\documentclass[a4paper,twoside,ngerman,english]{scrbook}
\newcommand{\texlive}{2009}
\usepackage{pibonn-thesis}
\usepackage{thesisDefinitions}
\ifthenelse {\texlive = 2009} {%
  \usepackage[backend=bibtex8,bibencoding=latin1,style=numeric-comp,sorting=none,block=ragged,firstinits=true]{biblatex}
}{%
  \usepackage[backend=biber,style=numeric-comp,block=ragged,firstinits=true]{biblatex}
}
\ifthenelse {\texlive = 2009} {%
  \usepackage{./biblatex/biblatex-num-v2009}
  \bibliography{./lcd2,bib/books,bib/lcd-notes,bib/proceedings,bib/other,bib/articles,bib/collections}
}{%
  \usepackage{./biblatex/biblatex-num-v2011}
  \addbibresource{./mythesis/thesis_refs.bib}
  \addbibresource{./refs/standard_refs-biber.bib}
}
\newcommand{\thesistitle}{Detector Optimization Studies and Light Higgs Decay into Muons at CLIC}
\newcommand{\thesisauthor}{Christian Grefe}
\newcommand{\thesistown}{L\"uneburg}

\newcommand{\thesisyear}{2013} 

\newcommand*{\thesisrefereeonetext}{1.\ Gutachter}
\newcommand*{\thesisrefereeone}{Prof.\ Dr.\ Klaus Desch}
\newcommand*{\thesisrefereetwotext}{2.\ Gutachterin}
\newcommand*{\thesisrefereetwo}{Prof.\ Dr.\ Ian Brock}
\newcommand{\thesissubmit}{14.05.2012}
\newcommand{\thesispromotion}{05.10.2012}
\graphicspath{%
  {./figures/},%
  {./figures/cover/},%
  {./figures/calorimetry/},%
  {./figures/geometry/},%
  {./figures/higgs/},%
  {./figures/mipcalibration/},%
  {./figures/software/},%
  {./figures/tracking/}%
}
\begin{document}
%
%
{\addtolength{\oddsidemargin}{1.0cm}\addtolength{\topmargin}{1.0cm}
%
%
 %
%


\title{\thesistitle}
\subtitle{\vspace*{4ex}
  \begin{otherlanguage}{ngerman}
    Dissertation\\
    zur\\
    Erlangung des Doktorgrades (Dr.\ rer.\ nat.)\\
    der\\
    Mathematisch-Naturwissenschaftlichen Fakultät\\
    der\\
    Rheinischen Friedrich-Wilhelms-Universität Bonn
  \end{otherlanguage}
}
\author{%
  von\\
  \LARGE \thesisauthor\\
  aus\\
  \thesistown
}
\date{}
\publishers{%
  Bonn, \thesissubmit
}
\lowertitleback{\normalsize
  \raggedright
  \begin{otherlanguage}{ngerman}
    Dieser Forschungsbericht wurde als Dissertation von der
    Mathematisch-Naturwissenschaftlichen Fakultät der Universität
    Bonn angenommen und ist auf dem
    Hochschulschriftenserver der ULB Bonn
    \url{http://hss.ulb.uni-bonn.de/diss_online} elektronisch
    publiziert.
  \end{otherlanguage}

  \vspace*{6ex minus 4ex}

  \begin{otherlanguage}{ngerman}
    Diese Dissertationsschrift wurde im Rahmen des Wolfgang-Gentner-Programms am europ\"aischen Zentrum f\"ur Teilchenphysik CERN angefertigt. Betreuerin: Dr.\ Lucie Linssen.
  \end{otherlanguage}

  \vspace*{10ex minus 4ex}

  \noindent
  \begin{otherlanguage}{ngerman}
    \begin{tabular}{ll}
      \thesisrefereeonetext: & \thesisrefereeone\\
      \thesisrefereetwotext: & \thesisrefereetwo\\[2ex]
      Tag der Promotion: & \thesispromotion\\
      Erscheinungsjahr: & \thesisyear
    \end{tabular}
  \end{otherlanguage}
}

\maketitle

%
%
%
%
%
%
}

\frontmatter
\pagestyle{scrplain}

\begin{otherlanguage}{english}

\begin{center}
 \emph{\textbf{\Huge{Abstract}}}
\end{center}

\noindent
The Compact Linear Collider (CLIC) is a concept for a future \epem linear collider with a center-of-mass energy of up to \unit[3]{TeV}. The design of a CLIC experiment is driven by the requirements related to the physics goals, as well as by the experimental conditions. For example, the short time between two bunch crossings of \unit[0.5]{ns} and the backgrounds due to beamstrahlung have direct impact on the design of a CLIC experiment. The Silicon Detector (SiD) is one of the concepts currently being discussed as a possible detector for the International Linear Collider (ILC). In this thesis we develop a modified version of the SiD simulation model for CLIC, taking into account the specific experimental conditions. In addition, we developed a software tool to investigate the impact of beam-related backgrounds on the detector by overlaying events from different simulated event samples. Moreover, we present full simulation studies, determining the performance of the calorimeter and tracking systems. We show that the track reconstruction in the all-silicon tracker of SiD is robust in the presence of the backgrounds at CLIC. Furthermore, we investigate tungsten as a dense absorber material for the hadronic calorimeter, which allows for the construction of a compact hadronic calorimeter that fulfills the requirements on the energy resolution and shower containment without a significant increase of the coil radius. Finally, the measurement of the decays of light Higgs bosons into two muons is studied in full simulation. We find that with an integrated luminosity of \unit[2]{\abinv}, corresponding to 4 years of data taking at CLIC, the respective Higgs branching ratio can be determined with a statistical uncertainty of approximately 15\%.
\end{otherlanguage}

\vfill

\begin{otherlanguage}{ngerman}

\begin{center}
 \emph{\textbf{\Huge{Zusammenfassung}}}
\end{center}

\noindent
Der Compact Linear Collider (CLIC) ist ein Konzept für einen zukünftigen \epem-Linearbeschleuniger mit einer Schwerpunktsenergie von bis zu \unit[3]{TeV}. Das Design eines CLIC-Experiments wird bestimmt durch die Anforderungen die sowohl aus den physikalischen Zielsetzungen als auch aus der experimentellen Umgebung herr\"uhren. Zum Beispiel haben die kurze Zeit zwischen zwei Strahlkreuzungen von \unit[0.5]{ns} und die Untergr\"unde durch die Beamstrahlung direkte Auswirkungen auf den Aufbau eines CLIC-Experiments. Der Silicon Detector (SiD) ist eines der Detektorkonzepte, die momentan als mögliche Experimente für den International Linear Collider (ILC) diskutiert werden. In dieser Arbeit entwickeln wir eine modifizierte Version des SiD-Simulationsmodels, wobei die CLIC-spezifischen Anforderungen berücksichtigt werden. Au{\ss}erdem haben wir ein Computerprogramm entwickelt, um die Effekte der strahlinduzierten Untergründe auf den Detektor zu untersuchen, indem simulierte Ereignisse aus verschiedenen Simulationen überlagert werden. Zudem stellen wir Simulationsstudien vor, in denen die Leistungsfähigkeit der Kalorimeter und der Spurdetektoren bestimmt werden. Wir zeigen, dass die Spurrekonstruktion in den in SiD vorgesehenen Siliziumspurdetektoren nicht durch die bei CLIC vorhandenen Untergr\"unde beeinflusst wird. Des Weiteren untersuchen wir Wolfram als m\"ogliches dichtes Absorbtionsmaterial f\"ur das hadronische Kalorimeter, was die Konstruktion eines kompakten hadronischen Kalorimeters erlaubt das sowohl die Anforderungen an die Energieaufl\"osung als auch an die m\"oglichst vollst\"andige Absorbtion der Schauer erf\"ullt, ohne einen signifikant gr\"osseren Spulenradius zu ben\"otigen. Schlie{\ss}lich wird eine Simulationsstudie vorgestellt, in der Higgsbosonzerf\"alle in zwei Myonen gemessen werden. Mit einer gesammelten Luminosität von \unit[2]{\abinv}, was in etwa 4 Jahren Datennahme bei CLIC entspricht, kann das entsprechende Verzeigungsverhältnis mit einer statistischen Unsicherheit von ca.\xspace 15\% bestimmt werden.
\end{otherlanguage}

%
%
\chapter*{Acknowledgements}
\label{cha:Acknowledgements}
I would like to thank Prof. Dr. Klaus Desch for being my supervisor and thus giving me the possibility to write this thesis. I would also like to thank Dr. Lucie Linssen for the supervision at CERN during the past years, many useful discussions and especially for being very supportive at all times.

I would like to thank the CERN LCD group as a whole for making the last few years extremely enjoyable. I am especially thankful to Peter Speckmayer and Phillip Rohloff with whom I had the pleasure to share an office and who were always available to discuss problems. I would like to thank Dominik Dannheim for many discussions on the vertex detector and the beam-related backgrounds. In addition, I would like to thank Martin Killenberg for the discussions concerning the tracking. I am also indebted to Stephane Poss who provided endless support in fixing problems with the Grid and providing the \whizard files. I am thankful to Andr\'e Sailer for giving useful tips concerning \geant and providing many insights into the beam-related backgrounds. I would also like to thank Jan Strube for many discussions especially on the Higgs branching ratio analysis. I also would like to thank our secretary Kate Ross for helping with travel arrangements and other administrative tasks.

At SLAC I would like to thank Norman Graf and Jeremy McCormick, who helped me in many ways with the simulation and reconstruction software, as well as Richard Partridge who helped me understand the track reconstruction algorithm.

I am very grateful to those that helped improving this thesis by reading parts of it and providing many useful comments: Lucie Linssen, Dominik Dannheim, Dieter Schlatter, Phillip Bechtle and my brother Michael.

Finally, I would like to express my deep gratitude to my parents and my brother who are always there to support me.

I acknowledge the support by the Bundesministerium f\"ur Forschung und Bildung through a Wolfgang-Gentner-Stipendium.

\tableofcontents

\mainmatter
\pagestyle{scrheadings}

%
%
\tikzset{
    photon/.style={decorate, decoration={snake}, draw=black},
    vectorboson/.style={decorate, decoration={snake}, draw=black},
    higgs/.style={densely dashed, draw=black},
    basefermion/.style={draw=black},
    fermion/.style={draw=black, postaction={decorate},
        decoration={markings,mark=at position .55 with {\arrow[draw=black]{>}}}},
    antifermion/.style={draw=black, postaction={decorate},
        decoration={markings,mark=at position .55 with {\arrow[draw=black]{<}}}},
    gluon/.style={decorate, draw=black,
        decoration={coil,amplitude=4pt, segment length=5pt}} 
}

\chapter{Introduction}
\label{cha:Introduction}

The \ac{SM} of particle physics was formulated almost half a century ago~\cite{Glashow:1961tr,Weinberg:1967tq,Salam:1968rm,Englert:1964et,Higgs:1964pj,Guralnik:1964eu,Gross:1973id,Politzer:1973fx,Fritzsch:1973pi,Fritzsch:1972jv}. Since then it has emerged as one of the most successful theories in modern physics. Not only did it successfully predict the existence of the heavier quarks and the massive gauge bosons of the electroweak interaction, it also succeeded in explaining all of the precision measurements of the interactions of fundamental particles performed in many collider experiments. Nevertheless, the \ac{SM} is still incomplete, since it predicts the existence of a massive scalar boson---the Higgs boson---that is required to explain electroweak symmetry breaking and thus the masses of the \PW and \PZ bosons. In addition, the existence of this boson could also explain the observation of massive fermions, which would otherwise contradict the theory.

The search for the Higgs boson is one of the main tasks of the \acs{ATLAS} and \acs{CMS} experiments operated at the \ac{LHC} at \acs{CERN}. With the \ac{LHC} performing above expectations there is good hope to be able to proof or falsify the existence of the Higgs boson by the end of this year, at least if its interactions are as predicted by the \ac{SM}. Nevertheless, this will not be the end of the journey. On one hand, if the Higgs is not found, this will require an alternative explanation of electroweak symmetry breaking. On the other hand, even if the Higgs boson is found, its true nature will still remain unclear for quite some time. All the predicted parameters of the Higgs boson including its couplings to the other particles in the \ac{SM} will have to be tested.

Besides the elusive Higgs boson there are many other open questions in \ac{HEP}. In the past decades, several theoretical limitations of the \ac{SM} have motivated a wide range of models that extend the current theory and might lead to a deeper understanding of nature (see~\cite{Lykken:2010mc} for an overview). One direct motivation for extending the \ac{SM} comes from cosmological observations. While the now established Big Bang theory (see \cite{Trodden:2004st} for an introduction) is another example of the success of the \ac{SM}, since it consistently explains the cosmic microwave background~\cite{Penzias:1965wn,Komatsu:2008hk} and the primordial abundances of the different chemical elements in the universe~\cite{Alpher:1948ve,Steigman:2005uz}, it also leads to questions that cannot be answered directly. The apparent asymmetry in the amount of matter and anti-matter in the universe is one of the observations that cannot be explained by the \ac{SM}. Another problem is the observation that the largest fraction of the matter in the universe exists in the form of dark matter~\cite{Bertone:2004pz}. This would require the existence of particles with properties that do not correspond to any particle in the \ac{SM}. Some of the possible extensions of the \ac{SM} introduce new particles that could well form an explanation for the dark matter in the universe. Depending on the concrete model these new particles could be in the energy range accessible to the \ac{LHC} and the next generation of particle colliders.

There is broad agreement in the \ac{HEP} community that the next collider at the energy frontier should be an \epem linear collider with a center-of-mass energy between several hundreds of GeV and a few TeV. A lepton collider will complement the measurements at the \ac{LHC} by allowing precise measurements of the properties of those particles that may be discovered in the coming years. In addition, a lepton collider would extend the discovery potential of the \ac{LHC} for several hypothetical models, especially for electroweak states. One concept of a future linear collider that is currently being discussed is the \ac{ILC}~\cite{Brau:2007zza} with a center-of-mass energy of up to \unit[1]{TeV}. Another possible future linear collider is the \ac{CLIC}. The present thesis focuses on this linear collider, with a center-of-mass energy of up to \unit[3]{TeV}. The \ac{CLIC} accelerator concept foresees normal conducting cavities with field gradients of approximately \unitfrac[100]{MV}{m} to limit its total length to less than \unit[50]{km}. The \ac{RF} power for accelerating the particles in the main beam is generated by a second particle beam---the drive beam---whose particles are decelerated in this process. This allows for almost lossless distribution of the power along the accelerator and is more efficient than the generation of \ac{RF} power using klystrons. These and other aspects of the \ac{CLIC} accelerator scheme are studied at the \ac{CTF3}~\cite{Corsini:2001ku} and its predecessors. The feasibility of most technological challenges have been demonstrated and the current status will be reported in the accelerator volume of the \ac{CLIC} \ac{CDR}, which will be published in the coming months~\cite{cdrvol1}.

Two detector concepts, the \ac{ILD}~\cite{ILD:2009} and the \ac{SiD}~\cite{Aihara:2009ad} concepts, have been developed as possible experiments at the \ac{ILC}. They are based on the particle flow paradigm. Since the requirements on the basic observables are very similar for an experiment at \ac{CLIC}, these two detector concepts have been chosen as a starting point for developing a possible detector for \ac{CLIC}. The main differences between the \ac{ILC} and \ac{CLIC} from a detector point of view are the available center-of-mass energy, leading to more energetic particles and jets at \ac{CLIC}, the amount of beam-related backgrounds, mostly due to beamstrahlung, as well as the much shorter time between two bunch crossings at \ac{CLIC}. In combination, the latter two differences result in higher occupancies in the subdetectors at \ac{CLIC}, which have to be taken into account when designing the experiment. The physics and detector volume of the \ac{CLIC} \ac{CDR}, discussing the physics potential of \ac{CLIC} and the performance of the two modified detector concepts \clicild and \clicsid, has been published recently~\cite{cdrvol2}. Some of the simulation studies of the \clicsid concept that contributed to the \ac{CDR} are presented in detail in this thesis.


This thesis is structured as follows. In \cref{cha:SM} we give a brief overview of the \ac{SM} and the Higgs mechanism, the limitations of the \ac{SM}, as well as examples of possible extensions and their implications for the Higgs sector. Afterwards we give an introduction to the \ac{CLIC} accelerator concept in \cref{cha:CLIC} where we also discuss the beam-related backgrounds and their implications for an experiment. In \cref{cha:DetectorDesign} we recapitulate the basic principles of momentum and energy measurements in \ac{HEP} experiments and discuss the design goals for an experiment at \ac{CLIC}. \Cref{cha:SiD} gives an overview of the \ac{SiD} concept as proposed for \ac{CLIC}, including a description of the simulation model \clicsid, which is used for the studies presented in this thesis as well as those presented in the \ac{CLIC} \ac{CDR}. The simulation and reconstruction of collisions at \ac{CLIC} are discussed in \cref{cha:Software}, including a detailed introduction to the concept of overlaying events to realistically simulate the effect of beam-related backgrounds, as well as the track reconstruction algorithm used later in the thesis. The performance of the all-silicon tracking detectors in \clicsid is investigated in \cref{cha:Tracking}, taking into account the most important beam-related background. The basic performance is studied in simulated single muon events and the efficiency and fake rate are studied in simulated di-jet events. In \cref{cha:Calorimetry} we show a systematic optimization study for the \ac{HCal} layout, comparing the performances of sampling calorimeters using steel and tungsten absorber plates. A performance study of the energy reconstruction in the calorimeter systems in \clicsid is presented in \cref{cha:Calorimetry_PFA}. In addition to the studies of the fundamental observables, a set of detector benchmark analyses has been defined for CLIC at $\sqrt{s} = \unit[3]{TeV}$~\cite{lcd:2011-016}, where each of the processes is especially demanding in one or more of the basic observables. This allows to study the simulated detector performance in the context of a realistic analysis. One of these benchmark analyses, the simulation of the measurement of the Higgs branching ratio into two muons for a light \ac{SM} Higgs, is presented in \cref{cha:Higgs}. This measurement requires excellent momentum resolution, good muon identification and efficient electron tagging also in the forward region to reject \ac{SM} backgrounds. 
Finally, the results of this thesis are summarized in \cref{cha:Conclusion}.

Some supplementary information is given in the appendices. \cref{app:Convention} defines the units and conventions used throughout this thesis. The \acs{RMS}$_{90}$, an alternative measure of the spread of a distribution, is introduced in \cref{App:RMS90}. The track parametrization used in the tracking studies is described in \cref{App:trackParametrization} together with a derivation of the uncertainties of several physical quantities from the uncertainties of the track fit. \cref{cha:Appendix_ttTracking} shows the tracking performance in simulated \ttbar events, which is a different event topology compared to the di-jet events studied in \cref{cha:Tracking}.

We have implemented the \clicsid detector model as a modification of an existing simulation model of \ac{SiD}, providing all necessary calibration files. The calculation of the total material budget in the tracking detectors was done by Astrid M\"unnich. A detailed description of this simulation model has been published in~\cite{lcd:grefemuennich2011}. 

The software tool to overlay background events, which is essential for a realistic simulation of collisions at \ac{CLIC}, has been developed in collaboration with Peter Schade who implemented the corresponding software tool in the \ac{ILD} reconstruction framework. Its functionality has been documented in~\cite{lcd:grefesidoverlay2011}.

The results from the study of the tracking performance have been partially published in~\cite{cdrvol2}. The optimization study of the \ac{HCal} layout has been performed in collaboration with Peter Speckmayer and the results have already been published in~\cite{SpeckmayerGrefe2010}.

The simulation study of the Higgs branching ratio into two muons has already been published in~\cite{lcd:grefeHmumu2011}. In addition, some of the results of the Higgs study have been published in~\cite{cdrvol2,Grefe:2012eh}. 

\chapter{Standard Model of Particle Physics}
\label{cha:SM}

The interactions of fundamental particles can be described in the framework of relativistic quantum field theory by the \acl{SM} of particle physics, which combines the observed strong and electroweak interactions. The \ac{SM} is a gauge invariant quantum field theory based on the $SU(3)_C \times SU(2)_L \times U(1)_Y$ gauge symmetry group, where $SU(3)_C$ is the color group describing the strong interaction and $SU(2)_L \times U(1)_Y$ describes the electroweak interaction which is spontaneously broken by the Higgs mechanism.

Many detailed discussions of quantum field theory and the \ac{SM} can be found in the literature, \eg~\cite{Griffiths:111880,Peskin:257493}. We want to briefly recapitulate some of these basic principles to motivate the interest in the Higgs sector of the \ac{SM}. \Cref{sec:QFT} illustrates the concept of gauge theories in the example of \ac{QED}. The particles and interactions in the \ac{SM} are introduced in \cref{sec:SM_SM} together with the Higgs mechanism, which is required to explain the observation of massive vector bosons. \cref{sec:BSM} highlights some of the shortcomings of the \ac{SM} and discusses possible extensions of it.

The notations and conventions used here are defined in \cref{app:Convention}. We follow the ideas presented in \cite{Hollik:2010id,Peskin:257493}.

\section{Quantum Field Theory}
\label{sec:QFT}
In \ac{QFT} particles are described by fields, which act as the creation and annihilation operators of the respective particles. The field theory approach intrinsically allows the description of multi-particle interactions and also guarantees causality, which is not the case in a particle quantization. The dynamics of a system that consists of the set of fields $\Phi$ is usually described in the Lagrangian formalism, which is invariant under Lorentz transformations. The classical Lagrangian density $L$, usually just referred to as the Lagrangian, depends on the discrete coordinates $q_i$ and the velocities $\dot{q}_i$, which have to be replaced by the continuous fields $\Phi(x)$ and their derivatives $\partial_\mu \Phi(x)$, respectively. The action is then given by the space-time integral of the field theoretical Lagrangian $\lagrange$:
\begin{equation}
 S(\Phi) = \int \lagrange\left(\Phi(x), \partial_\mu \Phi(x)\right) \text{d}^4x.
\end{equation}
The equations of motion follow from Hamilton's principle, $\delta S = 0$, and are given by the Euler-Lagrange equations
\begin{equation}
 \partial_\mu \frac{\partial\lagrange}{\partial(\partial_\mu \Phi)} - \frac{\partial \lagrange}{\partial \Phi} = 0.
\end{equation}

\abs
The \ac{SM} contains three types of fields, spin-0 particles described by scalar fields $\phi(x)$, spin-1 particles described by vector fields $A_\mu(x)$ and spin-$\frac{1}{2}$ particles described by 4-component spinor fields $\psi(x)$.

The dynamics of a free scalar field $\phi(x)$ with the mass $m$ is described by
\begin{equation}
 \lagrange = \frac{1}{2}(\partial_\mu \phi)^2 - \frac{m^2}{2}\phi^2.
\end{equation}
The corresponding equation of motion is the Klein-Gordon equation
\begin{equation}
 (\partial^\mu \partial_\mu + m)^2\phi = 0.
\end{equation}
Its solutions are linear combinations of plane waves of the form $e^{\pm ikx}$.

\abs
The Lagrangian of a free vector field $A_\mu(x)$ is given by
\begin{equation}
 \lagrange = - \frac{1}{4}F_{\mu\nu}F^{\mu\nu} - \frac{m}{2} A_\mu A^\mu,
 \label{eq:vector_lagrangian}
\end{equation}
where $F_{\mu\nu}$ is the field strength tensor $F_{\mu\nu} = \partial_\mu A_\nu - \partial_\nu A_\mu$. The resulting field equation is the Proca equation
\begin{equation}
 \left[\left(\partial^\mu \partial_\mu + m^2\right)g^{\mu\nu} - \partial^\mu \partial^\nu \right] A_\nu = 0,
\end{equation}
which can be solved by linear combinations of planar waves, spanned by three linearly independent polarization vectors. 

\abs
Fermions with a spin of $\frac{1}{2}$ are described by a 4-component spinor field $\psi(x)$. The dynamics of a free spinor field is described by the Dirac Lagrangian
\begin{equation}
 \lagrange = \overline{\psi} \left(i \gamma^\mu \partial_\mu - m\right) \psi,
\label{eq:dirac_lagrangian}
\end{equation}
where $m$ is the fermion mass and $\overline{\psi}$ is the adjoint spinor with $\overline{\psi} = \psi^\dag \gamma^0$. The corresponding equation of motion is the Dirac equation
\begin{equation}
 \left(i \gamma^\mu \partial_\mu - m\right) \psi = 0,
\label{eq:dirac_equation}
\end{equation}
which has two solutions of the form $u(p)e^{-ipx}$ and $v(p)e^{ipx}$, corresponding to the particle and anti-particle wave functions.

\abs
The interaction of fermions and vector fields can be derived by requiring that the theory is invariant under a certain group of gauge transformations. In \ac{QED} we assume that the Dirac Lagrangian is invariant under global phase transformations, $\psi \to \psi' = e^{i\alpha} \psi$, where $\alpha$ is real. We can even assume the invariance under local phase transformations $\alpha(x)$, $\psi \to \psi' = e^{i\alpha(x)} \psi$, which define the abelian group of unitary transformations $U(1)$. This requires the presence of a vector field $A_\mu$ which has to transform as $A_\mu(x) \to A^\prime_\mu(x) = A_\mu(x) + \frac{1}{e}\partial_\mu \alpha(x)$ in order to preserve invariance of the Lagrangian under this transformation, since the derivative in \cref{eq:dirac_lagrangian} has to be replaced by the covariant derivative $D_\mu = \partial_\mu - i e A_\mu$. The Lagrangian then reads
\begin{equation}
 \lagrange = \overline{\psi} \left(i \gamma^\mu \partial_\mu - m\right) \psi + e \overline{\psi} \gamma^\mu \psi A_\mu.
\end{equation}
If we identify $A_\mu$ as the photon field and $j^\mu = e \overline{\psi} \gamma^\mu \psi$ as the electromagnetic current with the coupling $e$, the electron charge, this Lagrangian describes the interaction of electron and positron fields with photon fields. The kinetic term of the photon field $- \frac{1}{4}F_{\mu\nu}F^{\mu\nu}$ has to be added according to \cref{eq:vector_lagrangian} in order to obtain the full Lagrangian of \ac{QED}. The mass term in \cref{eq:vector_lagrangian} can be omitted since the photon is massless.

The concept of invariance under certain gauge transformations can be applied similarly to non-abelian symmetries by replacing the derivative in \cref{eq:dirac_lagrangian} by the covariant derivative and introducing several vector fields. The number of vector fields introduced is given by the number of generators of the symmetry group. Similarly to the case of \ac{QED} the kinetic terms of the vector fields have to be added. The vector fields in such a theory are necessarily massless since the mass terms are not invariant under non-abelian gauge transformations. The non-abelian nature also introduces self-couplings of the vector fields through the gauge transformations, which are not present in \ac{QED}. These vector fields are thus carriers of their respective charge.

\section{The Standard Model}
\label{sec:SM_SM}
As mentioned above, the \ac{SM} is described by the symmetry group $SU(3)_C \times SU(2)_L \times U(1)_Y$. \ac{QCD}~\cite{Gross:1973id,Politzer:1973fx,Fritzsch:1973pi,Fritzsch:1972jv} describes the strong interaction. It is a gauge theory invariant under local gauge transformations in the three dimensional color space, described by the color group $SU(3)_C$. The $SU(3)$ group has 8 generators which can be described by the Gell-Mann matrices, resulting in the introduction of 8 vector fields, the gluons. In the \ac{SM}, only the quarks are affected by the strong interaction. They are described by a color triplet of fermion fields that transform under color transformations. The other fermions are color singlets and are not affected by the strong interaction.

The electroweak interactions~\cite{Glashow:1961tr,Weinberg:1967tq,Salam:1968rm} are described by the symmetry group $SU(2)_L \times U(1)_Y$. The generators of this group are the three isospin operators $I_{1,2,3}$ and the hypercharge $Y$. These result in 4 vector fields, the triplet $W^{1,2,3}_\mu$ and the singlet $B_\mu$. The electric charge $Q$ is connected to the isospin and the hypercharge by the relation
\begin{equation}
 Q = I_3 + \frac{Y}{2}.
\end{equation}

Transformation of fermions under $SU(2)$ in the \ac{SM} depend on the chirality of the fermion field. One distinguishes between left-handed fermions $\psi\LH = \frac{1 - \gamma_5}{2}\psi$ and right-handed fermions $\psi\RH = \frac{1 + \gamma_5}{2}\psi$. Only the left-handed fermions (or right-handed anti-fermions), which carry an isospin of $I = \frac{1}{2}$, are affected by $SU(2)$ transformations and form isospin doublets. The right-handed fermions are isospin singlets and have an isospin of $I = 0$. In the \ac{SM} this is realized by having two different covariant derivatives for the two chiral projections, \ie the derivative for the other projection is 0. It should be noted that only for the case of massless fermions the left-handed and right-handed projections of \cref{eq:dirac_lagrangian} are decoupled and can be written independently. This means that massive fermions would violate gauge invariance under these $SU(2)$ transformations, which can be solved by the Higgs mechanism introduced below.

An overview of the fermions in the \ac{SM} is given \cref{tab:SM_particles} together with their quantum numbers $I_3$, $Y$ and $Q$. In general there could be more generations of particles, although they must be significantly heavier than the first three generations since they have not been observed yet. The strongest constraint on the existence of a fourth generation neutrino comes from the measurement of the \PZ boson decay width at the \ac{LEP}, which is only compatible with three neutrino generations~\cite{ALEPH:2005ab}. That means that a fourth generation neutrino, if it exists, has to have at least half the \PZ boson mass, such that it could not be produced at \ac{LEP}. This would be in strong contrast to the other three neutrinos which are assumed to be massless in the basic \ac{SM}, or are at least of very small mass as discussed in \cref{sec:SM_extensions}.

\begin{table}
 \caption[Fermions in the \acl{SM}.]{Fermions in the \acl{SM}. Given is the isospin $I_3$, the hypercharge $Y$ and the electrical charge $Q$. The anti-particles are the charge conjugates of the fermions and are not listed in the table. The quark fields are color triplets, while the lepton fields are color singlets.}
 \label{tab:SM_particles}
 \centering
 \begin{tabular}{ccccccc}
  \toprule
                          & \multicolumn{3}{c}{Generation}   &        &      &     \\
                          & I        & II        & III       & $I_3$  & $Y$  & $Q$ \\
  \midrule
 \multirow{3}{*}{Leptons} & \multirow{2}{*}{$\begin{pmatrix}\PGne \\ \Pe\end{pmatrix}\LH$}  &  \multirow{2}{*}{$\begin{pmatrix}\PGnGm \\ \PGm\end{pmatrix}\LH$}  &  \multirow{2}{*}{$\begin{pmatrix}\PGnGt \\ \PGt\end{pmatrix}\LH$} & $+1/2$ & $-1$ & $0$ \\
                          &          &           &           & $-1/2$ & $-1$ & $-1$ \\
                          & $\Pe\RH$ & $\PGm\RH$ & $\PGt\RH$ & $0$    & $-2$ & $-1$ \\
  \midrule
 \multirow{4}{*}{Quarks}  & \multirow{2}{*}{$\begin{pmatrix}\PQu \\ \PQd\end{pmatrix}\LH$}  &  \multirow{2}{*}{$\begin{pmatrix}\PQc \\ \PQs\end{pmatrix}\LH$}  &  \multirow{2}{*}{$\begin{pmatrix}\PQt \\ \PQb\end{pmatrix}\LH$} & $+1/2$ & $+1/3$ & $+2/3$ \\
                          &          &           &           & $-1/2$ & $+1/3$ & $-1/3$ \\
                          & $\PQu\RH$ & $\PQc\RH$ & $\PQt\RH$ & $0$   & $+4/3$ & $+2/3$ \\
                          & $\PQd\RH$ & $\PQs\RH$ & $\PQb\RH$ & $0$   & $-2/3$ & $-1/3$ \\
  \bottomrule
 \end{tabular}
\end{table}

\subsection{The Higgs Mechanism}
\label{sec:SM_Higgs}
As mentioned in \cref{sec:QFT}, vector fields in non-abelian gauge theories are necessarily massless. This contradicts the observations of the massive gauge bosons $\PWpm$ and $\PZ$ of the weak interaction. Massive vector fields can be introduced in the theory by spontaneously breaking the $SU(2)_L \times U(1)_Y$ symmetry with the Higgs mechanism~\cite{Englert:1964et,Higgs:1964pj,Guralnik:1964eu}, while leaving the electromagnetic subgroup $U(1)_\text{em}$ unbroken.

We introduce a new isospin doublet of two complex scalar fields with an isospin of $\frac{1}{2}$ and a hypercharge of $Y = 1$
\begin{equation}
 \phi(x) = \begin{pmatrix}
            \phi^+ \\
            \phi^0
           \end{pmatrix},
\end{equation}
where $+$ and $0$ denote the electrical charge.
This adds another term to the Lagrangian:
\begin{equation}
 \lagrange_H = \left(D_\mu \phi\right)^\dag \left(D^\mu \phi\right) - V(\phi),
\label{eq:Higgs_lagrangian}
\end{equation}
where the interaction with the gauge fields enters through the covariant derivative $D_\mu = \partial_\mu - ig_2 \frac{\sigma_a}{2} W^a_\mu + i \frac{g_1}{2} B_\mu$, with the two couplings of the isospin and the hypercharge $g_{1,2}$ and the Pauli matrices $\sigma_a$. The self-interaction is described by the potential $V(\phi)$. The Higgs potential is postulated as
\begin{equation}
 V(\phi) = - \mu^2 \phi^\dag\phi + \frac{\lambda}{4} \left(\phi^\dag\phi\right)^2,
\label{eq:Higgs_potential}
\end{equation}
where $\mu^2$ and $\lambda$ are constants. If $\mu^2$ and $\lambda$ are positive, the minimum of this potential is not found for the vacuum state $\phi = 0$. Instead, the minimum of $V(\phi)$ is reached when $\phi^\dag \phi = \frac{2\mu^2}{\lambda}$. The Higgs doublet $\phi(x)$ has four degrees of freedom since it is complex. We chose a gauge transformation where the charged component vanishes and only the neutral component is left, referred to as the unitary gauge. The resulting vacuum expectation value of the Higgs doublet is then
\begin{equation}
 <\phi_0> = \frac{1}{\sqrt{2}}\begin{pmatrix}
                               0 \\
                               v
                              \end{pmatrix},
\end{equation}
with $v = \frac{2\mu}{\sqrt{\lambda}}$. Expanding the field in the real component around the vacuum state yields
\begin{equation}
  \phi(x) = \frac{1}{\sqrt{2}}\begin{pmatrix}
                               0 \\
                               v + H(x)
                              \end{pmatrix},
\label{eq:Higgs_unitary}
\end{equation}
where $H(x)$ is a real scalar field. The imaginary component is unphysical since it can be eliminated by a suitable gauge transformation. Using this representation of the Higgs doublet, the Higgs potential in \cref{eq:Higgs_potential} is given by
\begin{equation}
 V = \mu^2 H^2 + \frac{\mu^2}{v} H^3 + \frac{\mu^2}{4v^2} H^4.
\end{equation}
The Higgs field $H(x)$ is thus a massive scalar field with a mass of $m_H = \mu \sqrt{2}$ and with triple and quartic self-couplings that are both proportional to $m_H^2$.

Using this representation of the Higgs doublet in the kinetic term of \cref{eq:Higgs_lagrangian} results in mass terms for the four vector bosons of the standard model, as well as triple and quartic couplings of the $H$ field with the vector fields. The mass terms are given by
\begin{equation}
 \frac{1}{2} \left( \frac{g_2 v}{2} \right)^2 \left( (W_1)^2 + (W_2)^2\right) + \frac{1}{2} \left( \frac{v}{2}\right)^2 \left(W^3_\mu, B_\mu \right) \begin{pmatrix}
                                                                                                                                                      g^2_2 & g_1 g_2 \\
                                                                                                                                                      g_1 g_2 & g^2_1
                                                                                                                                                     \end{pmatrix} \begin{pmatrix}
                                                                                                                                                                     W^{3\,\mu} \\
                                                                                                                                                                     B^\mu
                                                                                                                                                                   \end{pmatrix}.
\end{equation}
The vector fields can be transformed into a basis that corresponds to the physical observed fields using
\begin{equation}
 \PW^\pm_\mu = \frac{1}{\sqrt{2}}\left(W^1_\mu \mp W^2_\mu\right)
\end{equation}
and
\begin{equation}
 \begin{pmatrix}
  \PZ_\mu \\ A_\mu
 \end{pmatrix} = \begin{pmatrix} \cos \theta_w & \sin \theta_w \\  -\sin \theta_w & \cos \theta_w \end{pmatrix} \begin{pmatrix} W^\mu_3 \\ B_\mu \end{pmatrix},
\end{equation}
with the weak mixing angle $\cos\theta_W = \frac{g_2}{\sqrt{g^2_1 + g^2_2}}$. The resulting mass terms are
\begin{equation}
 \frac{1}{2} \left( \frac{g_2 v}{2} \right)^2 \PW^+_\mu \PW^{-\,\mu} + \frac{1}{2} \left( \frac{v}{2}\right)^2 \left( A_\mu, \PZ_\mu \right) \begin{pmatrix}
                                                                                                                                                      0 & 0 \\
                                                                                                                                                      0 & 2\left(g^2_1 + g^2_2\right)
                                                                                                                                                     \end{pmatrix} \begin{pmatrix}
                                                                                                                                                                     A^\mu \\
                                                                                                                                                                     \PZ^\mu
                                                                                                                                                                   \end{pmatrix},
\end{equation}
where the mass of the \PW boson is given by $m_{\PW} = \frac{v}{2}g_2$ and the \PZ boson mass is given by $m_{\PZ} = \frac{v}{2}\sqrt{g^2_1 + g^2_2}$. The masses of the \PW and \PZ bosons are directly connected through the weak mixing angle by $m_{\PW} = m_{\PZ} \cos\theta_w$. The photon field stays massless as desired. A common interpretation of the Higgs mechanism is that the three vanishing degrees of freedom of $\phi$ are absorbed by the three massive gauge bosons in the \ac{SM} to give them their masses.

\subsubsection{Fermion Masses}
The fermion masses in the \ac{SM} cannot be generated by a mass term of the form $m_f(\overline{\psi}^f\LH \psi^f\RH + \overline{\psi}^f\RH \psi^f\LH)$ as in \cref{eq:dirac_lagrangian} since this would mix left-handed and right-handed states and thus violate gauge invariance. Instead, the fermions can acquire a mass by a postulated Yukawa coupling to the Higgs field with the coupling constant $g_f$. The Yukawa term in the Lagrangian for the fermion $f$ is given by
\begin{equation}
 \lagrange^f_Y = - g_f\ \overline{\psi}\LH\ \phi\ \psi\RH + \text{h.c.}
\end{equation}
Using \cref{eq:Higgs_unitary} for $\phi$ results in the Lagrangian
\begin{equation}
 \lagrange^f_Y = - g_f \frac{v}{\sqrt{2}} \left(\overline{\psi}\LH\ \psi\LH + \overline{\psi}\RH\ \psi\RH \right) - g_f \frac{1}{\sqrt{2}}\left(\overline{\psi}\LH\ \psi\LH H + \overline{\psi}\RH\ \psi\RH H \right).
\label{eq:yukawa_lagrangian}
\end{equation}
This includes a mass term that is identical for left-handed and right-handed fermions with $m_f = g_f\frac{v}{\sqrt{2}}$ and predicts a coupling to the scalar Higgs field, proportional to the particle mass. The choice of the coupling constant $g_f$ is free, and thus, the fermion masses are free parameters in the \ac{SM}.

In general, the mass eigenstates in \cref{eq:yukawa_lagrangian} are not necessarily the flavor eigenstates of the left-handed fermions under the transformations of the $SU(2)_L$ and thus flavor mixing can occur. In the \ac{SM} this is not realized for the leptons and the lepton number is conserved. The left-handed quarks in the \ac{SM} are, on the other hand, not mass eigenstates, which, if formulated as diagonal mass states, leads to flavor mixing in the charge-changing weak interactions. The mixing of the left-handed quarks in the \ac{SM} is described by the unitary $3\times3$ Cabibbo-Kobayashi-Maskawa (CKM) matrix $V_\text{CKM}$~\cite{Cabibbo:1963yz,Kobayashi:1973fv}. The entries of $V_\text{CKM}$ can be complex, which, constrained by the unitarity, leads to three mixing angles and one $CP$ violating complex phase as additional free parameters of the \ac{SM}.

It should be noted that the choice of the Higgs potential in \cref{eq:Higgs_potential} determines the self-interactions of the Higgs field and has to be experimentally verified by measuring the Higgs self-coupling. The choice of the gauge in \cref{eq:Higgs_unitary} determines the mass terms of the vector bosons. Choosing the gauge which leaves the charged component of the Higgs field would, for example, result in a massive photon field and would break $U(1)_\text{em}$.

\subsection[Higgs Production in \epem Collisions and Higgs Decays]{Higgs Production in \boldmath$\epem$\unboldmath\xspace Collisions and Higgs Decays}
\label{sec:SM_higgsProduction}
As we have seen, while allowing the introduction of a mass term for all massive particles in the \ac{SM}, the Higgs mechanism also introduces couplings of all of these particles to the scalar Higgs boson. If this mechanism is in fact the explanation for electroweak symmetry breaking one thus expects that the Higgs boson can be produced and observed in collider experiments.

\begin{figure}
 \begin{subfigure}[]{0.49\textwidth}
  \centering
  \begin{tikzpicture}[thick, >=latex]
   \node (positron) at (-2.0,2.0) {\Pep};
   \node (electron) at (-2.0,-2.0) {\Pem};
   \coordinate (eeZ) at (-0.5, 0.0);
   \coordinate (ZZH) at (1.5, 0.0);
   \node (higgs) at (3.0,2.0) {\PH};
   \node (Z) at (3.0,-2.0) {\PZ};

   \draw[antifermion] (positron) to (eeZ);
   \draw[vectorboson] (eeZ) to node[above] {$\PZ^*$} (ZZH);
   \draw[vectorboson] (ZZH) to (Z);
   \draw[higgs] (ZZH) to (higgs);
   \fill[black] (ZZH) circle (1.5pt);
   \draw[fermion] (electron) to (eeZ);
  \end{tikzpicture}
  \caption{$\epem \to \zhsm$}
  \label{fig:Higgsstrahlung}
 \end{subfigure}
 \hfill
 \begin{subfigure}[]{0.49\textwidth}
  \centering
  \begin{tikzpicture}[thick, >=latex]
   \node (positron) at (-2.0,2.0) {\Pep};
   \node (antineutrino) at (1.0,2.0) {\PAGne};
   \node (electron) at (-2.0,-2.0) {\Pem};
   \node (neutrino) at (1.0,-2.0) {\PGne};
   \coordinate (e1vW) at (-0.5, 1.0);
   \coordinate (e2vW) at (-0.5, -1.0);
   \coordinate (higgsLeft) at (0.0,0.0);
   \node (higgsRight) at (1.5,0.0) {\PH};

   \draw[antifermion] (positron) to (e1vW);
   \draw[antifermion] (e1vW) to (antineutrino);
   \draw[vectorboson] (e1vW) to node[right] {\PWp} (higgsLeft);
   \draw[vectorboson] (e2vW) to node[right] {\PWm} (higgsLeft);
   \fill[black] (higgsLeft) circle (1.5pt);
   \draw[higgs] (higgsLeft) to (higgsRight);
   \draw[fermion] (electron) to (e2vW);
   \draw[fermion] (e2vW) to (neutrino);
  \end{tikzpicture}
  \caption{$\epem \to \PH\nuenuebar$}
  \label{fig:WWfusion}
 \end{subfigure}
\caption{Feynman diagrams of the dominant Higgs production processes at an \epem collider.}
\label{fig:Higgsstrahlung_WWfusion}
\end{figure}
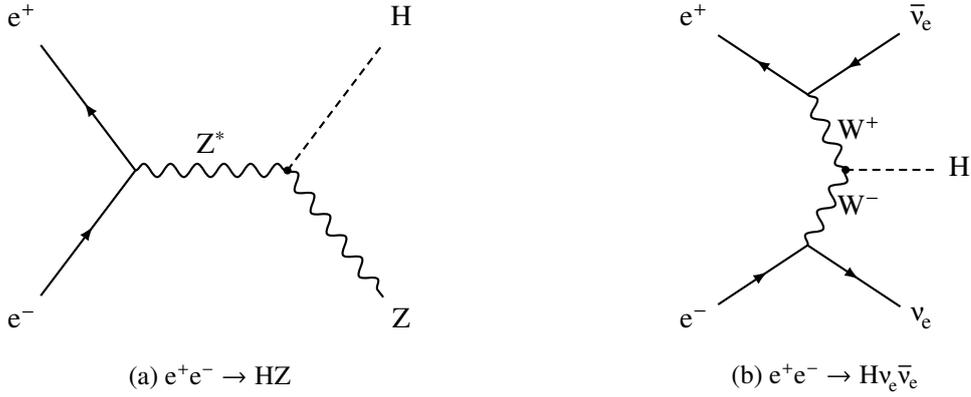

Due to the low mass of the electron, its coupling to the Higgs boson is also very small (see \cref{eq:yukawa_lagrangian}). The coupling is especially small compared to the coupling of the electron to $\PZ$ and $\PGg$. It would thus require a huge amount of statistics to identify a significant signal of the resonant Higgs production, $\epem \to \PH \to f\bar{f}$, at an \epem collider over the $\epem \to \PGg^*/\PZ^* \to f\bar{f}$ background. The most relevant Higgs production processes at an \epem collider are, depending on $\sqrt{s}$, the Higgsstrahlung process, $\epem \to \zhsm$, shown in \cref{fig:Higgsstrahlung} and the \PW{}\PW fusion process, $\epem \to \PH\nuenuebar$, shown in \cref{fig:WWfusion}. The cross section of the \zz fusion process, $\epem \to \PH\epem$, is suppressed approximately by one order of magnitude compared to the \PW{}\PW fusion process, due to the smaller coupling.

\begin{figure}
 \begin{subfigure}[]{0.49\textwidth}
  \includegraphics[width=\textwidth]{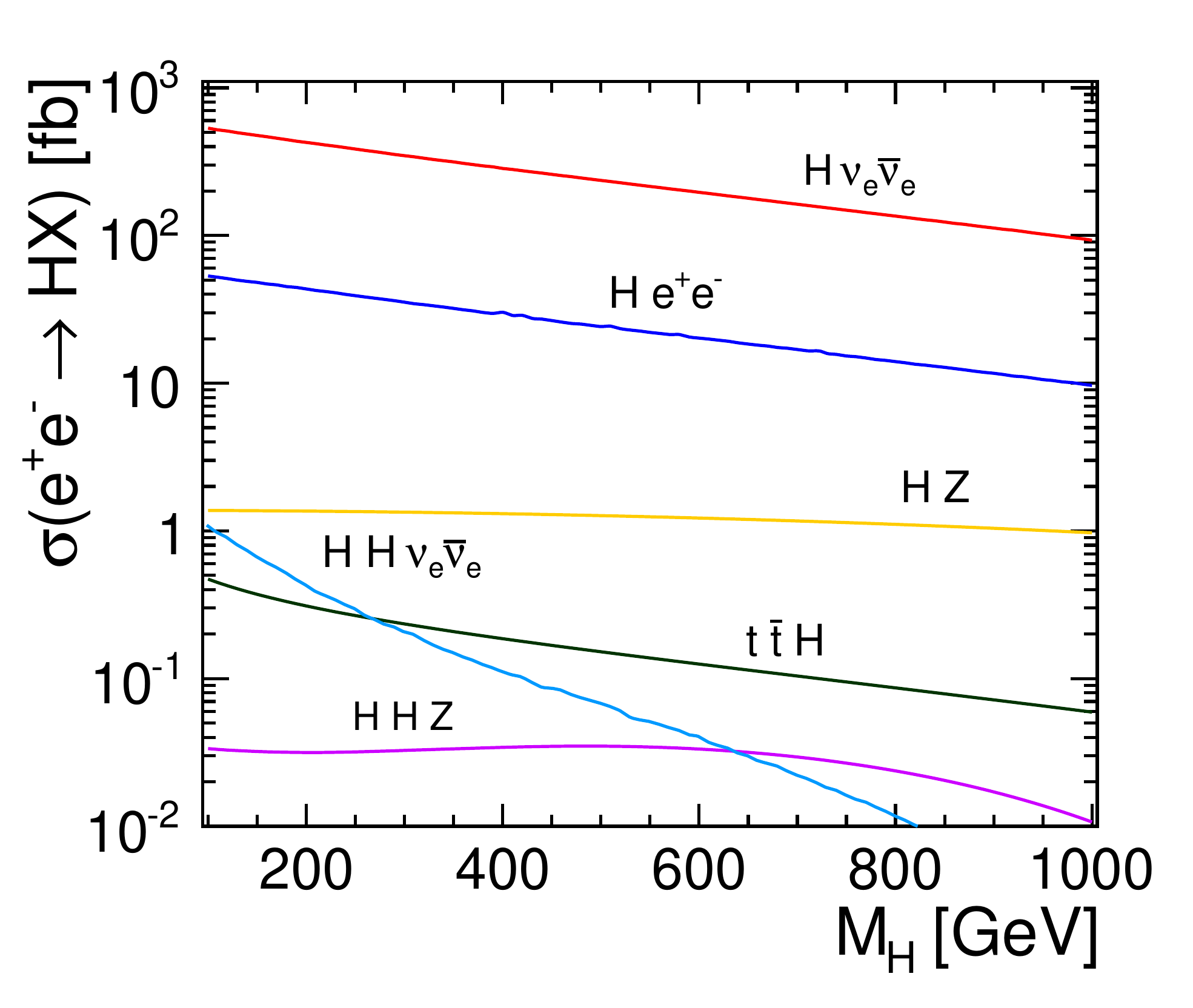}
 \end{subfigure}
 \hfill
 \begin{subfigure}[]{0.49\textwidth}
  \includegraphics[width=\textwidth]{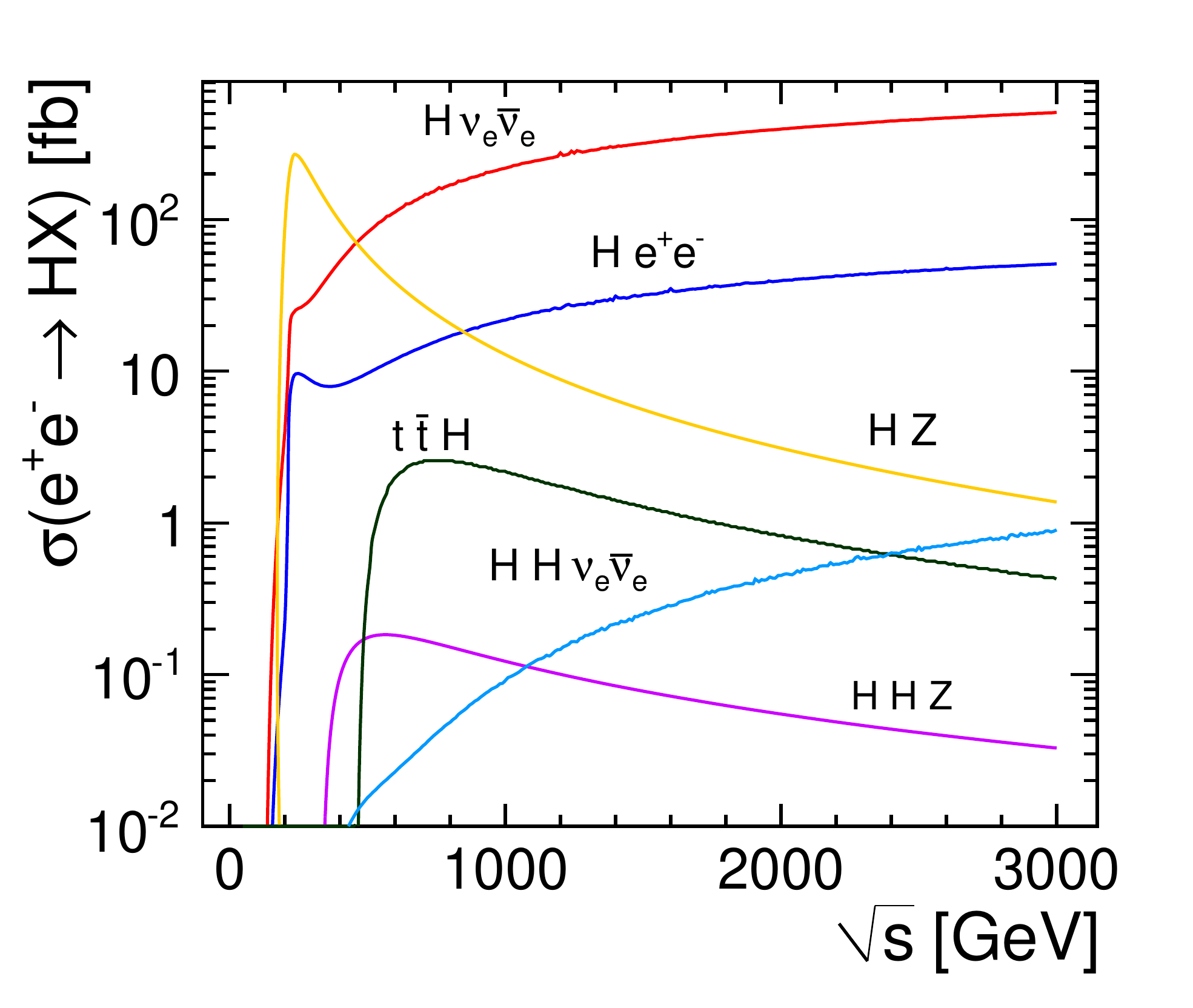}
 \end{subfigure}
 \caption[Higgs production cross section in the \acl{SM}.]{Higgs production cross section in the \ac{SM} depending on the Higgs mass $m_{\PH}$ for a center-of-mass energy $\sqrt{s}$ of \unit[3]{TeV}~(left) and depending on $\sqrt{s}$ for $m_{\PH} = \unit[120]{GeV}$~(right). Cross sections calculated with \whizard~\cite{Kilian:2007gr}. Figures taken from~\cite{cdrvol2}.}
 \label{fig:HiggsProduction}
\end{figure}

At tree-level, the total cross section of the Higgsstrahlung process is given by~\cite{Kilian:1995tr}
\begin{equation}
 \sigma_{\epem \to \zhsm} = \frac{G_F^2 m_{\PZ}^4}{96 \pi s}\left(\left(4\sin^2\theta_w -1\right)^2+1\right) \sqrt{\lambda} \frac{\lambda + 12m_{\PZ}^2 / s}{\left(1-m_{\PZ}^2/s \right)^2},
\label{eq:Higgsstrahlung}
\end{equation}
where $\lambda = \left(1 - (m_{\PH} + m_{\PZ})^2/s\right)\left(1 - (m_{\PH} + m_{\PZ})^2/s\right)$ and $G_F$ is the Fermi constant
\begin{equation}
 G_F = \frac{\sqrt{2}\ g^2_2}{8\ m^2_{\PW}}.
\end{equation}
The cross section rises sharply when \zhsm can be produced on-shell, \ie $\sqrt{s} = m_{\PH} + m_{\PZ}$, and drops with $1/s$ for high center-of-mass energies. The total cross section for the \ww fusion process can be approximated by~\cite{Kilian:1995tr}
\begin{equation}
 \sigma_{\epem \to \nuenuebar \PH} \approx \frac{G_F^3 m_{\PW}^4}{4\pi^3\sqrt{2}} \left(\left(1 + \frac{m_{\PH}^2}{s}\right) \log \frac{s}{m_{\PH}^2} - 2 - 2\frac{m_{\PH}^2}{s}\right).
\label{eq:WWfusion}
\end{equation}
This cross section scales with $\log(s)$ and although it is small at low $\sqrt{s}$ it eventually becomes the dominating Higgs production process at high center-of-mass energies. Both cross sections receive important radiative corrections~\cite{Kniehl:1991hk,Denner:2003iy} and interferences have to be taken into account if the \PZ decays into neutrinos~\cite{Kilian:1995tr}. Another important process is the Higgs production through radiation off a top quark $\epem \to \ttbar \to \ttbar \PH$~\cite{Djouadi:1991tk}. Although this process does not have a large cross section it is the only way to study the Yukawa coupling to top quarks, unless $m_{\PH} > 2m_{\PQt}$. The double Higgs production is induced through all processes listed above, where the Higgs decays into two Higgs bosons. There, the Higgsstrahlung process, $\epem \to \PH\PH\PZ$, and the \ww fusion process, $\epem \to \PH\PH\nuenuebar$, have the highest prospects of being measurable. \Cref{fig:HiggsProduction} gives an overview of all relevant Higgs production cross sections.

\begin{figure}
 \begin{subfigure}[]{0.53\textwidth}
  \includegraphics[width=\textwidth]{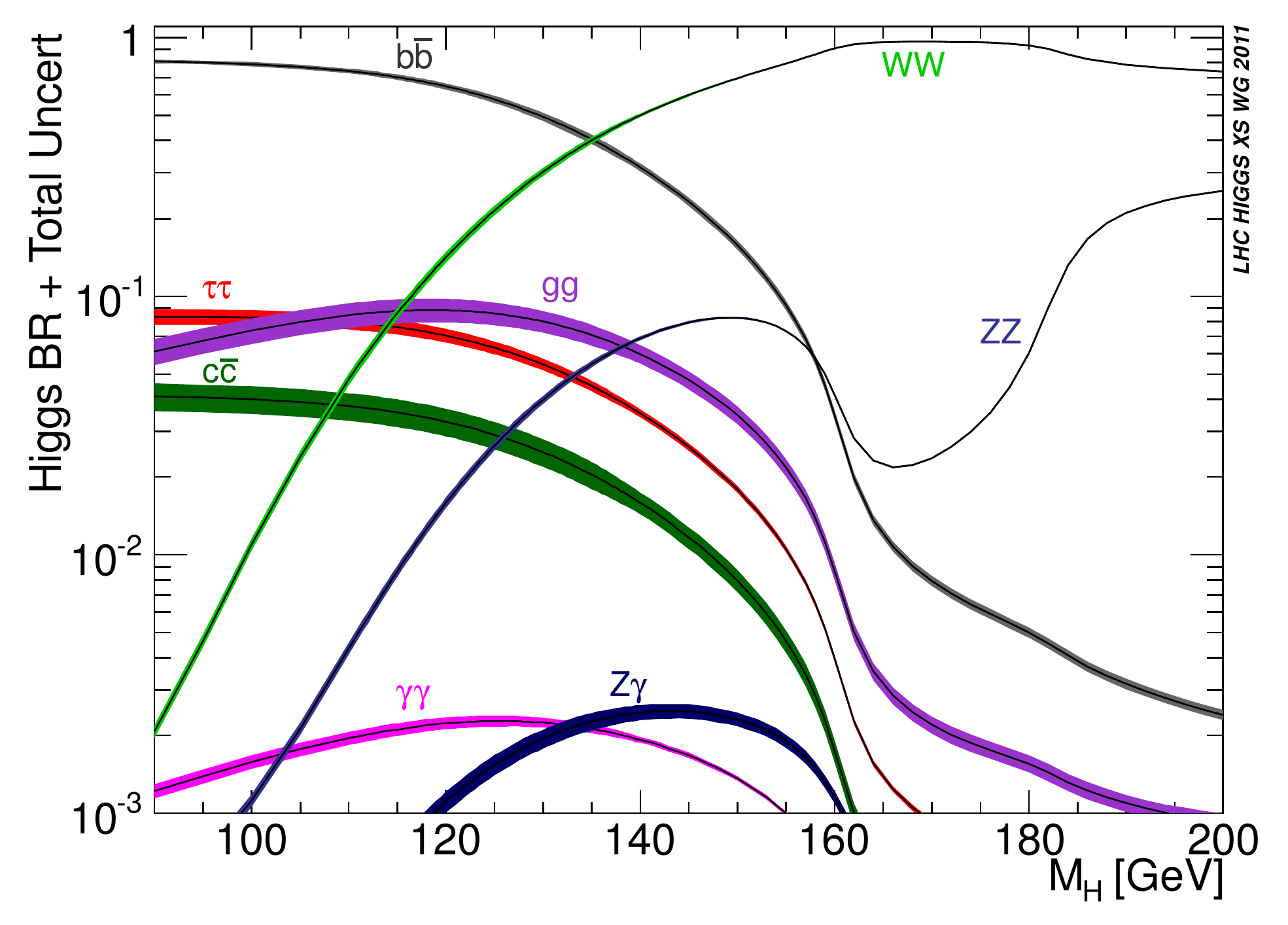}
 \end{subfigure}
 \hfill
 \begin{subfigure}[]{0.45\textwidth}
  \includegraphics[width=\textwidth]{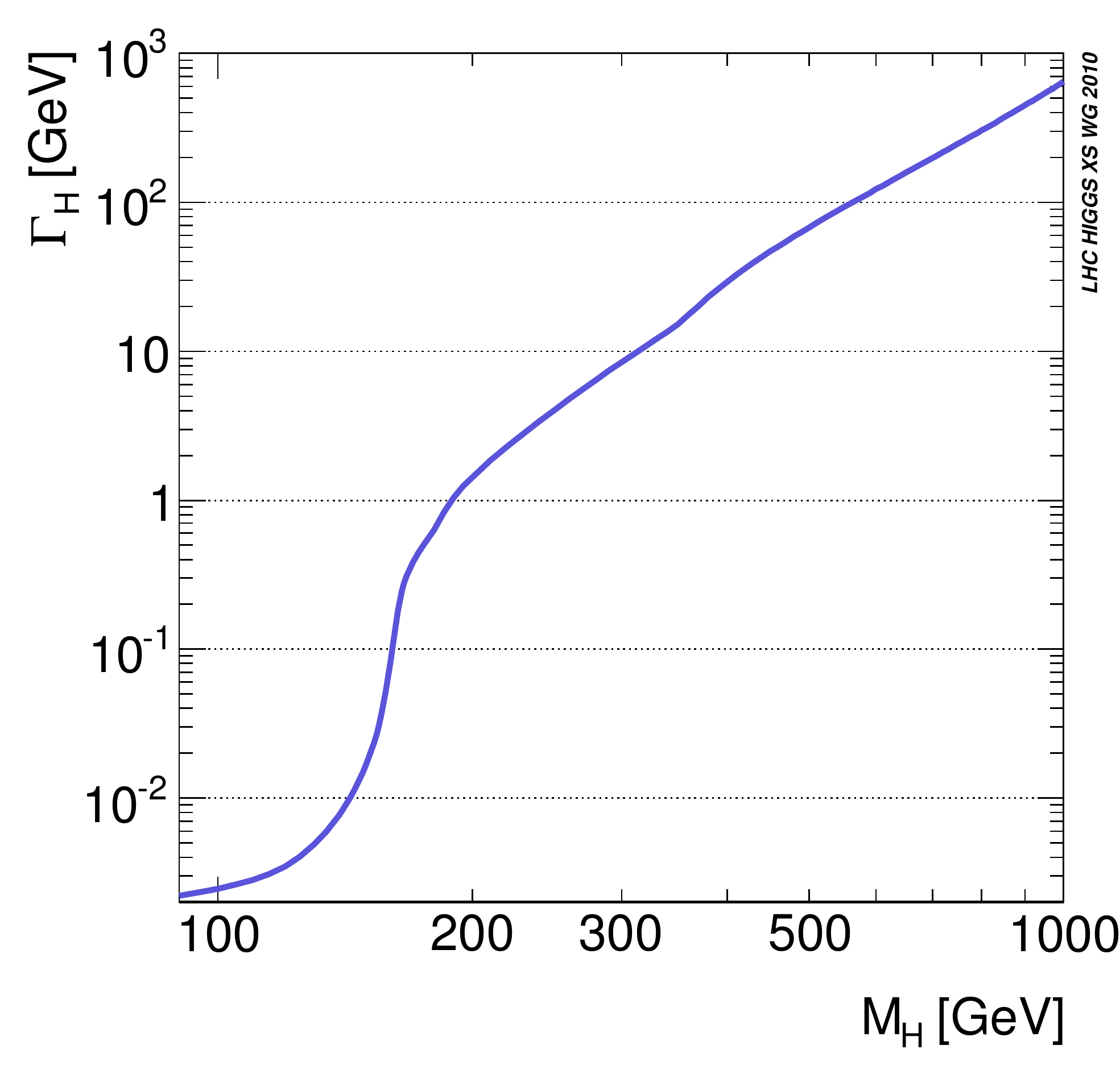}
 \end{subfigure}
 \caption[Higgs branching ratios in the \acl{SM}.]{Higgs branching ratios depending on the Higgs mass, $m_{\PH}$, including uncertainties from theory calculations~(left), taken from~\cite{Denner:2011mq}, and total width of the Higgs boson depending on $m_{\PH}$~(right), taken from~\cite{LHCHiggsCrossSectionWorkingGroup:2011ti}.}
 \label{fig:HiggsDecay}
\end{figure}

The partial widths of the different Higgs decays into fermions are given by
\begin{align}
 \Gamma_{\PH \to f\bar{f}} = \frac{G_F m^2_{f} m_{\PH} N_c}{4\pi\sqrt{2}}\left(1-m^2_f/m^2_{\PH}\right)^{3/2},
 \label{eq:higgsBranchinRatio}
\end{align}
where $N_c$ is the number of possible color charges, which is 1 for leptons and 3 for quarks. In addition, the Higgs can decay into the massive vector bosons as well as into two gluons or two photons. The latter two are induced through quark loops since they do not couple directly to the Higgs. The partial widths of all of these processes as well as the corresponding radiative corrections can be found for example in~\cite{Djouadi:2005gi} and references therein. The total width of the Higgs boson $\Gamma_{\PH}$ is then the sum of all partial widths that are kinematically allowed and the branching ratio of a certain decay is given by $\text{BR}_{\PH \to X} = \Gamma_{\PH \to X} / \Gamma_{\PH}$. The dependence of the individual branching ratios and the total width of the Higgs boson on its mass are shown in \cref{fig:HiggsDecay}.

It should be noted, that due to its kinematics, the Higgsstrahlung process has the highest prospects of the most precise measurement of the Higgs mass through the measurement of the recoil of the $\PZ$:
\begin{equation}
 m_{\PH}^2 = s - 2\sqrt{s}E_{\PZ} + m_{\PZ}^2.
\end{equation}
For ultimate precision a precise knowledge of $\sqrt{s}$ is required, which is only possible at a lepton collider. For example, a simulation study of this recoil mass measurement of a \ac{SM} Higgs with a mass of \unit[120]{GeV} at the \ac{ILC}, assuming a center-of-mass energy of \unit[250]{GeV} and an integrated luminosity of \unit[250]{\fbinv}, shows that an uncertainty on the Higgs mass of \unit[30]{MeV} is achievable~\cite{Li:2012ta}.
In addition, if only information of the $\PZ$ system is used in the analysis, this process allows for the determination of the total width of the Higgs boson in a model-independent way, including possible invisible decay modes.

\subsection{Determining the Parameters of the \acl{SM}}
\label{sec:SM_tests}
In total, the \ac{SM} has 19 free parameters that have to be determined from observations. These are 9 fermion masses, 3 angles and one $CP$ violating phase of the CKM matrix, the $CP$ violating phase of \ac{QCD}, the three gauge coupling constants $g_{1,2,3}$ and the parameters of the Higgs potential $\mu$ and $\lambda$. The latter five parameters can also be expressed in parameters that are directly accessible to experimental observation: the fine structure constant $\alpha$, the strong coupling constant $\alpha_s$, and the masses $m_{\PZ}$, $m_{\PW}$ and $m_{\PH}$.

Another important parameter is the Fermi constant $G_F$, which can be directly determined from the muon lifetime. It provides a relation between the weak boson masses and the fine structure constant
\begin{equation}
 G_F = \frac{\pi\ \alpha}{\sqrt{2}}\left(m^2_{\PW} - \frac{m^4_{\PW}}{m^2_{\PZ}}\right)^{-1},
 \label{eq:GF_mwmz}
\end{equation}
and is a direct measure of the vacuum expectation value of the Higgs potential $v^{-1} = \sqrt{2}G_F$. \Cref{eq:GF_mwmz} is one of many possible consistency tests of the \ac{SM}, since all of the parameters can be measured. Since the Higgs boson contributes to loop corrections also at lower energies, the combination of all observables, including the precise measurements of \PW and \PZ cross sections and branching ratios performed at \ac{LEP} and \acs{SLD}, allows to perform a global fit of the \ac{SM} and thus allows to put limits on the Higgs mass. Two discussions of electroweak fits and their results can be found in~\cite{ALEPH:2010aa,Baak:2011ze}. The latest published indirect limit on the \ac{SM} Higgs mass from June 2011 is $m_{\PH} = \unit[80^{+30}_{-23}]{GeV}$~\cite{Baak:2011ze}. The same fit including the additional limits from the direct Higgs searches at \ac{LEP} and the \acs{Tevatron} yields $m_{\PH} = \unit[116.4^{+18.3}_{-1.3}]{GeV}$~\cite{Baak:2011ze}.

In the mean time, the first results from Higgs searches at the \ac{LHC} have put even more stringent limits on the possible Higgs mass in the \ac{SM}. Earlier in 2012 \acs{CMS} reported that a \ac{SM} Higgs is excluded at the 95\% confidence level for masses of $m_{\PH} < \unit[114.4]{GeV}$ and $m_{\PH} > \unit[127.5]{GeV}$~\cite{Moriond2012CMS}. The results reported by \acs{ATLAS} put stronger limits for lower Higgs masses. They report that in the \ac{SM} the only Higgs masses that are not excluded at a confidence level of 95\% are $\unit[117.5]{GeV} < m_{\PH} < \unit[118.5]{GeV}$ and $\unit[122.5]{GeV} < m_{\PH} < \unit[129]{GeV}$~\cite{Moriond2012ATLAS}. It is expected that by the end of 2012 enough data are collected at the \ac{LHC} to either exclude the \ac{SM} Higgs over its full mass range or to claim its discovery with a significance of at least $5\sigma$.

\section{Beyond the Standard Model}
\label{sec:BSM}
Despite its great success in describing fundamental particle interactions, the \ac{SM} has several limitations. For example, the Higgs mass receives quadratic corrections from fermion loops, where the heaviest fermion, the top quark, has the largest contribution. This correction can be written as 
\begin{equation}
 \Delta m_{\PH} = - \frac{g_f^2}{8\ \pi^2} \Lambda^2 + \dots,
\end{equation}
where $\Lambda$ is the ultra-violet cut-off of the theory. From measurements of the weak boson masses as well as $G_F$ we know $v$ and thus also that $m_{\PH}$ has to be of the order of \unit[100]{GeV}. If $\Lambda$ is very high compared to the electroweak scale, \eg around the reduced Planck scale of $\unit[\sim10^{18}]{GeV}$, the bare Higgs mass has to be of the order of $\Lambda^2$ and fine-tuned to arrive at the observed Higgs mass. This is also referred to as the Hierarchy problem.


Although the \ac{SM} offers a mechanism to assign mass terms for the massive gauge bosons and the fermions it does not explain the actual size of the different masses. Similarly the different gauge couplings, the quark mixing angles, the Higgs mass and the Higgs vacuum expectation value are free parameters of the theory. The number of generations in the \ac{SM} is also not explained, as well as the observed pattern in the values of the electrical charge and the hypercharge of the fermions. In addition, the theory does not offer a mechanism to include gravity.

A direct motivation for the incompleteness of the \ac{SM} comes from cosmological observations. There is strong evidence for instance through the observations of rotational velocities of stars in galaxies~\cite{Rubin:1980zd}, the observations of gravitational lensing of galaxy clusters~\cite{Wu:1998ju}, the observation of the fluctuations in the cosmological microwave background~\cite{Komatsu:2008hk} and---probably the most compelling evidence---the observation of the mass distribution of the colliding galaxy clusters in the bullet cluster~\cite{Clowe:2006eq}, that there is a large discrepancy between the observed gravitating matter and the expected amount of matter from observations of stars and gas in galaxies. This can be explained if a large fraction of the mass of the universe is dark matter~\cite{Bertone:2004pz}, \ie matter that only interacts gravitationally or gravitationally and weak. The only particles in the \ac{SM} that fulfill this requirement are the neutrinos, but even when assuming massive neutrinos, the required dark matter largely exceeds the amount of neutrinos expected in the big bang theory. From simulations of galaxy formation processes it is evident that dark matter has to be cold~\cite{Springel:2005nw}, \ie has to consist of non-relativistic particles. This would not be the case if neutrinos are the main constituent of dark matter. Another important cosmological observation is the absence of anti-matter in the universe~\cite{Steigman:2005uz}. This asymmetry can only be explained through $CP$ violating processes that removed the anti particles from the universe after the big bang. The $CP$ violation allowed in the \ac{SM} is not sufficient to explain the necessary asymmetry to explain the observations.

All these are hints that the \ac{SM} might be only an effective field theory which has to be extended by a suitable model.

\subsection{Extensions of the \acl{SM}}
\label{sec:SM_extensions}
The observation of neutrino oscillations~\cite{Fukuda:1998fd,Ahmad:2001an} more than a decade ago constitutes already an extension of the basic \ac{SM} discussed in \cref{sec:SM_SM}, although it does not address any of the shortcomings of the \ac{SM} stated above. Similarly to the quark mixing mentioned in \cref{sec:SM_Higgs}, mixing of the neutrino states implies that the flavor eigenstates are different from the mass eigenstates. The observation of neutrino mixing is only sensitive to mass differences but not absolute masses. Since mixing of all neutrino flavors has been observed, it can be immediately concluded that at least two of the three neutrinos are massive. This leads to two possible modifications of the \ac{SM}. If the neutrinos are massive Dirac particles, it requires the existence of previously undetected right-handed neutrinos. On the other hand, if neutrinos would acquire their mass like Majorana particles, they would be their own anti-particle, which would imply violation of lepton number conservation.

\subsubsection{Supersymmetry}
\label{sec:SM_SUSY}
An attractive extension of the \ac{SM} is \ac{SUSY}, reviewed for example in~\cite{Martin:1997ns}. These models postulate a global symmetry between fermions and bosons. This symmetry introduces supersymmetric partners to all particles in the \ac{SM}, also referred to as sparticles. In the \ac{MSSM}, every particle has exactly one superpartner, where the superpartners of the fermions, the sleptons and squarks, have a spin of $0$ and the super partners of the gauge bosons, the gauginos or neutralinos and charginos, have a spin of $\frac{1}{2}$. In the Higgs sector, at least four additional scalar particles are predicted. The five scalar particles are two $CP$ even neutral particles, the $\PSh$ and $\PSH$ particles, the neutral pseudoscalar particle $\PSA$ and two charged particles $\PSHpm$. In many \ac{SUSY} models the $\PSh$ is very similar to the Higgs boson in the \ac{SM}, with a light mass of around \unit[120]{GeV}. In most of these cases the four other Higgs particles have almost identical and rather high masses. Unfortunately \ac{SUSY} introduces new couplings that violate both lepton number and baryon number, which are conserved in the \ac{SM}. These conservation laws are very well tested and have stringent limits from the minimum lifetime of the proton. This problem can be solved by postulating a new conserved quantity, the $R$-parity, with $P_R = (-1)^{2s+3B+L}$, where $s$ is the spin, $B$ is the baryon number and $L$ is the lepton number. As a result all particles in the \ac{SM} have an $R$-parity of 1 and all the superpartners have an $R$-parity of $-1$, thus the proton and also the \ac{LSP} are stable.

\ac{SUSY} naturally solves several of the limitations of the \ac{SM} discussed above. For example, the additional particles would introduce new loop corrections to the Higgs self-coupling that automatically cancel out the divergent terms and thus resolve the hierarchy problem. Similarly, the loop corrections change the dependence of the three effective gauge couplings on the transferred energy $Q$, allowing to unify at an energy of around \unit[$10^{16}$]{GeV}. This would allow a further unification of the strong and electroweak gauge interactions, similar to the unification of the weak and electromagnetic interactions in the \ac{SM}.
In addition, if the $R$ parity is conserved and if the \ac{LSP} were neutral, it would be a natural candidate for dark matter since it would be stable~\cite{Jungman:1995df}
. Even if $R$-parity would be slightly violated, the LSP could be an explanation for dark matter if its lifetime is large enough.

Since \ac{SUSY} particles have not been observed yet, the masses of the superpartners have to be significantly higher than the masses of the \ac{SM} particles. This requires that the symmetry has to be broken. In addition, the breaking of \ac{SUSY} should not re-introduce quadratic divergences in the loop corrections to the Higgs self-coupling, but at most logarithmic divergences. This is referred to as soft \ac{SUSY} breaking.

If \ac{SUSY} is introduced as a local gauge symmetry instead of a global gauge symmetry, it requires the inclusion of gravity. A new \ac{SUSY} doublet, the spin-2 graviton, which is the mediator of the gravitational force, and its super partner the spin-$\frac{3}{2}$ gravitino have to be introduced. The resulting theory is called \ac{SUGRA}~\cite{Freedman:1976xh,Deser:1976eh}
 and would be a natural combination of the theory of gravity and the current understanding of the other fundamental forces.


\subsubsection{Importance of the Higgs Sector}
\label{sec:SM_higgsImplications}
With the introduction of new particles, like for example in the case of \ac{SUSY}, new couplings to the Higgs boson(s) are predicted and thus the branching ratios of its decay modes should differ from the \ac{SM}. Similarly, models that include an alternative explanation of electroweak symmetry breaking naturally predict branching ratios that are different from those in the \ac{SM}. It is thus of utmost importance to not only measure the relative branching ratios of the different decay modes predicted in the \ac{SM}, which leaves one ignorant to other decays, but also measure the absolute branching ratios and the total width of the Higgs boson. This will be very difficult at the \ac{LHC} and is thus one of the most compelling reasons for the construction of a high-energy lepton collider. Finally, to decide between different concurring models it is important to measure all branching ratios as precisely as possible.

Extensive discussions of the Higgs sector and how to distinguish between different models can be found in \cite{Gunion:322177,Djouadi:2005gi,Djouadi:2005gj}.

\chapter{The Compact Linear Collider}
\label{cha:CLIC}

The \acl{CLIC} is a concept for a future \epem linear collider with a possible center-of-mass energy of several TeV. It will allow precise measurements of the properties of the Higgs boson, if it exists, and any other new particle that could be discovered at the \ac{LHC} if it falls within the energy reach of \ac{CLIC}. If it exists, the \ac{LHC} experiments will only be sensitive to the couplings of the Higgs to the gauge bosons (through top loops in case of photons and gluons) and possibly to the couplings to \PQt and \PQb through associated production. In the lepton sector only its coupling to \PGt and maybe the coupling to \PGm will be accessible. A lepton collider like \ac{CLIC} or the \ac{ILC} will be able to measure all of these couplings with a much higher precision and also allows the measurement of the Higgs self coupling. While the \ac{LHC} has very good chances of discovering new strongly interacting particles like squarks, a high energy lepton collider has a significantly larger discovery potential for particles in the electroweak sector, \eg gauginos, sleptons or a $\Zprime$. Measurements at \ac{CLIC} would thus be complementary to those performed at the \ac{LHC}.

This interplay of hadron and lepton colliders has proven beneficial throughout the history of particle physics. For example, the \PW and \PZ bosons have been first discovered in proton-antiproton collisions at the \ac{SPS} by the \acs{UA1}\cite{Arnison:1983mk,Arnison:1983rp} and \acs{UA2}~\cite{Banner:1983jy,Bagnaia:1983zx} experiments. Their properties have later been measured with much greater precision in the \epem collisions by the experiments at \ac{LEP}~\cite{ALEPH:2005ab,ALEPH:2010aa}. The properties of the \PZ boson were also measured in \epem collisions at the \ac{SLD} operated at the \ac{SLC}, the longest linear accelerator to date.
The reason why proton-proton or proton-antiproton colliders are often referred to as discovery machines is that they more easily allow for a much higher collision energy. When charged particles are forced on a circular trajectory they lose energy through synchrotron radiation. In a circular collider, the energy loss through synchrotron radiation per turn scales with
\begin{equation}
 \Delta E_\text{synchrotron} \propto \frac{E_\text{beam}^4}{m^4\ \rho}
\label{eq:synchrotron}
\end{equation}
where $\rho$ is the radius of the collider and $m$ is the mass of the particle. It is thus evident that protons, which are almost 2000 times heavier than electrons, experience a significantly lower energy loss per turn. The drawback of hadron colliders is that only one quark or one gluon of each of the colliding hadrons takes part in the hard interaction. Thus, only a fraction of the center-of-mass energy of the two hadrons is available in the hard interaction and the energy and momentum of the two partons is intrinsically unknown. Lepton colliders on the other hand collide fundamental particles, which means that $\sqrt{s}$ of the collision is precisely known through the beam energy.

In this chapter we will introduce the \ac{CLIC} accelerator concept and discuss the conditions for experiments at \ac{CLIC} including the time structure of the beam, the machine-related backgrounds and the luminosity spectrum. The information summarized in this chapter is extracted from the accelerator volume of the \ac{CLIC} \ac{CDR}~\cite{cdrvol1}.

\section{The \acs{CLIC} Concept}
\label{sec:CLIC_concept}

The aim of the next $\epem$ collider is a high center-of-mass energy, covering at least the energy scale of new physics \ac{BSM} (if it exists), while providing high luminosity of the order of $\unit[10^{34}]{cm^{-2}s^{-1}}$ to allow for precise measurements of the properties of all newly discovered phenomena.

As shown in \cref{eq:synchrotron}, the energy loss through synchrotron radiation in a circular collider rises with the fourth power of the beam energy. On the other hand, the energy consumption of a linear collider scales linearly with the beam energy as well as the luminosity. Already for beam energies slightly higher than those achieved at \ac{LEP} a linear collider becomes more efficient than a circular collider. A linear collider requires extremely high field gradients in the acceleration structures to limit the required total length of the accelerator. The \ac{ILC} proposes superconducting cavities with a gradient of \unitfrac[35]{MV}{m} to achieve a beam energy of \unit[250]{GeV} along a half-length of the collider of approximately \unit[14]{km}. The maximum gradient achievable with superconducting cavities is limited and thus, if much higher beam energies are desired, normal conducting cavities are the only option without requiring an excessively long accelerator.

\subsection{Field Gradient}
\label{sec:CLIC_gradient}
\begin{figure}
 \includegraphics[width=\textwidth]{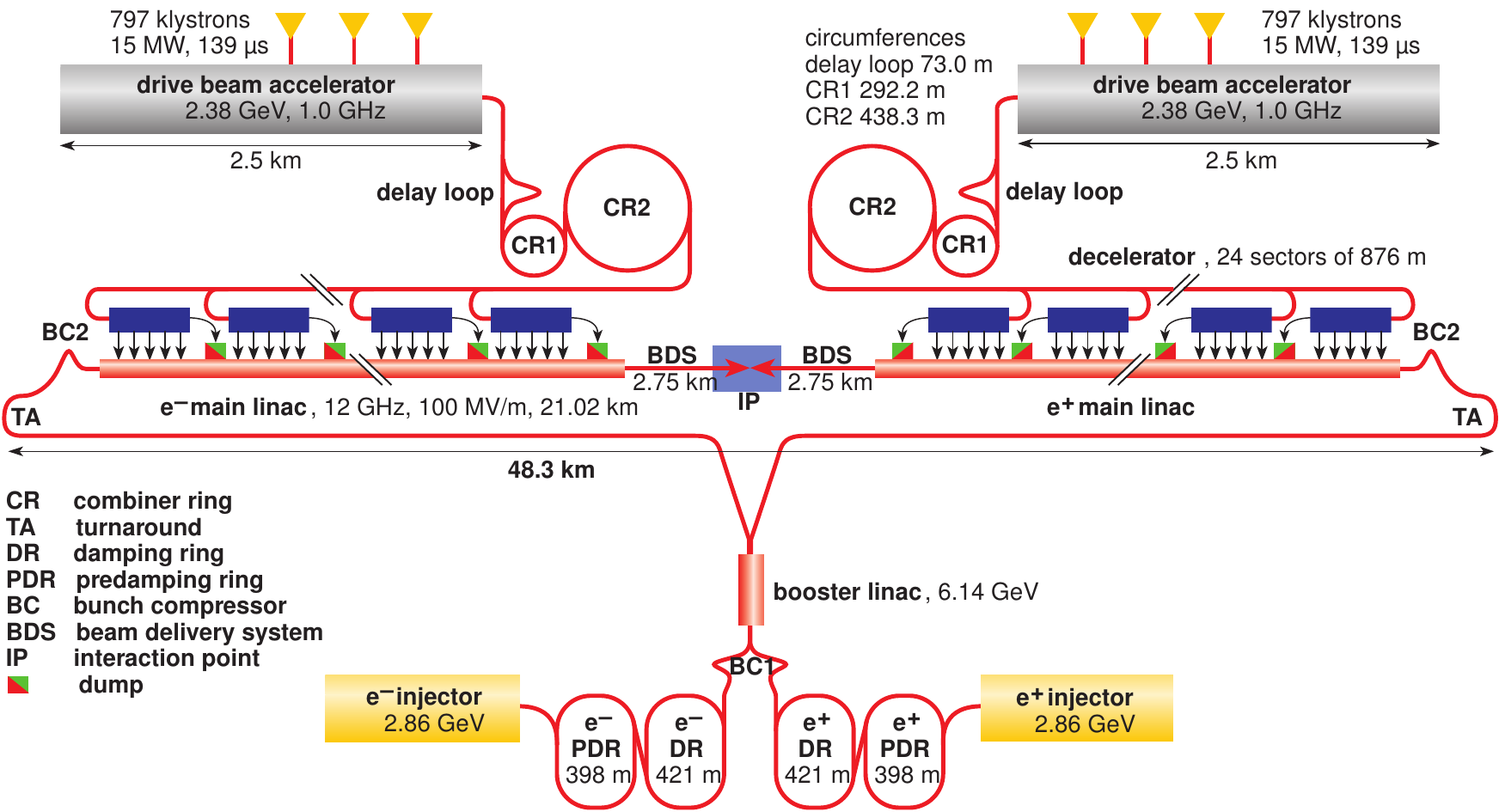}
 \caption[Schematic view of the \ac{CLIC} layout for a center-of-mass energy of $3\ \text{TeV}$.]{Schematic view of the \ac{CLIC} layout for a center-of-mass energy of \unit[3]{TeV}. Figure taken from~\cite{cdrvol1}.}
 \label{fig:CLIC_overview}
\end{figure}

For \ac{CLIC}, acceleration cavities with a gradient of \unitfrac[100]{MV}{m} and a frequency of \unit[12]{GHz} are foreseen, which allows to achieve a beam energy of \unit[1.5]{TeV} within an accelerator of \unit[21]{km} length. The total length of \ac{CLIC}, including both accelerators as well as the \ac{BDS}, is \unit[48.3]{km}. A schematic view of the default layout for $\sqrt{s} = \unit[3]{TeV}$ is shown in \cref{fig:CLIC_overview}. To achieve a center-of-mass energy of \unit[500]{GeV}, a total length of only \unit[13]{km} is required.

In addition to the high gradient in the cavities, also the breakdown rate of the \ac{RF} power in the accelerating cavities has to be sufficiently low to allow for an efficient operation of \ac{CLIC}. The design goal is that the probability of a breakdown happening anywhere in the accelerator is less than 1\%. This translates into a maximum breakdown rate within the cavities of $\unit[3\times10^{-7}]{m^{-1}pulse^{-1}}$.

\subsection{Two-Beam Acceleration}
\label{sec:CLIC_twoBeam}
The second crucial point is the efficient generation of the \ac{RF} power required to accelerate the particles in the beam. The high number of klystrons 
that would be required to provide the necessary power, as well as their low efficiency is prohibitive. Instead, a two beam acceleration scheme is proposed to generate and distribute the \ac{RF} power at \ac{CLIC}. The drive beam that provides the power is a low-energy electron beam, with an electron energy of \unit[2.38]{GeV}, at a very high intensity of \unit[100]{A} within the bunch train period. This beam is decelerated in a series of \acp{PETS}, where the energy is extracted to provide the accelerating field in the cavities of the main beam. The drive beam is decelerated over a length of \unit[876]{m} and dumped afterwards. In total there are 24 of these sectors in each of the two accelerators. The high beam intensity of the drive beam, required to produce sufficient power, is achieved by producing a beam with very short bunch spacings of \unit[2.4]{cm} that is split and overlaid using a delay loop followed by two combiner rings.

The distribution of the power transport in a particle beam is almost loss-less. The most important losses are introduced in the power extraction and transfer in the \ac{PETS}, which has to be very efficient ($\gg 90\%$) to provide enough power for the acceleration of the main beam.

\subsection{Beam-Beam Effects and Beam-Related Backgrounds}
\label{sec:CLIC_machineInduced}
The high space charge inside the two bunches leads to interactions between them when they cross each other at the interaction point. Since the bunches are of opposite charge at \ac{CLIC}, the particles see an attractive force which leads to an increase in luminosity. This effect is called pinch-effect, which is discussed for example in~\cite{Schulte:331845}. As in the case of synchrotron radiation, the change of the trajectory due to an electromagnetic field leads to the creation of photons. The photons of this so-called beamstrahlung can interact and lead to the creation of secondary particles, mostly \epem pairs~\cite{Chen:242895}.

\begin{figure}
 \begin{subfigure}[b]{0.49\textwidth}
  \centering
  \begin{tikzpicture}[thick, >=latex]
   \node (iPhoton) at (-2.0,-1.5) {\PGg};
   \node (fElectron1) at (1.5,-1.0) {\Pe};
   \node (fElectron2) at (1.5,1.0) {\Pe};
   \node[align=left] (field) at (-2.0,1.5) {Macroscopic\\Field};
   \coordinate (ge1) at (0.0, 0.0);

   \draw[photon] (iPhoton) to (ge1);
   \draw[basefermion] (fElectron1) to (ge1) to (fElectron2);
   \draw[decorate, decoration={snake, amplitude=3mm}] (field) to (ge1);
   \fill[black] (ge1) circle (2.0pt);
  \end{tikzpicture}
  \caption{Coherent Pair Production}
  \label{fig:CoherentPair}
 \end{subfigure}
 \hfill
 \begin{subfigure}[b]{0.49\textwidth}
  \centering
  \begin{tikzpicture}[thick, >=latex]
   \node (iElectron) at (-2.5,-2.0) {\Pe};
   \node (fElectron1) at (0.5,-2.0) {\Pe};
   \node (fElectron2) at (1.5,-1.0) {\Pe};
   \node (fElectron3) at (1.5,1.0) {\Pe};
   \node[align=left] (field) at (-2.0,1.5) {Macroscopic\\Field};
   \coordinate (ge1) at (-1.0, -1.0);
   \coordinate (ge2) at (0.0, 0.0);

   \draw[basefermion] (iElectron) to (ge1) to (fElectron1);
   \draw[photon] (ge1) to (ge2);
   \draw[basefermion] (fElectron2) to (ge2) to (fElectron3);
   \draw[decorate, decoration={snake, amplitude=3mm}] (field) to (ge2);
   \fill[black] (ge2) circle (2.0pt);
  \end{tikzpicture}
  \caption{Trident Pair Production}
  \label{fig:TridentPair}
 \end{subfigure}
 \caption{Feynman diagrams of the coherent pair production and trident pair production processes.}
 \label{fig:CoherentTridentPair}
\end{figure}
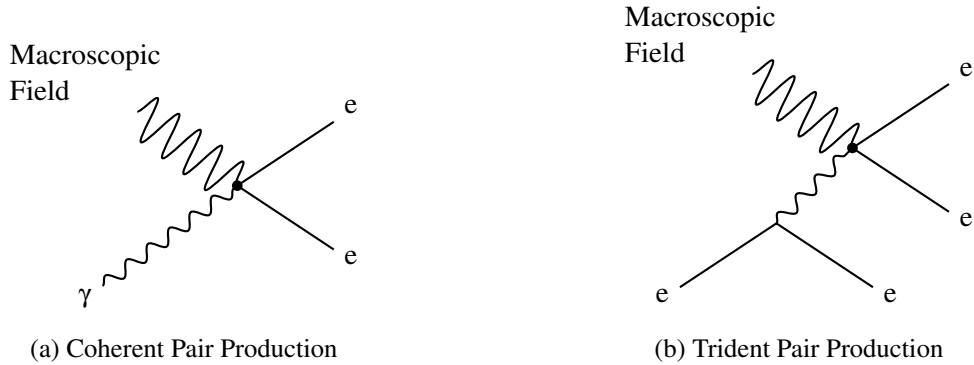

The most likely process is the conversion of one of the photons into an \epem pair. The photon conversion is only possible in a strong electromagnetic field, which in this case is the coherent field of all other particles seen by the photon. The \epem pairs are thus called coherent pairs (see \cref{fig:CoherentPair} for the Feynman diagram). Since the photons are created mostly collinear with the beam, the pairs are also created with very low transverse momentum. It is important to choose the opening angle of the outgoing beam pipe large enough, such that the detector is unaffected by the large flux of these high-energetic particles.

A related effect is the creation of trident pairs shown in \cref{fig:TridentPair}, where a virtual photon converts into an \epem pair in presence of the macroscopic field. The angular distribution of these pairs is very similar to that of the coherent pairs, but they are in general of lower energy. This means that they are of no additional concern for the design of the detector in the forward region.

\begin{figure}
 \begin{subfigure}[b]{0.30\textwidth}
  \centering
  \begin{tikzpicture}[thick, >=latex]
   \node (photon1) at (-1.5,1.5) {\PGg};
   \node (electron1) at (1.5,1.5) {\Pe};
   \node (photon2) at (-1.5,-1.5) {\PGg};
   \node (electron2) at (1.5,-1.5) {\Pe};
   \coordinate (ge1) at (0.0, 0.75);
   \coordinate (ge2) at (0.0, -0.75);

   \draw[photon] (photon1) to (ge1);
   \draw[photon] (photon2) to (ge2);
   \draw[basefermion] (electron1) to (ge1) to (ge2) to (electron2);
  \end{tikzpicture}
  \caption{Breit-Wheeler}
  \label{fig:Breit-Wheeler}
 \end{subfigure}
 \hfill
 \begin{subfigure}[b]{0.30\textwidth}
  \centering
  \begin{tikzpicture}[thick, >=latex]
   \node (iElectron) at (-2.5,2.0) {\Pe};
   \node (fElectron1) at (0.5,2.0) {\Pe};
   \coordinate (photon1) at (-1.0,1.5);
   \node (fElectron2) at (1.5,1.5) {\Pe};
   \node (photon2) at (-1.5,-1.5) {\PGg};
   \node (fElectron3) at (1.5,-1.5) {\Pe};
   \coordinate (ge1) at (0.0, 0.75);
   \coordinate (ge2) at (0.0, -0.75);

   \draw[basefermion] (iElectron) to (photon1) to (fElectron1);
   \draw[photon] (photon1) to node[below left]{\PGg} (ge1);
   \draw[photon] (photon2) to (ge2);
   \draw[basefermion] (fElectron2) to (ge1) to (ge2) to (fElectron3);
  \end{tikzpicture}
  \caption{Bethe-Heitler}
  \label{fig:Bethe-Heitler}
 \end{subfigure}
 \hfill
 \begin{subfigure}[b]{0.30\textwidth}
  \centering
  \begin{tikzpicture}[thick, >=latex]
   \node (iElectron1) at (-2.5,2.0) {\Pe};
   \node (fElectron1) at (0.5,2.0) {\Pe};
   \coordinate (photon1) at (-1.0,1.5);
   \node (fElectron2) at (1.5,1.5) {\Pe};
   \coordinate (photon2) at (-1.0,-1.5);
   \node (fElectron3) at (1.5,-1.5) {\Pe};
   \coordinate (ge1) at (0.0, 0.75);
   \coordinate (ge2) at (0.0, -0.75);
   \node (iElectron2) at (-2.5,-2.0) {\Pe};
   \node (fElectron4) at (0.5,-2.0) {\Pe};

   \draw[basefermion] (iElectron1) to (photon1) to (fElectron1);
   \draw[photon] (photon1) to node[below left]{\PGg} (ge1);
   \draw[photon] (photon2) to node[above left]{\PGg} (ge2);
   \draw[basefermion] (fElectron2) to (ge1) to (ge2) to (fElectron3);
   \draw[basefermion] (iElectron2) to (photon2) to (fElectron4);
  \end{tikzpicture}
  \caption{Landau-Lifshitz}
  \label{fig:Landau-Lifshitz}
 \end{subfigure}
\caption{Feynman diagrams of the most important processes that contribute to the incoherent pair production.}
\label{fig:IncoherentPair}
\end{figure}
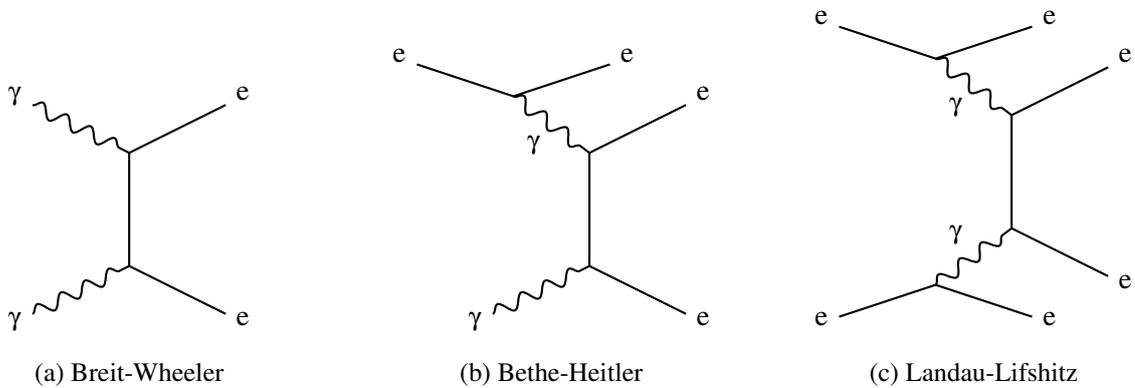

More important for the design of the detector are the incoherent \epem pairs, created from the interaction of two real or virtual photons. \Cref{fig:IncoherentPair} shows the three dominating processes contributing to their production. Although they are in general less energetic than the two previously mentioned pair backgrounds, they extend to larger polar angles. This requires a careful design of the vertex detector to avoid large occupancies as well as the forward calorimeters to avoid the creation of large amounts of backscattering particles. The strong solenoid field of the central detector helps in confining them to low radii.

All of these processes are of course also possible for all other charged fermions. Their production cross section is, however, largely reduced, due to their higher mass compared to the electron mass. The leptonic processes (other than \epem production) are thus of no concern in terms of detector occupancy, although they can fake physics signatures and have to be considered when looking for a corresponding final state. The production of quark pairs is of more concern since they hadronize and create mini-jets also at large polar angles~\cite{Chen:561345}. This background is referred to as \gghad.

\begin{figure}
 \includegraphics[width=0.49\textwidth]{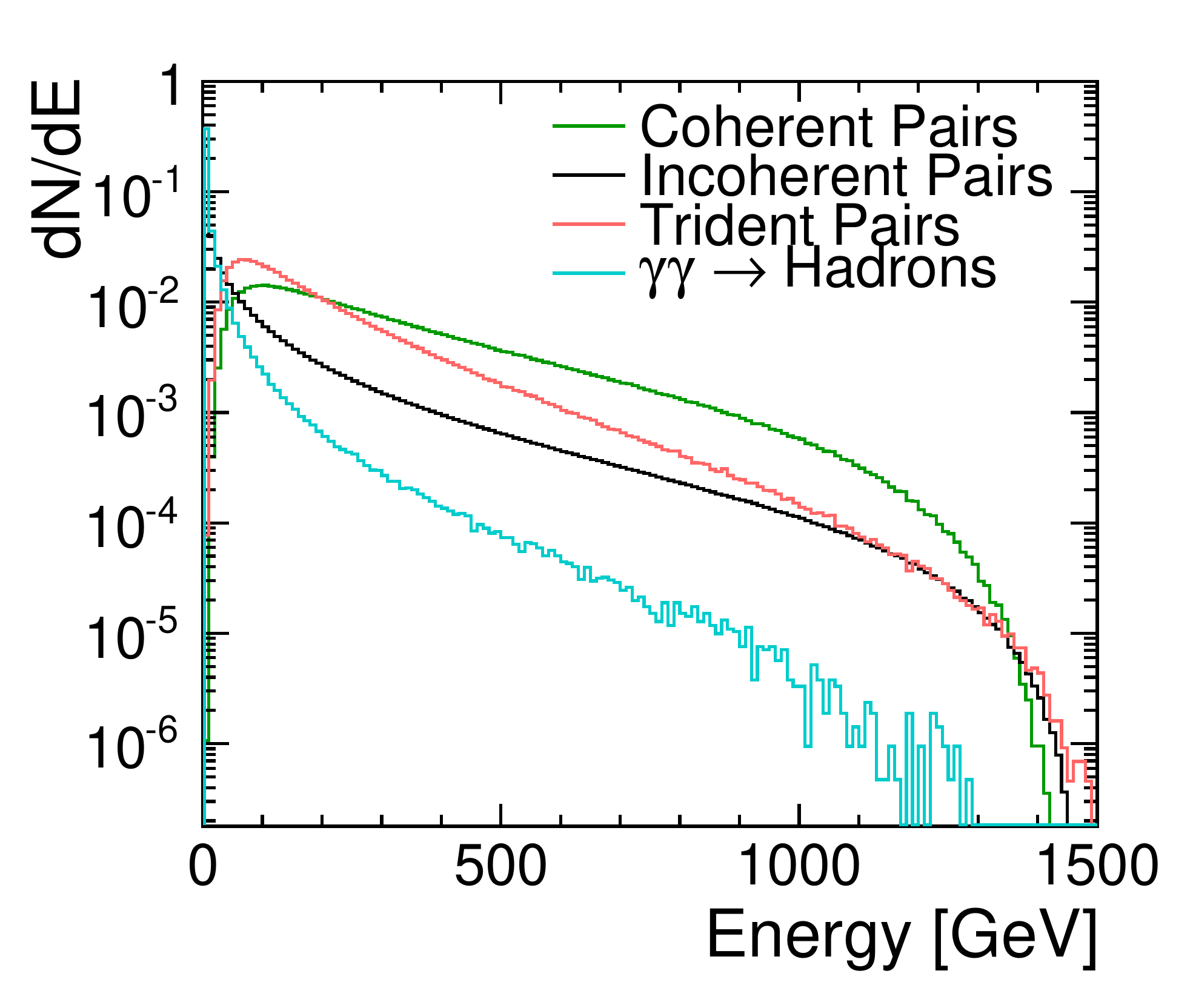}
 \hfill
 \includegraphics[width=0.49\textwidth]{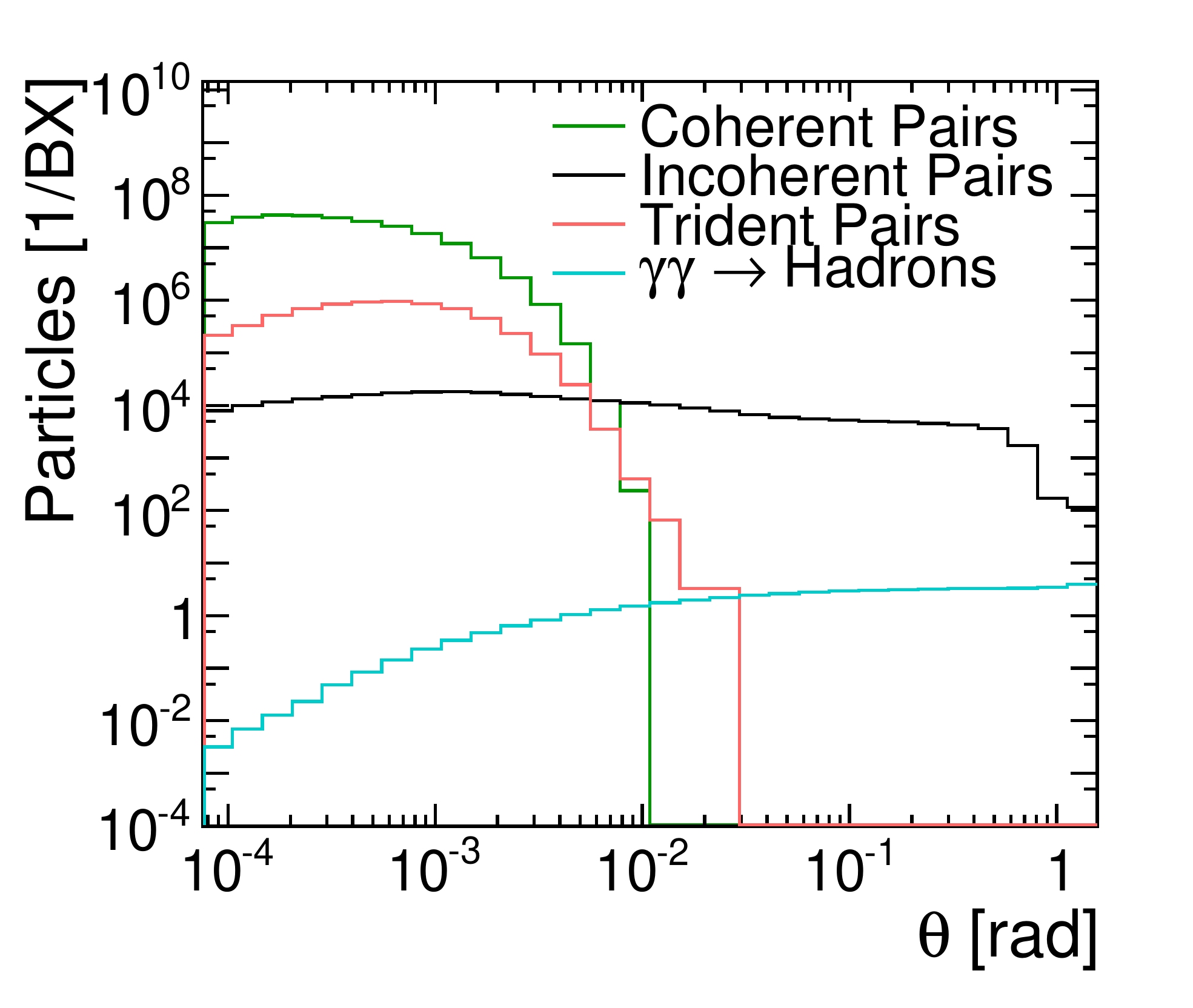}
\caption[Distribution of the energy and the polar angle of the particles from beam-induced backgrounds.]{Distribution of the particle energy in the beam-induced backgrounds~(left) and distribution of the polar angle of the background particles~(right). The polar angle is given with respect to the outgoing beam-axis. Figures taken from~\cite{cdrvol2}.}
\label{fig:CLIC_backgroundDistributions}
\end{figure}

All of these processes are simulated with the program \guineapig~\cite{Schulte:331845,Schulte:1999tx}, which is also discussed in \cref{sec:Software_beamBeam}. The resulting energy and polar angle distributions of the particles created in these processes for \ac{CLIC} at \unit[3]{TeV} are shown in \cref{fig:CLIC_backgroundDistributions}. The particle energies extend up to the full beam energy, although they peak at low energies. The coherent pairs are emitted only within \unit[10]{mrad} and the trident pairs only within \unit[30]{mrad}. The incoherent pairs and the particles from \gghad are created at all polar angles. The maximum polar angle of the coherent pairs also determines the minimum opening angle of the very forward calorimeter, \ie the \ac{BeamCal}, of \unit[10]{mrad}.

\subsection{Luminosity}
\label{sec:CLIC_luminosiy}
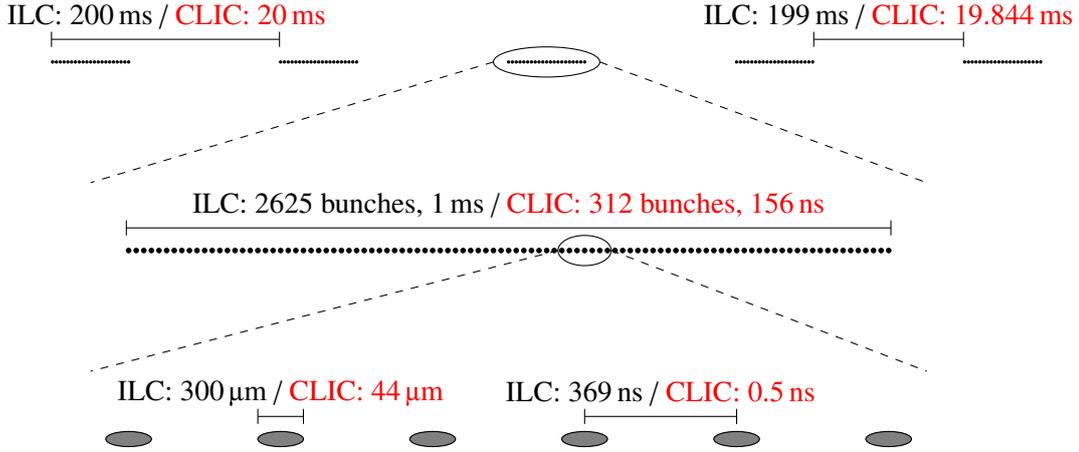
\begin{figure}
 \centering
 \begin{tikzpicture}
  \def\trainsy{5.0}
  \def\trainy{2.5}
  \def\bunchy{0.0}
  \draw (1.5,\trainsy+0.3) node[above] {ILC: \unit[200]{ms} / \color{red}CLIC: \unit[20]{ms}};
  \draw (-0.015,\trainsy+0.3) -- (2.985,\trainsy+0.3);
  \draw (-0.015,\trainsy+0.2) -- (-0.015,\trainsy+0.4);
  \draw (2.985,\trainsy+0.2) -- (2.985,\trainsy+0.4);

  \draw (11,\trainsy+0.3) node[above] {ILC: \unit[199]{ms} / \color{red}CLIC: \unit[19.844]{ms}};
  \draw (10.015,\trainsy+0.3) -- (11.985,\trainsy+0.3);
  \draw (10.015,\trainsy+0.2) -- (10.015,\trainsy+0.4);
  \draw (11.985,\trainsy+0.2) -- (11.985,\trainsy+0.4);
  \foreach \trainx in {0,3,6,9,12}
  {
   \foreach \x in {0,0.05,...,1.05}
   {
    \filldraw[black] (\trainx+\x,\trainsy) circle (0.015);
   }
  }

  \draw (6.5,\trainsy) ellipse (0.7 and 0.2);
  \draw[dashed] (6.5-0.7,\trainsy) -- (0.5, \trainy+0.9);
  \draw[dashed] (6.5+0.7,\trainsy) -- (11.5, \trainy+0.9);
  \draw (6,\trainy+0.3) node[above] {ILC: 2625 bunches, \unit[1]{ms} / \color{red}CLIC: 312 bunches, \unit[156]{ns}};
  \draw (0.97,\trainy+0.3) -- (11.03,\trainy+0.3);
  \draw (0.97,\trainy+0.2) -- (0.97,\trainy+0.4);
  \draw (11.03,\trainy+0.2) -- (11.03,\trainy+0.4);
  \foreach \bunchx in {1,1.1,...,11.1}
  {
   \filldraw[black] (\bunchx, \trainy) circle (0.03);
  }

  \draw (7,\trainy) ellipse (0.35 and 0.2);
  \draw[dashed] (7-0.35,\trainy) -- (0.5, \bunchy+0.9);
  \draw[dashed] (7+0.35,\trainy) -- (11.5, \bunchy+0.9);
  \draw (3,\bunchy+0.3) node[above] {ILC: \unit[300]{\micron} / \color{red}CLIC: \unit[44]{\micron}};
  \draw (2.7,\bunchy+0.3) -- (3.3,\bunchy+0.3);
  \draw (2.7,\bunchy+0.2) -- (2.7,\bunchy+0.4);
  \draw (3.3,\bunchy+0.2) -- (3.3,\bunchy+0.4);
  \draw (8,\bunchy+0.3) node[above] {ILC: \unit[369]{ns} / \color{red}CLIC: \unit[0.5]{ns}};
  \draw (7,\bunchy+0.3) -- (9,\bunchy+0.3);
  \draw (7,\bunchy+0.2) -- (7,\bunchy+0.4);
  \draw (9,\bunchy+0.2) -- (9,\bunchy+0.4);
  \foreach \bunchx in {1,3,...,11}
  {
   \filldraw[gray] (\bunchx, \bunchy) ellipse (0.3 and 0.1);
   \draw[black] (\bunchx, \bunchy) ellipse (0.3 and 0.1);
  }
  
 \end{tikzpicture}
\caption[Time structure of the beams at \ac{ILC} and \ac{CLIC}.]{Schematic view of the time structure of a beam at the \ac{ILC} at $\sqrt{s} = \unit[500]{GeV}$ (black numbers) and at \ac{CLIC} at $\sqrt{s} = \unit[3]{TeV}$ (red numbers). The beam is split into trains with a large gap in between (top). Each train consists of several bunches as indicated in the lower two sketches. Pictures are not to scale. Numbers for \ac{CLIC} from~\cite{cdrvol1}. Numbers for \ac{ILC} correspond to the nominal design in~\cite{Brau:2007zza}. }
\label{fig:CLIC_beamStructure}
\end{figure}

The luminosity of a collider with symmetric colliding beams can be written as~\cite{Schulte:331845}
\begin{equation}
 \lumi = \frac{N^2 f}{4\pi\sigma_x\sigma_y},
\end{equation} 
where $N$ is the number of particles in each bunch, $f$ is the frequency of the \acp{BX}, and $\sigma_x$ and $\sigma_y$ are the lateral extends of each bunch. In case of a linear collider that is operated in pulsed mode, \ie the bunches arrive in trains that are short with respect to the distance between them, the frequency has to be replaced by $f = f_\text{train} \cdot n_\text{bunches}$, where $f_\text{train}$ is the frequency of the trains and $n_\text{bunches}$ is the number of bunches in each train. A high luminosity thus requires both, a very high bunch charge, \ie a large number of particles in each beam, as well as very collimated beams. The total length of each train is limited by the choice of the normal conducting technology, since the electric field in the cavities dissipates quickly. The pulse length generated by the \ac{PETS} is \unit[176]{ns}. The total length of each train at \ac{CLIC} has to be shorter and is chosen to be \unit[156]{ns}, where each train consists of 312 bunches. For the \ac{ILC} on the other hand, much longer bunch trains of \unit[1]{ms} are proposed. \Cref{fig:CLIC_beamStructure} shows a schematic view of the typical beam structure at a linear collider and gives the numbers for the \ac{ILC} and \ac{CLIC} to highlight the very different time structures of the two proposed accelerator concepts.

While the luminosity scales with $1/\left(\sigma_x\sigma_y\right)$, the amount of energy lost in beamstrahlung scales with $1/\left(\sigma_x +\sigma_y\right)^2$~\cite{Schulte:331845}. This means that an asymmetric shape of the bunch in $xy$-direction with very different $\sigma_x$ and $\sigma_y$ is the optimal choice. The minimum size in any of these directions is limited by the feasible emittance of the beam, as well as the achievable stability of the beam position, due to the stability of the focusing elements in the \ac{BDS}, especially that of the \ac{QD0}. The parameters chosen for the default layout of \ac{CLIC} for a center-of-mass energy of \unit[3]{TeV} are given in \cref{tab:CLIC_parameters}.

The short spacing between two bunches leads to interactions between incoming and outgoing bunches, especially if the beams are not perfectly aligned. This and the desired placement of the \ac{QD0} as close as possible to the \ac{IP} infers a minimum crossing angle of the accelerator of \unit[20]{mrad}~\cite{Schulte:2001aw}.

\begin{table}[htb]
\caption[Parameters of the default \acs{CLIC} layout for $\sqrt{s} = 3\ \text{TeV}$.]{Parameters of the default \ac{CLIC} layout for $\sqrt{s} = \unit[3]{TeV}$. Given is the crossing angle $\theta_c$, the repetition rate of the bunch trains $f_\text{train}$, the number of bunches in each train $n_\text{bunches}$, the time between two bunch crossings $\Delta t$, the number of particles in each bunch $N$, the orthogonal bunch sizes $\sigma_{x,y,z}$, the total luminosity \lumi, the luminosity within the highest 1\% of the nominal beam energy $\lumi_{1\%}$, the number of beamstrahlung photons per beam particle $n_{\PGg}$, the fraction of the beam energy lost in beamstrahlung $\Delta E/ E$, the number of coherent pair particles per \ac{BX} $N_\text{coh}$, the total energy of all coherent pair particles per \ac{BX} $E_\text{coh}$, the number of incoherent pair particles per \ac{BX} $N_\text{incoh}$, the total energy of all incoherent pair particles per \ac{BX} $E_\text{incoh}$, and the number of \gghad events per \ac{BX} $n_{had}$ with a minimum transferred energy of \unit[2]{GeV}. Table adapted from~\cite{cdrvol2}. A full set of machine parameters for \ac{CLIC}, also for other $\sqrt{s}$, can be found in~\cite{cdrvol1}.}
\label{tab:CLIC_parameters}
\centering
\begin{tabular}{c c}
\toprule
 Parameter & Value \\
\midrule
 $\theta_c$ & \unit[20]{mrad} \\
 $f_\text{train}$ & \unit[50]{Hz} \\
 $n_\text{bunches}$ & \unit[312]{} \\
 $\Delta t$ & \unit[0.5]{ns} \\
 $N$ & $3.72\times10^{9}$ \\
 $\sigma_x$ & $\unit[\sim 45]{nm}$ \\
 $\sigma_y$ & $\unit[\sim 1]{nm}$ \\
 $\sigma_z$ & \unit[44]{\micron} \\
 $\lumi$ & $\unit[5.9\times10^{34}]{cm^{-2}s^{-1}}$ \\
 $\lumi_{1\%}$ & $\unit[2.0\times10^{34}]{cm^{-2}s^{-1}}$ \\
 $n_{\PGg}$ & 2.1 \\
 $\Delta E/ E$ & 0.28 \\
 $N_\text{coh}$ & $\unit[6.8\times10^8]{}$\\
 $E_\text{coh}$ & $\unit[2.1\times10^8]{TeV}$ \\
 $N_\text{incoh}$ & $\unit[3.0\times10^5]{}$\\
 $E_\text{incoh}$ & $\unit[2.3\times10^4]{TeV}$ \\
 $n_\text{had}$ & $\unit[3.2]{}$ \\
\bottomrule
\end{tabular}
\end{table}

\subsection{Luminosity Spectrum}
\label{sec:CLIC_lumiSpectrum}
\begin{figure}
 \centering
 \includegraphics[width=0.49\textwidth]{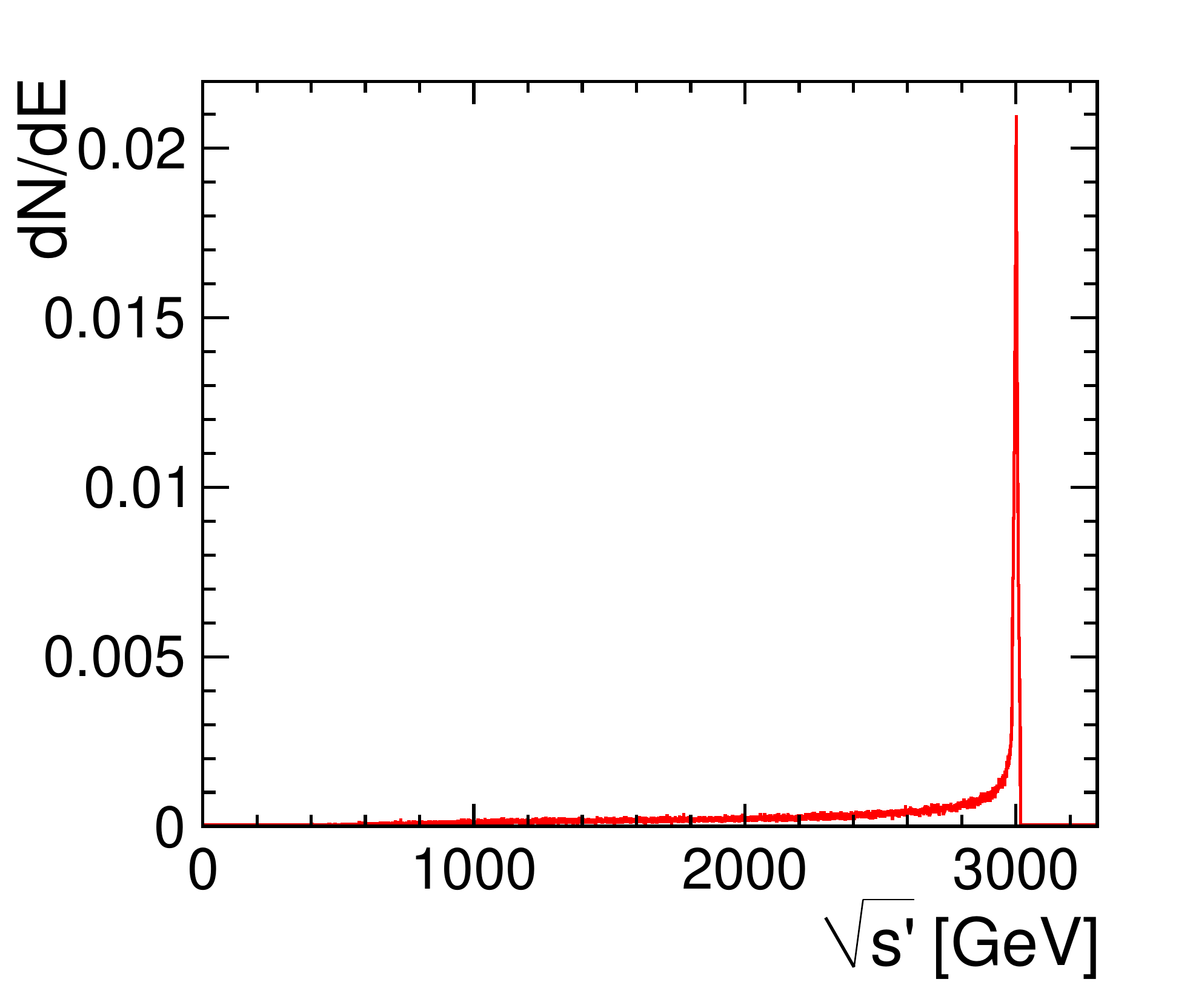}
\caption[Luminosity spectrum at \ac{CLIC} for a nominal center-of-mass energy of $3\ \text{TeV}$.]{Distribution of the effective center of mass energy $\sqrt{s'}$ at \ac{CLIC} for a nominal center-of-mass energy of \unit[3]{TeV}. Figure taken from~\cite{cdrvol2}.}
\label{fig:CLIC_lumiSpectrum}
\end{figure}

In any particle accelerator the energy of the colliding particles is subject to an intrinsic spread. At \ac{CLIC} this energy spread is expected to be 0.35\% around the nominal beam energy of \unit[1.5]{TeV}. In addition, the mean beam energy is expected to fluctuate by approximately 0.1\%~\cite{cdrvol2}. The biggest effect on the center-of-mass energy at \ac{CLIC} originates from the beamstrahlung introduced above. The potentially large energy loss of one or both of the colliding particles at the interaction point leads to long tails to low energies in the distribution of the effective center-of-mass energy: the luminosity spectrum. This effect is illustrated in \cref{fig:CLIC_lumiSpectrum}. For the nominal \ac{CLIC} parameters, shown in \cref{tab:CLIC_parameters}, the fraction of collisions within the highest 1\% of the nominal energy corresponds only to 35\% of the total luminosity. This mostly affects the measurement of processes with production thresholds close to the nominal center-of-mass energy. On the other hand, processes which can be produced at lower $\sqrt{s}$ will benefit from a significantly larger fraction of the total luminosity. In any case this luminosity spectrum has to be taken into account when calculating production cross sections at \ac{CLIC}.

We want to stress that the long tail in the luminosity spectrum is mostly caused by the beam-beam effects which can not be avoided if a high total luminosity is desired. A \ac{CLIC} accelerator at lower center-of-mass energies of $\sqrt{s} = \unit[500]{GeV}$ would have a much narrower luminosity spectrum with almost 75\% of the luminosity within the highest 1\% of the energy, but with a lower total luminosity of only  $\unit[2.0\times10^{34}]{cm^{-2}s^{-1}}$~\cite{cdrvol2}.

\subsection{Staged Construction}
\label{sec:CLIC_staging}
The configuration of the accelerator and especially the beam delivery system is chosen to optimize the available luminosity for the nominal center-of-mass energy. Although the accelerator can also be operated at lower $\sqrt{s}$ this will result in a significantly lower total luminosity. For certain scenarios, \eg a threshold scan, the accelerator will need to be operated far from its nominal energy. It is thus beneficial to construct \ac{CLIC} in several stages with increasing center-of-mass energies. The chosen energy stages will depend strongly on the new physics scenarios discovered at the \ac{LHC}. One possible scenario involving three energy stages is investigated in~\cite{cdrvol3}. There, the first stage is designed for a nominal center-of-mass energy of \unit[500]{GeV}, which allows to precisely measure the Higgs mass and some of its couplings in $\PH\PZ$ production as well as to precisely measure the top mass in a \ttbar threshold scan. The second stage is then chosen to be at $\sqrt{s} = \unit[1.4]{TeV}$ based on the mass spectrum in the \ac{SUSY} scenario that was chosen for this particular study. The final stage will then be the \unit[3]{TeV} configuration presented here.

\subsection{Demonstration of Feasibility}
\label{sec:CLIC_feasibility}
Most of the key technological challenges have been achieved in prototypes. For example, the generation of a high intensity beam using a delay loop followed by one combiner ring has been demonstrated at \ac{CTF3}. Similarly, the efficient power extraction using a series of \acp{PETS} demonstrated at \ac{CTF3}, which then has been used to demonstrate the concept of two-beam acceleration. In addition, a gradient of \unitfrac[106]{MV}{m} in combination with a maximum breakdown rate of $\unit[3.1\times10^{-7}]{m^{-1}pulse^{-1}}$ has been achieved in individually tested acceleration structures~\cite{cdrvol1}.



\chapter{Detector Design for Collider Experiments}
\label{cha:DetectorDesign}
High energy collisions are studied by measuring the four momenta, \ie the energies and the momenta, of ideally all particles that were created in the collision. All other observables are derived from these two quantities and the quality of their measurement determines the success of an experiment.

In this chapter we discuss the basic concepts of momentum measurement (see \cref{sec:SiD_momentumMeasurement}) and energy measurement (see \cref{sec:SiD_energyMeasurement}) in \ac{HEP} collider experiments, and motivate the formulae typically used to parametrize the detector performance. Subsequently the detector requirements for an experiment at \ac{CLIC} are discussed in \cref{sec:SiD_requirements}. The particle flow paradigm, which is driving most detector designs for future collider experiments is introduced in \cref{sec:SiD_particleFlow}.


\section{Track Momentum and Impact Parameter Measurement}
\label{sec:SiD_momentumMeasurement}

Charged particles are deflected in a magnetic field through the Lorentz force. In a homogeneous magnetic field the resulting particle trajectory is a helix defined only by the particle momentum $\vec{p}$, its charge $q$ and the magnetic field $\vec{B}$. This allows the measurement of the particle momentum in a known magnetic field. The momentum measurement can be divided into the measurement of the transverse momentum in the plane perpendicular to the magnetic field and the measurement of the polar angle $\theta$:
\begin{equation}
 p = \frac{\pT}{\sin\theta}.
\end{equation}
An illustration of the two track projections and more information on the track parameters can be found in \cref{App:trackParametrization}.

\subsection{Curvature Measurement}
\label{sec:SiD_momentum_curvature}
Assuming a perfect solenoid field parallel to the $z$-axis, the magnetic field inside the solenoid is $\vec{B} = (0,0,B)$. The circular path of the particle with a charge $q$ in the $r\phi$-plane is then given by the centripetal force introduced by the Lorentz force
\begin{equation}
 \frac{mv_\mathrm{T}^2}{\rho} = q\ v_\mathrm{T}\ B,
\end{equation}
where $\rho$ is the radius, $m$ is the particle mass and $v_\mathrm{T}$ is the absolute value of the particle velocity projected to the $r\phi$-plane. Using $mv_\mathrm{T} = \pT$ the radius and the curvature $\kappa$ are given by
\begin{equation}
 \frac{1}{\kappa} \equiv \rho = \frac{\pT}{qB}.
\label{eq:SiD_trackMeasurement1}
\end{equation}
Charged particles that are living long enough to form a measurable track carry unit charge, $\pm e$. Omitting the sign of the charge, which defines the orientation of the helix, and using units commonly used in \ac{HEP}, the transverse momentum can be calculated as
\begin{equation}
 \pT = \frac{0.3\ B}{\kappa} \frac{\mathrm{GeV}}{\mathrm{T\ m}}.
\label{eq:SiD_trackMeasurement2}
\end{equation}
The arc length $S$ of the projected track in the $r\phi$-plane can be calculated from the azimuthal angle $\varphi$ between the start point and the end point of the helix:
\begin{equation}
 S = \rho\varphi = \frac{\pT\ \varphi}{0.3\ B} \frac{\mathrm{T\ m}}{\mathrm{GeV}}.
 \label{eq:SiD_projectedTrackLength}
\end{equation}
The uncertainty of the curvature measurement using $N$ individual measurement points with an uncertainty of $\sigma(r\phi)$ and which are equally distributed along $S$ can be estimated using the Gl\"uckstern formula~\cite{Gluckstern:1963ng}
\begin{equation}
 \sigma(\kappa) \approx \sqrt{\frac{720}{N+4}} \frac{\sigma(r\phi)}{S^2}.
\label{eq:Glueckstern}
\end{equation}
This formula is derived using a parabola as an approximation of the circular track projection and only holds for small curvatures and $N \gg 3$. An exact solution for circular tracks can be found in~\cite{Karimaki:1997ff}.

Using \cref{eq:Glueckstern,eq:SiD_trackMeasurement2} the relative uncertainty of the transverse momentum due to the point resolution can be approximated as
\begin{equation}
 \frac{\sigma(\pT)}{\pT}\bigg|^\mr{track} \approx \sqrt{\frac{720}{N+4}} \frac{\sigma(r\phi) \pT}{0.3\ B\ S^2} \frac{\mathrm{T\ m}}{\mathrm{GeV}}.
 \label{eq:resolutionTrack_pT}
\end{equation}
For almost straight tracks, $S$ is equivalent to the distance between the innermost and outermost measurement point. This means that the radius of the detector is the most powerful variable to improve the momentum resolution for high momentum tracks.

\subsection{Polar Angle Measurement}
\label{sec:SiD_momentum_angle}
The projection of the track in the curved $Sz$-plane is a straight line described by
\begin{equation}
 z = z_0 + S\cot(\theta),
\end{equation}
where $z_0$ is the $z$-position of the track origin in the $xy$-plane. For tracks with a small curvature the track projection in the $rz$-plane can be approximated by a straight line. The uncertainty on the measurement of the polar angle from $N$ measurements with a point resolution of $\sigma(\xi)$ which are equally distributed along a straight line of the length $L$ is given by~\cite{Gluckstern:1963ng}
\begin{equation}
 \sigma(\theta) \approx \frac{\sigma(\xi)}{L}\sqrt{\frac{12(N-1)}{N(N+1)}}.
 \label{eq:GlucksternLambda}
\end{equation}
The point resolution $\sigma(\xi)$ is perpendicular to the track. This has to be translated into the detector point resolution which is parallel to the detector layer. For barrel detectors which are parallel to $z$ this results in $\sigma(\xi) = \sigma(z)\sin(\theta)$. For disk detectors which are perpendicular to $z$ this results in $\sigma(\xi) = \sigma(r)\cos(\theta)$. It should be noted that the length $L$ used here is the total length of the track along the helix
\begin{equation}
 L = \frac{S}{\sin(\theta)} = S\sqrt{1+\cot(\theta)}.
\end{equation}

\subsection{Multiple Scattering}
\label{sec:SiD_momentum_multipleScattering}
A charged particle that traverses material is deflected by multiple Coulomb scattering. The distribution of the scattering angle is described by Moli\`{e}re's theory of multiple scattering~\cite{Moliere:1948zz}. For small scattering angles $\alpha$ within a plane parallel to the original direction of the particle, the distribution of $\alpha$ can be described by a Gaussian distribution centered around $\alpha = 0\degrees$. For particles with $q = \pm e$, its width $\sigma(\alpha)$ is then given by the approximation~\cite{Highland:1975pq,Lynch:1990sq}
\begin{equation}
 \sigma(\alpha) \approx \frac{\unit[13.6]{MeV}}{\beta\ p}\sqrt{\frac{X}{\radlen}} \left( 1 + 0.038 \ln \left( \frac{X}{\radlen} \right) \right),
\label{eq:Highland}
\end{equation}
where $X$ is the thickness of the material along the particle trajectory and \radlen is the radiation length of the material. Since the second term in parentheses is only a small correction it will be neglected below.

For an estimation of the impact of multiple scattering on the track measurement we treat the scattering in $\theta$ and $\phi$ directions independently. The transverse momentum measurement is only affected by the multiple scattering in $\phi$ which introduces an additional uncertainty in $\varphi$. Using \cref{eq:SiD_projectedTrackLength} we find
\begin{equation}
\frac{\sigma(\pT)}{\pT} = \frac{\sigma(\varphi)}{\varphi} = \frac{\sigma(\varphi)\ \pT}{0.3\ B\ S}\frac{\mathrm{T\ m}}{\mathrm{GeV}}.
\end{equation}
In order to use \cref{eq:Highland} to describe the uncertainty in $\varphi$ due to multiple scattering we have to project the scattering angle into the $r\phi$-plane which gives $\alpha_{r\phi}$ = $\alpha/\sin\theta$. We also assume a homogeneous material distribution along the track such that $X = L$. The resulting uncertainty on the transverse momentum due to multiple scattering is then
\begin{equation}
\frac{\sigma(\pT)}{\pT}\bigg|^\mathrm{MS} \approx \frac{0.045\ \pT\ \mr{T}\ \mr{m}}{\beta\ p\ B\ S\ \sin\theta} \sqrt{\frac{L}{\radlen}} = \frac{0.045\ \mr{T}\ \mr{m}}{\beta B\ \sin\theta\ \sqrt{L\ \radlen}}.
\label{eq:SiD_multipleScatteringCurvature}
\end{equation}
It should be noted that for multiple individual layers the scattering angles introduced by different layer are correlated. In addition, when calculating the traversed length $L'$ of track in a layer, the direction in $\phi$ in which a layer is crossed has to be considered as well as the angle in $\theta$.

Using \cref{eq:resolutionTrack_pT} and \cref{eq:SiD_multipleScatteringCurvature} it is evident that the relative transverse momentum resolution can be parametrized as
\begin{equation}
 \frac{\sigma(\pT)}{\pT} =  a\pT \oplus \frac{b}{\sin\theta}.
\label{eq:resolutionParametrization_pT}
\end{equation}

The impact of multiple scattering on the measurement of the polar angle can be estimated similarly. Like above we assume a homogeneous material distribution along $L$. The additional uncertainty on $\theta$ is then directly given by \cref{eq:Highland}
\begin{equation}
 \sigma(\theta)|^\mathrm{MS} \approx \frac{\unit[13.6]{MeV}}{\beta\ p}\sqrt{L/\radlen}.
\end{equation}

As shown in~\cite{Valentan:2009zz} the total momentum resolution is given by
\begin{equation}
 \frac{\sigma(p)}{p} \approx \sqrt{\left(\frac{\sigma(\pT)}{\pT}\right)^2 - \cot^2\theta\ \sigma^2(\theta)}.
\label{eq:resolutionParametrization_p}
\end{equation}

\subsection{Impact Parameter Measurement}
\label{sec:SiD_momentum_impactParameter}
The impact parameter $d_0$ is defined as the distance of closest approach to the \ac{IP} of the helix projection in the $xy$-plane. It defines the origin $S = 0$ of the helix projection. The $z$-position of $P_0$ is called $z_0$.

The measurement of $d_0$ is the extrapolation of the azimuthal measurement to the point of closest approach, which we can approximate by a parabola for sufficiently small curvatures~\cite{Valentan:2009zz}
\begin{equation}
 d_0 \approx R\phi - R\varphi + \frac{\kappa}{2}R^2,
\end{equation}
where $R$ is the radius of the measurement layer. The track curvature has to be small compared to $R$, otherwise the track would not pass through multiple layers. We can thus neglect the last term, and the resulting uncertainty on $d_0$ is given by
\begin{equation}
 \sigma(d_0) \approx \sigma(R\phi) \oplus R\sigma(\varphi),
\end{equation}
when also neglecting correlations. We can identify $\sigma(R\phi)$ with the point resolution $\sigma(r\phi)$ in the measurement layer. In addition, the term $\sigma(\varphi)$ recieves a contribution from multiple scattering. Using \cref{eq:Highland} projected into the $r\phi$-plane for the additional uncertainty on $\varphi$ results in 
\begin{equation}
 \sigma(d_0) \approx \sigma(r\phi) \oplus R\sigma(\varphi) \oplus \frac{R\ \unit[13.6]{MeV}}{\beta\ p \sin\theta}\sqrt{\frac{X}{\radlen}}.
\end{equation}
For high track momenta the multiple scattering term and the uncertainty on the angular measurement should be small compared to $\sigma(r\phi)$, which then determines the impact parameter resolution. For low momenta only the multiple scattering term is relevant. In a barrel like detector with constant thickness $d = X\sin\theta$ we can write
\begin{equation}
 \sigma(d_0) \approx \sigma(r\phi) \oplus \frac{R\ \unit[13.6]{MeV}}{\beta\ p \sqrt{\sin\theta}}\sqrt{\frac{d}{\radlen}},
\end{equation}
which corresponds to the canonical parametrization
\begin{equation}
 \sigma(d_0) \approx a \oplus \frac{b}{p\ \sqrt{\sin\theta}}.
 \label{eq:resolutionParametrization_d0}
\end{equation}

The measurement of $z_0$ is a linear extrapolation of the track slope from the innermost measurement position $z$
\begin{equation}
 z_0 = z - S(z)\ \cot\theta.
\end{equation}
Neglecting correlations, the uncertainty of the $z_0$ measurement is given by
\begin{equation}
 \sigma(z_0) \approx \sigma(z) \oplus \frac{S\ \sigma(\theta)}{\sin(\theta)}.
\label{eq:z0res}
\end{equation}
Like in the case of $d_0$ only the second term recieves corrections due to multiple scattering. Using \cref{eq:Highland} to estimate the uncertainty on the polar angle yields
\begin{equation}
 \sigma(z_0) \approx \sigma(z) \oplus \frac{S\ \sigma(\theta)}{\sin(\theta)} \oplus \frac{S}{\sin(\theta)} \frac{\unit[13.6]{MeV}}{\beta\ p}\sqrt{\frac{X}{\radlen}}.
\end{equation}
If the distance to the \ac{IP} is sufficiently small, $\sigma(z_0)$ is determined by $\sigma(z)$ in the central region and by $\sigma(\theta)$ for lower $\theta$. For low momenta $\sigma(\theta)$ is dominated by multiple scattering which then also dominates the impact parameter measurement. Assuming a constant material budget $d = X\sin(\theta)$ we get the usual parametrization
\begin{equation}
 \sigma(z_0) =  a \oplus \frac{b}{p\ \sqrt{\sin\theta}},
\end{equation}
which is similar to the one for $d_0$.

\section{Energy Measurement}
\label{sec:SiD_energyMeasurement}

The particle energy is measured in the calorimeters. In this process the particle loses its energy through various interactions until all its energy is deposited in the calorimeter. These interactions usually lead to the creation of secondary particles. In this way, a shower cascade is created in the calorimeter.

The basic mechanisms of the shower development and the most important influences on the energy resolution are discussed below following the detailed discussions in~\cite{Wigmans2000}.

\subsection{Electromagnetic Showers}
\label{sec:SiD_energy_emShowers}

The dominant electromagnetic processes that lead to energy loss of a charged particle that passes through matter are either ionization of the nuclei in the passed material or creation of Bremsstrahlung photons for high particle energies. The transition of these two regimes is characterized by the critical energy of the particle $\epsilon_\mr{c}$ at which the energy loss of these two processes is equivalent. $\epsilon_\mr{c}$ depends both on the electron density in the absorber and the mass of the particle
\begin{equation}
 \epsilon_\mr{c} \propto \frac{m^2}{Z},
\end{equation}
where $Z$ is the atomic number of the material.
The critical energy for electrons is well below \unit[100]{MeV} in most materials, while for muons the critical energy is typically of the order of several hundred GeV. Coulomb scattering, introduced in \cref{sec:SiD_momentum_multipleScattering}, does affect the particle direction and thus contributes to the lateral expansion of the shower, but does not contribute to the energy loss.

The mean energy loss of charged particles heavier than electrons through ionization of the surrounding material can be approximated by the Bethe formula~\cite{PDG}
\begin{equation}
 \frac{dE}{dx} = Kz^2\frac{A}{Z}\frac{1}{\beta^2}\left(\frac{1}{2}\ln\frac{2 m_e c^2 \beta^2 \gamma^2 E_\mr{max}}{I^2} - \beta^2 - \frac{\delta(\beta \gamma)}{2}\right),
\end{equation}
where $A$ is the atomic number of the absorber material, $I$ is the mean excitation energy, $E_\mr{max}$ is maximum transferred energy for each interaction and $\delta(\beta \gamma)$ is a density dependent correction factor. With the electron mass $m_e$ and the classical electron radius $r_e = \dfrac{e^2}{4 \pi \epsilon_0 m_e c^2}$ the constant $K$ is given by $K = 4 \pi N_A r_e^2 m_e c^2$. The Bethe formula is a good approximation in the range $0.1 < \beta \gamma < 1000$. Particles with an energy loss around the minimum of the Bethe formula are called \acp{MIP}. The energy loss of a \ac{MIP} is described by a Landau distribution~\cite{Landau:1944if} which is asymmetric and has a long tail towards high energies. If the energy transfer in an ionization process is high, the resulting free electron is called a $\delta$ electron. Due to its high energy it will lead to more ionization along its path.

The dominating interactions of photons with matter are the photoelectric effect for very low energies (hundreds of GeV and less), Compton scattering for intermediate energies and \epem-pair production for high energies (several MeV and more). For positrons, the annihilation with an electron is another process of energy deposition and leads to the creation of a secondary photon. 

If a highly energetic particle enters material, its energy is successively reduced by the processes listed above until it is eventually absorbed. The secondary particles that are created, i.e. photons and \epem-pairs, also lose their energies through these processes and produce more secondary particles such that the energy of the initial particle is distributed over a shower of secondary particles. The shower builds up to a maximum until the average energy of the secondary particles drops below the energies where pair production or Bremsstrahlung are dominating, such that the number of particles does not multiply further.

The longitudinal shower development can be described in a material independent way by the radiation length \radlen. The radiation length is defined as the average pathlength in an absorber material after which the initial energy of an electron or positron is reduced by approximately 63.2\% ($E = E_0/e$). The radiation length of any given material can be approximated by~\cite{PDG}
\begin{equation}
 \radlen = \frac{716.4\ A}{Z(Z+1)\ln(287/\sqrt{Z})}\frac{\mr{g}}{\mr{cm}^2}.
\end{equation}
In the limit of very high energies, the radiation length also describes the mean free path length of photons, which can be approximated as $\dfrac{9}{7}\radlen$.

The lateral shower size can be characterized similarly in a material independent way. The Moli\`ere radius, which is defined as the radius of the cylinder around the shower axis that contains on average 90\% of the shower energy, is given by~\cite{Wigmans2000}
\begin{equation}
 \rho_\mr{M} = m_e\ c^2 \sqrt{\frac{4\pi}{\alpha}} \frac{\radlen}{\epsilon_\mr{c}} \approx \unit[21.2]{MeV} \frac{\radlen}{\epsilon_\mr{c}}.
\end{equation}

\subsection{Hadronic Showers}
\label{sec:SiD_energy_hadShowers}

Hadrons will lose their energy in strong interactions with the nuclei of the absorber material, when passing through matter. Charged hadrons also lose some of their energy through ionization or pair production, as explained above, but will eventually enter a nuclear interaction.

The inelastic scattering of a highly energetic hadron with a nucleus usually leads to spallation of the nucleus. In the first stage of the nuclear reaction, a cascade of intranuclear reactions leads to the release of high energetic mesons and nucleons in direction of the incident particle. In the slower evaporation phase of the spallation, low energetic nucleons, \PGa-particles and photons are released isotropically from the excited nucleus. The fraction of a hadronic shower energy that is lost in the binding energy through the release of fragments of a nucleus is intrinsically invisible.

The spallation protons lose a large part of their energy through ionization. The neutrons on the other hand can lose their energy only through nuclear interactions that can produce secondary \PGa-particles and photons. For kinetic energies below \unit[1]{MeV}, elastic scattering is the dominant interaction for neutrons. The maximum transferred energy in these elastic scattering processes is proportional to $1/A$. This is the reason that an active material that contains hydrogen is especially sensitive to the neutron component of a hadronic shower.

A hadronic shower always contains also an electromagnetic component. This is due to the creation of \PGpz and \PGh-particles, which both decay into two photons. The average fraction of the shower energy that is deposited via electromagnetic processes depends on the absorber material as well as the energy and the type of the initial hadron. Proton showers do not produce $\PGpz$s in the initial hadronic interaction and thus have a considerable smaller electromagnetic component.

The longitudinal development of hadronic showers is characterized by the mean free path length of a hadron in a material, also referred to as nuclear interaction length \lambdaint. In addition to the absorber material, the interaction length depends on the particle type and can be up to 50\% longer for pions than for protons. In this thesis \lambdaint always refers to the interaction length of a proton.

\subsection{Energy Resolution}
\label{sec:SiD_energy_resolution}
The resolution of a calorimeter is influenced by a large number of factors. These factors depend differently on the particle energy. Since they are mostly uncorrelated the energy resolution can be parametrized as
\begin{equation}
 \frac{\sigma(E)}{E} = \frac{s}{\sqrt{E}} \oplus \frac{n}{E} \oplus c,
\label{eq:energyResolution}
\end{equation}
where $s$ is the \emph{sampling} term, $n$ is the \emph{noise} term and $c$ is the \emph{constant} term. The effects contributing to these three terms are briefly discussed below.

\subsubsection{Sampling Term}
\label{sec:SiD_energy_sampling}

The energy that is measured in a calorimeter is deposited discretely, e.g. in ionization electrons or \v{C}erenkov photons. The number of energy depositions $N$ is proportional to the energy, $N = kE$. Ideally, the calorimeter response is also proportional to number of individual energy deposits, such that the only uncertainty originates from the Poisson fluctuations of the number of created energy deposits. This leads to the stochastic uncertainty in the energy resolution
\begin{align}
 \frac{\sigma(N)}{N} &= \frac{1}{\sqrt{N}} \\
 \Rightarrow \frac{\sigma(E)}{E}\bigg|^\mr{sampling} &= \frac{a}{\sqrt{E}},
\label{Eq:Energy_stochasticTerm}
\end{align}
where $a$ depends on $k$ convolved with the signal collection efficiency of the detector.

In case of a sampling calorimeter only a fraction of the deposited energy is directly measured. The average measured energy in an active layer is proportional to the total deposited energy, while each individual energy deposition fluctuates following the Poisson statistics. This introduces another uncertainty described by \cref{Eq:Energy_stochasticTerm}, where the factor $a$ depends by the ratio of material thicknesses between active and passive layers, the sampling fraction.

In addition to these two effects which apply to electromagnetic and hadronic showers, the uncertainty due to the fluctuation of the fraction of invisible energy, i.e. energy lost in binding energy, in hadronic showers also scales with $\sigma(E)/E \propto 1/\sqrt{E}$. Similarly, the fluctuation of the electromagnetic shower content in a hadronic shower fluctuates from event to event and introduces another uncertainty only observed in hadronic showers. This uncertainty scales with $\sigma(E)/E \propto E^{-j}$, where $j$ is typically less than 0.5. The value of $j$ of a given calorimeter depends on the different response of the calorimeter to electromagnetic and hadronic energy deposits. The deterioration of the resolution due to this effect can be avoided if the calorimeter is compensating, which is discussed in \cref{sec:SiD_energy_compensation}.

Due to these additional effects, the energy resolution of hadronic showers is necessarily worse than the resolution for electromagnetic showers. In addition, the sampling fluctuations in hadronic showers are much larger than in electromagnetic showers since the energy deposited through ionization by charged hadrons as well as slow spallation protons is usually very different from the minimum ionization energy.

\subsubsection{Noise Term}
\label{sec:SiD_energy_noise}

Electronic noise inevitably adds some energy to the energy collected from the shower. This amount of noise depends on the detector technology, thresholds, amplification factors and the number of channels used to calculate the energy sum of the shower. It is independent of the shower energy and thus results in an uncertainty on the energy measurement of
\begin{equation}
 \frac{\sigma(E)}{E}\bigg|^\mr{noise} = \frac{n}{E}.
\label{Eq:Energy_noiseTerm}
\end{equation}

\subsubsection{Constant Term}
\label{sec:SiD_energy_constant}

If the shower is not fully contained in the calorimeter only a part of the shower energy is measured. The amount of leakage depends on the particle energy, which fluctuates strongly for each event, since it strongly depends on the position of the first hard interaction as well as the random shower development. The leakage thus leads to an energy dependent uncertainty that translates to a constant term in the relative energy resolution
\begin{equation}
 \frac{\sigma(E)}{E}\bigg|^\mr{constant} = c.
\label{Eq:Energy_constantTerm}
\end{equation}

\subsubsection{Compensation}
\label{sec:SiD_energy_compensation}

As described in \cref{sec:SiD_energy_sampling}, a different response to electromagnetic and hadronic showers in a calorimeter will lead to an additional uncertainty in the energy measurement. This can be avoided in two ways. If the calorimeter is designed in a way that these responses are identical it is called a compensating calorimeter. This can be tested for example by comparing the response of the calorimeter to electron and pion showers. 
The effect can also be reduced by applying different weighting factors depending on the particle type, e.g. proton or pion. An effective correction can only be achieved by determining the electromagnetic shower content on an event by event basis, since the electromagnetic shower content is fluctuating.

\subsection{Digital Calorimetry}
\label{sec:SiD_energy_digital}
Since most of the energy of a shower is deposited through ionization which are roughly of similar energy one can also estimate the shower energy by counting these energy deposits, referred to as digital calorimetry. This is only possible if the cells in the calorimeter are small compared to the typical shower size. This technology significantly reduces the amount of data required to read out from each calorimeter cell, which is an essential prerequisite for having very small cell sizes that are read out individually. The option of a digital \ac{HCal} has been first proposed for a \acs{TESLA} detector assuming $\unit[1\times1]{cm^2}$ cells~\cite{Behnke:2001qq}. 

This technology can be refined by using one or more additional thresholds to separate between high and low energetic deposits. This is referred to as semi-digital calorimetry~\cite{Laktineh:2009zz}.

\subsection{Linearity}
\label{sec:SiD_energy_linearity}

The energy measurement of a calorimeter is only useful if its response is proportional to the deposited energy. Saturation effects can lead to a non-linear behavior at high energies and are more common in electromagnetic showers which deposit the energy in a smaller volume. Energy dependent weighting factors can be used to correct for this and can improve the linearity.

\subsection{Jet Energy Measurement}
\label{sec:SiD_energy_jets}
The measurement of the energy of a jet, which consists of several particles that enter the calorimeter in a narrow region, strongly depends on the type of particles in the jet. This composition fluctuates strongly on an event by event basis due the randomness of the hadronization of the original quark(s). If the jet can not be resolved into individual calorimeter clusters for the individual particles in the jet, the achievable energy resolution is necessarily worse than the resolution for a single hadronic particle, since an additional uncertainty on the electromagnetic content of the shower is introduced.

\section{Detector Requirements for a \acs{CLIC} Experiment}
\label{sec:SiD_requirements}

The design of any future experiment is directly motivated by the requirements imposed by its physics goals. The physics case for a future linear collider experiment is the precision measurement of various observables of the \ac{SM} and investigations of possible extensions of the \ac{SM} or deviations from it, as discussed in \cref{cha:SM}. The implications for the detector design of a future linear collider experiment have been formulated for \acs{TESLA}~\cite{AguilarSaavedra:2001rg}, the \ac{ILC}~\cite{Brau:2007zza,ILD:2009,Aihara:2009ad} and recently for \ac{CLIC}~\cite{cdrvol2}. Despite differences in the available center-of-mass energy and thus in the physics reach of the experiments at these colliders, the resulting detector requirements are rather similar. The requirements on the basic observables are summarized from~\cite{cdrvol2} together with their physics motivation.

\begin{description}
 \item [Track Momentum Resolution] The precise measurement of leptonic final states requires excellent momentum resolution, $\sigma(\pT)/\pT^2$, of the order of \unit[$2\times10^{-5}$]{GeV$^{-1}$} or better. This requirement ensues from the Higgs boson mass measurement through Higgsstrahlung process discussed \cref{sec:SM_higgsProduction}. Similar requirements are imposed by the measurement of the cross section times branching ratio of Higgs into two muons, as discussed in \cref{sec:MomentumResolution}. 
 
 \item[Jet Energy Resolution] The jet energy resolution $\sigma(E)/E$ has to be around 3.5\%-5\% in the range from \unit[1]{TeV} to \unit[50]{GeV}, to precisely identify hadronic final states. Many \ac{BSM} scenarios predict multi-jet final states, like the chargino pair production process, $\epem \to \chargino{+}\chargino{-} \to \ww\neutralino{1}\neutralino{1}$, in supersymmetric models. These require the identification and separation of hadronic decays of \PW, \PZ and \PSh bosons to supress background processes~\cite{LCD:2011-037}. 

 \item[Impact Parameter Resolution] The precise reconstruction of secondary vertices requires an excellent impact parameter resolution and is necessary for efficient flavor tagging. This is for example important for identifying $\PQb$ jets like in the measurement of the Higgs coupling to \PQb quarks~\cite{LCD:2011-036} or the measurement of the triple Higgs coupling, $\epem \to \PSh\nuenuebar \to \PSh\PSh\nuenuebar$. Using the parametrization for the $d_0$ resolution given in \cref{eq:resolutionParametrization_d0} the desired resolution is $a \leq \unit[5]{\micron}$ for the point resolution and $b \leq \unit[15]{\micron \cdot GeV}$ for the term depending on the material budget.
\end{description}

In addition, the particle identification capabilities, especially for leptons, should be excellent to correctly identify event topologies. This goes together with the required excellent coverage of the detector also in the forward region to correctly identify event topologies with particles created at low polar angles.

\section{The Particle Flow Paradigm}
\label{sec:SiD_particleFlow}

Traditionally the jet energy is measured as the energy sum of the energy deposits in the calorimeters in a region around the jet axis. The desired jet energy resolution of 3.5\% to 5\% (see \cref{sec:SiD_requirements}) translates into a energy resolution of $\sigma(E)/E \approx 30\%/\sqrt{E/\mr{GeV}}$~\cite{Thomson:2009rp}, which is significantly better than what has been achieved in previous experiments using sampling calorimeters and thus novel approaches are required to reach this energy resolution.

One solution to this problem might be dual-readout calorimetry~\cite{Akchurin:2005eu,Akchurin:2005an} which aims for ultimate hadronic energy resolution. By individually measuring the response to the electromagnetic and hadronic shower components (see \cref{sec:SiD_energy_hadShowers}) the largest uncertainty in the energy measurement of hadronic showers and jets is removed. In this case, a compensating calorimeter, as explained in \cref{sec:SiD_energy_compensation}, is not required and the calorimeter can be optimized for the best intrinsic sampling resolution.

The second approach is particle flow~\cite{Brient:2002gh,Morgunov:2001cd}, which aims at identifying the showers of the individual particles that constitute the jet and measure their individual four momenta.
This idea has evolved from the concept of energy flow which has for example been used in the \acs{ALEPH}~\cite{Buskulic:1994wz} experiment to successfully improve its jet energy resolution.

The average particle content in a jet has been measured at \acs{LEP}~\cite{Knowles:1997dk} and consists approximately of 62\% charged particles, 27\% photons, 10\% neutral hadrons and 1.5\% neutrinos. If the showers of these particles can be resolved, the total jet energy resolution can be significantly improved. Especially the energy of charged particles can be determined from the track momentum measurement with far greater accuracy than achievable in the calorimeters. For photons the energy resolution in the \ac{ECal} applies, which leaves only a small fraction of the jet energy, the neutral hadrons, that has to be measured in the \ac{HCal} directly.

The main goal of particle flow is thus not the actual energy measurement but the correct identification of the individual showers. This requires a highly segmented calorimeter (laterally and longitudinally) as well as a sophisticated pattern recognition. The largest source of uncertainty in a particle flow energy reconstruction comes from confusion. For example, if a shower from a neutral particle is not resolved from that of a neighboring charged particle and the energy of the charged particle determined from its momentum is used, the total reconstructed energy will be too low. On the other hand, if a subcluster of a charged particle shower is identified as an individual neutral shower, the total energy will be overestimated. This shows that, despite not being used for most of the direct energy measurement, a good energy resolution of the calorimeter is still important to identify discrepancies between the energy determined from the track momentum and the energy of the corresponding calorimeter shower. Significant leakage will also contribute to the confusion if the algorithm compensates longitudinal energy loss by identifying neighboring neutral clusters as parts of a charged cluster instead.

It should be noted that $\sigma(\pT)/\pT$ scales with $\pT$, as shown in \cref{eq:resolutionParametrization_pT}, while $\sigma(E)/E$ scales mostly with $\sqrt{E}$. The energy measurement of an energetic charged particles can thus ultimately be more precise in the calorimeters if the contribution of the constant term, i.e. leakage, is small enough. It has been demonstrated in simulation that even at typical jet energies at a \ac{CLIC} experiment, particle flow will perform better than a pure calorimetric energy measurement~\cite{Thomson:2009rp}.

In addition to the requirements on the segmentation of the calorimeters, which depend mostly on $\rho_\mr{M}$, \radlen and \lambdaint of the absorber materials used, the inner radius of the calorimeters and the magnetic field strength are of importance for particle flow. Charged particles will be deflected from their original trajectory by the magnetic field, and will thus be separated from the neutral particles in a jet and reduce the possible confusion. The strength of the magnetic field and the path length, i.e. the inner radius of the \ac{ECal}, determine the typical separation and can thus also affect the choice of the lateral segmentation. It has been shown that for ultimate particle flow performance, a larger radius is preferable over a larger magnetic field~\cite{Thomson:2009rp}. This has been driving the design decisions of the \ac{ILD} concept. The \ac{SiD} concept, which aims at a more cost effective solution, deviates from that and prefers using a higher magnetic field with a smaller inner radius of the calorimeters. The desired lateral segmentation of the calorimeter has been determined in simulation studies to be \unit[5]{mm} for the \ac{ECal} and \unit[30]{mm} for the \ac{HCal}~\cite{Thomson:2009rp}.

A definitive test of the particle flow paradigm can only be achieved in a full sized experiment since both momentum and energy measurements have to be performed in jet events. Nevertheless, there are efforts to test parts of the concept using test beam data. For example, two events taken in a beam test can be merged offline by displacing them in the calorimeter and adding their signals. This allows to test the effectiveness of a \ac{PFA} using real data as demonstrated in~\cite{Collaboration:2011ha}. The results indicate that there is no noticeable difference in the performance of \pandora using simulated events or test beam data.

Particle flow event reconstruction is also used in the \acs{CMS} experiment, which was not designed for that purpose, to improve the jet energy resolution significantly in several event topologies~\cite{Beaudette:2010zz}.

\chapter{The \clicsid Detector Model}
\label{cha:SiD}

The \acl{SiD}~\cite{Aihara:2009ad} is a concept for an experiment at a future \epem collider. It was originally designed for the \ac{ILC}~\cite{Brau:2007zza}.
It is an example of a typical $4\pi$ multi-purpose detector for high energy collider experiments with highly granular calorimeters according to the particle flow paradigm introduced in \cref{sec:SiD_particleFlow}. Despite originally being designed for a center-of-mass energy of \unit[500]{GeV} the concept is an excellent starting point for a detector concept used at collision energies of several TeV. Nevertheless, we have introduced several modifications to the detector concept to account for the different experimental conditions at \ac{CLIC}.
\begin{itemize}
 \item The vertex detector layout had to be adapted to avoid high occupancies due to beam induced background.
 \item The forward region had to be changed completely due to the placement of the final focusing quadrupole and the requirements on its stability.
 \item The hadronic calorimeters had to be made significantly deeper to avoid leakage due to the higher jet energies.
\end{itemize}
While the first two points are specific to the \ac{CLIC} accelerator, the last point is purely due to the higher center-of-mass energy of the collisions compared to the default parameters of the \ac{ILC}.


This chapter describes the \clicsid model as it is implemented in the \geant simulations used in the \ac{CLIC} \ac{CDR} studies and the studies presented in this thesis. Although the simulation model contains some details, as for example cabling and support structures in the tracking detectors, it is still a simplified model. For example, an engineering model has been developed of \clicsid for machine-detector integration studies~\cite{LCD:2011-011}. In that model, the size of the yoke has been increased to provide sufficient shielding for the magnetic field, which thus yields a more realistic estimate of the overall dimensions.

\begin{figure}[htpb]
  \centering
  \begin{subfigure}[b]{0.49\textwidth}
   \centering
   \begin{tikzpicture} [scale=0.528]
     \draw[scale=1.0] (-0.2,0.0) node[anchor=south east]{\includegraphics[width=\textwidth, trim=41 364 241 165 ,clip=true]{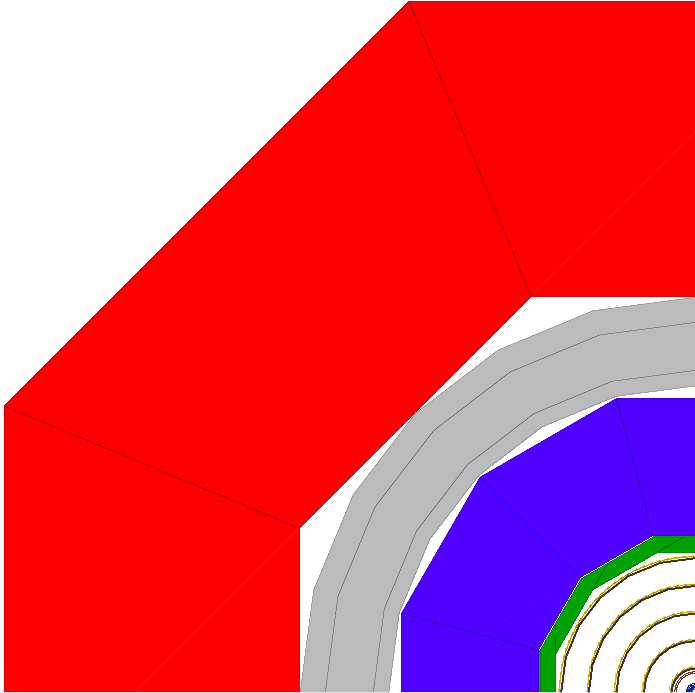}};
     \scriptsize
     \draw[scale=1.0] (-0.46,-0.4)  node[below left]  {0}    -- (-0.46,0.0) ; 
     \draw[scale=1.0] (-3.422,-0.4)  node[below right] {1265} -- (-3.422,0.0) ; 
     \draw[scale=1.0] (-3.783,-0.4)  node[below left]  {1419} -- (-3.783,0.0) ; 
     \draw[scale=1.0] (-6.947,-0.4)  node[below] {2770} -- (-6.947,0.0) ; 
     \draw[scale=1.0] (-8.847,-0.4)  node[below]  {3581} -- (-8.847,0.0) ; 
     \draw[scale=1.0] (-15.10,-0.4) node[below right] {6251} -- (-15.10,0.0) ; 
   \end{tikzpicture}
   \caption{$xy$}\label{fig:sidoverviewXY}
  \end{subfigure}
  \begin{subfigure}[b]{0.49\textwidth}
   \centering
   \begin{tikzpicture} [scale=0.528]
     \draw[scale=1.0] (0.2,0.0) node[anchor=south west]{\includegraphics[width=\textwidth, trim=241 364 41 165 ,clip=true]{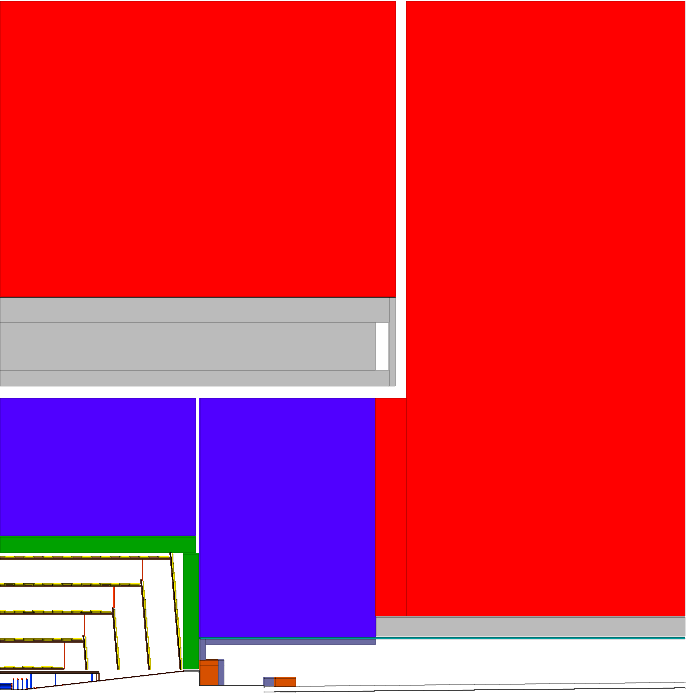}};
     \scriptsize
     \draw[scale=1.0] (0.46,-0.4)  node[below right] {0}    -- (0.46,0.0) ; 
     \draw[scale=1.0] (4.363,-0.4)  node[below left]  {1657} -- (4.363,0.0) ; 
     \draw[scale=1.0] (4.712,-0.4)  node[below right] {1805} -- (4.712,0.0) ; 
     \draw[scale=1.0] (8.458,-0.4)  node[below left]  {3395} -- (8.458,0.0) ; 
     \draw[scale=1.0] (9.118,-0.4)  node[below right] {3675} -- (9.118,0.0) ; 
     \draw[scale=1.0] (14.97,-0.4) node[below left]  {6195} -- (14.97,0.0) ; 
     \draw[scale=1.0] (4.5,11.5) node {\LARGE{Yoke}};
     \draw[scale=1.0] (4.5,7.6) node {\Large{Coil}};
     \draw[scale=1.0] (2.5,5.0) node {\Large{HCal}};
     \draw[scale=1.0] (2.5,3.4) node {\tiny{ECal}};
     \draw[scale=1.0] (8.7,4.2) node[rotate=270] {\small{Yoke Plug}};
     \draw[scale=1.0] (11.7,1.7) node {\tiny{Anti-Solenoid}};
   \end{tikzpicture}
   \caption{$zx$}\label{fig:sidoverviewZX}
  \end{subfigure}
  \caption[One quadrant of the \clicsidcdr detector model.]{One quadrant of the \clicsidcdr detector model in the $xy$-plane \protect\subref{fig:sidoverviewXY} and in the $zx$-plane \protect\subref{fig:sidoverviewZX}. Values are given in millimeters.}
  \label{fig:sidoverview}
\end{figure}

With an overall length of \unit[12.39]{m} and a total height of \unit[12.50]{m}, the detector is rather compact and smaller than for example \acs{CMS}. An overview of the parameters of the main detector components is given in \cref{tab:sidoverview}. Figure~\ref{fig:sidoverview} shows one quadrant of the detector model in the $xy$-plane as well as in the $xz$-plane.

\begin{table}[tphb]
  \centering
  \caption[Parameters for the main elements of the \clicsidcdr detector
    model.]{Parameters for the main elements of the \clicsidcdr detector
    model. The parameters $z_{\mathrm{min/max}}$ are the beginning and end in one half of the
    detector. While $r_{\mathrm{min/max}}$ are the radii of the inscribing
    circles for the polygons given in the last column. The polygon column gives the
    number of corners for polygonal shaped detector elements. All other elements are cylindrical.}
  \label{tab:sidoverview}
  \begin{threeparttable}
  \begin{tabular}{l  *4{R{4}{0}} *1{R{2}{0}} }
    \toprule
                     & \tabt{$z\mm{min}$}   & \tabt{$z\mm{max}$} & \tabt{$r\mm{min}$} & \tabt{$r\mm{max}$} & \tabt{Polygon} \\\midrule
ECal Barrel          & 0                    & 1765               & 1265               & 1403.5             & 12             \\
HCal Barrel          & 0                    & 1765               & 1419               & 2656.5             & 12             \\
Coil                 & 0                    & 3575               & 2770.21            & 3571.21            & \tabempty      \\
Yoke Barrel          & 0                    & 3575               & 3581.21            & 6251.21            & 8              \\
ECal Endcap          & 1657                 & 1795.5             & 210                & 1250               & 12             \\
HCal Endcap          & 1805                 & 3395               & 500                & 2656.5             & 12             \\
Yoke Plug            & 3395                 & 3675               & 690                & 2656.5             & 12             \\
Yoke Endcap          & 3675                 & 6195               & 690                & 6251.21            & 8              \\
LumiCal              & 1805                 & 1975.65            & 64                 & 240                & \tabempty      \\
BeamCal\tnote{A}     & 2485.65              & 2671.15            & 0                  & 130                & \tabempty      \\
    \bottomrule
  \end{tabular}
  \begin{tablenotes}
    \item[A] The BeamCal is centered around the detector axis. The holes for the incoming and outgoing beam pipes, which are not in the center, are described in \cref{sec:SiD_model_beamCal}.
  \end{tablenotes}
  \end{threeparttable}
\end{table}

The tracking system is described in \cref{sec:SiD_model_tracking}. It consists of the vertex detector with five barrel layers in the central region and four disk layers of silicon pixel detectors in the endcaps. The vertex detector is surrounded by the main tracking system, consisting of five barrel layers of silicon strip detectors and four disks of silicon stereo-strip detectors on each side. The tracking system is completed by three additional silicon pixel disks in the forward region. A solenoidal field of \unit[5]{T} is provided by a superconducting coil outside of the calorimeters and is described in \cref{sec:SiD_model_coil}.

Between the coil and the tracking systems are the calorimeters which are discussed in \cref{sec:SiD_model_calorimetry}. The silicon-tungsten \ac{ECal} and the \ac{HCal} are highly granular sampling calorimeters following the particle flow paradigm. The \ac{HCal} is instrumented with scintillator plates and steel is used as the absorber material in the endcaps. Tungsten is used as the absorber material in the \ac{HCal} barrel in order to keep the size of the coil feasible.

The iron yoke, which surrounds the whole detector, is instrumented with \acp{RPC}. It helps with the identification of muons and serves as a tail-catcher for the \ac{HCal} and is described in \cref{sec:SiD_model_yoke}.

\Cref{sec:SiD_model_forward} describes the far forward region of the detector. There, the \ac{LumiCal} and the \ac{BeamCal} complement the coverage for electromagnetic showers provided by the \ac{ECal}. The design of the forward region is driven by the requirements for the \ac{QD0}. The \ac{QD0} is placed behind the \ac{BeamCal} and has to be supported by a large support tube from the tunnel walls in order to achieve the required stability.

The occupancies due to the machine-induced backgrounds in the tracking detectors are discussed in \cref{sec:SiD_occupancies}.

\section{Tracking System}
\label{sec:SiD_model_tracking}

\begin{figure}[htpb]
  \centering
  \includegraphics[width=.7\linewidth]{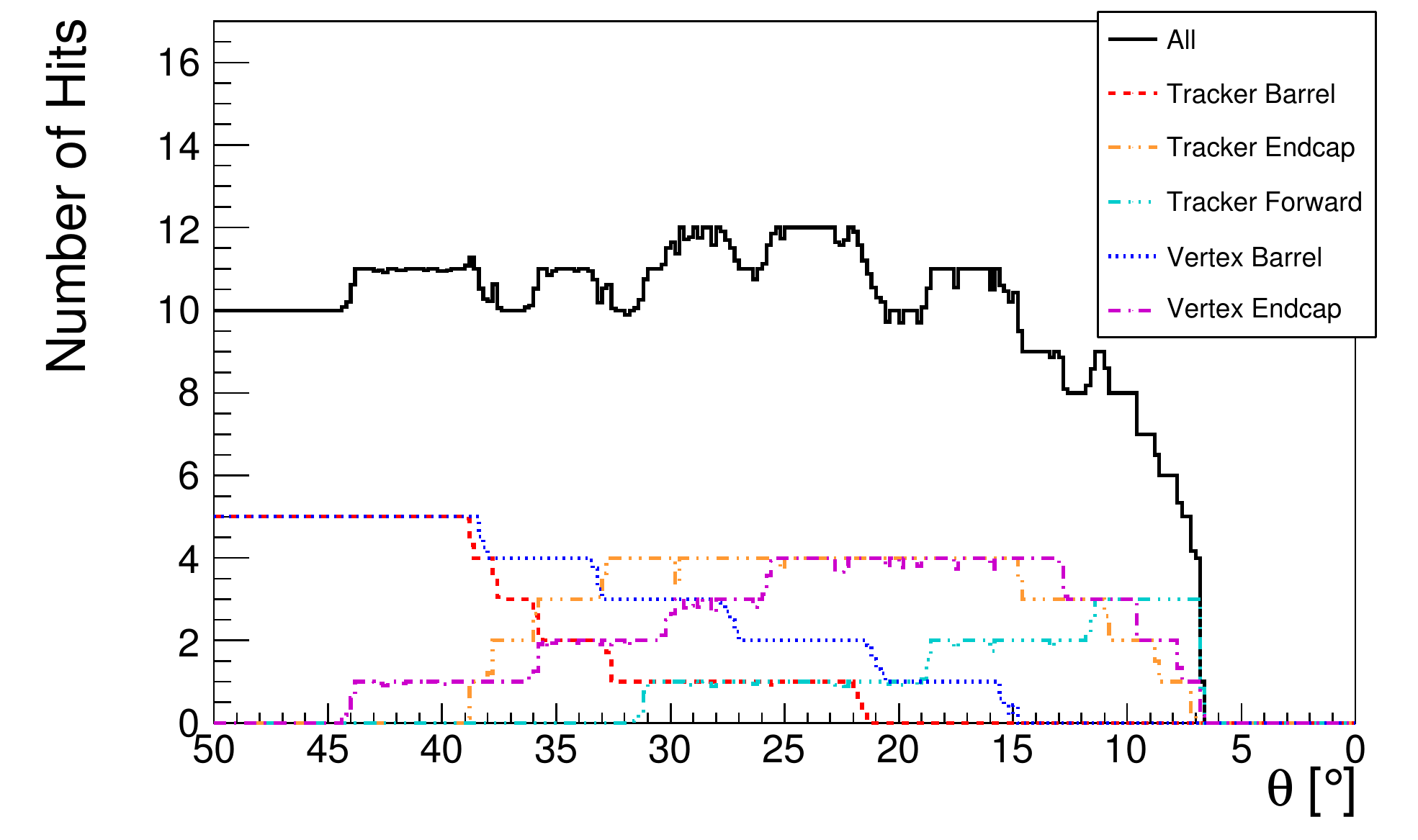}
  \caption[Coverage of the tracking systems with respect to the polar angle $\theta$.]{Coverage of the tracking systems with respect to the polar angle $\theta$. Shown is the average number of hits created by a \unit[500]{GeV} muon in full simulation. At least six hits are measured for all 
  tracks with a polar angle down to about 8\degrees.}
  \label{fig:trackingcoverage}
\end{figure}

\begin{figure}[htpb]
  \centering
  \begin{subfigure}[htbp]{0.415\textwidth}
   \centering
   \begin{tikzpicture} [scale=0.528]
     \draw[scale=1.0] (-0.2,0.0) node[anchor=south east]{\includegraphics[width=\textwidth, trim=65 364 241 165 ,clip=true]{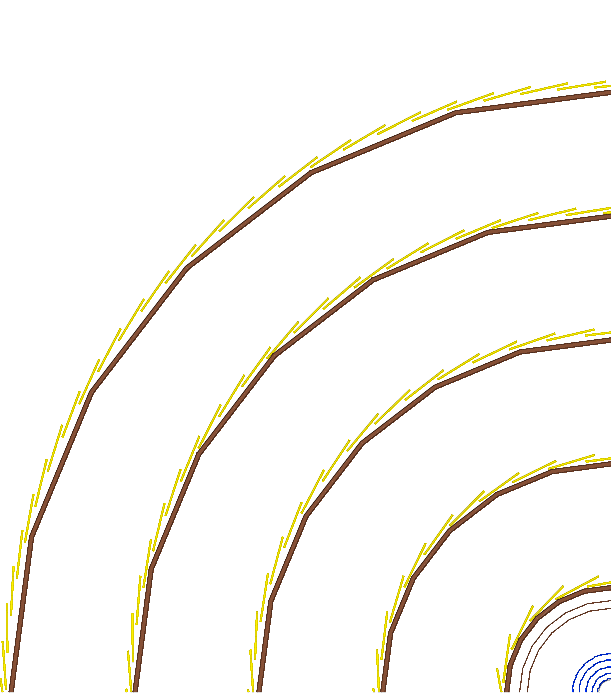}};
     \scriptsize
     \draw[scale=1.0] (-0.47,-0.4)  node[below left]  {0}    -- (-0.47,0.0) ; 
     \draw[scale=1.0] (-2.552,-0.4)  node[below] {206} -- (-2.552,0.0) ; 
     \draw[scale=1.0] (-5.079,-0.4)  node[below]  {456} -- (-5.079,0.0) ; 
     \draw[scale=1.0] (-7.606,-0.4)  node[below] {706} -- (-7.606,0.0) ; 
     \draw[scale=1.0] (-10.133,-0.4)  node[below]  {956} -- (-10.133,0.0) ; 
     \draw[scale=1.0] (-12.66,-0.4) node[below right] {1206} -- (-12.66,0.0) ; 
   \end{tikzpicture}
   \caption{$xy$}\label{fig:sidtrackerXY}
  \end{subfigure}
  \begin{subfigure}[htbp]{0.568\textwidth}
   \centering
   \begin{tikzpicture} [scale=0.528]
     \draw[scale=1.0] (0.2,0.0) node[anchor=south west]{\includegraphics[width=\textwidth, trim=241 364 0 165 ,clip=true]{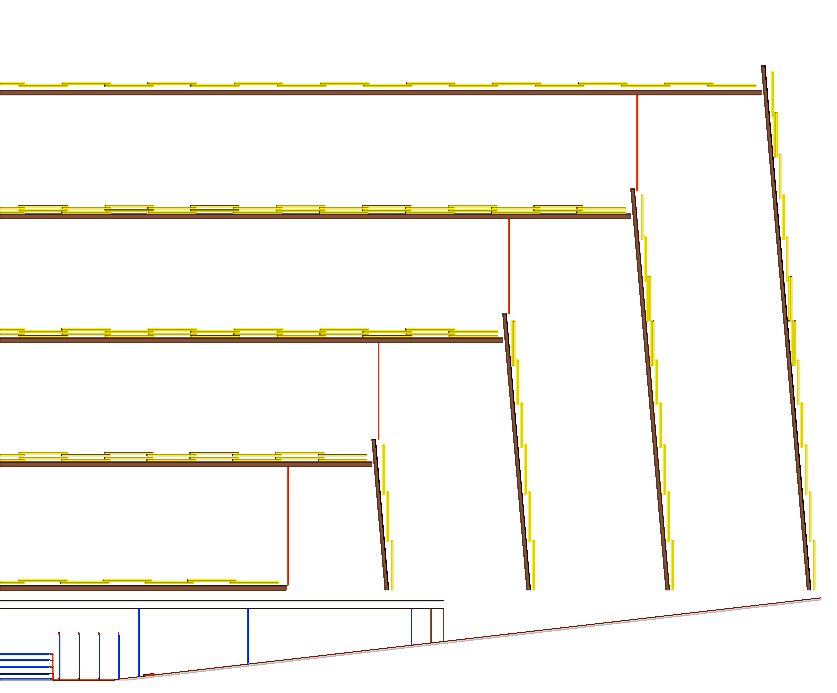}};
     \scriptsize
     \draw[scale=1.0] (0.46,-0.4)  node[below right] {0}    -- (0.46,0.0) ; 
     \draw[scale=1.0] (6.295,-0.4)  node[below]  {577} -- (6.295,0.0) ; 
     \draw[scale=1.0] (8.313,-0.4)  node[below] {777} -- (8.313,0.0) ; 
     \draw[scale=1.0] (11.207,-0.4)  node[below]  {1063} -- (11.207,0.0) ; 
     \draw[scale=1.0] (14.042,-0.4)  node[below] {1344} -- (14.042,0.0) ; 
     \draw[scale=1.0] (16.927,-0.4) node[below left]  {1629} -- (16.927,0.0) ; 

   \end{tikzpicture}
   \caption{$zx$}\label{fig:sidtrackerZX}
  \end{subfigure}
  \caption[Layout of the tracking system in \clicsid.]{Layout of the tracking system in \clicsid in the $xy$-plane \protect\subref{fig:sidtrackerXY} and in the $zx$-plane \protect\subref{fig:sidtrackerZX}.
  The main tracker modules are displayed in yellow and support structures are displayed in brown. A detailed view of the vertex detector can be found in \cref{fig:sidvertexXY} and \cref{fig:sidvxdendcap}. Values are given in millimeters.}
  \label{fig:sidtracker}
\end{figure}
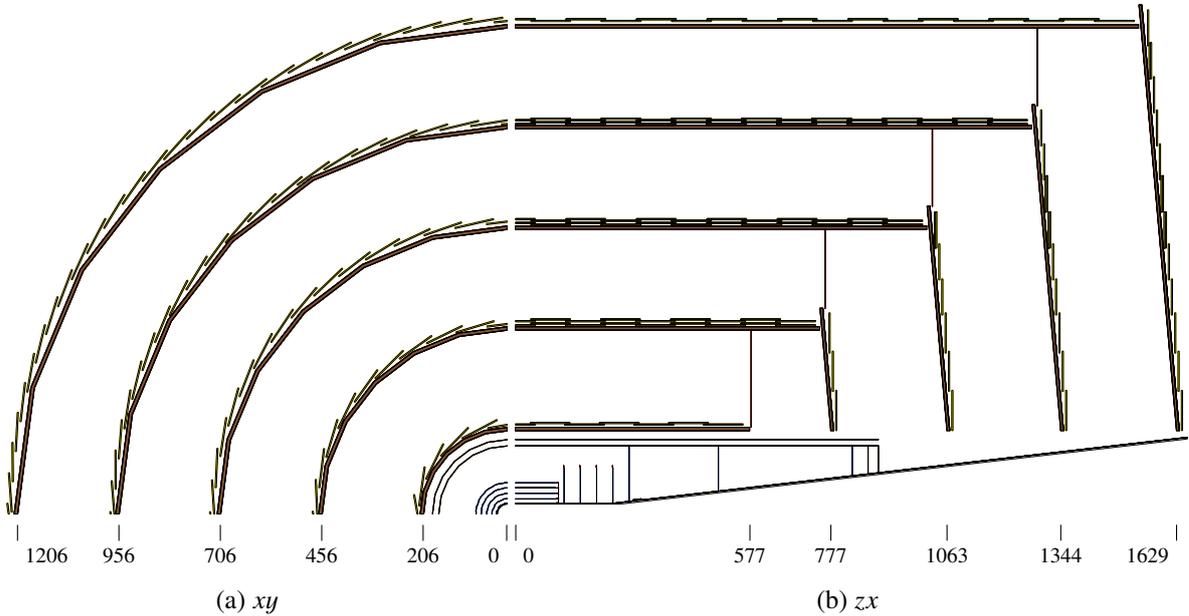

The all-silicon tracking system in \clicsid is designed to provide excellent point resolution combined with low material budget. The pixel detectors (see \cref{sec:SiD_model_vertex}) and the main tracking system (see \cref{sec:SiD_model_mainTracking}) form an integrated system that provides at least ten precisely measured points for all tracks
down to a polar angle of about 15\degrees and at least six measured points down to a polar angle of about 8\degrees, as shown in \cref{fig:trackingcoverage}. A cut through the tracking region of \clicsid is shown in \cref{fig:sidtracker}.

The layers of the tracking detectors are made up of individual modules including realistic overlaps. Beyond that, the modules are not segmented in the simulation model, instead, the segmentation into strips and pixels is applied during the digitization step of the reconstruction, as discussed in Section~\ref{sec:Software_trackerDigitizationSegmentation}. Nevertheless, the assumed pitches for the strip and pixel detectors are given here for completeness.

\subsection{The Vertex Detector}
\label{sec:SiD_model_vertex}

The layout of the vertex detector system has been changed significantly with respect to the default \ac{SiD} layout to account for the background from incoherent pairs. For example, the radius of the first layer has been increased from \unit[15]{mm} to \unit[27]{mm}. All other radii are changed accordingly. The length of the barrel section has been increased from \unit[62.5]{mm} to \unit[100]{mm} to provide a similar coverage in the polar angle. The disk detectors have been moved out accordingly and the distances between the disks, especially in case of the forward disks, has been modified to a more equidistant layout. A systematic fast simulation study of various vertex detector layouts including varying assumptions on the material budget can be found in~\cite{lcd:2011-DannheimVosLayoutOptNote}.

\subsubsection{Vertex Detector Barrel}
\label{sec:SiD_model_vertexBarrel}
The vertex barrel detector consists of five concentric layers. Each layer is made up by several modules as described in \cref{tab:sidvxdbarrel}. Each module consists of \unit[50]{\micron} of silicon (\unit[$\approx0.053\%$]{\radlen}) followed by \unit[130]{\micron} of carbon (\unit[$\approx0.061\%$]{\radlen}) as support. The silicon is segmented into \unit[20$\times$20]{\micron$^2$} pixels. Figure~\ref{fig:sidvertexXY} shows the arrangement of the layers and the modules in the vertex barrel detector.

The vertex detector for \ac{CLIC} will have to combine high spatial resolution with a low material budget. While these requirements are similar to those of an \ac{ILC} vertex detector, \ac{CLIC} will require also high time resolution to mitigate the high occupancy due to beam induced backgrounds. There is currently no pixel technology available that fulfills all requirements of the \ac{CLIC} vertex detector. In terms of material budget, the key issue will be the use of power pulsing, \ie switching off the detector between two bunch trains, to reduce the power consumption. In the current design no cooling except for air cooling is foreseen. The alternative would be  evaporative cooling using \ce{CO2}, which would increase the material budget due to the required pipes. For the pixel technologies multiple integrated pixel solutions are pursued, like for example designs based on the TimePix chip~\cite{Llopart:1066001}, the MIMOSA sensors~\cite{Turchetta:2001dy}, the CHRONOPIXEL sensors~\cite{Baltay:2011hk} or fully integrated solutions based on Silicon On Insulator~\cite{Ikeda:2007zza} or 3D~architectures~\cite{Parker:1996dx}. For a detailed discussion on the possible pixel technologies we refer to~\cite{cdrvol2}.

\begin{minipage}{\textwidth}
 \begin{minipage}[]{0.30\textwidth}
   \begin{tikzpicture} [scale=0.352]
     \draw[scale=1.0] (0.0,0.0) node[anchor=south west]{\includegraphics[width=\textwidth, trim=229.8 364 241 353.1 ,clip=true]{figures/geometry/clic_sid_cdr_vis_tracking_cut_xy-eps-converted-to.pdf}};
     \scriptsize
     \draw[scale=1.0] (0.556,-0.4)  node[below]  {77}    -- (0.556,0.0) ; 
     \draw[scale=1.0] (2.792,-0.4)  node[below]  {64}    -- (2.792,0.0) ; 
     \draw[scale=1.0] (5.028,-0.4)  node[below]  {51}    -- (5.028,0.0) ; 
     \draw[scale=1.0] (7.264,-0.4)  node[below]  {38}    -- (7.264,0.0) ; 
     \draw[scale=1.0] (9.156,-0.4)  node[below left]  {27}    -- (9.156,0.0) ; 
     \draw[scale=1.0] (9.5,-0.4)  node[below right]  {25}    -- (9.5,0.0) ; 
     \draw[scale=1.0] (13.8,-0.4)  node[below] {0} -- (13.8,0.0) ; 
  \end{tikzpicture}
  \figcaption[Layout of the vertex barrel detector in \clicsid in the $xy$-plane.]{Layout of the vertex barrel detector in the $xy$-plane. Values are given in millimeters.}
  \label{fig:sidvertexXY}
 \end{minipage}
 \hfill
 \begin{minipage}[b]{0.55\textwidth}
  \vspace{20pt}
  \centering
  \tabcaption[Parameters for the vertex detector barrel layers.]{Parameters for the vertex detector barrel layers. The number of
    modules $N$ in the layer, the mean radius $r$ of the layer, the half-length $z$ of the module and
    width $w$ of the module are given. Each module consists of a layer of \unit[50]{\micron} of silicon followed by \unit[130]{\micron} of carbon fiber.}
  \label{tab:sidvxdbarrel}
  \begin{tabular}{c c R{2}{1} R{2}{1}  R{2}{1}}\toprule
    Layer & $N$ & \tabt{$r\mm{}$} & \tabt{$z\mm{}$}       & \tabt{$w\mm{}$} \\\midrule
    1     & 18  & 27.0            & 98.5                  & 9.8             \\
    2     & 18  & 38.0            & 98.5                  & 13.8            \\
    3     & 24  & 51.0            & 98.5                  & 13.8            \\
    4     & 30  & 64.0            & 98.5                  & 13.8            \\
    5     & 36  & 77.0            & 98.5                  & 13.8            \\\bottomrule
  \end{tabular}
 \end{minipage}
\end{minipage}

\subsubsection{Vertex Detector Endcap and Forward Tracking Disks}
\label{sec:SiD_model_forwardTracking}

There are seven pixel disks covering the forward and far-forward region of the detector. The first four disks are close together and are
considered part of the vertex detector. The three forward tracking disks extend the coverage of the tracker endcap to the beam pipe. All pixel
disks consist of several trapezoidal modules as described in \cref{tab:sidvxdendcap}. Like in the vertex barrel each module consists of 
\unit[50]{\micron} of silicon (\unit[$\approx0.053\%$]{\radlen}) followed by \unit[130]{\micron} of carbon (\unit[$\approx0.061\%$]{\radlen}), with the silicon being segmented into \unit[20$\times$20]{\micron$^2$}
pixels. For the pixel technology the same applies as for the vertex barrel detector.

\begin{figure}[htpb]
  \centering
   \tikzstyle{line} = [draw]
   \begin{tikzpicture} [scale=0.528]
     \draw[scale=1.0] (0.0,0.0) node[anchor=south west]{\includegraphics[width=0.9\textwidth, trim=241 364 110 335 ,clip=true]{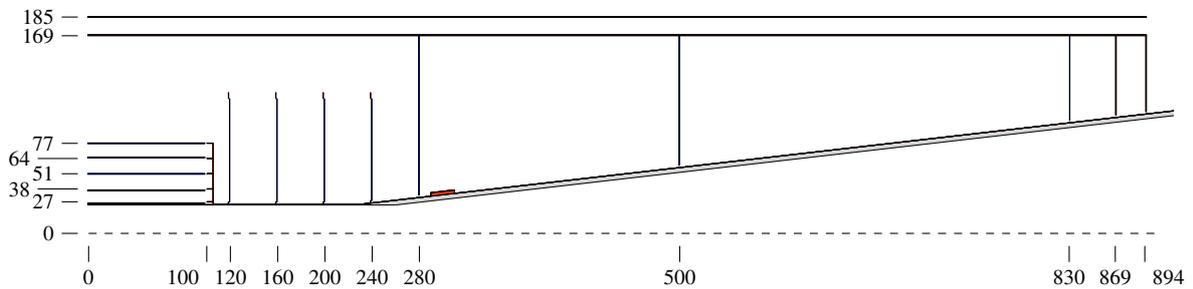}};
     \scriptsize
     \draw[scale=1.0] (-0.4,0.317)  node[left] {0}    -- (0.0,0.317) ; 
     \draw[scale=1.0] (-0.4,1.11)  node[left] {27}    -- (0.0,1.11) ; 
     \draw[scale=1.0] (-1.0,1.433)  node[left] {38}    -- (0.0,1.433) ; 
     \draw[scale=1.0] (-0.4,1.815)  node[left] {51}    -- (0.0,1.815) ; 
     \draw[scale=1.0] (-1.0,2.197)  node[left] {64}    -- (0.0,2.197) ; 
     \draw[scale=1.0] (-0.4,2.578)  node[left] {77}    -- (0.0,2.578) ; 
     \draw[scale=1.0] (-0.4,5.28)  node[left] {169}    -- (0.0,5.28) ; 
     \draw[scale=1.0] (-0.4,5.75)  node[left] {185}    -- (0.0,5.75) ; 

     \draw[scale=1.0] (0.27,-0.4)  node[below] {0}    -- (0.27,0.0) ; 
     \draw[scale=1.0] (3.214,-0.4)  node[below left]  {100} -- (3.214,0.0) ; 
     \draw[scale=1.0] (3.803,-0.4)  node[below]  {120} -- (3.803,0.0) ; 
     \draw[scale=1.0] (4.981,-0.4)  node[below]  {160} -- (4.981,0.0) ; 
     \draw[scale=1.0] (6.158,-0.4)  node[below]  {200} -- (6.158,0.0) ; 
     \draw[scale=1.0] (7.336,-0.4)  node[below]  {240} -- (7.336,0.0) ; 
     \draw[scale=1.0] (8.513,-0.4)  node[below]  {280} -- (8.513,0.0) ; 
     \draw[scale=1.0] (15.0,-0.4)  node[below]  {500} -- (15.0,0.0) ; 
     \draw[scale=1.0] (24.706,-0.4)  node[below]  {830} -- (24.706,0.0) ; 
     \draw[scale=1.0] (25.854,-0.4)  node[below]  {869} -- (25.854,0.0) ; 
     \draw[scale=1.0] (26.59,-0.4)  node[below right]  {894} -- (26.59,0.0) ; 

     \path[line,dashed] (0.27,0.317) -- (27.0,0.317);
   \end{tikzpicture}
  \caption[Layout of the vertex region in \clicsid in the $zy$-plane.]{Layout of the vertex region in \clicsid in the $zx$-plane. Shown are the vertex barrel layers, the vertex endcap disks and the forward tracking disks together with the vertex support, cabling and the central beam pipe. All values are given in millimeters.}
  \label{fig:sidvxdendcap}
\end{figure}

\begin{table}[htbp]
  \centering
  \caption[Parameters for the vertex endcap and forward tracking disks.]{Parameters for the vertex endcap and forward tracking disks. The number $N$ of trapezoidal
    modules, the inner radius $r_{\mathrm{in}}$, the outer radius $r_{\mathrm{out}}$, the
    inner width $w_{\mathrm{in}}$, the outer width $w_{\mathrm{out}}$
    and the position in $z$ of the modules are given. Each module consists of a layer of \unit[50]{\micron} of silicon followed by \unit[130]{\micron} of carbon fiber.}
\label{tab:sidvxdendcap}
  \begin{tabular}{c c R{2}{1} R{3}{1} R{1}{1} R{2}{1} R{3}{1} c c }\toprule
    Disk  & $N$ & \tabt{$r\mm{in}$} & \tabt{$r\mm{out}$} & \tabt{$w\mm{in}$} & \tabt{$w\mm{out}$} & \tabt{$z\mm{}$} \\\midrule
    1     & 16  & 27.0              & 115                & 10.81             & 45.149             & 120             \\
    2     & 16  & 27.0              & 115                & 10.81             & 45.149             & 160             \\
    3     & 16  & 27.0              & 115                & 10.81             & 45.149             & 200             \\
    4     & 16  & 28.147            & 115                & 11.27             & 45.149             & 240             \\
    5     & 16  & 32.785            & 168.7              & 13.13             & 66.232             & 280             \\
    6     & 16  & 58.241            & 168.7              & 23.32             & 66.232             & 500             \\
    7     & 16  & 96.425            & 168.7              & 38.61             & 66.232             & 830             \\\bottomrule
  \end{tabular}
\end{table}

\subsubsection{Vertex Detector Support and Cabling}
\label{sec:SiD_model_vertexSupport}

The vertex detector and the forward tracking disks are supported by a double-walled carbon fiber tube with an inner radius of \unit[168.7]{mm}, an outer radius
of \unit[184.7]{mm} and a half length of \unit[894.8]{mm}. This tube is supported from the beam pipe by two carbon fiber disks at each end of the tube. The first one is placed at \unit[868.8]{mm} in $z$
with an inner radius of \unit[100.9]{mm} and the second one is placed at \unit[894.3]{mm} in $z$ with an inner radius of \unit[103.9]{mm}. The outer radius for both disks is
\unit[168.7]{mm}. The thickness of each carbon fiber wall and disk is \unit[0.5]{mm}. The support tube fits inside the main tracker, as shown in \cref{fig:sidtrackerZX} and the vertex detector can thus be removed together with the beam pipe when opening the detector.

Some estimation of required cabling is included in the simulation model. Copper disks are placed at the end of the barrel layers connecting down to the beam pipe. A copper layer around the beam pipe represents the connection to the outside of the detector. Routing the cables along the support tube instead of the beam pipe might be beneficial for the impact parameter in the far forward region since it removes some material in front of the first measurement layer.

\subsection{Main Tracking System}
\label{sec:SiD_model_mainTracking}

The main tracking system is unchanged with respect to the \ac{SiD} concept since the requirement on the momentum resolution are very similar. The segmentation is sufficient to provide robust pattern recognition also in presence of the \gghad background at \ac{CLIC} as shown in \cref{cha:Tracking}.

\subsubsection{Main Tracker Barrel}
The main tracking barrel detector consists of five layers of silicon strip detectors. The overall dimensions of the layers are given in \cref{tab:sidtkrbarrel}.
Each layer is made from several square modules with a size of \unit[97.8$\times$97.8]{mm$^2$}. Each module consists of \unit[0.3]{mm}
of sensitive silicon (\unit[$\approx0.32\%$]{\radlen}) and \unit[2.6]{mm} of support and electronics (\unit[$\approx0.19\%$]{\radlen}). A detailed description of the material in each module can be found in \cref{tab:sidtkrbarrellayer}. The readout assumed for the strip detectors is the KPiX \acs{ASIC}~\cite{Freytag:2008zz} developed for the \ac{SiD} concept.  It is designed to provide time resolution of several hundred nanosecons, which is sufficient to resolve individual bunch crossings at the \ac{ILC}. For \ac{CLIC} this has to be improved to a time resolution of the order of \unit[5]{ns}.

The silicon strips run along the $z$-direction and, with a length of \unit[92.03]{mm}, span almost the whole module. Their pitch is \unit[25]{\micron}, with only every second strip being read out directly, while the intermediate strip is read out via capacitive coupling (see \cref{sec:Software_trackerDigitizationSegmentation}). The readout pitch is thus \unit[50]{\micron}. 

\begin{table}[htbp]
  \centering
  \caption[Parameters for the barrel tracker layers.]{Parameters for the barrel tracker layers. The number $N$ of
    modules in the layer in $z$ and $\phi$, the mean radius of the layer $r$ and the half-length of the layer
    $z$ are given. Each module consists of \unit[300]{\micron} of silicon and \unit[2.6]{mm} of support material (see \cref{tab:sidtkrbarrellayer}).}
  \label{tab:sidtkrbarrel}
  \begin{tabular}{c c c R{4}{1} R{4}{1}}\toprule
    Layer & $N_{z}$ & $N_{\phi}$ & \tabt{$r\mm{}$} & \tabt{$z\mm{}$}\\\midrule
    1     & 13      & 20         & 230             & 578            \\
    2     & 17      & 38         & 483             & 749.8          \\
    3     & 23      & 58         & 725.5           & 1013.9         \\
    4     & 29      & 80         & 988.5           & 1272.3         \\
    5     & 35      & 102        & 1239            & 1535.7         \\\bottomrule
  \end{tabular}
\end{table}

\begin{table}[htbp]
  \caption[Materials in the tracker barrel modules.]{Materials and corresponding thicknesses $d$ in each of the modules in the tracker barrel \protect\subref{tab:sidtkrbarrellayer} and
in the tracker endcap \protect\subref{tab:sidtkrendcaplayer}, ordered as seen from the interaction point. The carbon fiber and \rohacell used in the main tracker modules have only half of their nominal densities and thus twice the radiation length.}
  \label{tab:sidtkrlayer}
  \begin{subtable}[t]{0.45\linewidth}
    \centering
    \caption{Tracker barrel module}
    \label{tab:sidtkrbarrellayer}
    \begin{tabular}[ht]{l *1{R{4}{1}}}
      \toprule
      Material  & \tabt{$d\mic{}$} \\\midrule
      Copper &              3.8    \\
      Kapton &             38.0    \\
      Silicon (passive) &   4.8    \\
      Silicon (sensitive) &   300.0   \\
      Carbon Fiber (50\%)& 160.0   \\
      Epoxy     &          175.0   \\
      \rohacell (50\%)&   1800.0   \\
      Carbon Fiber (50\%)& 160.0   \\
      PEEK      &          200.0   \\
      \bottomrule
    \end{tabular}
  \end{subtable}
  \hfill
  \begin{subtable}[t]{0.54\linewidth}
    \centering
    \caption{Tracker endcap module}
    \label{tab:sidtkrendcaplayer}
    \begin{threeparttable}
    \begin{tabular}[ht]{l r}
      \toprule
      Material  & \tabt{$d\mic{}$}\\\midrule
      Copper &             5.2\tnote{A}\ \ /  7.9\tnote{B} \\
      Kapton &            51.0\tnote{A}\ \ / 78.0\tnote{B} \\
      Silicon (passive) &  4.8    \\
      Silicon (sensitive) &   300.0  \\
      Carbon Fiber (50\%)& 160.0  \\
      Epoxy     &          175.0  \\
      \rohacell (50\%)&    1800.0 \\
      Carbon Fiber (50\%)&  160.0 \\
      Silicon (sensitive) &   300.0  \\
      Silicon (passive) &  4.8.0  \\
      Kapton &            51.0\tnote{A}\ \ / 78.0\tnote{B} \\
      Copper &             5.2\tnote{A}\ \ /  7.9\tnote{B} \\
      \bottomrule
    \end{tabular}
    \begin{tablenotes}
     \item{A} Material in the inner three rings.
     \item{B} Material in the outer rings.
    \end{tablenotes}
    \end{threeparttable}
  \end{subtable}
\end{table} 

\subsubsection{Main Tracker Disks}
The tracker endcap consist of four silicon stereo strip layers. Each layer consists of several rings of trapezoidal modules.
The rings are arranged to follow a conical shape, as shown in \cref{fig:sidtrackerZX}).
A detailed description of the dimensions of the individual disks is given in \cref{tab:sidtkrendcap}.

Each module in the tracker endcap has two layers of \unit[0.3]{mm} of
sensitive silicon, since they are modelled as stereo strip detectors, amounting to about \unit[0.64\%]{\radlen}, and other material representing support and electronics, as described in detail in \cref{tab:sidtkrendcaplayer}. The modules used in the inner three rings
have a radial extent of \unit[100.1]{mm} and the modules in the outer rings have a radial extent of \unit[89.8]{mm}.

Like for the tracker barrel, the silicon strips have a pitch of \unit[25]{\micron}, while the readout pitch is \unit[50]{\micron}. The readout is assumed to be the KPiX chip mentioned above. The strips
in the first sensitive layer are perpendicular to one side of the trapezoid, while the strips in the second sensitive layer are perpendicular to the
other side of the trapezoid, which then automatically defines the stereo angle. The layout of the inner and outer modules is chosen, such that in both cases the stereo angles between the two sensitive layers
is 12\degrees.

\begin{table}[htbp]
  \centering
  \caption[Parameters for the tracker endcap disks in \clicsid.]{Parameters for the tracker endcap disks, which are made of trapezoidal modules that are arranged in rings. The number of
    modules $N$, the inner radius $r_{\mathrm{in}}$, the outer radius $r_{\mathrm{out}}$, the two widths of the trapezoid $w_{\mathrm{in/out}}$ and the position in $z$ of the modules within the ring are given. The material budget of the modules is described in \autoref{tab:sidtkrendcaplayer}.}
\label{tab:sidtkrendcap}
  \begin{tabular}{c c R{4}{1} R{4}{1} R{2}{1} R{3}{1} R{4}{1} }\toprule
    Disk  & $N$ & \tabt{$r\mm{in}$} & \tabt{$r\mm{out}$} & \tabt{$w\mm{in}$} & \tabt{$w\mm{out}$} & \tabt{$z\mm{}$}\\\midrule
    1     & 24  & 206.716           & 306.716            & 72.22             & 93.27              & 787.105        \\
          & 32  & 303.991           & 403.991            & 72.22             & 93.27              & 778.776        \\
          & 40  & 399.180           & 499.180            & 72.22             & 93.27              & 770.544        \\
    2     & 24  & 206.716           & 306.716            & 72.22             & 93.27              & 1073.293       \\
          & 32  & 303.991           & 403.991            & 72.22             & 93.27              & 1064.966       \\
          & 40  & 399.180           & 499.180            & 72.22             & 93.27              & 1056.734       \\
          & 40  & 493.520           & 583.520            & 90.49             & 109.36             & 1048.466       \\
          & 48  & 580.654           & 670.654            & 90.49             & 109.36             & 1041.067       \\
          & 54  & 658.666           & 748.666            & 90.49             & 109.36             & 1033.725       \\
    3     & 24  & 206.716           & 306.716            & 72.22             & 93.27              & 1353.786       \\
          & 32  & 303.991           & 403.991            & 72.22             & 93.27              & 1345.457       \\
          & 40  & 399.180           & 499.180            & 72.22             & 93.27              & 1337.225       \\
          & 40  & 493.520           & 583.520            & 90.49             & 109.36             & 1328.957       \\
          & 48  & 580.654           & 670.654            & 90.49             & 109.36             & 1321.558       \\
          & 54  & 658.666           & 748.666            & 90.49             & 109.36             & 1314.217       \\
          & 58  & 748.448           & 848.448            & 90.49             & 109.36             & 1306.828       \\
          & 64  & 829.239           & 919.239            & 90.49             & 109.36             & 1299.486       \\
          & 68  & 913.364           & 1003.364           & 90.49             & 109.36             & 1292.189       \\
    4     & 24  & 206.716           & 306.716            & 72.22             & 93.27              & 1639.164       \\
          & 32  & 303.991           & 403.991            & 72.22             & 93.27              & 1630.835       \\
          & 40  & 399.180           & 499.180            & 72.22             & 93.27              & 1622.603       \\
          & 40  & 493.520           & 583.520            & 90.49             & 109.36             & 1614.335       \\
          & 48  & 580.654           & 670.654            & 90.49             & 109.36             & 1606.936       \\
          & 54  & 658.666           & 748.666            & 90.49             & 109.36             & 1599.595       \\
          & 58  & 748.448           & 841.448            & 90.49             & 109.36             & 1592.206       \\
          & 64  & 829.239           & 919.239            & 90.49             & 109.36             & 1584.864       \\
          & 68  & 913.364           & 1003.364           & 90.49             & 109.36             & 1577.567       \\
          & 72  & 995.970           & 1095.970           & 90.49             & 109.36             & 1570.222       \\
          & 78  & 1079.167          & 1169.167           & 90.49             & 109.36             & 1562.916       \\
          & 84  & 1161.937          & 1251.937           & 90.49             & 109.36             & 1555.647       \\\bottomrule
  \end{tabular}
\end{table}

\subsubsection{Main Tracker Support and Cabling}
The barrel modules are mounted on carbon fiber cylinders. The endcap modules are mounted on conical disks made of carbon fiber as indicated in \cref{fig:sidtrackerZX}. In addition copper disks are added at the end of each of the barrel layers to represent cables. Like for the vertex detector only air cooling is assumed and no material is added for other cooling mechanisms. Thus, also here power pulsing with the repetition rate of the bunch trains is an essential requirement.

\subsection{Tracker Material Budget}
\begin{figure}[htpb]
  \centering
  \includegraphics[width=0.5\textwidth]{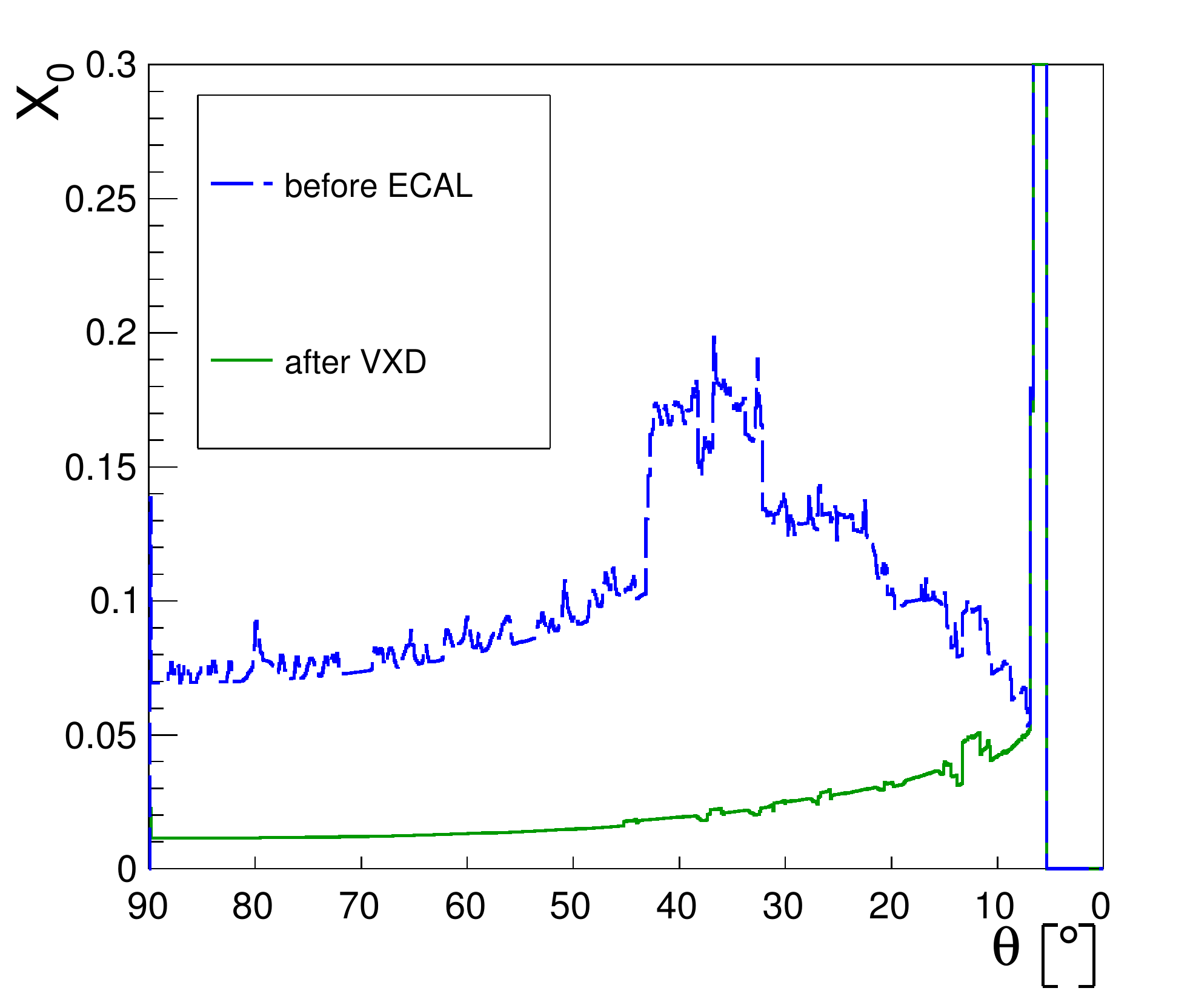}
  \caption[The material budget in the tracking region of \clicsid.]{The material budget in the tracking region given in radiation lengths \radlen, with respect to the polar angle $\theta$. The conical beam pipe
    causes the sharp peak at $\theta\approx7\degrees$. Figure kindly provided by Astrid M\"unnich.}
  \label{fig:materialtracking}
\end{figure}

The total amount of material in the vertex detector amounts to approximately \unit[1\%]{\radlen}, including the beam pipe discussed in \cref{sec:SiD_model_beamPipe} but excluding the vertex detector support tube. The whole tracking region amounts to approximately \unit[7\%]{\radlen} at a polar angle of 90\degrees. This value slowly increases towards lower polar angles, as shown in \cref{fig:materialtracking}. At $\theta \approx 42\degrees$ the material budget rises sharply because of the forward disks which are traversed at a shallow angle. For even lower polar angler the total material budget decreases in steps, depending on how many of the main tracker disks are traversed. The conical part of the beam-pipe is designed to be pointing (see \cref{sec:SiD_model_beamPipe}) and thus results in a huge material budget at its opening angle of $\theta \approx 7\degrees$. Otherwise the total material budget of the tracking region never exceeds \unit[20\%]{\radlen}.

\section{Calorimeters}
\label{sec:SiD_model_calorimetry}

As discussed in \cref{sec:SiD_particleFlow}, the main goal of particle flow calorimetry is to minimize confusion by providing high lateral and longitudinal segmentation. This paradigm also drives the design of \ac{SiD} and \clicsid.

\subsection{Electromagnetic Calorimeter}
\label{sec:SiD_model_ecal}

\begin{figure}[htbp]
 \begin{subfigure}[]{0.49\textwidth}
  \includegraphics[width=\textwidth, trim=50 300 50 110 ,clip=true]{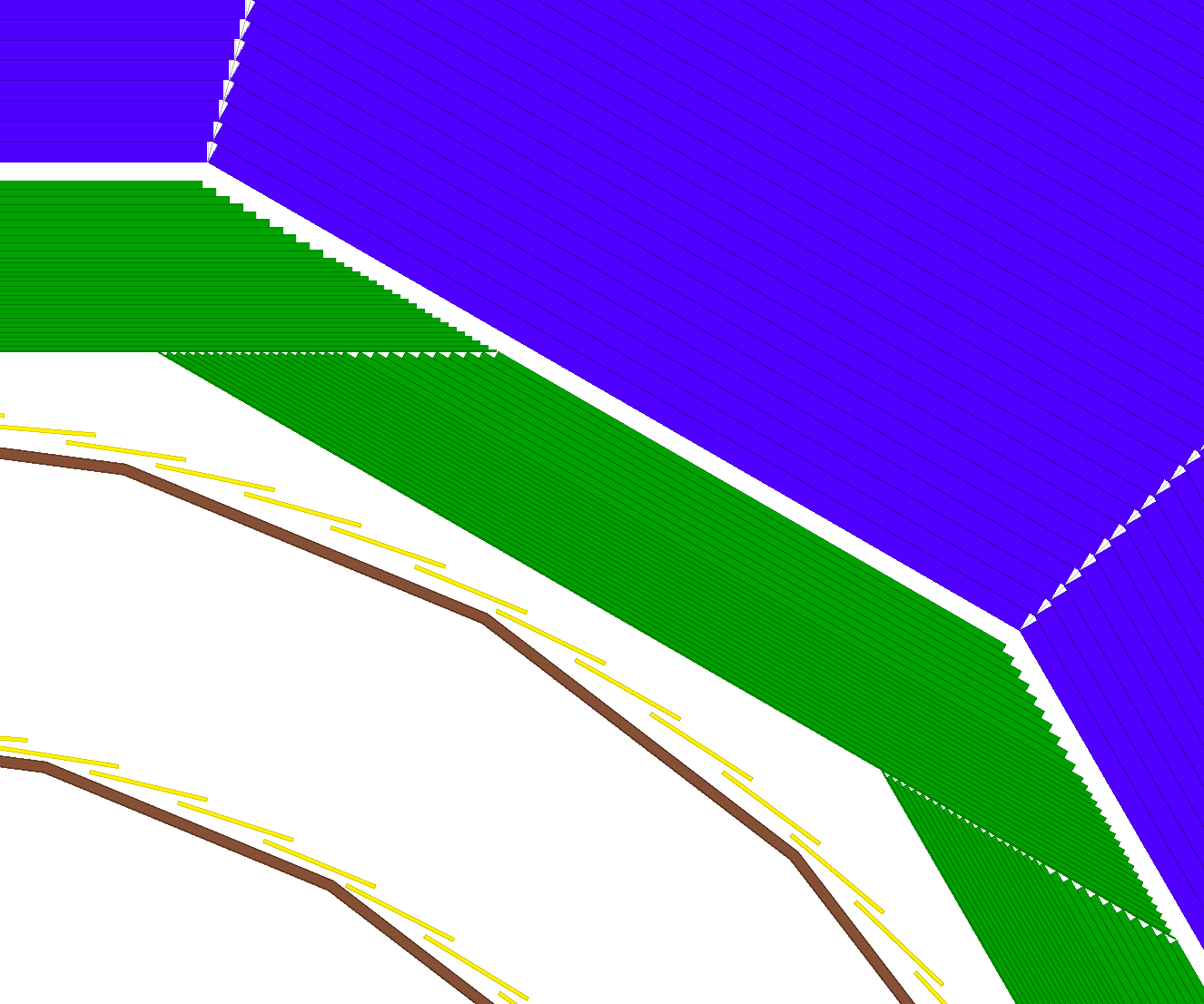}
  \caption{}
  \label{fig:sidecallayout}
 \end{subfigure}
 \begin{subfigure}[]{0.49\textwidth}
  \includegraphics[width=\textwidth, trim=60 230 150 272 ,clip=true]{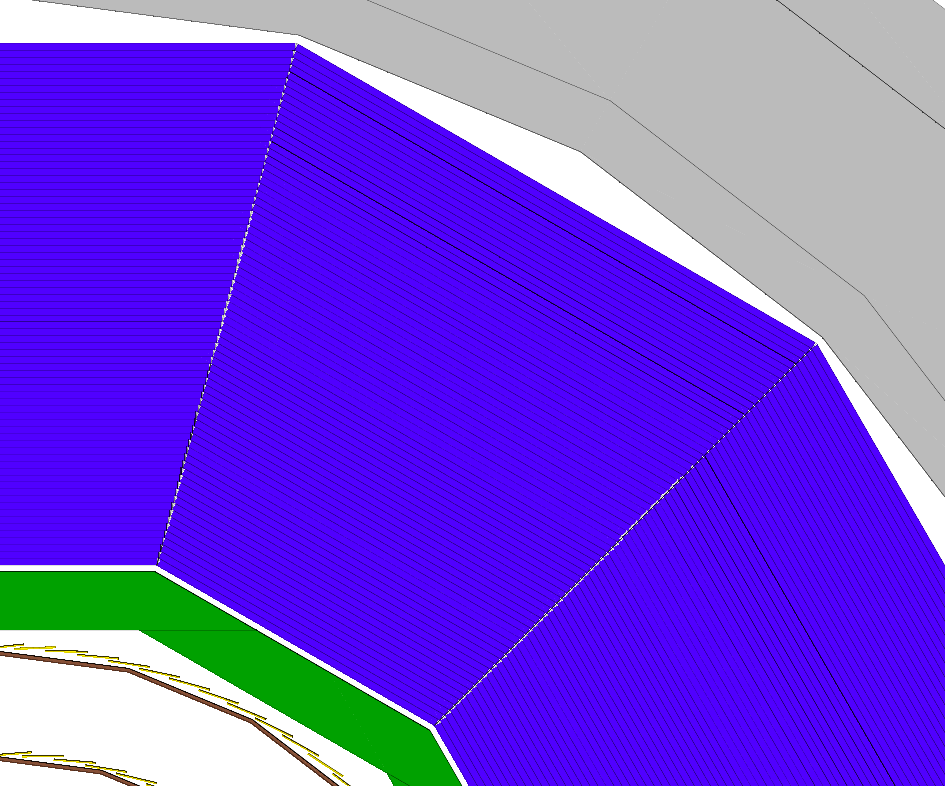}
  \caption{}
  \label{fig:sidhcallayout}
 \end{subfigure}
 \caption[Cut through the barrel region of the calorimeter in \clicsid.]{Cut through the barrel region of the electromagnetic calorimeter \protect\subref{fig:sidecallayout} and the hadronic calorimeter \protect\subref{fig:sidhcallayout}, showing the layers and modules.}
 \label{fig:sidcalolayout}
\end{figure}

The \ac{ECal} is a highly segmented tungsten sampling calorimeter with silicon as the sensitive material.
The innermost layer of the ECal consists only of a sensitive layer. It is followed by 20 layers with an absorber thickness
of \unit[2.5]{mm} and another 10 layers with an absorber thickness of \unit[5]{mm}. The absorber material is tungsten.
Each layer consists of the absorber material followed by an air gap of \unit[0.25]{mm}, \unit[0.32]{mm} of
silicon (sensitive), \unit[0.05]{mm} of copper, \unit[0.3]{mm} of Kapton and another air gap of \unit[0.33]{mm}.
The cell size in all layers is \unit[3.5$\times$3.5]{mm$^2$}.
The total material corresponds to approximately \unit[1.0]{\nuclen} or \unit[25.7]{\radlen}. The \ac{ECal} barrel consists of twelve modules with non-pointing gaps, illustrated in \cref{fig:sidecallayout}.

This layout is identical to that forseen in the \ac{SiD} model. The total depth is sufficient for most electromagnetic showers also at \ac{CLIC} and the highly granular \ac{HCal} works as a tail catcher. The baseline technology for the \ac{ECal} in \ac{SiD} is hexagonal silicon pixels with an area of $\unit[13]{mm^2}$ which we approximate here using square pixels. The readout has to be fully embedded to minimize the gap sizes. It is forseen to use the KPiX chip~\cite{Freytag:2008zz}.

\subsection{Hadronic Calorimeter}
\label{sec:SiD_model_hcal}

The \ac{HCal} is a sampling calorimeter with polystyrene plates as the sensitive material.
The polystyrene plates have a thickness of \unit[5]{mm} and are segmented into \unit[30$\times$30]{mm$^2$} cells.
In addition to the sensitive layer and the absorber plates described below, there is an air gap of \unit[1.5]{mm} in each layer.
There is no material included to represent electronics. This does not affect the overall material budget significantly.
The total gap size of \unit[6.5]{mm} should be sufficient to accommodate sensitive material and electronics for all sampling calorimeter technologies currently being discussed.

There are 75 layers in the barrel with \unit[10]{mm} tungsten absorber plates arranged in twelve modules with pointing gaps (see \cref{fig:sidhcallayout}). The total material corresponds to \unit[7.9]{\nuclen}.
This amount of material does provide sufficient absorption for hadronic showers at \ac{CLIC} energies, and the sampling provides the best energy resolution in the given space as shown in \cref{cha:Calorimetry}.
In the endcaps there are 60 layers with \unit[20]{mm} steel plates as absorber material, which corresponds to \unit[7.6]{\nuclen}. There is no need to replace the absorber material in the endcap, since only the size of the \ac{HCal} barrel is constrained by the coil.

There are many different technologies being considered for the \ac{HCal} in \ac{SiD}. Several prototypes have been built and operated in beam tests. These options include scintillators with analog readout using \acp{SiPM}~\cite{collaboration:2010hb} or \acp{RPC} with pads that are read out digitally~\cite{Bilki:2010xe}. Alternative gaseous detector options are using \acs{MPGD} like GEM foils~\cite{Yu:1900zza} or MicroMegas~\cite{Adloff:2009fs} instead of the pad readout. The baseline for \ac{SiD} is the digital \ac{RPC} option with $\unit[1\times1]{cm^2}$ pads. For \clicsid the analog scintillator option has been chosen since the \pandora algorithm which was chosen as default particle flow algorithm was only validated for analog calorimeter signals.

Another important difference compared to the \ac{SiD} concept is the increased inner radius of the \ac{HCal} endcap of \unit[50]{cm} to accommodate the support tube for the \ac{QD0} as discussed in \cref{sec:SiD_model_acceleratorComponents}. This results in a significantly reduced acceptance.

\section{Solenoid and Magnetic Field}
\label{sec:SiD_model_coil}

The superconducting coil is modelled as an aluminium cylinder, which represents the conductor and the mandrel. It is surrounded by a
vacuum layer and a steel layer, which represent the vacuum vessel. The vaccum vessel is closed by steel at its ends. The total thickness
of the coil corresponds to approximately \unit[1.5]{\nuclen}. An overview of the parameters of the coil is shown in \autoref{tab:coil}.

The solenoidal magnetic field is a homogeneous field of \unit[5]{T} parallel to the detector axis throughout the volume inside of the coil.
The field outside of the coil is \unit[1.5]{T} pointing in the opposite direction of the inner field. There is no field beyond the end of
the coil in $z$-direction.

The coil radius is increased by \unit[18]{cm} with respect to the \ac{SiD} design to allow calorimeters of similar depth in the \clicild and \clicsid models.
This radius corresponds approximately to that of the \acs{CMS} coil~\cite{Acquistapace:1997fm} which has a nominal field strength of \unit[4]{T}. Constructing a coil of similar size and with \unit[5]{T} field will require novel technologies like reinforced conductors~\cite{LCD:2009-001,LCD:2011-007}.

\begin{table}[htpb]
  \caption[Parameters of the coil elements.]{Parameters of the coil elements. For all elements the material, the longitudinal extent in one half of
the detector $z_{\mathrm{min/max}}$ and the radial extent $r_{\mathrm{min/max}}$ are given.}
  \label{tab:coil}
  \centering  
  \begin{tabular}{l *4{R{4}{0}}}
    \toprule
    Material  & \tabt{$z\mm{min}$} & \tabt{$z\mm{max}$} & \tabt{$r\mm{min}$} & \tabt{$r\mm{max}$}   \\\midrule
    Steel     &                0   &            3515.0  &          2770.21   &           2800.21    \\
    Vacuum    &                0   &            3515.0  &          2800.21   &           2910.21    \\
    Aluminium &                0   &            3395.0  &          2910.21   &           3344.21    \\
    Vacuum    &                0   &            3515.0  &          3344.21   &           3531.21    \\
    Steel     &                0   &            3515.0  &          3531.21   &           3571.21    \\
    Vacuum    &           3395.0   &            3515.0  &          2910.21   &           3344.21    \\
    Steel     &           3515.0   &            3575.0  &          2770.21   &           3571.21    \\
    \bottomrule
  \end{tabular}
\end{table}

\section{Instrumented Yoke}
\label{sec:SiD_model_yoke}

The coil is surrounded by an iron return yoke, which shields the experimental area from the magnetic field as well as from radiation.
It is instrumented with double-layers of \acp{RPC} to help identifying muons.

Each layer consists of two \ac{RPC} layers, followed by an air gap of \unit[10]{mm} and the absorber material.
There are 18 layers in the barrel and the endcap with \unit[100]{mm} of iron as absorber material for each layer.
The first barrel layer has only \unit[50]{mm} of iron to give a finer sampling directly after the coil.
The second and the last absorber layer in the barrel are twice as thick, in order to take the stress.
Unlike the barrel, the yoke endcap begins with an absorber layer and ends with a sensitive layer. The 18 endcap layers have iron absorber plates of \unit[100]{mm} thickness and are instrumented like the barrel layers.

In addition, a yoke plug fills the gap between HCal and yoke in the endcap region inside of the coil.
It is instrumented with an \ac{RPC} double-layer between two iron layers of \unit[150]{mm} and \unit[90]{mm} thickness.

\section{Forward Region and Beam Pipe}
\label{sec:SiD_model_forward}
\begin{figure}[htpb]
  \centering
   \begin{tikzpicture} [scale=0.528]
     \draw[scale=1.0] (0.0,0.0) node[anchor=south west]{\includegraphics[width=0.9\textwidth, trim=270 352 100 330 ,clip=true]{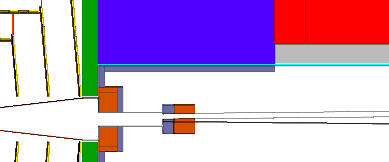}};
     \scriptsize
     \draw[scale=1.0] (-0.4,1.608)  node[left] {$-210$}    -- (0.0,1.608) ; 
     \draw[scale=1.0] (-0.4,3.22)  node[left] {0}    -- (0.0,3.22) ; 
     \draw[scale=1.0] (-0.4,4.832)  node[left] {210}    -- (0.0,4.832) ; 

     \draw[scale=1.0] (9.5,1.377)  node[right] {$-240$}    -- (9.1,1.377) ; 
     \draw[scale=1.0] (14.4,2.222)  node[right] {$-130$}    -- (14.0,2.222) ; 

     \draw[scale=1.0] (5.985,-0.4)  node[below left]  {1657} -- (5.985,0.0) ; 
     \draw[scale=1.0] (7.14,-0.4)  node[below]  {1805} -- (7.14,0.0) ; 
     \draw[scale=1.0] (8.45,-0.4)  node[below]  {1976} -- (8.45,0.0) ; 
     \draw[scale=1.0] (11.582,-0.4)  node[below left]  {2386} -- (11.582,0.0) ; 
     \draw[scale=1.0] (12.350,-0.4)  node[below]  {2486} -- (12.350,0.0) ; 
     \draw[scale=1.0] (13.77,-0.4)  node[below]  {2671} -- (13.77,0.0) ; 

     \draw[scale=1.0] (24.1,9.8) node {\LARGE{Yoke}};
     \draw[scale=1.0] (23.7,7.7) node {\large{Anti-Solenoid}};
     \draw[scale=1.0] (13.0,9.1) node {\LARGE{HCal}};
     \draw[scale=1.0] (11.0,5.0) node[above] {\Large{LumiCal}} -- (7.8,4.2);
     \draw[scale=1.0] (16.0,4.6) node[above] {\Large{BeamCal}} -- (13.1,3.9);
     \draw[scale=1.0] (6.55,8.0) node[rotate=270] {\Large{ECal}};
   \end{tikzpicture}
  \caption[Layout of the forward region in \clicsid.]{Layout of the forward region in \clicsid. The BeamCal and the LumiCal are shown together with the forward part of the beam pipe, the support tube and several shielding and mask elements. Not included are the \ac{QD0} which will be placed around the beam pipes inside of the anti-solenoid, a vacuum valve which will be placed between \ac{LumiCal} and \ac{BeamCal} and beam monitoring equipment which will be placed behind the \ac{BeamCal}. All values are given in millimeters.}
  \label{fig:sidforwardregion}
\end{figure}
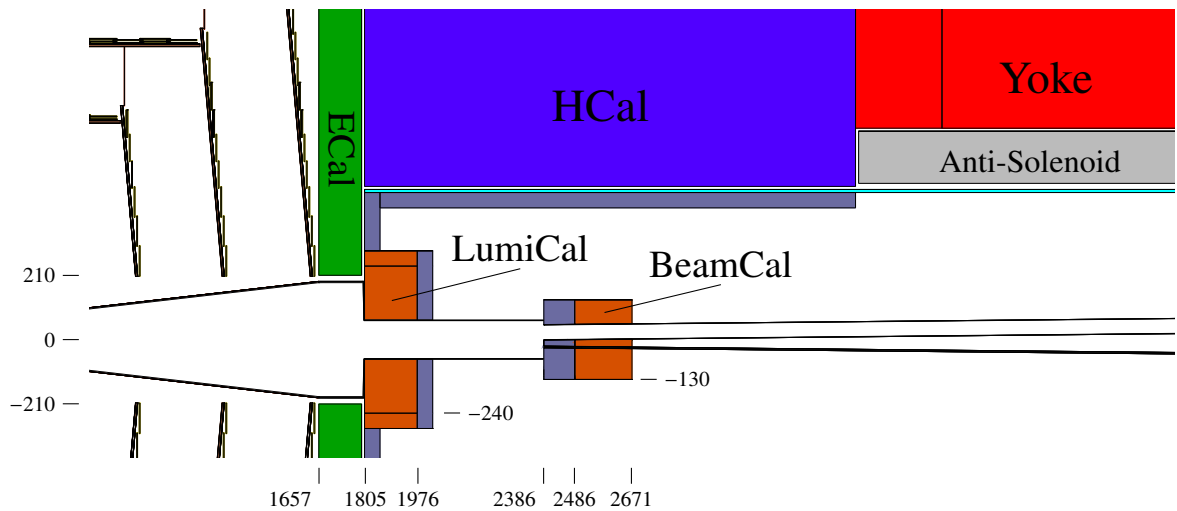
\subsection{LumiCal}
\label{sec:SiD_model_lumiCal}

The \ac{LumiCal} extends the coverage for the identification of electromagnetic showers down to \unit[64]{mm}, which corresponds to an opening angle of about \unit[35]{mrad}.
It is placed behind the ECal at a $z$-position of \unit[1805]{mm}. The outer radius of \unit[240]{mm} creates enough overlap with the \ac{ECal} to avoid a gap in the coverage.
Its main purpose is the measurement of the luminosity spectrum through the measurement of Bhabha scattering events but is also important for identifying other forward processes as shown in the analysis in \cref{cha:Higgs}.

The instrumentation is very similar to the one of the \ac{ECal} described in \cref{sec:SiD_model_ecal}.
It also consists of two sections with different absorber thicknesses.
The first 20 layers have a tungsten thickness of \unit[2.71]{mm} and the last 15 layers have a tungsten thickness of \unit[5.43]{mm}.
Each layer consists of the absorber material followed by \unit[0.32]{mm} of
silicon (sensitive), \unit[0.05]{mm} of copper, \unit[0.3]{mm} of Kapton and an air gap of \unit[0.33]{mm}.
The cell size is \unit[3.5$\times$3.5]{mm$^2$} and the total material corresponds to \unit[1.4]{\nuclen} or \unit[34.8]{\radlen}.
A tube of G10 with a thickness of \unit[50]{mm} is placed around the LumiCal as a placeholder for readout electronics.

In the simulation model the \ac{LumiCal} is placed symmetric to the detector axis. This is a simplification since it should be placed centered around the outgoing beam axis. 

\subsection{BeamCal}
\label{sec:SiD_model_beamCal}

The \ac{BeamCal} completes the forward coverage of the electromagnetic calorimeters. It is has an outer radius of \unit[130]{mm} and it starts at a $z$ of \unit[2486]{mm}. The BeamCal has two holes for the incoming and outgoing beam pipes. The radius of the hole for the incoming beam pipe is \unit[2.55]{mm} and the radius for the outgoing beam pipe is \unit[24.91]{mm}, which corresponds to an opening angle of \unit[10]{mrad}. This opening angle is enough to avoid the majority of the coherent pair background (see \cref{sec:CLIC_machineInduced}).

The \ac{BeamCal} consists of 50 layers of \unit[2.71]{mm} tungsten, \unit[0.32]{mm} of
silicon (sensitive), \unit[0.05]{mm} of copper, \unit[0.3]{mm} of Kapton and an air gap of \unit[0.33]{mm}.
Like the other electromagnetic calorimeters, the sensitive layer is segmented into \unit[3.5$\times$3.5]{mm$^2$} cells.
Due to the high radiation dose at these low angles, the sensitive detectors have to be radiation hard like for example diamond sensors~\cite{Grah:2009zz}.

A graphite layer of \unit[100]{mm} is placed in front of the BeamCal in order to absorb backscattered particles~\cite{dannheimsailer2011,sailerphd}. The position in $z$ of the BeamCal including the graphite mask is thus \unit[2386]{mm}.

\subsection{Beam Pipe}
\label{sec:SiD_model_beamPipe}

The central part of the beam pipe is made of beryllium because of its low radiation length. It has an outer radius of \unit[25]{mm} and a thickness of \unit[0.5]{mm}. The conical parts and the elements in the far forward region are made of steel. The beam pipe becomes conical at a $z$ of \unit[260]{mm} with an opening angle of 6.6\degrees. The thickness of the conical beam pipe is \unit[4]{mm} in radial direction. It has been shown that this amount of steel effectively removes most of the backscattered particles that are created from the beam-related backgrounds in the far-forward region of the detector~\cite{dannheimsailer2011}. The beam pipe reaches a radius of \unit[190]{mm} at the front face of the ECal and then becomes cylindrical again. This radius is small enough to allow an opening scenario of the detector where the vertex detector is removed together with the beam pipe.

The thickness of the beam pipe in front of LumiCal is \unit[1]{mm}. A cylindrical beam pipe with an outer radius of \unit[63]{mm} and a thickness of \unit[1]{mm} is inside of the LumiCal and continues to the front face of the BeamCal. Two cylindrical beam pipes which are centered around the incoming and outgoing beam axis are placed inside of the BeamCal holes and continue until the end of the detector volume. The outgoing beam pipes are simplified in this model and should instead be conical with an opening angle of \unit[10]{mrad}.

\subsection{Accelerator Components}
\label{sec:SiD_model_acceleratorComponents}

The \ac{QD0} has to placed as close to the \ac{IP} as possible to provide the necessary beam focusing for high luminosity. Alternatively the \ac{QD0} would have to be placed outside of the detector which would result in a significantly reduced luminosity~\cite{cdrvol1}. The \ac{QD0} has to be stable to around \unit[1]{nm} during a bunch train which can not be achieved if it is supported from the detector. All of the far-forward detectors and the \ac{QD0} will thus be supported from a support tube which is mounted on the tunnel wall to fulfill the stability requirements on the \ac{QD0}. This support tube is simulated as a iron cylinder inside the HCal and yoke endcaps.
The \ac{QD0}, which would be placed around the incoming beam pipe behind the \ac{BeamCal} is not represented in the detector simulation.

In addition there are several shielding elements to avoid backscatters from the \ac{BeamCal} and other far-forward elements to reach the calorimeters.
These are modelled as a layer of \unit[50]{mm} of tungsten behind the \ac{LumiCal} and the \ac{ECal} and right inside of the support tube within the \ac{HCal} endcap.

An anti-solenoid placed right outside of the support tube in the region of the yoke endcap.
It is there to protect the \ac{QD0} magnet, which includes a permanent magnet, from the field of the central solenoid. The anti-solenoid is also modelled as an iron cylinder.

\section{Occupancies from Beam-Related-Backgrounds}
\label{sec:SiD_occupancies}
The occupancies from beam-related backgrounds have been studied in \clicild~\cite{dannheimsailer2011,sailerphd}. The main differences between the \ac{ILD} and \ac{SiD} concepts are the solenoid field strength, \ie \unit[4]{T} for \ac{ILD} and \unit[5]{T} which affect the angular distributions of charged background particles, as well as the tracking systems, which will be the focus of this section. However, we want to mention one important result from~\cite{dannheimsailer2011}: the train occupancies in the endcap calorimeters can reach up 60\% in the \ac{ECal} mostly from the \gghad background and up to 1100\% at the innermost radii in the \ac{HCal} from incoherent pairs interacting in the outgoing beampipes. This shows that the highly granular calorimeters serve an important purpose at \ac{CLIC} beyond their intended use for particle flow calorimetry. A highly granular calorimeter gives the possibility to disentangle energy deposits from beam-related backgrounds and physics. For more details on the occupancies in the calorimeters and estimations of radiation doses we refer to those two studies.

\subsection{Simulation Samples}
We study the occupancy using two background samples generated from \guineapig~\cite{Schulte:331845,Schulte:1999tx} for the nominal \ac{CLIC} parameters at a center-of-mass energy of \unit[3]{TeV}. The incoherent pairs corresponding to one full bunch train and another for the \gghad background, corresponding to approximately 10000 bunch trains. Each bunch crossing of incoherent pair background contains approximately 300000 electrons and positrons and can thus not be simulated as one event. Instead, each particle is passed through the full detector simulation individually. The resulting tracker hits are merged together into events corresponding to the signal from 1000 incoherent pair particles, \ie $\sim1/300$ of a bunch crossing, before passing it through the hit digitization. For the \gghad background we simulate and digitize each event individually, while we expect about 3.2 events in each bunch crossing (see \cref{tab:CLIC_parameters}). The steps involved in the simulation and hit digitization are explained in \cref{cha:Software}. Consequently each hit represents a reconstructed cluster, \ie several pixel or strip hits. The resulting hit rates using full digitization are comparable to the rates obtained from counting all simulation hits that deposit more energy than $1/5$ of a \ac{MIP} as done in~\cite{dannheimsailer2011}. However, using digitized hits allows to directly estimate the occupancy based on resulting cluster sizes.


\subsection{Hit Densities}
\begin{figure}[tbh]
 \begin{subfigure}[]{0.49\textwidth}
  \includegraphics[width=\textwidth]{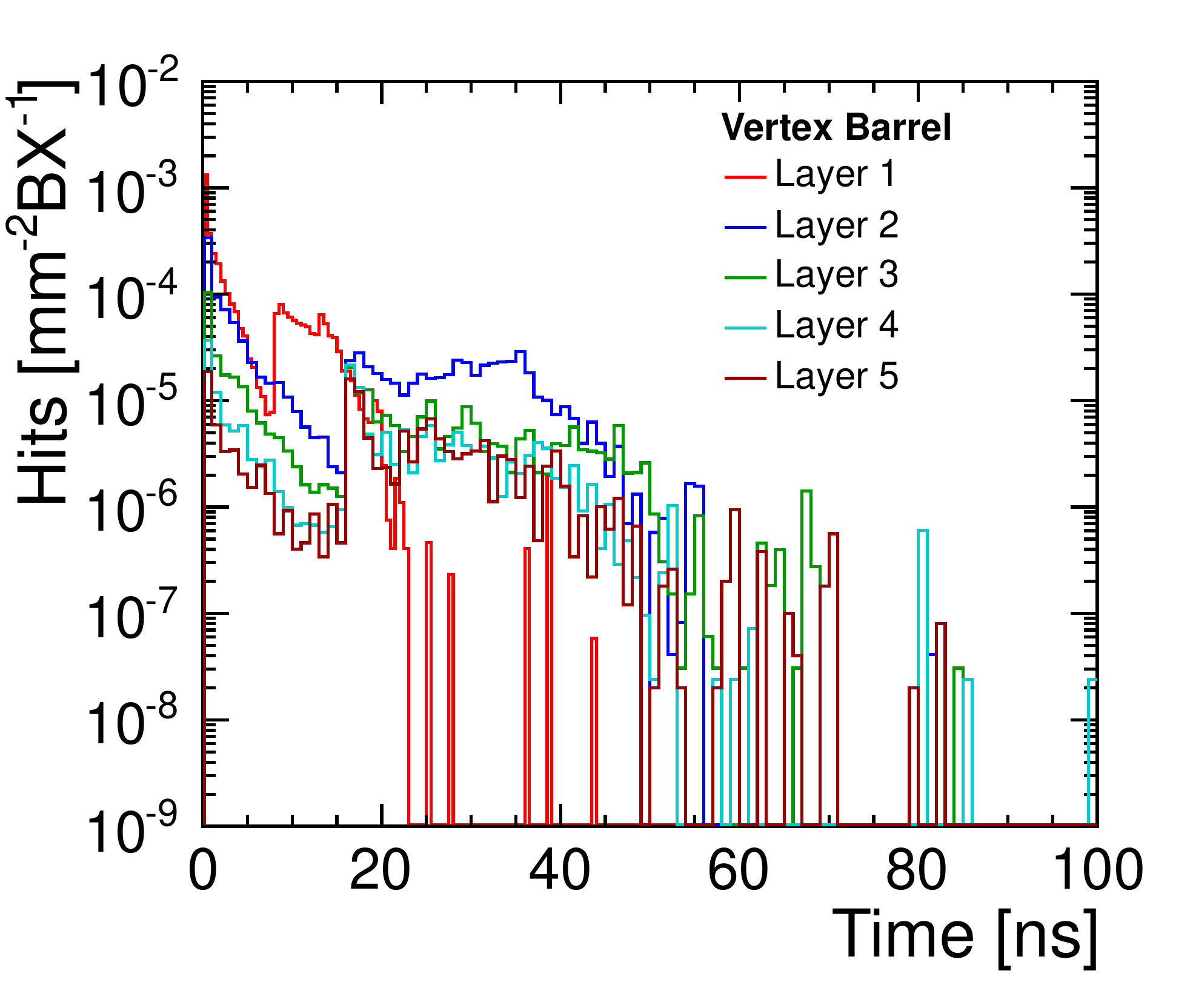}
  \caption{Time structure of hits from a single BX}\label{fig:SiD_hitDensitiesVertexBarrel_t_incoh}
 \end{subfigure}
 \hfill
 \begin{subfigure}[]{0.49\textwidth}
  \includegraphics[width=\textwidth]{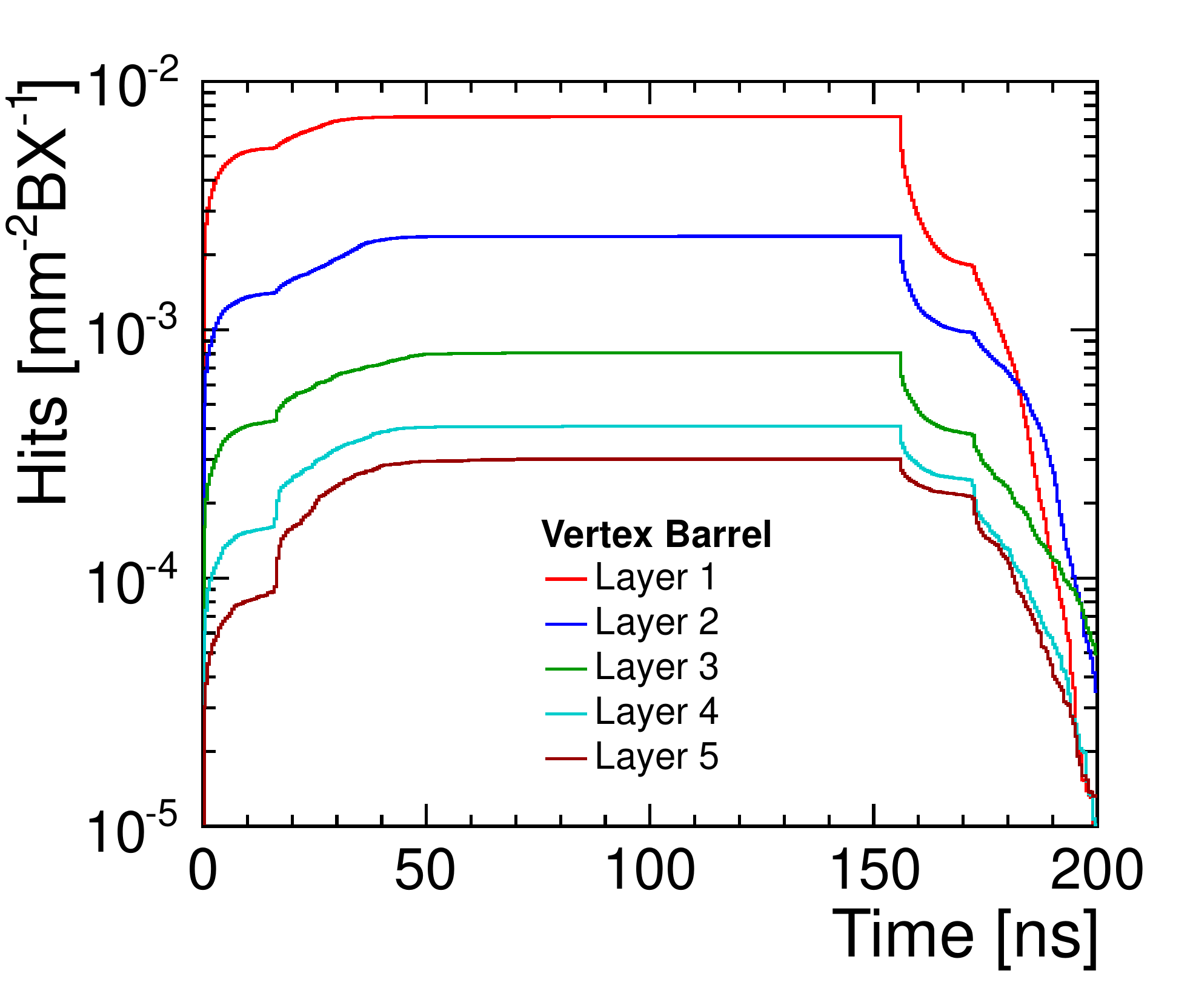}
  \caption{Time structure of hits in a full bunch train}\label{fig:SiD_hitDensitiesVertexBarrel_t_full}
 \end{subfigure}
 \caption[Time dependence of the hit densities in the vertex barrel layers from incoherent pairs.]{Time dependence of the hit densities in the five vertex barrel layers from incoherent pairs. Shown is the time structure of the hits from a single BX \subref{fig:SiD_hitDensitiesVertexBarrel_t_incoh} and for a full bunch train \subref{fig:SiD_hitDensitiesVertexBarrel_t_full}.}
 \label{fig:SiD_hitDensitiesVertexBarrel_t}
\end{figure}

The hits originating from the incoherent pair background originate in parts from backscatters that are significantly displaced in time as shown in \cref{fig:SiD_hitDensitiesVertexBarrel_t_incoh}. These are mostly created in the very forward region, \ie the \ac{BeamCal} as well as the outgoing beam pipe. The time delay between these direct and indirect hits is given by the distance of the \ac{BeamCal} from the \ac{IP}. A careful design of the \ac{BeamCal} and its mask is essential to keep these backscatters at manageable levels. A thick conical beam pipe is similarly important to shield the central detectors from these backscatters~\cite{dannheimsailer2011}. A thinner beam pipe, as foreseen in the \ac{ILC} detectors, results in an increase of hits due to backscatters of one order of magnitude. They would thus constitute the dominating contribution to the total occupancy~\cite{dannheimsailer2011}. The resulting time structure for the full bunch train is shown in \cref{fig:SiD_hitDensitiesVertexBarrel_t_full}. We obtained the distribution by repeatedly displacing the average time distribution of a single BX and thus, no statistical fluctuations are visible. In the first \unit[15]{ns} of a bunch train the vertex barrel detectors only experience direct hits. Afterwards backscatters start to arrive in time with the particles from the \ac{IP}. The highest occupancy levels are reached after \unit[30--40]{ns}. As a result of this time structure there are hits visible in the detector until \unit[40]{ns} after the end of a bunch train. 

Low \pT particles are confined to small radii by the magnetic field. This does not necessarily apply for backscatters. Thus, for large radii the importance of backscatters increases. The maximum hit density within a bunch train, on the other hand, decreases significantly with larger radii from $\unit[7\times10^{-3}]{Hits/mm^2/BX}$ in the first barrel layer of the vertex detector to $\unit[2.5\times10^{-4}]{Hits/mm^2/BX}$ in the fifth layer.

\begin{figure}[tbh]
 \begin{subfigure}[]{0.30\textwidth}
  \includegraphics[width=\textwidth]{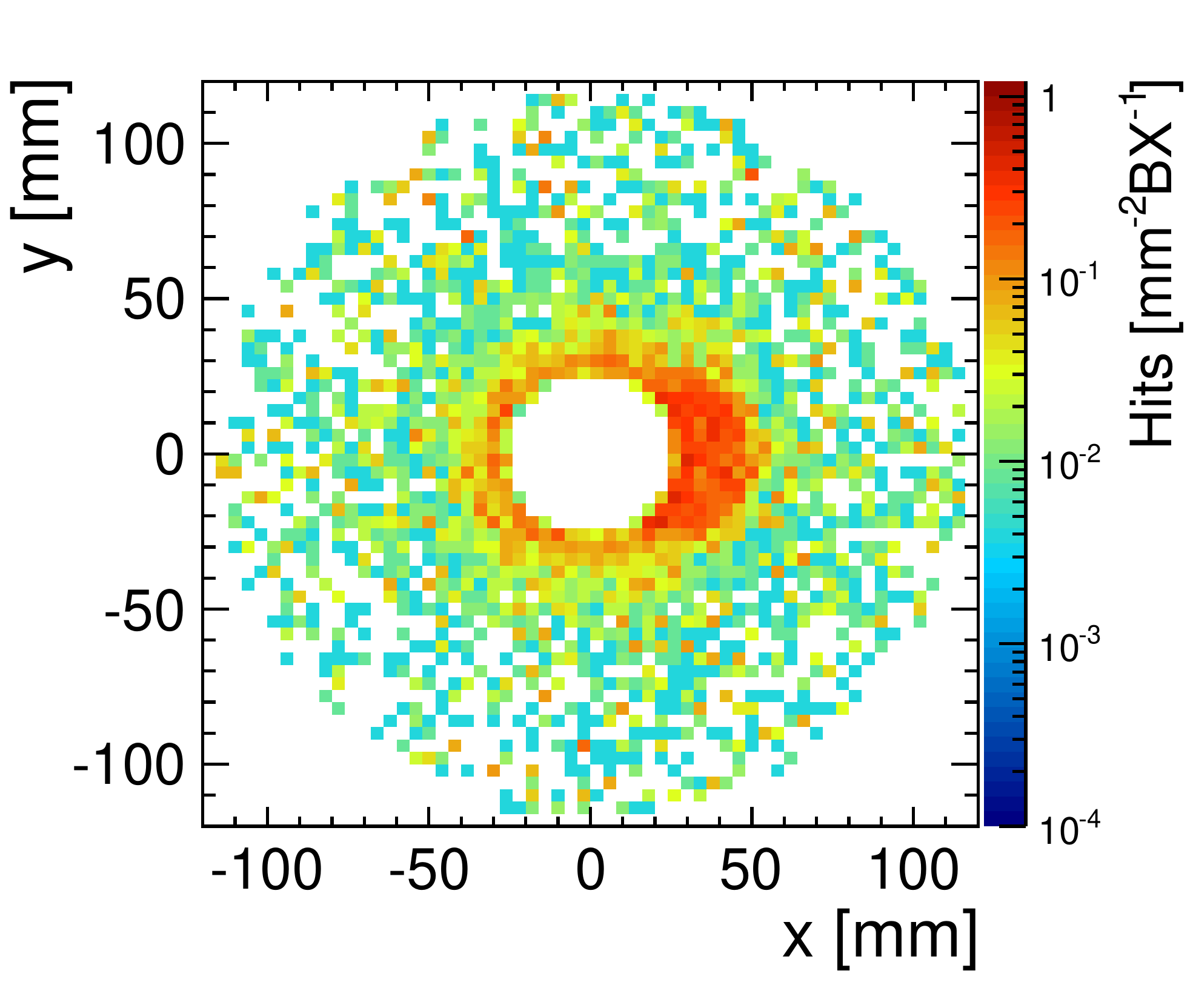}
  \caption{All hits from one BX}\label{fig:SiD_hitDensities_incoh_xy_all}
 \end{subfigure}
 \hfill
 \begin{subfigure}[]{0.30\textwidth}
  \includegraphics[width=\textwidth]{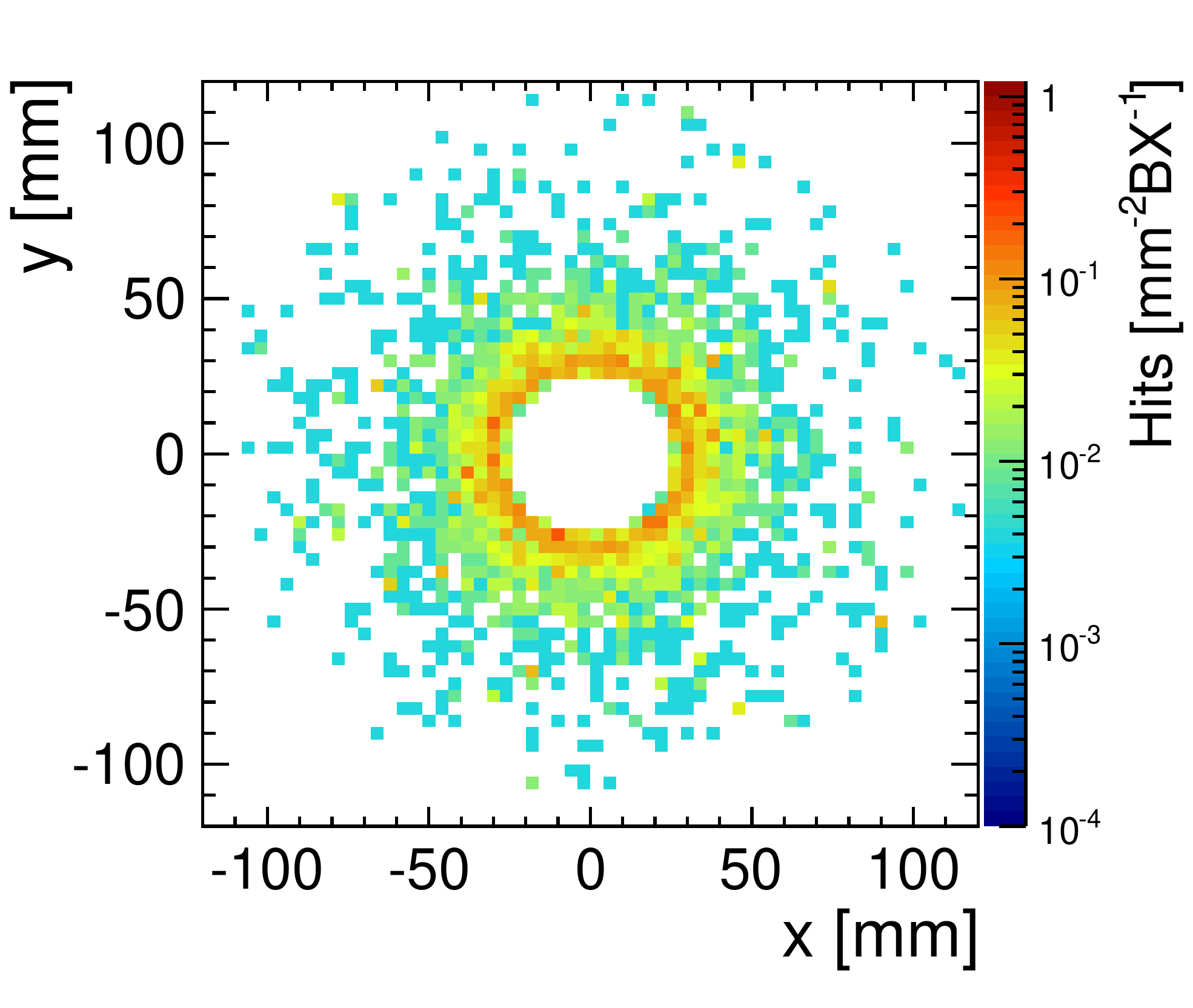}
  \caption{$t < \unit[15]{ns}$}\label{fig:SiD_hitDensities_incoh_xy_t1}
 \end{subfigure}
 \hfill
 \begin{subfigure}[]{0.30\textwidth}
  \includegraphics[width=\textwidth]{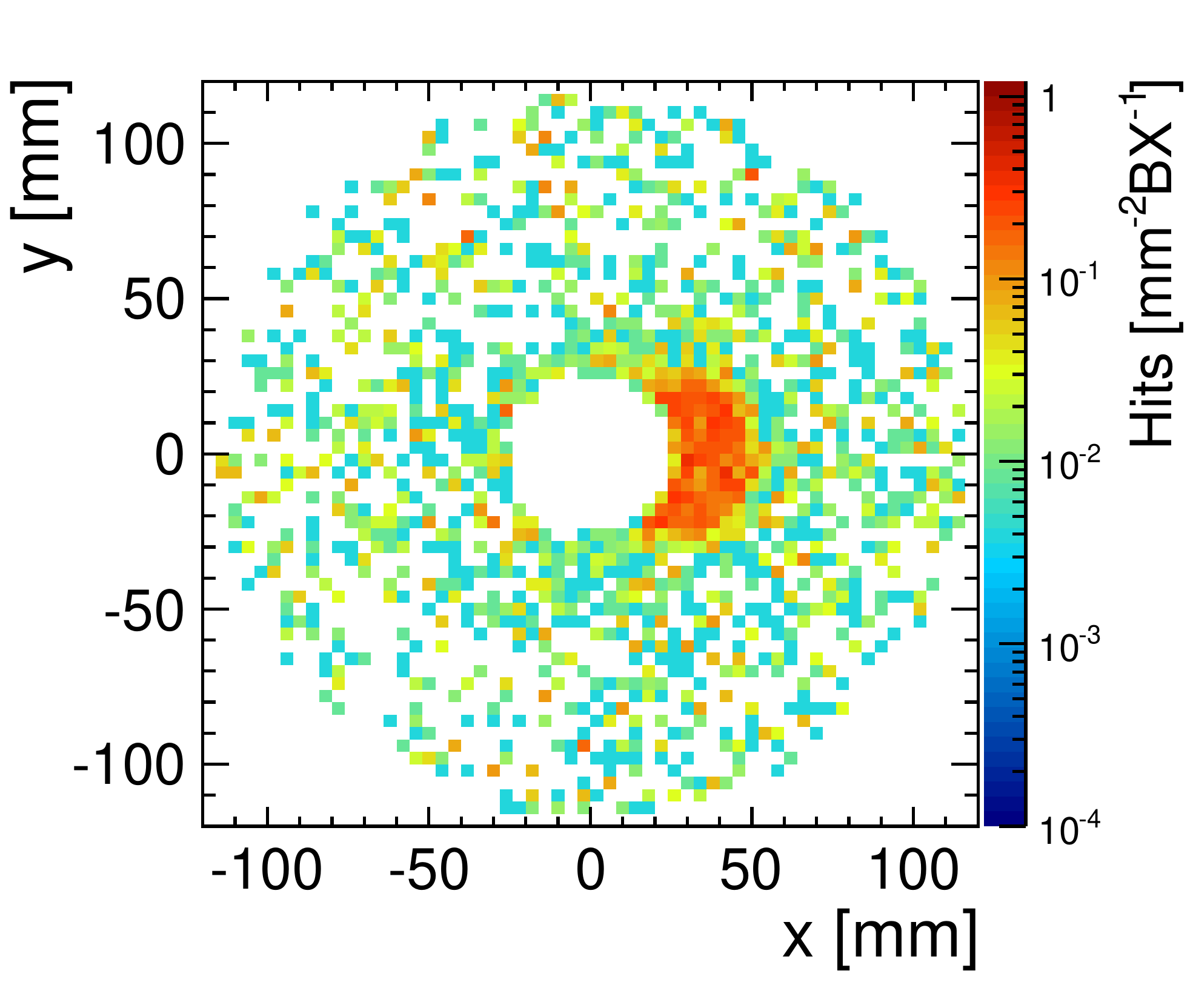}
  \caption{$t > \unit[15]{ns}$}\label{fig:SiD_hitDensities_incoh_xy_t2}
 \end{subfigure}
 \caption[Hit densities from incoherent pair background in the innermost vertex endcap layer depending on $x$ and $y$.]{Hit densities from incoherent pair background in the innermost vertex endcap layer depending on $x$ and $y$.}
 \label{fig:SiD_hitDensities_incoh_xy}
\end{figure}

The hits that originate from backscatters are not distributed uniformly. They have a strong dependency on the azimuthal angle $\phi$ originating from the asymmetric shape of the forward region. For example, backscatters created in the outgoing beam pipe can re-enter the detector only through the opening of the \ac{BeamCal}. This results in hot spots in the tracking detectors from the projection of this opening as shown in \cref{fig:SiD_hitDensities_incoh_xy}.

\begin{figure}
 \begin{subfigure}[]{0.49\textwidth}
  \includegraphics[width=\textwidth]{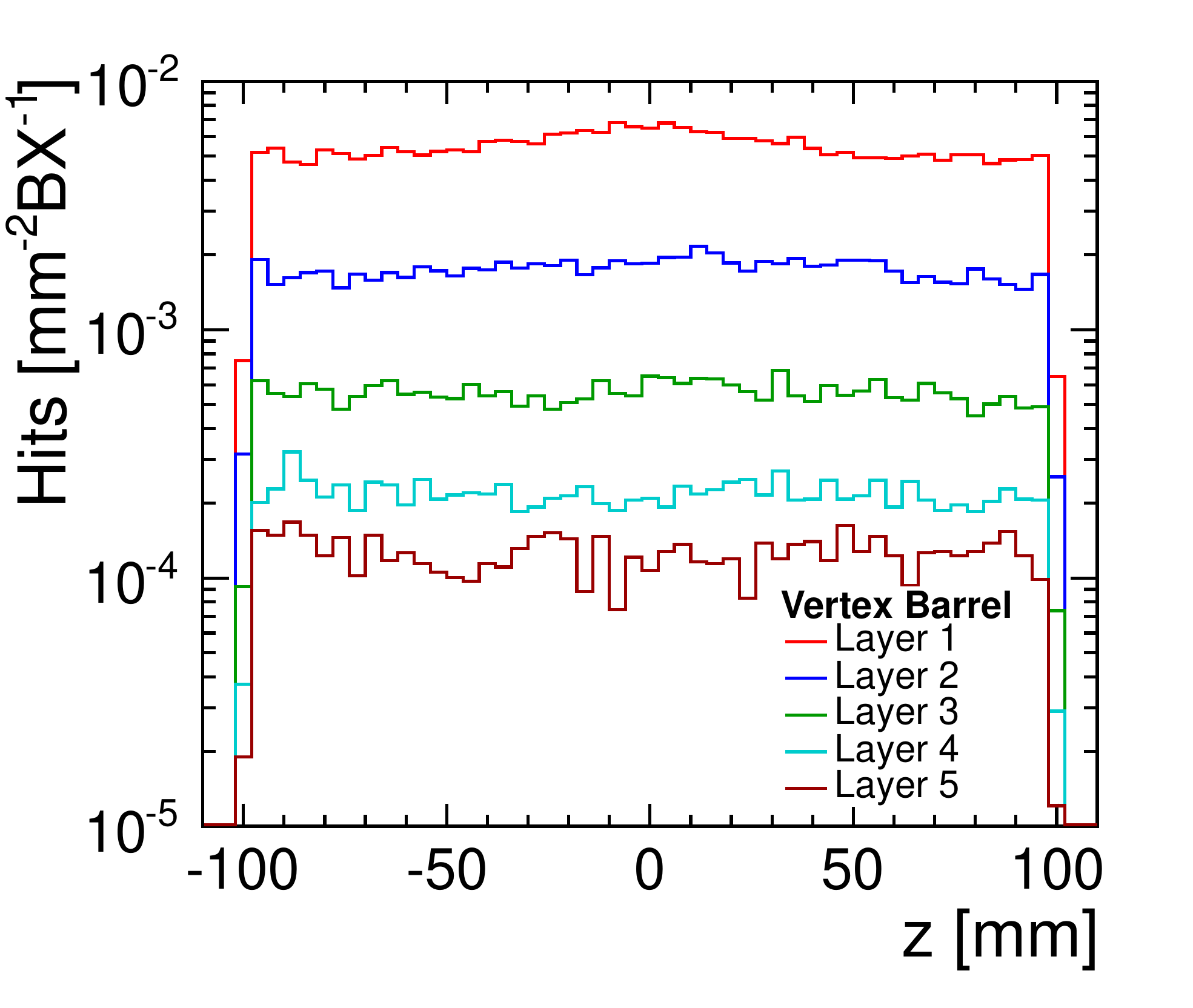}
  \caption{Incoherent pairs}\label{fig:SiD_hitDensitiesVertexBarrel_z_incoh}
 \end{subfigure}
 \hfill
 \begin{subfigure}[]{0.49\textwidth}
  \includegraphics[width=\textwidth]{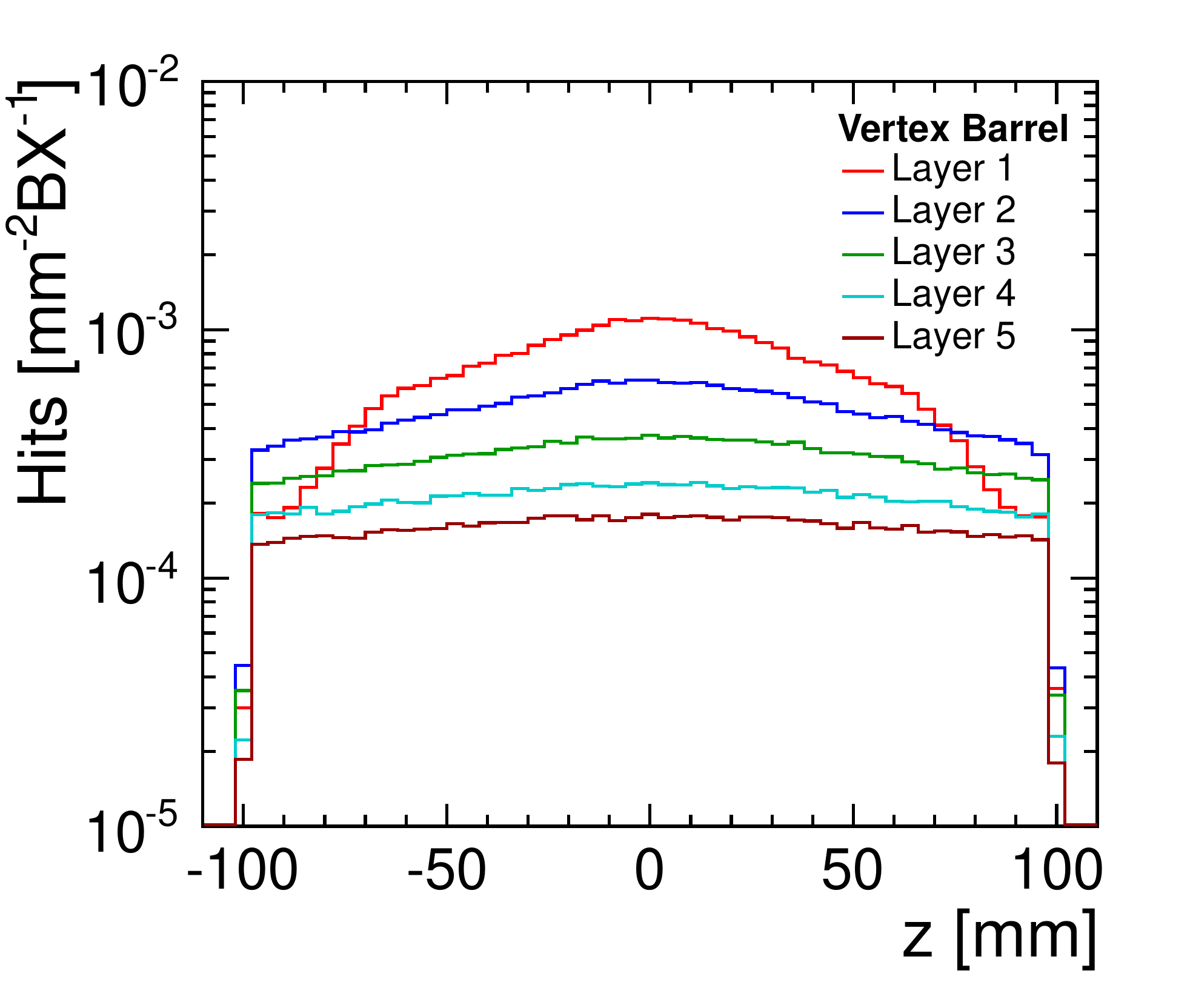}
  \caption{\gghad}\label{fig:SiD_hitDensitiesVertexBarrel_z_gghad}
 \end{subfigure}
 \caption[Hit densities in the vertex barrel layers from incoherent pairs and the \gghad background depending on $z$.]{Hit densities in the five vertex barrel layers from incoherent pairs \subref{fig:SiD_hitDensitiesVertexBarrel_z_incoh} and the \gghad background \subref{fig:SiD_hitDensitiesVertexBarrel_z_gghad} depending on $z$.}
 \label{fig:SiD_hitDensitiesVertexBarrel_z}
\end{figure}

\begin{figure}
 \begin{subfigure}[]{0.49\textwidth}
  \includegraphics[width=\textwidth]{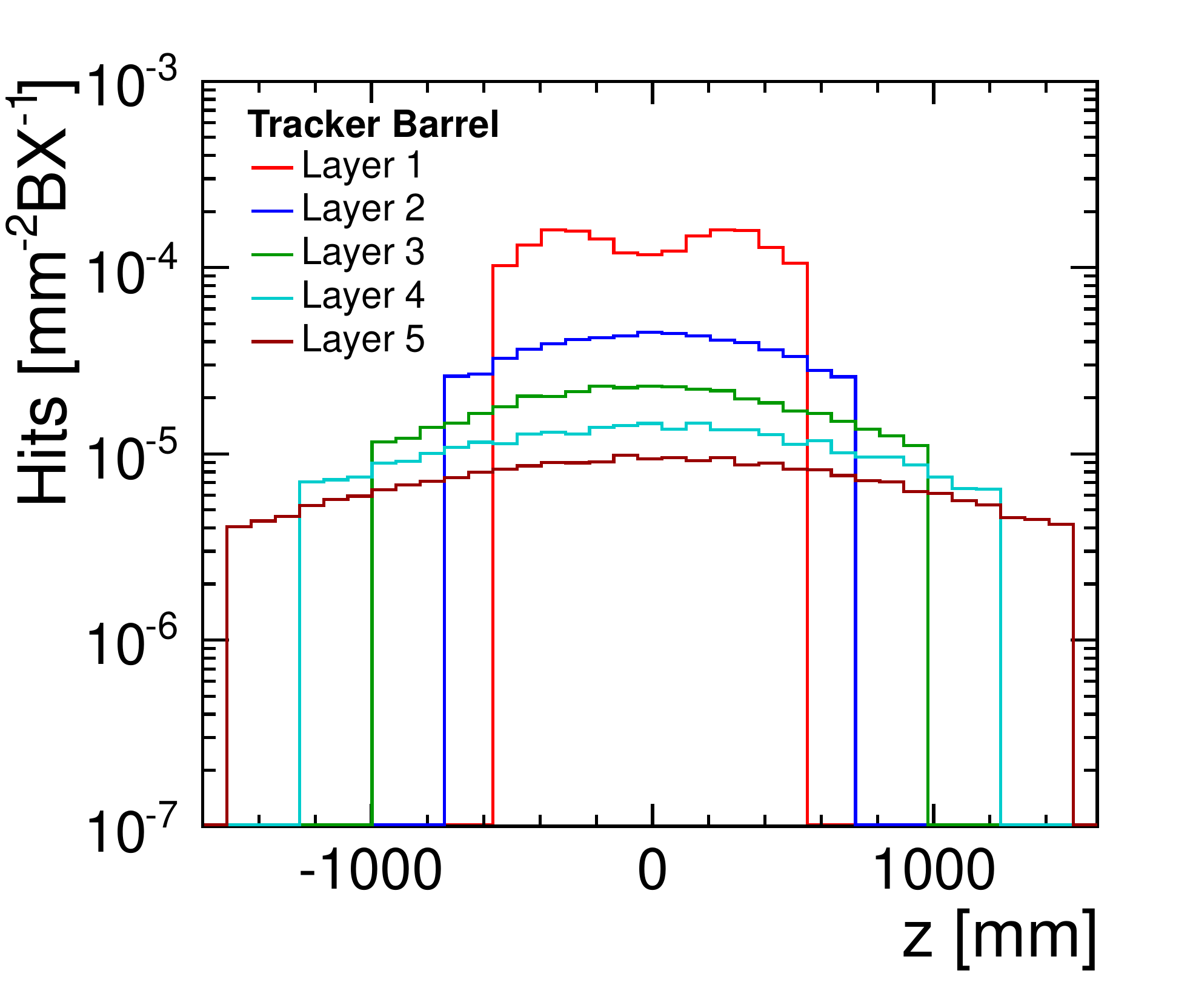}
  \caption{Incoherent pairs}\label{fig:SiD_hitDensitiesTrackerBarrel_z_incoh}
 \end{subfigure}
 \hfill
 \begin{subfigure}[]{0.49\textwidth}
  \includegraphics[width=\textwidth]{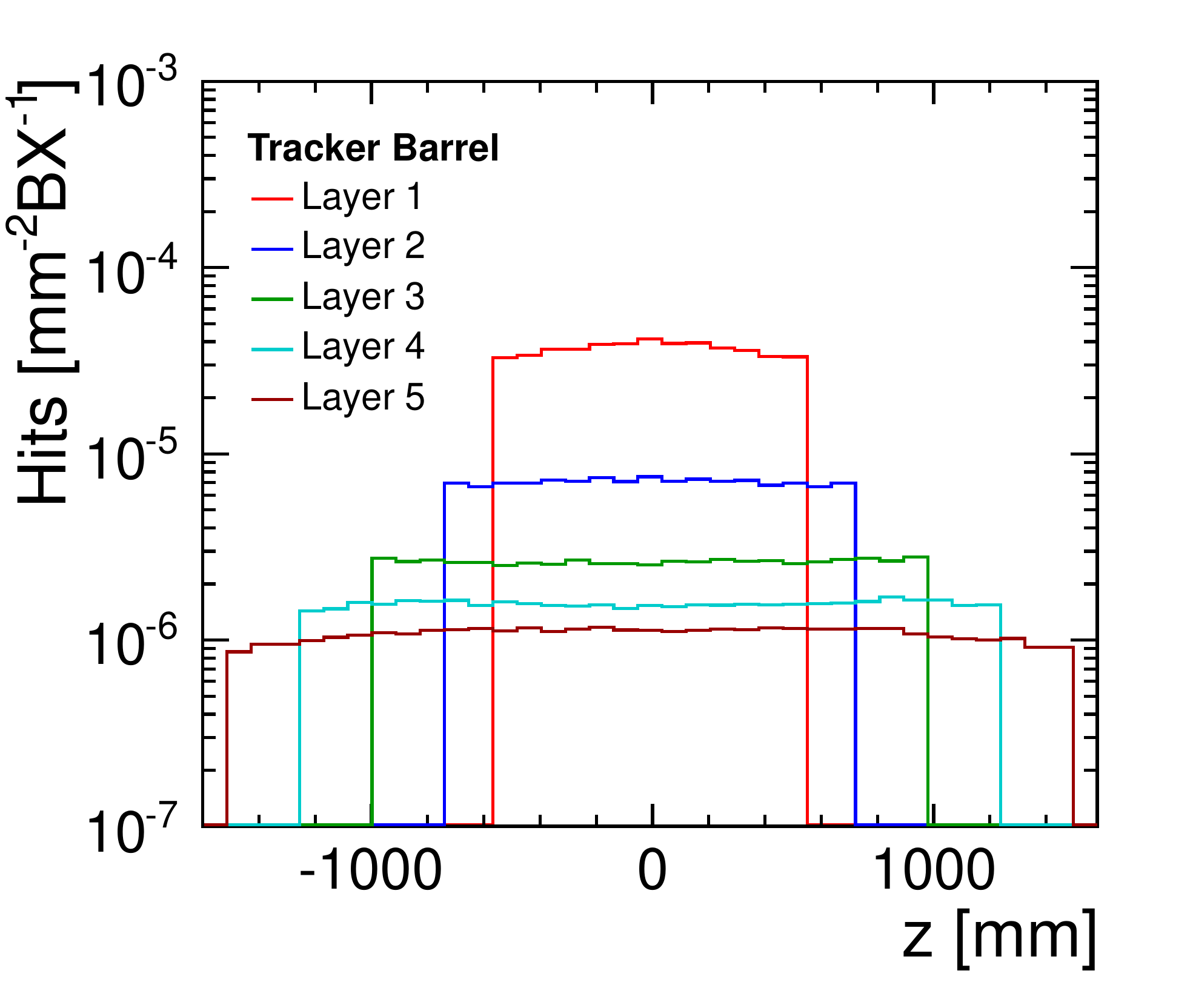}
  \caption{\gghad}\label{fig:SiD_hitDensitiesTrackerBarrel_z_gghad}
 \end{subfigure}
 \caption[Hit densities in the main tracker barrel layers from incoherent pairs and the \gghad background depending on $z$.]{Hit densities in the five main tracker barrel layers from incoherent pairs \subref{fig:SiD_hitDensitiesTrackerBarrel_z_incoh} and the \gghad background \subref{fig:SiD_hitDensitiesTrackerBarrel_z_gghad} depending on $z$.}
 \label{fig:SiD_hitDensitiesTrackerBarrel_z}
\end{figure}

\begin{figure}
 \begin{subfigure}[]{0.49\textwidth}
  \includegraphics[width=\textwidth]{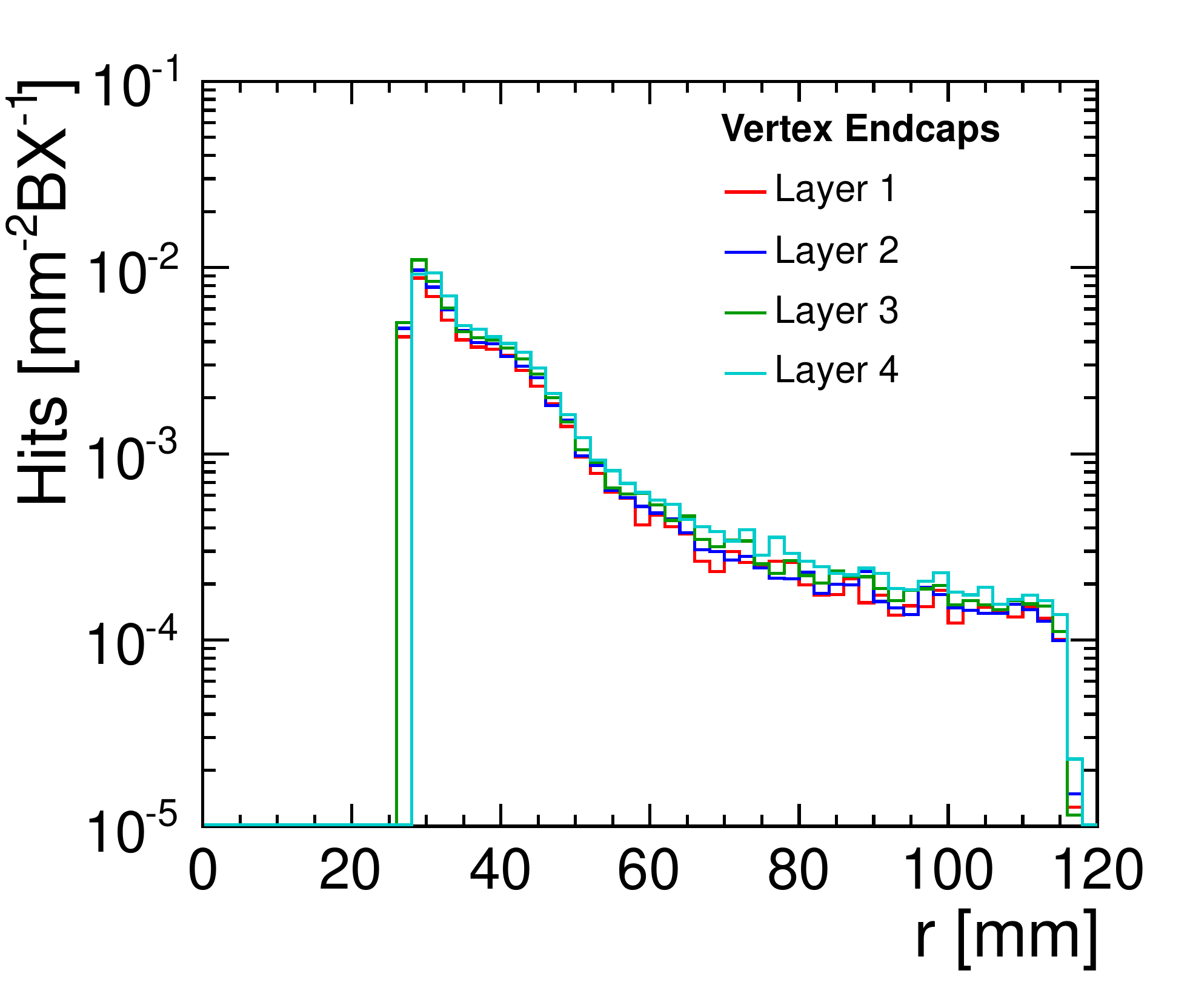}
  \caption{Incoherent pairs}\label{fig:SiD_hitDensitiesVertexEndcap_r_incoh}
 \end{subfigure}
 \hfill
 \begin{subfigure}[]{0.49\textwidth}
  \includegraphics[width=\textwidth]{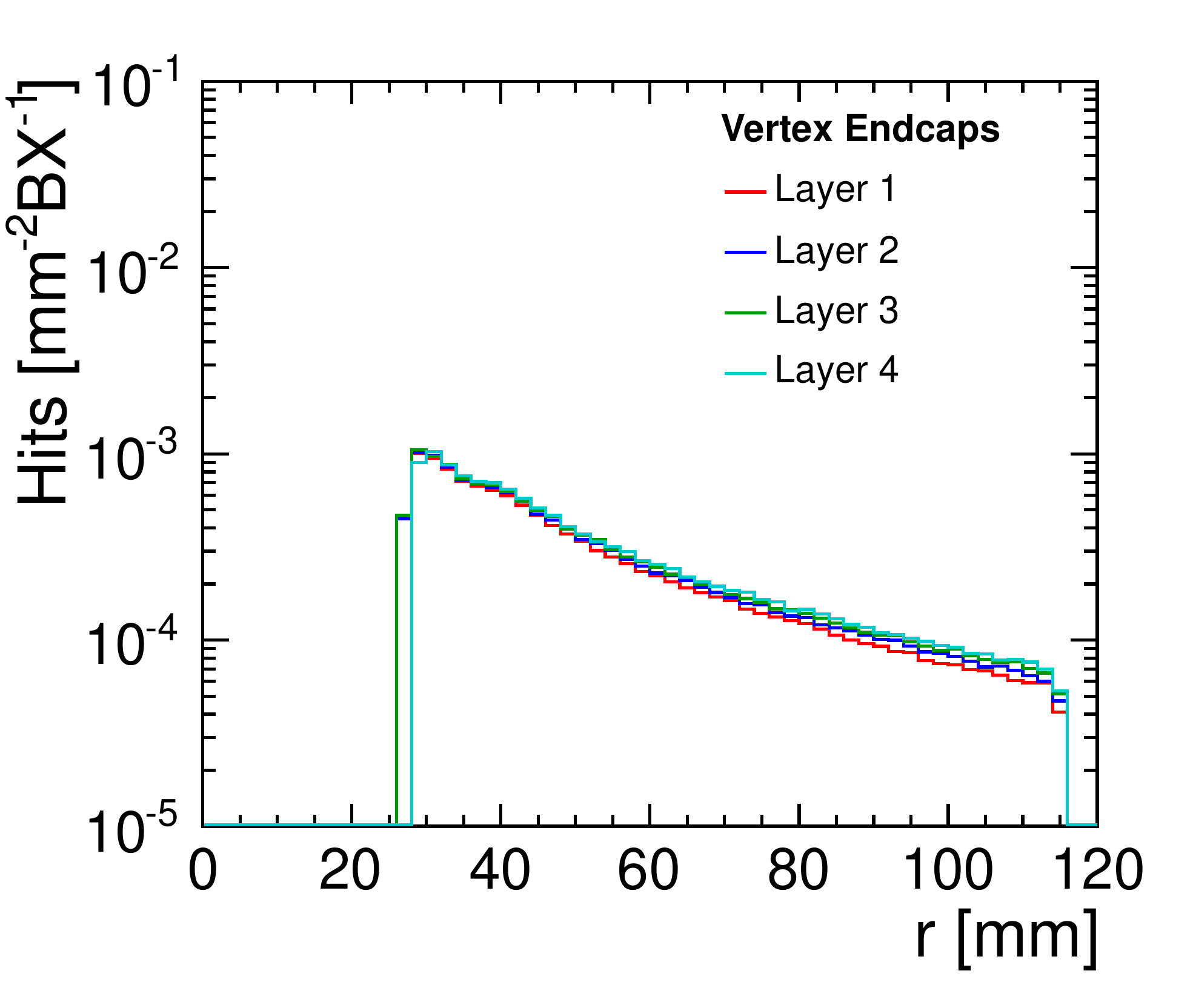}
  \caption{\gghad}\label{fig:SiD_hitDensitiesVertexEndcap_r_gghad}
 \end{subfigure}
 \caption[Hit densities in the vertex endcap layers from incoherent pairs and the \gghad background depending on $r$.]{Hit densities in the four vertex endcap layers from incoherent pairs \subref{fig:SiD_hitDensitiesVertexEndcap_r_incoh} and the \gghad background \subref{fig:SiD_hitDensitiesVertexEndcap_r_gghad} depending on $r$.}
 \label{fig:SiD_hitDensitiesVertexEndcap_r}
\end{figure}

\begin{figure}
 \begin{subfigure}[]{0.49\textwidth}
  \includegraphics[width=\textwidth]{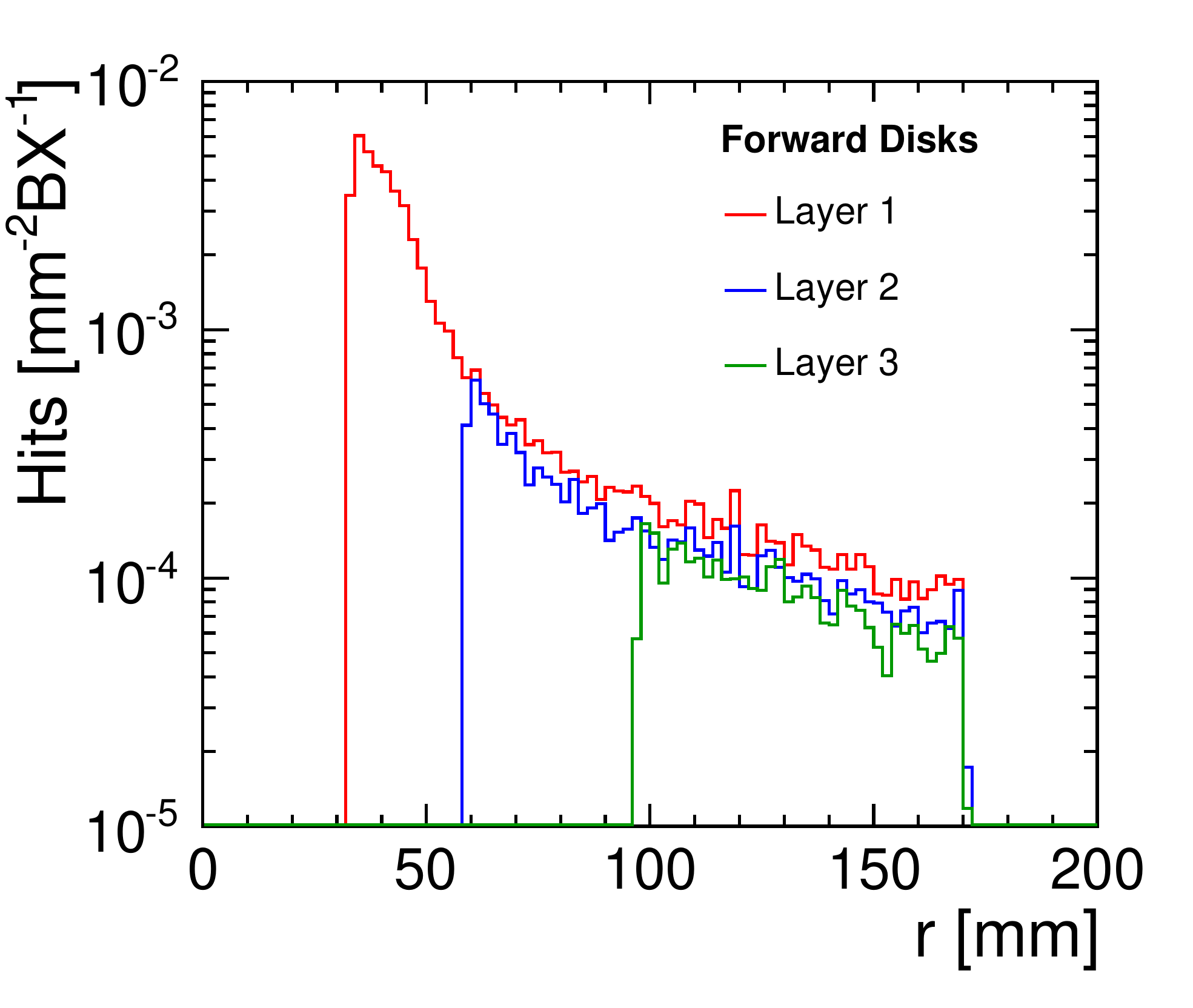}
  \caption{Incoherent pairs}\label{fig:SiD_hitDensitiesForwardDisks_r_incoh}
 \end{subfigure}
 \hfill
 \begin{subfigure}[]{0.49\textwidth}
  \includegraphics[width=\textwidth]{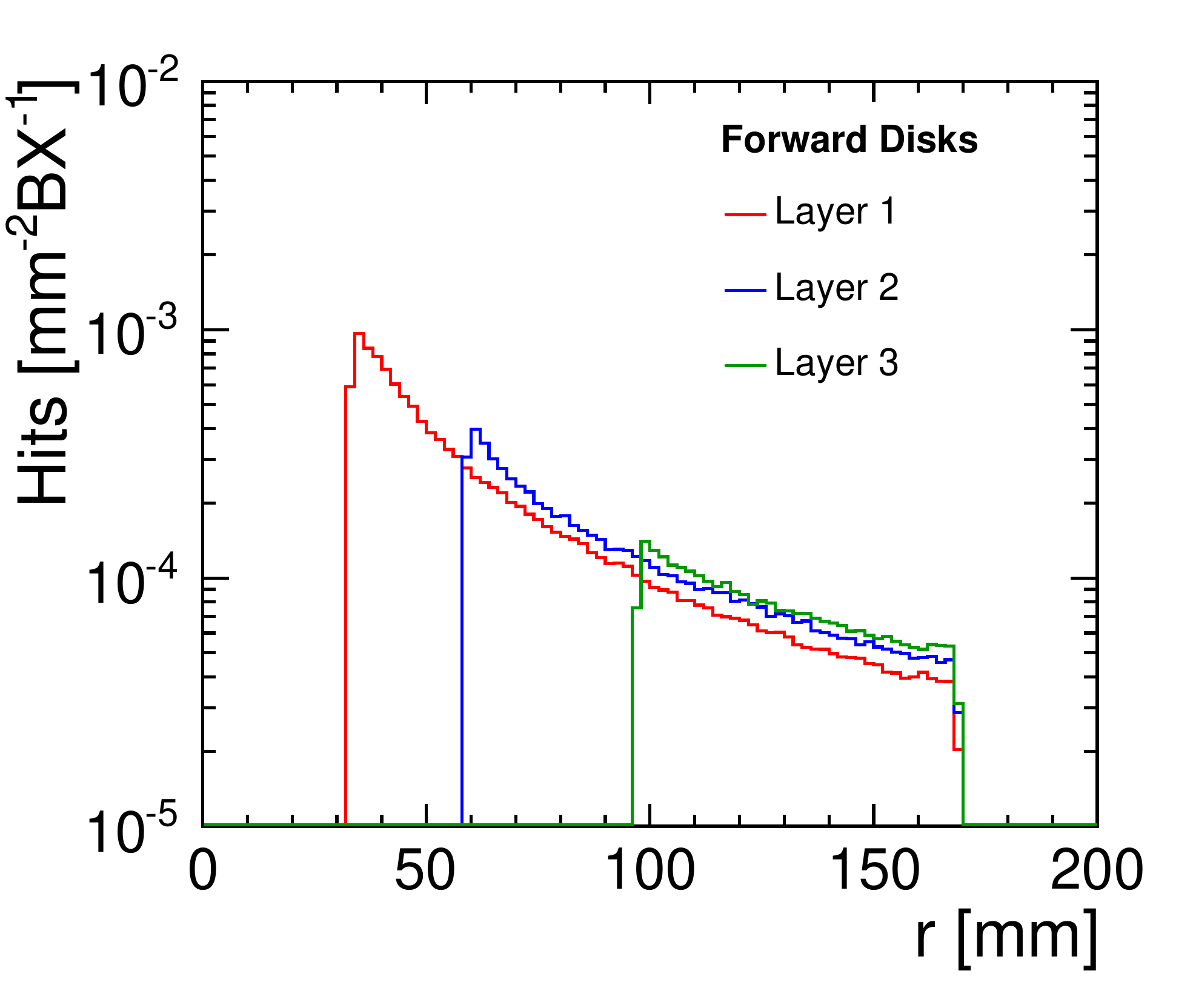}
  \caption{\gghad}\label{fig:SiD_hitDensitiesForwardDisks_r_gghad}
 \end{subfigure}
 \caption[Hit densities in the forward tracking pixel disks from incoherent pairs and the \gghad background depending on $r$.]{Hit densities in the three forward tracking pixel disks from incoherent pairs \subref{fig:SiD_hitDensitiesForwardDisks_r_incoh} and the \gghad background \subref{fig:SiD_hitDensitiesForwardDisks_r_gghad} depending on $r$.}
 \label{fig:SiD_hitDensitiesForwardDisks_r}
\end{figure}

\begin{figure}
 \begin{subfigure}[]{0.49\textwidth}
  \includegraphics[width=\textwidth]{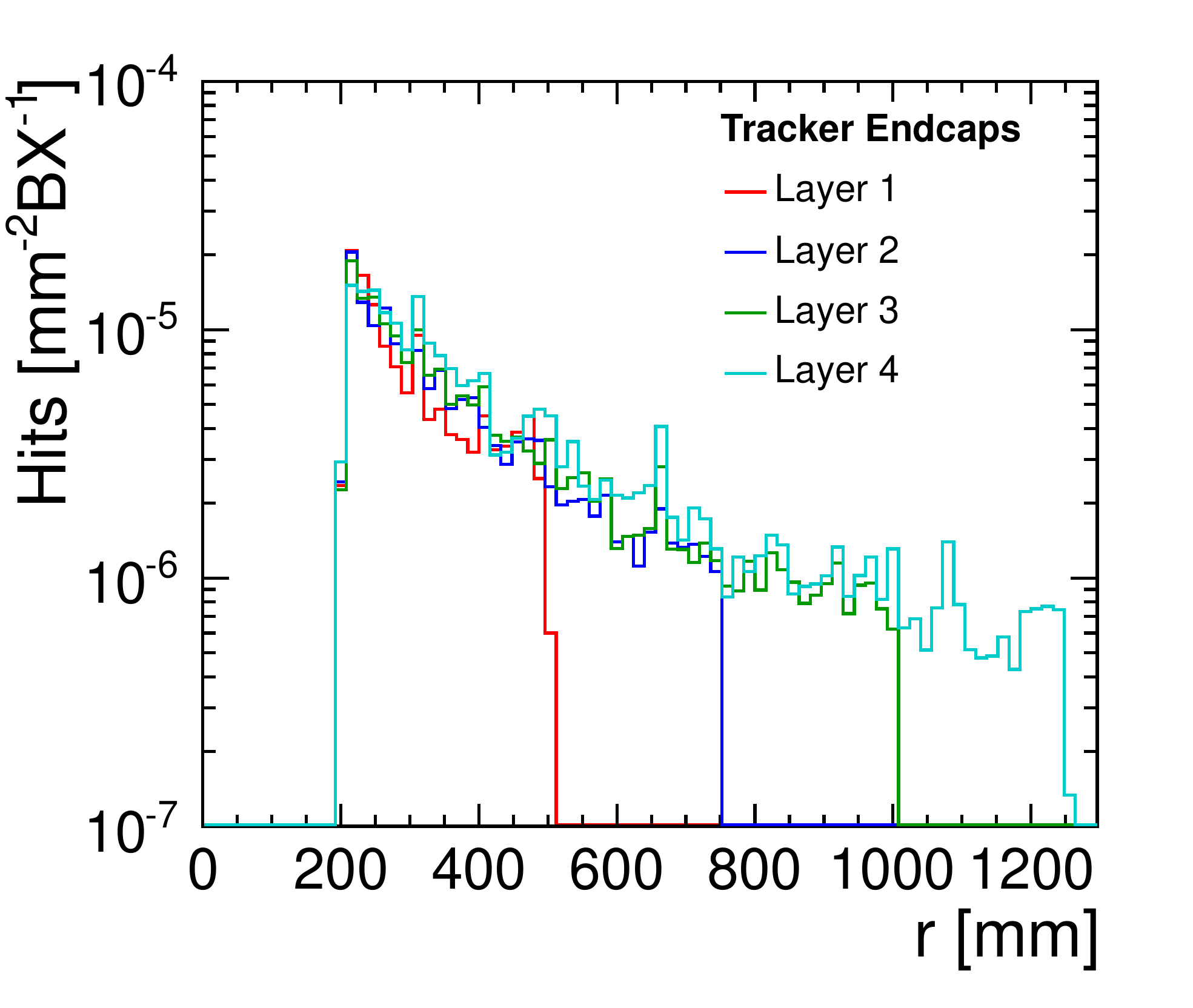}
  \caption{Incoherent pairs}\label{fig:SiD_hitDensitiesTrackerEndcap_r_incoh}
 \end{subfigure}
 \hfill
 \begin{subfigure}[]{0.49\textwidth}
  \includegraphics[width=\textwidth]{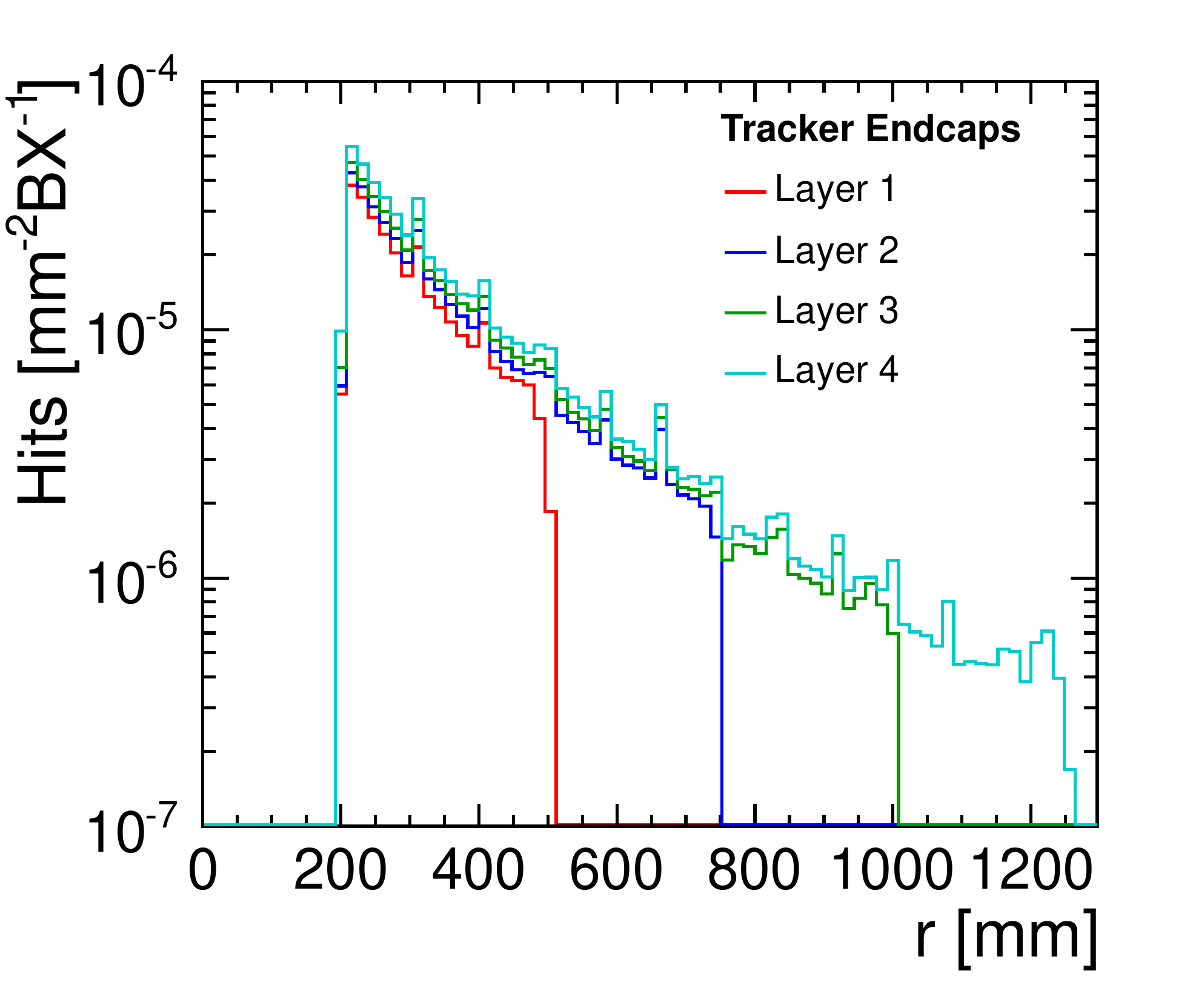}
  \caption{\gghad}\label{fig:SiD_hitDensitiesTrackerEndcap_r_gghad}
 \end{subfigure}
 \caption[Hit densities in the main tracker endcap layers from incoherent pairs and the \gghad background depending on $r$.]{Hit densities in the four main tracker endcap layers from incoherent pairs \subref{fig:SiD_hitDensitiesTrackerEndcap_r_incoh} and the \gghad background \subref{fig:SiD_hitDensitiesTrackerEndcap_r_gghad} depending on $r$.}
 \label{fig:SiD_hitDensitiesTrackerEndcap_r}
\end{figure}

The structure of the hits from the \gghad events is quite different from those from the incoherent pairs. The particles are created not as forward and no strong azimuthal dependence of the backscatters is present. The hit densities in all tracking detectors are shown for both backgrounds in \cref{fig:SiD_hitDensitiesVertexBarrel_z,fig:SiD_hitDensitiesTrackerBarrel_z,fig:SiD_hitDensitiesVertexEndcap_r,fig:SiD_hitDensitiesForwardDisks_r,fig:SiD_hitDensitiesTrackerEndcap_r}. In the innermost vertex barrel layers the amount of hits from incoherent pairs exceeds the number of hits for \gghad events by one order of magnitude. For higher radii the hit densities are comparable. This is also visible in the distributions of the hit densities in the vertex endcaps (see \cref{fig:SiD_hitDensitiesVertexEndcap_r}) and the forward tracking disks (see \cref{fig:SiD_hitDensitiesForwardDisks_r}) which are similar at all radii greater than \unit[50]{mm}. At lower radii the backscatters from the incoherent pair background are the dominating background. In the main tracker endcaps the \gghad background is dominating, as shown in \cref{fig:SiD_hitDensitiesTrackerEndcap_r}. While the incoherent pair background consists of individual particles the hadronic background consists of low energetic jets. Higher local densities lead to more ghost hits in the stereo strip detectors of the endcaps and thus result in the higher hit rates due to the \gghad background.

\clearpage

\subsection{Occupancies}
\label{sec:SiD_occupancies2}

\begin{figure}
 \centering
 \includegraphics[width=0.49\textwidth]{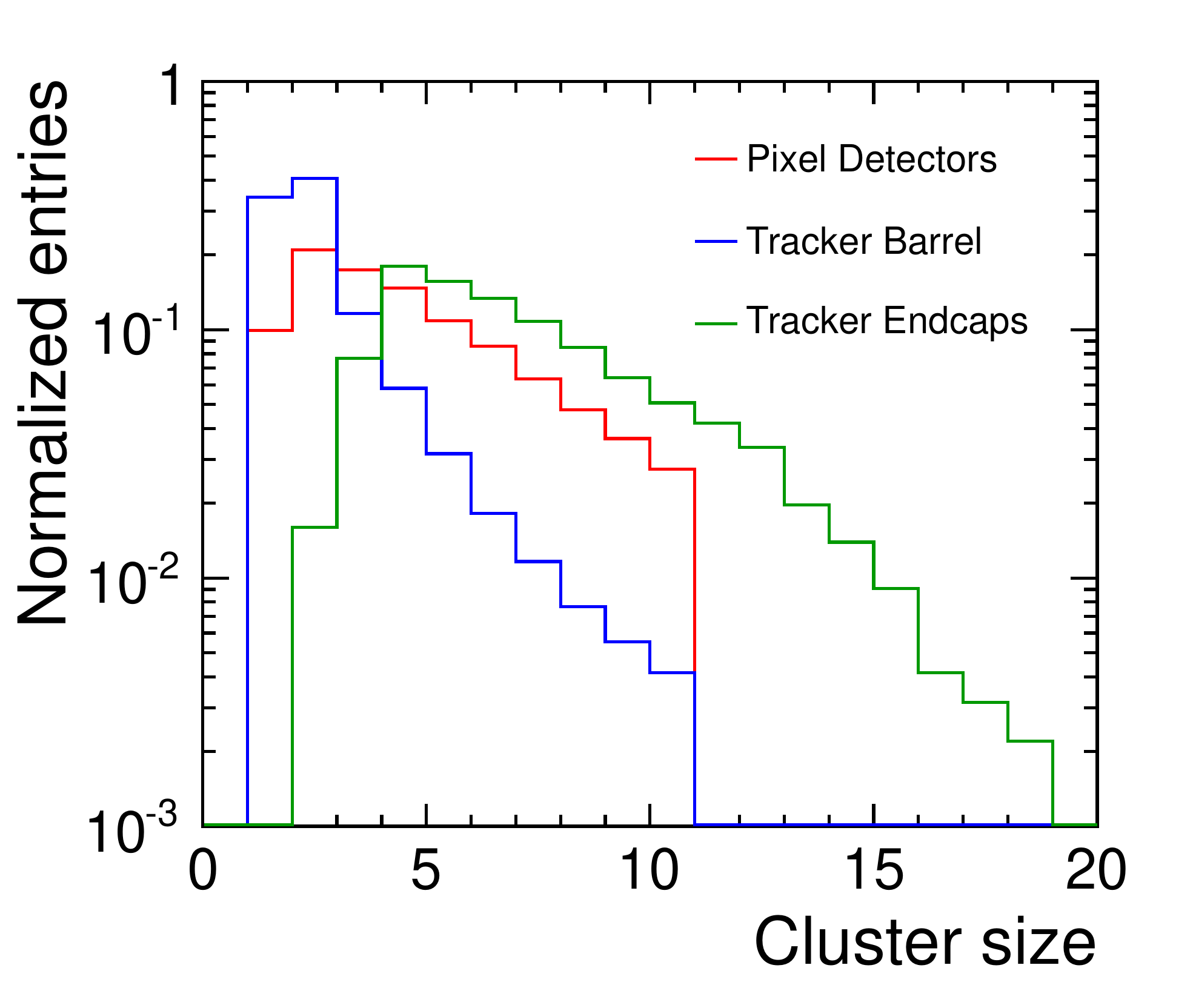}
 \caption{Distribution of the sizes of the reconstructed clusters in the tracking detectors.}
 \label{fig:SiD_clusterSize}
\end{figure}

From the hit densities shown above we can estimate the expected occupancies for the different subdetectors. We assume the nominal segmentation of the subdetectors, \ie $\unit[20\times20]{\micron^2}$ for the pixel detectors and $\unit[50]{\micron}\times\unit[985]{mm}$ for the strip detectors, and use the average reconstructed cluster sizes. The distribution of the reconstructed cluster sizes in the tracking detectors is shown in \cref{fig:SiD_clusterSize}. If a pixel is hit in the pixel detectors, all directly neighboring pixels are usually hit as well, resulting in an average cluster size of 4.6. Similarly, in the strip detectors the two neighboring strips are hit resulting in an average cluster size of 2.7. The clusters in the stereo strip detectors of the main tracker endcap detectors are more than twice that size. The number of hits seen in a stereo strip layer is the multiplication of the number of the hits in the two strip layers. The average cluster size in the stereo strip detectors is thus 7.2. For the occupancy we count each stereo layer as a single layer, we thus have to use 3.6 as the average cluster size. We apply an additional safety factor of 5 to the occupancy from incoherent pairs to account for uncertainties and fluctuations in the background rates. The amount of backscatters estimated from the full simulation constitutes the largest source of uncertainty and can not be easily verified. For the \gghad background we use a safety factor of 2, since it is causing significantly less amounts of backscatters.

Using these assumptions, the highest occupancies are reached in the innermost layer of the tracker barrel which experiences a maximum occupancy of 200--300\% from incoherent pairs, depending on the azimuthal angle and the position in $z$, when intergrating over a full bunch train. An occupancy of 15--30\% resulting from the \gghad background has to be added to this value. Similarly critical are the innermost regions of the main tracker endcaps which have an occupancy of up to 35\% from \gghad hits and an additional 30\% from incoherent pairs over a full bunch train. This occupancy drops quickly for larger radii and is below 5\% for $r > \unit[600]{mm}$. The pixel detectors are of much less concern. The innermost vertex barrel layer has an occupancy of 1.5--2.0\% for a full bunch train, depending on the azimuthal angle. The numbers for the pixel disks are similar and drop below 0.5\% for radii greater than \unit[50]{mm}.

Since we simulated and digitized only small fractions of a bunch train simultaneously, the numbers for the stereo strip detectors are underestimated. For occupancies as high as stated above a significant amount of additional ghost hits will be reconstructed. This problem does not occur for the other detectors where we overestimate the occupancy since some of the clusters would overlap.

It is evident that at least for the innermost main tracker barrel layer and the inner parts of the main tracker endcaps multi-hit readout capability is necessary to keep occupancies at a manageable level. Alternatively one could replace the modules in that region with pixel detectors, which could have larger pixels than those used in the vertex detectors. The desired time resolution for all sub detectors at \ac{CLIC} is of less than \unit[10]{ns}~\cite{cdrvol2} which would reduce the occupancies by a factor of approximately 15. It should be noted that the numbers obtained here agree very well with the study performed for \clicild for those detector regions which are dominated by direct background hits. For regions which are dominted by hits from backscattered particles we find significantly higher hit rates. Several optimizations of the layout of the forward region were applied to the \clicild model~\cite{sailerphd}, which are not present in the \clicsid model. This suggests that the hit rates at high radii are probably overestimated.

\subsection{Other Sources Of Backgrounds}
As discussed in \cref{sec:CLIC_machineInduced}, the coherent pair background and the trident pair background are only relevant at very low angles and do not have any direct impact on the tracking systems. Their impact on the tracking system through backscatters remains to be studied but is assumed to be covered within the safety factors. Another source of background particles are muons that can be created from the beam halo during the collimation in the \ac{BDS}. This background can be reduced by placing magnetized iron spoilers along the \ac{BDS} to deflect the muons. Simulation studies suggest that it is realistic to assume an average of one muon per \unit[20]{BX} traversing the detector volume~\cite{cdrvol1,cdrvol2}. This rate is completely negligible compared to the other backgrounds.

\section{Summary}
We have presented the \clicsid simulation model which was used for the \ac{CLIC} \ac{CDR} simulation studies as well as the studies presented in this thesis. The most important changes with respect to the \ac{SiD} detector model are the layout of the vertex region to avoid large amounts of direct hits from the beam-related backgrounds, the replacement of the \ac{HCal} barrel absorber material with a denser material together with an increase in the inner radius of the coil to accommodate a sufficiently deep calorimeter and the redesign of the forward region due to the required stability of the \ac{QD0}.

We have also studied the occupancies in the tracking detectors due to particles from incoherent pairs and the \gghad background. With the current design of the tracking systems the most critical regions are the innermost layer of the main tracker barrel and the inner modules of the main tracker endcaps. In both regions occupancies of 100\% or more are reached when integrating over a full bunch train and when including a safety factor of 5. In addition to the desired $\unit[<10]{ns}$ time resolution this could be mitigated by increasing the segmentation in these parts of the detectors.

\chapter{Simulating High Energy Collisions and Detector Response}
\label{cha:Software}

A problem including many convolved effects, which is probably even impossible to solve analytically, can be solved using the Monte~Carlo method. This method allows to determine the result of a complicated chain of steps if the probability distributions of the possible results for each step are known. Generating a random outcome for each of the steps yields a final result. Repeating this exercise will result in the probability distribution for the original problem.

Determining the response of a complicated detector model is an ideal example of the Monte~Carlo method. All possible interactions of the particles with matter are usually well understood and the problem arises only from the convolution of these countless effects. By repeatedly simulating the random passage of some particle through the model we can estimate the expected distribution of an observable. Another application of the Monte~Carlo method in \ac{HEP} is the event generation where it is used to determine the phase space distribution of particle interactions. Finally, a simulation offers the unique ability to disentangle effects in cases which are not directly accessible to a measurement, \eg a hypothesis can be tested by simulating the outcome of the experiment and comparing it with the observation. 

In this chapter we want to briefly discuss the different steps that are necessary for a realistic detector simulation and event reconstruction. First, we introduce the event generation step \cref{sec:Software_eventGeneration} which produces the samples used in all other steps. Afterwards, we discuss full detector simulations in \geant in \cref{sec:Software_Simulation}. The major part is the description of the event reconstruction in \cref{sec:Software_digitizationReconstruction}, where we put emphasis on the event overlay which is necessary for the simulation of \ac{CLIC} experimental conditions as well as the tracker hit digitization and the track reconstruction used later in \cref{cha:Tracking} 

\section{Event Generation}
\label{sec:Software_eventGeneration}
The first step in each physics simulation is the event generation, \eg the creation of a list of particles with all relevant attributes which are required to be properly propagated through the detector simulation.

\subsection{Physics Samples}
\label{sec:Software_physicsSampleGeneration}
Event samples that represent high energy particle collision processes have to be generated with correct distributions and correlations for all variables of interest, like angles, momentum and energy. These samples are usually used to compare measurements from experiments with expectations from physics models. They are also useful to estimate the performance of a future experiment to measure a certain quantity of an assumed model, like the possible measurement of the $\PSh \to \mpmm$ branching ratio at CLIC, presented in \cref{cha:Higgs}.

The event generator \whizard~\cite{Kilian:2007gr} is used for generating the jet samples used in the tracking studies in \cref{cha:Tracking}, the calorimetry studies in \cref{cha:Calorimetry_PFA} as well as for producing the physics samples for the $\PSh \to \mpmm$ branching ratio analysis. \whizard is optimized for calculating cross sections of \epem collisions. The matrix elements are calculated on tree-level and generated by \OMega~\cite{Moretti:2001zz}. The effect of the luminosity spectrum (see \cref{sec:CLIC_lumiSpectrum}) and the initial state radiation (ISR) are taken into account in the event generation. Both effects lead to lower energy of the colliding electrons and modified cross sections. The ISR photons are forced to be collinear with the outgoing beam. The hadronization of the quarks is handled in \pythia~\cite{Sjostrand2006}, which also takes care of the generation of final state radiation (FSR). The hadronization parameters in \pythia are tuned to the results of the OPAL experiment~\cite{berggren-pythia-tuning2010}. The decays of $\uptau$ leptons of some of the simulated background processes for \cref{cha:Higgs} are performed using TAUOLA~\cite{tauola}.

The file format used for the physics sample is the binary format \stdhep~\cite{stdhep}.

\subsection{Beam-Beam Interactions}
\label{sec:Software_beamBeam}
The generation of the samples of beam-induced backgrounds, which we discussed in \cref{sec:CLIC_machineInduced}, require several simulation steps. First the particles are tracked through the \ac{LINAC} and the \ac{BDS} as described in~\cite{Schulte:572820} to determine the beam parameters at the interaction point. Afterwards, the beam-beam interactions are simulated in \guineapig~\cite{Schulte:331845,Schulte:1999tx}, which produces the event samples for the produced electron pairs, photons and hadrons as well as a file containing the luminosity spectrum. The hadronic pairs have a minimum invariant mass of \unit[2]{GeV}. After the event generation they have to be run through \pythia for the fragmentation of the hadrons. More details on the simulation of the \gghad background are given in~\cite{barklow_bg}.

\subsection{Individual Particles}
\label{sec:Software_particleGun}
Fundamental detector parameters like resolutions and efficiencies are usually obtained using dedicated samples of certain particles with a well defined angular and energy distribution. The \ac{GPS} provided by \geant (see below) allows to generate various particle types at fixed angles and energies or following user provided distributions. Event samples generated by the general particle source are used for example in the tracking studies in \cref{cha:Tracking}, the calorimeter layout optimization in \cref{cha:Calorimetry} and the calorimeter calibration in Chapter~\ref{cha:Calorimetry_PFA}.

\section{Detector Simulation}
\label{sec:Software_Simulation}
The simulation of the interaction of the particles with the detector model is performed using the program \slic~\cite{Graf:2006ei,slic}. \slic is a thin wrapper around \geant~\cite{Agostinelli2003,Allison2006} that allows to easily define a detector geometry in an \acs{XML} format. The initial particles for the simulation can either be defined by a file in the \stdhep format or by using \geant \ac{GPS} interface. The output files containing the simulated events are produced in the \lcio~\cite{Gaede:2003ip,lcio} file format. All simulations presented in this thesis use \slic version $2.9.8$ together with \geant version $9.3.2$.

Since the events are generated in \whizard as head-on collisions the crossing angle of the beams at the interaction point has to be taken into account in the detector simulation. All initial particles are boosted according to the crossing angle of \unit[20]{mrad} at the beginning of the simulation.

\subsection{Particle Transport in \geant}
\label{sec:Software_geantParticleTransport}
The particles are propagated by \geant through the detector volume in discrete steps. The length of each step is the shortest value of either the mean free path length in the respective material, the distance to the boundary of the respective volume or a defined maximum step length. The interaction of the particle within each step is then randomly selected from a list of possible processes defined in the physics list (see below), according to their probability. The energy loss in this step is then calculated accordingly and the particle momentum and position after the step are set. If one or more secondary particles are created, e.g by Bremsstrahlung or pair production, they are propagated individually through the detector afterwards, starting from the position of their production. If the energy of a particle is below a certain minimum energy, all its energy is deposited in the step and the particle is stopped. More precise results are achieved for shorter step lengths and lower minimum energy limits. This will result in longer computing time required to simulate the event such that the parameters are always a trade-off between precision and performance of the simulation.

For those detector volumes that are declared as active detectors, the energy deposited by a particle traversing the volume is written out as a hit. In addition to the energy, the simulation hit also stores the time and a reference to the particle(s) that created the hit. Each energy deposit in the tracking detectors creates an individual hit, since the volumes are not segmented into strips or pixels within the simulation model. The calorimeters, on the other hand, are segmented into individual cells and each hit is a list of all energy deposits within the respective cell during the simulation of the event.

\subsection{Physics Lists in \geant}
\label{sec:Software_physicsList}
\geant offers various models to describe the different interactions of particles with matter. Reference physics lists offer validated combinations of these models and the choice of the physics list depends on the focus of the simulation. The \qgspbert physics list is the reference physics list used for the simulation for the \ac{LHC} experiments. It has been shown that this list offers good agreement between simulations and data obtained from beam tests for the energy and the shower shape, \eg in~\cite{Piperov:2008ui}.

The models can be grouped into models for electromagnetic interactions and hadronic interactions. The electromagnetic models are fairly well understood and the same combination of models is used in almost all of the reference physics lists. These models include descriptions for energy loss through ionization, multiple Coulomb scattering, Bremsstrahlung, pair production and annihilation of anti-electrons. The models for photons include the photoelectric effect, Compton scattering and conversion into $\epem$ pairs. Details about the electromagnetic models can be found in~\cite{Ivanchenko:2011} and references therein.

For the inelastic scattering of hadrons a wide range of models exist in \geant. The \qgspbert physics list uses the Bertini cascade model~\cite{Bertini1963,Binary2004} for energies below \unit[9.9]{GeV} and the quark-gluon string model~\cite{Toneev:1989pu,Folger2003} for energies higher than \unit[12]{GeV}. As a transition, a low energy parametrization model is used from \unit[9.5--25]{GeV} and in regions where more than one model is defined, one is randomly chosen for each interaction. The elastic scatterings involving hadrons are modeled using the \ac{CHIPS}~\cite{Degtyarenko:2000ks,Degtyarenko:2000er,Degtyarenko:2000es}. In addition, models for neutron capture and neutron induced fission are included in \qgspbert. All details concerning this and the other hadronic physics lists are given in~\cite{geant4physicsManual}.

Most physics lists only differ in details and the level of agreement between data and simulation depends on which observable is of interest. For example, the jet energy resolution achieved with \pandora (see \cref{sec:Software_particleFlow}) is not very sensitive to the choice of the physics list~\cite{Thomson:2009rp}. In addition, these differences only matter when comparing simulations with data, especially in high precision measurements, but usually do not apply to the simulation studies currently being performed for future linear colliders. One notable exception is the \qgspberthp physics list. It offers precise neutron tracking down to lowest energies and is used in the \ac{HCal} study in \cref{cha:Calorimetry}.

\section{Event Digitization and Reconstruction}
\label{sec:Software_digitizationReconstruction}
The hits created during the detector simulation have to be passed through various pattern recognitions to reconstruct the original event. These include track reconstruction and identification of clusters in the calorimeters which are then combined into reconstructed particles. To simulate a realistic event reconstruction, the hits have to be passed through a digitization step that models the response of an actual detector to the deposited energy. This can include threshold effects, noise, cross-talk of neighboring channels, smearing of the signal and conversion into a digital signal with limited precision.

The digitization and reconstruction is performed in several steps using the \lcsim~\cite{lcsim} framework and the slicPandora software package. The individual steps that are performed are explained below.



\subsection{Overlay of Beam-Induced Background}
\label{sec:Software_BackgroundOverlay}
An essential part of realistic detector simulations for CLIC is the inclusion of the high levels of beam-induced backgrounds. This allows to estimate the impact of hits and energy deposits from background particles on the pattern recognition in the tracking system or the reconstruction of clusters in the calorimeters. The background overlay has to be performed before the digitization step of the event reconstruction to correctly account for a realistic energy deposit within each channel. This is particularly important when incorporating measurement thresholds and saturation effects.

For the CLIC detector studies the background events are simulated independently and then merged with the physics events at a later stage. This approach removes the overhead in computing resources used when simulating background particles together with a physics event. It also allows to study the impact of the background on an event by event basis since each event can be reconstructed with and without the background particles.

We have developed a software tool for \lcsim that allows for overlaying events from different simulated events in a flexible way. The \textit{OverlayDriver}~\cite{lcd:grefesidoverlay2011} allows to merge a user defined number of background events with a signal event by merging the individual simulated hits. Signal and background events are shifted in time with respect to each other if desired, thus, effectively recreating the beam structure as seen by the detector. The time of each hit is shifted depending on when the background event happened with respect to the signal event. In addition, a readout time window can be defined for every collection\footnote{In the simulation each subdetector writes its hits to a different collection. A collection in \lcio is an ordered list of arbitrary objects which can have meta data attached to it.}.

\subsubsection{Time Structure and Amount of Background}
\label{sec:Software_BackgorundOverlayBackgroundWeight}

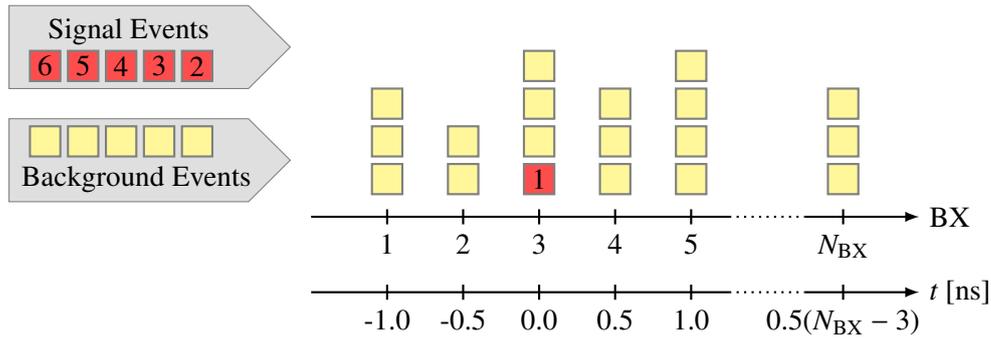
\begin{figure}[htpb]
 \centering
 \begin{tikzpicture}
    [signal/.style={rectangle, draw=gray, fill=red!70, thick, inner sep=0pt, minimum size=4mm},
     background/.style={rectangle, draw=gray, fill=yellow!50, thick, inner sep=0pt, minimum size=4mm},
     >=latex]

  \node at (-3.4,1.75) [shape=single arrow, draw=gray, fill=gray!25, minimum height=3.7cm, minimum width=1.1cm, single arrow head extend=0pt] (signal arrow) {};
  \node[above] at (-3.4,1.65) {Signal Events};

  \node at (-4.5,1.5) [signal] {6};
  \node at (-4.0,1.5) [signal] {5};
  \node at (-3.5,1.5) [signal] {4};
  \node at (-3.0,1.5) [signal] {3};
  \node at (-2.5,1.5) [signal] {2};

  \node at (-3.4,0.25) [shape=single arrow, draw=gray, fill=gray!25, minimum height=3.7cm, minimum width=1.1cm, single arrow head extend=0pt] (signal arrow) {};
  \node[below] at (-3.3,0.30) {Background Events};

  \node at (-4.5,0.5) [background] {};
  \node at (-4.0,0.5) [background] {};
  \node at (-3.5,0.5) [background] {};
  \node at (-3.0,0.5) [background] {};
  \node at (-2.5,0.5) [background] {};

  \draw[thick] (-1, -0.5) -- (4.5, -0.5);
  \draw[thick, dotted] (4.5, -0.5) -- (5.5, -0.5);
  \draw[thick, ->] (5.5, -0.5) -- (7, -0.5);
  \node[anchor=west] at (7, -0.5) {BX};

  \draw[thick] (0.0, -0.4) -- (0.0,-0.6);
  \draw[thick] (1.0, -0.4) -- (1.0,-0.6);
  \draw[thick] (2.0, -0.4) -- (2.0,-0.6);
  \draw[thick] (3.0, -0.4) -- (3.0,-0.6);
  \draw[thick] (4.0, -0.4) -- (4.0,-0.6);
  \draw[thick] (6.0, -0.4) -- (6.0,-0.6);
  \node[below] at (0.0,-0.6) {1};
  \node[below] at (1.0,-0.6) {2};
  \node[below] at (2.0,-0.6) {3};
  \node[below] at (3.0,-0.6) {4};
  \node[below] at (4.0,-0.6) {5};
  \node[below] at (6.0,-0.6) {$N_\mathrm{BX}$};

  \draw[thick] (-1, -1.5) -- (4.5, -1.5);
  \draw[thick, dotted] (4.5, -1.5) -- (5.5, -1.5);
  \draw[thick, ->] (5.5, -1.5) -- (7, -1.5);
  \node[anchor=west] at (7, -1.5) {$t$ [ns]};

  \draw[thick] (0.0, -1.4) -- (0.0,-1.6);
  \draw[thick] (1.0, -1.4) -- (1.0,-1.6);
  \draw[thick] (2.0, -1.4) -- (2.0,-1.6);
  \draw[thick] (3.0, -1.4) -- (3.0,-1.6);
  \draw[thick] (4.0, -1.4) -- (4.0,-1.6);
  \draw[thick] (6.0, -1.4) -- (6.0,-1.6);
  \node[below] at (0.0,-1.6) {-1.0};
  \node[below] at (1.0,-1.6) {-0.5};
  \node[below] at (2.0,-1.6) { 0.0};
  \node[below] at (3.0,-1.6) { 0.5};
  \node[below] at (4.0,-1.6) { 1.0};
  \node[below] at (6.0,-1.6) {$0.5 (N_\mathrm{BX}-3)$};

  \node at (0,0.0) [background] {};
  \node at (0,0.5) [background] {};
  \node at (0,1.0) [background] {};

  \node at (1,0.0) [background] {};
  \node at (1,0.5) [background] {};

  \node at (2,0.0) [signal] {1};
  \node at (2,0.5) [background] {};
  \node at (2,1.0) [background] {};
  \node at (2,1.5) [background] {};

  \node at (3,0.0) [background] {};
  \node at (3,0.5) [background] {};
  \node at (3,1.0) [background] {};

  \node at (4,0.0) [background] {};
  \node at (4,0.5) [background] {};
  \node at (4,1.0) [background] {};
  \node at (4,1.5) [background] {};

  \node at (6,0.0) [background] {};
  \node at (6,0.5) [background] {};
  \node at (6,1.0) [background] {};
 \end{tikzpicture}
 \caption[Example of a bunch train structure created by the \emph{OverlayDriver}.]{Example of a bunch train structure created by the \emph{OverlayDriver} assuming a time of \unit[0.5]{ns} between two bunch crossings. One signal event is placed at a random bunch crossing. A random number of background events is added to each bunch crossing in the train. The signal event determines the reference time.}
 \label{fig:Software_BackgroundOverlayScheme}
\end{figure}

The beam structure is simulated in the \textit{OverlayDriver} as follows. First, the time between two bunch crossings and the total length of a train are defined. The signal event is then either placed randomly at one of the bunch crossings or at a user defined bunch crossing. By default the time of the collision in the signal event, $t_\mathrm{signal}$, is chosen as the reference time $t_0 = \unit[0]{ns}$. Afterwards, for each bunch crossing in the train a number is drawn from a Poisson distribution to determine the number of background events that occur at that bunch crossing.
The corresponding number of background events is then merged successively with the signal event, shifting the times of all hits and particles according to the time of the bunch crossing. A schematic display of the resulting time structure is shown in \cref{fig:Software_BackgroundOverlayScheme}.

A bunch train of 60 bunch crossings with a bunch spacing of \unit[0.5]{ns} was used for the background overlay in the \ac{CLIC} \ac{CDR} and the studies presented in this thesis. This corresponds only to approximately one fifth of the bunch train at \ac{CLIC} but is sufficient to have realistic background levels throughout the readout time windows of \unit[10]{ns}, as described below. In these studies only the \gghad background is included systematically with an expectation value of 3.2 events per bunch crossing. The signal event is placed at the tenth bunch crossing within the bunch train to guarantee realistic amounts of background throughout the whole time window.

\subsubsection{Read-out Time Windows}
\label{sec:Software_BackgroundOverlayTimeWindows}
In a \ac{CLIC} experiment the readout will be triggerless. Each subdetector will read out continuously for the duration of the train and a time stamp of a certain resolution will be applied to each signal. The reconstruction software will then have to reconstruct the full event in sliding time windows by looking at different combinations of adjacent readout cycles to identify the hard interaction. This procedure is not implemented in the current reconstruction software. Read-out time windows can be defined instead for each subdetector, assuming that the hard interaction can be identified and constrained to the corresponding time window. These time windows present the subsequent digitization and reconstruction software following the \emph{OverlayDriver} with a single event.

The readout time windows have a defined length for each of the collections present in the event, $\Delta t_\mathrm{collection}$. All hits outside the respective time window are removed. The starting point of the time windows is defined by $t_\mathrm{signal}$. This simplification assumes that the correct time window of the signal has been identified and avoids removing late parts of the developing showers. In addition, a time-of-flight correction based on the hit position $\vec{x}_\mathrm{hit}$ is applied, assuming that the particles move with the speed of light $c$ along the line-of-sight from the interaction point. The time window for a certain hit is defined by its limits $t_\mathrm{min}$ and $t_\mathrm{max}$ which can be written as
\begin{eqnarray}
 t_\mathrm{min} &=& t_\mathrm{signal} + \frac{|\vec{x}_\mathrm{hit}|}{c}, \\
 t_\mathrm{max} &=& t_\mathrm{min} + \Delta t_\mathrm{collection}.
\end{eqnarray}

\begin{figure}
 \includegraphics[width=0.32\textwidth]{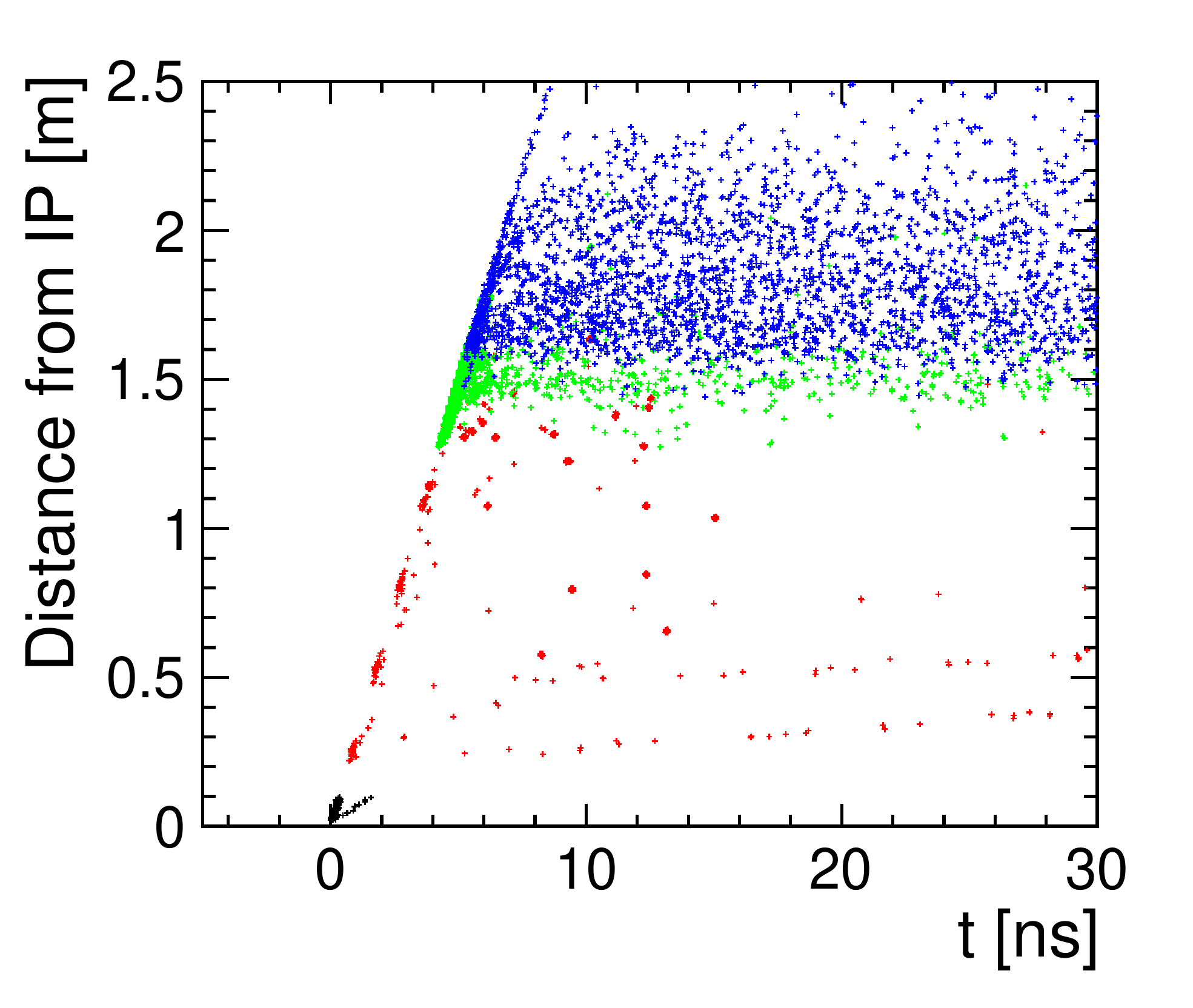}
 \hfill
 \includegraphics[width=0.32\textwidth]{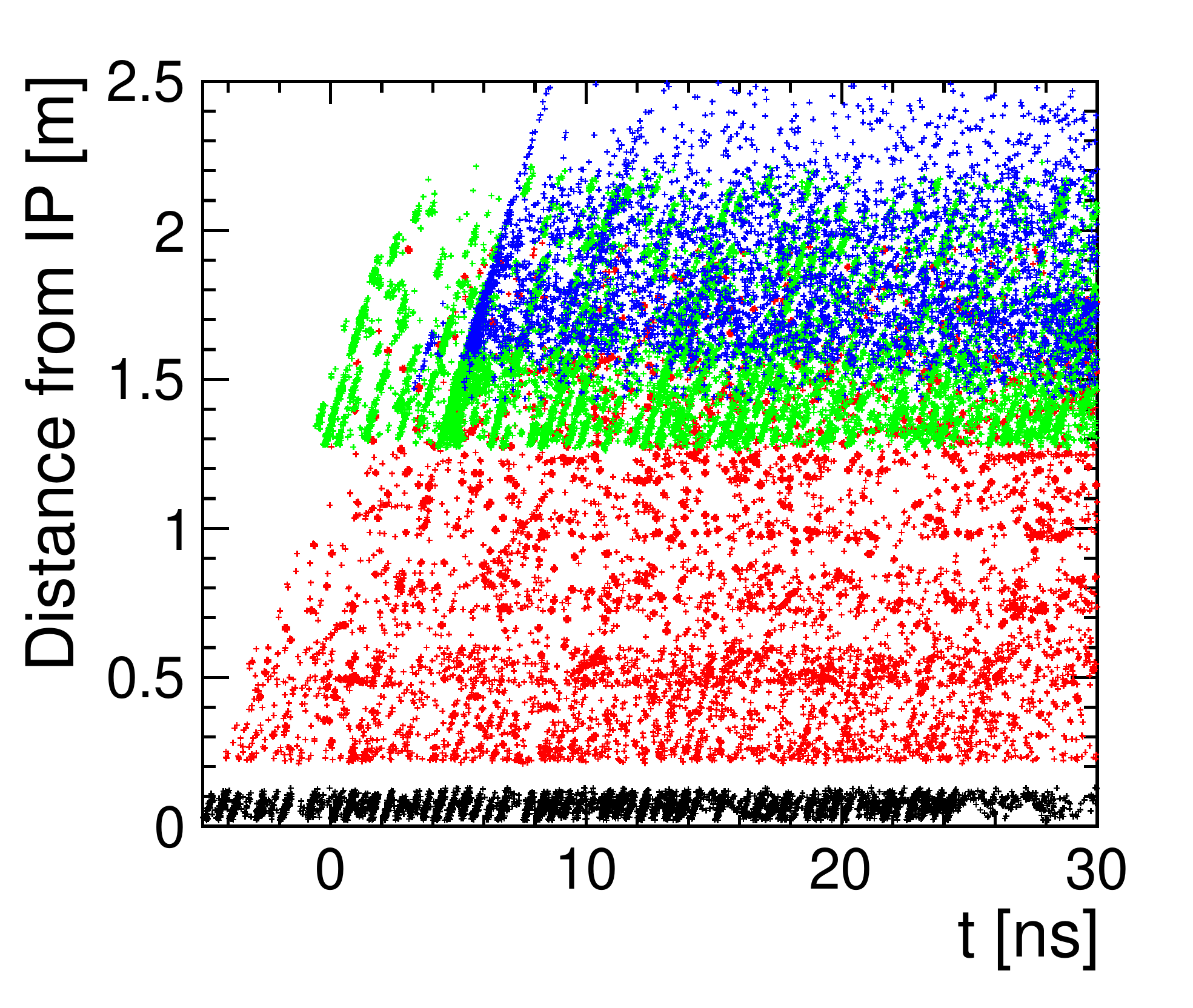}
 \hfill
 \includegraphics[width=0.32\textwidth]{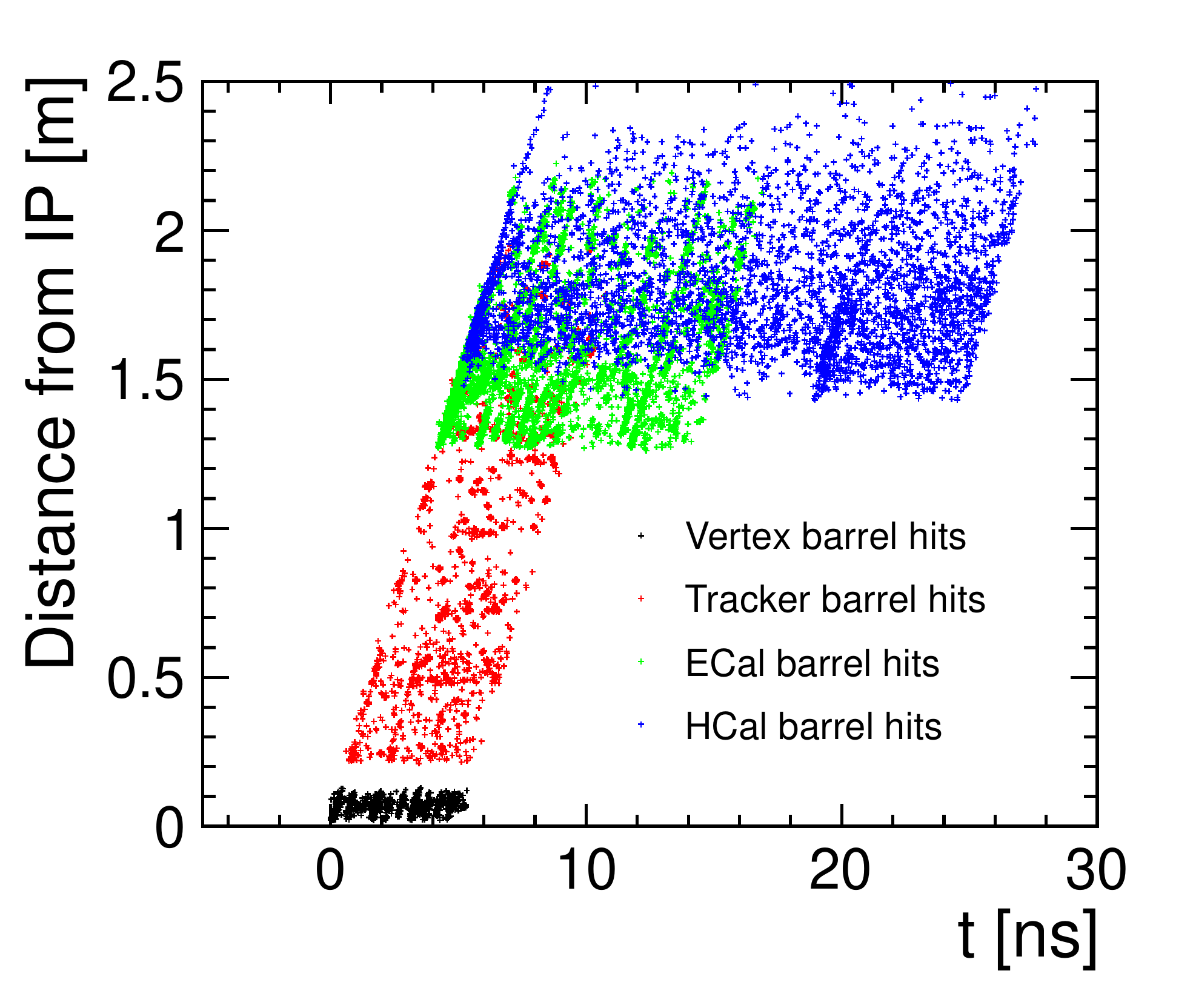}
 \caption[Time structure of the hits in the barrel detectors with and without applying readout time windows.]{Time structure of the hits in the barrel detectors in the signal event (left), together with a bunch train of \gghad events (center) and after introducting readout time windows (right). The readout time windows used here are \unit[5]{ns} for the vertex detector (black) and the main tracker (red), \unit[10]{ns} for the \ac{ECal} (green) and \unit[20]{ns} for the \ac{HCal} (blue).}
 \label{fig:Software_BackgroundOverlayTimeWindows}
\end{figure}

\Cref{fig:Software_BackgroundOverlayTimeWindows} shows the time structure of the hits in an arbitrary event with and without addition of \gghad background events. While most hits in the tracking systems are produced shortly after the expected time from the time-of-flight, the shower development in the calorimeter leads to much later hits.

For the \ac{CLIC} \ac{CDR} a readout time window of \unit[10]{ns} is assumed throughout all subdetectors, except for the \ac{HCal} barrel which uses tungsten as absorber material. Hadronic showers in tungsten have a larger neutron component compared to steel absorbers, which tend to have a later calorimetric response. The time window of the HCal barrel is thus set to \unit[100]{ns}, which is feasible since the occupancy from the beam induced backgrounds at these radii is low compared to the forward detectors.

\subsubsection{Treatment of Tracker Hits}
\label{sec:Software_BackgorundOverlayTrackerHits}
The tracking detectors are segmented into modules in the simulation model as described in \cref{sec:SiD_model_mainTracking}. Individual strips or pixels are not implemented in the simulation. Thus, each tracker hit represents a localized energy deposit and not a pixel or strip hit. These pixel or strip hits are only created during the digitization step taking into account all energy deposits that would belong to the same readout channel. This leaves the \textit{OverlayDriver} only with the task of shifting the hits in time according to their bunch crossing and removing all hits that do not fall within the defined time window.

\subsubsection{Treatment of Calorimeter Hits}
\label{sec:Software_BackgorundOverlayCalorimeterHits}
Calorimeter hits represent the total energy deposition within a certain readout cell of the calorimeter. They can consist of several individual energy depositions that occurred during the simulation within the corresponding volume. Each of those energy deposits has a time stamp and the information of which particle it originates from. The time window cuts are applied to the individual energy deposits instead of the calorimeter hit as a whole, since he time of the hit is simply defined as the time of the earliest of its energy depositions. Multiple calorimeter hits corresponding to the same cell are merged into a single calorimeter hit by merging all of their energy depositions within the respective time window.

\subsubsection{Treatment of Monte~Carlo Particles}
\label{sec:Software_BackgorundOverlayMcParticles}
The list of all Monte~Carlo particles in the signal event is kept to maintain the full true event information. The number of background particles can be very high and they are usually only of limited interest. Only those background particles that are associated with a hit that passed the time window selection are kept and placed into the list of Monte~Carlo particles together with the full hierarchy of their respective ancestors. Two lists of pointers are created to identify which of the particles originated from the signal event and which ones are background particles. The production times of all particles are modified according to the time of the bunch crossing at which they occur.

\subsubsection{Multiple Sources of Backgrounds}
\label{sec:Software_BackgorundOverlayMultiple}
The \textit{OverlayDriver} was intended from the beginning to allow overlaying different types of background from different sources. For this, multiple instances can be run one after the other using different parameters. A label can be set to distinguish the individual lists of background Monte~Carlo particles. The parameters defining the placement of the signal event have to be set carefully to avoid inconsistencies. For example, if the signal event should be placed randomly within the bunch train, the beginning of the bunch train has to be selected as $t_0$ in the first overlay step. Then, for all following instances of the \textit{OverlayDriver} the signal has to be forced to the first bunch crossing since the output of the first overlay has set its time structure relative to the first bunch crossing in the train.

\subsection{Tracker Hit Digitization and Clustering}
\label{sec:Software_trackerDigitization}
The digitization of the tracker hits transforms energy deposits from the \geant simulation into signals as they would be seen from a detector readout, i.e. \ac{ADC} counts for each readout channel. Since the simulation model only defines modules and not individual strips or pixels the respective modules have to be segmented virtually and the individual tracker hit positions have to be identified with their respective channel. Afterwards, the charge build-up in each channel due to the simulated energy deposition has to be simulated and translated into the resulting measurable signal height. Finally, a clustering algorithm is used to combine signals from neighboring channels and identify the most probable hit position.

The digitization code that is used is the \emph{SiSim} package~\cite{SiSim}
implemented in \lcsim. The algorithm is explained in some detail below since no written documentation exists to date.

\subsubsection{Segmentation of Tracker Modules}
\label{sec:Software_trackerDigitizationSegmentation}
The segmentation of the tracking modules are defined for each subdetector individually by setting the desired sensor pitch. For pixel sensors values are set for the pitches in $x$ and $y$-direction, which are the local dimensions parallel to the sensor plane. For strip detectors only one pitch has to be defined for the strip width since the strip length is defined by the module extent. The direction of the strips in the barrel detector is parallel to the $z$-axis. In case of the stereo strip detectors in the tracker endcaps, the strips are placed perpendicular to the sides of the trapezoidal modules (see Section~\ref{sec:SiD_model_mainTracking}). The chosen stereo angle is 12\degrees for all stereo strip modules. This step also sets the transfer efficiencies for all subdetectors. They represent the efficiency of charge transport within the silicon as well as from the surface of the sensitive material to the readout and reduce the signal height as explained below.

It is forseen that only every other strip in the strip detectors is read out directly to lower the number of required readout channels. The intermediate strips are connected via capacitive coupling to their neighboring strips which allows to maintain a high spatial resolution of the hit position as shown in~\cite{Krammer:1997ke}. This is taken into account in the silicon simulation by defining a readout pitch of twice the sensor pitch. The charge collected in the intermediate strips is added to both neighboring strips that are read out directly using a lower transfer efficiency.

In \clicsid the sensitive and readout pitch for all pixel detectors are set to \unit[20]{\micron} in $x$ and $y$ and the transfer efficiency is set to 1. For all strip detectors the sensitive pitch is set to \unit[25]{\micron}, while the readout pitch is set to \unit[50]{\micron}. For the strips that are read out directly a transfer efficiency of $\sim99\%$ is assumed. The intermediate strips, which are read out via capacitive coupling by both neighboring strips is set to a value of $\sim40\%$, which is significantly lower than half of the efficiency assumed the directly read out strips.

\subsubsection{Silicon Simulation}
\label{sec:Software_trackerDigitizationSiSim}
The simulation of the charge deposition and distribution is implemented for a generic silicon sensor device. Generic variables like the bias and depletion voltages, the doping concentration, as well as the electron and hole concentrations are used to describe the sensor material. For the basic principles of semi-conductor sensors we refer to~\cite{Knoll:1300754}. The readout device is similarly defined in a generic way by its threshold and noise level.

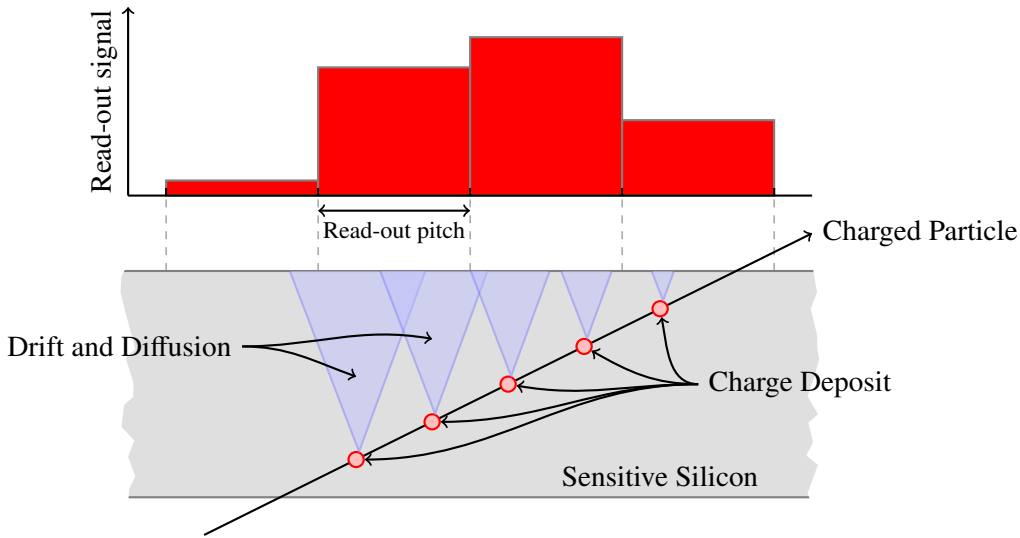
\begin{figure}
 \centering
 \begin{tikzpicture}[thick,
                     charge/.style={ellipse, draw=red, fill=red!25, inner sep=0pt, minimum size=2mm}]

  \begin{scope}[draw=gray,
                fill=red]
    \filldraw (-4.0,2.5) rectangle (-2.0,2.7);
    \filldraw (-2.0,2.5) rectangle (-0.0,4.2);
    \filldraw (0.0,2.5) rectangle (2.0,4.6);
    \filldraw (2.0,2.5) rectangle (4.0,3.5);
  \end{scope}

  \draw (-4.5,2.5) -- (4.5,2.5);
  \foreach \i in {-4,-2,...,4} {
    \draw (\i,2.5) -- (\i,2.6);
    \draw[gray, dashed, thin] (\i,1.5) -- (\i,2.5);
  }
  \draw[<->] (-2.0,2.3) -- (0.0,2.3);
  \node[below, font=\footnotesize] at (-1.0,2.3) {Read-out pitch};

  \draw[->] (-4.5,2.5) -- (-4.5, 5.0);
  \node[above,rotate=90] at (-4.5, 3.75) {Read-out signal};

  \begin{scope}[fill=gray!25,
                decoration={random steps, segment length=6pt, amplitude=3pt}]
  \path[clip] (-4.5,-1.6) decorate{ -- (-4.5,1.6)} -- (4.5,1.6) decorate{ -- (4.5,-1.6)} -- cycle;
  \fill[draw=gray] (-5.0,-1.5) rectangle (5.0,1.5);
  \node[above] at (2.5,-1.5) {Sensitive Silicon};
  \end{scope}

  \draw[->] (-3.5,-2.0) -- (4.5,2.0) node[right] {Charged Particle};

  \node (charge1) at (-1.5,-1) [charge] {};
  \node (charge2) at (-0.5,-0.5) [charge] {};
  \node (charge3) at (0.5,0.0) [charge] {};
  \node (charge4) at (1.5,0.5) [charge] {};
  \node (charge5) at (2.5,1.0) [charge] {};

  \node[right] (chargeText) at (3.0,0.0) {Charge Deposit}
    edge[->, out=180, in=0] (charge1)
    edge[->, out=180, in=0] (charge2)
    edge[->, out=180, in=350] (charge3)
    edge[->, out=180, in=330] (charge4)
    edge[->, out=180, in=280] (charge5);

  \begin{scope}[draw=blue!50, fill=blue!25, opacity=0.5]
    \path[clip] (-5.0,-1.5) rectangle (5.0,1.5);
    \filldraw (charge1) -- ($(charge1) + (70:5cm)$) -- ($(charge1) + (110:5cm)$) -- cycle;
    \filldraw (charge2) -- ($(charge2) + (70:5cm)$) -- ($(charge2) + (110:5cm)$) -- cycle;
    \filldraw (charge3) -- ($(charge3) + (70:5cm)$) -- ($(charge3) + (110:5cm)$) -- cycle;
    \filldraw (charge4) -- ($(charge4) + (70:5cm)$) -- ($(charge4) + (110:5cm)$) -- cycle;
    \filldraw (charge5) -- ($(charge5) + (70:5cm)$) -- ($(charge5) + (110:5cm)$) -- cycle;
  \end{scope}
  \node[left] (driftText) at (-3.0,0.5) {Drift and Diffusion}
    edge[->, out=0, in=160] (-1.5,0.1)
    edge[->, out=0, in=170] (-0.5,0.6);

 \end{tikzpicture}
 \caption[Illustration of the charge deposition in the sensitive material of silicon detectors as simulated in the tracker hit digitization.]{Illustration of the charge deposition in the sensitive material, charge drift to the electrode and conversion into signals in the readout cell as simulated by the tracker hit digitization code. See text for details.}
 \label{fig:Software_SiliconSimulation}
\end{figure}

First all simulated tracker hits that deposited energy in the respective sensitive volume are identified. A simulated hit is defined by its position in the center of the sensitive medium, its energy deposit and the two points where the particle that it originated from entered and left the sensitive plane. For a realistic charge simulation a single energy deposition in the center of the medium is not sufficient. Instead, the energy is evenly distributed in small steps throughout the material along the path of the particle as shown in \cref{fig:Software_SiliconSimulation}. The particle trajectory is determinded from the two endpoints and divided into small track segments. Each segment has a maximum length of 1/10 of the sensor pitch. The local deposited energy for each segment is then converted into the corresponding number of electron-hole pairs that are created in the doped silicon. The amount of charge is then corrected for its collection inefficiency due to local charge trapping. Afterwards, the diffusion of the charge during its drift to the electrodes is calculated taking into account the electric field created by the bias and depletion voltages. The impact of the magnetic field is neglected. As a result the charge is distributed in a 2 dimensional Gaussian distribution on the electrode surface.

Once the charge distribution of all track segments of all hits in a module has been calculated, the charge collected within the surface of each readout channel is combined. The transfer efficiency is applied to calculate the visible signal. Randomly generated noise can be added to the signal of each channel, which was not done for the \ac{CLIC} \ac{CDR} studies. Finally, the charge in the cell is converted to an integer value which represents the digital signal of the \ac{ADC}.
If the resulting signal is above the desired threshold it is added to the list of digitized hits. The time of the hit is not used during the digitization. The only constraint on the time of the hits comes from the background overlay time windows introduced in \cref{sec:Software_BackgroundOverlayTimeWindows}.

\subsubsection{Clustering}
\label{sec:Software_trackerDigitizationClustering}
The digitized hits are clustered together using a nearest neighbor algorithm. Two threshold values have to be defined for each subdetector. The first value defines the minimum signal height for a hit to be used as a seed, while the second threshold defines the minimum signal height for a neighboring cell to be clustered together with a seed.

The identified clusters are converted into tracker hits to be used by the track finding algorithm explained in \cref{sec:Software_trackReconstruction}. The position of the tracker hit is calculated as the signal weighted center of the cluster. The uncertainties of the hit position are calculated from the readout pitch of the cell, $w_\mathrm{cell}$, and a correction factor $f$ which depends on the cluster width, $w_\mathrm{cluster}$, in the relevant direction
\begin{equation}
 \sigma = w_\mathrm{cell} \cdot f(w_\mathrm{cluster}).
\end{equation}
The correction factors used are given in \cref{tab:Software_trackerDigitizationFactors}.

\begin{table}
 \centering
 \caption[Correction factors used for the calculation of the tracker hit uncertainty.]{Correction factors $f$ used for the calculation of the tracker hit uncertainty depending on the cluster width $w_\mathrm{cluster}$.}
 \label{tab:Software_trackerDigitizationFactors}
 \begin{tabular}{c c}
  \toprule
  $w_\mathrm{cluster}$ & $f$           \\
  \midrule
  1                    & $1/\sqrt{12}$ \\
  2                    & $1/5$ \\
  3                    & $1/3$ \\
  4                    & $1/2$ \\
  5                    & $1$ \\
  \bottomrule
 \end{tabular}
\end{table}

\subsection{Removal of Tracker Hits}
\label{sec:Software_trackerHitRemoval}
The layers of the tracking system in the \clicsid detector model are made of individual sensitive modules, which are partially overlapping. This leads to a more realistic material budget in the full simulation, but it also creates multiple tracker hits within a single layer when a particle passes through such overlapping modules. The track reconstruction algorithm, described in  \cref{sec:Software_trackReconstruction}, was originally written for a simpler geometry description using cylindrical and disk tracking layers. In those geometries overlaps are not existing and only a single hit from every layer is allowed to be assigned to a track. The additional hits in the overlapping region can lead to a performance drop, since more combinations are possible, or, in the worst case, lead to the creation of fake tracks.

This shortcoming of the tracking algorithm is circumvented by removing those hits before the track finding is run. All layers are checked for multiple tracker hits belonging to the same Monte~Carlo particle and every hit except for the one that occurred first is removed. This driver is the only one using Monte~Carlo truth information throughout the reconstruction chain, which was done for simplicity. A similar result could be obtained by implementing a hit removal that is based on the pure geometrical overlap of the individual modules.

Ideally the additional information should be used in the pattern recognition as a short vector defining the track direction in the overlap region.

\subsection{Removal of Muon Chamber Hits}
\label{sec:Software_calorimeterHitRemoval}
The return yoke in the \clicsid detector model is modelled with 18 equidistant sensitive layers for muon identification (see \cref{sec:SiD_model_yoke}). It has been shown, that an efficient muon identification can be achieved with a significantly smaller number of layers, i.e. three groups of three instrumented layers throughout the yoke thickness~\cite{kraaij11}. This is taken into account in the event reconstruction by ignoring all hits in the return yoke except of those in the layers 1--3, 8--10 and 16--18.

\subsection{Track Reconstruction}
\label{sec:Software_trackReconstruction}
The tracks are reconstructed from the digitized tracker hits with the \emph{SeedTracker} algorithm~\cite{SeedTracker}
implemented in \lcsim. This algorithm is a generic tracking algorithm intended for silicon based tracking detectors with only a small number of hits along a track. The geometry information of the detector is decoupled from the algorithm by relying on a set of strategies that encapsulate the required geometry information. In addition, it uses generalized three dimensional tracker hits for the track fitting and does not rely on certain hit types. Thus, this algorithm is especially suited for detector design studies where the detector geometry is frequently changed.

\subsubsection{Track Finding Steps}
\label{sec:Software_trackingFinding}
The track finding is controlled by strategies. Each strategy assigns a certain role to each of the layers in the tracking system that should be considered. Three layers are defined as seed layers, one layer is defined as the confirmation layer and usually all other layers are defined as possible extension layers. There is no requirement on the order of seed and confirmation layers, i.e. inside-out and outside-in tracking strategies are possible. There is also no requirement to chose adjacent layers as seed and confirmation layers although this is beneficial for the track finding.

In addition to the roles of each tracking layer, the strategy defines a minimum number of hits that are required to form a track candidate, $N_{\mathrm{min}}$, a maximum total $\chi^2$ for the track fit, $\chi^2_{\mathrm{max}}$, a constraint on the minimum transverse momentum for the track, $p_{\mathrm{T, min}}$, and a constraint on the distance of closest approach of the track to the \ac{IP} in $r\phi$ and $z$, referred to as $d_{0, \mathrm{max}}$ and $z_{0, \mathrm{max}}$, respectively. The vertex constraint is required to reduce the number of fake tracks in case of a low number of required hits and should be chosen rather loosely to avoid that tracks from displaced vertices fail the reconstruction. Secondary decays that occur far away from the \ac{IP} can not be found by this algorithm due to this vertex constraint. It has been shown that these tracks can be found efficiently by a second tracking algorithm that uses the track stubs created by \acp{MIP} in the \ac{ECal} as seeds~\cite{Onoprienko:2007wt}. The high granularity of the \ac{ECal} allows a sufficiently precise determination of the track parameters to extrapolate the trajectory back into the tracking region and identify the associated hits. This calorimeter assisted tracking algorithm was not used in the tracking studies presented in this thesis since it is incompatible with the digitization code used.


For tracking systems consisting of multiple subdetectors, several strategies are required to cover the necessary layer combinations of seed and confirmation layers to be able to find all the tracks within the acceptance of the detector. This is usually the case, since barrel and endcap detectors are treated as individual subdetectors in the simulation. The set of strategies then has to cover all necessary combinations of seed layers from different subdetectors to also find tracks in the transition region. The training procedure for obtaining a full set of strategies is explained below.

The track finding is performed for each of the strategies and results in a single list of reconstructed tracks. It is performed in steps corresponding the three different roles that are assigned to the different layers in that strategy. During these steps hits are successively added to a track candidate by checking its consistency with a helical fit explained below. The hit is only added to the track candidate if the fit succeeded and the total $\chi^2$ of the fit does not exceed the $\chi^2_{\mathrm{max}}$.
\begin{description}
 \item[Seeding] The hits in the seed layers defined by the strategy are used to form the initial track candidates. This list contains all possible combinations of hits from the three different layers. An initial helix fit is performed to determine the track parameters of each track candidate.
 \item[Confirming] All track candidates resulting from the seeding are tested for a successful helix fit with any of the hits in the confirmation layer fulfilling $\chi^2 \leq \chi^2_{\mathrm{max}}$. It is possible to create multiple confirmed seeds from a single track candidate.
 \item[Extending] The list of confirmed seeds is tested if they can be extended to a full track. For each track candidate all the hits in each of the extension layers are tested if they pass a helix fit ($\chi^2 \leq \chi^2_{\mathrm{max}}$). Once the number of remaining layers to be tested drops below the number of hits still required to form a track, the candidate is discarded. Each resulting track candidate is checked if it shares more than one hit with any of the previously reconstructed tracks. If this is the case, the only the best track candidate is kept. The best track candidate is defined as the one with the most hits associated and in case of equal number of hits the one with the smallest $\chi^2$.
\end{description}
The hits in each layer are grouped into sectors in $\phi$ and $z$ which allows to check only the hits in the sectors consistent with the current set of helix parameters when confirming and extending the track seeds. While hits have to be present in the seed and confirmation layers to find a track, extension layers can be skipped if there are sufficient hits in other extension layers to fulfill the $N_{\mathrm{min}}$ criterion.

\subsubsection{Helix Fitting}
\label{sec:Software_trackingFit}
The track fit used throughout the \emph{SeedTracker} algorithm assumes a helical track path. This helix fit consists of a non-iterative circle fit~\cite{Karimaki:1990zz} in the $r\phi$-plane followed by a straight line fit in the $zS$-plane, where $S$ is the arc length from the $z$ position of the point of closest approach in the $xy$-plane, $z_0$, to the position of interest along the helix. A definition of the coordinates can be found in \cref{App:trackParametrization}.
The hit uncertainties in $r\phi$ and $z$ are a assumed to be uncorrelated. For stereo strip hits the hit position and covariance matrix is depending on the direction at which the layer is passed. The track parameters from the previous fit are used to calculate the corrected covariance matrix.

If possible, \ie at least three pixel hits are available, only those hits are used in the linear fit of the slope, since those hits have a far better $z$-resolution compared to the long strips\footnote{This is always possible in the \clicsid detector}. 
A $\chi^2$ penalty is added if the resulting track parameters exceed the limits for $p_{T, \mathrm{min}}$, $d_{0, \mathrm{max}}$ or $z_{0, \mathrm{max}}$. The total $\chi^2$ of the track fit is the combination of the individual fit $\chi^2$ and the different penalties,
\begin{equation}
 \chi^2 = \chi^2_\text{circle} + \chi^2_\text{line} + \frac{(\kappa - \kappa_\text{max})^2}{\sigma_\kappa}\bigg|_{\kappa > \kappa_\text{max}} + \frac{(d_0 - d_{0, \text{max}})^2}{\sigma_{d_0}}\bigg|_{(d_0 - d_{0, \text{max}}} + \frac{z_0 > z_{0, \text{max}})^2}{\sigma_{z_0}}\bigg|_{z_0 > z_{0, \text{max}}},
\end{equation}
where $\chi^2_\text{circle}$ and $\chi^2_\text{line}$ are the $\chi^2$ of the circle and the line fit, respectively.

\subsubsection{Multiple Scattering}
\label{sec:Software_trackingScattering}
The material budget is taken into account for each helix fit beyond the initial fit performed during the seed step. This is done by combining the uncertainty due to multiple scattering with the uncertainty of the single point resolution of each hit.

First, the width of the scattering angle distribution introduced by each layer is calculated taking into account the angle at which the layer is crossed and the track candidate momentum. This requires some track parameters and is thus only possible after the initial fit has been performed. The layers considered for the multiple scattering estimation are the physical layers present in the tracking volume which includes support material in addition to the sensitive layers.

The uncertainty due to the scattering angle is estimated for each layer passed using \cref{eq:Highland}. Although the incident angle of the particle is taken into account when calculating the pathlength within the material, the curvature of the particle trajectory within the material is ignored. These multiple scattering uncertainties are then extrapolated to the position of the candidate hit for all layers that are closer to the \ac{IP} than the hit. The uncertainties for all layers are extrapolated individually and added in quadrature. This neglects any correlations between the multiple scattering uncertainties introduced by the individual layers which underestimates the total uncertainty. Finally, the total multiple scattering uncertainty is added in quadrature to the uncertainty of the hit position.

Once a hit is added to a track candidate the helix parameters are changed based on the new fit and the calculation of the multiple scattering explained above has to be repeated. Since confirmation and extension steps are done per layer and only one hit is added per layer this calculation has to be done once per layer and track candidate.

\subsubsection{Strategy Training}
\label{sec:Software_trackingStrategyTraining}
Since the number of strategies required to cover the full geometry can be very high, depending on the number of different layers in the detector and the desired minimum number of hits to form a track, a software tool is available to generate the list of necessary strategies.

A sample of training events covering the desired angular and momentum range is used to map out the possible combinations of layers that can be hit by individual particles. A minimum transverse momentum can be set to select only those particles that are desired to be found. Then, the set of particles is reduced to those that created at least the minimum number of hits required to form a track. All combinations of 4 different layers are generated from the available tracking layers, representing the three seed layers and the confirmation layer of the strategies. This set of strategy candidates is then used to obtain the smallest set of strategy candidates required to find all of the particles in the training sample.

Each layer can have a weight assigned and in case of redundant strategy candidates only the one with the highest combined weight is used in the final strategy list. By default, all layers which are not used as seed layers are defined as extension layers for that strategy. If desired, the strategy builder enforces symmetry between strategies involving one of the endcap layers.


\subsection{Calorimeter Clustering, Particle Flow and Particle Identification}
\label{sec:Software_particleFlow}
The identification of the calorimeter clusters and the creation of \acp{PFO}, i.e. reconstructed particles, is performed in the \pandora framework~\cite{Thomson:2009rp,PandoraPFANew}. The client application \slicPandora takes care of the reading and writing of the event, digitizes the hits before passing them to \pandora and provides the detector geometry.

\subsubsection{Calorimeter Hit Digitization}
\label{sec:Software_particleFlowDigitization}
The calorimeter hit digitization is very basic and only applies a threshold cut to the hit energy. In addition, only hits within the first \unit[100]{ns} of the event are taken into account, modeling the read-out cycle of the calorimeters. The energy of the hits in the muon chambers, which are instrumented with a digital \ac{RPC} readout, is set to 1 if the deposited energy is above threshold. All other hits are ignored. 


\subsubsection{Pandora Particle Flow Algorithm}
\label{sec:Software_particleFlowAlgorithm}
The \pandora framework offers a wide range of clustering algorithms that can be used for the identification of clusters. A short overview of the steps performed in \pandora is given below. Details of these steps can be found in~\cite{Thomson:2009rp,PandoraPFANew}.

\begin{itemize}
 \item The energy of the hits is corrected by the sampling fraction of the respective calorimeter layer. Different sampling fractions for electromagnetic and hadronic showers are foreseen and both hypothesis for the hit energy are calculated. The final decision on which sampling fraction to use for calculating the shower energy is done after the particle identification step.
 \item Individual hits that are most likely caused by slow neutrons are removed, since they are hard to attribute to specific showers.
 \item Initial clusters are formed from the calorimeter hits using a cone-based algorithm starting in the first layer of the calorimeter. The direction of tracks that reach the calorimeters are used as seeds. This step tends to produce small clusters which can be merged at a later stage.
 \item Neighboring clusters are merged based on their topology, i.e. the observed pattern matches a typical topology of two subclusters. Clusters that are consistent with a purely electromagnetic shower hypothesis are not merged with neighboring clusters.
 \item Clusters are merged with reconstructed tracks based on the topology. If the energies of the track and the cluster do not match, several re-clustering algorithms try to improve the match by breaking up the cluster or merging it with neighboring clusters.
 \item \acp{PFO} are created from the clusters. If the \ac{PFO} has a track, its momentum is directly taken from the track and the energy is calculated assuming the particle mass from the particle identification, as explained below. For neutral hadrons and photons the energy is set to the cluster energy. The momentum is calculated from the energy and the direction of the cluster. A mass of 0 is used for the calculation of the momentum since no proper identification of neutral hadrons is performed.
\end{itemize}

\subsection{Particle Identification in \pandora}
\label{sec:Software_particleID}
\pandora offers a basic particle identification. If a track is associated with the \ac{PFO}, the particle is identified as a charged pion or, if the cluster size resembles the size of a purely electromagnetic shower, it is identified as an electron. Similarly, if no track is associated with the cluster it is identified as a neutron or, if the shower is a purely electromagnetic shower, it is identified as a photon.

A dedicated algorithm to identify muons~\cite{lcd:2011-pfamuon} is used as the first step in \pandora. All hits associated with a muon are removed before any of the other algorithms are run.

\subsection{\acs{PFO} Selection and Truth Linking}
\label{sec:TimingTruth}
The final part of the reconstruction is again performed in \lcsim. It applies several selection cuts to the reconstructed particles in order to reduce the amount of beam-induced background. It also creates the truth links between Monte~Carlo and reconstructed particles required for performance studies.

\subsubsection{\acs{PFO} Selection Cuts and Truth Links}
\label{sec:PFOSelection}
The readout windows applied to collections, as described in \cref{sec:Software_BackgroundOverlayTimeWindows}, partially remove the beam-induced background by reducing the number of bunch crossings that are reconstructed together with the signal event. Once the reconstructed particles are available, a more sophisticated background rejection is possible using the combined time information from reconstructed clusters~\cite{PFOSelector}.

It is assumed that the calorimeter readout will provide time information for individual calorimeter hits on the order of \unit[1]{ns}, which is possible when using for example fast \acp{SiPM}~\cite{Simon:2010yj}. The truncated mean time of a reconstructed calorimeter cluster is known to sub-nanosecond precision. This mean cluster time is then propagated back to the interaction point, either along the line of sight for neutral particles or along the helix of the reconstructed track for a charged particle to calculate a precise time of the particle creation. The time of flight calculation is taking into account the mass of the particle based on its particle type. Using this procedure it is possible to select particles that were created within a certain time window around the signal event that is much narrower than the original readout time windows. It thus allows to efficiently identify and select only those particles that originate from the physics event together with those that are produced in-time with the physics event.

Different \ac{PFO} selection cuts are defined for each type of particle---charged particles, photons and neutral hadrons---and depend on whether the particle was reconstructed in the forward region of the detector, which is experiencing a much higher background occupancy. Two cuts on the transverse momentum define which particles are affected by the selection cuts. Particles below the limit are always rejected and particles with a higher \pT are always kept. The \emph{loose} selection cut defined for the \ac{CLIC} \ac{CDR} studies is given as an example set of \ac{PFO} selection cute in \cref{tab:PFOselection}. This set of cuts was used in the Higgs analysis in \cref{cha:Higgs}.

\begin{table}
\caption[One set of cuts used for the \ac{PFO} selection at \ac{CLIC}.]{One set of cuts used for the \ac{PFO} selection at \ac{CLIC}. Given is the minimum transverse momentum $p_\text{T, min}$ for particles that are kept, the maximum transverse momentum for particles that are removed $p_\text{T, max}$, and the maximum reconstructed production time of the \ac{PFO} at the \ac{IP}. These selction cuts depend on the reconstructed polar angle range and the particle type.}
\label{tab:PFOselection}
 \centering
 \begin{tabular}{c c c c c}
 \toprule
           Particle type & Polar angle region                       & $p_\text{T, min}$ [GeV] & $p_\text{T, max}$ [GeV] & $t_\text{max}$ [ns] \\\midrule
\multirow{4}{*}{Photons} & \multirow{2}{*}{$\cos\theta \leq 0.975$} &  0.75                  &                    4.0 &                 2.0 \\
                         &                                          &  0.0                   &                   0.75 &                 2.0 \\
                         & \multirow{2}{*}{$\cos\theta > 0.975$}    &  0.75                  &                    4.0 &                 2.0 \\
                         &                                          &  0.0                   &                   0.75 &                 1.0 \\\midrule
\multirow{4}{*}{Neutral hadrons} & \multirow{2}{*}{$\cos\theta \leq 0.975$} &  0.75          &                    8.0 &                 2.5 \\
                         &                                          &  0.0                   &                   0.75 &                 1.5 \\
                         & \multirow{2}{*}{$\cos\theta > 0.975$}    &  0.75                  &                    8.0 &                 2.5 \\
                         &                                          &  0.0                   &                   0.75 &                 1.5 \\\midrule
\multirow{2}{*}{Charged particles} & \multirow{2}{*}{All angles}    &  0.75                  &                    4.0 &                 3.0 \\
                         &                                          &  0.0                   &                   0.75 &                 1.5 \\
\bottomrule
 \end{tabular}
\end{table}


\subsubsection{Truth Linking}
\label{sec:Software_TruthLinking}
Reconstructed high level objects like tracks, calorimeter clusters and \acp{PFO} have no intrinsic link to the Monte~Carlo particles that they originated from. This connection is required to asses the quality of the event reconstruction and to quantify reconstruction efficiencies. We have implemented a truth linker in \lcsim to provide these links by creating weighted relations between Monte~Carlo particles and the reconstructed objects mentioned above~\cite{TruthLinker}. These links are used to identify the true particles for the tracking performance studies in \cref{cha:Tracking}.


The relations for tracks and clusters are evaluated based on the truth information of the contributing hits. The weight of the relation is the fraction of hits that a Monte~Carlo particle contributed to a track, or, in case of a calorimeter cluster, the fraction of energy that it contributed to the total true energy. The relations for the reconstructed particles are based on the contributing tracks in case of charged particles and the contributing clusters for neutral particles. This resembles how the information is used in the \ac{PFA}, which only uses the track momentum to calculate the particle energy, if available, because it has usually much higher accuracy.

\chapter{Tracking Performance in \clicsid}
\label{cha:Tracking}

In this chapter the performance of the tracking system of the \clicsid model is studied. First, the performance in determining the fundamental observables is studied in single muon events in \cref{sec:Tracking_momentumResolution}. In \cref{sec:Tracking_eventSamples} we introduce the samples used to study the performance in jet events. Furthermore we motivate our definition of the track finding efficiency and the fake rate in \cref{sec:Tracking_efficiency}. Afterward we present a parameter scan used to determine the optimal track finding efficiency strategies in \cref{sec:Tracking_strategy}. Finally, in \cref{sec:Tracking_jets}, we determine the performance in di-jet events from track finding efficiency and fake rates including the impact of the \gghad background.

\section{Track Resolution}
\label{sec:Tracking_momentumResolution}
\subsection{Momentum Resolution}
As we have discussed in \cref{cha:DetectorDesign} the momentum is one of the main observables in a \ac{HEP} experiment. In case of particle flow, a precise measurement of the track momentum is also essential for the energy measurement of charged particles. It is especially important for channels that involve only leptonic final states like for example the analysis of the Higgs decay into muons in \cref{cha:Higgs}.

\begin{figure}[htpb]
 \includegraphics[width=0.49\textwidth]{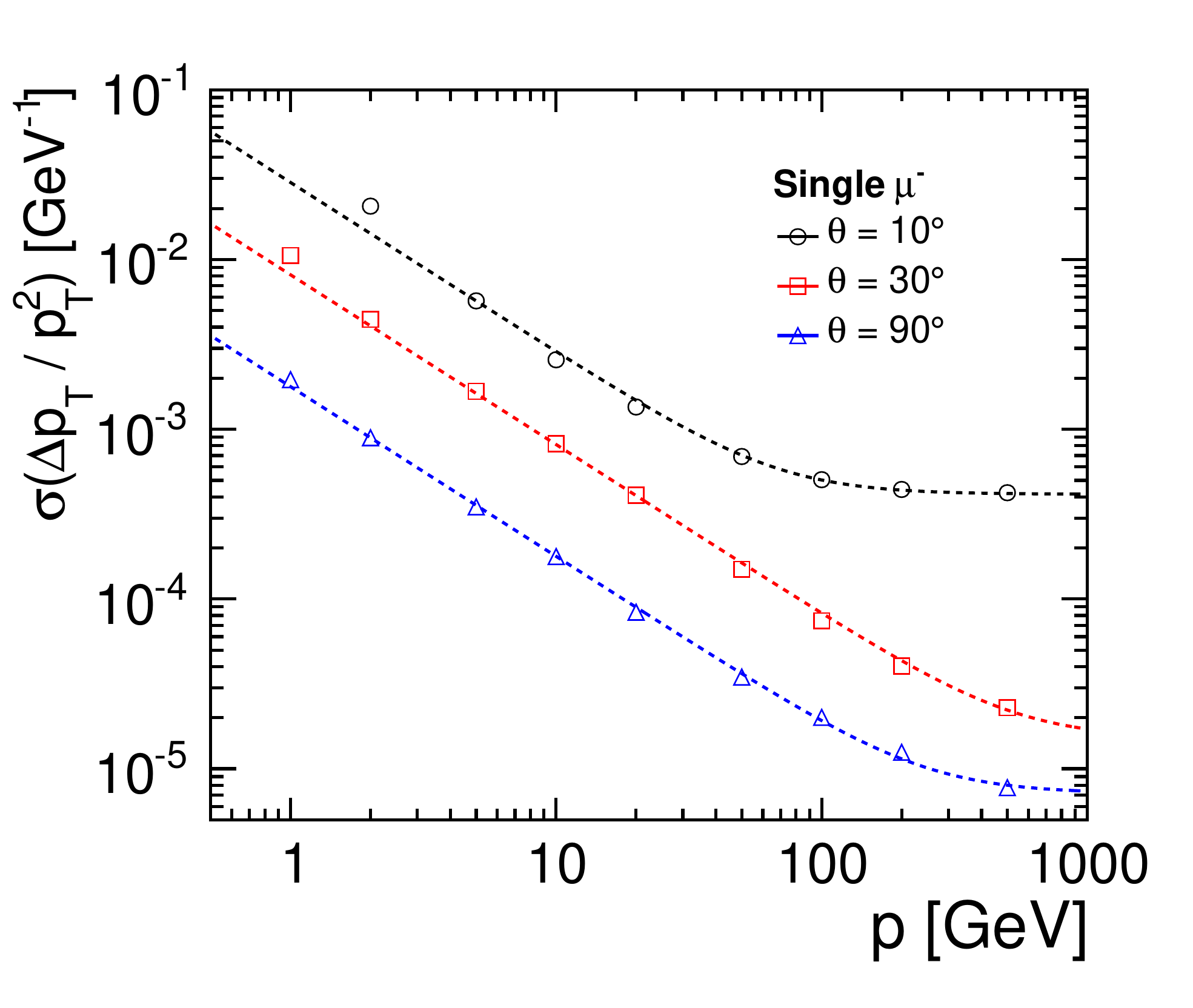}
 \hfill
 \includegraphics[width=0.49\textwidth]{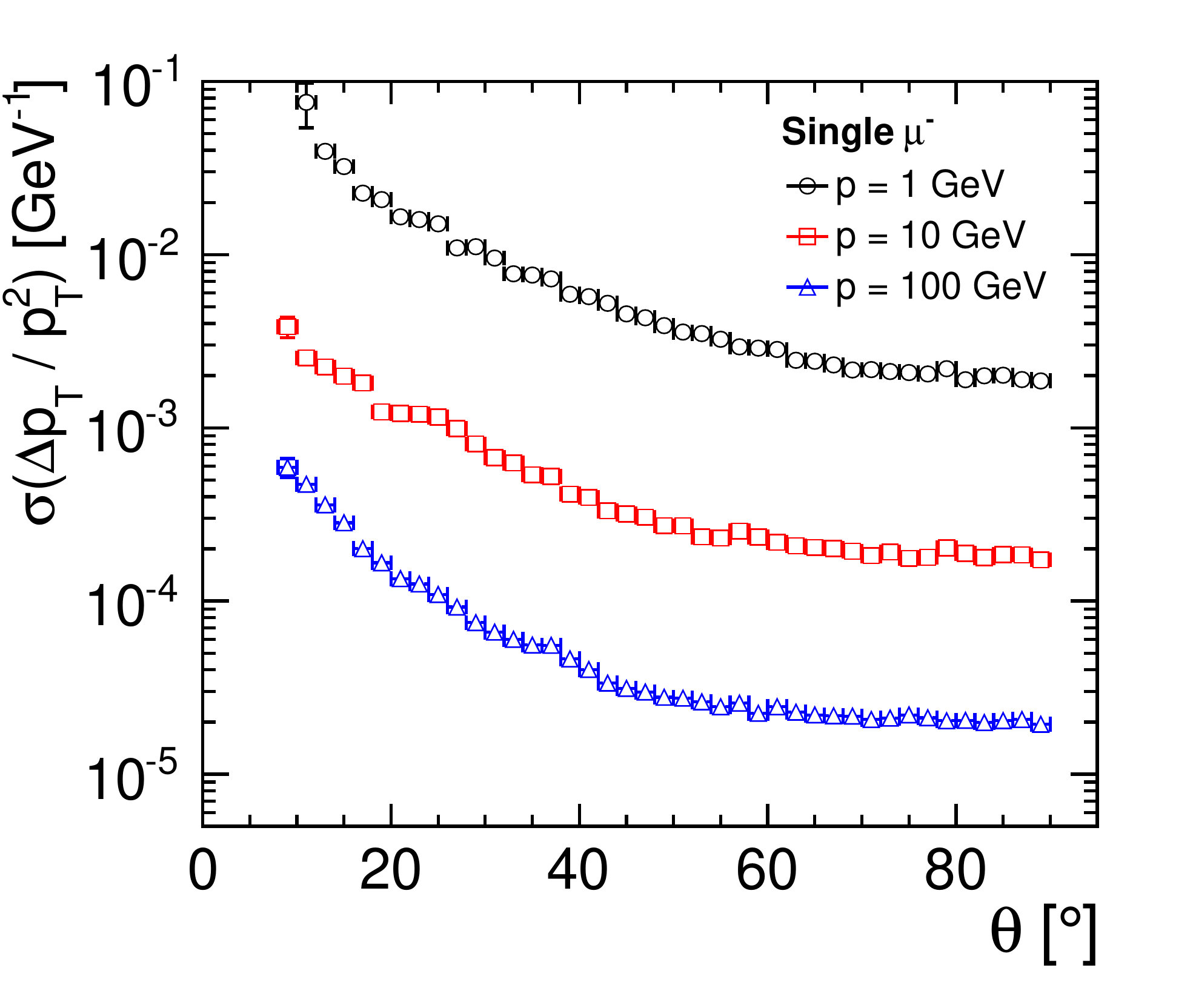}
\caption[Momentum resolution of single muons depending on the momentum and the polar angle.]{Transverse momentum resolution depending on the momentum $p$ for single muons at polar angles of 10\degrees, 30\degrees and 90\degrees (left). Transverse momentum resolution depending on the polar angle $\theta$ for single muons of \unit[1]{GeV}, \unit[10]{GeV} and \unit[100]{GeV} (right).}
\label{fig:tracking_TransverseMomentumResolution}
\end{figure}

\Cref{fig:tracking_TransverseMomentumResolution}~(left) shows the transverse momentum resolution of the \clicsid detector versus the momentum for single muons at different polar angles $\theta$. The points show the result obtained from a Gaussian fit on 10000 fully simulated and reconstructed tracks at the respective angle and momentum. The dashed lines shows a fit to \cref{eq:resolutionParametrization_pT} for fixed $\theta$. The resulting fit parameters are given in \cref{tab:tracking_momentumResolution}. The desired momentum resolution of $\sigma_{\pT}/\pT^2 \approx \unit[2\cdot10^{-5}]{GeV^{-1}}$ is achieved for central tracks with a momentum above \unit[100]{GeV}. For forward tracks with a polar angle of 30\degrees this resolution is only reached for very energetic particles of \unit[500]{GeV} or more. The parameter $b$ is increasing towards lower $\theta$, which is expected due to the higher material budget in the forward region (see \cref{fig:materialtracking}). Although the material budget assumed in the simulation model is rather ambitious it is still the limiting factor for the momentum resolution up to very high momenta. The dependence of the transverse momentum resolution on the polar angle is shown in \cref{fig:tracking_TransverseMomentumResolution}~(right) for different track momenta. The strong $\theta$-dependence originates from the dependence of \pT on $\theta$.

\begin{table}[htbp]
\centering
\caption[Transverse momentum resolution in \clicsid for single muons.]{Transverse momentum resolution $\sigma (\Delta \pT / \pT^2)$ in \clicsid for single muons,
parametrized by Equation~\ref{eq:resolutionParametrization_pT}.}
\label{tab:tracking_momentumResolution}
\begin{tabular}{r r r}\toprule
\tabt{$\theta$ \footnotesize{$\left[ ^{\circ}\right]$}} & \tabt{$a$ \footnotesize{$\left[\mathrm{GeV}^{-1}\right]$}} & \tabt{$b$}\\\midrule
90 & $7.3\cdot10^{-6}$ & $2.0\cdot10^{-3}$ \\
30 & $1.9\cdot10^{-5}$ & $3.8\cdot10^{-3}$ \\
10 & $4.0\cdot10^{-4}$ & $5.3\cdot10^{-3}$ \\\bottomrule
\end{tabular}
\end{table}

\begin{figure}[htpb]
 \includegraphics[width=0.49\textwidth]{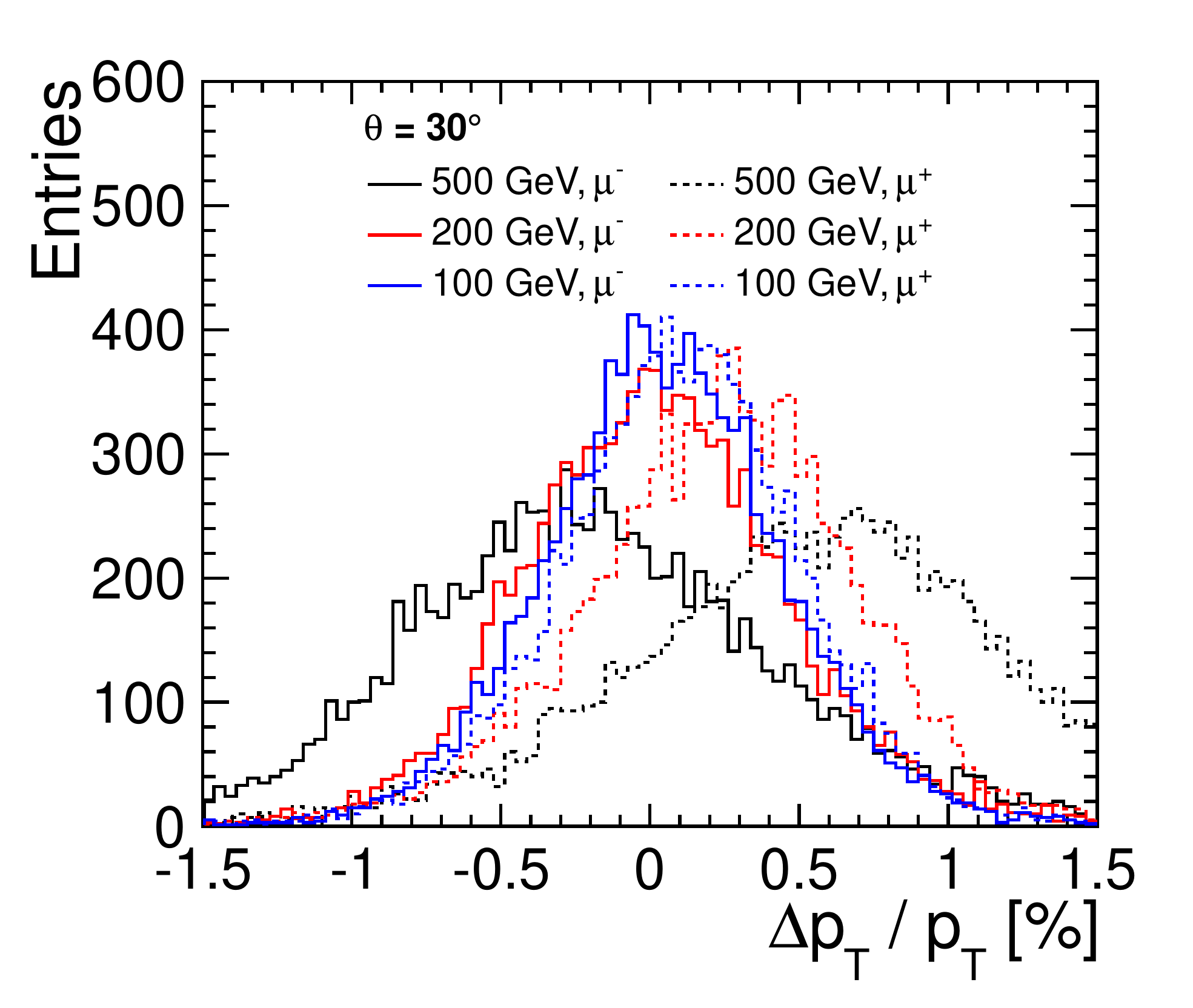}
 \hfill
 \includegraphics[width=0.49\textwidth]{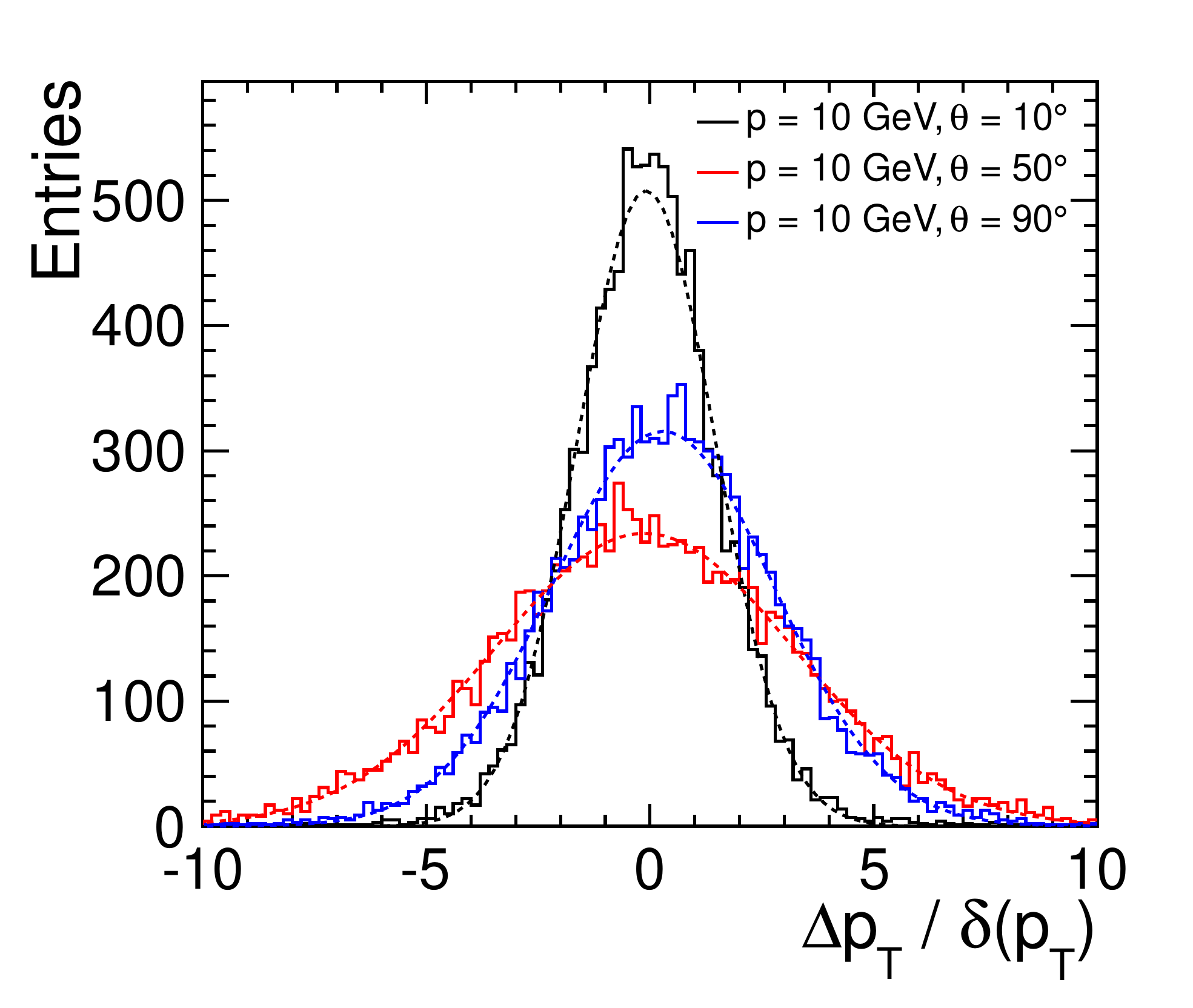}
\caption[$\Delta \pT$-distribution and pull distribution for single muons.]{$\Delta \pT$-distribution for high energetic muons at a polar angle of 30\degrees (left). Pull-distribution of the transverse momentum measurement for muons of \unit[10]{GeV} (right).}
\label{fig:tracking_momentumpull}
\end{figure}

The $\Delta \pT$-distribution of the reconstructed momenta follows a Gaussian distribution for all simulated samples. This is expected, since there is no confusion when reconstructing single tracks which could induce errors beyond the normal distributed errors due to multiple scattering and the measurement of the curvature. For high momentum forward tracks of $\theta \leq 30\degrees$ the reconstructed momentum is slightly biased, depending on the charge, as shown in \cref{fig:tracking_momentumpull}~(left). This shows that the curvature fit is biased towards one direction in $\phi$ for these tracks, resulting in too low curvatures for negatively charged tracks and too high curvatures for positively charged tracks. It remains to be understood if this effect is due to the detector geometry, \eg the fact that the modules in the disk detectors are overlapping in one direction, or is caused by the algorithm itself. Since only a very specific group of tracks is affected, this problem was not considered problematic for the \ac{CDR} studies.

The pull of the transverse momentum is defined as $\Delta \pT/\delta \pT$, where $\delta \pT$ is the uncertainty calculated by the track fit. It allows to test the uncertainty calculated by the track fit. Ideally, the pull-distribution follows a Gaussian distribution with a width of 1. To calculate the pull we first need to propagate of the uncertainty from the track parametrization to the desired physical quantity which is shown in~\cref{App:trackParametrization}. The resulting pull distribution is shown in \cref{fig:tracking_momentumpull} for muons of \unit[10]{GeV} at three different polar angles. The widths of all distributions are significantly larger than 1 which means that the uncertainties are underestimated by the fit. This was expected, as the correlations between the multiple scattering corrections introduced by each traversed layer are neglected. While the error is underestimated by approximately 50\% for the muons at 10\degrees the difference for muons at 90\degrees is more than a factor of 3.

\begin{figure}[htpb]
 \includegraphics[width=0.49\textwidth]{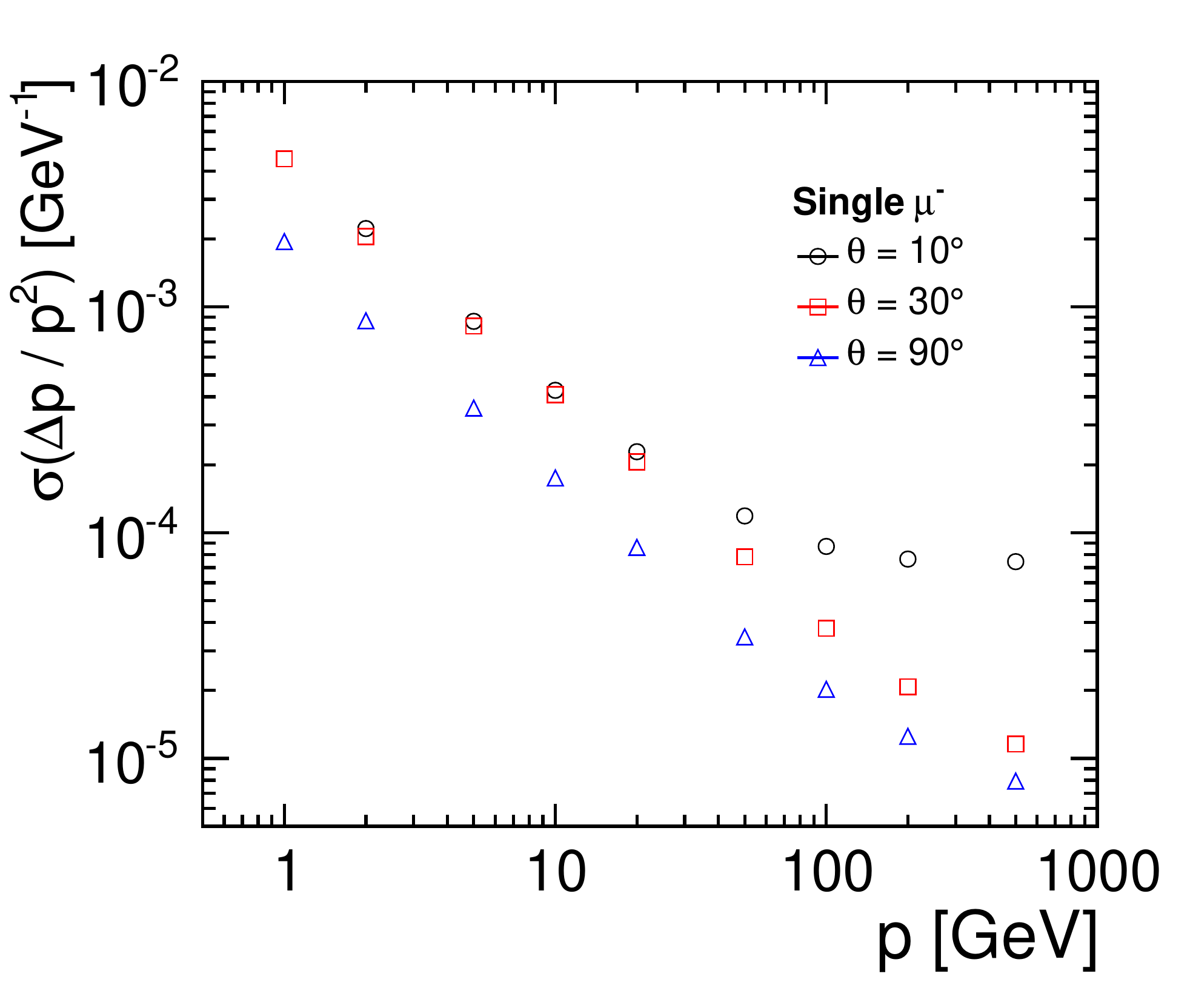}
 \hfill
 \includegraphics[width=0.49\textwidth]{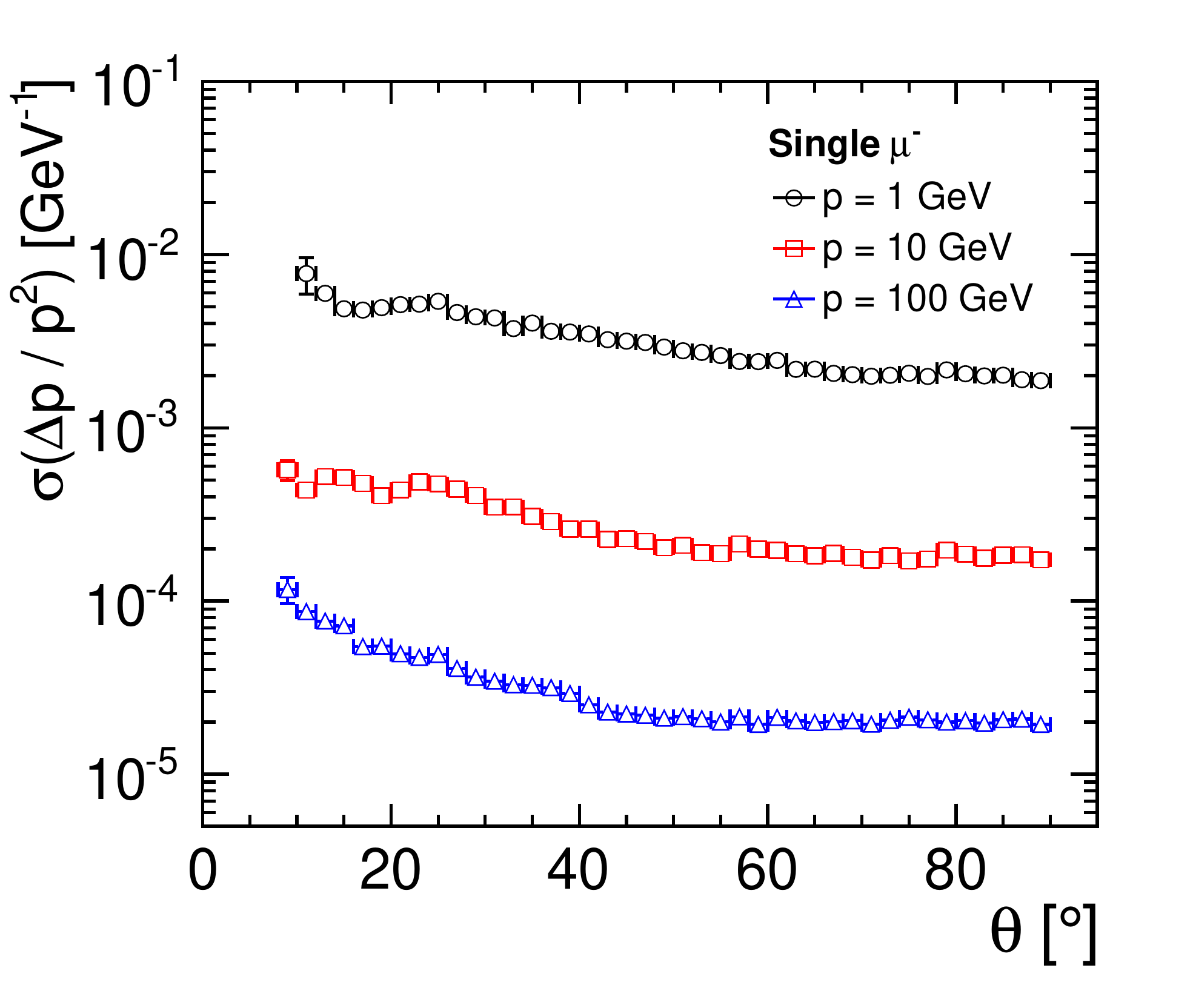}
\caption[Momentum resolution of single muons depending on the momentum and the polar angle.]{Momentum resolution depending on the momentum $p$ for single muons at polar angles of 10\degrees, 30\degrees and 90\degrees (left). Momentum resolution depending on the polar angle $\theta$ for single muons of \unit[1]{GeV}, \unit[10]{GeV} and \unit[100]{GeV} (right).}
\label{fig:tracking_momentumResolution}
\end{figure}

The momentum resolutions for single muons is shown in \cref{fig:tracking_momentumResolution}. It is much less dependend on the polar angle than the transverse momentum resolution, which dominates the momentum resolution only in the central region. The polar angle resolution, which we discuss below, dominates the momentum resolution in the forward region (see \cref{eq:resolutionParametrization_p}). 

\subsection{Angular Resolution}

\begin{figure}[htpb]
 \includegraphics[width=0.49\textwidth]{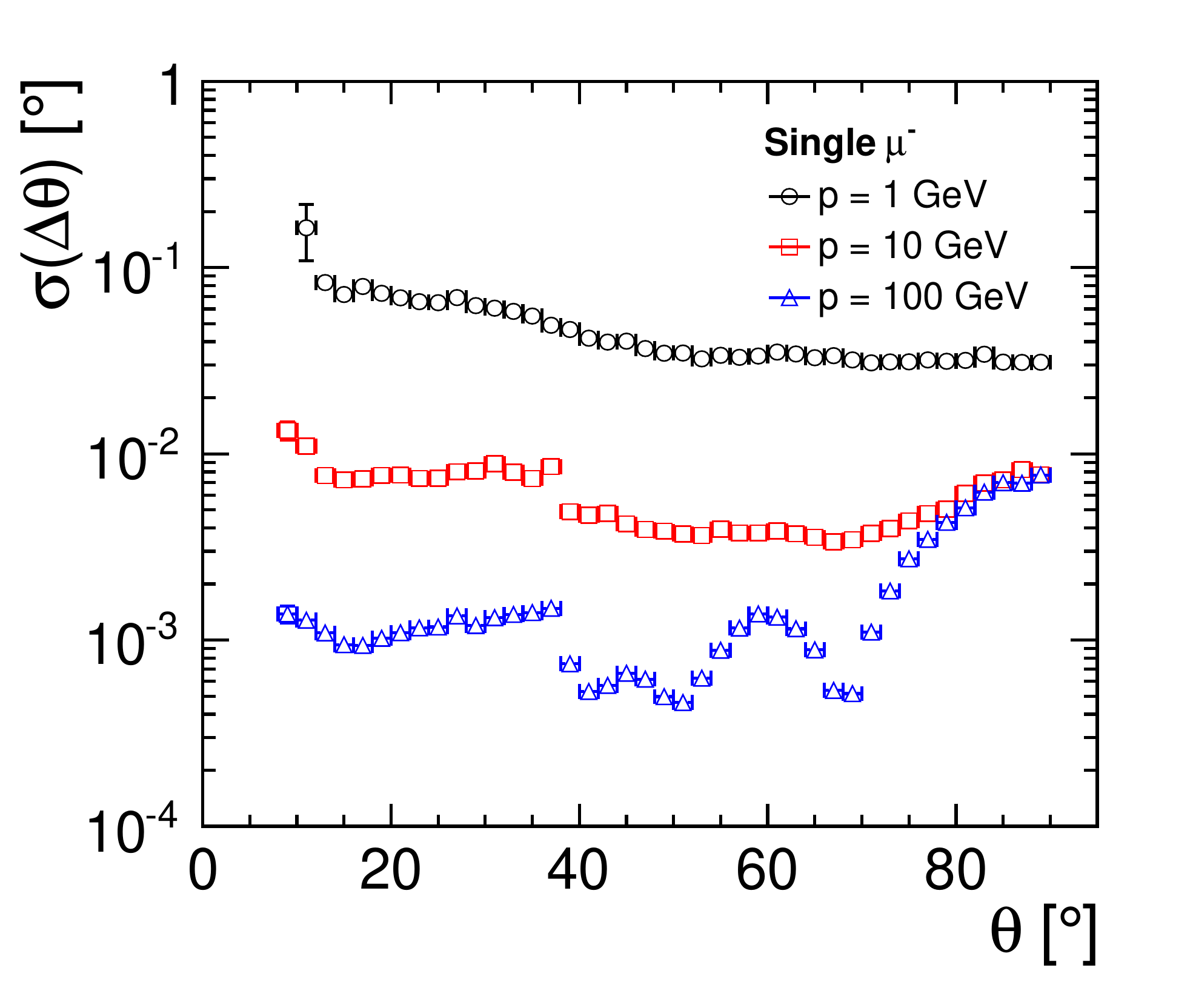}
 \hfill
 \includegraphics[width=0.49\textwidth]{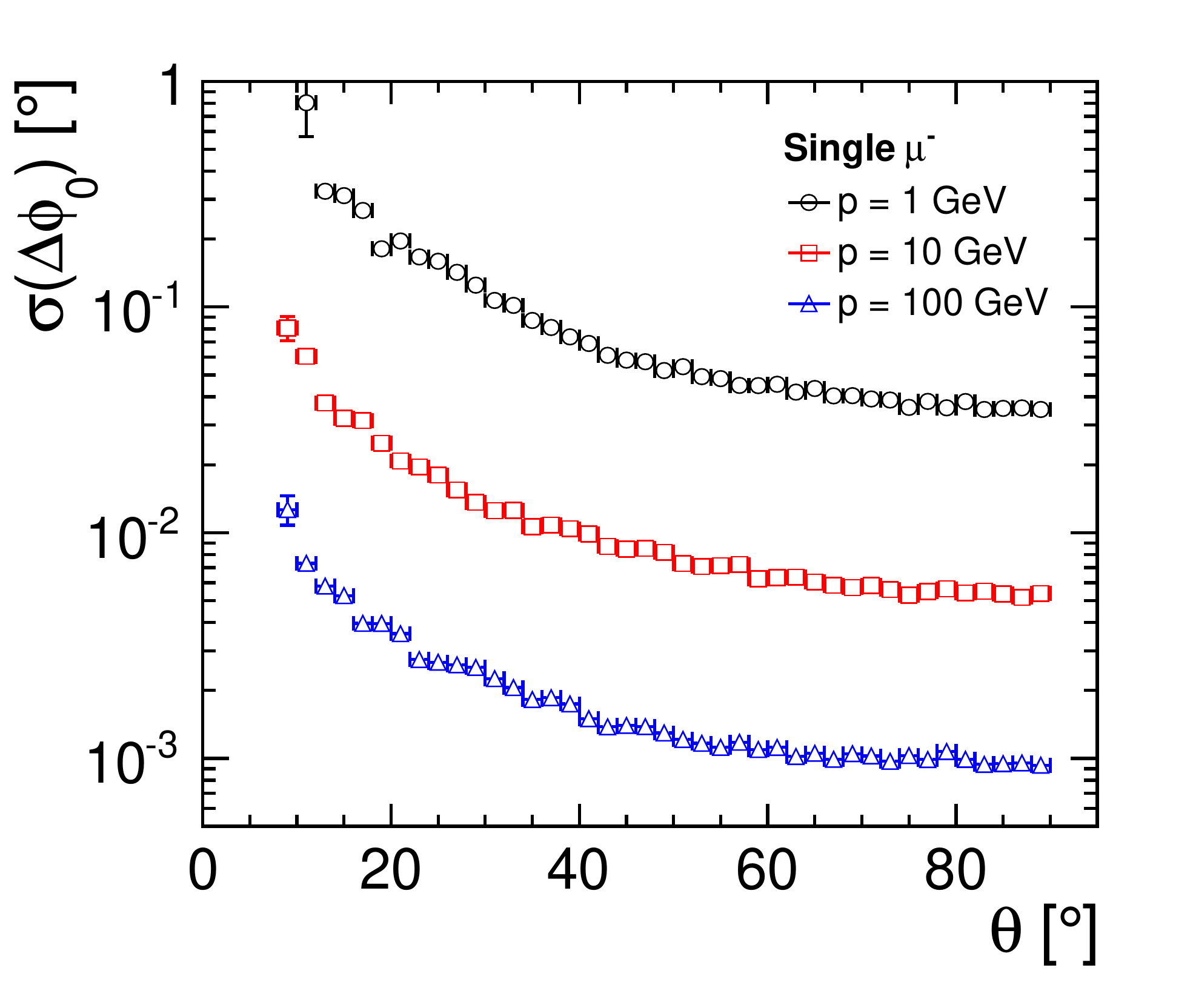}
\caption[Angular track resolutions for single muons.]{Polar angle resolution depending on the polar angle for single muons~(left). Azimuthal angle resolution depending on the polar angle for single muons~(left).}
\label{fig:tracking_thetaResolution}
\end{figure}

The dependence of the polar angle resolution on the polar angle is shown in \cref{fig:tracking_thetaResolution}~(left). For low energies the resolution dependency is almost flat and is given by the multiple scattering contribution. For high track momenta, on the other hand, it has some distinct features. These originate from the fact that only pixel hits are used in the fit of the slope (see \cref{sec:Software_trackingFit}). This results in a very short lever arm for the fit of the slope in the central region. Using \cref{eq:GlucksternLambda} we can estimate the expected polar angle resolution for a given angle. For example, using $\sigma(\xi) \approx \unit[0.00577]{mm}$, $L \approx \unit[50]{mm}$ and $N = 5$ yields a resolution of approximately $\sigma(\theta) \approx 0.0065\degrees$ which is slightly better than the number in \cref{fig:tracking_thetaResolution} since we have neglected the multiple scattering term. The complete angular dependency of the $\sigma(\theta)$-resolution for high momentum tracks can be explained from \cref{eq:GlucksternLambda}, as $\sigma(\xi)$, $L$, $N = 5$ vary with the polar angle. The resolution can not be easily improved by using the strip hits in the fit due to their large point resolution in $z$. This effects has no impact on the momentum resolution as shown in \cref{fig:tracking_momentumResolution}.

The dependency of the azimuthal angle resolution on the polar angle, shown in \cref{fig:tracking_thetaResolution}~(right), is similar to the behavior of the transverse momentum resolution. This is expected since they are directly connect via \cref{eq:SiD_projectedTrackLength}.

\subsection{Impact Parameter Resolution}
\label{sec:Tracking_impactParameter}
\begin{figure}[htpb]
 \includegraphics[width=0.49\textwidth]{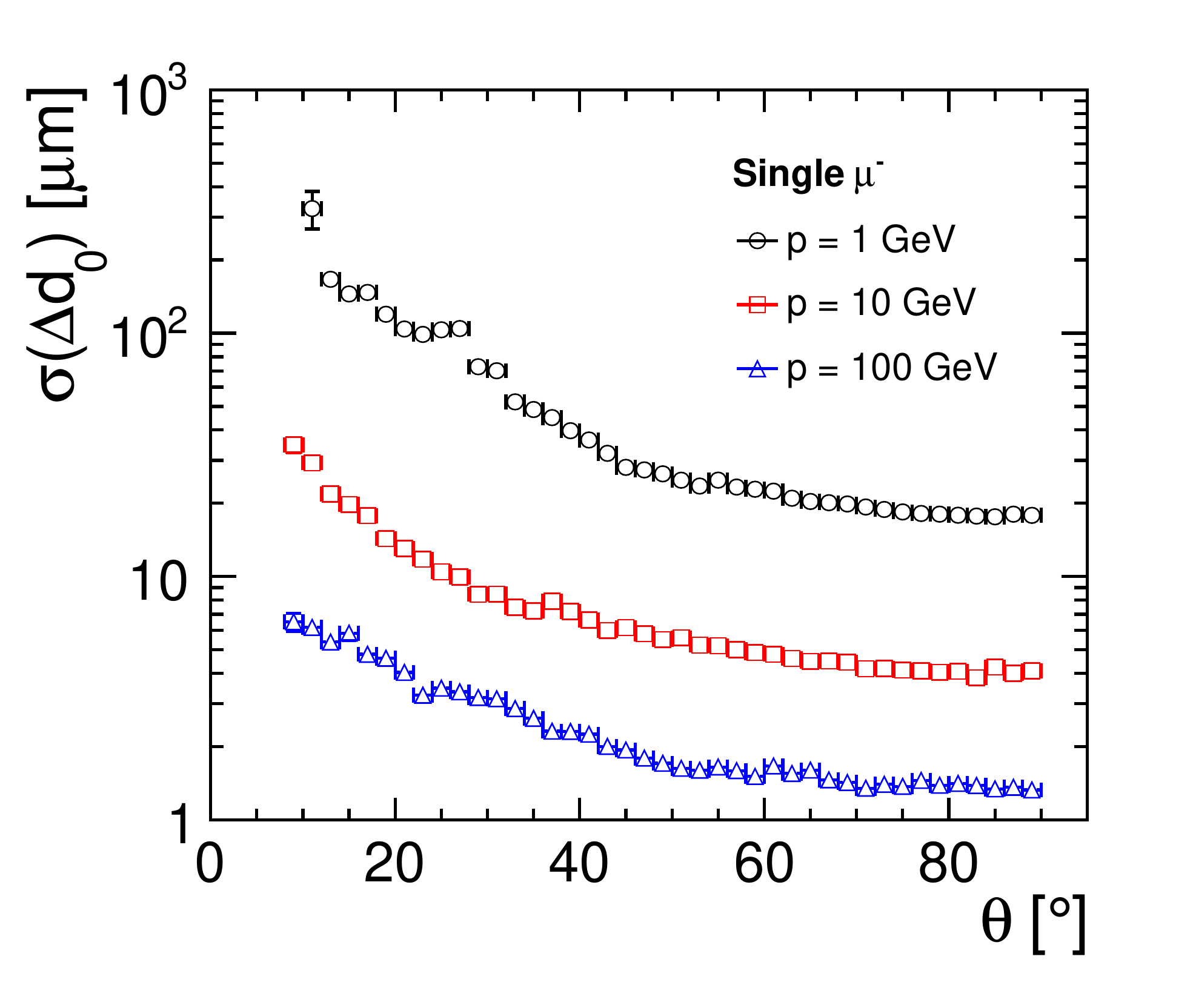}
 \hfill
 \includegraphics[width=0.49\textwidth]{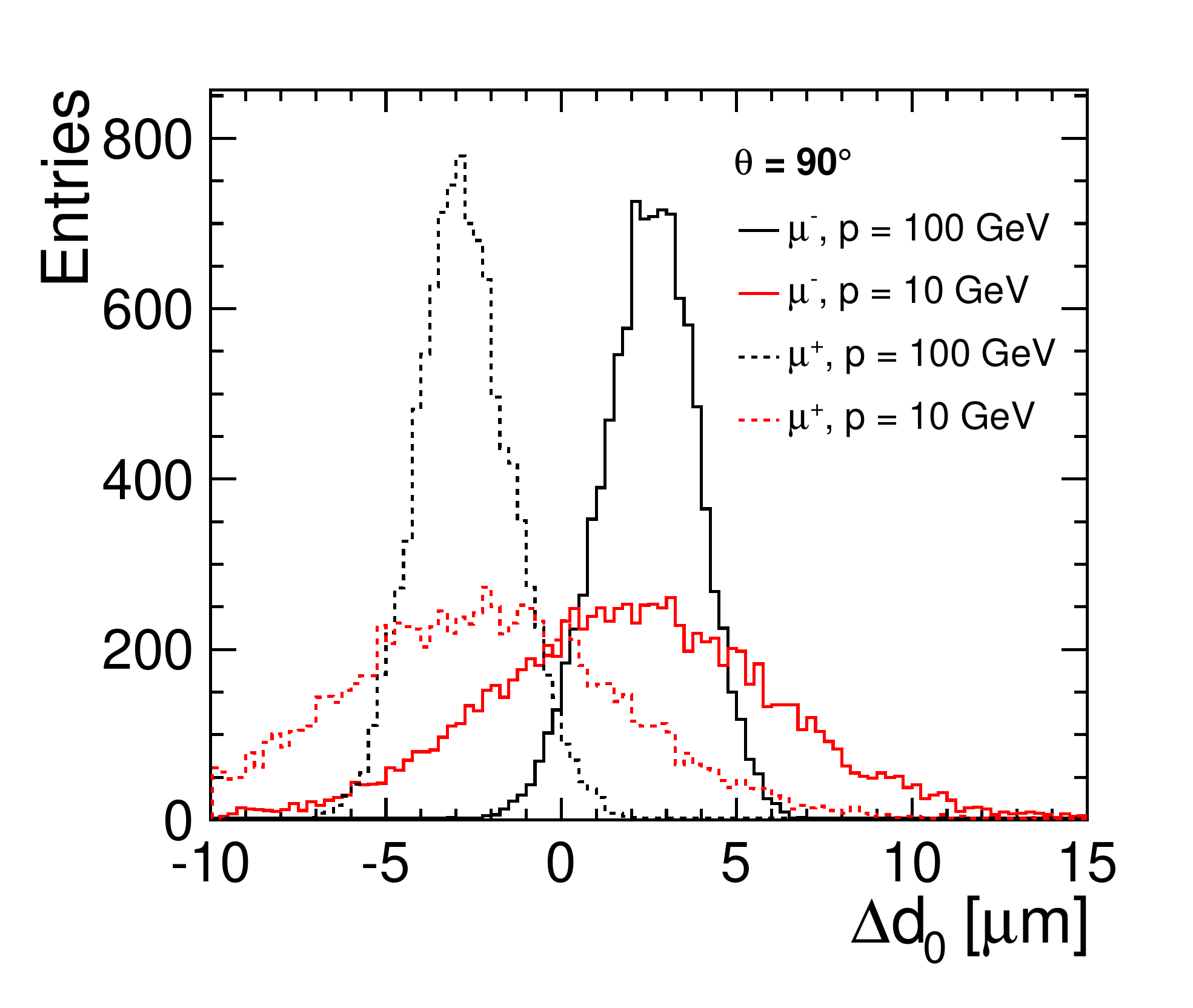}
\caption[Impact parameter $d_0$ resolution for single muons.]{Impact parameter $d_0$ resolution depending on the polar angle $\theta$ for single muons of different energies (left). $\Delta d_{0}$-distribution for \PGmp and \PGmm of \unit[10]{GeV} and \unit[100]{GeV} at a polar angle $\theta$ of 90\degrees (right).}
\label{fig:tracking_d0Resolution}
\end{figure}

\Cref{fig:tracking_d0Resolution}~(left) shows the impact parameter resolution for $d_0$. As expected, the $d_{0}$-resolution dependency on the polar angle is similar to the azimuthal angle dependency and reaches a few \micron for high momentum tracks. The reconstructed $d_{0}$ is significantly biased depending on the track charge for tracks with a momentum above \unit[10]{GeV}. This effect, which can be seen in \cref{fig:tracking_d0Resolution}~(right), shows that the track fitting is currently underestimating the curvature of very straight tracks.

\begin{figure}[htpb]
 \centering
 \includegraphics[width=0.49\textwidth]{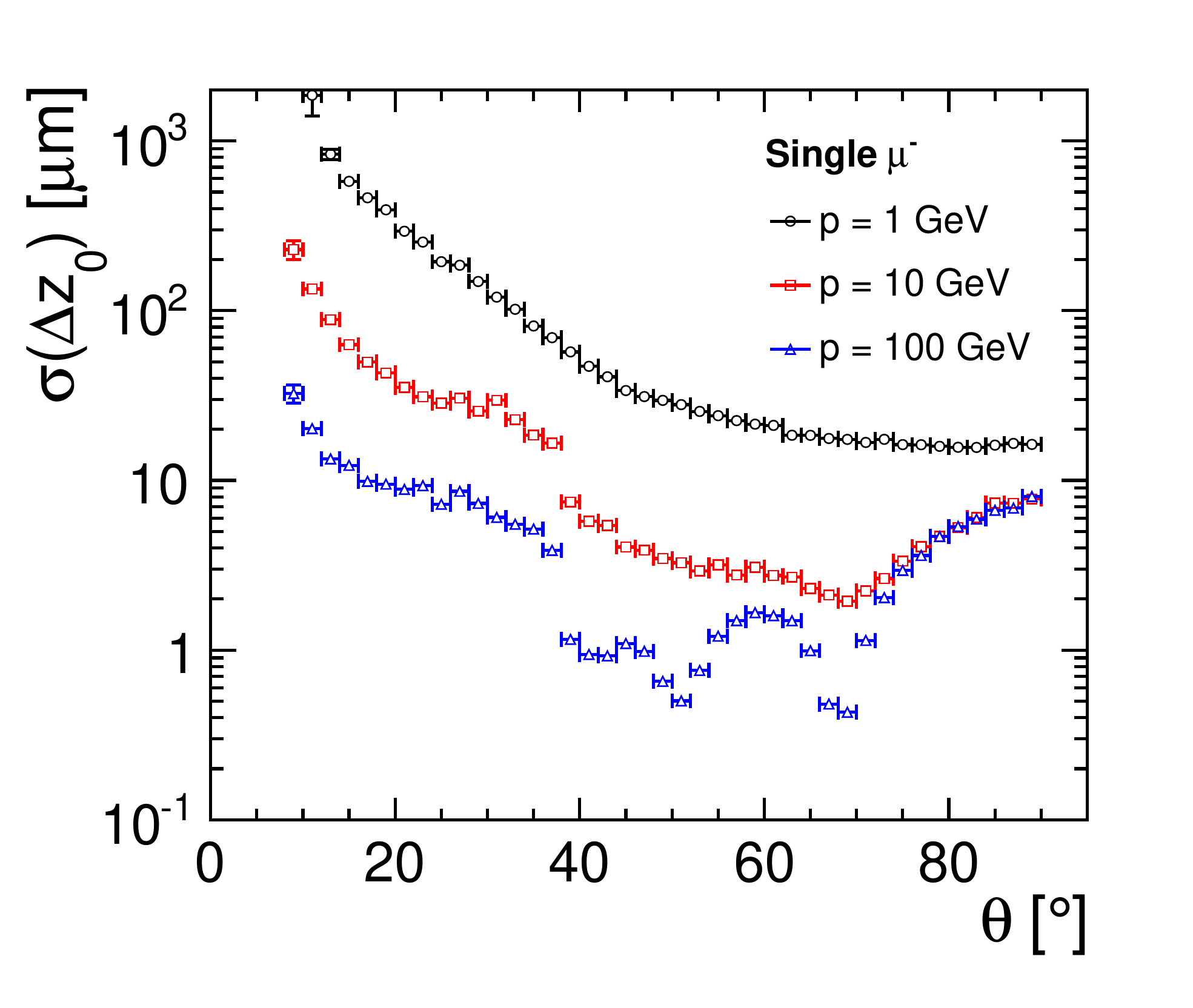}
\caption[Impact parameter $z_0$ resolution for single muons depending on the polar angle.]{Impact parameter $z_0$ resolution depending on the polar angle $\theta$ for single muons of different energies.}
\label{fig:tracking_z0Resolution}
\end{figure}

The $z_0$ resolution, shown in \cref{fig:tracking_z0Resolution}, is clearly dominated by the polar angle resolution with an added $1/\sin\theta$ dependence according to \cref{eq:z0res}. This limits the $z_0$ resolution in the central region to values significantly larger than \unit[1]{\micron}.

\section{Jet Event Samples}
\label{sec:Tracking_eventSamples}



The tracking performance in jets is studied in di-jet events from the decay of a hypothetical $\PZ'$ with a mass of \unit[3]{TeV} which has been produced at rest and is presented in Section~\ref{sec:Tracking_jets}.
These events offer two highly energetic jets with high local track densities which are particularly challenging for the pattern recognition. 
The performance in jet events was also studied in \ttbar events with a center-of-mass energy of \unit[3]{TeV} which can be found in \cref{cha:Appendix_ttTracking}. These events offer quite a different topology with two to six jets, depending on the decay mode of the two \PW bosons. The \ttbar events have a more complex topology compared to the di-jet events but are slightly less challenging for the pattern recognition, which is limited by local occupancy instead of global occupancy.

\begin{figure}[htpb]
 \begin{subfigure}[t]{0.49\textwidth}
 \includegraphics[width=\textwidth]{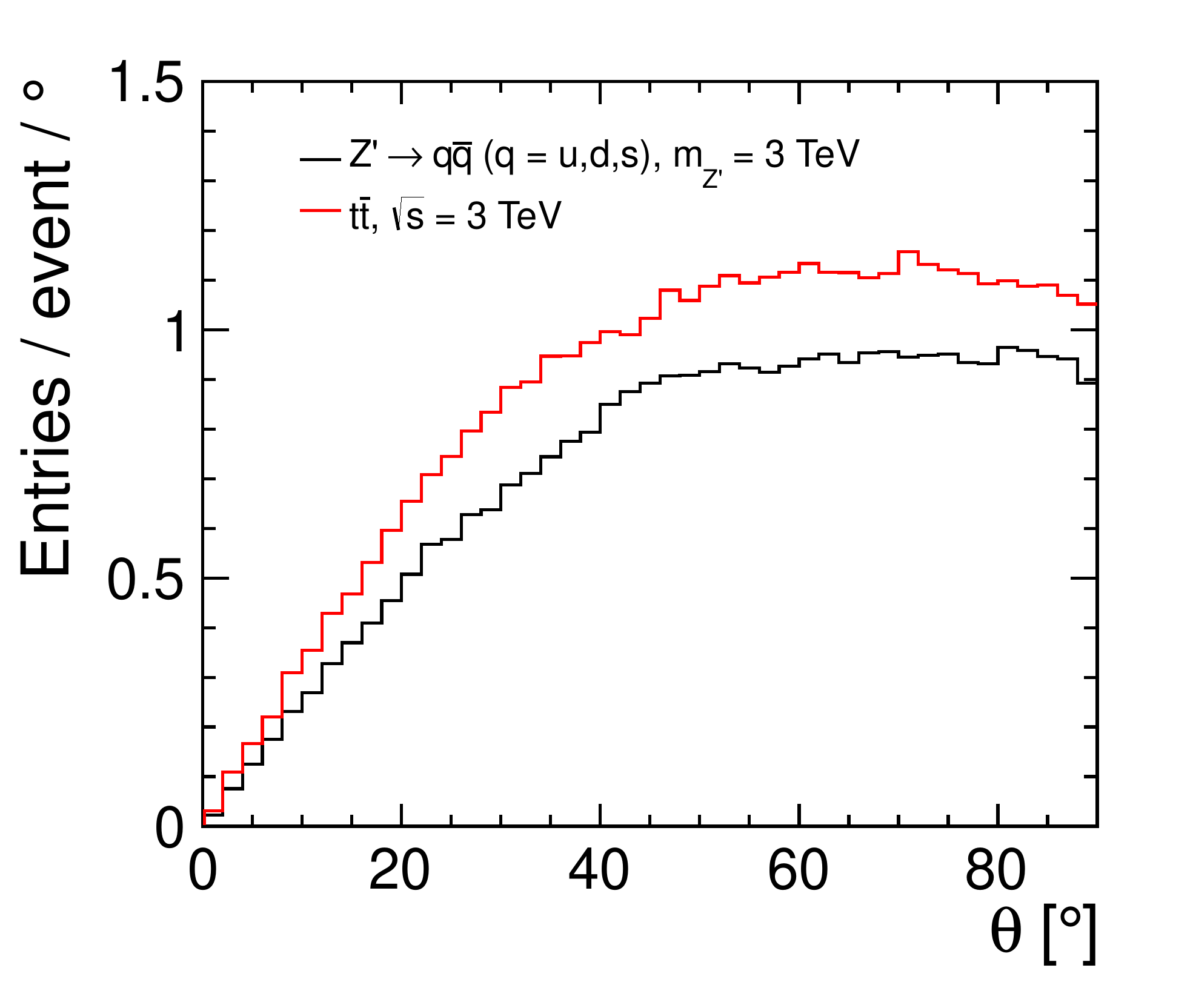}
 \end{subfigure}
 \begin{subfigure}[t]{0.49\textwidth}
 \includegraphics[width=\textwidth]{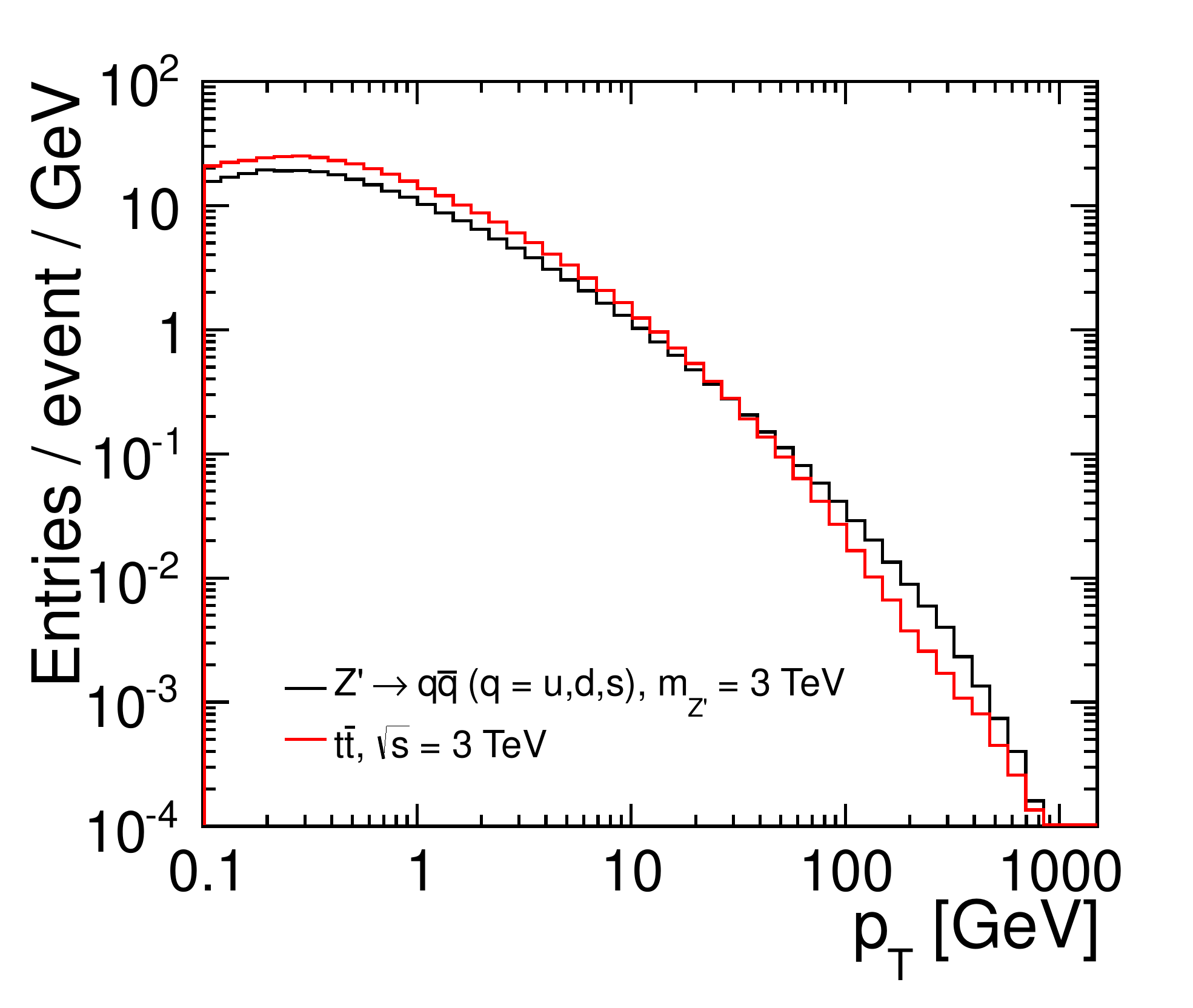}
 \end{subfigure}
\caption[Distribution of polar angles and of transverse momenta of findable particles in jet events.]{Distribution of polar angles $\theta$ (left) and distribution of transverse momenta \pT (right) of findable particles in $\PZ' \rightarrow qq (q = uds)$ and \ttbar events.}
\label{fig:Tracking_jetDistributions}
\end{figure}

The distributions of the polar angle and the transverse momentum of the findable particles in the jet events are shown in \cref{fig:Tracking_jetDistributions} (see \cref{sec:Tracking_efficiency} for the definition of findable particles). While the angular distribution is similar in both cases with most of the tracks inside the barrel region of $\theta > 40\degrees$, the transverse momentum distribution extends to higher momenta in case of the di-jet events. This is due to the fact that the same amount of energy is distributed over fewer tracks.

\begin{figure}[htpb]
 \begin{subfigure}[t]{0.49\textwidth}
 \includegraphics[width=\textwidth]{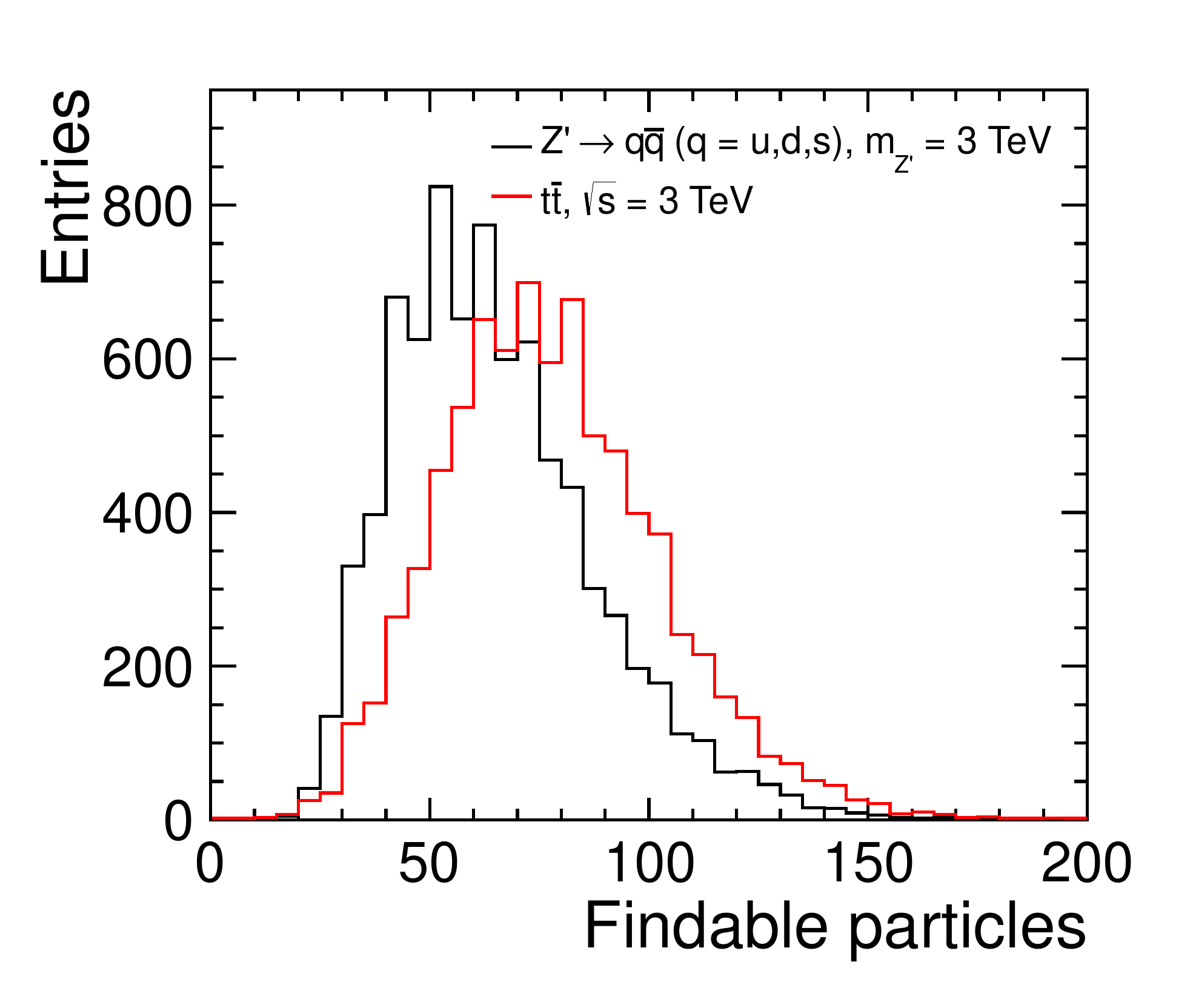}
 \end{subfigure}
 \begin{subfigure}[t]{0.49\textwidth}
 \includegraphics[width=\textwidth]{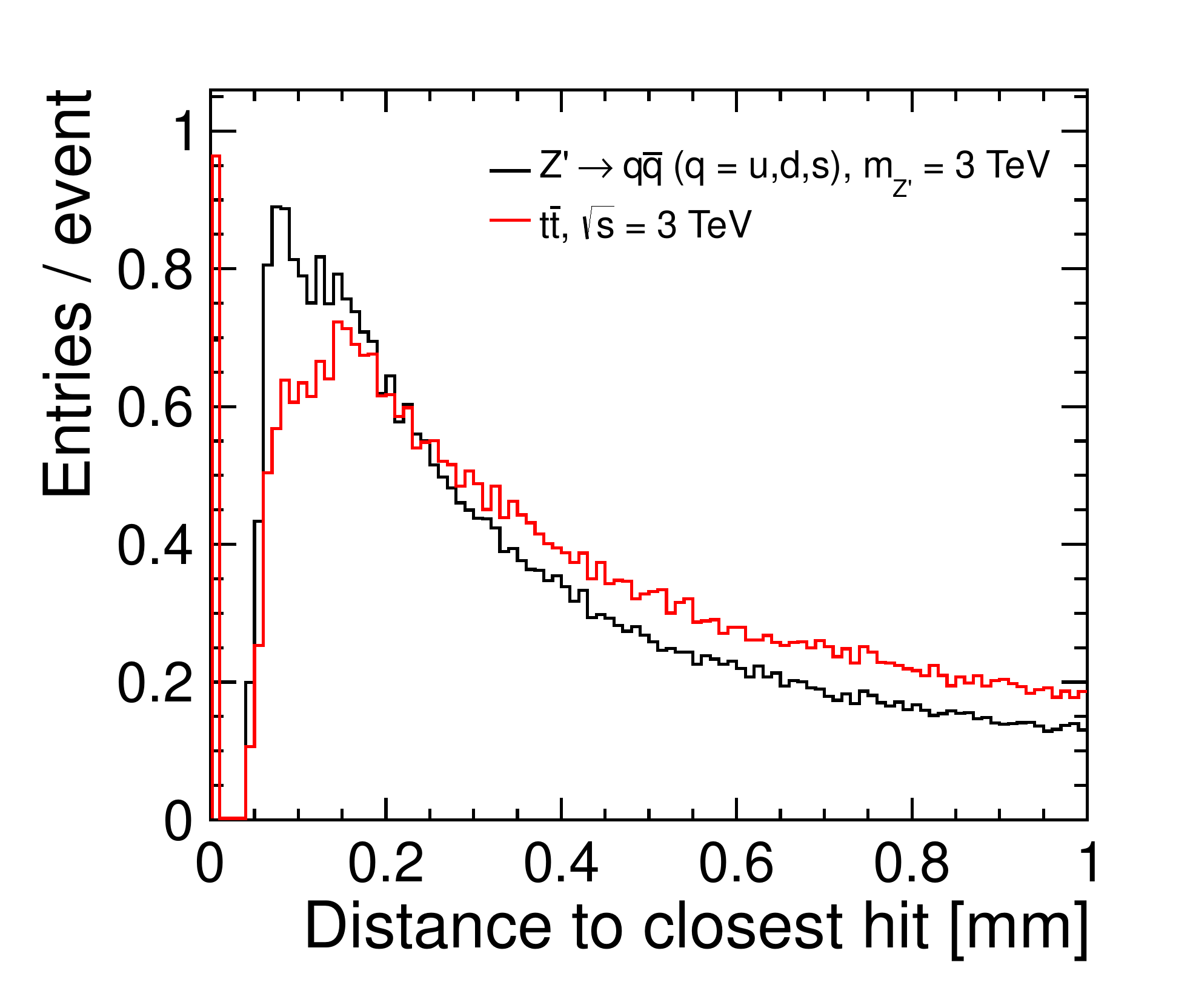}
 \end{subfigure}
\caption[Distribution of number of findable particles and the distance to the closest hit in jet events.]{Distribution of number of findable particles (left) and the distance to the closest hit not originating from these particle (right) in $\PZ' \rightarrow qq (q = uds)$ and \ttbar events.}
\label{fig:Tracking_jetDistributions2}
\end{figure}

The track multiplicity is shown in \cref{fig:Tracking_jetDistributions2}~(left). The average number of findable tracks in di-jet events is 64.6, while it is 77.6 in \ttbar events. Despite the higher track multiplicity, the local occupancy is lower in \ttbar events since the tracks are distributed over up to six jets and those jets are less pencil like due to the lower track momentum. This can be seen in \cref{fig:Tracking_jetDistributions2}~(right) which shows the closest distance from any of the track hits to any other hit. Any hit closer than \unit[40]{\micron} will not results in a separate digitized cluster and can thus not be resolved. In both types of events a large number of tracks has at least one hit within its direct vicinity. The peak for \ttbar events is around \unit[190]{\micron}, while for di-jet events it is below the threshold that can be resolved.

\section{Tracking Efficiency and Fake Rate Definition}
\label{sec:Tracking_efficiency}

An important aspect of the tracking performance is the efficiency of reconstructing a particle track.
First the reconstructed tracks have to be matched to Monte~Carlo particles. This is done using the truth linking explained in \cref{sec:Software_TruthLinking}. The particle contributing to the majority of hits that have been associated to the track during the track reconstruction is considered the true particle. In addition, if this particle contributed to multiple reconstructed tracks it is only considered the true particle for the track to which it contributed most of its hits.
The tracking efficiency can then be expressed as
\begin{equation}
 \epsilon = \frac{N_{\mathrm{reconstructed}}}{N_{\mathrm{findable}}},
\end{equation}
where $N_{\mathrm{reconstructed}}$ is the number of findable particles that have been matched to a reconstructed track and $N_{\mathrm{findable}}$ is the number of particles that are supposed to be found.
The definition of what to call a findable particle is strongly dependend on which type of tracking efficiency is looked at.
Throughout this Chapter two types of tracking efficiencies are used.
\begin{description}
 \item[Algorithmic tracking efficiency] is a measure of how efficient the tracking algorithm finds tracks of particles that fulfill all requirements that are intrinsic to the algorithm: the minimum transverse momentum, maximum impact parameter, minimum number of total hits and sufficient number of hits in the seed layers. This tracking efficiency is neither sensitive to the geometric coverage nor the coverage by the chosen tracking strategies.
 \item[Total tracking efficiency]---further referred to as just tracking efficiency---includes inefficiencies due to the algorithm and the geometric coverage and represents the efficiency relevant to any realistic physics analysis. Nevertheless, some selection cuts have to be applied to the particles in order to select only those that are actually relevant for a physics analysis. First, only those charged particles that travel at least \unit[5]{cm}, measured along their helical path, are considered findable particles to suppress those intermediate particles short life times. In addition, only those particles originating from within a sphere with a radius of \unit[5]{cm} from the interaction point are taken into account. This cut does not reject particles originating from typical secondary vertices but removes particles created from bremsstrahlung in the tracking volume.

 Two additional selection cuts are applied depending on the variable that is plotted if not stated otherwise. For efficiency plots versus $\theta$ a minimum transverse momentum of \unit[1]{GeV} is required. Similarly, for plots versus \pT a minimum polar angle of 15\degrees is implied. This is done to disentangle the effects of detector acceptance and low efficiency for low momentum tracks.
\end{description}

\begin{figure}[htpb]
 \includegraphics[width=0.49\textwidth]{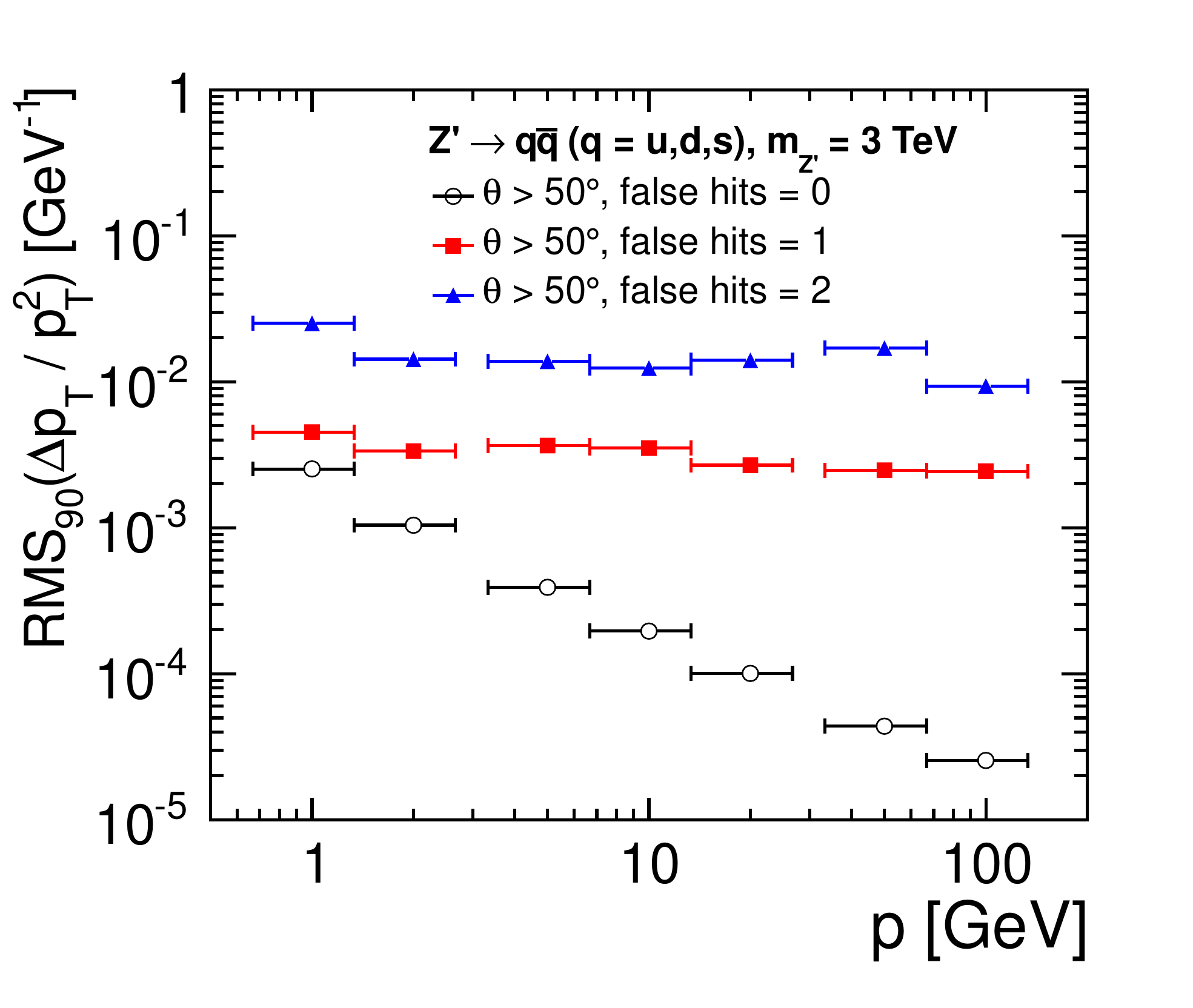}
 \hfill
 \includegraphics[width=0.49\textwidth]{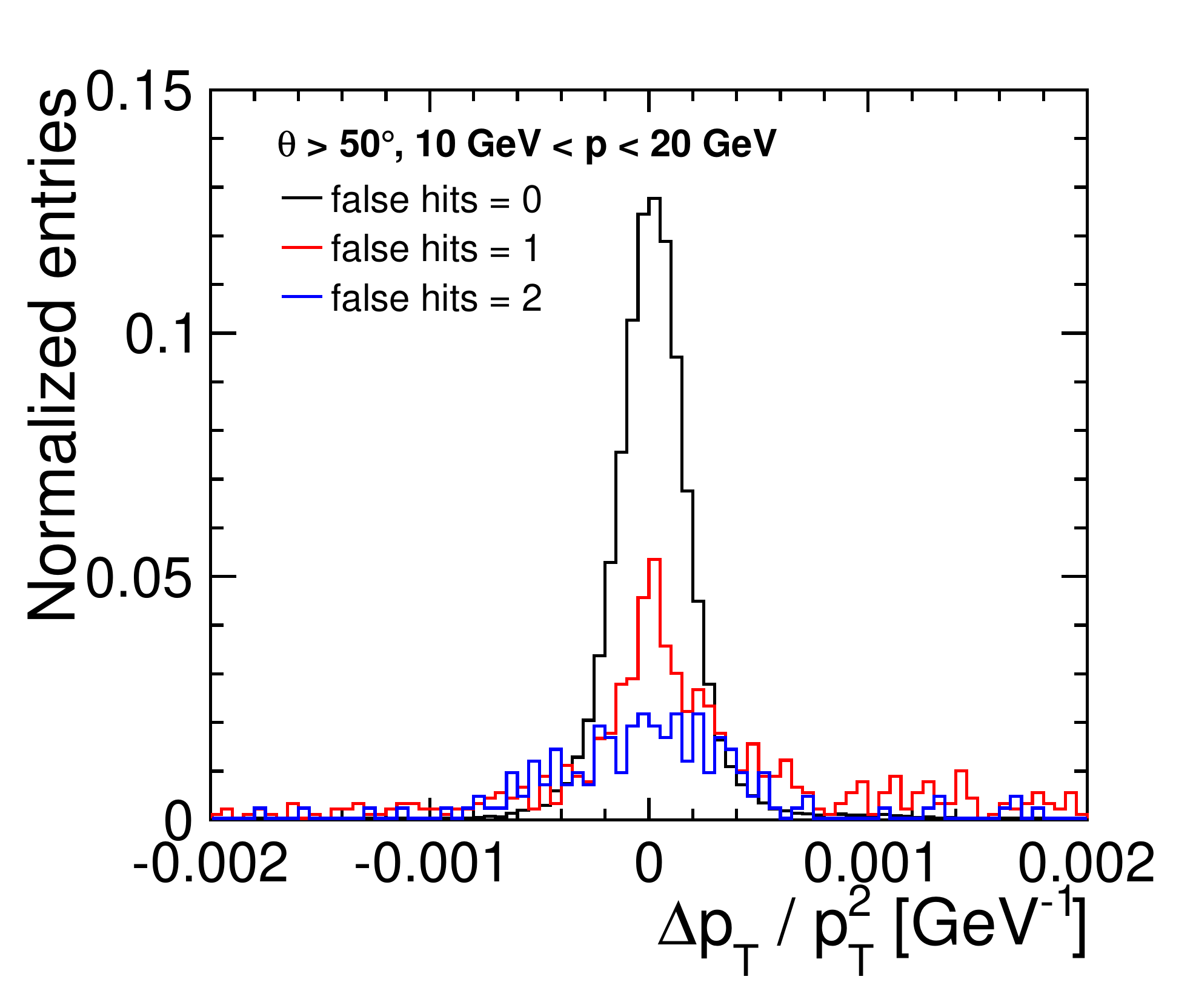}
\caption[Momentum resolution in di-jet events for different numbers of false hits.]{Momentum resolution depending on the track momentum (left) and the $\frac{\Delta\pT}{\pT^2}$-distribution for $\unit[10]{GeV} < p < \unit[20]{GeV}$ (right) for reconstructed tracks in the central region of the detector with different numbers of false hits in di-jet events.}
\label{fig:Tracking_resolutionPt2FalseHits}
\end{figure}
\begin{figure}[htpb]
 \includegraphics[width=0.49\textwidth]{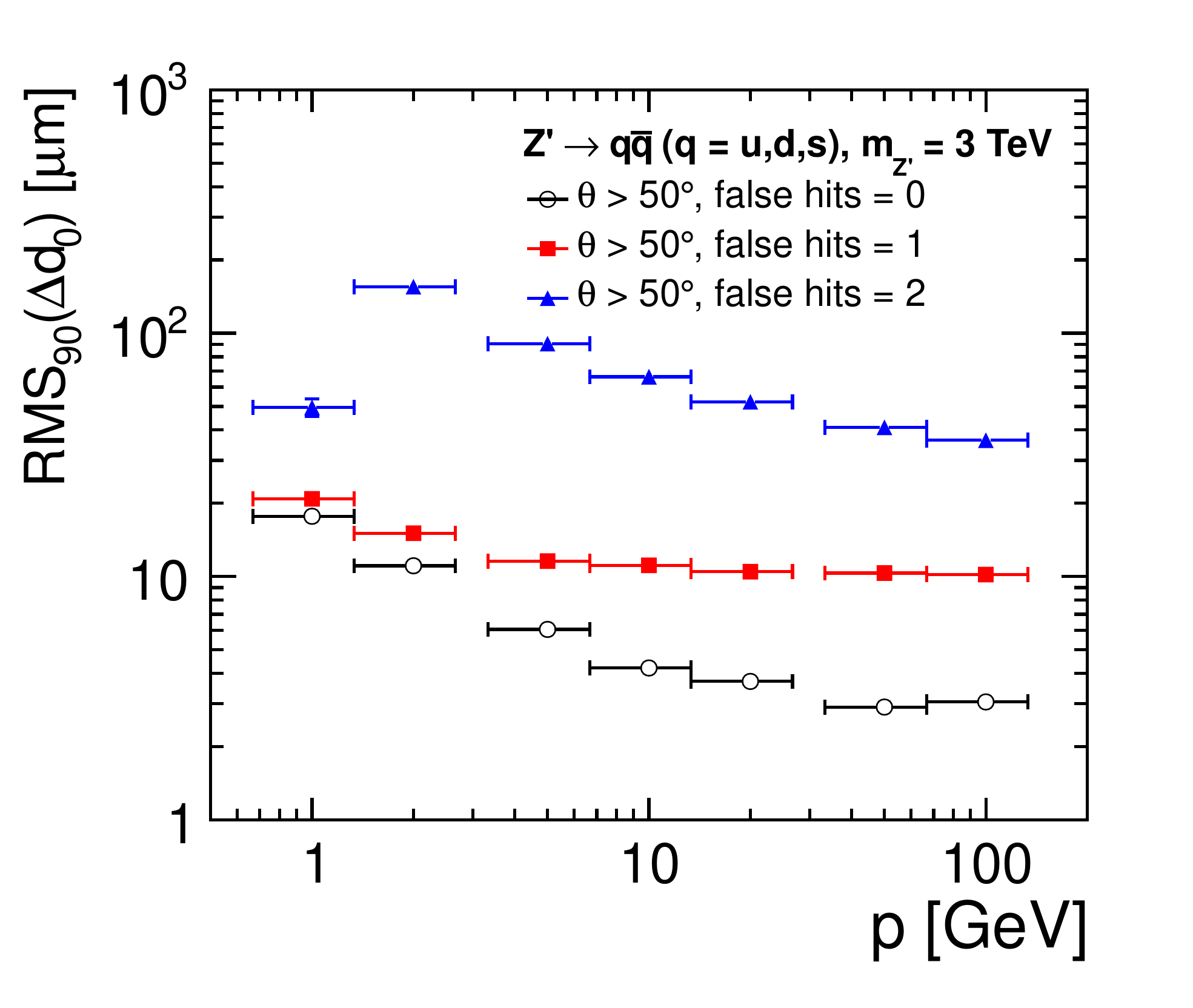}
 \hfill
 \includegraphics[width=0.49\textwidth]{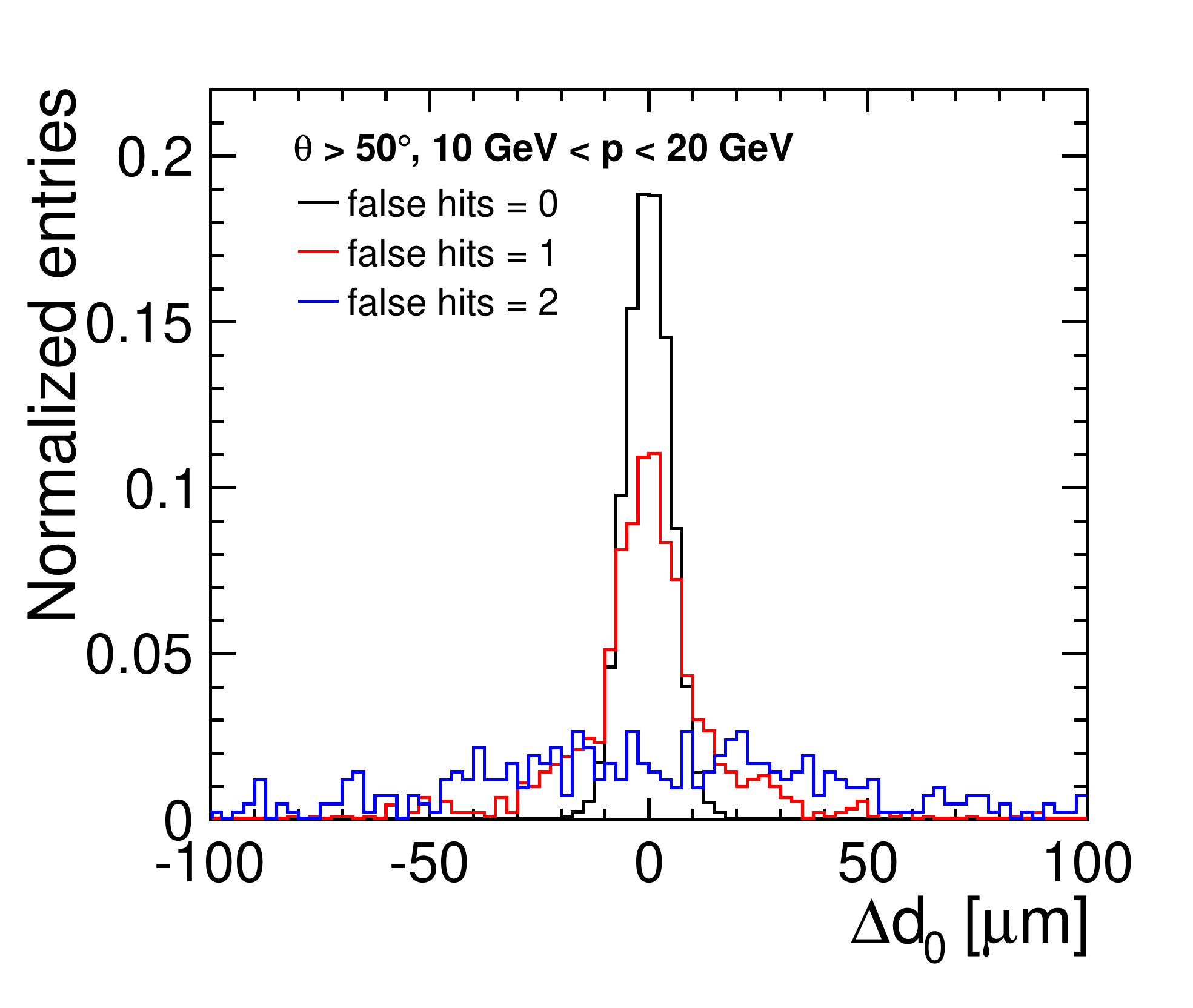}
\caption[Impact parameter resolution in di-jet events for different numbers of false hits.]{The $d_0$ impact parameter resolution depending on the track momentum (left) and the $\Delta d_{\mathrm 0}$ distribution depending on the polar angle (right) for reconstructed tracks with different numbers of false hits in di-jet events.}
\label{fig:Tracking_resolutionD0FalseHits}
\end{figure}

All selection cuts are applied to both the reconstructed and findable particles. An additional quality cut is applied to the reconstructed particles to ensure that the parameters of the reconstructed track resemble those of the original particle. This quality cut is based on the number of false hits that have been assigned to a reconstructed track. A hit is considered a false hit if the truth matched particle did not contribute to the hit. \cref{fig:Tracking_resolutionPt2FalseHits} shows the transverse momentum resolution depending on the track momentum and the $\frac{\Delta\pT}{\pT^2}$-distribution for central tracks in di-jet events with different numbers of false hits. \cref{fig:Tracking_resolutionD0FalseHits} shows the impact parameter resolution depending on the track momentum and the $\Delta d_{0}$-distribution for central tracks in di-jet events with different numbers of false hits. Both resolutions are degrading strongly if one or more false hits have been assigned to a track and large non-Gaussian tails can be seen in the distributions. For this reason, the $RMS_{90}$ method was preferred over a Gaussian fit to obtain the resolutions.

\begin{figure}[htpb]
 \includegraphics[width=0.49\textwidth]{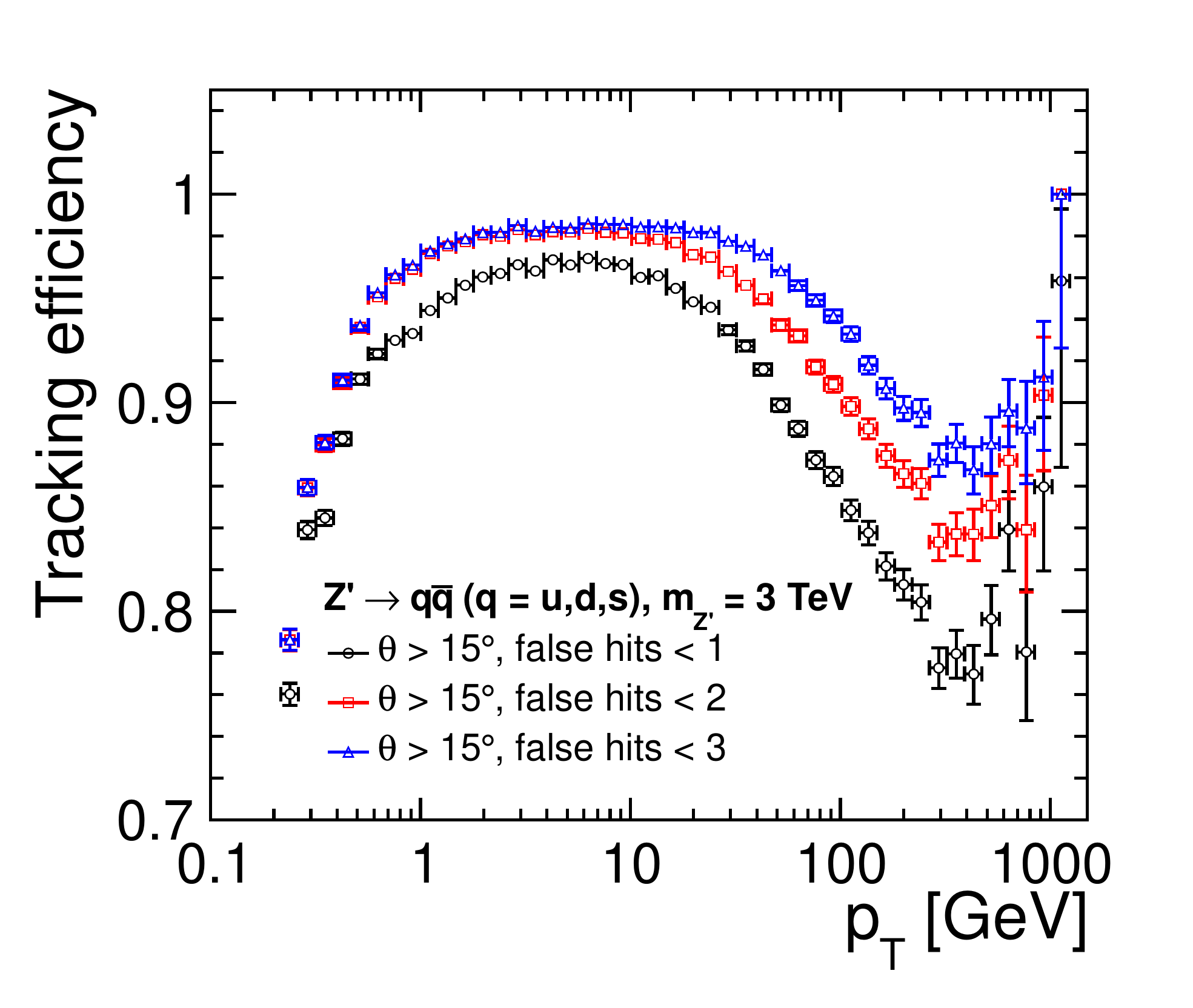}
 \hfill
 \includegraphics[width=0.49\textwidth]{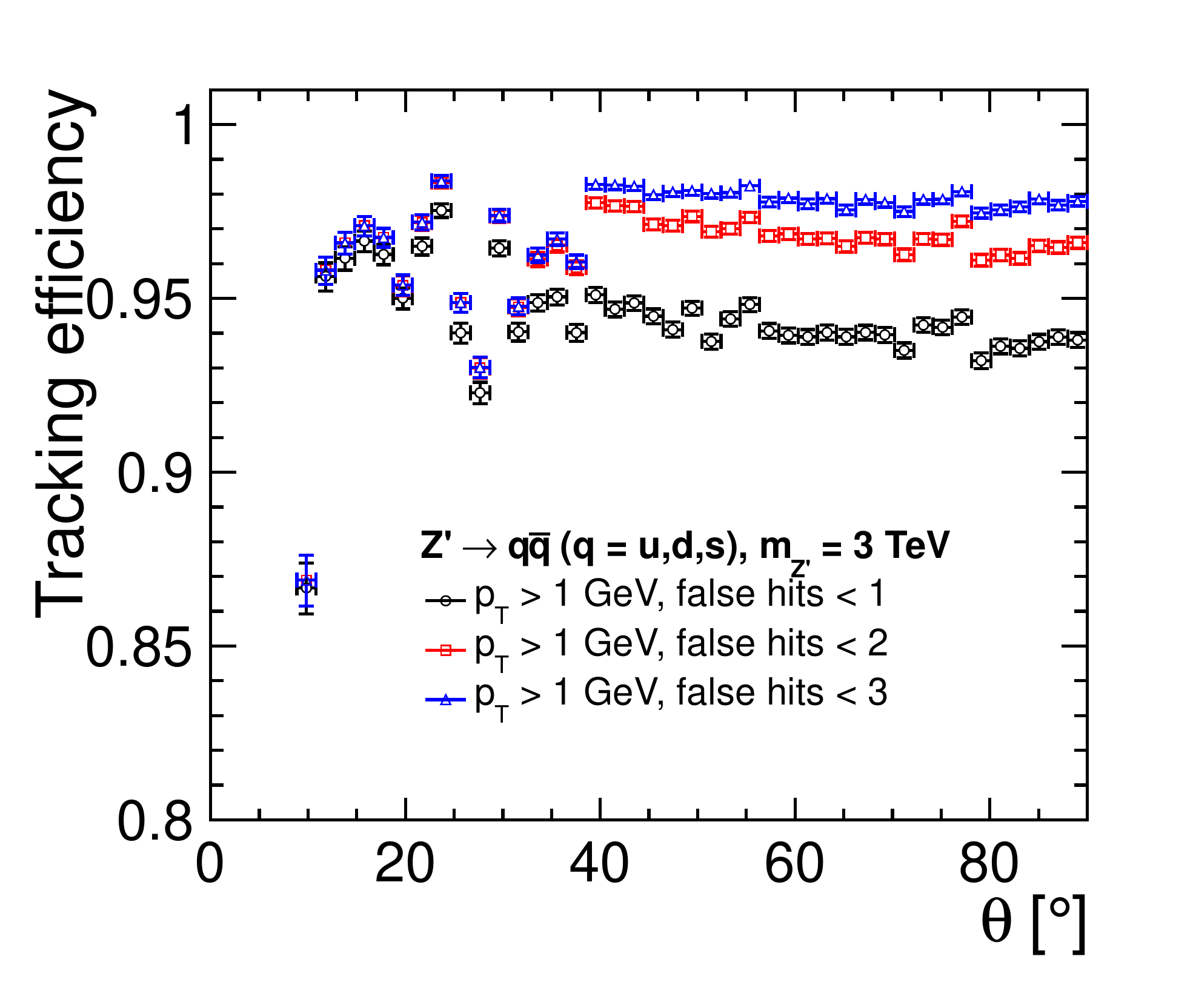}
\caption[Tracking efficiency for reconstructed tracks with different numbers of false hits in di-jet events.]{Tracking efficiency depending on the transverse momentum (left) and the polar angle (right) for reconstructed tracks with different numbers of false hits in di-jet events.}
\label{fig:Tracking_efficiencyFalseHits}
\end{figure}

Requiring a pure track with no false hits reduces the track reconstruction efficiency, as shown in Figure~\ref{fig:Tracking_efficiencyFalseHits}. This effect is most strongly seen in the barrel region, where due to the missing information on the $z$-coordinate of the strip detectors it is more likely to pick up a false hit. Thus, the definition of a good track is always a tradeoff between ultimate resolution and good efficiency. For the remainder of this chapter the maximum number of false hits on a track is one for those tracks that are counted as reconstructed tracks in the efficiency calculation.

The fake rate is a measure of the number of tracks that are reconstructed but that do not correspond to any of the true particles. It can be expressed as
\begin{equation}
 f = \frac{N_{\mathrm{fake}}}{N_{\mathrm{reconstructed}}},
\end{equation}
where $N_{\mathrm{fake}}$ is the number of fake tracks and $N_{\mathrm{reconstructed}}$ is the number of all reconstructed tracks.
One type of fake tracks is purely combinatorial fakes where a track is reconstructed from random hits. A bad track on the other hand is a track that correctly identifies the hits originating from a single particle but has picked up additional hits and results in track fit parameters far from those of the original particle. Both effects mostly depend on the occupancy in the tracking detectors. Here the second, much stricter, definition is used and the fake rate is given as the fraction of tracks that have more than one false hit.



\section{Strategy Optimization}
\label{sec:Tracking_strategy}

As explained in \cref{sec:CLIC_machineInduced}, the track finding is steered by a set of strategies which can be trained automatically. For this training, a dedicated event sample of 100000 single muon events is used. The muons are uniformly distributed in $p$ between \unit[0.1]{GeV} and \unit[50]{GeV}, in $\theta$ between 5\degrees and 175\degrees and between 0\degrees and 360\degrees in $\phi$. The minimum transverse momentum for the strategy training is set to \unit[0.2]{GeV}. The layer weights are set such that vertex layers are preferred for seeding with decreasing weights for layers further away from the interaction point, resulting in inside-out tracking strategies.

The incoherent pair background, introduced in \cref{sec:CLIC_machineInduced}, could not be included in the track reconstruction due to its huge number of particles. In order to not overestimate the track finding performance, the two innermost vertex layers are excluded as seed layers. Their occupancy is considerably higher than those of the outer layers, as shown in \cref{sec:SiD_occupancies}. Nevertheless, these layers are allowed as extension layers and thus their hits are used for the track fit.

\begin{figure}[htpb]
 \includegraphics[width=0.49\textwidth]{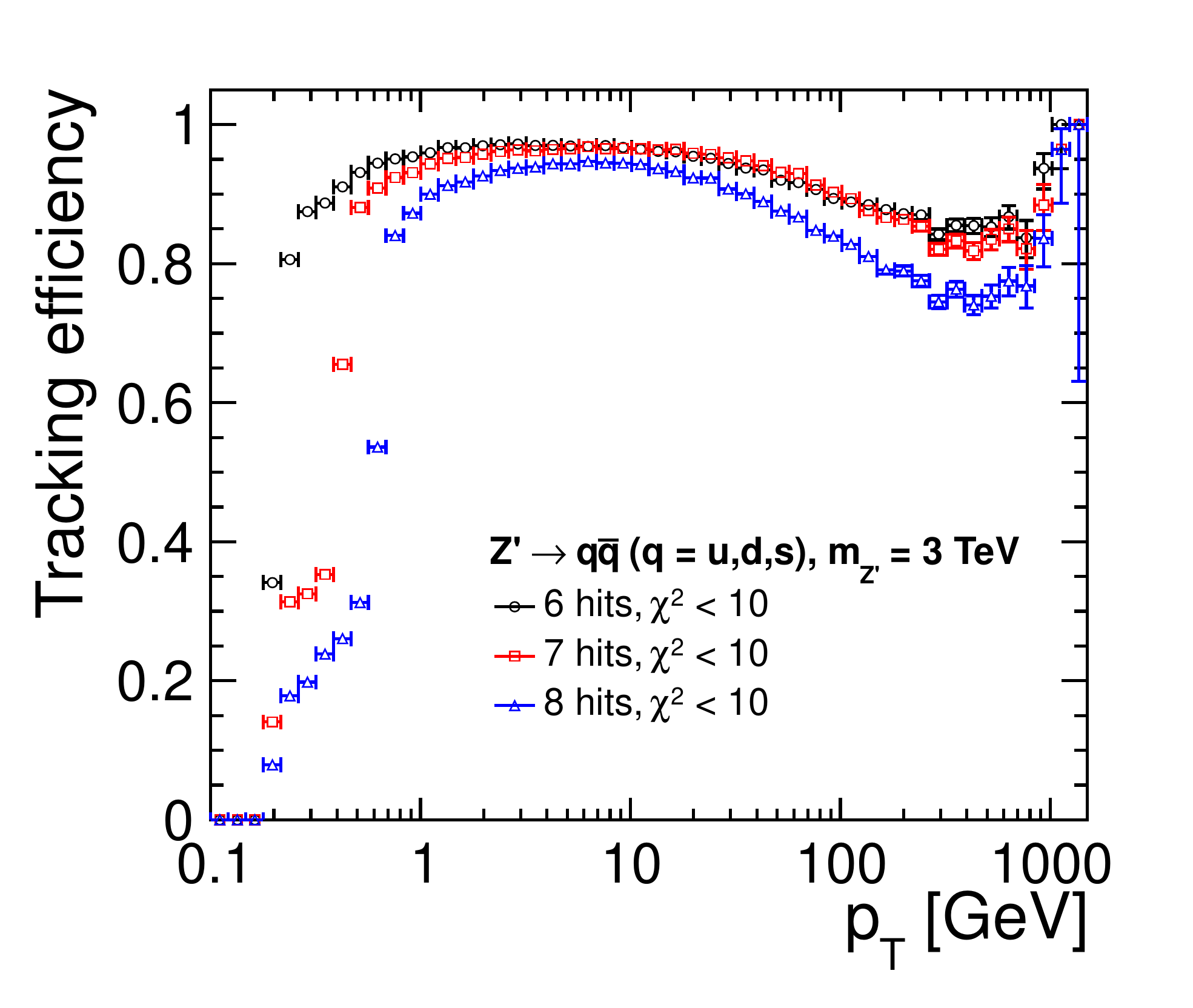}
 \hfill
 \includegraphics[width=0.49\textwidth]{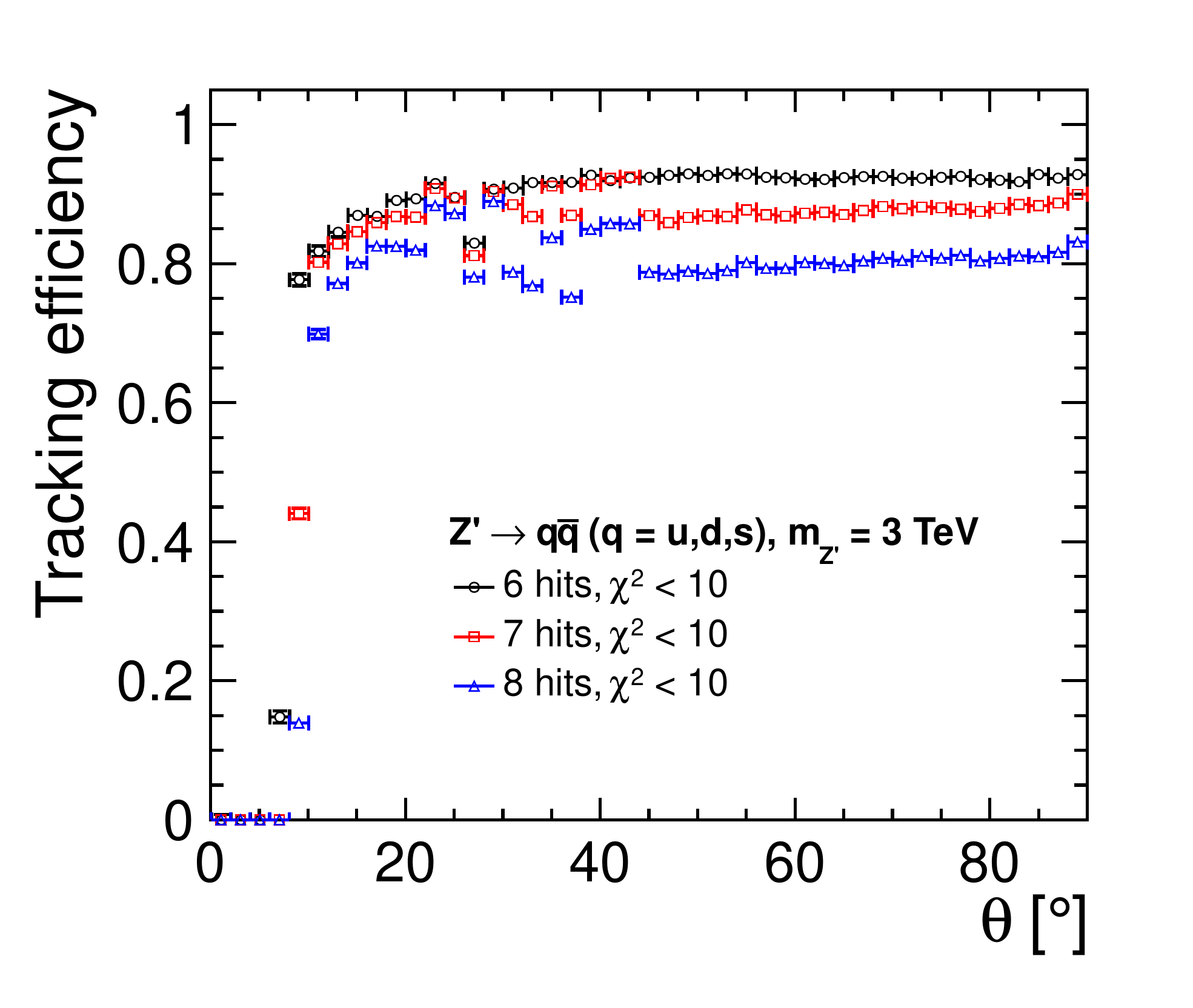}
\caption[Tracking efficiency for different minimum hits requirements in di-jet events.]{Tracking efficiency depending on the transverse momentum \pT (left) and depending on the polar angle $\theta$ (right) for various strategy sets requiring a minimum of 6, 7 or 8 hits for a reconstructed track in $\PZ' \rightarrow qq (q = uds)$ events.}
\label{fig:Tracking_strategyHits}
\end{figure}

\begin{figure}[htpb]
 \includegraphics[width=0.49\textwidth]{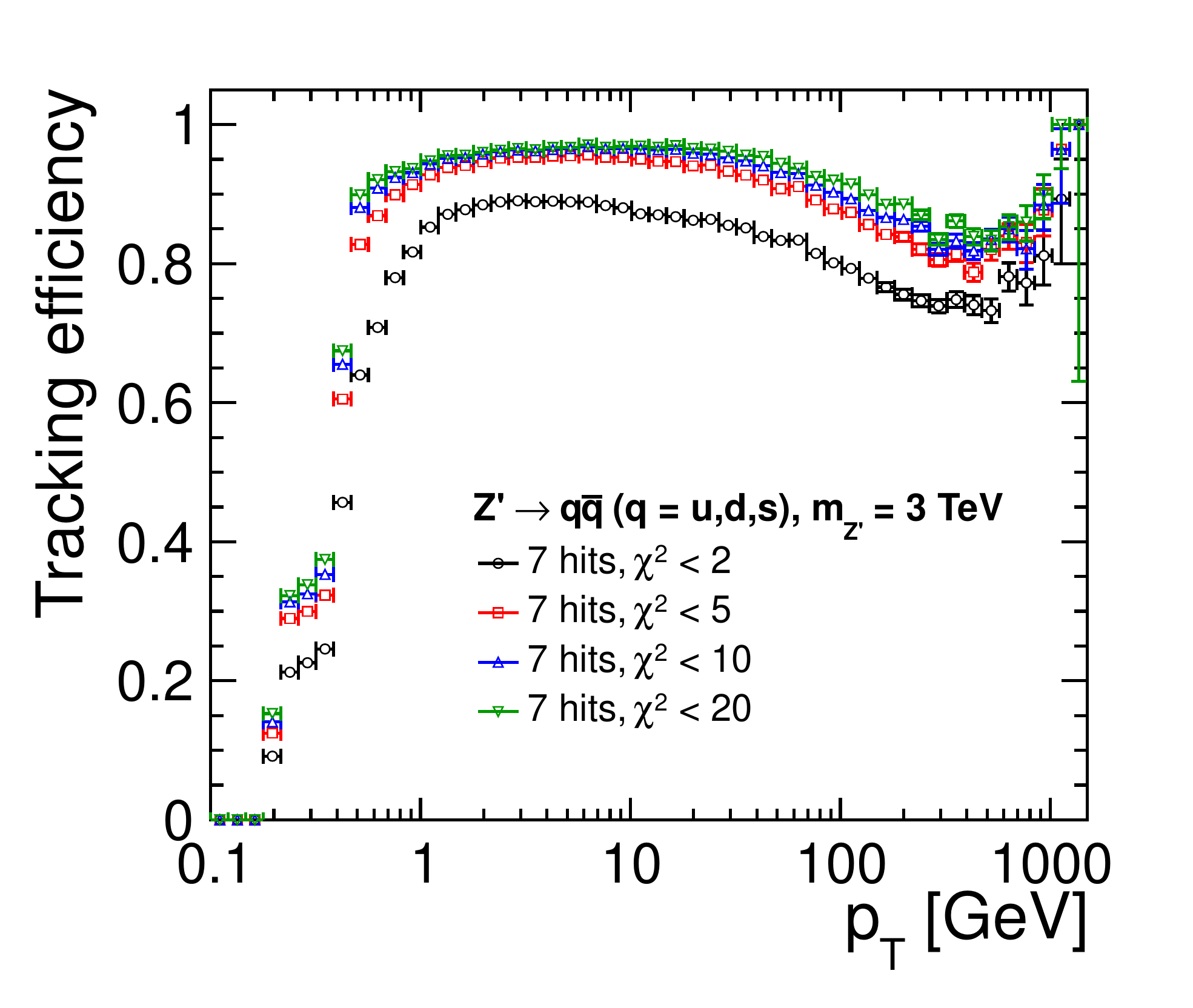}
 \hfill
 \includegraphics[width=0.49\textwidth]{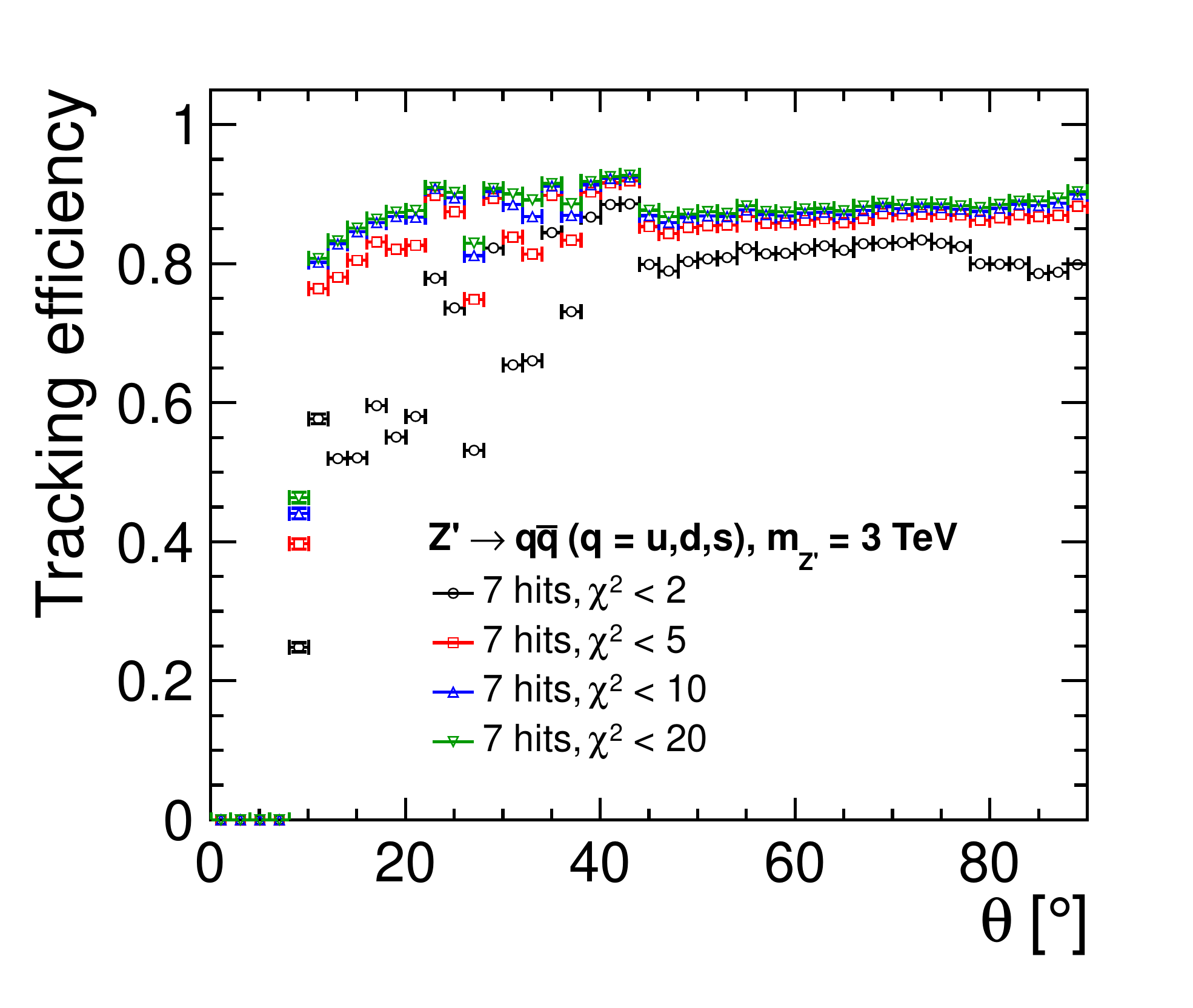}
\caption[Tracking efficiency for different $\chi^2_\mathrm{max}$ requirements in di-jet events.]{Tracking efficiency depending on the transverse momentum \pT (left) and depending on the polar angle $\theta$ (right) for various strategy sets with different $\chi^2_\mathrm{max}$ and a minimum of 7 hits for the track finding in $\PZ' \rightarrow qq (q = uds)$ events.}
\label{fig:Tracking_strategyChisq}
\end{figure}

The $\chi^2_\mathrm{max}$ and $N_\mathrm{min}$ (see \cref{sec:Software_trackingFinding}) for the final set of tracking strategies were chosen by a parameter scan and the resulting impact on the tracking efficiency in di-jet events. \cref{fig:Tracking_strategyHits} shows the tracking efficiency depending on \pT and $\theta$ for different $N_\mathrm{min}$ and a fixed $\chi^2_\mathrm{max}$. Requiring a large number of hits for a track results in lowered tracking efficiency for low momentum tracks, which are less likely to pass a sufficient number of layers, as well as a smaller acceptance in the very forward region of $\theta < 10\degrees$. The final strategy set thus requires a minimum of 7 hits with an exception of barrel only tracks which require only 6 hits to improve finding of central low momentum tracks. The impact of varying the $\chi^2_\mathrm{max}$ is shown in \cref{fig:Tracking_strategyChisq} where the minimum number of hits is kept at 7. For $\chi^2_\mathrm{max}$ below 10 the performance is degrading significantly, while for higher values the performance is almost constant. The final strategy set uses a $\chi^2_\mathrm{max}$ of 10 for all strategies.

\section{Performance in Jet Events}
\label{sec:Tracking_jets}

This section discusses the tracking performance in di-jet events. The performance in \ttbar events is slightly different due to the different topology and can be found in \cref{cha:Appendix_ttTracking}.

\subsection{Track Finding Efficiency}
\begin{figure}[htpb]
 \includegraphics[width=0.49\textwidth]{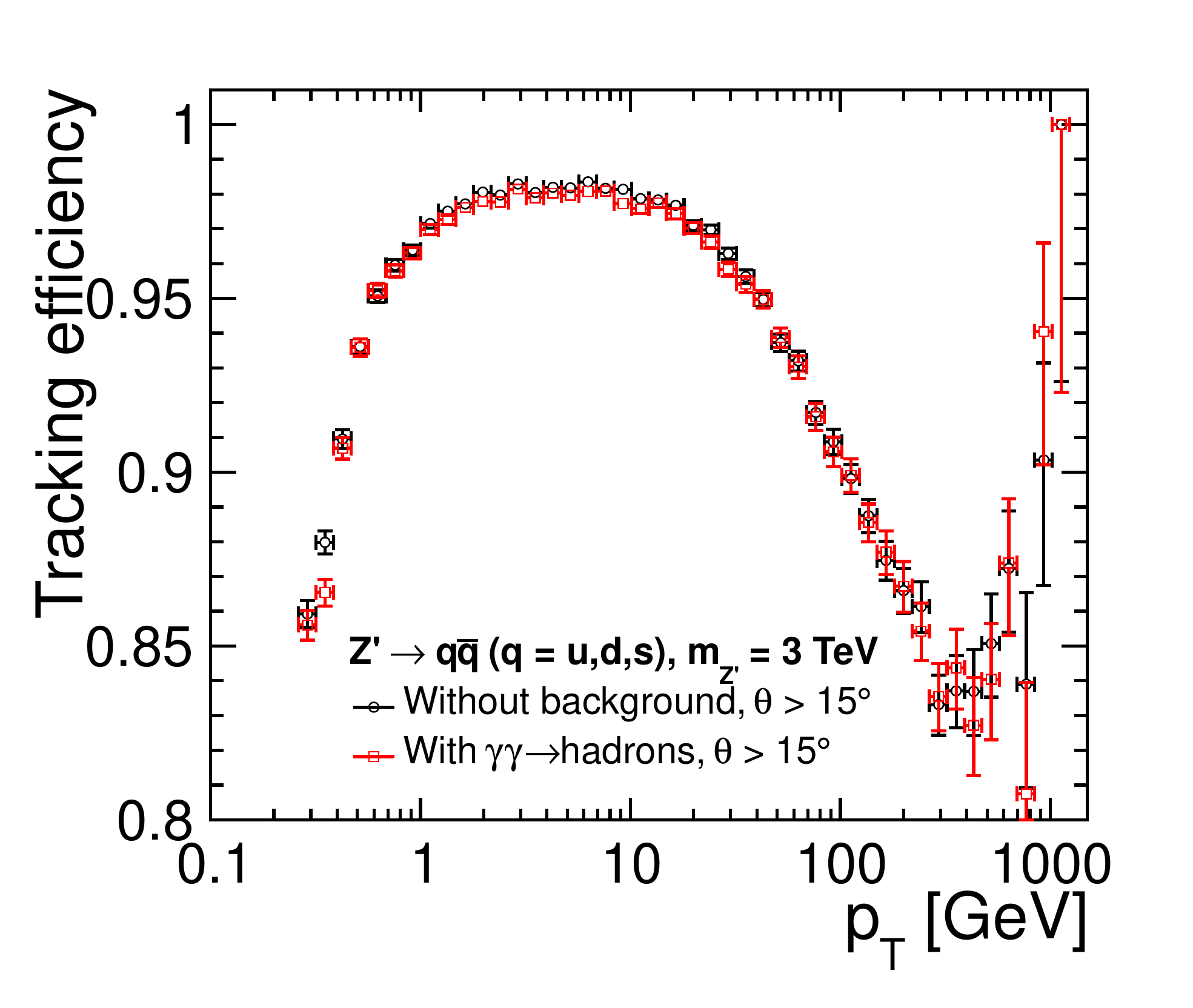}
 \hfill
 \includegraphics[width=0.49\textwidth]{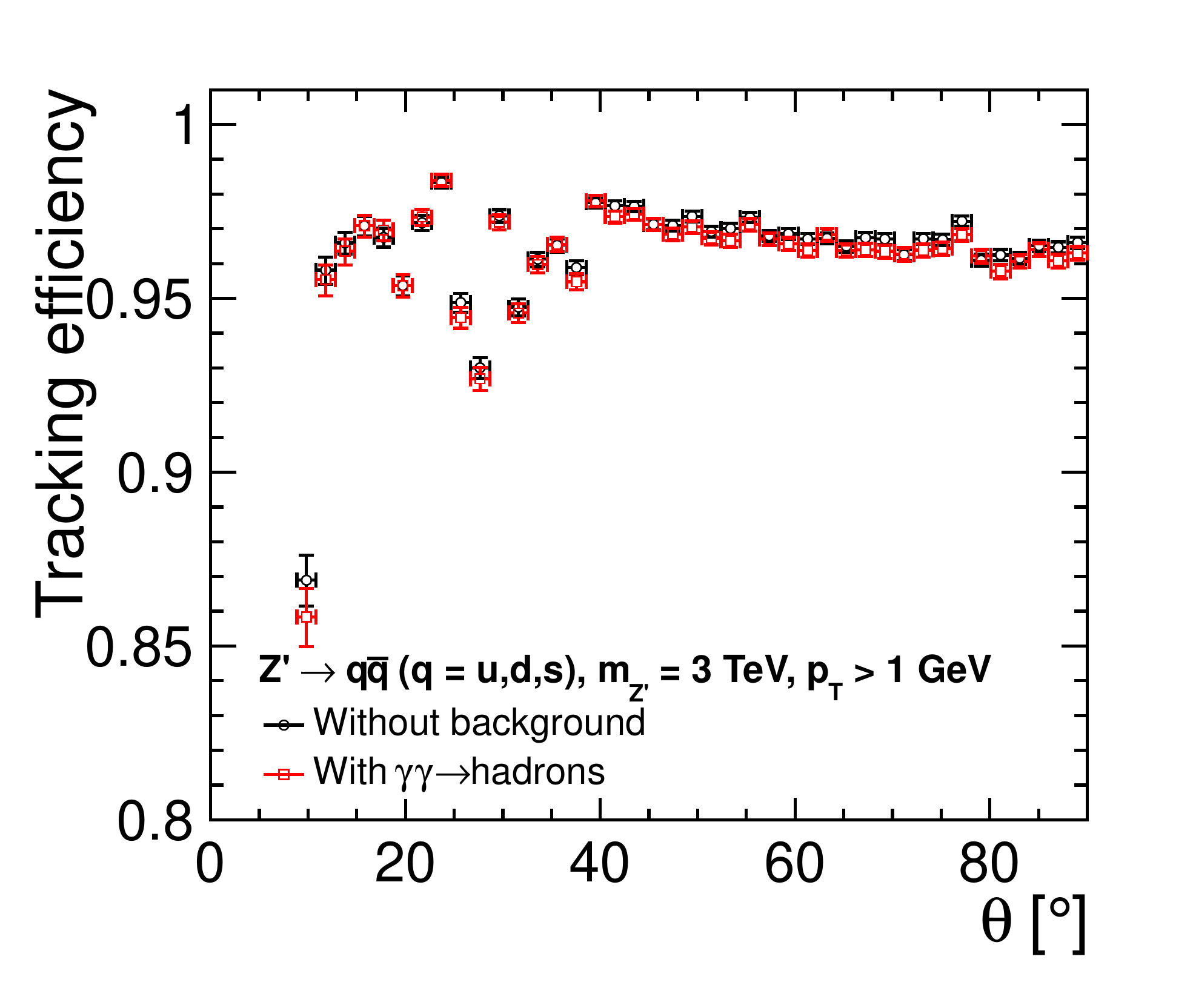}
\caption[Tracking efficiency in di-jet events with and without \gghad background.]{Tracking efficiency depending on the transverse momentum \pT (left) and the polar angle $\theta$ (right) in $\PZ' \rightarrow qq (q = uds)$ events with and without \gghad background.}
\label{fig:Tracking_dijetEfficiency}
\end{figure}

The tracking efficiency in di-jet events depending on the polar angle and the transverse momentum is shown in \cref{fig:Tracking_dijetEfficiency}. The highest tracking efficiency of 97\% to 98\% is achieved for tracks with a transverse momentum between \unit[1]{GeV} and \unit[20]{GeV} which constitutes the majority of the tracks. The efficiency drops to below 85\% for tracks with a high transverse momentum of around \unit[400]{GeV}. The efficiency increases for higher momentum tracks which have a higher probability of being isolated, since they carry a very large fraction of the jet energy. The efficiency depending on the polar angle is rather constant in the barrel region going from 96\% for very central tracks to about 98\% for tracks with a polar angle of 40\degrees. In the forward region, below 40\degrees, the tracking efficiency shows large fluctuations depending on the local material budget, resulting in efficiencies between 92\% and 98\%. The efficiency drops sharply for angles below 10\degrees due to the detector acceptance.

Overlaying typical amounts of \gghad background expected at CLIC has no significant effect on the track finding efficiency.

\begin{figure}[htpb]
 \includegraphics[width=0.49\textwidth]{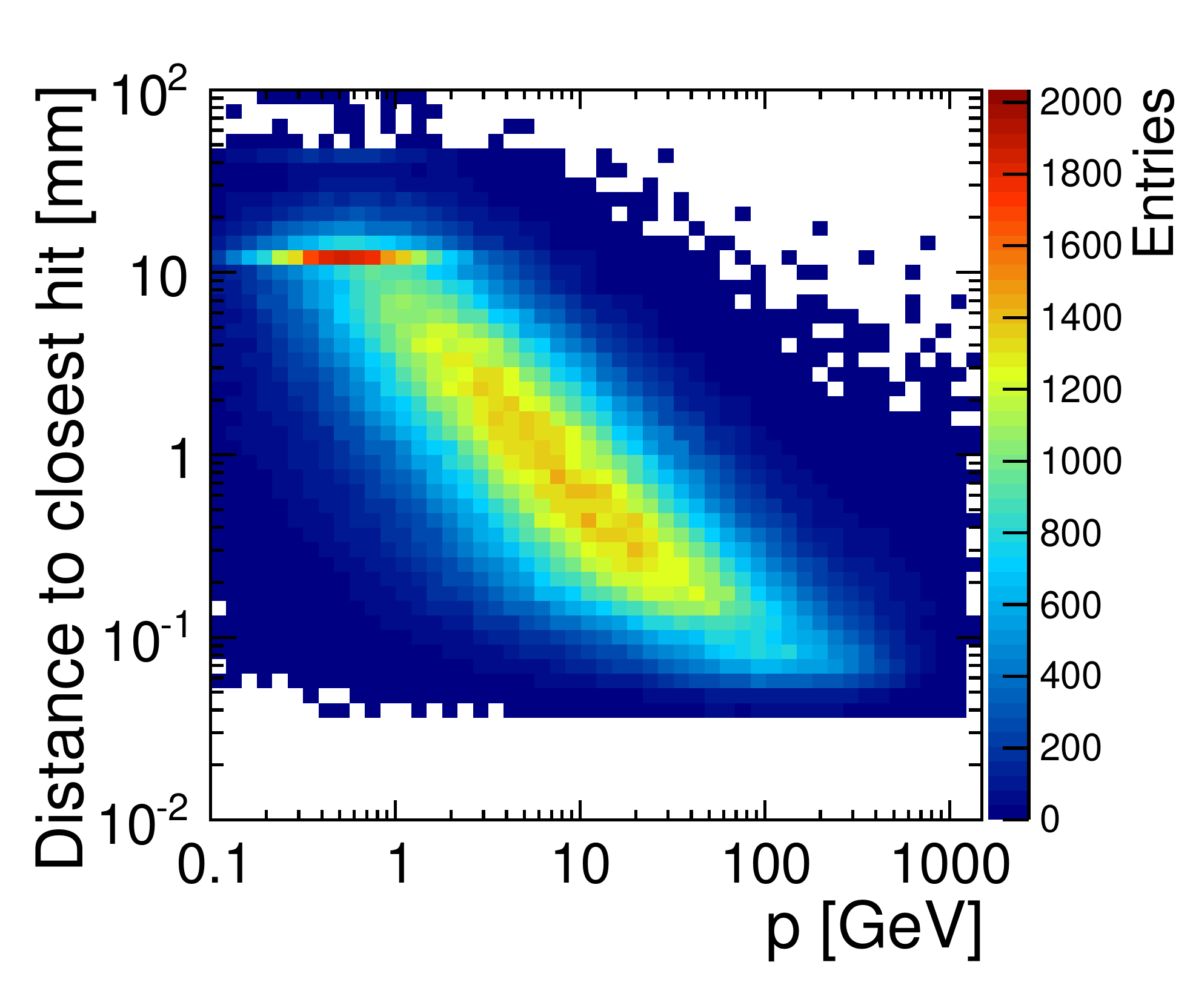}
 \hfill
 \includegraphics[width=0.49\textwidth]{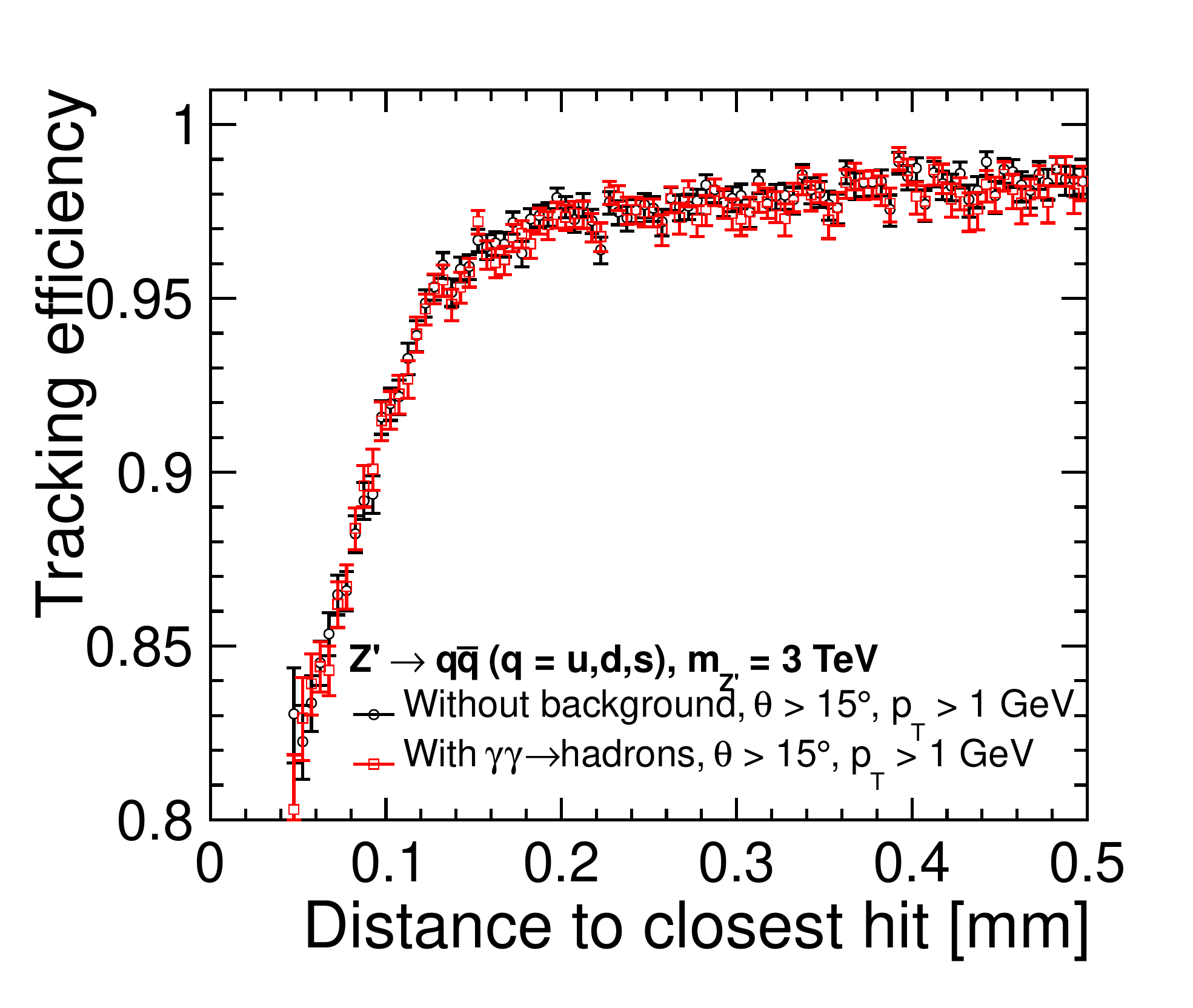}
\caption[Tracking efficiency depending on the distance to the closest hit.]{Correlation between track momentum $p$ and the distance to the closest hit in $\PZ' \rightarrow qq (q = uds)$ events (left). Tracking efficiency depending on the distance to the closest hit in $\PZ' \rightarrow qq (q = uds)$ events with and without \gghad background (right).}
\label{fig:Tracking_dijetEfficiency2}
\end{figure}

High momentum tracks are most likely in the center of the jet, as shown in \cref{fig:Tracking_dijetEfficiency2}~(left). The local occupancy is highest in the center of the jets which leads to a higher probability of confusing hits and not finding proper tracks. \cref{fig:Tracking_dijetEfficiency2}~(right) shows the efficiency depending on the distance to the closest hit along the track. The track finding efficiency is very high ($>97\%$) for tracks that do not have another hit within \unit[130]{\micron} but drops quickly for tracks that do have a very close hit from another particle.

\begin{figure}[htpb]
 \includegraphics[width=0.49\textwidth]{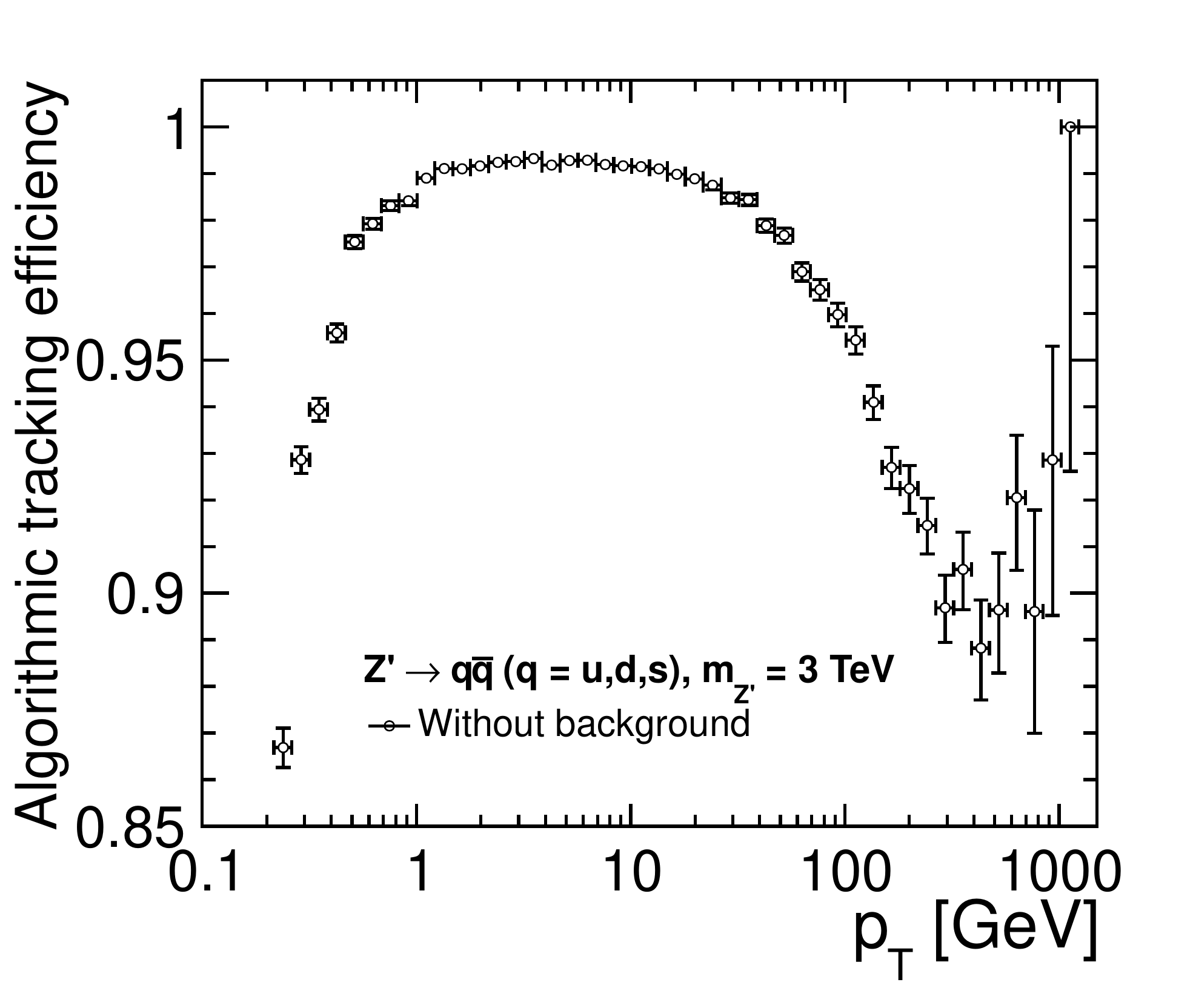}
 \hfill
 \includegraphics[width=0.49\textwidth]{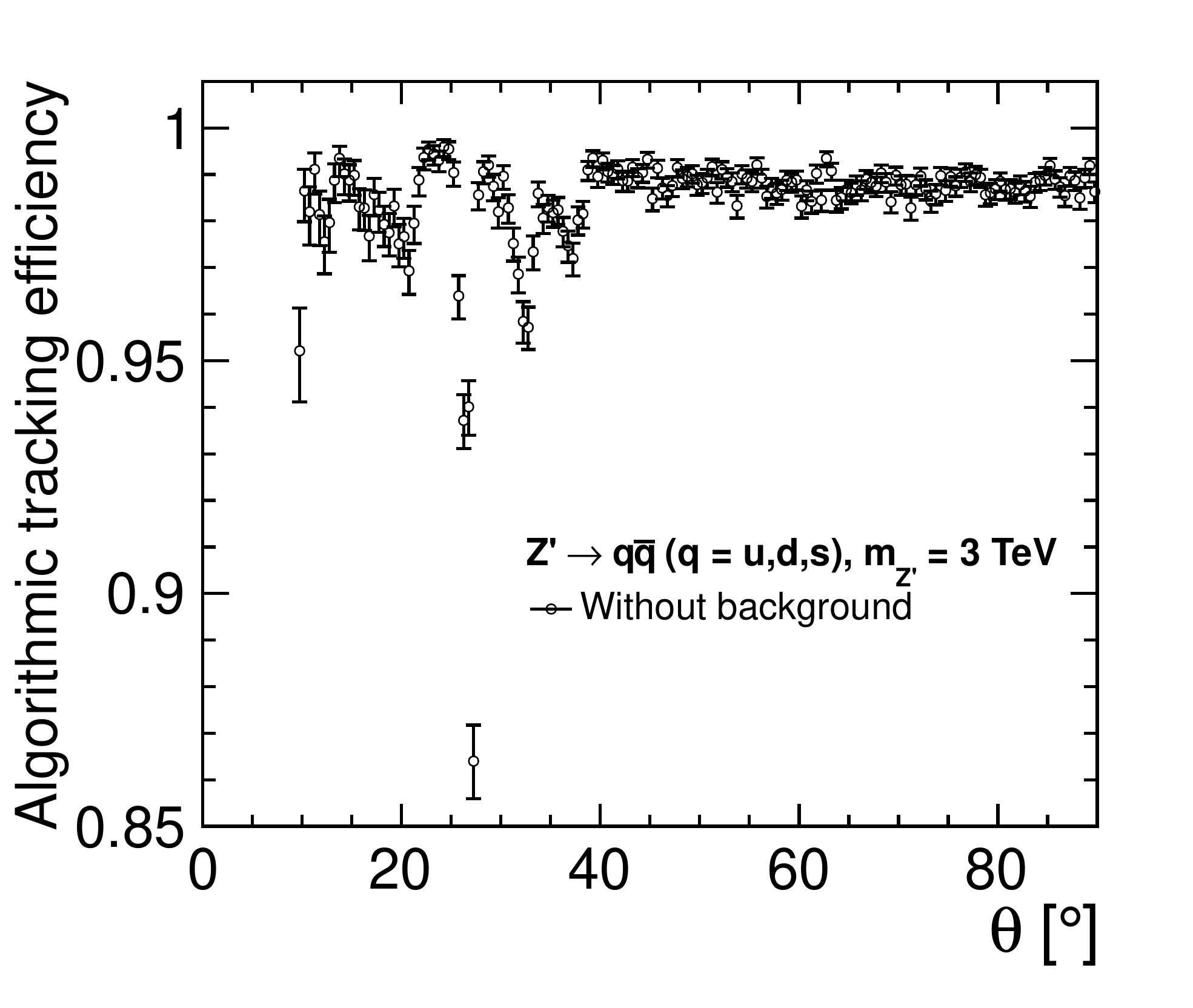}
\caption[Algorithmic tracking efficiency in di-jet events.]{Algorithmic tracking efficiency depending on the transverse momentum \pT (left) and the polar angle $\theta$ (right) in $\PZ' \rightarrow qq (q = uds)$ events.}
\label{fig:Tracking_dijetAlgorithmicEfficiency}
\end{figure}

The algorithmic tracking efficiency is shown in \cref{fig:Tracking_dijetAlgorithmicEfficiency}. In most regions the results are better by 1--2\% than the efficiencies found with the more general definition of the tracking efficiency shown in \cref{fig:Tracking_dijetEfficiency}. These 1--2\% of the tracks are not findable by the algorithm with the set of strategies that we used. The most important features of the $\theta$ and $\pT$ dependency are also visible in the algorithmic tracking efficiency, although less pronounced. As we already discussed, these effects are caused by high local occupancies which the algorithm can not resolve, although, from their parameters these tracks should be findable. In principle it should thus be possible to recover these tracks if the segmentation were sufficiently increased or the algorithm would be improved. One possible improvement could be an iterative approach to the track finding, where hits belonging to an identified track are removed from the list of available hits to allow for a second track finding step to recover tracks that were missed in the first iteration. 

\subsection{Fake Rate}

\begin{figure}[htpb]
 \includegraphics[width=0.49\textwidth]{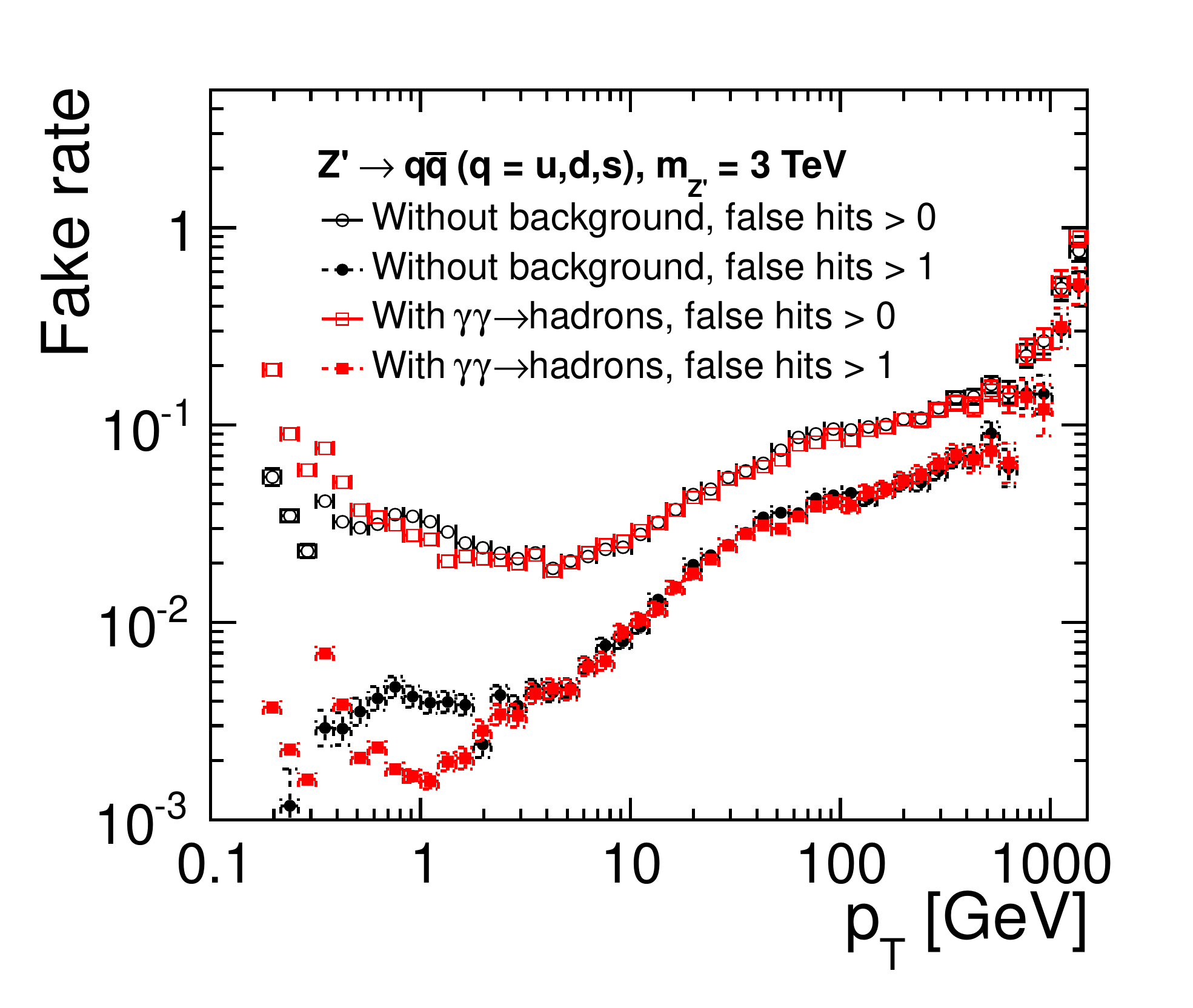}
\hfill
 \includegraphics[width=0.49\textwidth]{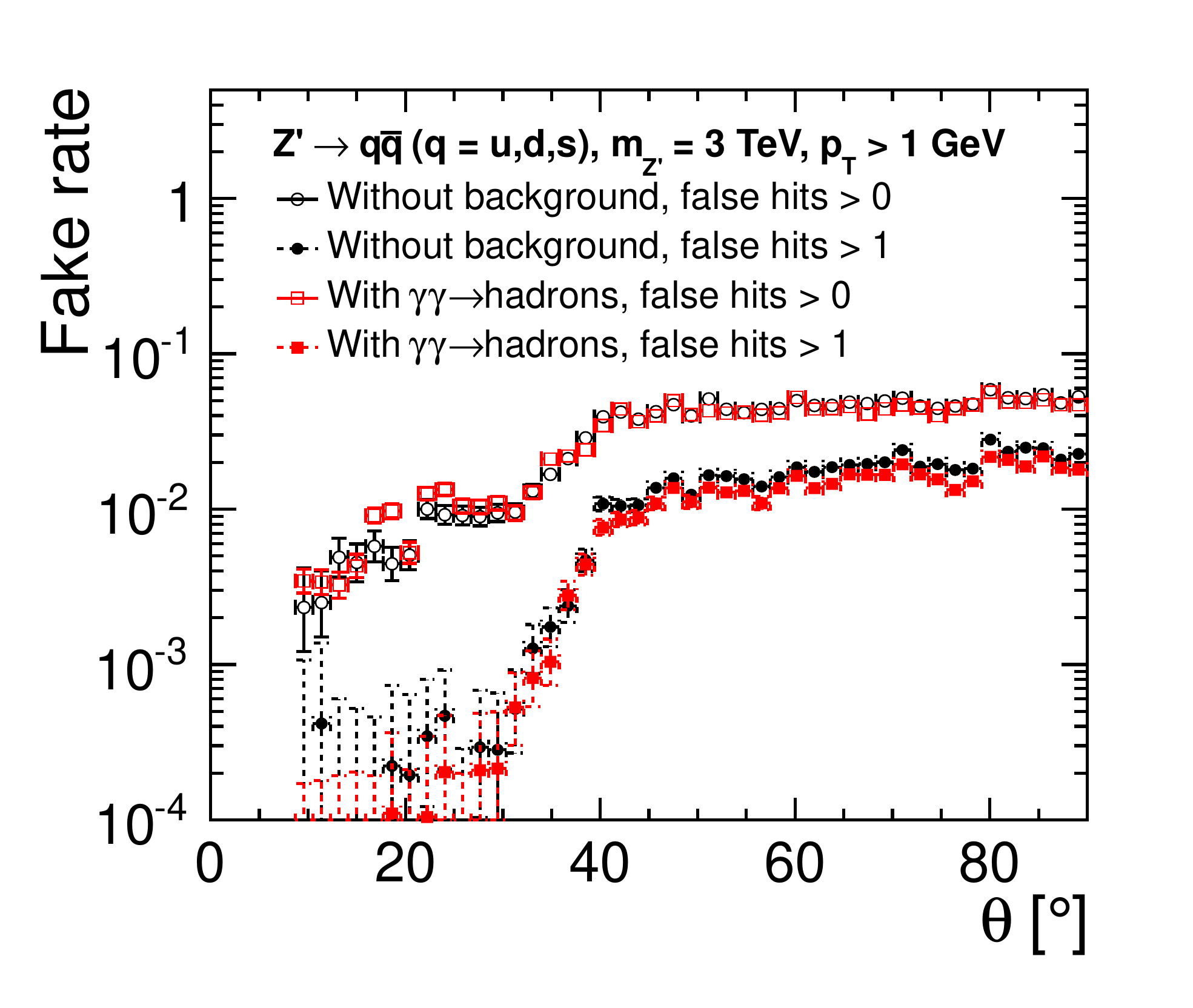}
\caption[Fake rate in di-jet events with and without \gghad background.]{Fake rate depending on the transverse momentum \pT (left) and the polar angle $\theta$ (right) in $\PZ' \rightarrow qq (q = uds)$ events with and without \gghad background.}
\label{fig:Tracking_dijetFake}
\end{figure}

The fake rate in di-jet events is shown in \cref{fig:Tracking_dijetFake}. It is highest in the barrel region with between 1\% and 2\% of fake tracks, when considering each track with two or more false hits as being a fake. This is caused by the low point resolution in $z$ due to the long strip length of the strip detectors in the main tracker barrel. It drops by more than a factor of ten in the forward region where picking up false hits is less likely due to the smaller point resolutions. High \pT tracks are especially likely to be fake tracks since they are in the center of the jets (see \cref{fig:Tracking_dijetEfficiency2}).

If one uses a stricter definition where each track that has a single false hit is also considered a fake, the fake rate increases by a factor of two to ten.

Like for the tracking efficiency, the fake rate is almost unaffected by the addition of the \gghad background. The only effect is a significantly lower fake rate for tracks around \unit[1]{GeV} since the background adds a large amount of tracks to the events with that transverse momentum that are mostly well reconstructed.


\subsection{Track Purity}

\begin{figure}[htpb]
 \includegraphics[width=0.49\textwidth]{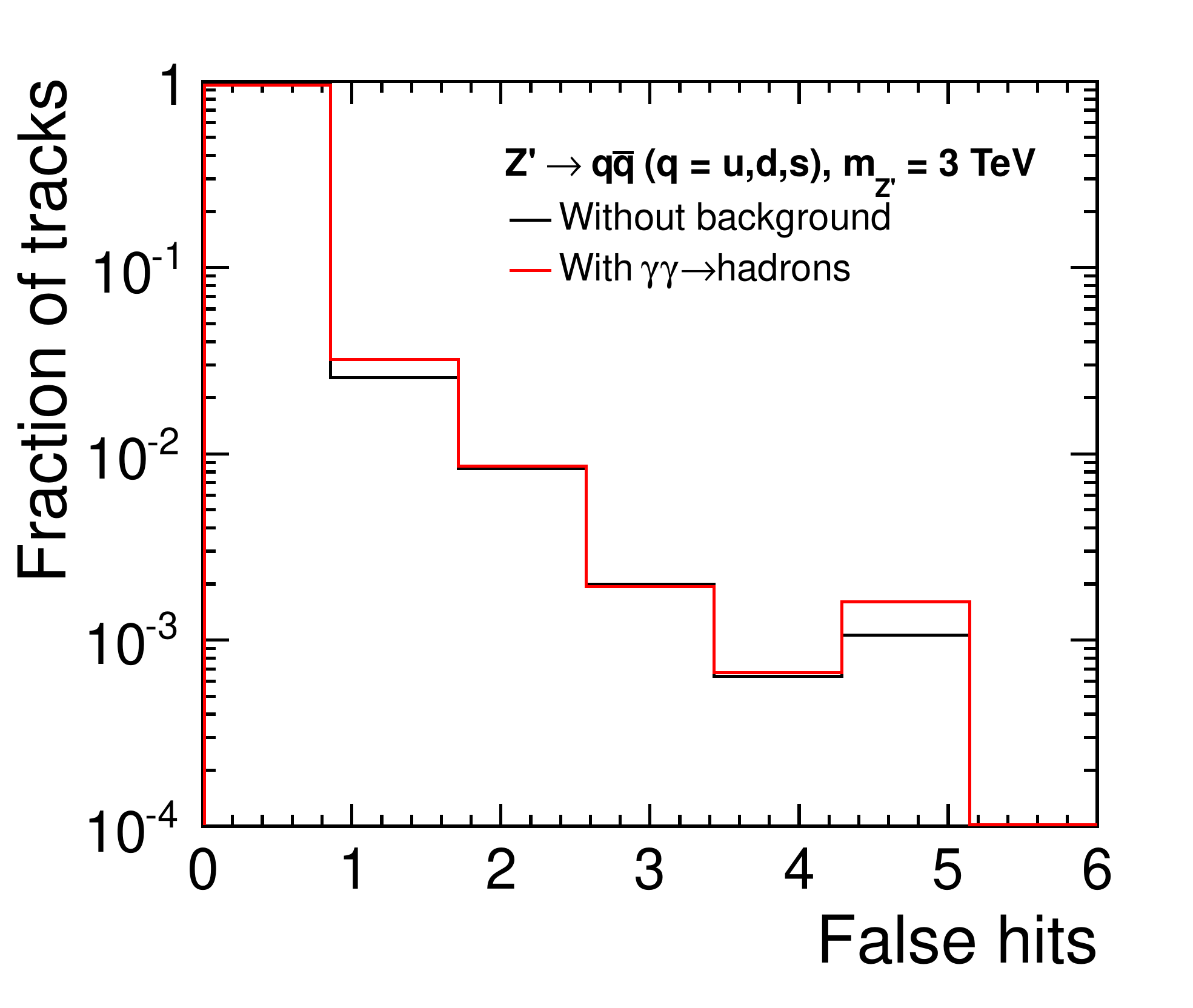}
 \hfill
 \includegraphics[width=0.49\textwidth]{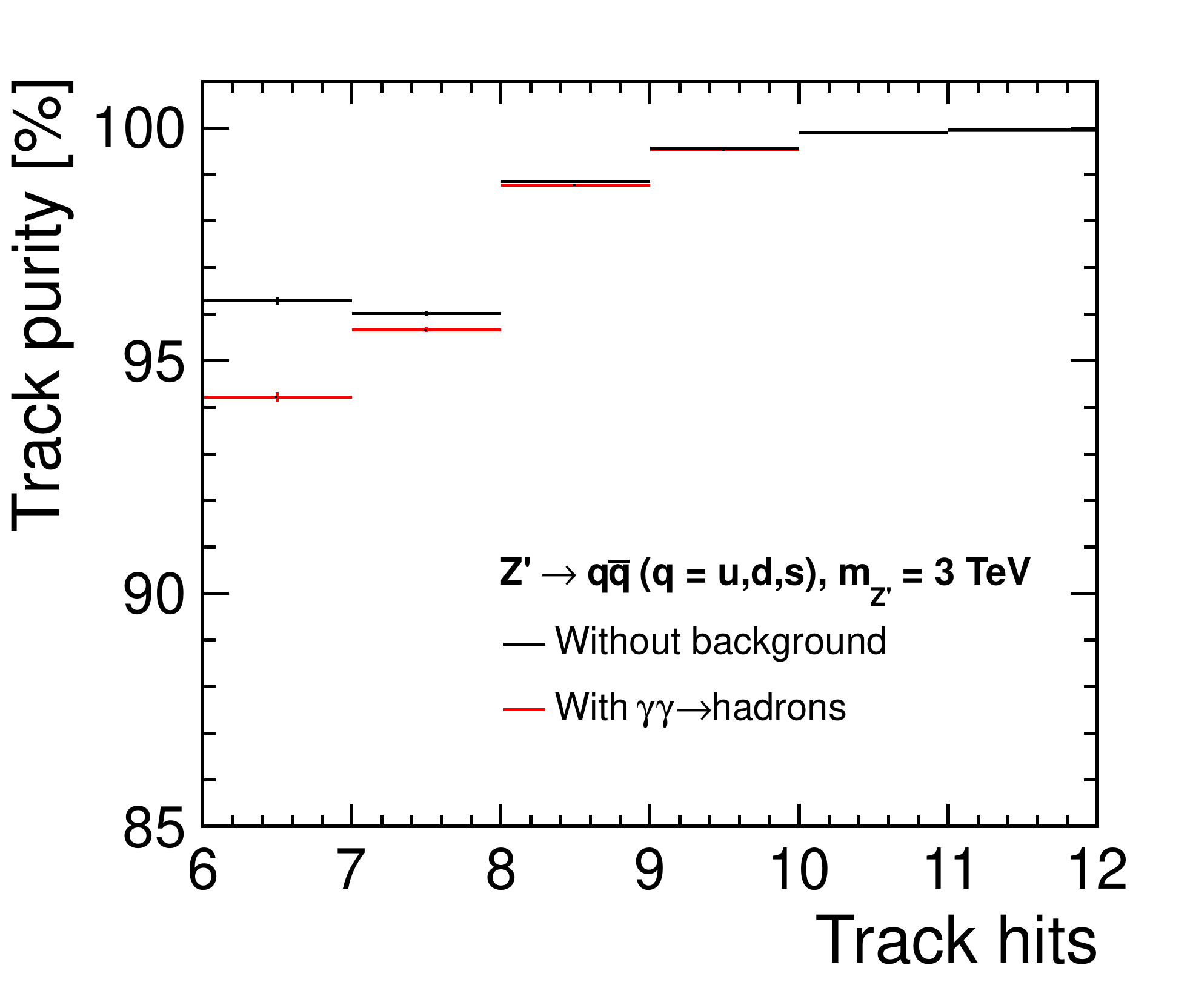}
\caption[Number of false hits on a track and track purity in di-jet events.]{Probability of reconstructing a track with a certain number of false hits~(left) and average track purity depending on the number of hits on a reconstructed track~(right) in $Z \rightarrow qq (uds)$ with and without $\gamma\gamma\rightarrow\mathrm{hadrons}$.}
\label{fig:Tracking_dijetPurity}
\end{figure}

The impact of the \gghad background on the track purity is illustrated in \cref{fig:Tracking_dijetPurity}. The probability of reconstructing a pure track drops from 96.2\% to approximately 95.5\% when adding the \gghad background. In turn, the probability to reconstruct a track with one false hit increases from 2.6\% to 3.2\%. The probability to reconstruct a track with two false hits is almost unchanged at around 0.8\%. The number of tracks with even higher numbers of false hits are negligible. In terms of the purity only the tracks with a low number of total hits are affected significantly. The average purity of tracks with 6 hits is reduced from 96.3\% to 94.2\%, while the purity for tracks with 7 hits is reduced to 95.7\% from 96\%.

\section{Summary}
\label{sec:Tracking_summary}

It has been shown that the all-silicon tracker designed for the SiD detector concept is also suitable as the main tracking device in a CLIC detector at \unit[3]{TeV}. It offers excellent momentum and impact $d0$ resolution and despite the low number of tracking layers achieves an excellent pattern recognition with high track finding efficiency, high purity and a low fake rate. The high purity is essential to the tracking performance in this layout with a minimal number of layers since already a single false hit deteriorates the track fit significantly. Some shortcomings of the current track fitting algorithm have been identified and remain to be investigated. In addition, the short lever arm in the vertex barrel detector was identified as the limiting factor in the $\theta$ resolution and thus also limits the $z_0$ resolution. If a better resolution is desired this would require at least one highly segmented layer at a large radius.

The amount of beam-related background assumed in these studies does not significantly affect the track reconstruction. Instead, the high local occupancy in multi-TeV jets is the most important source of inefficiencies in the track finding. This effect is most notable in the barrel region where the low $z$-resolution of the strip detectors can lead to confusion. The impact of increasing the longitudinal segmentation in one or more of the barrel tracker layers should thus be investigated in the future.


\chapter{\acs{HCal} Layout Optimization}
\label{cha:Calorimetry}

The design of the calorimeter system has to take into account several parameters. Most important is the desired energy resolution, which is driven by the sampling fraction, the choice of the absorber material as well as the active material and the readout technology. A good energy resolution also relies on a linear response over the full energy range as well as on minimal leakage, which means that the calorimeter depth has to be well adapted to the expected shower energies. Finally, cost and feasibility have to be taken into account as well.

The aim for excellent jet energy resolution as discussed in \cref{sec:SiD_requirements} requires that both the \ac{ECal} and the \ac{HCal} are placed inside of the solenoid to avoid large amounts of dead material in the region most important to the energy measurement---the shower maximum. The desired magnetic field in the inner detector of \unit[5]{T} for the \ac{SiD} concept implies a strong constraint on the maximum coil radius through cost and technical feasibility. This translates directly into the maximum radius of the calorimetric system. When modifying the \ac{ILD} and \ac{SiD} detector concepts for the \ac{CLIC} scenario it was decided to keep the inner and outer radii of the calorimeter systems and replace the iron absorber plates in the \ac{HCal} with a denser absorber material. This requires a re-assessment of the absorber thickness to achieve optimal performance. There is no need to modify the \ac{ECal} since the \ac{ILC} concepts already foresee dense tungsten absorber plates with a total depth of approximately \unit[25]{\radlen}. In addition, the highly granular \ac{HCal} offers sufficient performance to measure the tails of electromagnetic showers that might leak out of the \ac{ECal}.

This chapter presents a systematic simulation study to identify the optimal sampling fraction for the \ac{HCal} in a \ac{CLIC} detector, comparing different calorimeter configurations using steel and tungsten absorber plates. The different calorimeter geometries that were used in the simulations are explained in \cref{sec:Calorimetry_HCalOptimizationModel}. The energy reconstruction is performed using a \ac{NN}, which is discussed in \cref{sec:Calorimetry_HCalOptimizationNeuralNet}. \cref{sec:Calorimetry_HCalOptimizationResolutionLinearity} shows the intrinsic resolution that results from the different sampling frequencies and \cref{sec:Calorimetry_HCalOptimizationResolutionDepth} discusses the energy resolution that is achievable when a finite detector length and leakage are taken into account. The impact of additional calorimetric information from a tail catcher that is placed behind the coil is shown in \cref{sec:Calorimetry_HCalOptimizationTailCatcher}.







\section{Calorimeter Models}
\label{sec:Calorimetry_HCalOptimizationModel}

To optimize the sampling fraction of a calorimeter a simplified detector model is sufficient. The simulation models used in this study consist of simple \ac{HCal} stacks with varying absorber materials and absorber plate thicknesses. The sampling fraction is homogeneous within each stack and thus, no dedicated \ac{ECal} is present in the models. The absorber materials studied are tungsten and steel as well as combinations of both materials, where each absorber layer consists of two slices of each material, each with a thickness of half the total thickness. The models with both absorber materials were simulated in two configurations, such that each material is used as the first material in direction of the particle. The order of the material has an impact mostly on the electromagnetic shower component since the \radlen of steel and tungsten are very different. In total, 15 different configurations have been simulated. They are listed in \cref{tab:Calorimetry_HCalOptimizationParameters}.
The depth of each calorimeter stack model is chosen large enough ($>20\lambdaint$) such that longitudinal leakage is negligible. With a plate size of \unit[$5\times5$]{m$^2$} the lateral leakage can also be ignored. 

The tungsten alloy used in the simulation, simply referred to as tungsten, consists of 93\% W, 6.1\% Ni and 0.9\% Fe (by volume), which results in a radiation length of \unit[0.37]{cm} and the nuclear interaction length is \unit[10.16]{cm} For the steel we assume a typical structural steel (\emph{steel 235}) with 99.8\% Fe, 0.02\% Cu and a density of \unit[7.85]{g/cm$^3$}. The radiation length is \unit[1.73]{cm} and the nuclear interaction length is \unit[16.87]{cm}. The active material thickness and the readout is the same in all setups. Each active layer consists of \unit[5]{mm} of plastic scintillator (Polysterene), followed by \unit[2.5]{mm} of G10, a fibre-reinforced plastic that serves as a placeholder for the readout electronics and cabling present in each layer.
The layers are segmented into \unit[$1\times1$]{cm$^2$} cells with analog readout. The scope of this study is not the optimization of the lateral segmentation and only showers originating from single particles are simulated. In addition, only variables that are not strongly dependent on the lateral segmentation are used to describe the shower shape (see \cref{sec:Calorimetry_HCalOptimizationNeuralNet}).

The minimum energy of a hit to be considered in the reconstruction is \unit[250]{keV}. This represents a noise cut that would be applied in any real experiment. The simulation does not include noise hits. 

\begin{table}
  \caption[List the \ac{HCal} geometries used for calorimeter optimization.]{List of all simulated \ac{HCal} geometries. Given is the absorber material, the thickness of the absorber layer, the average nuclear interaction length $\lambdaint$ and the average radiation length $\radlen$ for each full layer including the passive and active material.}
  \label{tab:Calorimetry_HCalOptimizationParameters}
  \centering
    \begin{tabular}{cccc}
\toprule
absorber material & absorber thickness [mm] & $\lambdaint$ per layer & $\radlen$ per layer \\
\midrule
tungsten & 5 & 0.06 & 1.37 \\
tungsten & 10 & 0.11 & 2.72 \\
tungsten & 15 & 0.16 & 4.07 \\
tungsten & 20 & 0.21 & 5.42 \\
steel & 10 & 0.07 & 0.60 \\
steel & 15 & 0.10 & 0.89 \\
steel & 20 & 0.13 & 1.18 \\
steel & 25 & 0.16 & 1.47 \\
steel & 30 & 0.19 & 1.76 \\
tungsten-steel & 10 & 0.09 & 1.66 \\
tungsten-steel & 15 & 0.13 & 2.48 \\
tungsten-steel & 20 & 0.17 & 3.30 \\
steel-tungsten & 10 & 0.09 & 1.66 \\
steel-tungsten & 15 & 0.13 & 2.48 \\
steel-tungsten & 20 & 0.17 & 3.30 \\
\bottomrule
    \end{tabular}
\end{table}

The depth of the calorimeter is varied by ignoring all simulated hits in those layers that are not considered part of the detector. This also allows to study the impact of large inactive areas within the calorimeter and is used to study the impact of the coil position, as discussed in \cref{sec:Calorimetry_HCalOptimizationTailCatcher}.

\section{Event Samples}
\label{sec:Calorimetry_HCalOptimizationSample}

An event sample of single pions is used to study the hadronic energy measurement. A data set of 100000 \PGpp events was simulated in each of the detector configurations described above. Since a \ac{NN} is used to reconstruct the pion energy, a sample with a continuous energy distribution is used for the training in order to avoid overtraining. The particle energy is chosen to be in the range between \unit[1]{GeV} and \unit[300]{GeV} and follows an exponential distribution with
\begin{equation}
 N \propto 0.5^{E / \unit[300]{GeV}}
\end{equation}
to provide slightly more statistics at lower particle energies. This energy range represents the typical energy of individual particles in jets at the TeV scale, e.g. as shown in Figure~\ref{fig:Tracking_jetDistributions2}~(left). The pions are generated using the \geant particle gun and are shot perpendicular onto the center of the first calorimeter layer. A small angular spread of $\pm 1\degrees$ is added to the particle direction to avoid a bias due to the impact position of the particle. 

The neutron content of hadronic showers increases with the atomic mass of the materials traversed. The response of plastic scintillators is especially sensitive to signals from elastic neutron scattering due to the high fraction of hydrogen. The higher cross section for neutron capture in tungsten compared to steel leads to an enhanced slow shower component and a significantly different time structure of the signal. In combination, this means that the simulation of the neutron shower component requires special attention when simulating tungsten-scintillator sampling calorimeters. \geant offers special high precision physics list for low energetic neutrons that are denoted by an appended \texttt{HP} to the name of the physics list. These physics lists allow for tracking of neutrons with kinetic energies below \unit[20]{MeV} down to thermal energies~\cite{geant4lowEnergyNeutrons}. Unlike the other simulation studies presented in this thesis, which use the \qgspbert physics list (see \cref{sec:Software_Simulation}), the simulations presented in this chapter are performed using the \qgspberthp physics list. The differences of the two physics lists for the tungsten absorber case are discussed in~\cite{Speckmayer2010}. First results from a test beam campaign with a tungsten \ac{HCal} prototype~\cite{Simon:2011ua} indicate that simulations using the \qgspberthp physics list are in good agreement with the observed time development of hadronic showers~\cite{Simon:2011rm}.

\section{Energy Reconstruction with a \acl{NN}.}
\label{sec:Calorimetry_HCalOptimizationNeuralNet}

The standard approach of energy reconstruction in a sampling calorimeter is a linear model where the sampling fraction is used to calculate the particle energy from the sum of the visible energy. This often requires to apply ad-hoc corrections for leakage as well as weightings depending on the shower type to obtain a good energy resolution. A multivariate classifier allows to directly use the maximum information of the shower topology in the energy reconstruction, taking into account possible correlations of the variables. Simulation studies performed for various \ac{HEP} experiments, e.g. \acs{H1}~\cite{Hoeppner1997}, \acs{CMS}~\cite{Damgov2002} and \acs{ATLAS}~\cite{daSilva:2006pa}, have found that an energy reconstruction using \ac{NN} performs usually much better than the respective standard energy reconstruction of the experiments. For the \ac{HCal} optimization study presented here, the particle energy is therefore reconstructed from the energy deposited in the active scintillator material using the \acl{NN} classifier implemented in the \tmva~\cite{Hoecker2007}. Using shower shape information is especially beneficial when assessing the impact of the coil position and a possible tail catcher, which requires an estimation of the energy deposited in the dead region from adjacent layers.

The \acl{NN} is created with two hidden layers of $N+30$ and $N+20$ nodes, where $N$ is the number of variables used. The network is trained with the true particle energy as the target value on half of the simulated events. For each of the detector configurations an individual network is trained using the hits from a calorimeter depth of approximately \unit[3]{\lambdaint}, \unit[4]{\lambdaint}, \unit[5]{\lambdaint}, \unit[6]{\lambdaint}, \unit[7]{\lambdaint}, \unit[8]{\lambdaint}, \unit[9]{\lambdaint} and the full calorimeter stack. Specific \acp{NN} are also trained for the various tail catcher configurations. This way the leakage is implicitly taken into account in the training. The training events are not used in the analysis discussed in Sections~\ref{sec:Calorimetry_HCalOptimizationResolutionLinearity}, \ref{sec:Calorimetry_HCalOptimizationResolutionDepth} and \ref{sec:Calorimetry_HCalOptimizationTailCatcher}.

\subsection{\acl{NN} Variables}
Since only single particles are simulated, no clustering algorithm is necessary and all hits with an energy above the threshold of \unit[250]{keV} define the cluster. The following variables are used in the \ac{NN} to reconstruct the particle energy.
\begin{itemize}
 \item The total energy of the cluster
\begin{equation} \label{eq:clusterEnergy}
 E_\mr{cluster} = \sum_{i}^{N_\mr{cells}} E_{\mr{cell},i},
\end{equation}
 where $E_{\mr{cell},i}$ is the deposited energy in the cell with the index $i$ and with $E_{\mr{cell},i} > E_\mr{threshold}$.
 \item The energy density in the cluster
\begin{equation} \label{eq:clusterDensity}
 \rho_{\mr{E},{\mr{cluster}}} = \frac{E_\mr{shower}}{N_\mr{cells}}.
\end{equation}
 \item The energy weighted center of the cluster in $z$ direction
\begin{equation} \label{eq:clusterBaryCenterZ}
 \overline{z}_{\mr{cluster}} = \frac{1}{E_{\mr{cluster}}}\sum_{i}^{N_{\mr{cells}}} z_{\mr{cell},i} E_{\mr{cell},i},
\end{equation}
where $z_{\mr{cell},i}$ is the $z$ position of the cell with the index $i$.
 \item The energy weighted distance of the cluster from the $z$ axis
\begin{equation} \label{eq:clusterBaryCenterR}
 \overline{r}_{\mr{cluster}} = \frac{1}{E_{\mr{cluster}}}\sum_{i}^{N_{\mr{cells}}} r_{\mr{cell},i} E_{\mr{cell},i},
\end{equation}
where $r_{\mr{cell},i}$ is the distance from the $z$ axis of the cell with index $i$.
 \item The length of the cluster in $z$, estimated by the \ac{RMS} of the energy weighted $z$ position
\begin{equation} \label{eq:clusterBaryCenterZRMS}
 \mr{RMS}_{z,{\mr{cluster}}} = \frac{1}{E_{\mr{cluster}}} \sqrt{\sum_{i}^{N_{\mr{cells}}}\left(z_{\mr{cell},i}-\overline{z}\right)^2 E_{\mr{cell},i}^2}.
\end{equation}
 \item The width of the cluster, estimated by the \ac{RMS} of the energy weighted distance from the $z$ axis
\begin{equation} \label{eq:clusterBaryCenterRRMS}
 \mr{RMS}_{r,{\mr{cluster}}} = \frac{1}{E_{\mr{cluster}}} \sqrt{\sum_{i}^{N_{\mr{cells}}}\left(r_{\mr{cell},i}-\overline{r}\right)^2 E_{\mr{cell},i}^2}.
\end{equation}
 \item Three additional energy sums are calculated similarly to \cref{eq:clusterEnergy} for different regions of the calorimeter. The energy deposited within the first \lambdaint of the calorimeter, the last \lambdaint of the calorimeter and the energy deposited in between these two regions.
 \item In case of the tail catcher studies, the sum of the energy visible in the tail catcher is used as an additional variable. The tail catcher is defined as a region of the calorimeter that follows an inactive volume with a length of approximately \unit[2]{\lambdaint} which represents the coil.
\end{itemize}

\section{Energy Resolution and Linearity}
\label{sec:Calorimetry_HCalOptimizationResolutionLinearity}
The energy resolution is determined as the $\mr{RMS}_{90}(E_\mr{reco}/E_\mr{true})$ in 13 intervals of the true particle energy in the range between \unit[40]{GeV} and \unit[270]{GeV}. The intervals are chosen such that the ratio $E_\mr{reco}/E_\mr{true}$ is almost constant over the interval. The $\mr{RMS}_{90}$, which is explained in \cref{App:RMS90}, is preferred over a Gaussian fit to describe the width of the reconstructed energy. It takes into account possible non-Gaussian tails while rejecting strong outliers at the same time. The energy dependence of the resolution can be described by the parametrization given in \cref{eq:energyResolution}. The noise term can be ignored since noise is not part of the simulation. The parametrization used for the fit is thus
\begin{equation}
 \frac{\sigma(E)}{E} = \frac{s}{\sqrt{E}} \oplus c.
\label{eq:energyResolutionNoNoise}
\end{equation}

The energy resolution of a calorimeter is limited by its sampling fraction. The best resolution for a given active material is achieved in the limit of a homogeneous calorimeter without any absorber plates. 

\begin{figure}[htpb]
 \centering
 \includegraphics[width=0.8\textwidth]{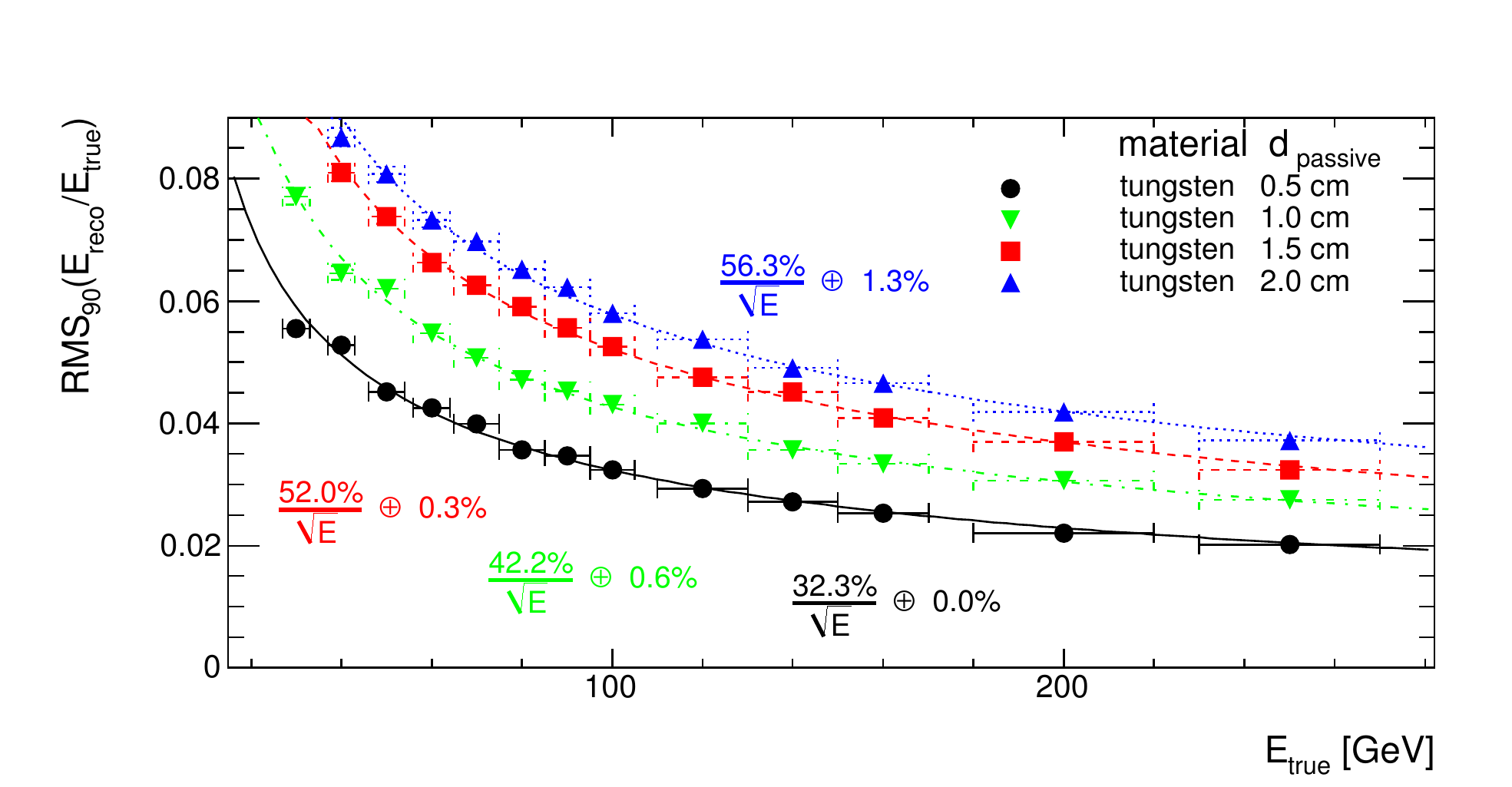}
 \caption[Energy resolution for \PGpp depending on the true particle energy for different tungsten absorber thicknesses.]{Energy resolution for \PGpp depending on the true particle energy for different tungsten absorber thicknesses, $d_\mr{passive}$, in the full calorimeter stack.}
\label{fig:calorimetry_resolutionTungstenInfinite}
\end{figure}

\begin{figure}[htpb]
 \centering
 \includegraphics[width=0.8\textwidth]{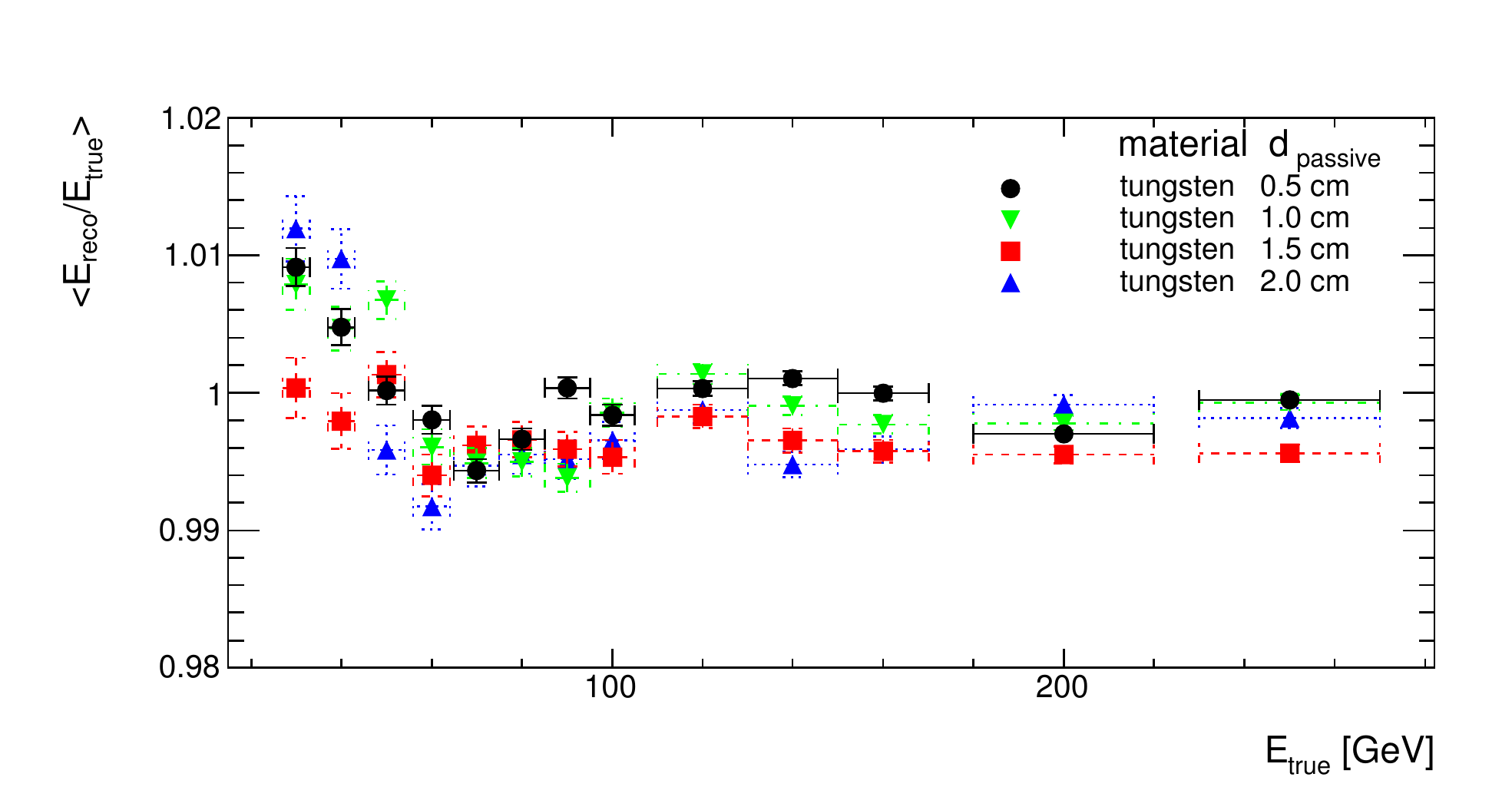}
 \caption[Linearity for \PGpp depending on the true particle energy for different tungsten absorber thicknesses.]{Linearity for \PGpp depending on the true particle energy for different tungsten absorber thicknesses, $d_\mr{passive}$, in the full calorimeter stack.}
\label{fig:calorimetry_linearityTungstenInfinite}
\end{figure}

Figure~\ref{fig:calorimetry_resolutionTungstenInfinite} shows the energy resolution for different tungsten absorber thicknesses in the full calorimeter stack where leakage is negligible. The best resolution of $32.3\%/\sqrt{E/\mr{GeV}}\oplus0.0\%$ is achieved for the finest sampling with an absorber thickness of \unit[5]{mm}. For coarser samplings the resolution function worsens to up to $56.6\%/\sqrt{E/\mr{GeV}}\oplus1.3\%$ for an absorber thickness of \unit[20]{mm}. The constant term is always very small since leakage is not an issue in these detector setups and non-uniformities of the cells are not simulated. The reconstructed energy stays within $\pm 1.5\%$ of the true energy over the whole energy range (see \cref{fig:calorimetry_linearityTungstenInfinite}), which confirms that the \ac{NN} provides a good calibration.

\begin{figure}[htpb]
 \centering
 \includegraphics[width=0.8\textwidth]{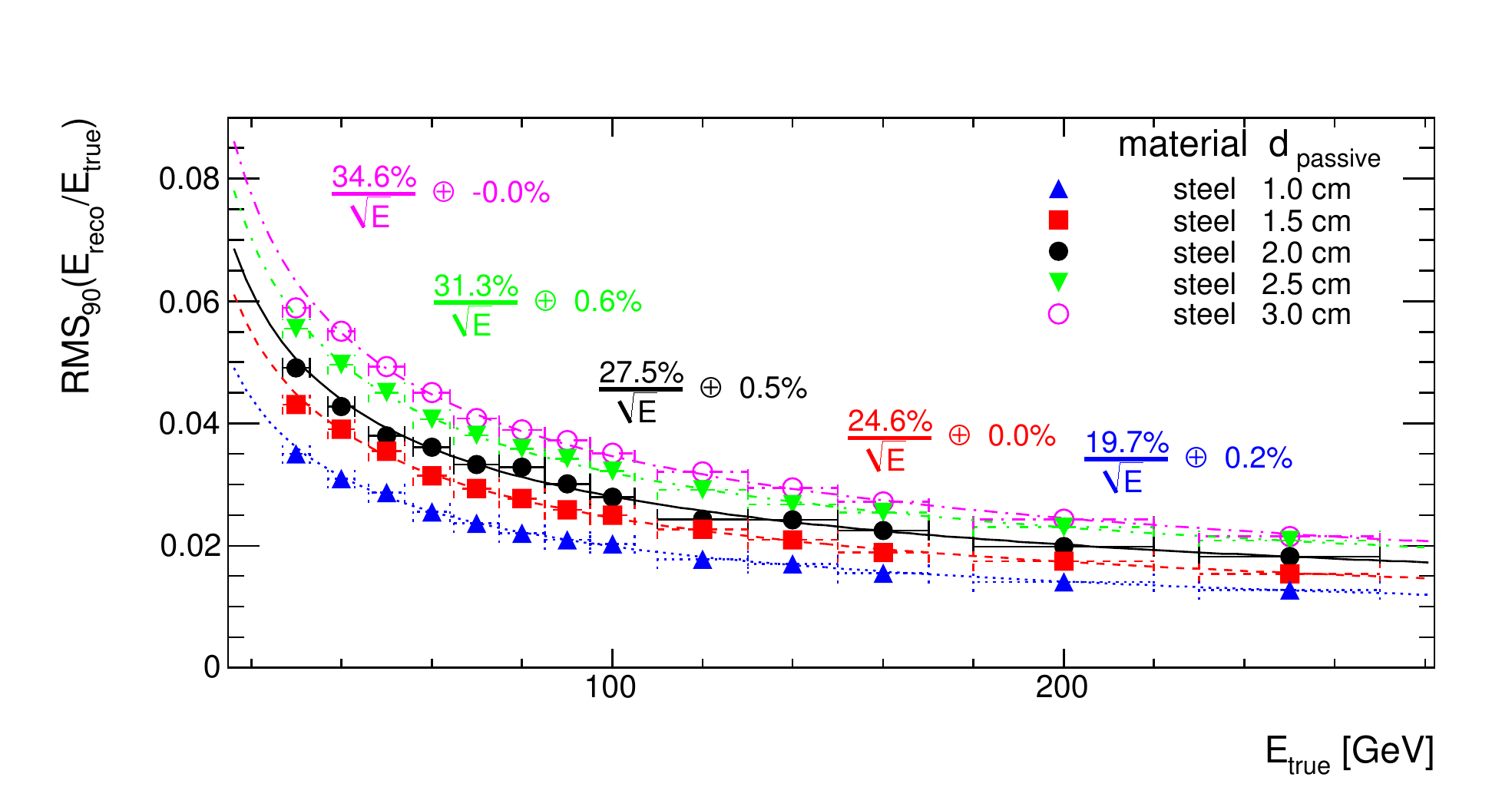}
 \caption[Energy resolution for \PGpp depending on the true particle energy for different steel absorber thicknesses.]{Energy resolution for \PGpp depending on the true particle energy for different steel absorber thicknesses, $d_\mr{passive}$, in the full calorimeter stack.}
\label{fig:calorimetry_resolutionSteelInfinite}
\end{figure}

The energy resolution of the steel calorimeter stacks is shown in \cref{fig:calorimetry_resolutionSteelInfinite}. The resolution is considerably better than in the tungsten case for similar interaction lengths per layer. For example, when comparing the tungsten \unit[15]{mm} configuration and the steel \unit[25]{mm} configuration with both approximately \unit[0.16]{\lambdaint} per layer, the resolution in the steel \unit[25]{mm} case is $31.3\%/\sqrt{E}\oplus0.6\%$ and $52.0\%/\sqrt{E}\oplus0.3\%$ in the tungsten \unit[15]{mm} configuration. This can be explained by the large differences in the radiation length of the two absorber materials: A steel calorimeter offers a much finer sampling of the electromagnetic shower content.

\begin{figure}[htpb]
 \centering
 \includegraphics[width=0.8\textwidth]{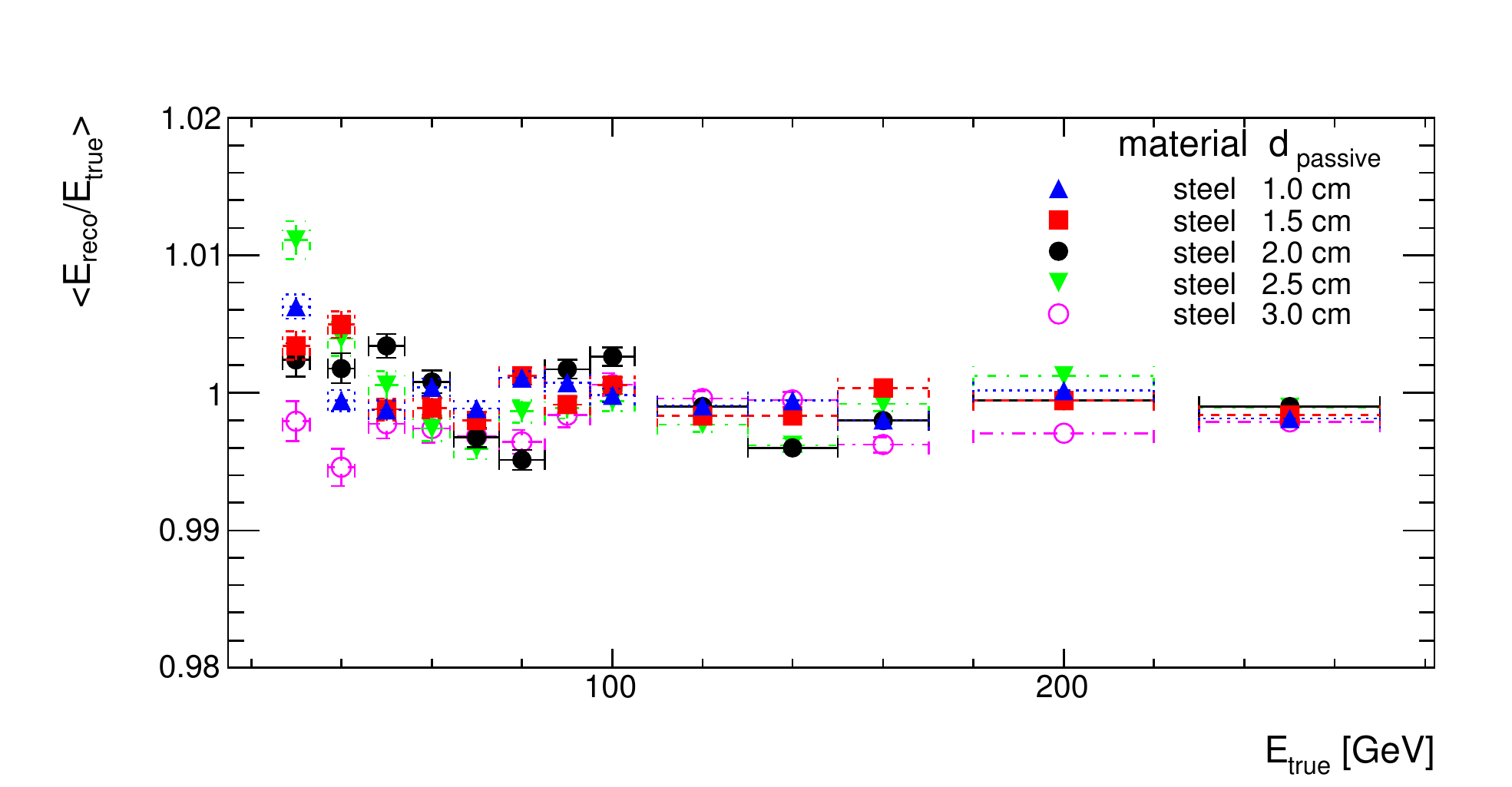}
\caption[Linearity for \PGpp depending on the true particle energy for different steel absorber thicknesses.]{Linearity for \PGpp depending on the true particle energy for different steel absorber thicknesses, $d_\mr{passive}$, in the full calorimeter stack.}
\label{fig:calorimetry_linearitySteelInfinite}
\end{figure}

The deviations from a linear energy response in the steel configurations, which is shown in \cref{fig:calorimetry_linearitySteelInfinite}, is similar to the tungsten case and never exceeds $\pm 1.5\%$.

The configurations with the mixed absorber plates, i.e. tungsten-steel and steel-tungsten, performed in between the pure tungsten and steel configurations, as expected from the relative density. No significant difference was observed when changing the order of the absorber materials.

\section{Energy Resolution Depending on HCal Depth}
\label{sec:Calorimetry_HCalOptimizationResolutionDepth}

The intrinsic resolution due to the sampling fraction discussed in \cref{sec:Calorimetry_HCalOptimizationResolutionLinearity} is only reached for an infinite calorimeter. In case the calorimeter depth is limited, leakage reduces the achievable resolution. For the \ac{SiD} concept the depth of the calorimeter is limited by the radius of the coil.

\begin{figure}[htpb]
 \centering
 \begin{tikzpicture}
  \draw[scale=1.0] (0.0,0.0) node[anchor=south west]{\includegraphics[width=0.8\textwidth]{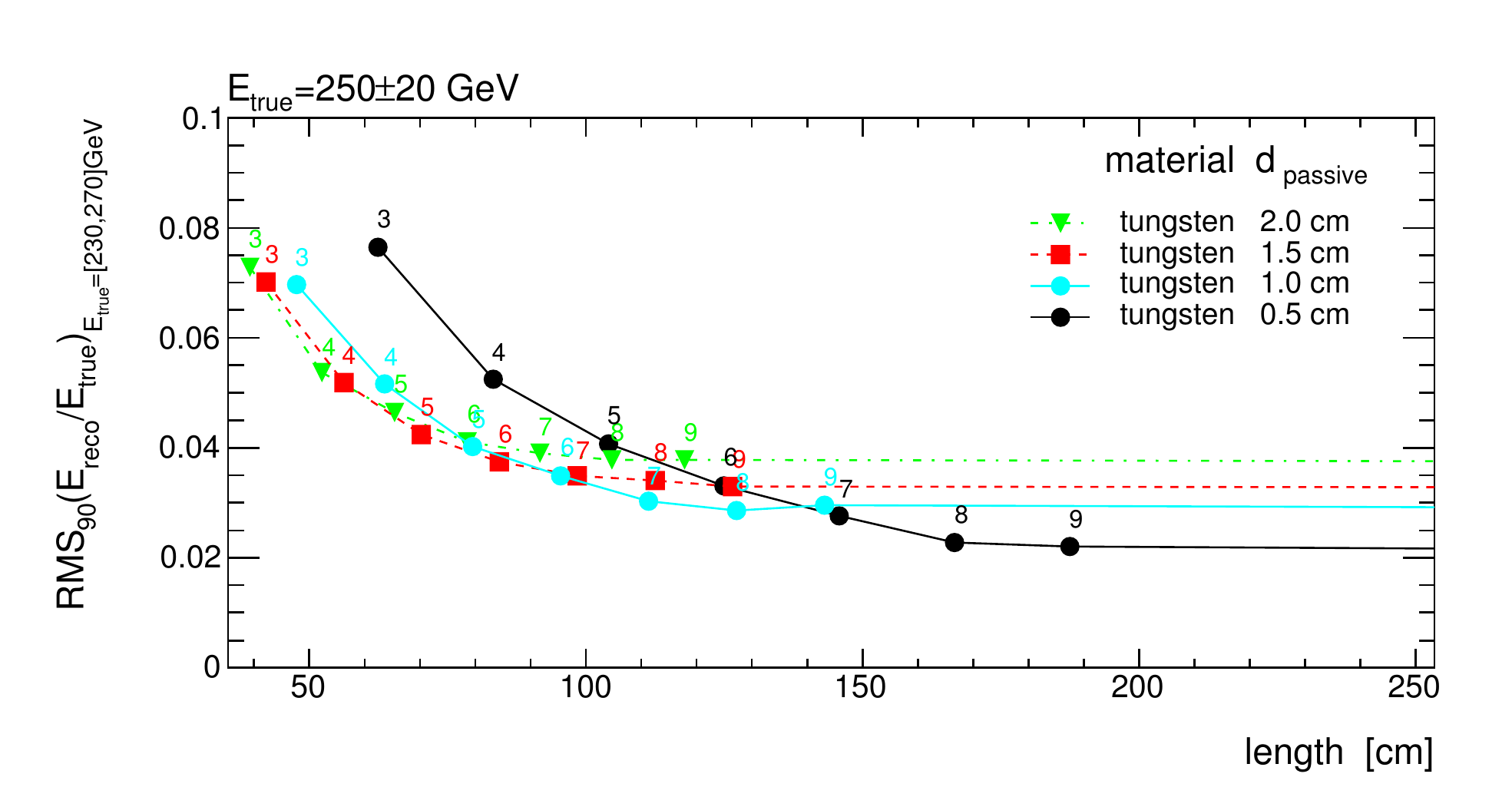}};
  \draw[red,thick,dashed] (6.0,1.5) -- (6.0,4.5);
 \end{tikzpicture}
\caption[Energy resolution for \PGpp with an energy of $250 \pm 20\ \text{GeV}$ depending on the calorimeter depth for tungsten calorimeters.]{Energy resolution for \PGpp with an energy of \unit[$250\pm20$]{GeV} depending on the calorimeter depth for tungsten calorimeters with different absorber thicknesses $d_\mr{passive}$. The numbers next to the data points denote the calorimeter thickness in interaction lengths. The dashed vertical line indicates the approximate size foreseen for the \ac{HCal}.}
\label{fig:calorimetry_resolutionTungstenDepth}
\end{figure}

\begin{figure}[htpb]
 \centering
 \begin{tikzpicture}
  \draw[scale=1.0] (0.0,0.0) node[anchor=south west]{\includegraphics[width=0.8\textwidth]{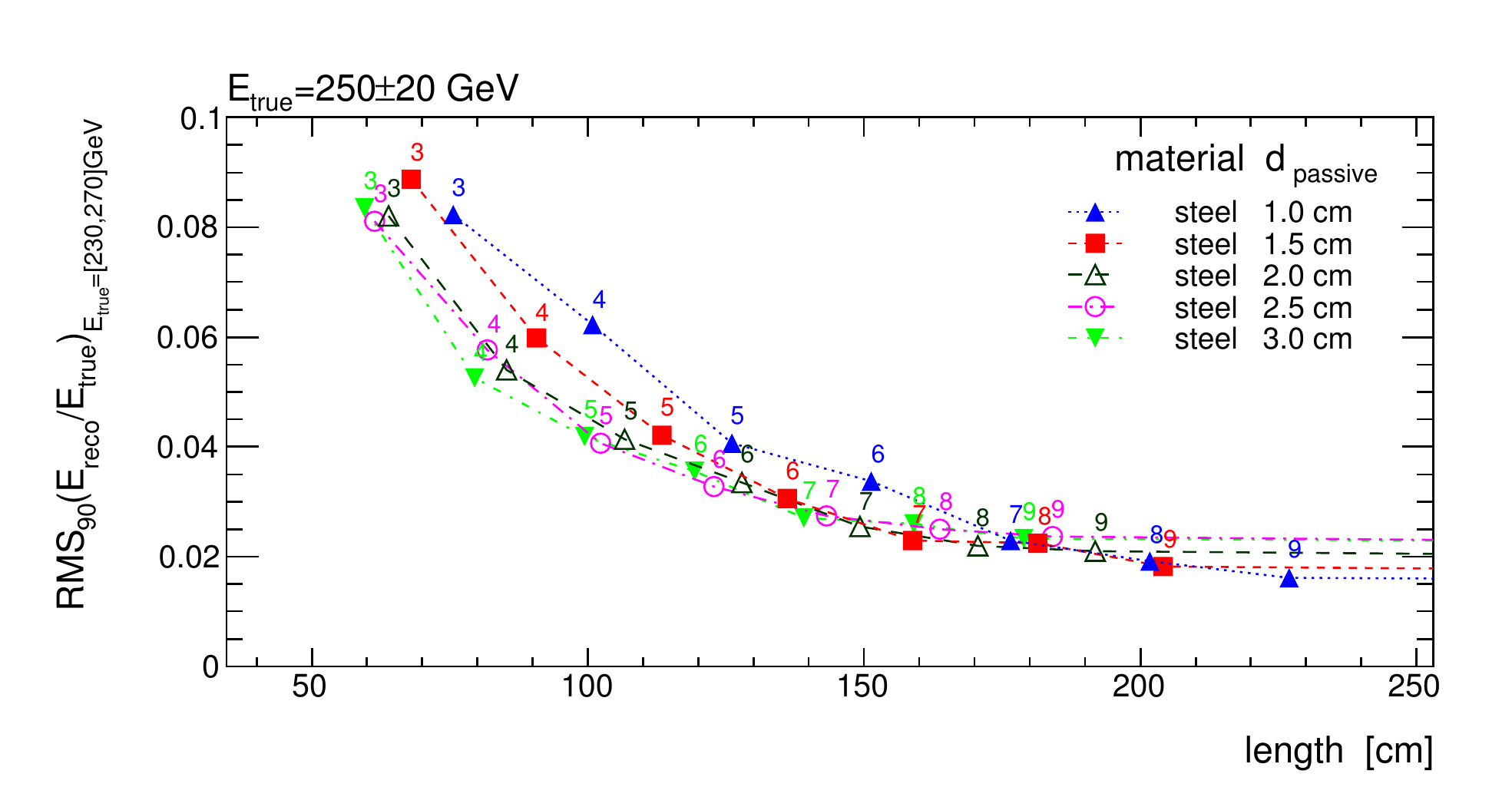}};
  \draw[red,thick,dashed] (6.0,1.5) -- (6.0,4.5);
 \end{tikzpicture}
\caption[Energy resolution for \PGpp with an energy of $250 \pm 20\ \text{GeV}$ depending on the calorimeter depth for steel calorimeters.]{Energy resolution for \PGpp with an energy of \unit[$250\pm20$]{GeV} depending on the calorimeter depth for steel calorimeters with different absorber thicknesses $d_\mr{passive}$. The numbers next to the data points denote the calorimeter thickness in interaction lengths. The dashed vertical line indicates the approximate size foreseen for the \ac{HCal}.}
\label{fig:calorimetry_resolutionSteelDepth}
\end{figure}

The energy resolution for high energy pions of \unit[$250\pm20$]{GeV} as a function of the total depth of the calorimeter is shown in \cref{fig:calorimetry_resolutionTungstenDepth} for tungsten and in \cref{fig:calorimetry_resolutionSteelDepth} for steel. The energy resolution is worst for very short calorimeters where the resolution is clearly dominated by leakage. For longer calorimeters the resolution improves until it reaches a plateau which is determined by the intrinsic resolution given by the sampling fraction of the configuration.

These figures allow to choose the optimal calorimeter configuration for a given calorimeter thickness. In case of the \ac{CLIC} detectors, the available space between the \ac{ECal} and the coil is approximately \unit[120]{cm}. For this thickness (and a gap size of \unit[7.5]{mm}) the optimal configuration is the tungsten calorimeter with an absorber thickness of approximately \unit[10]{mm}, which corresponds to an \ac{HCal} depth of approximately \unit[7.5]{\lambdaint}. In addition to offering the best resolution, the tungsten \unit[10]{mm} configuration is also in the plateau region of the resolution at a thickness of \unit[120]{cm}, which means that leakage is negligible. The energy resolution in the steel configurations, on the other hand, are dominated by leakage for this thickness. In case of a particle flow energy reconstruction, full shower containment is preferred over better intrinsic resolution, since most of the information that is used comes from the shower topology.

\section{Impact of a Tail Catcher on the Energy Resolution}
\label{sec:Calorimetry_HCalOptimizationTailCatcher}

The added value of a tail catcher is studied by using a slightly modified setup. The first \unit[8]{\lambdaint} of the calorimeter stack are considered as the main calorimeter, i.e. \ac{ECal} and \ac{HCal}, followed by a dead region of \unit[2]{\lambdaint} representing the coil, and a tail catcher of varying length. The tail catcher here uses the same sampling as the \ac{HCal} which overestimates the performance, since a typical tail catcher only consists of the sparsely instrumented return yoke outside of the coil.

\begin{figure}[htpb]
 \centering
 \includegraphics[width=0.8\textwidth]{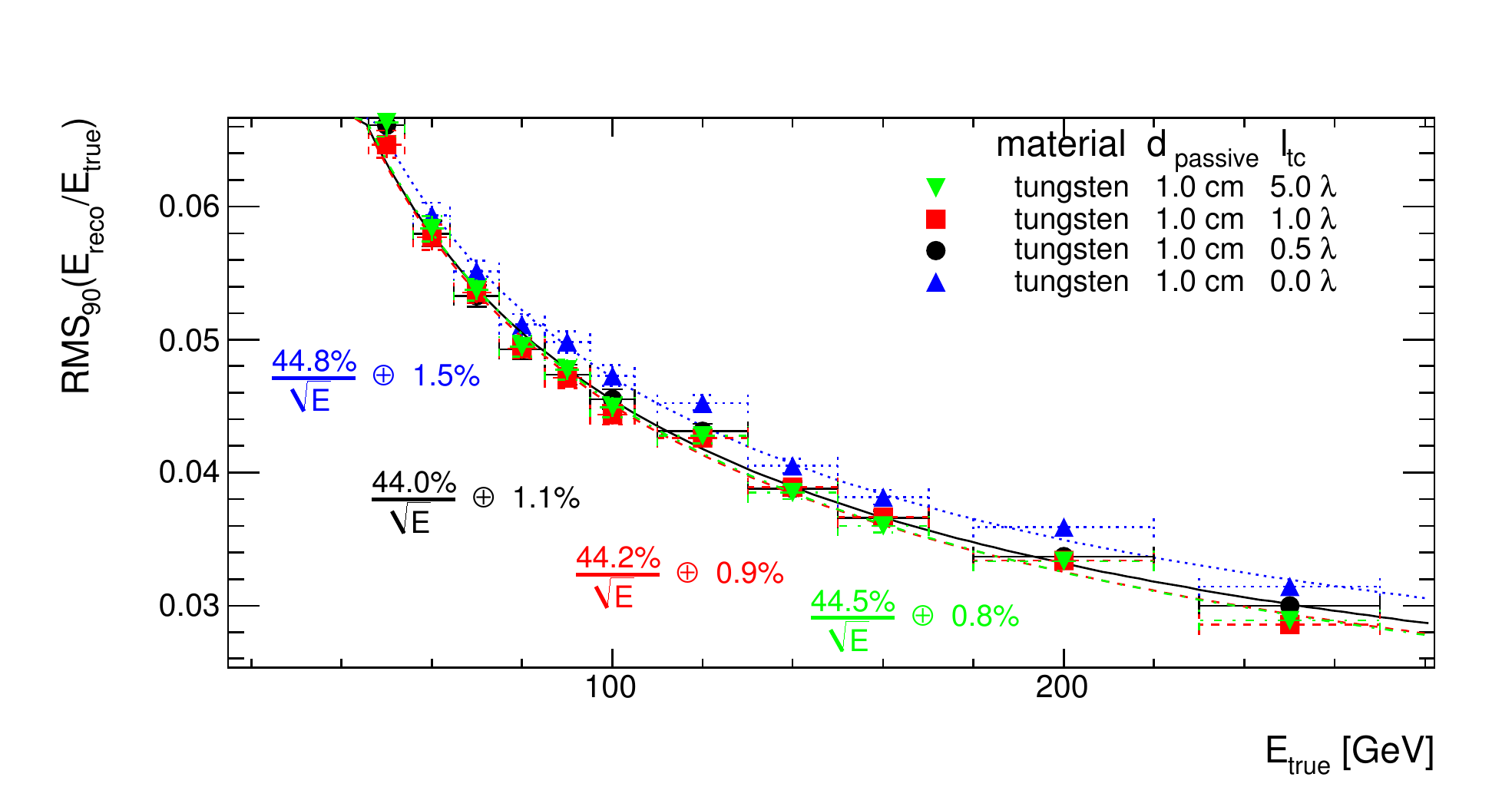}
\caption[Energy resolution for \PGpp depending on the true particle energy for a calorimeter with tail catcher.]{Energy resolution for \PGpp depending on the true particle energy for a calorimeter of \unit[8]{\lambdaint} using \unit[10]{mm} tungsten absorber plates followed by a coil of \unit[2]{\lambdaint} and a tail catcher of varying thickness, $l_\mr{tc}$. In case of $l_\mr{tc} = \unit[0]{\lambdaint}$ no tail catcher information is used.}
\label{fig:calorimetry_resolutionTungstenTail}
\end{figure}

The resolution for a calorimeter of \unit[8]{\lambdaint} with \unit[10]{mm} tungsten absorber plates and varying tail catcher configurations is shown in Figure~\ref{fig:calorimetry_resolutionTungstenTail}. The resolution improves by less than 1\% when information from a tail catcher is added. As expected, mainly the constant term is improved, since some of the leakage is recovered. The size of the tail catcher is almost irrelevant, as long as some information from the tail catcher is used. The large size of the calorimeter of \unit[8]{\lambdaint} provides sufficient depth to contain the majority of the shower and the leakage correction from the tail catcher only leads to a small improvement. In addition, the variables used for the energy reconstruction implicitely allow for longitudinal leakage correction without using tail catcher information by using the shower position, longitudinal extent as well as the amount of energy in the last \lambdaint of the calorimeter.

\section{Summary}
\label{sec:Calorimetry_HCalOptimizationSummary}

The aim of this study was to compare the relative calorimetric performance of steel and tungsten sampling calorimeters. For this we developed an energy reconstruction using a \ac{NN} with several variables to describe the shower shape. An individual \ac{NN} was trained for each of the calorimeter setups to guarantee optimal performance. There are several caveats when comparing the resolutions presented here with those achieved in other studies. First of all this is a simplified study where only the relative performance of different sampling ratios are studied. Detector effects like noise are not taken into account. In addition, the differences between $\mr{RMS}_{90}$, $\mr{RMS}$ or a Gaussian width have to be taken into account, as discussed in \cref{App:RMS90}. Nevertheless, using more information of a calorimeter shower than its total energy does certainly result in a significantly improved energy reconstruction. For example, it has been shown in test beam data from the CALICE analog \ac{HCal} prototype that the energy resolution can be improved significantly when the local shower densities in a highly granular calorimeter are used to correct for electromagnetic shower content~\cite{Chadeeva:2012pe}. This correction leads to relative improvement of the energy resolution by 20\%. Another recent study using data from the same prototype applies a leakage correction using the information from the high longitudinal segmentation~\cite{Marchesini2011}. The variables used for the leakage correction are the shower start and the energy sum in the last four layers, which is very similar to the variables that we use to describe the shower shape. This leakage correction results in a relative improvement of the energy resolution by 25\%.

The simulation study presented in this chapter shows that for the given space in the \ac{CLIC} detector concepts a tungsten sampling calorimeter performs better than a steel sampling calorimeter. The optimal performance is achieved for a tungsten absorber plate thickness of approximately \unit[10]{mm} when assuming a gap size of \unit[7.5]{mm} for each active layer. The simulation study assumes an analog scintillator readout but the result will be applicable to other readout technologies as long as the gap size does not deviate significantly from the number used here. It was found that for a calorimeter depth that corresponds to \unit[8]{\lambdaint}, information from a tail catcher that is placed behind the coil can slightly improve the energy measurement. The findings concerning optimal \ac{HCal} depth were later confirmed by an independent study using \pandora, where only the total depth of the \ac{HCal} was varied~\cite{cdrvol2}. In combination, these studies lead to the decision for the current \ac{HCal} layout used in the \clicild and \clicsid models.

A tungsten \ac{HCal} prototype has been built within the CALICE collaboration to verify the simulation results with test beam data. This prototype has been successfully operated in two test beam campaigns at the \acs{PS} and \acs{SPS} at CERN using analog scintillator detectors~\cite{Simon:2011ua}. Preliminary results show good agreement between simulation and data~\cite{CAN036}.

\chapter{Calorimeter Performance in \clicsid}
\label{cha:Calorimetry_PFA}

In this chapter we discuss the calorimetric performance of \clicsid after the modifications of the \ac{HCal}, which we motivated in \cref{cha:Calorimetry}.

The energy of a calorimeter cluster is calculated from the sum of the energy in hits associated with that cluster. Since we are using sampling calorimeters, a correction factor has to be applied to account for the energy deposited in the absorber layers, which depends on the sampling fraction of the respective layer. In addition, we use different sampling fractions depending on the shower type to account for the potentially different responses of electromagnetic and hadronic showers, as discussed in \cref{sec:SiD_energyMeasurement}.

First, we discuss the calibration procedure used to obtain the different sampling fractions in \cref{sec:Calorimetry_PFACalibration}. Afterwards, we discuss the performance of the reconstruction of showers from individual neutral particles in \cref{sec:Calorimetry_PFASingleParticles}, which allows us to deduce the intrinsic calorimetric resolution of \clicsid. The energy reconstruction in di-jet events using \pandora is discussed in \cref{sec:Calorimetry_PFAJets}. Since we are interested in the basic performance, all events are simulated without the \gghad background.

\section{Calorimeter Calibration Procedure}
\label{sec:Calorimetry_PFACalibration}

The sampling fractions for the different calorimeters are calculated from the response of simulated single particle events at several energies. The simulated events that are used, are single photon events to determine the electromagnetic sampling fractions and single \PKzL events for the hadronic sampling fractions. Both particle types are generated with the \geant particle gun and are simulated for kinetic energies of \unit[1]{GeV}, \unit[2]{GeV}, \unit[5]{GeV}, \unit[10]{GeV}, \unit[20]{GeV} and \unit[50]{GeV}. Samples are generated for polar angles of 90\degrees and 30\degrees to determine the sampling fractions for the barrel and endcap calorimeters independently. Each sample contains 10000 events.

The calorimeter consists of two \ac{ECal} sections with different longitudinal sampling and the \ac{HCal} section, which has an even coarser sampling. In addition, the absorber material and the sampling fraction of the \ac{HCal} in the barrel region differs from those in the endcap. The energy reconstruction has to account for all these differences by using different correction factors for each of these subdetectors. If we assume a linear response in all of the individual calorimeters the energy of the particle in one event is given by
\begin{equation}
 E = \sum\limits_{i=0}^N \frac{E_i}{c_i},
\end{equation}
where the index $i$ denotes the calorimeter compartment with the sampling fraction $c_i$ and the sum of the deposited energy in its active layers $E_i$. The sum of the deposited energy $E_i$ takes into account only those hits that are above the threshold of the respective calorimeter, as given in \cref{sec:Software_particleFlowDigitization}. In addition, a cone based clustering algorithm with a fixed opening angle of $0.5$ is used to identify the main cluster of each event. Hits not attributed to the cluster are ignored. $E$ is given by the true generated energy of the initial particle, taking into account the particle mass in case of the Kaons. Together with the calculated sums for all $E_i$, the optimal sampling fractions can be determined using the least-squares method.

Since the clustering in the actual event reconstruction is performed in \pandora which uses additional topological information to refine the clustering, a second iteration is required to determine the final sampling fractions. The calibration events are passed through \slicPandora using the initial sampling fractions. The resulting clusters from \pandora are then used to re-calculate the values for $E_i$ and the final sampling fractions are again determined with the least-squares method.

The resulting sampling fractions are given in \cref{tab:Calorimetry_samplingFractions}. As expected, the sampling fractions for the second \ac{ECal} are approximately half of the first \ac{ECal} section, since the absorber thickness differs by that amount (see \cref{sec:SiD_model_ecal}). For hadronic showers a slightly higher weight is put on the second \ac{ECal} section due to the fact that the first section usually lies before the shower start and mostly sees energy deposits from backscattering evaporation protons and neutrons. Especially the slow protons, which can deposit significantly more energy than a \ac{MIP} on a given path length, should in fact be weighted less. The sampling fractions for electromagnetic and hadronic showers are within 5\% of each other for the hadronic calorimeters and slightly further apart for the \acp{ECal}. This shows that the calorimters are almost compensating and only a limited degradation of the hadronic energy resolution due to this effect is expected.

\begin{table}
 \centering
 \caption{Electromagnetic and hadronic sampling fractions used for the different calorimeters in \clicsid.}
 \label{tab:Calorimetry_samplingFractions}
 \begin{tabular}{l l c c c}
  \toprule
   \multicolumn{2}{c}{Sampling fraction}       & \ac{ECal}$_1$ & \ac{ECal}$_2$ & \ac{HCal} \\
  \midrule
   \multirow{2}{*}{Barrel} & Electromagnetic   & 0.0168        & 0.00816       & 0.0485    \\
                           & Hadronic          & 0.0173        & 0.00939       & 0.0461    \\
   \multirow{2}{*}{Endcap} & Electromagnetic   & 0.0173        & 0.00845       & 0.0334    \\
                           & Hadronic          & 0.0156        & 0.00907       & 0.0350    \\
  \bottomrule
 \end{tabular} 
\end{table} 

\section{Single Particle Response}
\label{sec:Calorimetry_PFASingleParticles}

The basic performance of the calorimeter system was tested with several simulated single particle event samples. Only neutral particles can be used since the particle flow algorithm always uses the track momentum to determine the particle energy of a charged particle. 10000 events of single \PGg, \PGpz, \PKzS and \PKzL have been simulated for each of the kinetic energies of \unit[1]{GeV}, \unit[2]{GeV}, \unit[5]{GeV}, \unit[10]{GeV}, \unit[20]{GeV}, \unit[50]{GeV}, \unit[100]{GeV}, \unit[200]{GeV} and \unit[500]{GeV}. The polar angles of the particles have been restricted to $5\degrees < \theta < 175\degrees$ and the azimuthal angle was left free. The events have been passed through the standard event reconstruction, as explained in \cref{sec:Software_digitizationReconstruction}. In this case, \pandora is only used to determine the calorimeter clusters, which is trivial in the single particle case. The resulting energy resolution can be interpreted as the intrinsic calorimeter performance. 
For the analysis, the energy of the original particle, $E_\text{true}$, is compared with the total reconstructed energy, $E_\text{reco}$. The resolution is determined as the \ac{RMS} of the $E_\text{reco}/E_\text{true}$ distribution for each of the simulated energies. A cut on the polar angle of the initial particle of $\theta > 30\degrees$ was introduced to restrict the sample to fully contained showers. The linearity is determined as the mean of the $E_\text{reco}/E_\text{true} - 1$ distribution.

\begin{figure}[htpb]
 \includegraphics[width=0.49\textwidth]{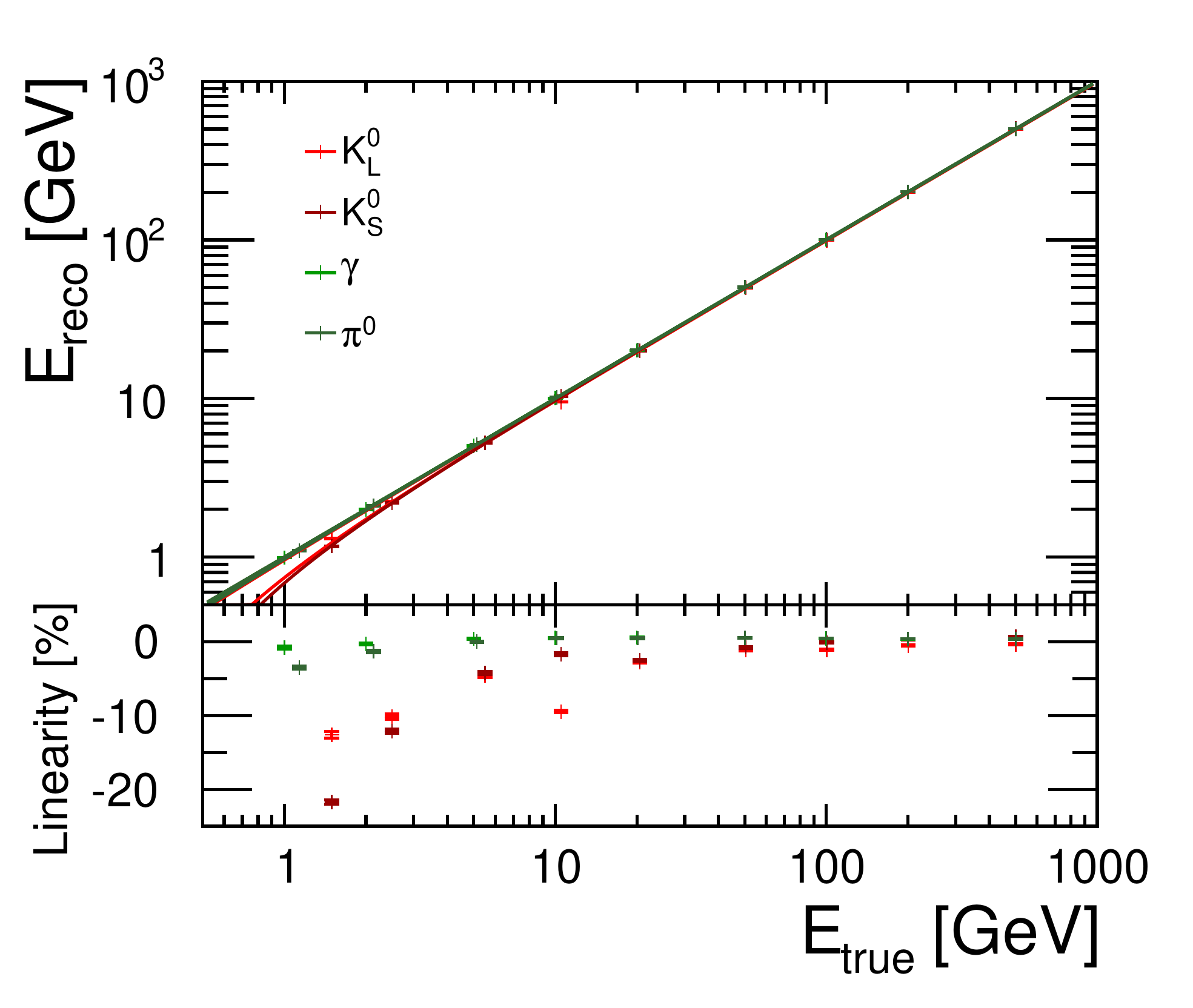}
 \hfill
 \includegraphics[width=0.49\textwidth]{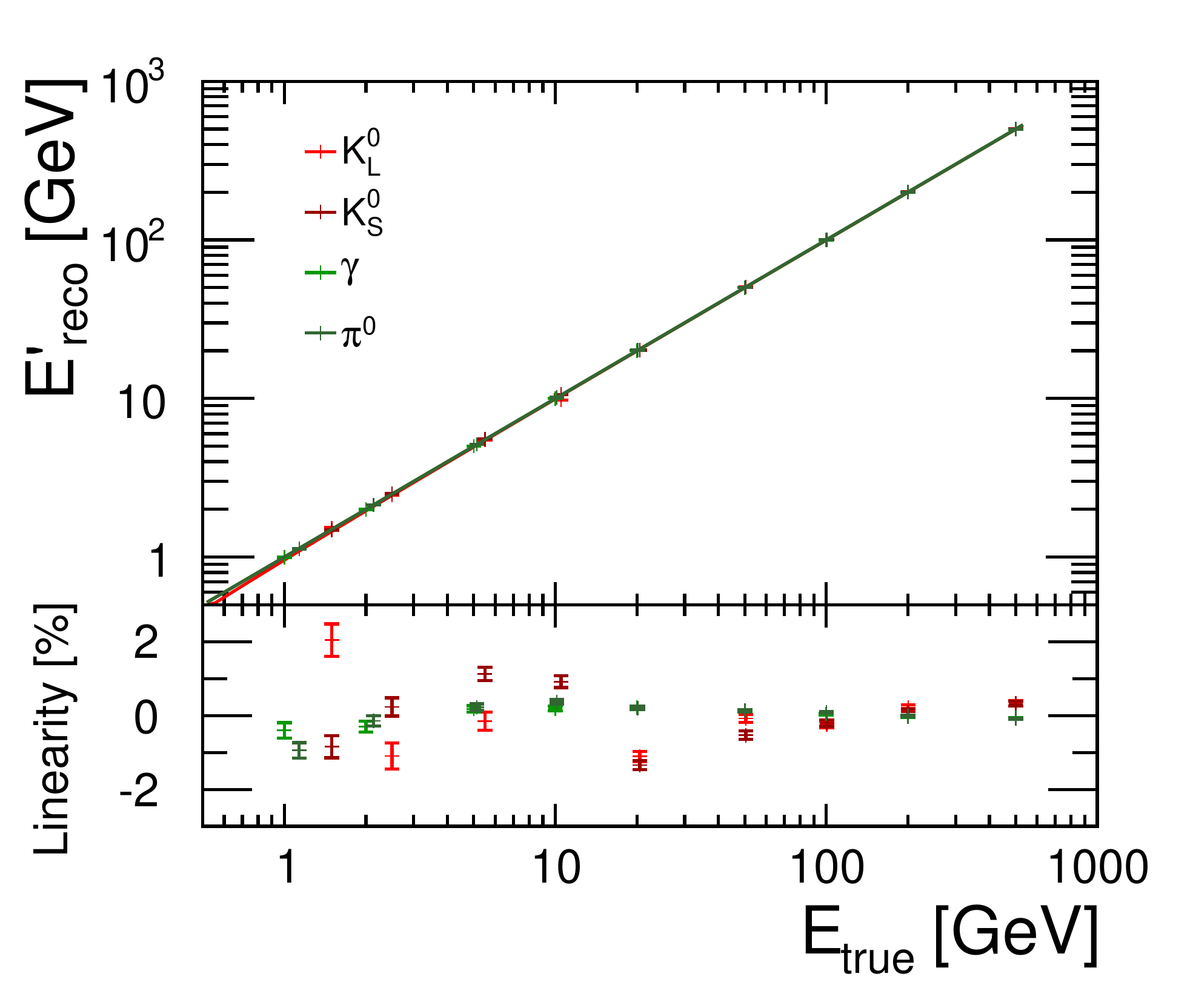}
\caption[Linearity for single neutral particles in the \clicsid detector]{Linearity for single neutral particles in the \clicsid detector. The left plot shows the linearity of the reconstructed energy obtained from the mean of the reconstructed energy for the respective sample. The right plot shows the improved linearity after a correction has been applied. The lines correspond to a linear fit as discussed in the text.}
\label{fig:Calorimetry_singleParticlesLinearity}
\end{figure}

The resulting linearity is shown in \cref{fig:Calorimetry_singleParticlesLinearity}~(left) over the full energy range. Especially for the hadronic showers the reconstructed energies deviates significantly from the true energy for very low energies. For example, the reconstructed energy for \unit[1]{GeV} \PKzL is approximately 20\% too low. For \PGpz particles of $E_\text{kin} = \unit[1]{GeV}$ the reconstructed energy is by approximately 4\% too low. For photons the difference never exceeds 1\%. For higher energies, the difference between true and reconstructed energy becomes significantly smaller and approaches asymptotically 0. An especially large deviation is observed for \PKzL of \unit[10]{GeV}, which is caused by the physics list, since this deviation occurs directly in the transition region of the models for the inelastic scattering of hadrons (see \cref{sec:Software_physicsList}). The effect of significantly reduced visible energy for hadrons with energies in the transition region of \qgspbert is discussed in~\cite{Speckmayer2010}. The \PKzS particles on the other hand almost never reach the calorimeters and thus the available energy of the hard interaction is only that of its decay products. This effect, although much less pronounced, can be seen for \PKzS of \unit[20]{GeV}.

The linearity was fitted with a linear equation
\begin{equation}
 E^{\prime}_\text{reco} \equiv a \cdot E_\text{reco} + b = E_\text{true} 
 \label{eq:linearityCorrection}
\end{equation}
to obtain correction factors that can be applied to the reconstructed energy. The point corresponding to \unit[10]{GeV} \PKzL was excluded from the fit to avoid a bias. The resulting parameters are given in \cref{tab:Calorimetry_singleParticlesLinearityCorrection}. Using these corrections a much better linearity is achieved as shown in \cref{fig:Calorimetry_singleParticlesLinearity}~(right). Even for low energetic Kaons the mean reconstructed energy does not deviate more than 2\% from the true energy.

Although this correction significantly improves the reconstructed energy of low energetic neutral hadrons it depends on the particle type and thus requires particle identification. Since there is no particle identification for neutral hadrons in the current reconstruction software, as mentioned in \cref{sec:Software_particleID}, this correction can not be used. Applying this correction should also improve the reconstructed energy of jets, which still has to be investigated. 

\begin{table}
 \centering
 \caption[Correction to the reconstructed particle energy to improve the linearity in \clicsid.]{Correction to the reconstructed particle energy to improve the linearity in \clicsid. Given is the particle type with the corresponding correction factors $a$ and $b$ according to \cref{eq:linearityCorrection}.}
 \label{tab:Calorimetry_singleParticlesLinearityCorrection}
 \setlength{\extrarowheight}{1pt}
 \begin{tabular}{l c c}
  \toprule
   Particle   & $a$ & $b$ [GeV] \\
  \midrule
   \PGg       & 0.996 & 0.009 \\
   \PGpz      & 0.996 & 0.034 \\
   \PKzS      & 0.996 & 0.351 \\
   \PKzL      & 1.008 & 0.299 \\
  \bottomrule
 \end{tabular} 
\end{table} 

\begin{figure}[htpb]
 \centering
 \includegraphics[width=0.49\textwidth]{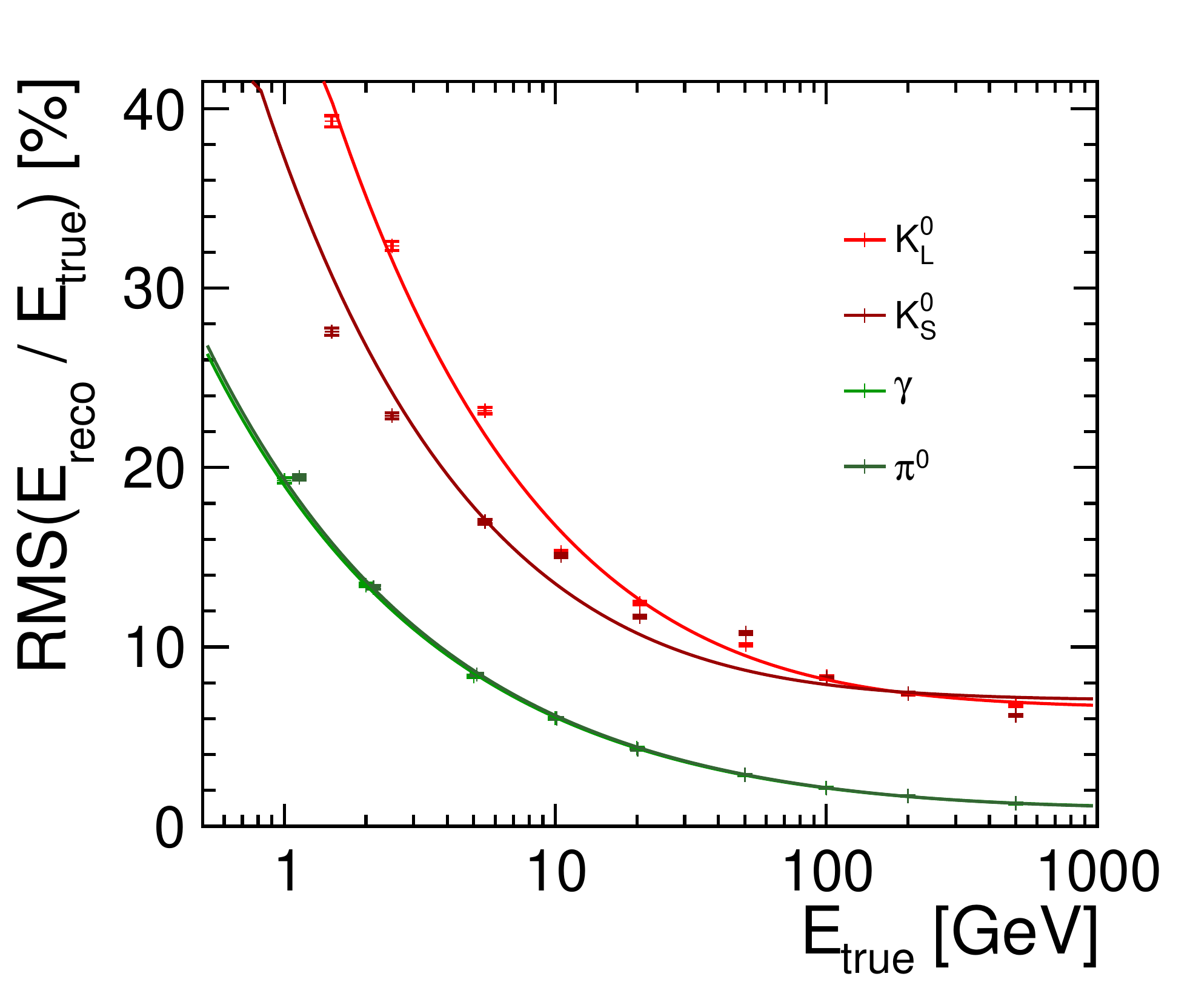}
\caption[Energy resolution for single neutral particles in the \clicsid detector.]{Energy resolution for single neutral particles in the \clicsid detector calculated from the \ac{RMS} of the reconstructed energy for the different test samples. The line shows the fit of the energy resolution as discussed in the text.}
\label{fig:Calorimetry_singleParticlesResolution}
\end{figure}

The energy resolution, which is shown in \cref{fig:Calorimetry_singleParticlesResolution}, has been fitted for all particles with the parametrization given in \cref{eq:energyResolutionNoNoise}.
The resulting parameters of the fit to \cref{eq:energyResolutionNoNoise} for all of the particle types are given in \cref{tab:Calorimetry_singleParticlesResolution}. The energy resolution achieved for photons and $\pi^0$s is almost the same, which is expected, since \PGpz particles decay almost exclusively into two photons. The calorimetric energy resolution for electromagnetic showers in \clicsid is thus approximately $19.1\%/\sqrt{E/\mr{GeV}} \oplus 1.0\%$. The calorimetric energy resolution for hadronic showers is given by the resolution for \PKzL, which is $50.3\%/\sqrt{E/\mr{GeV}} \oplus 6.5\%$. The situation for \PKzS is a bit more complex, since they most likely decay in flight before reaching the calorimeters. The two relevant decays are $\PKzS \to \PGpp\PGpm$ and $\PKzS \to \PGpz\PGpz$, with branching ratios of $\sim 69.2\%$ and $\sim 30.7\%$~\cite{PDG}, respectively. The later decay mode will further decay into 4 photons before reaching the calorimeters due to the short lifetime of the $\PGpz$s. Due to the late decay of the \PKzS, the tracks of the charged pions can not be reconstructed with the current tracking algorithm, as discussed in \cref{sec:Software_trackReconstruction}. The corresponding showers are thus reconstructed as neutral hadrons and the total resolution for \PKzS particles is between the electromagnetic and the hadronic resolution of the calorimeters.

The numbers for the energy resolution given here are obtained from the uncorrected reconstructed energy. Calculating the energy resolution for $E'$ gives very similar results.

\begin{table}
 \centering
 \caption[Energy resolution for single neutral particles in \clicsid.]{Energy resolution for single neutral particles in \clicsid. Given is the particle with the sampling term $s$ and the constant term $c$ according to Eq.~(\ref{eq:linearityCorrection}).}
 \label{tab:Calorimetry_singleParticlesResolution}
 \setlength{\extrarowheight}{1pt}
 \begin{tabular}{l c c}
  \toprule
   Particle   & $s$ [\%] & $c$ [\%] \\
  \midrule
   \PGg       & 19.1 & 1.0 \\
   \PGpz      & 19.3 & 0.9 \\
   \PKzS      & 38.7 & 6.9 \\
   \PKzL      & 50.3 & 6.5 \\
  \bottomrule
 \end{tabular} 
\end{table}

\section{Jet Response}
\label{sec:Calorimetry_PFAJets}

The performance of the energy reconstruction of jet events was studied using di-jet samples from \Zprime decaying at rest. These are the same event samples that are used to evaluate the tracking performance in jets. In addition to the sample with a \Zprime mass of \unit[3]{TeV} discussed in \cref{sec:Tracking_eventSamples} we also use samples with lower masses in the range from \unit[91]{GeV} to \unit[3]{TeV} to study the dependence of the jet energy resolution on the jet energy. No jet finding is used in the event reconstruction to avoid a bias depending on the choice of the algorithm and its parameters. Instead, the energy of all reconstructed particles, \ie \acp{PFO}, is summed up to find the reconstructed energy $E_\text{reco}$ of the di-jet system. The energy of each jet is assumed to be $E_\text{reco}/2$. The resolution is then calculated as the \rmsn of the distribution of $E_\text{reco}$ and the resolution for each individual jet is calculated as $\sigma(E_\text{jet}) = \sqrt{2}\sigma(E_\text{reco})$. Since no jet clustering is used, we define the polar angle of the jet as the polar angle of the original quark\footnote{The quarks are created back-to-back and can thus be identified with a single polar angle in the interval $[0\degrees,90\degrees]$}.

\begin{figure}[htpb]
 \includegraphics[width=0.49\textwidth]{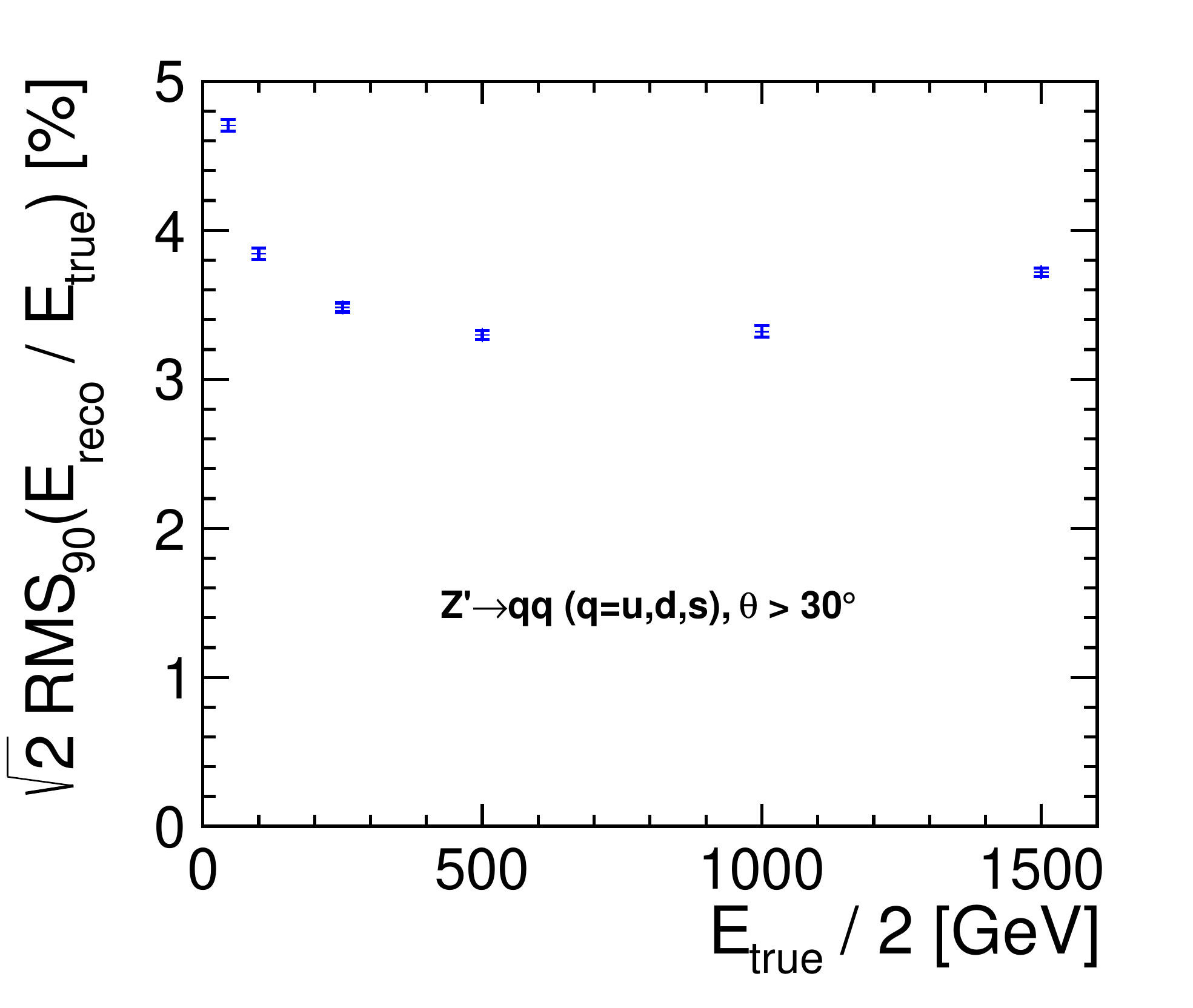}
 \hfill
 \includegraphics[width=0.49\textwidth]{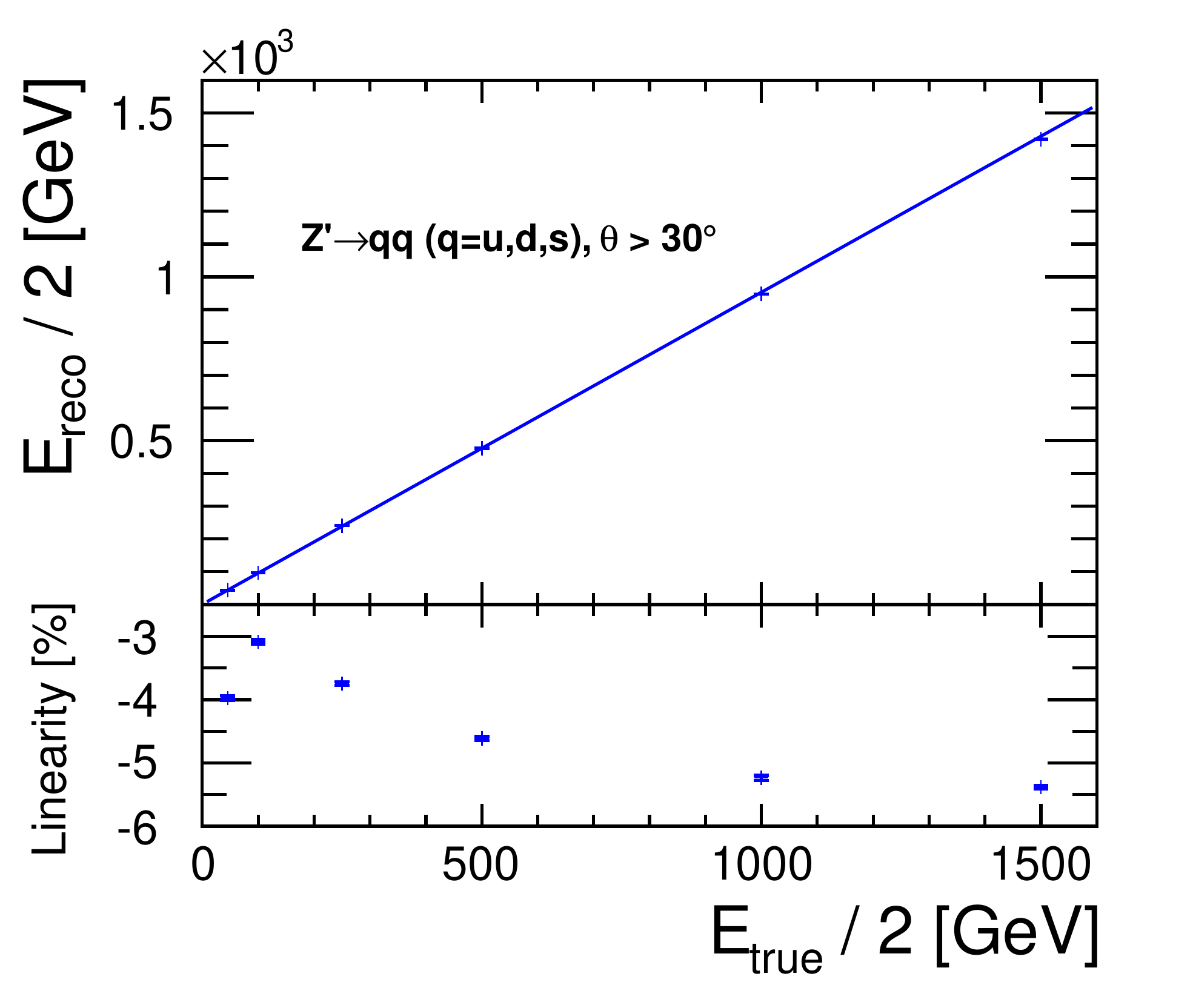}
\caption[\ac{PFA} response for single jets in di-jet decays depending on the jet energy.]{\ac{PFA} response for single jets in di-jet decays of a \Zprime at rest, depending on the jet energy. Shown is the jet energy resolution (left) and the linearity (right) obtained from the energy sum of all reconstructed \acp{PFO}.}
\label{fig:Calorimetry_jetPerformanceEnergy}
\end{figure}

The jet energy resolution for central jets, which is shown in \cref{fig:Calorimetry_jetPerformanceEnergy}~(left), is between 3\% and 5\%, depending on the jet energy and when using \rmsn. The best resolution is achieved for jet energies of approximately \unit[500]{GeV} and drops significantly for lower jet energies. For higher jet energies it degrades moderately. The energy is underestimated in the reconstruction by 3--5\%, as shown in \cref{fig:Calorimetry_jetPerformanceEnergy}. Since this effect is almost constant over the full energy range, the result can be improved by rescaling the reconstructed energy by a constant factor of 1.050 which was obtained from the slope of the linearity in \cref{fig:Calorimetry_jetPerformanceEnergy}~(right). It is assumed that this systematic bias in the jet energy reconstruction is caused by confusion when soft neutral showers are not resolved from large neighboring charged particle showers. It should be noted that the configuration of the algorithms in \pandora was originally tuned for \clicild and not specifically adapted to \clicsid. The smaller inner radius of the calorimeters in \clicsid, compared to \clicild might be the cause of this, but further studies are required.


\begin{figure}[htpb]
 \includegraphics[width=0.49\textwidth]{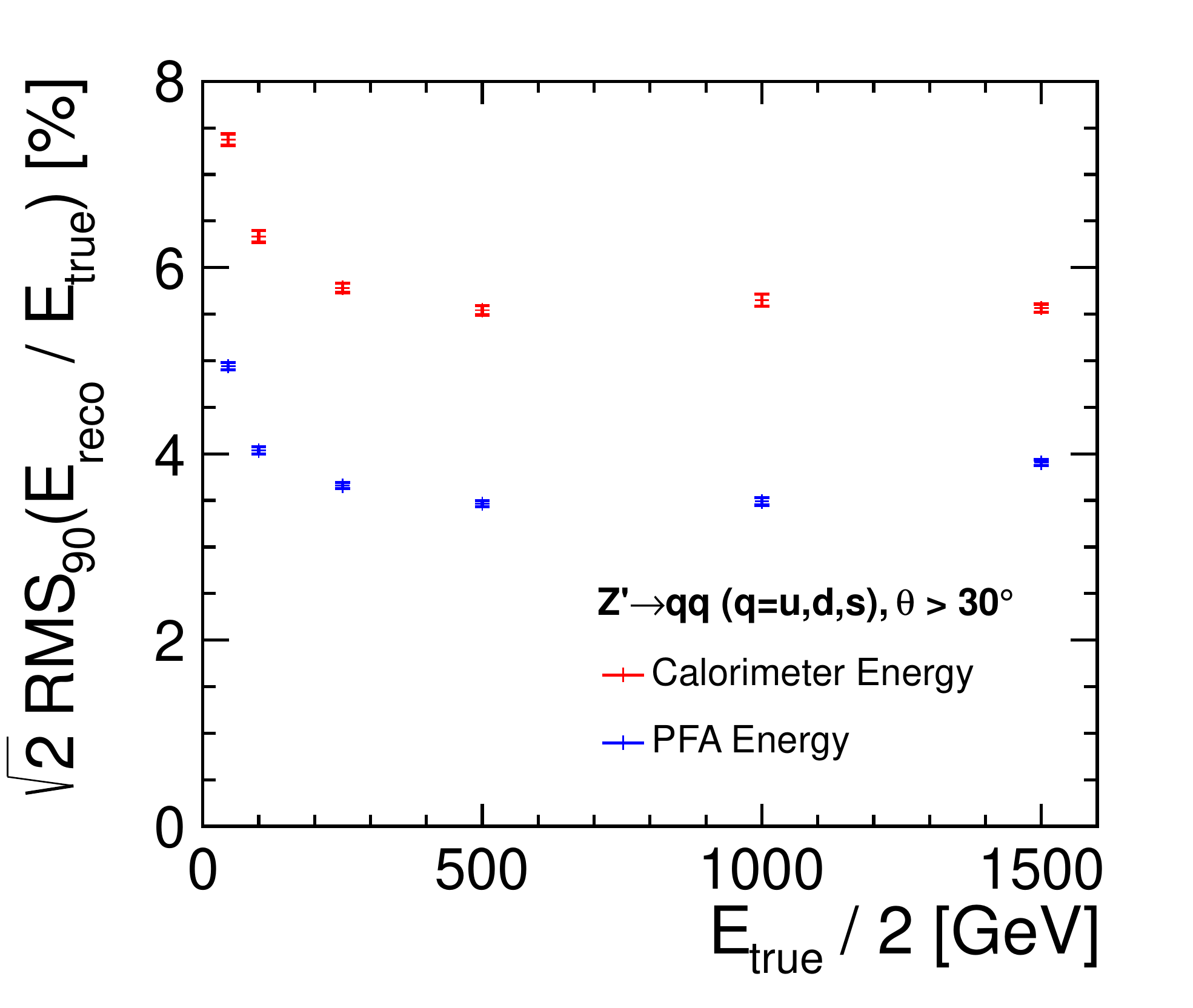}
 \hfill
 \includegraphics[width=0.49\textwidth]{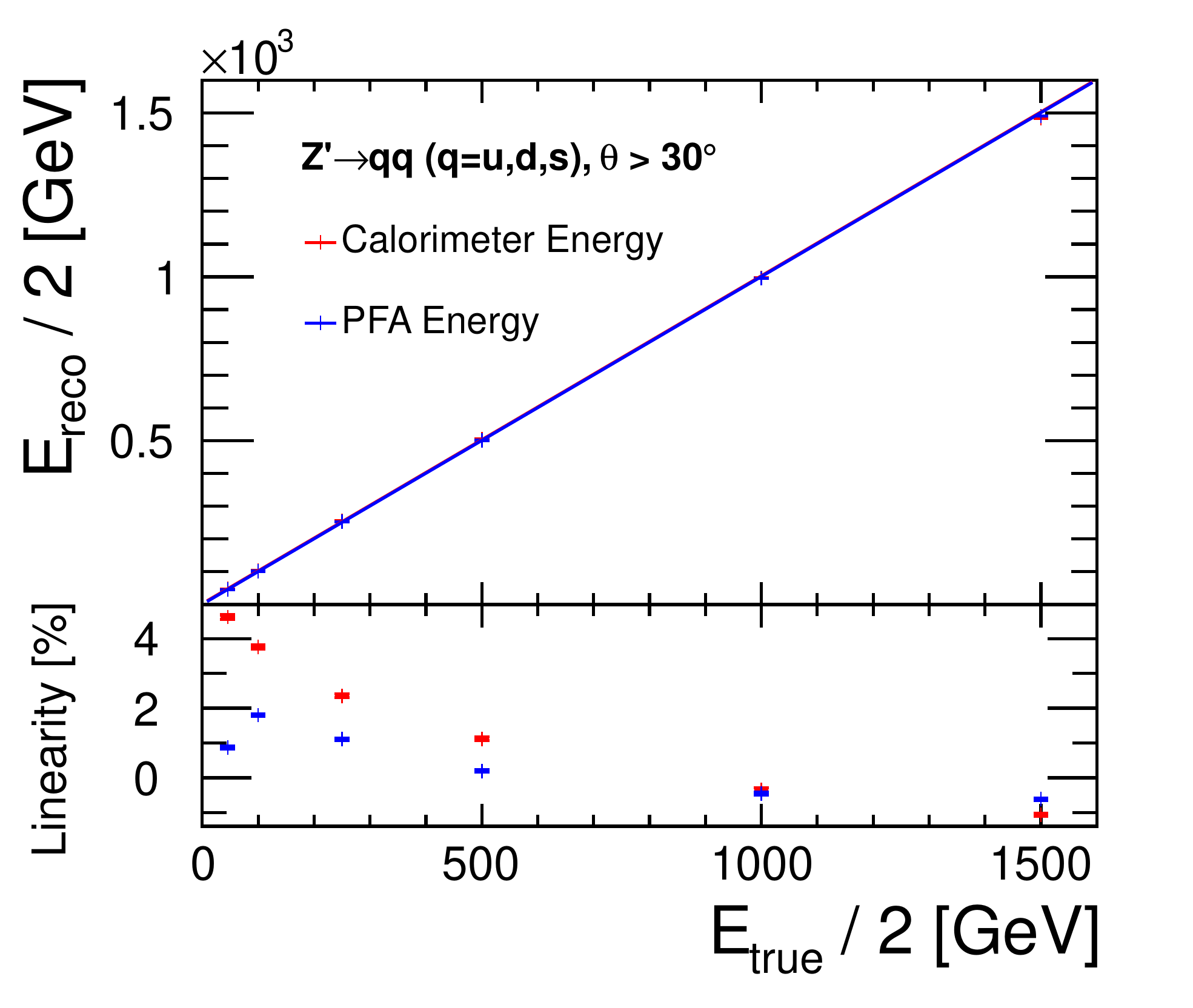}
\caption[Calorimeter response for single jets in di-jet decays depending on the jet energy.]{Calorimeter response for single jets in di-jet decays of a \Zprime at rest, depending on the jet energy. Shown is the jet energy resolution (left) and the linearity (right) using either the energy sum of all reconstructed \acp{PFO} or the energy sum of all corrected calorimeter hits.}
\label{fig:Calorimetry_jetPerformanceEnergyComparison}
\end{figure}

\begin{figure}[htpb]
 \begin{subfigure}[]{0.49\textwidth}
  \includegraphics[width=\textwidth]{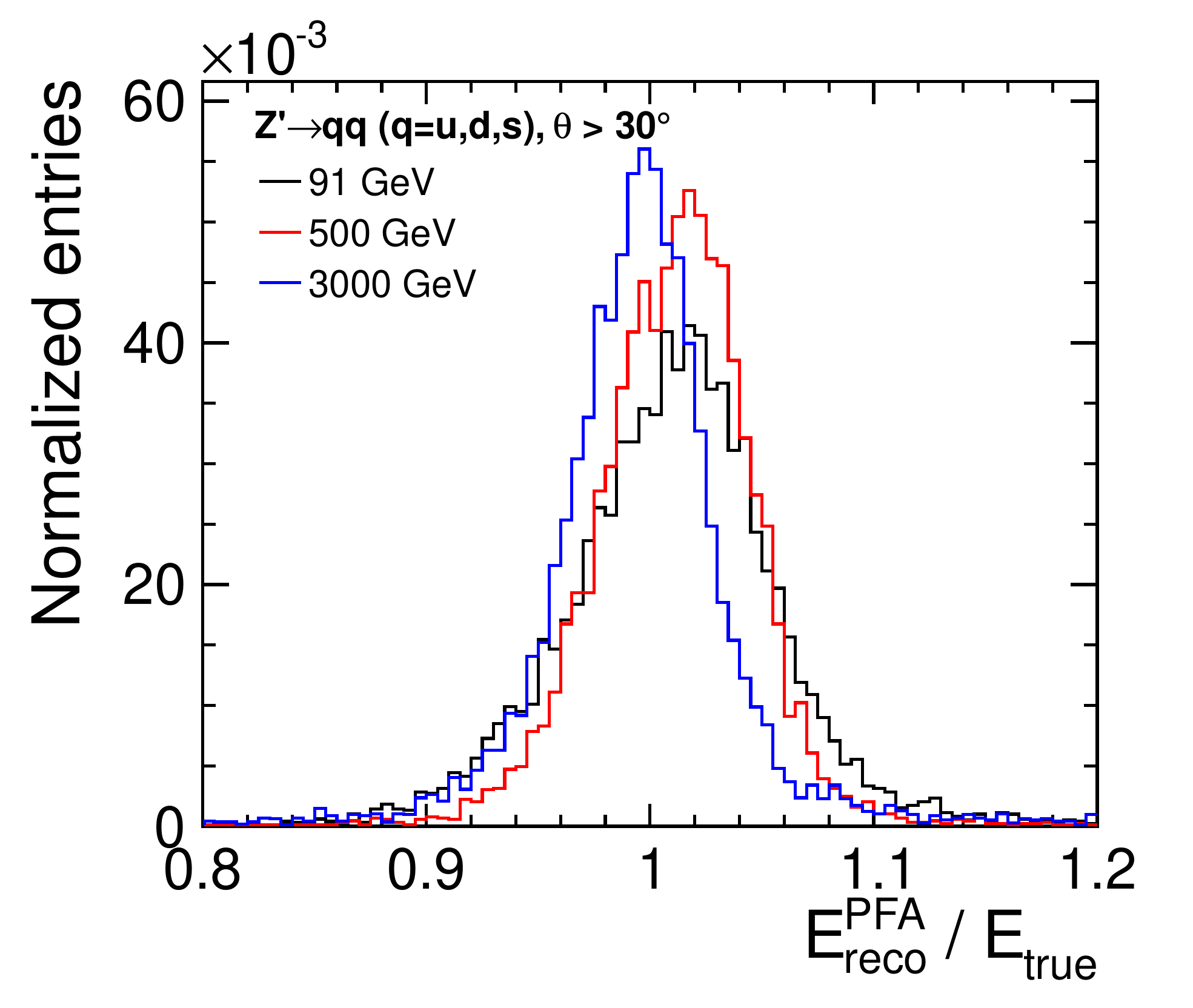}
  \caption{\ac{PFA} energy}
  \label{fig:Calorimetry_jetEnergyDistributionPFO}
 \end{subfigure}
 \hfill
 \begin{subfigure}[]{0.49\textwidth}
  \includegraphics[width=\textwidth]{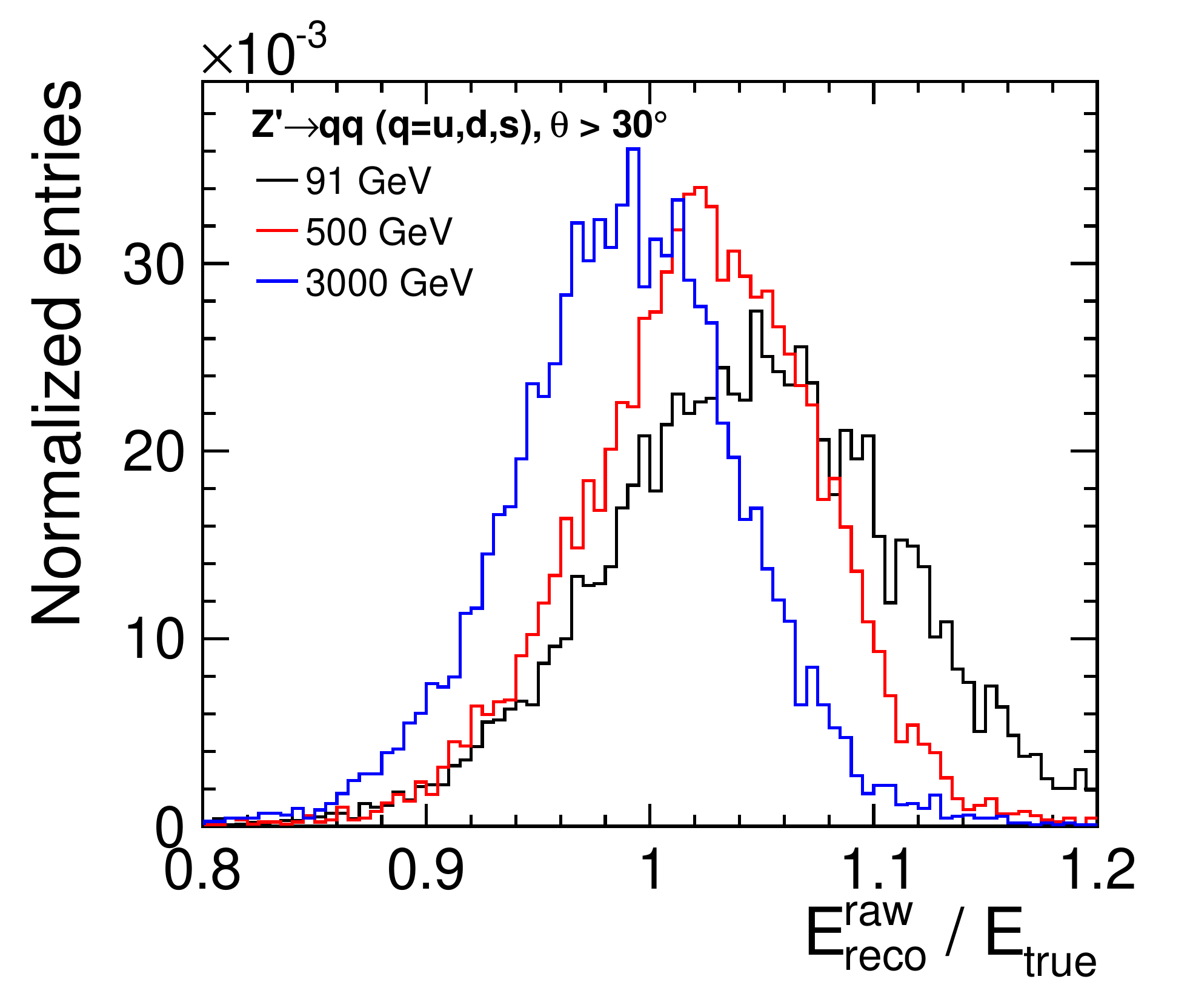}
  \caption{Cluster energy}
  \label{fig:Calorimetry_jetEnergyDistributionCluster}
 \end{subfigure}
\caption[Distribution of the reconstructed energy in di-jet decays depending on the jet energy.]{Distribution of the reconstructed energy in di-jet decays of a \Zprime at rest for different \Zprime masses, using the energy sum of all reconstructed \acp{PFO}~\subref{fig:Calorimetry_jetEnergyDistributionPFO} and using the energy sum of all calorimeter hits~\subref{fig:Calorimetry_jetEnergyDistributionCluster}.}
\label{fig:Calorimetry_jetEnergyDistribution}
\end{figure}

\Cref{fig:Calorimetry_jetPerformanceEnergyComparison} shows a comparison of the \ac{PFA} performance and a pure calorimetric energy measurement. The calorimeter energy is obtained as the sum of all hit energies corrected by the respective sampling fractions given in \cref{tab:Calorimetry_samplingFractions}, where the electromagnetic sampling fraction was used in the \ac{ECal} and the hadronic sampling fraction was used in the \ac{HCal}. In addition,  the total reconstructed energy is scaled such that the slope of the linearity is 1. Similarly, we applied the energy scaling of 1.050 to the \ac{PFA} energy as discussed above. The energy reconstruction using \ac{PFA} is significantly better than using just the energy measurement in the calorimeter over the full energy range. 
These results are consistent with the results found for an \ac{ILD}-like detector geometry, as discussed in~\cite{Thomson:2009rp}.
The distribution of the total reconstructed energies, shown in \Cref{fig:Calorimetry_jetEnergyDistribution}, also highlights the large advantage in using momentum information to reconstruct the jet energy. It also shows that the distributions have long asymmetric tails which motivates the use of \rmsn.

\begin{figure}[htpb]
 \begin{subfigure}[]{0.49\textwidth}
  \includegraphics[width=\textwidth]{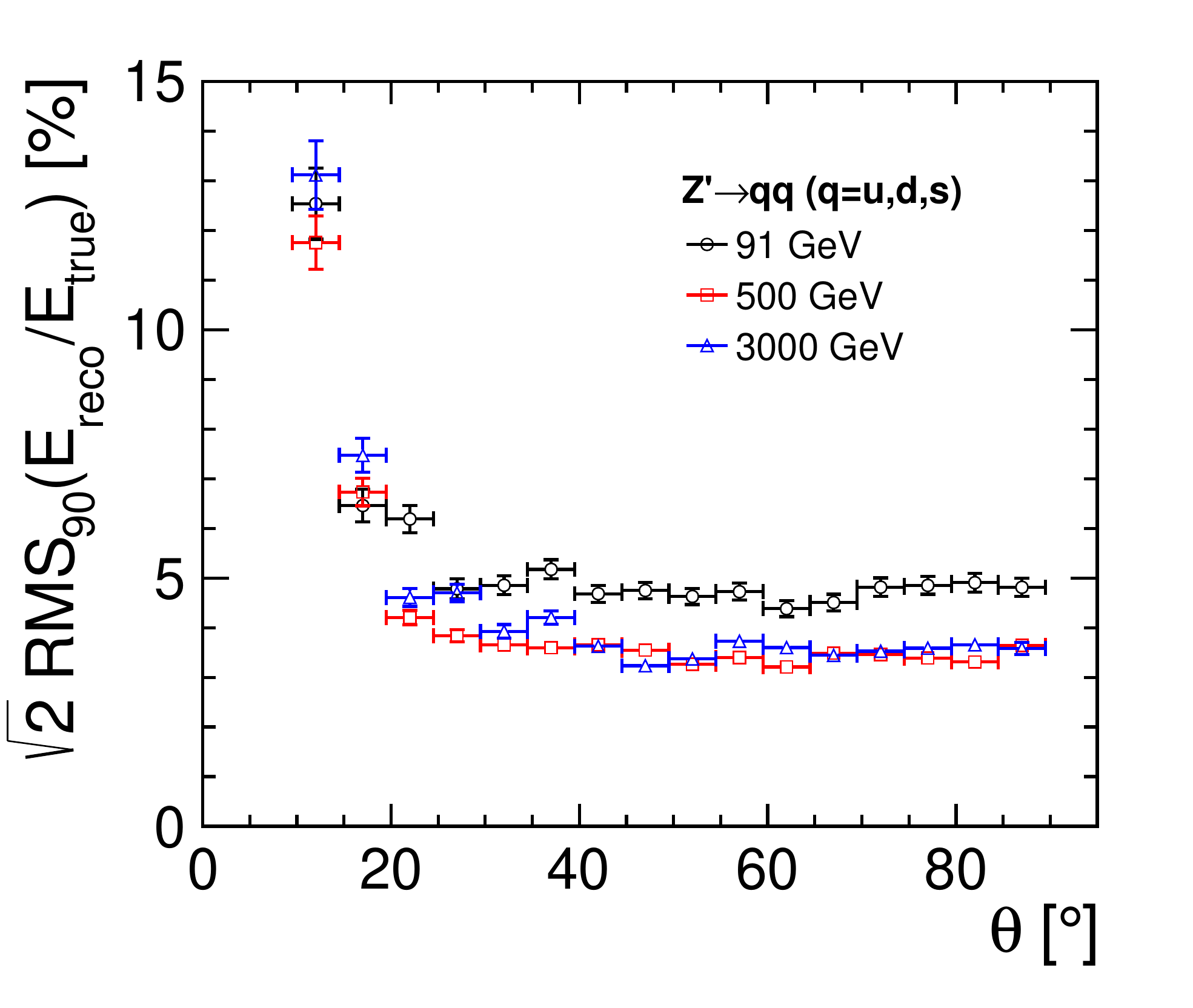}
  \caption{}\label{fig:Calorimetry_jetPerformanceThetaRes}
 \end{subfigure}
 \hfill
 \begin{subfigure}[]{0.49\textwidth}
  \includegraphics[width=\textwidth]{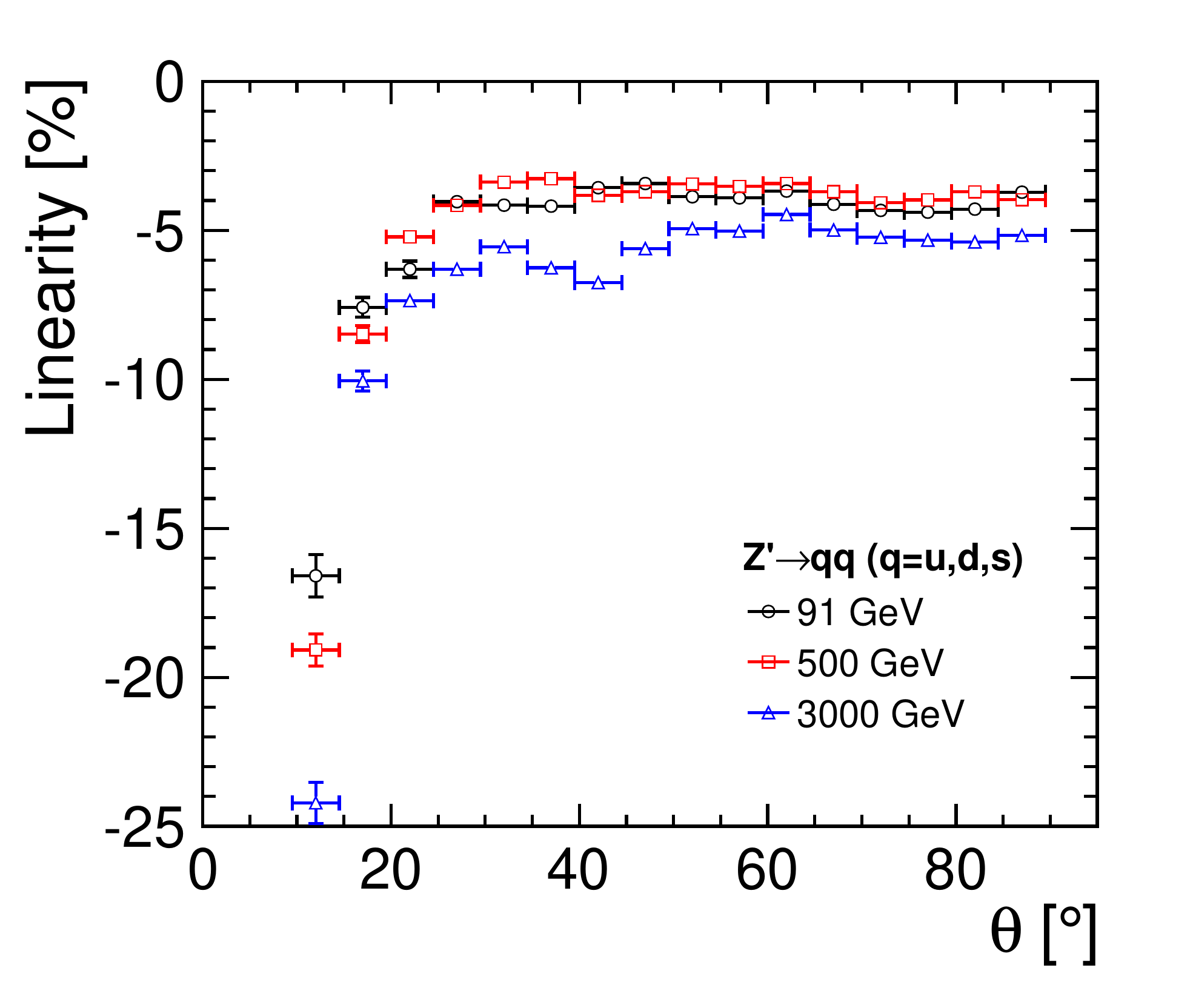}
  \caption{}\label{fig:Calorimetry_jetPerformanceThetaMean}
 \end{subfigure}
\caption[Calorimeter response for single jets in di-jet decays depending on the polar angle of the jet.]{Calorimeter response for single jets in di-jet decays of a \Zprime at rest, for different \Zprime masses, depending on the polar angle $\theta$ of the jet. Shown are the resolution (left) and the linearity (right) using the energy sum of the reconstructed \ac{PFO}.}
\label{fig:Calorimetry_jetPerformanceTheta}
\end{figure}

The angular dependence of the jet energy resolution can be seen in \cref{fig:Calorimetry_jetPerformanceThetaRes}. The jet energy resolution is almost constant for all polar angles $\theta$ throughout the central region of the detector and drops significantly for jet angles lower than 20\degrees. This drop in the performance is to some extent due to the degrading momentum resolution in the forward region but mostly due to the acceptance of the calorimeters in that region, where especially the \ac{HCal} acceptance is limited due to the support tube necessary for stabilizing the \ac{QD0} (see \cref{sec:SiD_model_acceleratorComponents}). The effect of the acceptance can be seen more directly in the dependence of linearity on the polar angle, which is shown in \cref{fig:Calorimetry_jetPerformanceThetaMean}. The energy is underestimated by 4--5\% in the central region, which increases to 10--20\% for polar angles of less than 20\degrees. While we have discussed possible explanations of this discrepancy in the central region, the showers in the very forward region are clearly not fully contained. These effects are more pronounced for high energy jets at all polar angles. In the transition region between barrel and endcap calorimeters at $\theta \approx 40\degrees$ the reconstructed energy is especially low for high jet energies. This shows that the energy loss in the gap between these two calorimeter regions is not correctly accounted for.

For completeness we summarize some of the findings presented in~\cite{lcd:2011-028} concerning the impact of the \gghad background. This background significantly increases the reconstructed energy in each event. Depending on the event topology a suitable jet clustering algorithm might be sufficient to retain the original event information. The $K_t$-algorithms~\cite{Catani:1993hr} have proven especially useful at \ac{CLIC} conditions, as for example shown in the squark mass measurement benchmark study~\cite{LCD:2011-27}. In more complex event topologies rejecting reconstructed particles based their reconstructed production time is necessary using the method explained in \cref{sec:PFOSelection}. Although this method is extremely efficient, tight selection cuts also degrade the jet energy resolution by removing particles from the original event. This effect is only noticeable for low energetic jets, since the individual particles in \gghad events are of low energy and thus, only low energetic particles are removed by the cuts. In these cases, the jet energy resolution can degrade by up to a factor of two.

One example of the effectiveness of these timing cuts is the gaugino and neutralino pair production benchmark analysis~\cite{LCD:2011-037}.

\section{Summary}
\label{sec:Calorimetry_PFASummary}

We have presented the method that is used to determine the sampling fractions for the \clicsid detector model. A set of simulated single photon and \PKzL particle events at various energies and two polar angles is used to determine the optimal sampling fractions of all subdetectors in a least-squares fit. The resulting calibration is tested using independent samples of single particles and di-jet events, not taking into account the \gghad background.

For single particles we found that, especially for very low energetic Kaons, the reconstructed energy is too low by up to 20\%. A linear correction can be used to improve this significantly but requires particle identification also for neutral hadrons which is not available in the current reconstruction software. The energy resolution for electromagnetic showers was found to be $19.1\%/\sqrt{E/\mr{GeV}} \oplus 1.0\%$. As expected, the hadronic energy resolution is significantly worse with $50.3\%/\sqrt{E/\mr{GeV}} \oplus 6.5\%$.

The jet energy resolution in di-jet events was found to be in between 5\% and 3.5\% (using \rmsn) for jet energies of \unit[45]{GeV} to \unit[1.5]{TeV}. We have shown that \ac{PFA} improves the jet energy resolution significantly. It remains to be studied if the performance of \pandora can be improved by tuning some of its algorithms specifically for \clicsid.

\chapter{Measurement of Light Higgs Decay into Muons}
\label{cha:Higgs}

As discussed in \cref{cha:SM}, the measurement of the Higgs branching ratios is an important test of the \ac{SM}.
The \ac{SM} makes precise predictions of the Higgs branching ratios with the mass of the Higgs boson being the only unknown parameter. A deviation from these predictions would be a strong indication of new physics.
Measuring the Higgs decay into muons is especially challenging because of its very small branching ratio of 0.028\% for a Higgs boson mass of \unit[120]{GeV}.


This chapter presents the cross section times branching ratio measurement for the \ww fusion process, $\epem \to \PSh \nuenuebar$, with the Higgs decaying into two muons, at \ac{CLIC} with a center-of-mass energy of \unit[3]{TeV}. This analysis is part of the detector benchmark analyses~\cite{lcd:2011-016} described in the \ac{CLIC} \ac{CDR}~\cite{cdrvol2} with the aim of demonstrating the detector performance in realistic physics analyses. This channel, requiring the precise reconstruction of two muons of medium energies, challenges the momentum resolution of the tracking system at all angles. The analysis is performed in the \clicsid detector model which is described in \cref{cha:SiD}. A total integrated luminosity $\cal L$ of \unit[2]{\abinv} is used for the analysis. This corresponds to about 4 years of data taking with an instantaneous luminosity of \unit[$5.9\times10^{34}$]{cm$^{-2}$s$^{-1}$}, assuming 200 days of operation per year with an efficiency of 50\%. This analysis has already been published in~\cite{lcd:grefeHmumu2011}.

We begin by describing the event samples which are used in the analysis in \cref{sec:Samples}. In \cref{sec:EventSelection} we introduce the event selection procedure used and in \cref{sec:MassFit} we discuss the invariant mass fit used to extract the number of signal events as well as the resulting uncertainties on the cross section times branching ratio measurement. In \cref{sec:BeamInducedBackground} we investigate the impact of the \gghad background on this analysis. The dependency of the uncertainty of the branching ratio measurement on the momentum resolution is studied in a fast simulation study presented in \cref{sec:MomentumResolution}. Moreover, the importance of efficient electron tagging in the forward calorimeters is demonstrated in \cref{sec:ElectronTagging}. Furthermore, the translation of the measured cross section branching ratio into the corresponding coupling constant $g_{\PGm}$ is discussed in \cref{sec:ElectronTagging}. Finally, we compare the achievable uncertainty at \ac{CLIC} with the potential for measuring this branching ratio at the \ac{ILC} and at the \ac{LHC} in \cref{sec:Higgs_otherPotential}.

\section{Event Samples}
\label{sec:Samples}

This section introduces the different event samples used throughout this analysis. An overview is given in \cref{tab:samples} showing the cross sections of all processes and the number of events that were simulated.

\begin{table}[htb]
 \centering
 \caption[List of processes considered for the $\PSh \to \mpmm$ analysis.]{List of processes considered for the $\PSh \to \mpmm$ analysis with their respective cross sections $\sigma$ and the number of simulated events $N_{\mathrm{events}}$.}
 \label{tab:samples}
\begin{threeparttable}
\begin{tabular}{p{6cm} r r l}
\toprule
Process & $\sigma$ [fb] & $N_{\mathrm{events}}$ & Short label \\\midrule

$\epem \to \PSh \nuenuebar$; $\PSh \to \mpmm$ (signal) & 0.120            &   21000 & $\PSh \to \mpmm$    \\\midrule
$\epem \to \mpmm \nunubar$                             & 132              & 5000000 & $\mpmm \nunubar$    \\
$\epem \to \mpmm \epem$                                & 346\tnote{A}     & 1350000 & $\mpmm \epem$       \\\midrule
$\epem \to \mpmm$                                      &  12\tnote{B}     &   10000 & $\mpmm$             \\
$\epem \to \tptm$                                      & 250              &  100000 & $\tptm$             \\
$\epem \to \tptm \nunubar$                             & 125              &  100000 & $\tptm \nunubar$    \\\midrule
$\gamgam \to \mpmm$ (generator level only)             & 20000\tnote{B}   & 1000000 & $\gamgam \to \mpmm$ \\
\bottomrule
\end{tabular}
\begin{tablenotes}
 \item[A] Including a cut of $\unit[100]{GeV} < M(\mumu) < \unit[140]{GeV}$ and requiring a minimum polar angle for both muons of 8\degrees.
 \item[B] Including a cut of $\unit[100]{GeV} < M(\mumu) < \unit[140]{GeV}$.
\end{tablenotes}
\end{threeparttable}
\end{table}

\subsection{Signal Sample}
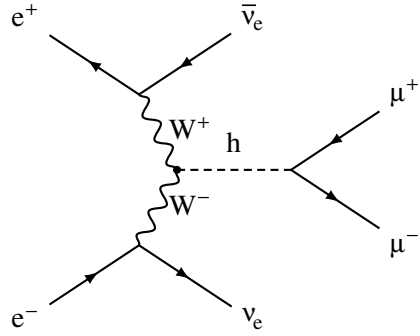
\begin{figure}[htb]
 \centering
 \begin{subfigure}[]{0.49\textwidth}
  \centering
  \begin{tikzpicture}[thick, >=latex]

   \node (positron) at (-2.0,2.0) {\Pep};
   \node (antineutrino) at (1.0,2.0) {\PAGne};
   \node (electron) at (-2.0,-2.0) {\Pem};
   \node (neutrino) at (1.0,-2.0) {\PGne};
   \coordinate (e1vW) at (-0.5, 1.0);
   \coordinate (e2vW) at (-0.5, -1.0);
   \coordinate (higgsLeft) at (0.0,0.0);
   \coordinate (higgsRight) at (1.5,0.0);
   \node (muon) at (3,-1) {\PGmm};
   \node (antimuon) at (3,1) {\PGmp};

   \draw[antifermion] (positron) to (e1vW);
   \draw[antifermion] (e1vW) to (antineutrino);
   \draw[vectorboson] (e1vW) to node[right] {\PWp} (higgsLeft);
   \draw[vectorboson] (e2vW) to node[right] {\PWm} (higgsLeft);
   \fill[black] (higgsLeft) circle (1.5pt);
   \draw[higgs] (higgsLeft) to node[above] {\PSh} (higgsRight);
   \draw[antifermion] (higgsRight) to (antimuon);
   \draw[fermion] (higgsRight) to (muon);
   \draw[fermion] (electron) to (e2vW);
   \draw[fermion] (e2vW) to (neutrino);
  \end{tikzpicture}
 \end{subfigure}

 \caption{Feynman diagram of the Higgs production through \ww fusion with the Higgs decaying into two muons. }
 \label{fig:higgsproduction}
\end{figure}

At a center of mass energy of \unit[3]{TeV} the most relevant production processes for a light Higgs are gauge boson fusion processes, as discussed in \cref{sec:SM_higgsProduction}. The Higgs production through \ww fusion, $\epem \to \PSh \nuenuebar$, has a cross section of \unit[422]{fb} when the \ac{CLIC} luminosity spectrum is taken into account. The corresponding Feynman diagram is shown in \cref{fig:higgsproduction}. The cross section of the \zz fusion process $\epem \to \PSh \epem$ is \unit[42.6]{fb} which is about 10 times smaller. The Higgs branching ratios depend on the Higgs boson mass and are calculated using \pythia. The BR$_{\PSh \to \mpmm}$ is only 0.028\% for a \unit[120]{GeV} \ac{SM} Higgs, which is compatible with latest theoretical calculations~\cite{Denner:2011mq}. For an integrated luminosity of \unit[2]{\abinv} one thus expects about 236 events from the \ww fusion process and only 24 events from the \zz fusion process.
Therefore, the Higgs boson production through \ww fusion is chosen for this study. The final state of the signal is thus $\mpmm \nuenuebar$, which will further be referred to as $\PSh \to \mpmm$. The event topology of interest comprises two muons and missing energy.

The number of events from the \zz fusion channel with this signature is very low. The selection efficiency is on the few per cent level, depending on the final cuts. Taking it into account in the signal and background hypothesis is not expected to have a noticeable effect on the final results.

The signal events used in this analysis have been created by generating events with a final state of $\PSh \nuenuebar$ using \whizard. The Higgs boson was forced to decay into two muons in \pythia. Since final state particles have to be stable in \whizard, the width of the Higgs boson is not taken into account. The width of a \acl{SM} Higgs with a mass of \unit[120]{GeV} is \unit[3.6]{MeV}, which is considerably smaller than the invariant mass resolution of the detector and is thus negligible.

For this analysis 21000 $\PSh \to \mpmm$ events were generated and simulated. They were reconstructed with and without overlaying the \gghad background in order to study the impact of this background on the analysis result.

\subsection{Main Backgrounds}

The two main background processes are the $\epem \to \mpmm \nunubar$ and the $\epem \to \mpmm \epem$ processes where the muons are not produced by a Higgs boson and, in the case of the second process, the electrons escape the detector through the beam pipe. Their total cross sections are \unit[132]{fb} and \unit[5.4]{pb}, respectively. Both were generated as inclusive final states in \whizard, where the diagrams including a Higgs boson were kinematically forbidden by artificially setting the Higgs mass to \unit[12]{TeV}.

Because of the large cross section of the $\epem \to \mpmm \epem$ process, several generator level cuts were introduced to limit the number of events that needed to be simulated and reconstructed. The invariant mass of the di-muon system, $M(\mumu)$, was limited to masses between \unit[100]{GeV} and \unit[140]{GeV} and the minimum polar angle of the muons was set to 8\degrees. The cross section of the $\epem \to \mpmm \epem$ process including these cuts is \unit[346]{fb}. These cuts introduce no bias, since the pre-selection requires $\unit[105]{GeV} < M(\mumu) < \unit[135]{GeV}$ and the detector acceptance to reconstruct muons is around 10\degrees, as discussed in \cref{sec:PreSelection}.

Because of its final state of only two muons, 5 million events were simulated and reconstructed for the $\epem \to \mpmm \nunubar$ background process, which corresponds to an integrated luminosity of almost \unit[40]{\abinv}. The simulation and reconstruction of the $\epem \to \mpmm \epem$ process is considerably slower and thus only 1.35 million events were generated for this background, corresponding to an integrated luminosity of approximately \unit[2]{\abinv}. In both cases no \gghad background was overlaid since it would have significantly increased the reconstruction time per event and the impact is only minor. For a discussion of this background see \cref{sec:BeamInducedBackground}.

\subsection{Additional Background Samples}

Several other two and four fermion final states were simulated to verify that they can be easily identified as background events. The $\epem \to \mpmm$ process has a total cross section of about \unit[350]{fb}. It was generated with an invariant mass of the di-muon system between \unit[100]{GeV} and \unit[140]{GeV} which reduces the cross section to about \unit[12]{fb}. The processes $\epem \to \tptm$ and $\epem \to \tptm \nunubar$ have cross sections of \unit[250]{fb} and \unit[125]{fb}, respectively. They can mimic the signal topology if both taus are decaying into muons, which happens in about 3\% of the cases. In addition to the large reduction by requiring two muons, the invariant mass of the di-muon system allows to distinguish these channels from the signal events.

Finally, a sample of beam-induced incoherent pair background was generated with \whizard, using a photon spectrum generated by \guineapig~\cite{Schulte:331845,Schulte:1999tx}. The $\gamgam \to \mpmm$ background has a large cross section of \unit[20]{pb} even after requiring the invariant mass of the di-muon system to be in the range from \unit[100]{GeV} to \unit[140]{GeV}. This sample was not put through full simulation since it became clear that all of its events can easily be rejected by cutting on the transverse momentum of the di-muon system (see \cref{sec:SelectionAdditionalBackground}).

\section{Event Selection}
\label{sec:EventSelection}

The event selection is done in two steps. First, the pre-selection reduces the samples to the relevant invariant mass region around the Higgs mass peak. Then, the final event selection using several kinematic variables in a boosted decision tree classifier is applied to separate signal and background events.

\subsection{Pre-Selection}
\label{sec:PreSelection}
For the analysis only events with at least two reconstructed muons are used. In addition to the \ac{PFO} selection cuts given in \cref{tab:PFOselection}, each reconstructed particle is required to have a transverse momentum of \unit[5]{GeV} or larger, which removes most of the particles from the \gghad background, see \cref{sec:BeamInducedBackground}. In case there are more than two muons reconstructed, the two most energetic muons are used. In this analysis, the most energetic muon is referred to as $\upmu_1$ and the second most energetic muon is referred to as $\upmu_2$. In addition, the invariant mass of the two muons, $M(\mumu)$, is required to be between \unit[105]{GeV} and \unit[135]{GeV}. This limits the background events used for the training of the classifier to the region relevant to the fit of the Higgs mass peak.

\begin{figure}[htb]
 \begin{subfigure}[]{0.49\textwidth}
  \includegraphics[width=\textwidth]{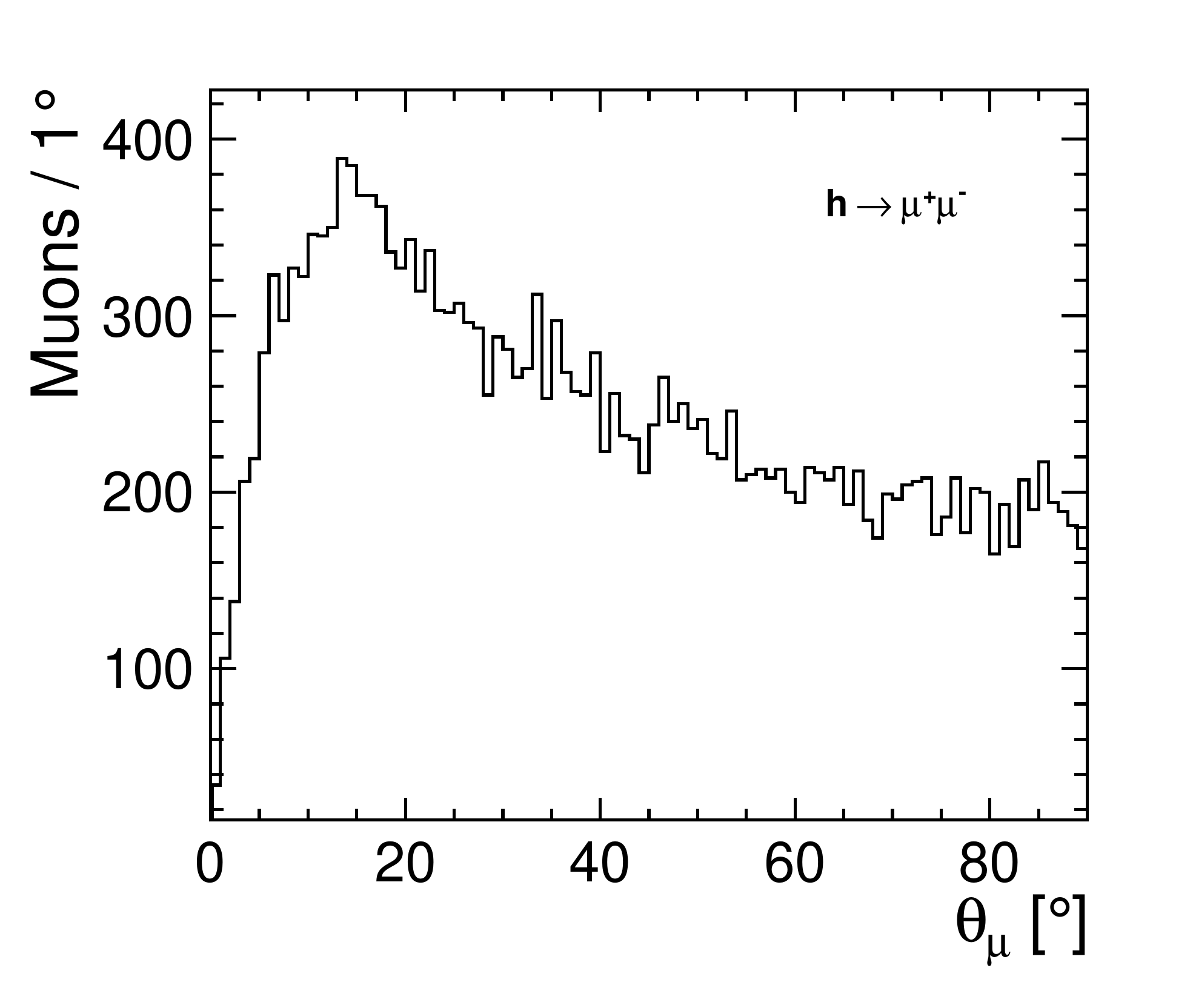}
 \end{subfigure}
\hfill
 \begin{subfigure}[]{0.49\textwidth}
    \includegraphics[width=\textwidth]{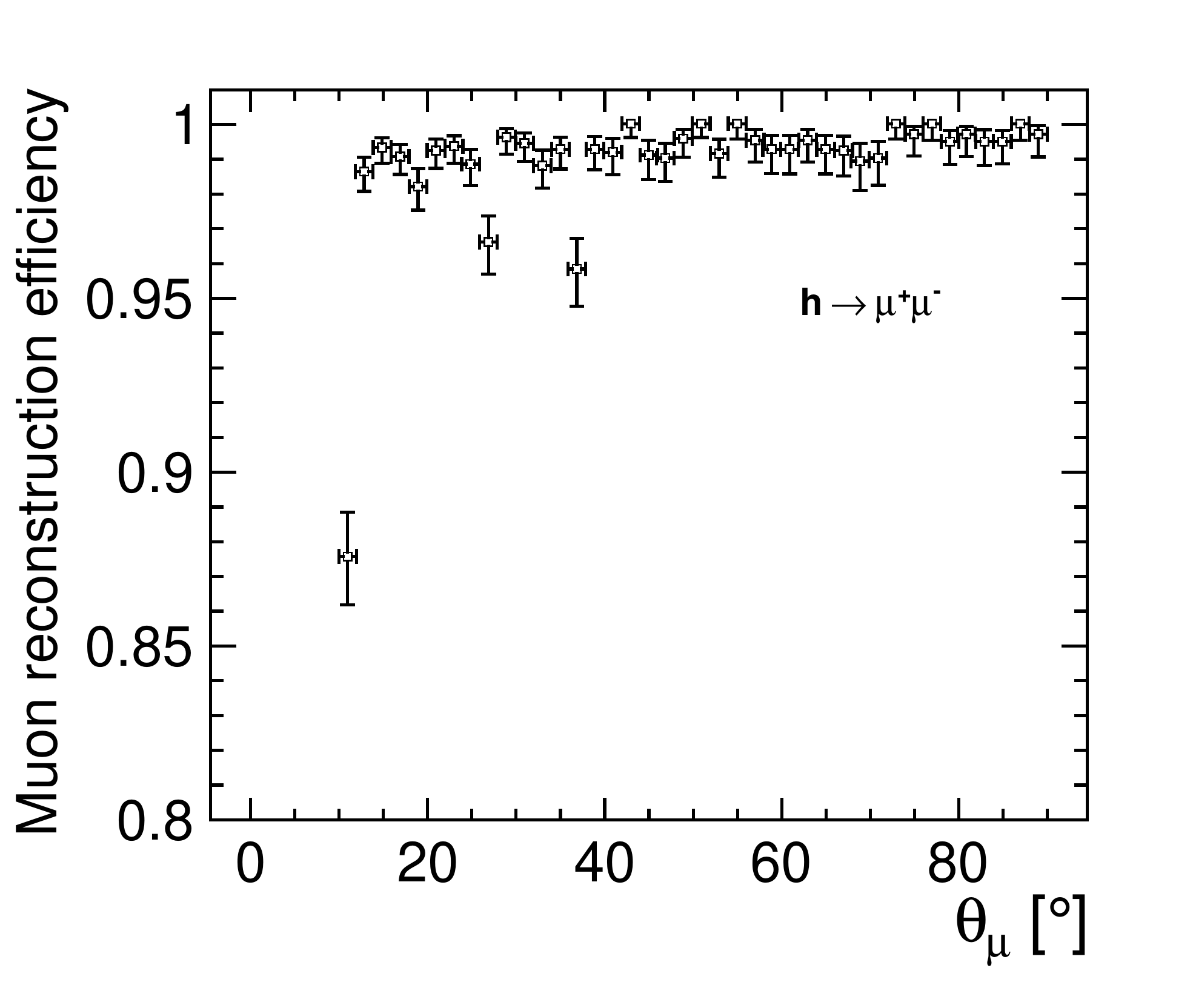}
 \end{subfigure}
\caption[Polar angle distribution and reconstruction efficiency of the muons in $\PH \to \mpmm$ events.]{Distribution of the true polar angles of the muons $\theta_{\upmu}$ (left) and the muon reconstruction efficiency depending on the true polar angle $\theta_{\upmu}$ (right) in $\PSh \to \mpmm$ events.}
\label{fig:muon_efficiency}
\end{figure}

The muon reconstruction efficiency is a combination of the track finding efficiency and the muon identification efficiency. As shown in \cref{fig:muon_efficiency}~(right) it is very high within the detector acceptance. There are two small regions with slightly lower efficiencies around 27\degrees and 37\degrees which are due to the barrel-endcap transition in the tracking system and the calorimeters, respectively. The average reconstruction efficiency for muons with a polar angle $\theta_{\upmu}$ larger than 10\degrees is 99.6\% in the $\PSh \to \mpmm$ sample. The efficiency drops quickly for polar angles below 10\degrees due to the tracker acceptance. The efficiency to reconstruct both muons is dominated by this acceptance and is 78.7\% for the $\PSh \to \mpmm$ sample, since a sizable amount of muons is produced at low polar angles, as shown in \cref{fig:muon_efficiency}~(left). The pre-selection cut on the invariant mass of the di-muon system reduces the selection efficiency further to 74.1\%. \cref{tab:preselection} summarizes the impact of the pre-selection cuts on all simulated samples.

\begin{table}[htb]
 \centering
 \caption[Selection efficiencies forthe pre-selection cuts.]{Selection efficiencies $\epsilon$ and number of selected events $N_{\mathrm{events}}$ in ${\cal L} = \unit[2]{\abinv}$ when applying the pre-selection cuts to the simulated samples.}
 \label{tab:preselection}

\begin{tabular}{l c c c c}
\toprule
& \multicolumn{2}{c}{Two muons reconstructed} & \multicolumn{2}{c}{$\unit[105]{GeV} < M(\mumu) < \unit[135]{GeV}$} \\
Process          & $\epsilon$ & $N_{\mathrm{events}}$ & $\epsilon$ & $N_{\mathrm{events}}$ \\\midrule
$\PSh \to \mpmm$ & 78.7\%     & 187.6                 & 74.1\%     & 176.6                 \\
$\mpmm\nunubar$  & 43.3\%     & 127700                &  1.0\%     & 2563                  \\
$\mpmm\epem$     & 81.4\%     & 564400                & 57.1\%     & 395600                \\
$\mpmm$          & 27.7\%     & 6638                  & 16.6\%     & 3974                  \\
$\tptm$          &  0.7\%     & 3368                  &  0.02\%     &  105.9               \\
$\tptm\nunubar$  &  1.4\%     & 3456                  &  0.007\%     &   17.79             \\\bottomrule
\end{tabular}

\end{table}

\subsection{\acl{BDT}}
The event selection is done using the boosted decision tree classifier implemented in \tmva~\cite{Hoecker2007}. The goal of the classification is to reject the $\epem \to \mpmm\epem$ background, while the $\epem \to \mpmm\nunubar$ background is indistinguishable from the signal except for the di-muon invariant mass distribution. The $\epem \to \mpmm$, $\epem \to \tptm$ and $\epem \to \tptm \nunubar$ samples are not used in the training and are only used to verify that they are removed by the final event selection. Thus, the \ac{BDT} training is performed using only events from the $\PSh \to \mpmm$ and $\epem \to \mpmm\epem$ samples. 10000 $\PSh \to \mpmm$ events and 500000 $\epem \to \mpmm\epem$ events were used for the training, which leaves sufficient statistics for the actual analysis.

The variables used for the event selection by the \ac{BDT} are chosen to distinguish resonant di-muon production as in case of the Higgs decaying into muons and radiative processes which are dominating in the $\mpmm\epem$ background.
The following variables are being used:
\begin{itemize}
 \item The visible energy excluding the two reconstructed muons, $E_{\mathrm{vis}}$.
 \item The scalar sum of the transverse momenta of the two muons, $\pT(\upmu_1) + \pT(\upmu_2)$.
 \item The helicity angle $\cos\theta^*(\mumu) = \frac{\vec{p}'(\upmu_1) \cdot \vec{p}(\mumu)}{|\vec{p}'(\upmu_1)| \cdot |\vec{p}(\mumu)|}$, where $\vec{p}'$ is the momentum in the rest frame of the di-muon system. Since the two muons are back-to-back in the rest frame of the di-muon system, there is no additional information to be gained from calculating a similar angle for $\upmu_2$.
 \item The velocity of the di-muon system, $\upbeta(\mumu)$, where $\upbeta = \frac{v}{c}$.
 \item The transverse momentum of the di-muon system, $\pT(\mumu)$.
 \item The polar angle of the di-muon system, $\theta(\mumu)$.
\end{itemize}

\begin{figure}
 \begin{subfigure}[]{0.49\textwidth}
  \includegraphics[width=\textwidth]{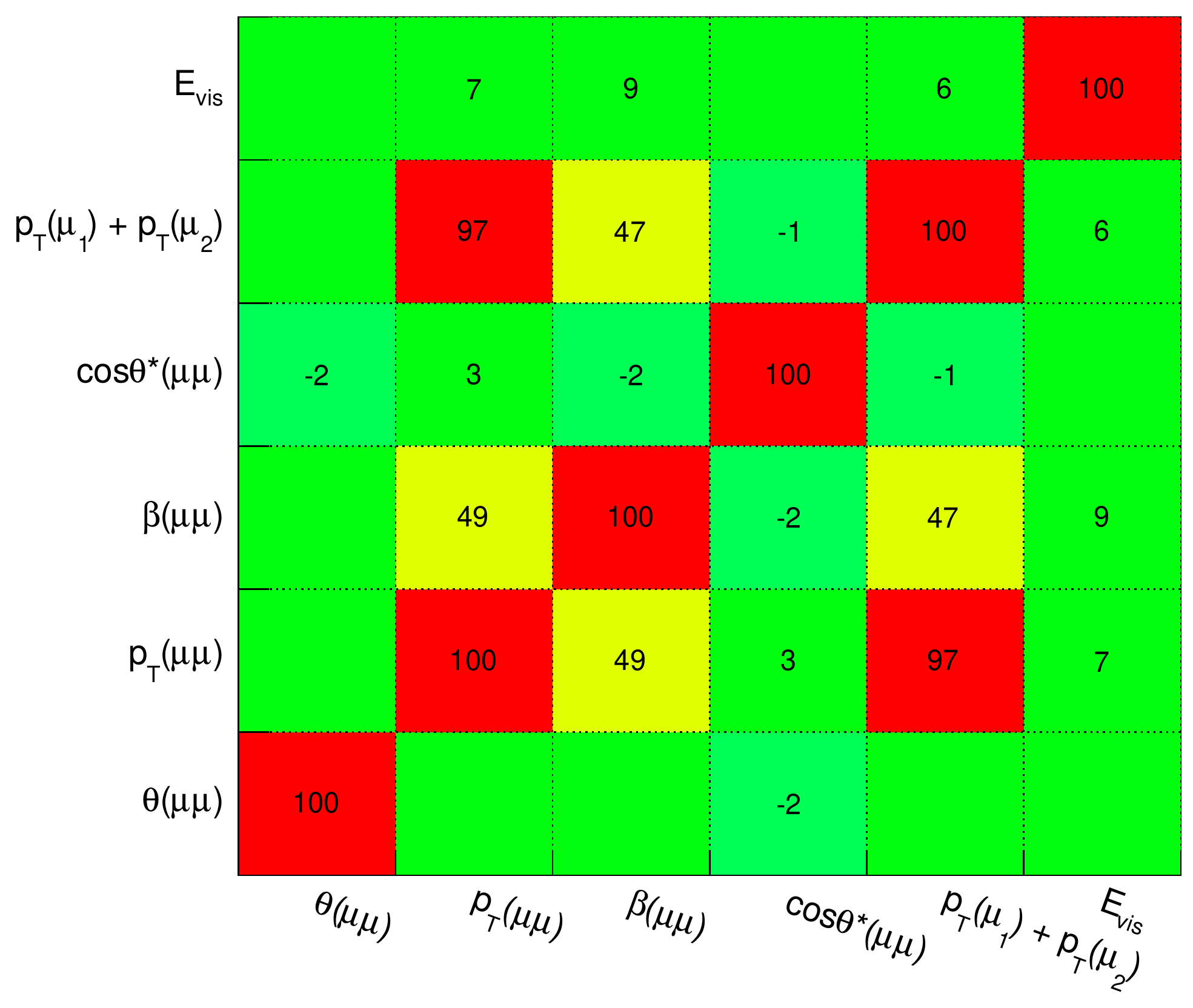}
 \end{subfigure}
\hfill
 \begin{subfigure}[]{0.49\textwidth}
    \includegraphics[width=\textwidth]{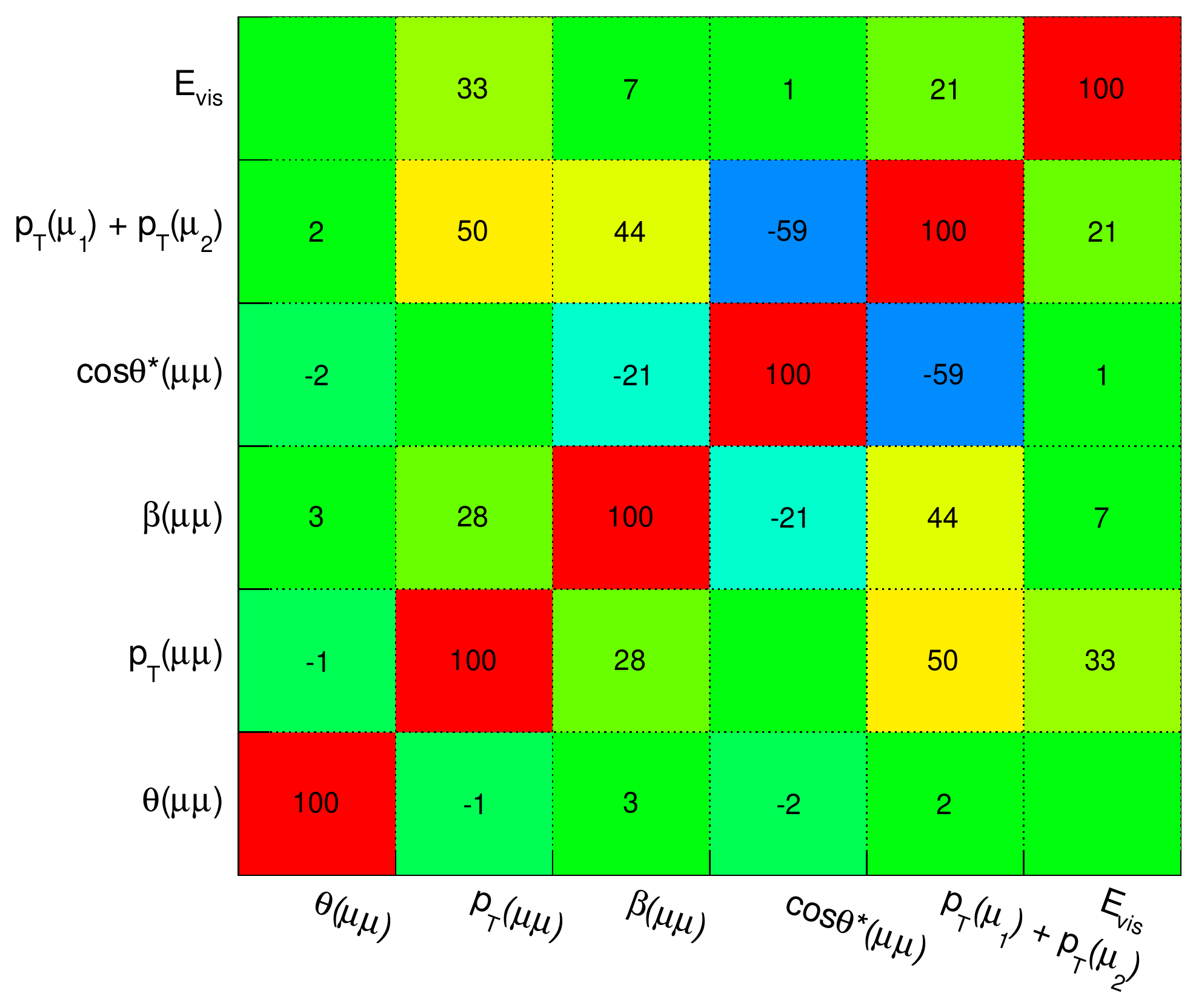}
 \end{subfigure}
\caption[Correlation matrices for the kinematic variables used in the boosted decision tree classifier.]{Correlation matrices for the kinematic variables used in the boosted decision tree classifier for $\PSh \to \mpmm$ events (left) and $\mpmm\epem$ events (right).}
\label{fig:variable_correlations}
\end{figure}

The most powerful variable is the visible energy whenever there is an electron within the detector acceptance. Otherwise the background can be rejected by the transverse momentum of the di-muon system or the sum of the two individual transverse momenta. According to the correlation matrices shown in \cref{fig:variable_correlations} the two latter variables are highly correlated. The level of correlation is different for signal and background samples, which means that both variables are useful for the background rejection.

\begin{figure}
 \begin{subfigure}[]{0.49\textwidth}
  \includegraphics[width=\textwidth]{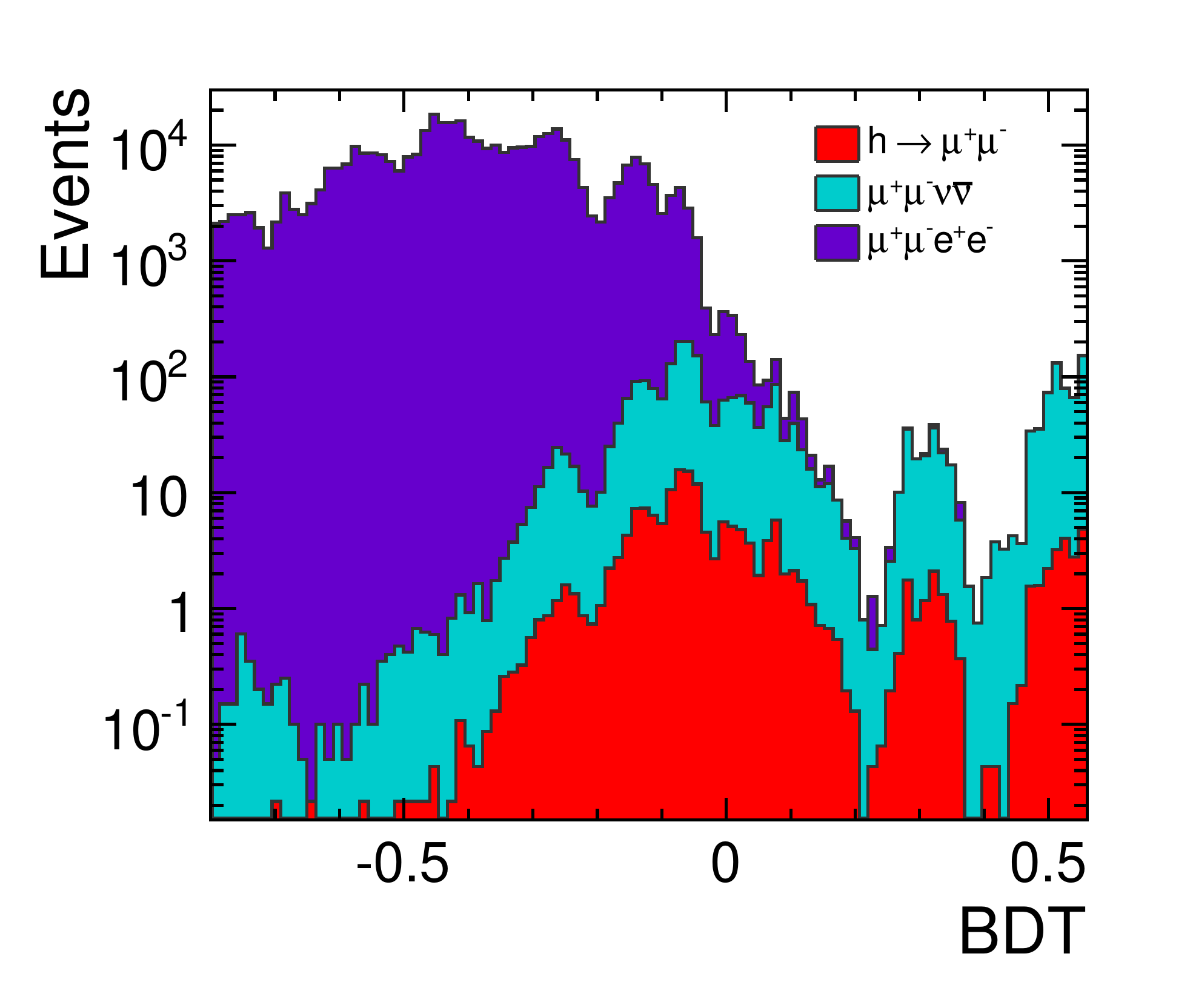}
 \end{subfigure}
\hfill
 \begin{subfigure}[]{0.49\textwidth}
    \includegraphics[width=\textwidth]{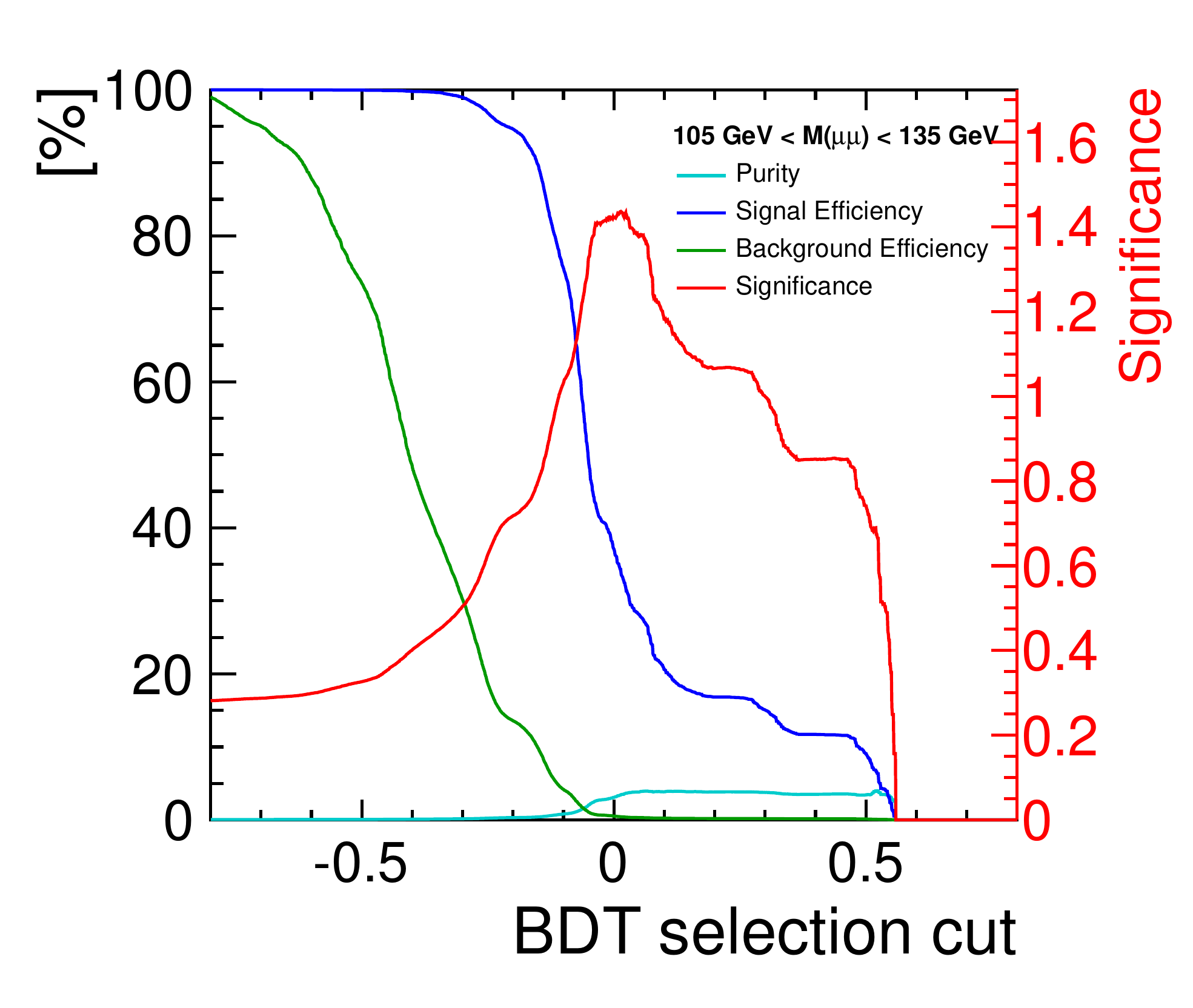}
 \end{subfigure}
\caption[Response of the boosted decision tree classifier and resulting significance, purity, signal efficiency and background efficiency]{Response of the boosted decision tree classifier for the signal and the two most important background processes (left) and the resulting significance, purity, signal efficiency and background efficiency (right). See text for details.}
\label{fig:variable_bdt}
\end{figure}

\cref{fig:variable_bdt} shows the response of the \ac{BDT} classifier for the different samples. It also shows the dependence of the signal selection efficiency $N_{\mathrm S}/N^{\mathrm{total}}_{\mathrm S}$, signal purity $N_{\mathrm S}/(N_{\mathrm S}+N_{\mathrm B})$, background efficiency $N_{\mathrm B}/N^{\mathrm{total}}_{\mathrm B}$ and significance\footnote{It should be noted that this is the significance over the arbitrarily chosen range of $\unit[105]{GeV} < M(\mumu) < \unit[135]{GeV}$ and should not be confused with the significance of the signal peak.} $N_{\mathrm S}/\sqrt{N_{\mathrm S} + N_{\mathrm B}}$ on the \ac{BDT} cut value, where $N_{\mathrm S}$ and $N_{\mathrm B}$ are the number of selected signal and selected background events, respectively. $N^{\mathrm{total}}_{\mathrm S}$ and $N^{\mathrm{total}}_{\mathrm B}$ are the number of events before the event selection for signal and background samples, respectively.

\begin{figure}
 \begin{subfigure}[]{0.49\textwidth}
  \includegraphics[width=\textwidth]{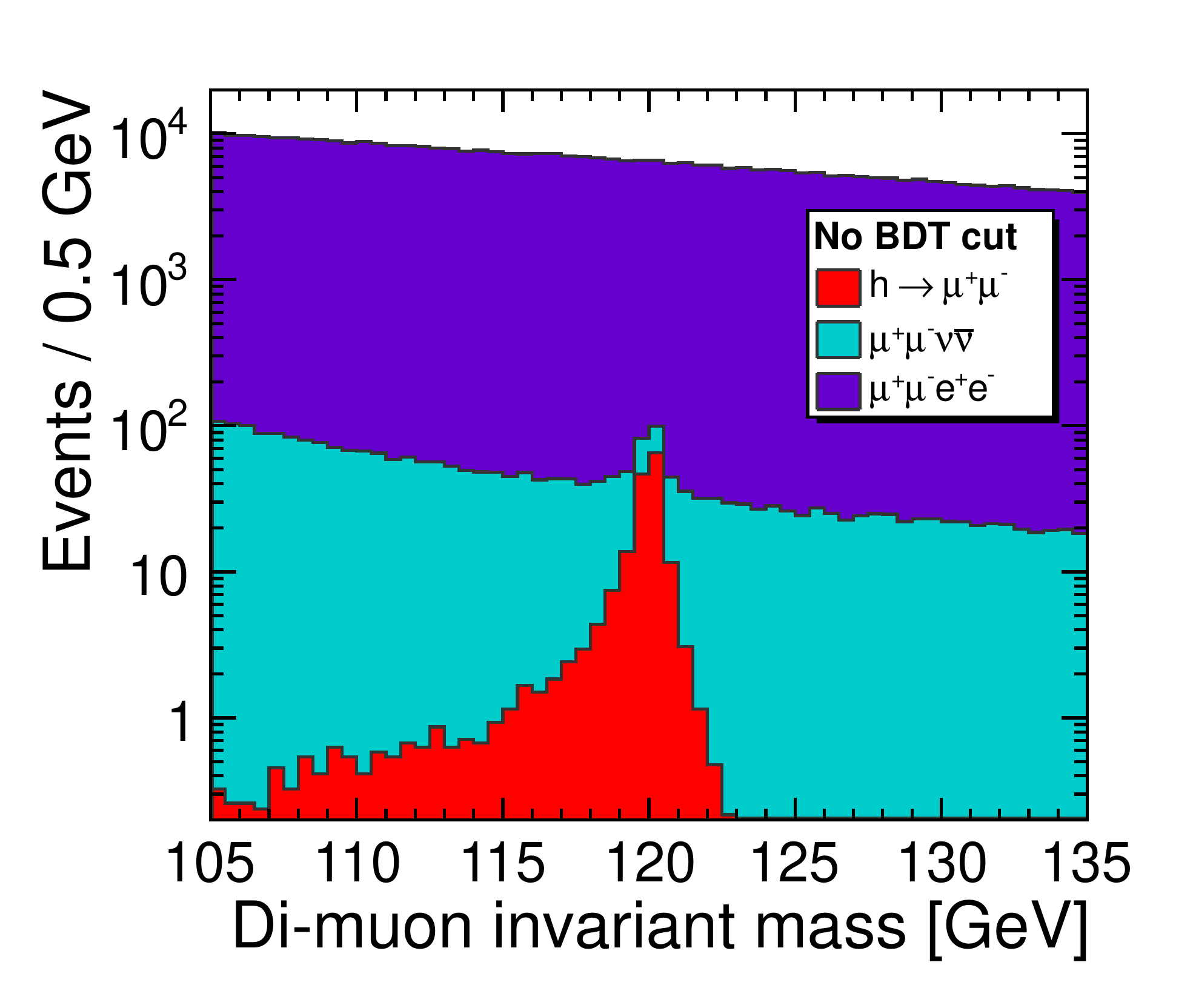}
 \end{subfigure}
\hfill
 \begin{subfigure}[]{0.49\textwidth}
    \includegraphics[width=\textwidth]{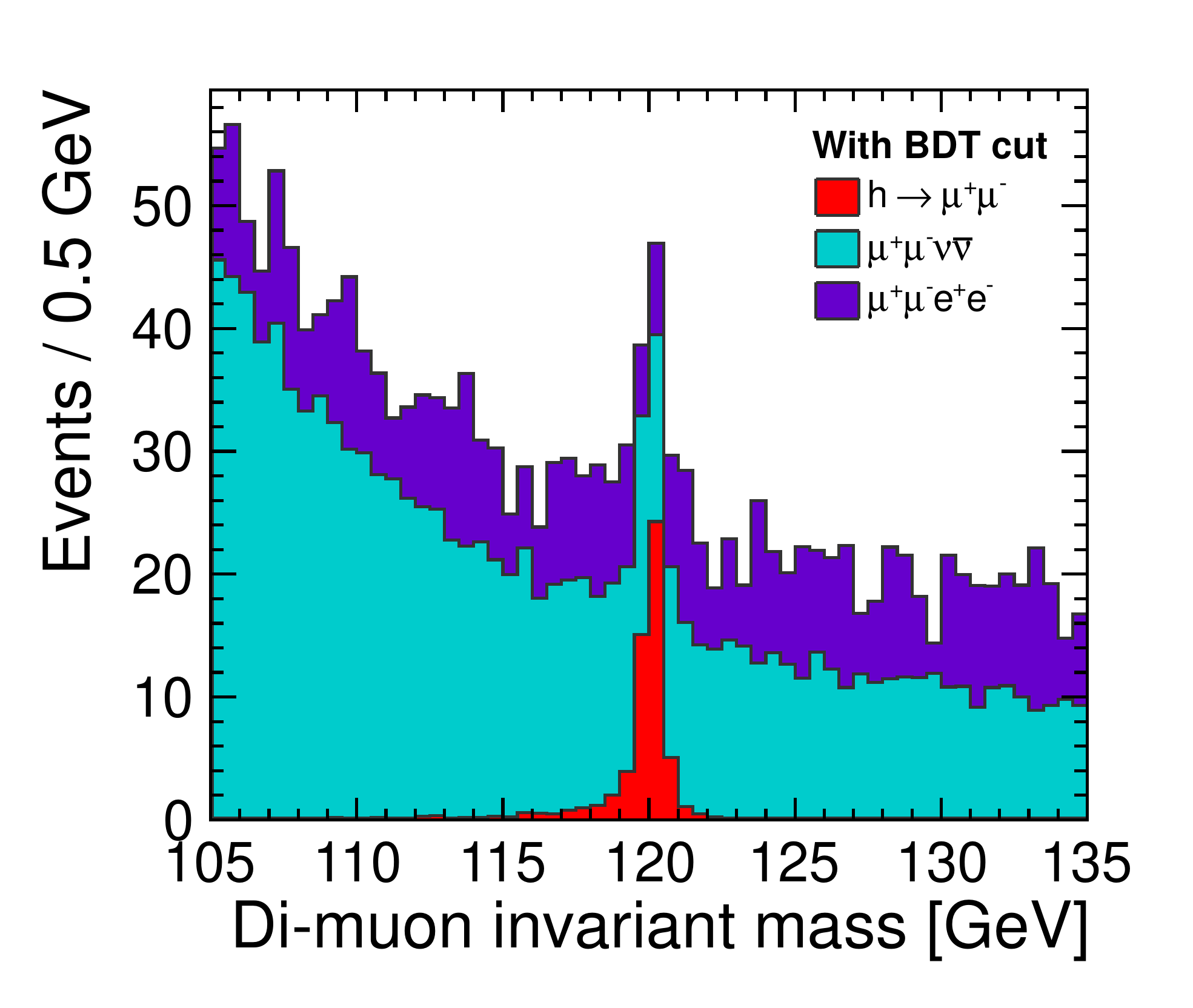}
 \end{subfigure}
\caption[Distribution of the invariant mass of the di-muon system before and after the event selection.]{Distribution of the invariant mass of the di-muon system for the signal and the two main background channels before the event selection (left) and after the event selection (right). Histograms are stacked and normalized to ${\cal L} = \unit[2]{\abinv}$.}
\label{fig:variable_mass}
\end{figure}

\begin{figure}
 \begin{subfigure}[]{0.49\textwidth}
  \includegraphics[width=\textwidth]{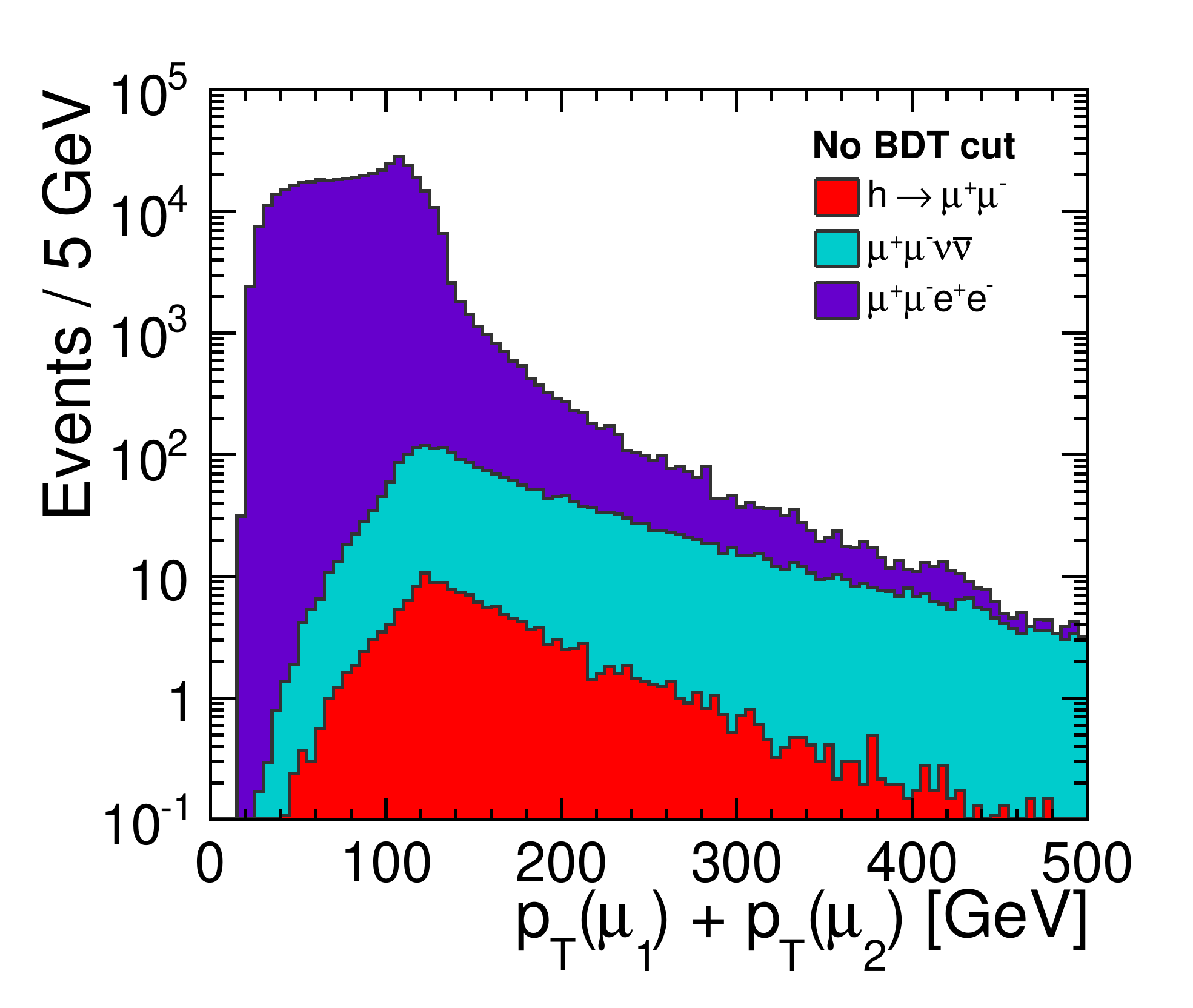}
 \end{subfigure}
\hfill
 \begin{subfigure}[]{0.49\textwidth}
    \includegraphics[width=\textwidth]{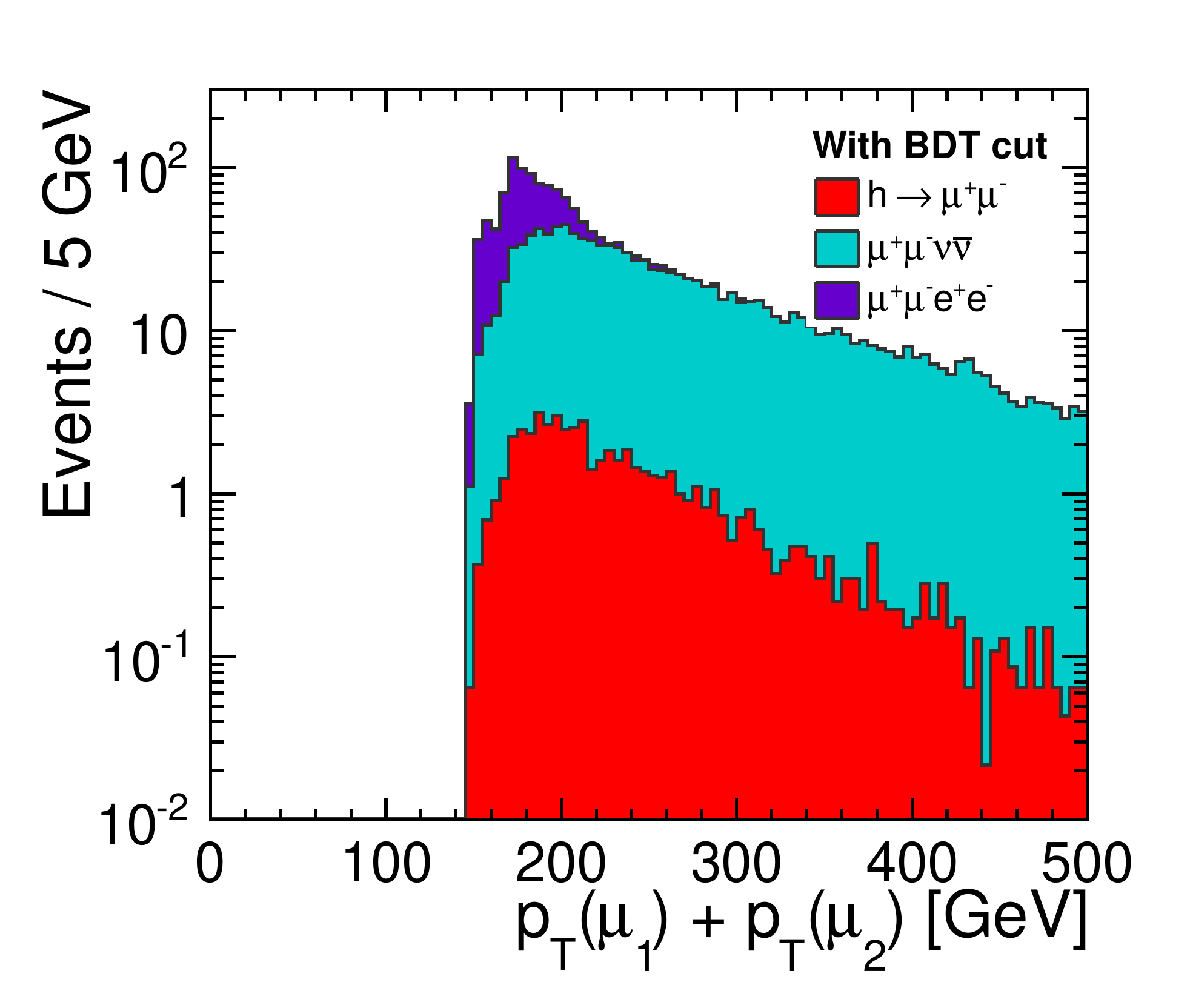}
 \end{subfigure}

 \begin{subfigure}[]{0.49\textwidth}
  \includegraphics[width=\textwidth]{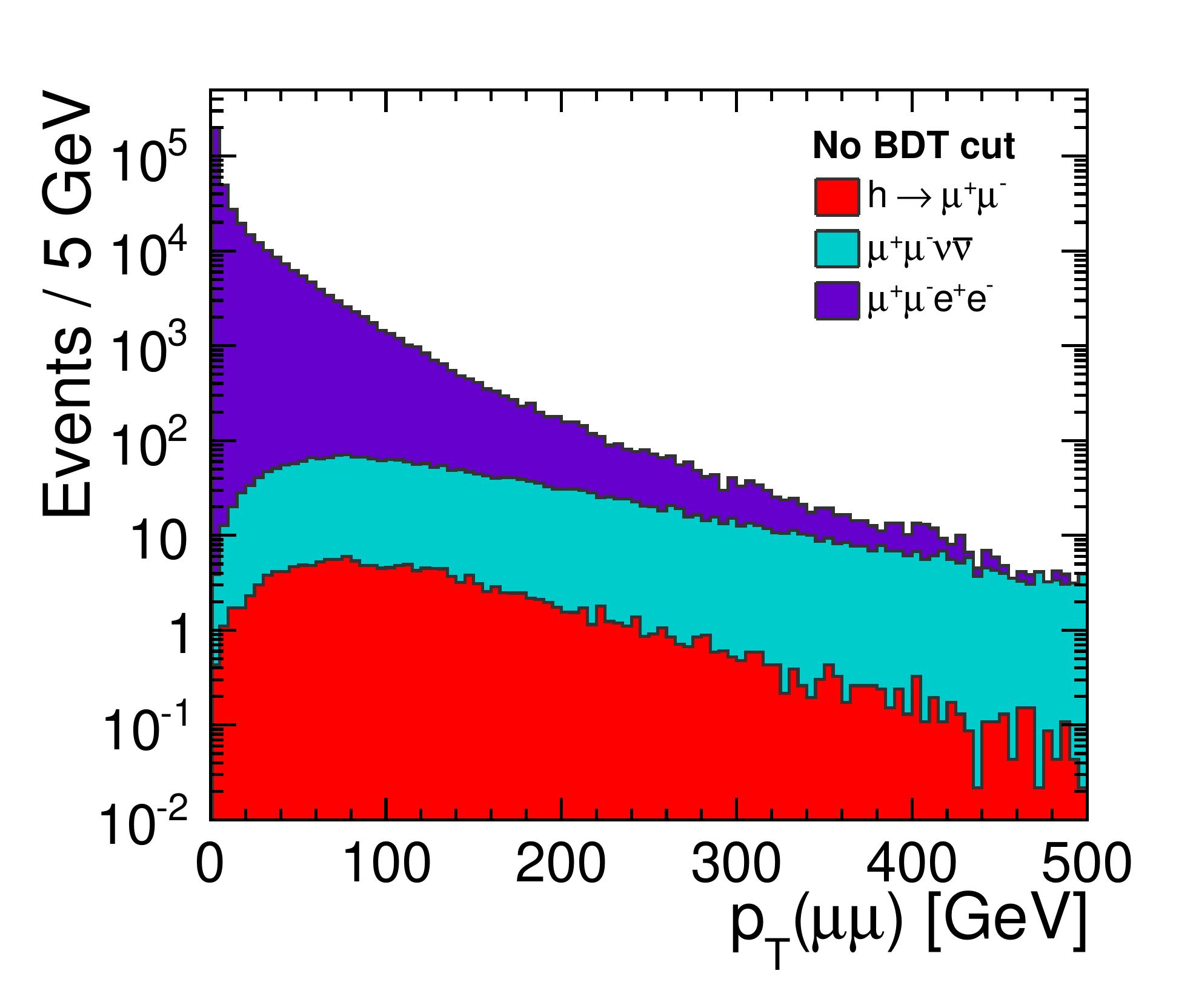}
 \end{subfigure}
\hfill
 \begin{subfigure}[]{0.49\textwidth}
    \includegraphics[width=\textwidth]{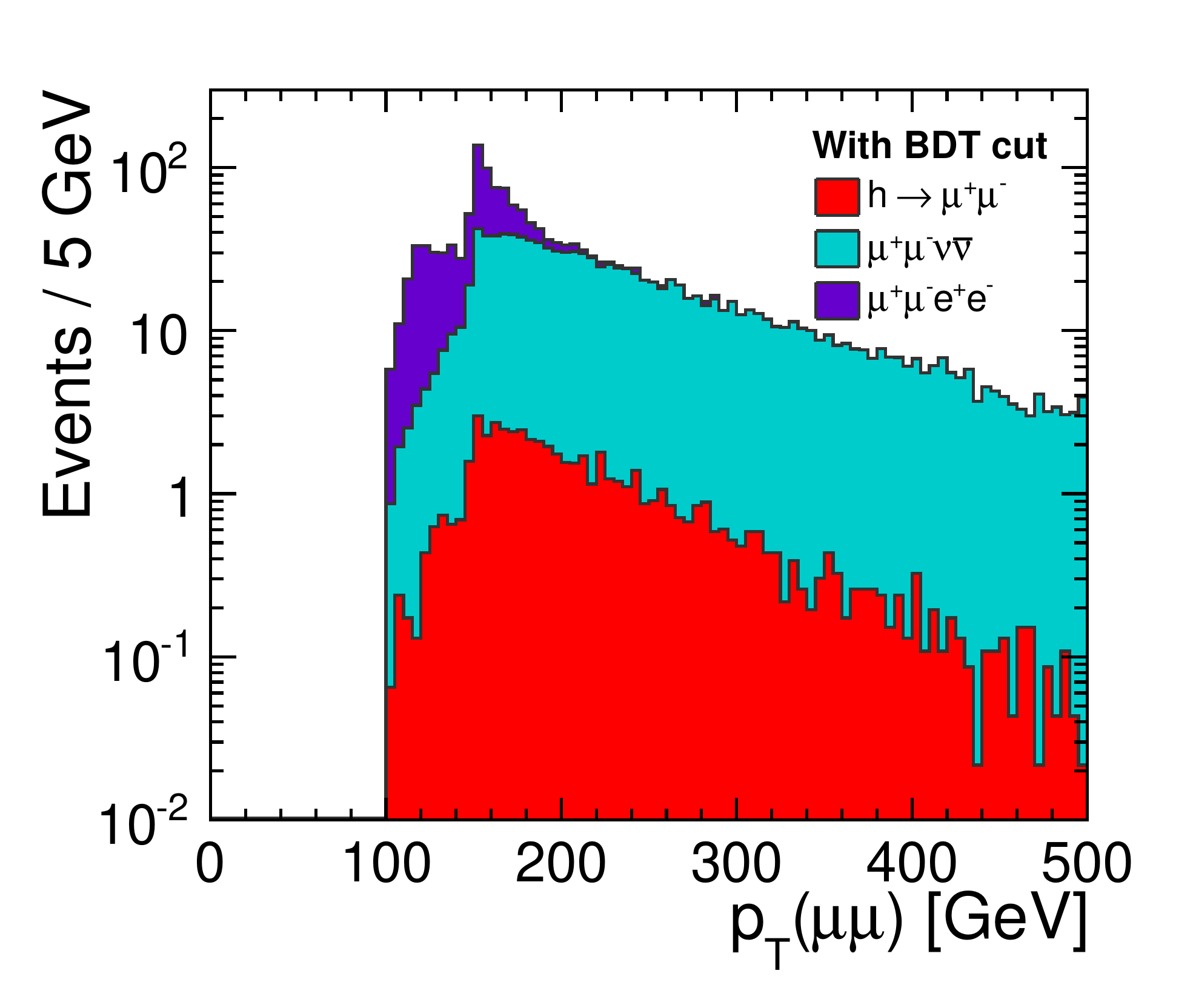}
 \end{subfigure}

 \begin{subfigure}[]{0.49\textwidth}
  \includegraphics[width=\textwidth]{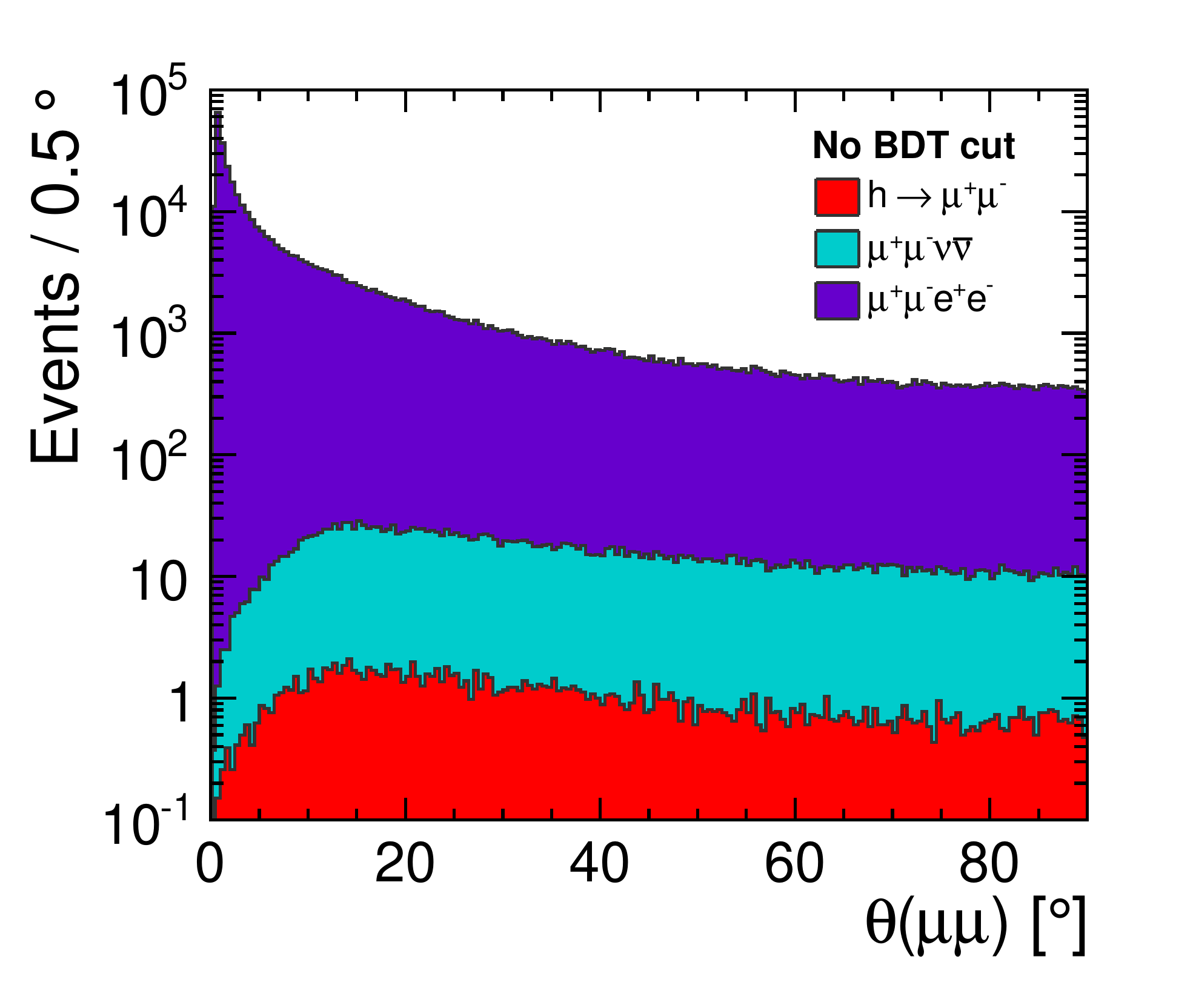}
 \end{subfigure}
\hfill
 \begin{subfigure}[]{0.49\textwidth}
    \includegraphics[width=\textwidth]{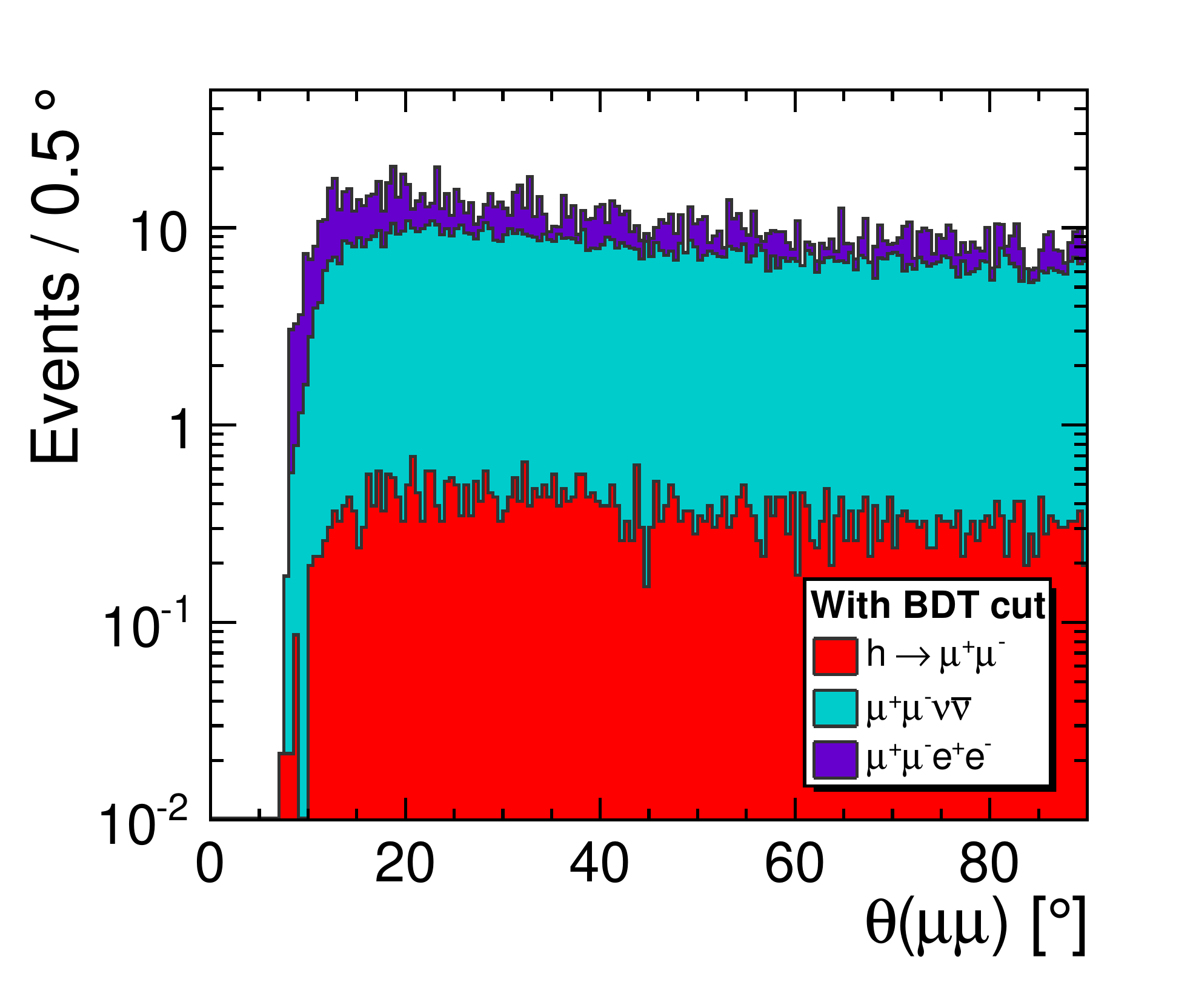}
 \end{subfigure}
\caption[Distributions of kinematic variables before and after the event selection.]{Distribution of the sum of the transverse momenta of the two muons $\pT(\upmu_1)+\pT(\upmu_2)$, the transverse momentum of the di-muon system $\pT(\mumu)$ and the polar angle of the di-muon system $\theta(\mumu)$ for the signal and the two main background channels before the event selection (left) and after the event selection (right). Histograms are stacked and normalized to ${\cal L} = \unit[2]{\abinv}$.}
\label{fig:kinematic_variables_1}
\end{figure}

\begin{figure}
 \begin{subfigure}[]{0.49\textwidth}
  \includegraphics[width=\textwidth]{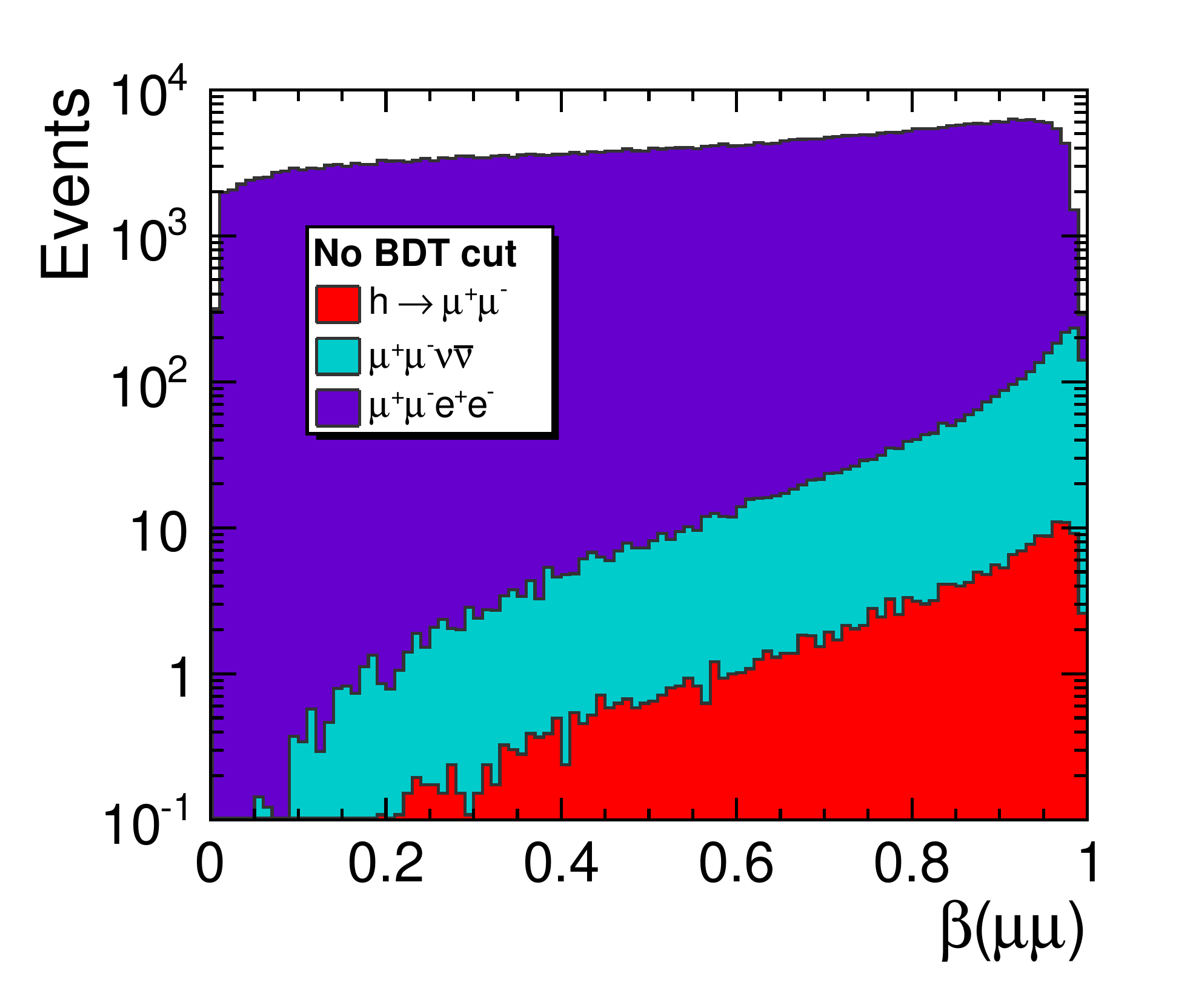}
 \end{subfigure}
\hfill
 \begin{subfigure}[]{0.49\textwidth}
    \includegraphics[width=\textwidth]{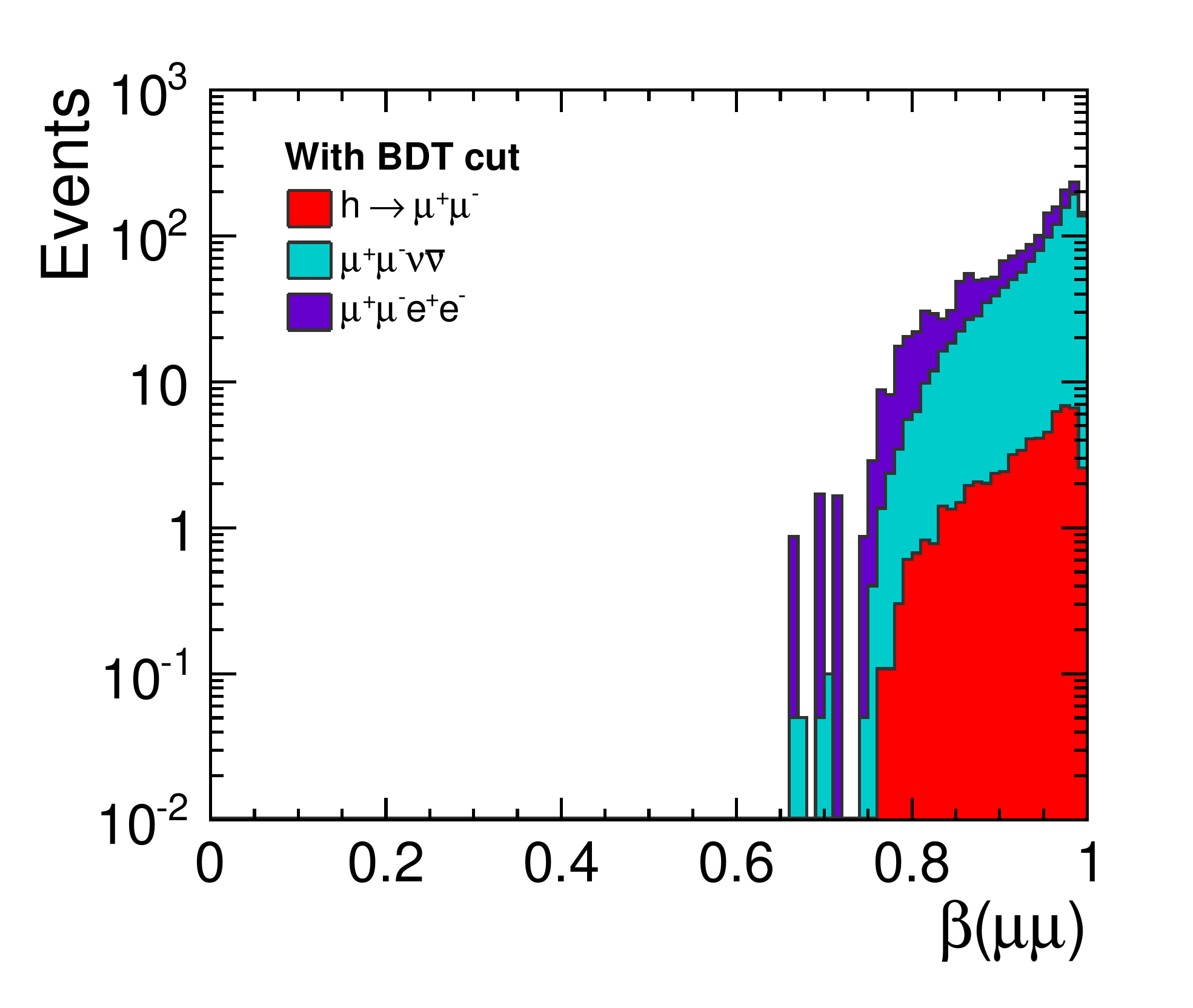}
 \end{subfigure}

 \begin{subfigure}[]{0.49\textwidth}
  \includegraphics[width=\textwidth]{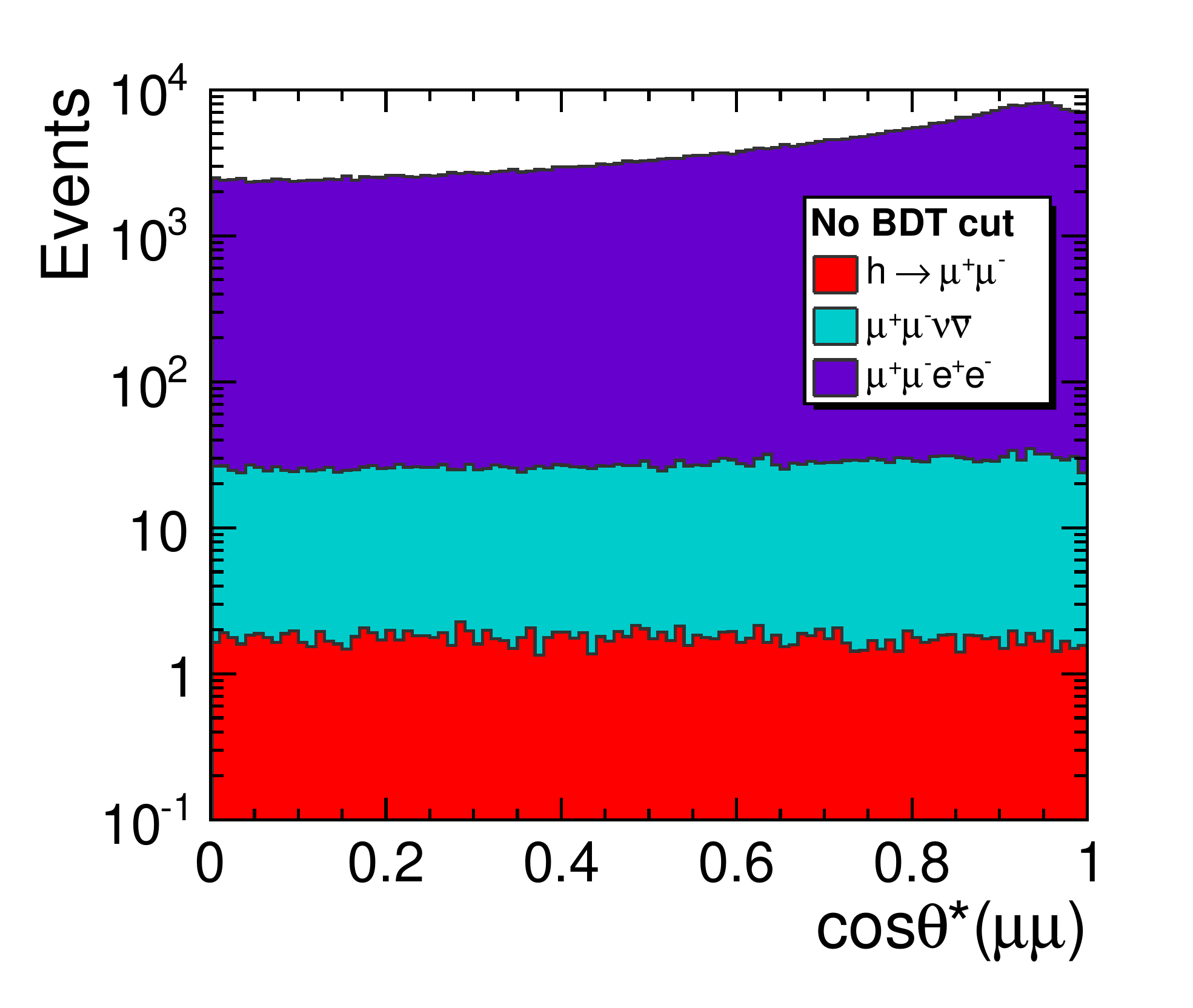}
 \end{subfigure}
\hfill
 \begin{subfigure}[]{0.49\textwidth}
    \includegraphics[width=\textwidth]{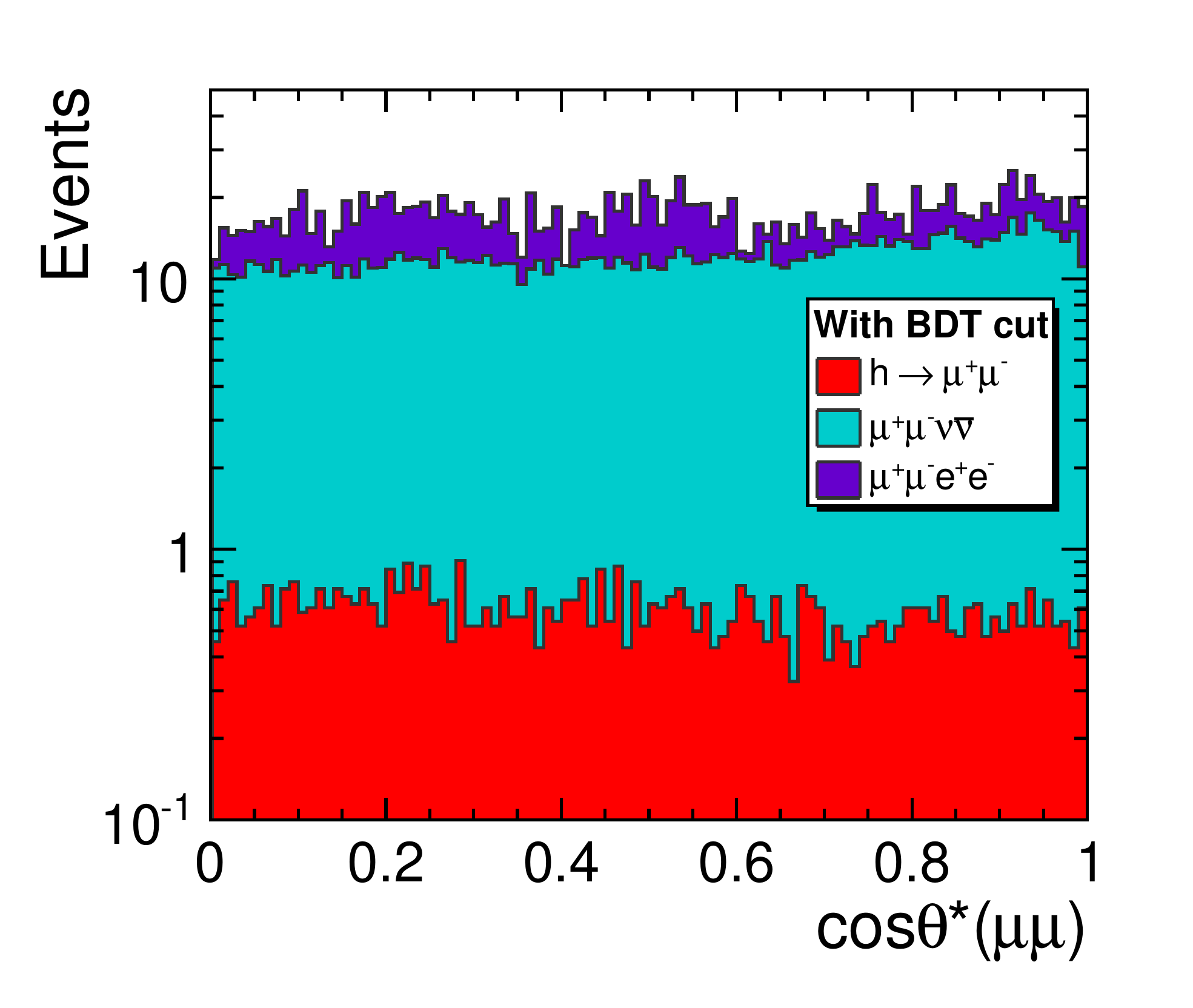}
 \end{subfigure}

 \begin{subfigure}[]{0.49\textwidth}
  \includegraphics[width=\textwidth]{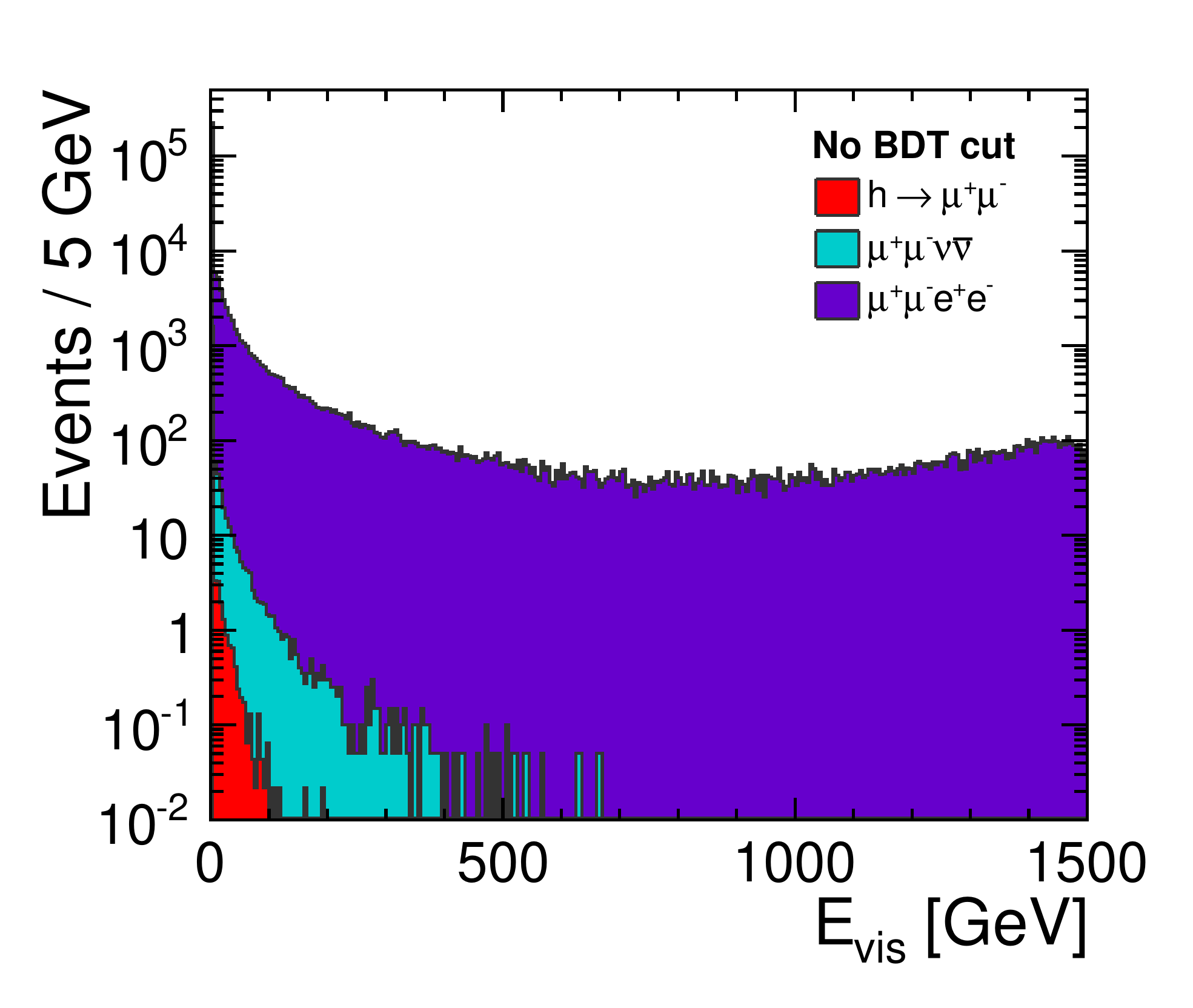}
 \end{subfigure}
\hfill
 \begin{subfigure}[]{0.49\textwidth}
    \includegraphics[width=\textwidth]{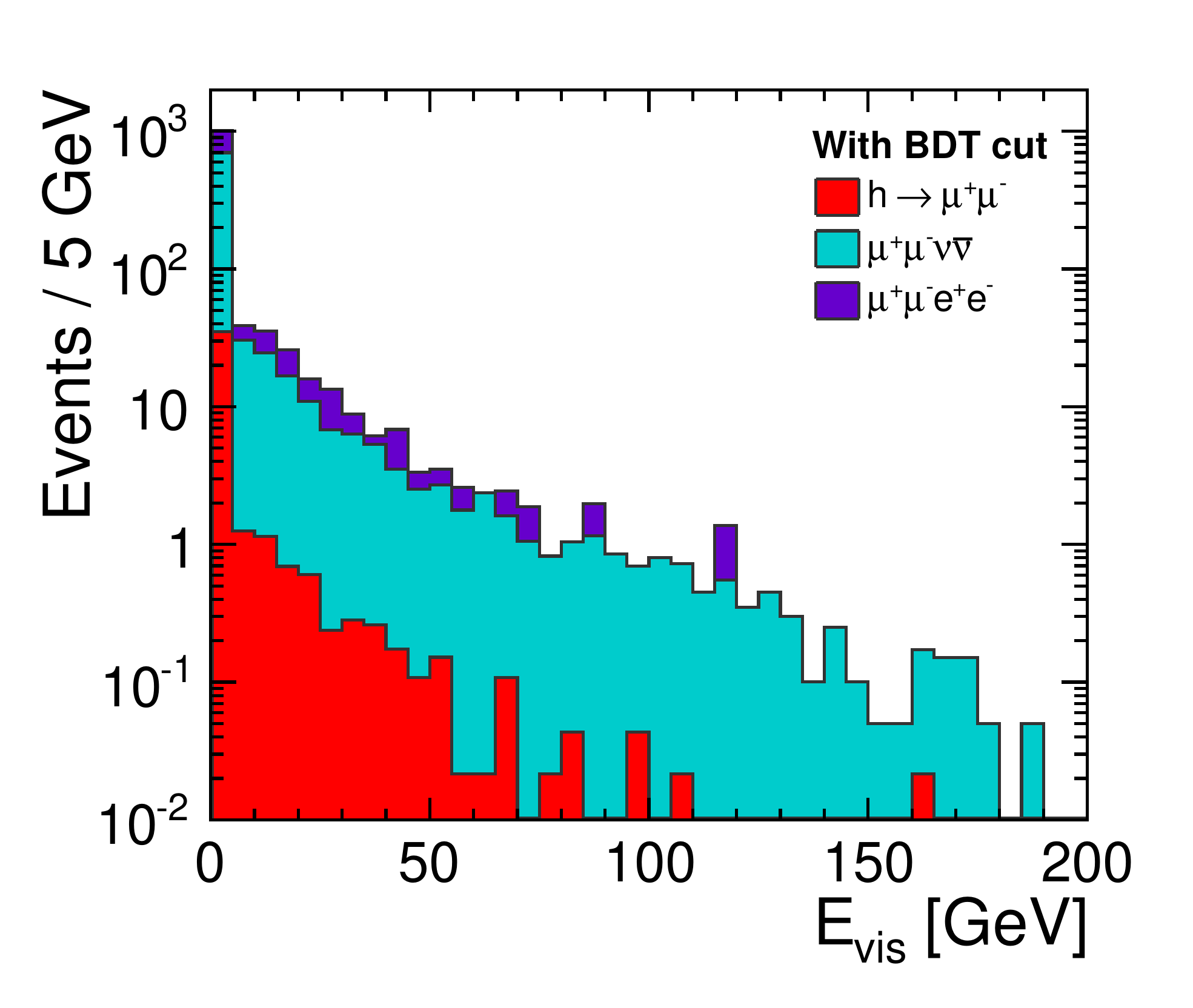}
 \end{subfigure}
\caption[Distributions of kinematic variables before and after the event selection.]{Distribution of the velocity $\beta$ of the di-muon system, the helicity angle $\cos\theta^\ast$ and the visible energy without the two muons $E_{\mathrm{vis}}$ for the signal and the two main background channels before the event selection (left) and after the event selection (right). Histograms are stacked and normalized to ${\cal L} = \unit[2]{\abinv}$.}
\label{fig:kinematic_variables_2}
\end{figure}

The final \ac{BDT} selection cut is chosen such that the significance is maximized, which is usually at a \ac{BDT} value of around 0. The distribution of all input variables before and after the BDT selection are shown in \cref{fig:kinematic_variables_1} and \cref{fig:kinematic_variables_2}. The distribution of the invariant mass of the di-muon system before and after the event selection is shown in \cref{fig:variable_mass}. The Higgs mass peak is clearly visible after the event selection.

\subsection{Additional Background Samples}
\label{sec:SelectionAdditionalBackground}

Although not used in the \ac{BDT} training, the background selection efficiency was checked for all background samples.

Only 4 events of the $\epem \to \tptm$ sample and 3 events of the $\epem \to \tptm\nunubar$ sample pass the final event selection. Assuming a total integrated luminosity of \unit[2]{\abinv} this corresponds to approximately 20 selected events from the $\epem \to \tptm$ channel and 8 selected events from the $\epem \to \tptm\nunubar$ channel with an invariant di-muon mass between \unit[105]{GeV} and \unit[135]{GeV}. It is reasonable to assume that their distribution is flat within that invariant mass region, which implies that their influence on the fit of the Higgs peak is negligible.

None of the $\epem \to \mpmm$ background events passes the final event selection. In fact, these events can be effectively rejected just by removing events with $\pT(\mumu) < \unit[25]{GeV}$ which is well below the selection cut introduced by the \ac{BDT}, as shown in \cref{fig:kinematic_variables_1}.

\begin{figure}
\centering
 \begin{subfigure}[]{0.49\textwidth}
    \includegraphics[width=\textwidth]{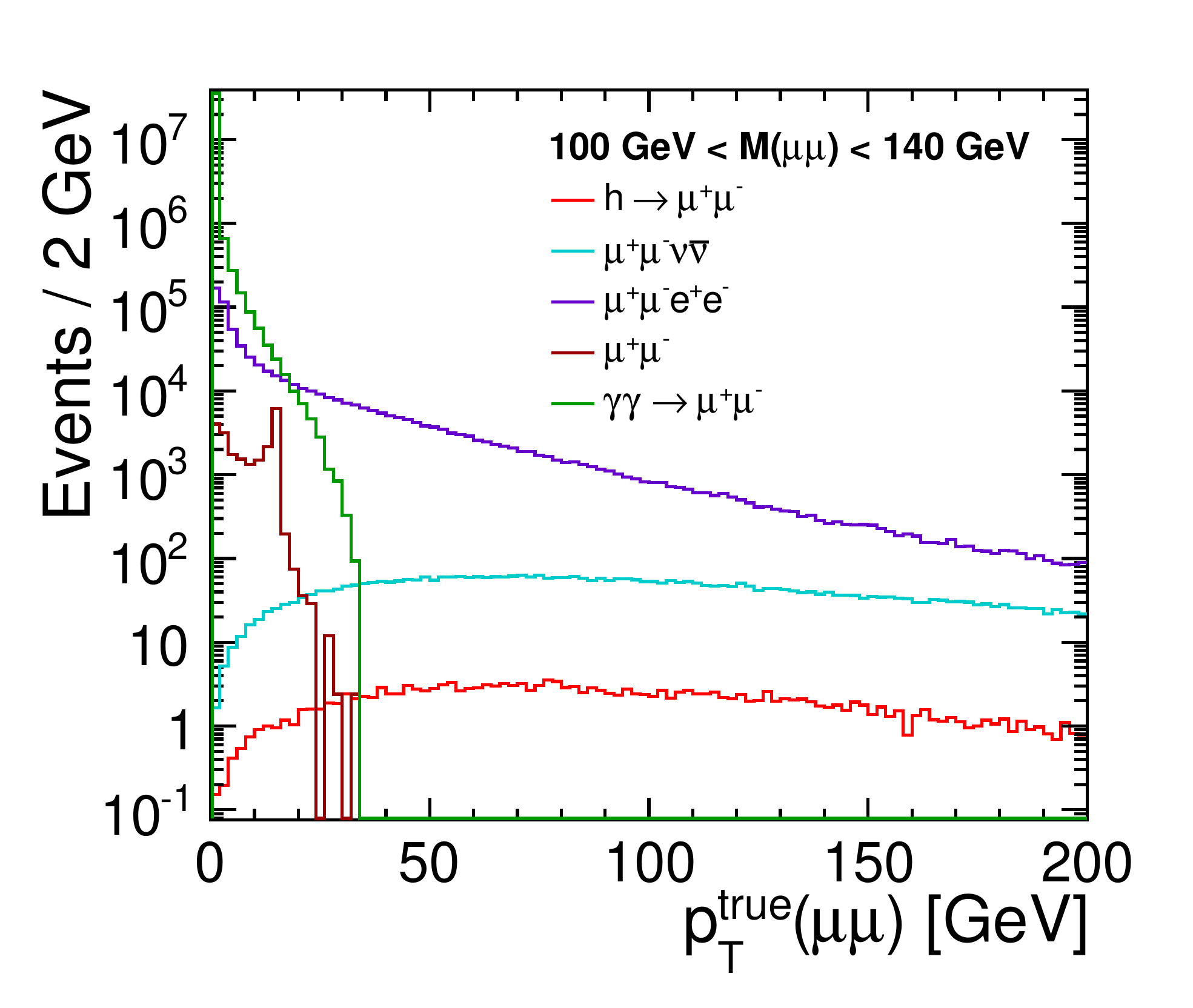}
 \end{subfigure}
\caption[Distribution of the true transverse momentum of the di-muon system.]{Distribution of the true transverse momentum of the di-muon system for all channels normalized to ${\cal L} = \unit[2]{\abinv}$. No detector acceptance effects are taken into account.}
\label{fig:additional_samples}
\end{figure}

The $\gamgam \to \mpmm$ sample was studied at generator level. \cref{fig:additional_samples} shows the $\pT(\mumu)$ distribution of all samples at generator level and---despite the large event weights and low statistics in the tail---it can be seen that this variable can be used to effectively reject the $\gamgam \to \mpmm$ background, well below the selection cut introduced by the \ac{BDT} (see \cref{fig:kinematic_variables_1}).

\section{Invariant Mass Fit}
\label{sec:MassFit}
The number of signal events is obtained by fitting the expected shape of signal and background to the invariant mass distribution. This is a viable procedure since the Higgs mass can be obtained from other measurements with much higher precision. All the fits that are described in the following section are unbinned likelihood fits performed with the \roofit package~\cite{RooFit:2003}.

\subsection{Invariant Mass Shapes}
\label{sec:Shapes}
In order to fit the invariant mass, the shapes of the invariant mass distributions for all channels are fitted individually first, using the full statistics available. The resulting \acp{PDF} then serve as templates to fit the simulated data. They also serve to generate more, statistically independent data sets for the background channels to perform multiple toy Monte~Carlo experiments. The normalization is irrelevant for determining the shape and is thus not part of the PDFs.

\begin{figure}
\begin{subfigure}[]{0.32\textwidth}
 \includegraphics[width=\textwidth]{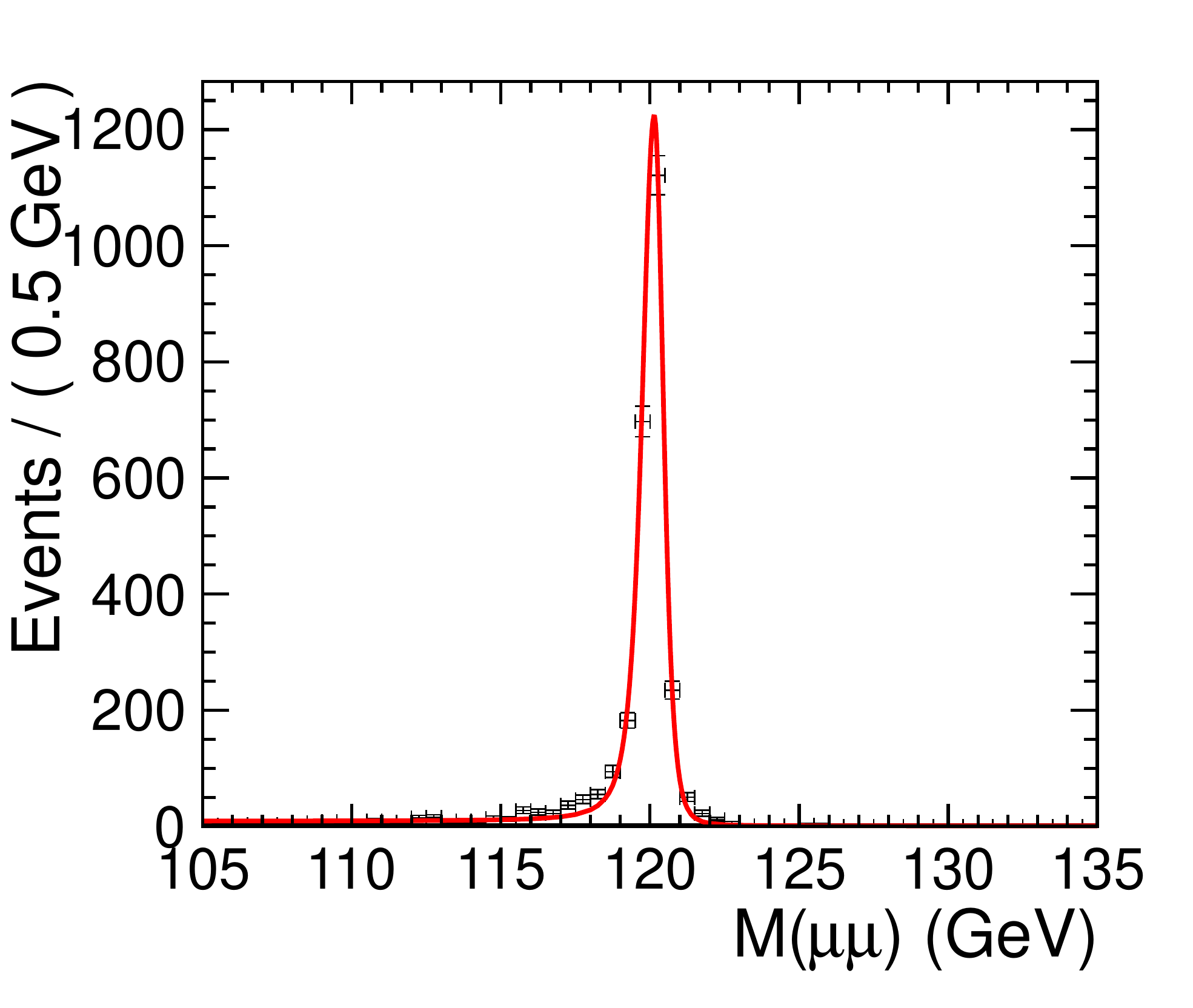}
\end{subfigure}
\hfill
\begin{subfigure}[]{0.32\textwidth}
 \includegraphics[width=\textwidth]{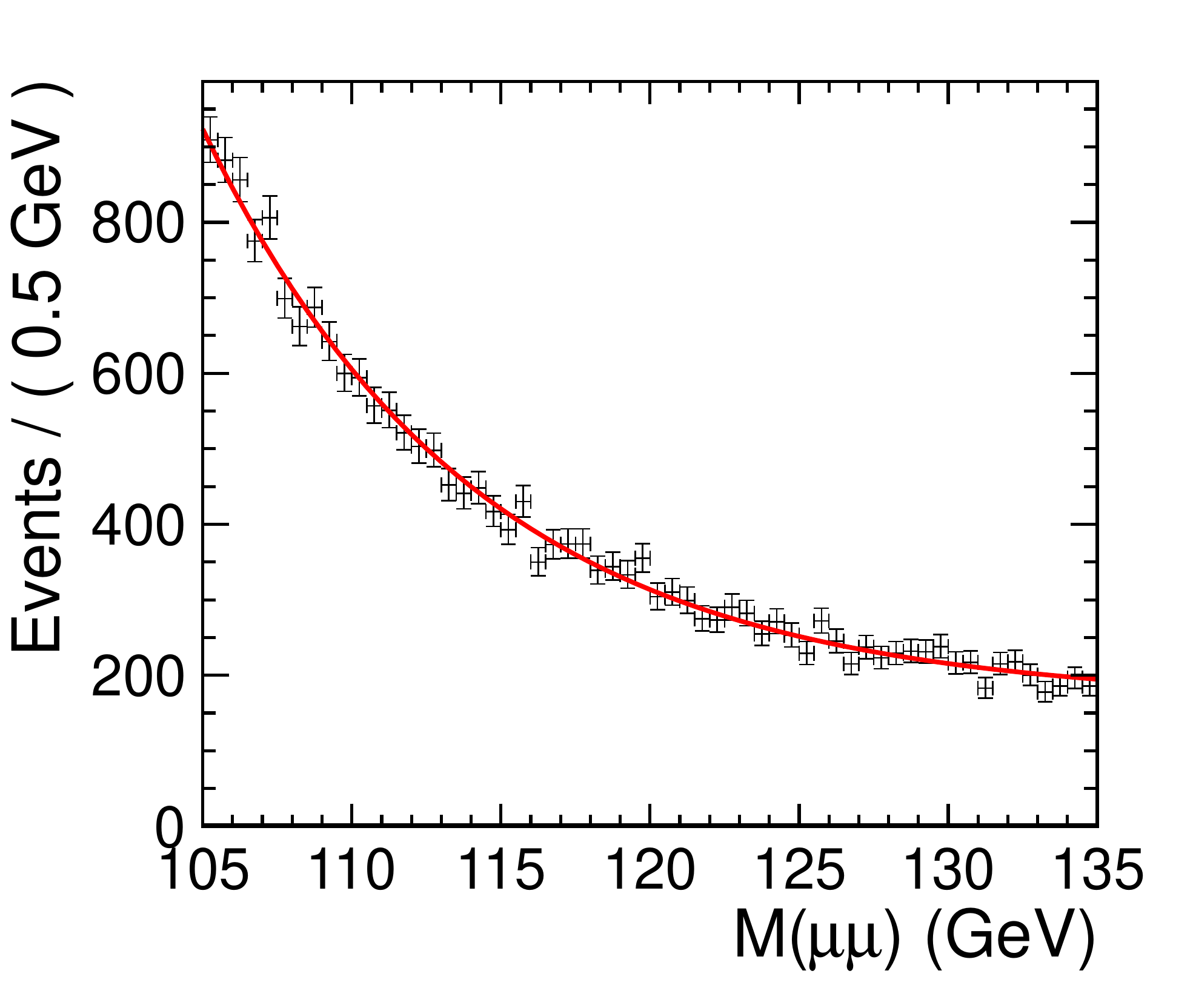}
\end{subfigure}
\hfill
\begin{subfigure}[]{0.32\textwidth}
 \includegraphics[width=\textwidth]{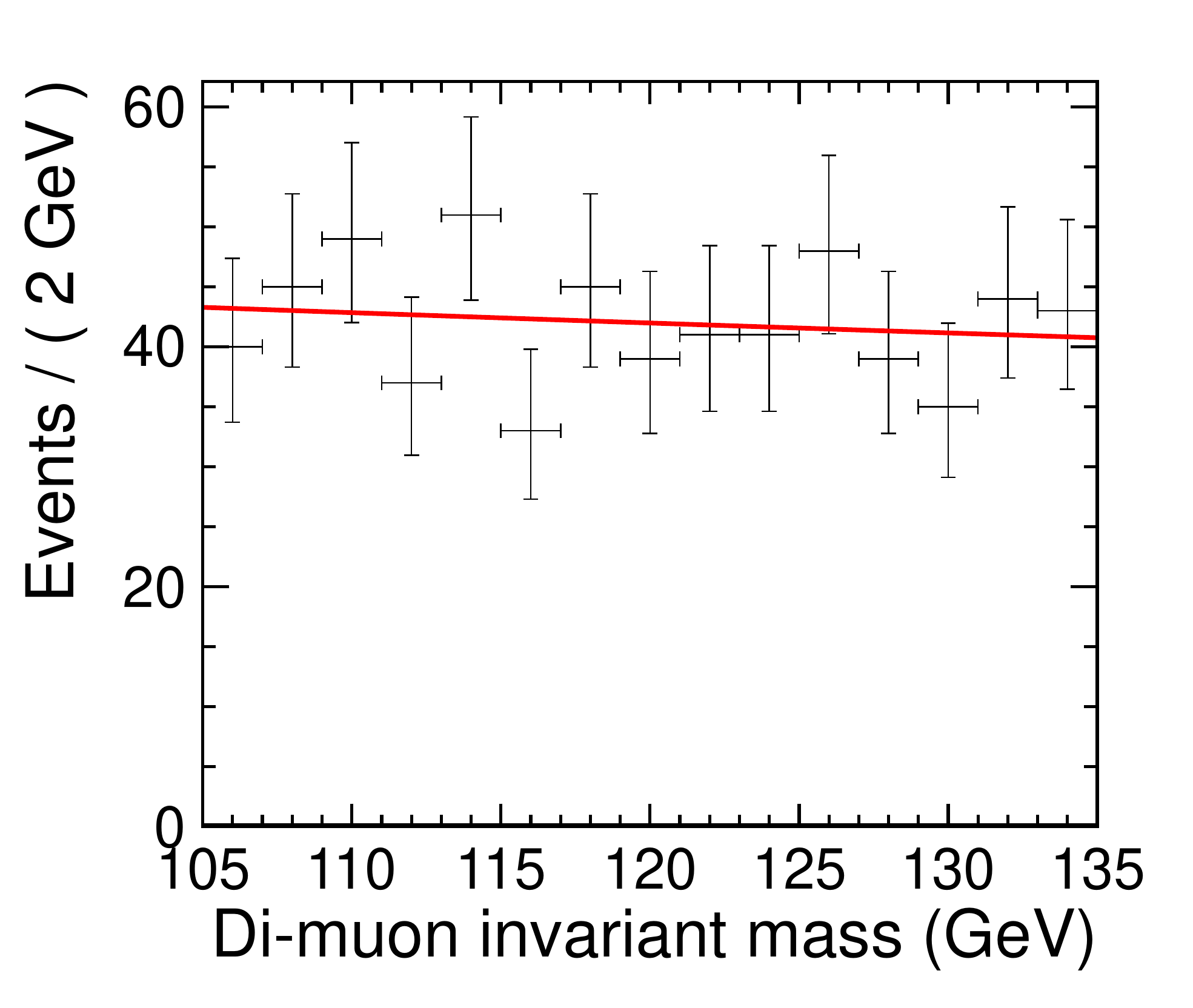}
\end{subfigure}
\caption[Invariant mass distributions for signal and background events with fitted \acp{PDF}.]{The invariant mass distribution of the signal events after event selection fitted with the function given in \cref{eq:MassFit_hmumu} (left). The invariant mass distribution of the $\epem \to \mpmm\nunubar$ events (middle) and the $\epem \to \mpmm\epem$ events (right) after event selection, both fitted with a combination of a flat and an exponential probability density function.}
\label{fig:MassFits}
\end{figure}

The reconstructed Higgs mass peak can be described by a Gaussian distribution, which is determined completely by the momentum resolution of the detector. The detector resolution is in fact a convolution of several Gaussian distributions, since the momentum resolution is strongly depending on the transverse momentum and the polar angle of the reconstructed muon (see \cref{sec:MomentumResolution}).
The distribution has a tail towards lower masses because of final state radiation. In principle, this can be corrected for by identifying and matching reconstructed photons with the muons. This correction was not applied in this analysis because the overall effect on the shape is rather small.
Instead, the shape is described well by two half Gaussian distributions with an exponential tail. Together with the mean value this results in five free parameters in the fitted function, which can be written as
\begin{equation}
 f(x) = n \left\{ \begin{array}{rl}
                 e^{\frac{-(x - m_0)^2}{2\sigma_L^2 + \alpha_L(x - m_0)^2}}, x \leq m_0 \\
                 e^{\frac{-(x - m_0)^2}{2\sigma_R^2 + \alpha_R(x - m_0)^2}}, x > m_0
                \end{array} \right .
 \label{eq:MassFit_hmumu}
\end{equation}
where $m_0$ is the mean of both Gaussian distributions, $\sigma_L$ and $\sigma_R$ are the widths, and $\alpha_L$ and $\alpha_R$ are the tail parameters of the left and the right Gaussian distribution, respectively. The normalization $n$ is required to normalize the total integral to 1 and transform the function into a probability distribution. The best fit of \cref{eq:MassFit_hmumu} to the invariant mass distribution after event selection is shown in \cref{fig:MassFits}~(left). The values for the best fit are $m_0 = \unit[120.1]{GeV}$, $\sigma_L = \unit[0.379]{GeV}$, $\sigma_R = \unit[0.293]{GeV}$, $\alpha_L = 0.203$ and $\alpha_R = 0.133$.


Both main background channels can be described by a combination of an exponential \ac{PDF} and a flat \ac{PDF}
\begin{equation}
 f(x) = r\ n_{0}\ e^{\lambda x} + (1 - r)\ n_{1},
 \label{eq:MassFit_background}
\end{equation}
where the two free parameters are $\lambda$, describing the exponential behavior, and the fraction $r$, describing the amount of the exponential and the flat contributions. The two normalization constants $n_0$ and $n_1$ are required to normalize each individual probability distribution to 1.

As shown in \cref{fig:MassFits} the distribution of the di-muon invariant mass in the $\epem \to \mpmm\nunubar$ events after the event selection using the \ac{BDT} is mostly described by an exponential drop with $\lambda = -0.109$ and $r = 0.573$. The di-muon invariant mass distribution of the $\epem \to \mpmm\epem$ events is almost flat after the event selection. The best fit is $\lambda = -0.004$ and $r = 0.493$.

After determining the shapes of the three contributing channels, the three \acp{PDF} are added into a combined \ac{PDF} which describes the signal plus background hypothesis in case of a \ac{SM} Higgs decaying into two muons. The parameters defining the shape of the individual \acp{PDF} are fixed and three free parameters for the number of events contributed by the individual channels are introduced.

\subsection{Toy Monte Carlo}

The fit of the Higgs mass peak is done on a very small event sample and upward or downward fluctuations of the number of signal events can strongly influence the measurement. Therefore, it is useful to perform the measurement on several independent event samples to estimate the average expected measurement accuracy. Since the available statistics from full simulation, especially for the $\mpmm\epem$ sample, is not sufficient to create enough independent samples corresponding to an integrated luminosity of \unit[2]{\abinv}, the toy Monte~Carlo method is used instead. For this, the \acp{PDF} for the different channels determined in \cref{sec:Shapes} are used to generate random event samples corresponding to the desired integrated luminosity. The number of events to be generated for each process is drawn from a Poisson distribution with a most probable value of $N = \sigma \cdot \epsilon \cdot {\cal L}$, where $\epsilon$ is the total selection efficiency for the respective process. Since the $\PSh \to \mpmm$ sample has sufficient statistics, those events are drawn from the fully simulated sample instead. This method allows to generate a large number of independent samples that each correspond to a realistic measurement. It also allows to generate data sets corresponding to larger integrated luminosities than possible with the original samples, provided the original sample is large enough to extract the shape.

\begin{figure}
 \centering
 \begin{subfigure}[]{0.49\textwidth}
  \includegraphics[width=\textwidth]{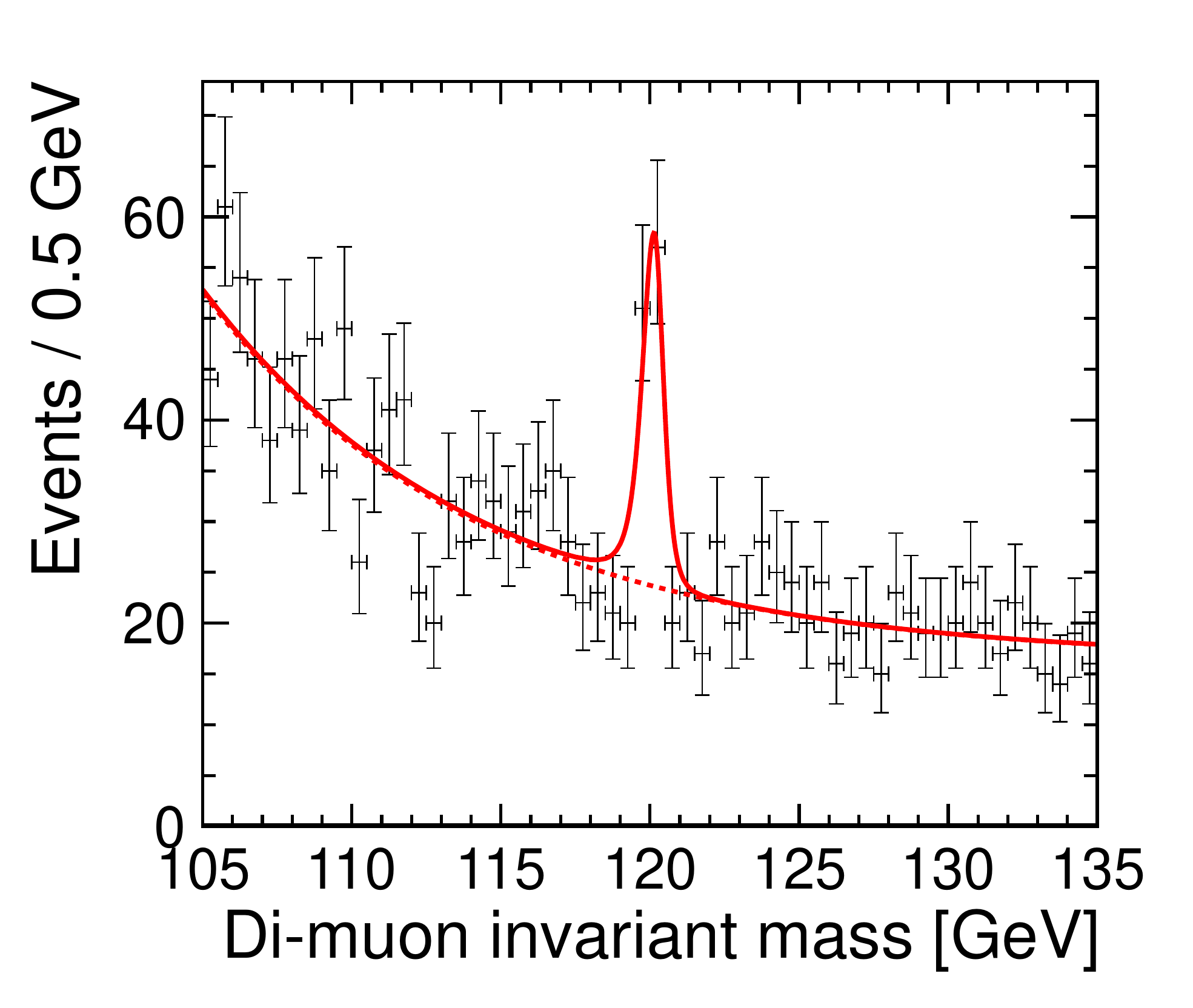}
 \end{subfigure}
 \hfill
 \begin{subfigure}[]{0.49\textwidth}
  \includegraphics[width=\textwidth]{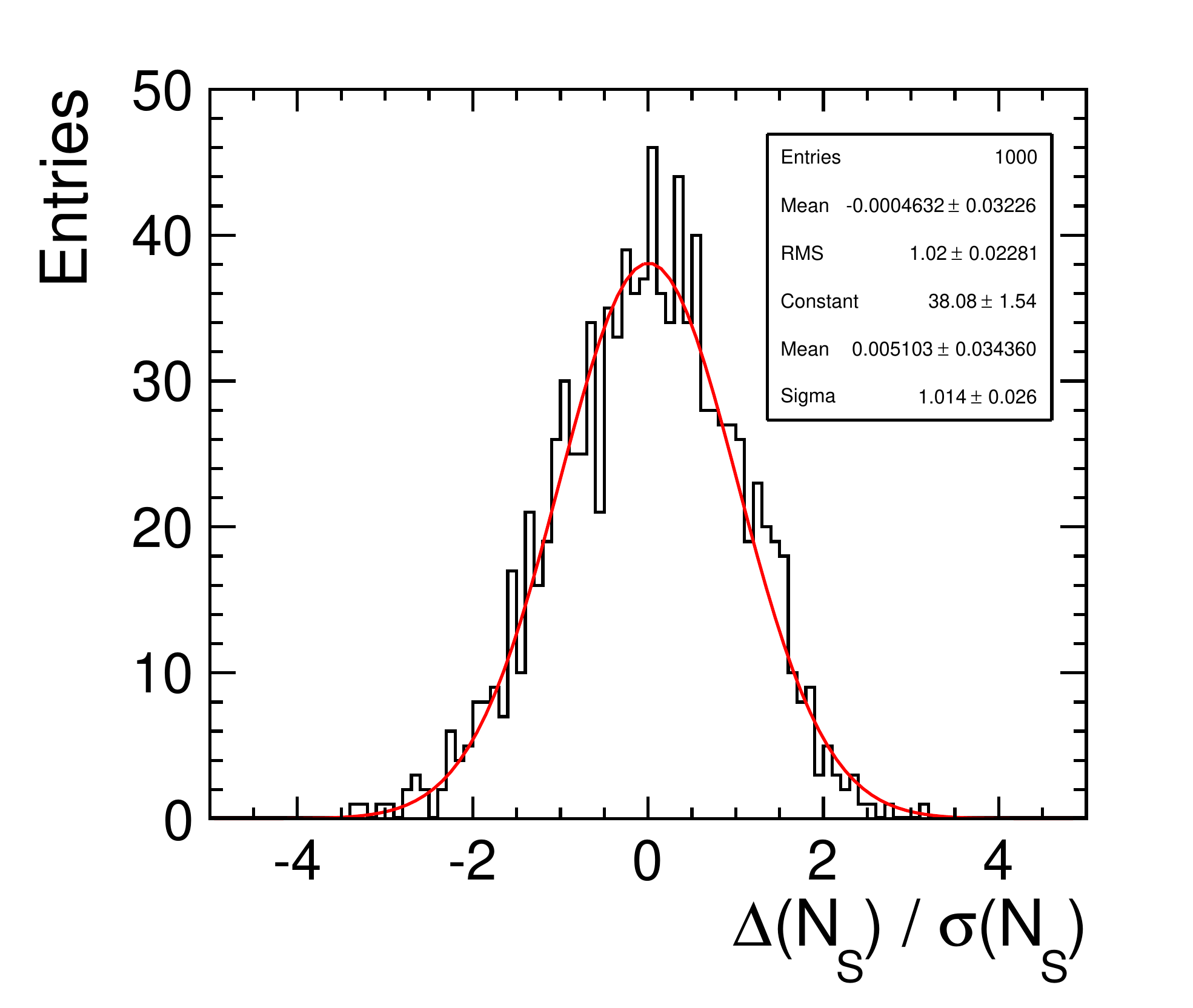}
 \end{subfigure}
\caption[Fit of the Higgs mass peak and pull distribution of the toy Monte~Carlo.]{One example of the Higgs mass peak fit for a data set corresponding to ${\cal L} = \unit[2]{\abinv}$ (left). Pull distribution for 1000 toy Monte~Carlo experiments extracting the number of signal events $N_{\mathrm S}$ assuming a total integrated luminosity of \unit[2]{\abinv} and fitted with a Gaussian distribution (right).}
\label{fig:pulldistribution}
\end{figure}

The randomly generated event sample is then fitted with the combined \ac{PDF} using an unbinned likelihood fit to obtain the number of signal events. An example of this fit for an integrated luminosity of \unit[2]{\abinv} is shown in \cref{fig:pulldistribution}~(left). The cross section times branching ratio is then determined by
\begin{equation}
 \sigma_{\PSh\nuenuebar} \times \text{BR}_{\PSh \to \mpmm} = \frac{N_\text{S}}{{\cal L}\ \epsilon_\text{S}},
\end{equation}
where $\epsilon_\text{S}$ is the total signal selection efficiency

The large number of samples that can be generated with the toy Monte~Carlo method also allows to test the shape assumed for the signal distribution, since the $\PSh \to \mpmm$ events are drawn from the full simulation sample. If the shape describes the distribution well, the resulting pull distribution should follow a Gaussian distribution centered around 0 with a width of 1. The pull is defined as $\Delta N_\text{S} / \sigma(N_\text{S})$, where $\Delta N_\text{S}$ is the difference between the number of signal events extracted from the respective fit and the mean expected signal events. $\sigma(N_\text{S})$ is the uncertainty of the number of signal events for that fit. \cref{fig:pulldistribution}~(right) shows the pull distribution obtained from 1000 toy Monte~Carlo experiments which has been fitted with a Gaussian distribution. It can be seen that the invariant mass distribution is well described by the assumed shapes.

\subsection{Result}
\label{sec:result}

In order to estimate the achievable precision, 100 toy Monte~Carlo experiments are performed. The resulting cross section times branching ratio of $\unit[0.122]{fb}$ is the average obtained from those 100 fits. The relative statistical uncertainty of 23\% is the average of the relative uncertainties of the individual measurements. The results are summarized in \cref{tab:results}. Here, we assumed a \ac{SM}-like cross section times branching ratio. If the true value differs from the \ac{SM} prediction by a factor $a$, then the achievable uncertainty scales with a factor $1/\sqrt{a}$ due to the different amount of available signal events. The observed signal would correspond to a significance of $4.3\sigma$, which would be slightly to low to actually claim a discovery of the muon decay channel.

There are several systematic uncertainties that have to be taken into account in such an experiment. The expected uncertainty of the peak luminosity is currently being studied but is estimated to be around 1\% or less. The systematic uncertainties originating from detector effects are difficult to estimate and a detailed study goes beyond the scope of this thesis. The most important detector effects will be the uncertainty on the momentum resolution, the uncertainty on the resolution of the track angle and the uncertainty on the muon identification efficiency directly influencing the uncertainty of the signal selection efficiency and the expected signal shape.

An estimate of the detector related systematic uncertainties can be found from the experience at the \ac{LEP} experiments. For the measurement of $\sigma_{\PZ \to \mpmm}$ at \ac{LEP}, the systematic uncertainty due to detector effects was between 0.1 and 0.4\%, depending on the experiment~\cite{ALEPH:2005ab}. Together with the estimate on the uncertainty due to the luminosity measurement, one can expect that the systematic uncertainty of this analysis will be negligible compared to the statistical uncertainty.

\begin{table}
 \centering
 \caption[Summary of the results for the $\PSh \to \mpmm$ branching ratio measurement.]{Summary of the results for the $\PSh \to \mpmm$ branching ratio measurement for an integrated luminosity of $\unit[2]{\abinv}$.}
 \label{tab:results}
\begin{tabular}{l r r r}
\toprule
Signal events                                                 & $62 \pm 14$              \\
Signal efficiency                                             & 25.2\%                     \\
$\sigma_{\PSh\nuenuebar} \times \mathrm{BR}_{\PSh \to \mpmm}$ & \unit[0.122]{fb}         \\
Stat. uncertainty                                             & 23.3\%                     \\\bottomrule
\end{tabular}

\end{table}

\section{Impact of Beam-Induced Background}
\label{sec:BeamInducedBackground}

The beam-induced electron pair background with particles at very low angles and low transverse momenta leads to high occupancies in the vertex detector and the forward calorimeters (see \cref{sec:CLIC_machineInduced,sec:SiD_occupancies}). This has to be considered when designing those sub-detectors but is of no concern for this analysis. The \gghad background, on the other hand, introduces particles at higher \pT and has to be dealt with at the analysis level (see also \cref{sec:CLIC_machineInduced}). The default event overlay settings for the \gghad background are used, as defined in \cref{sec:Software_BackgroundOverlay}. The impact of this background is studied by adding the background only to the signal events due to the large statistics of the background samples.

\begin{figure}
 \begin{subfigure}[]{0.49\textwidth}
  \includegraphics[width=\textwidth]{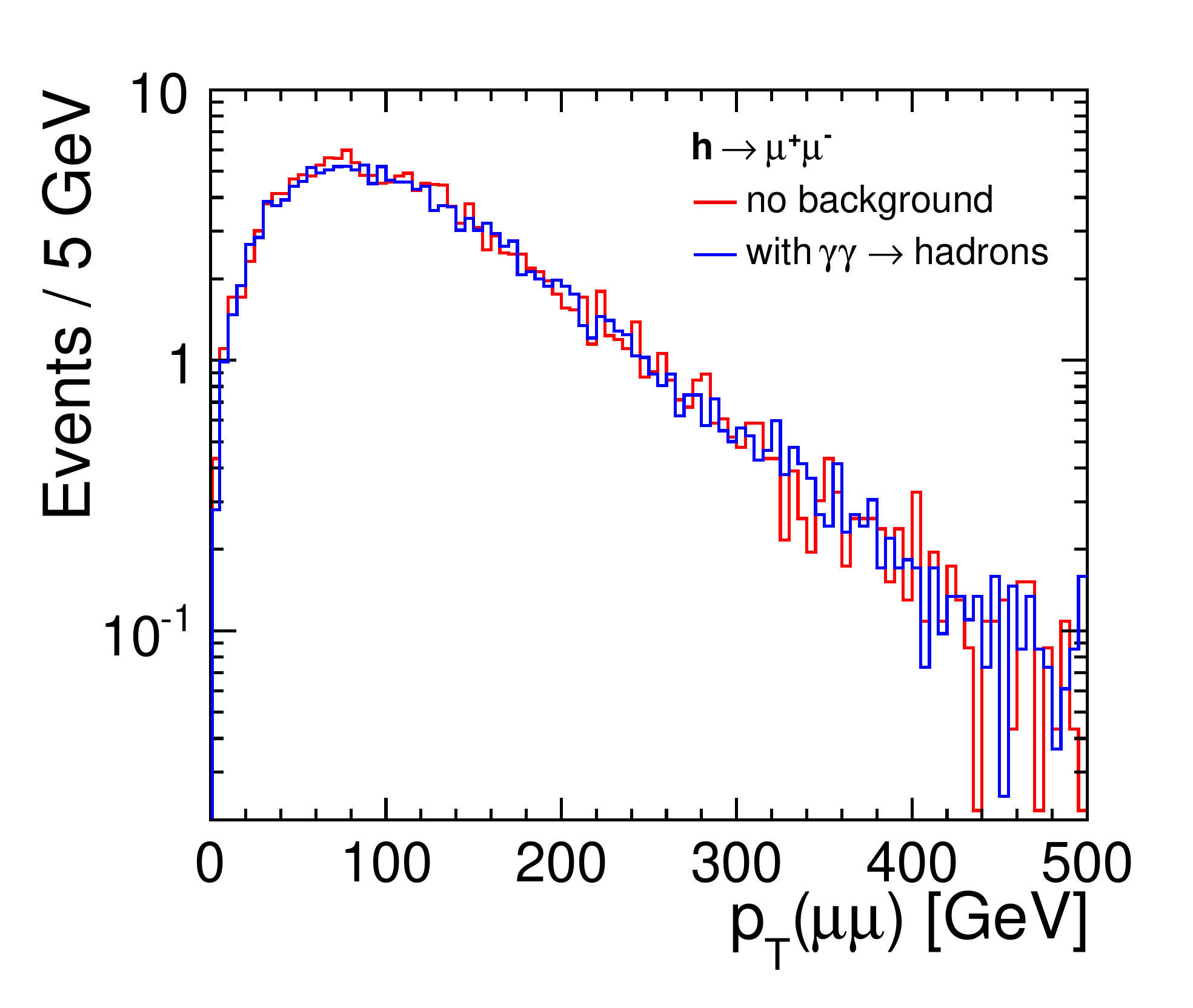}
 \end{subfigure}
\hfill
 \begin{subfigure}[]{0.49\textwidth}
  \includegraphics[width=\textwidth]{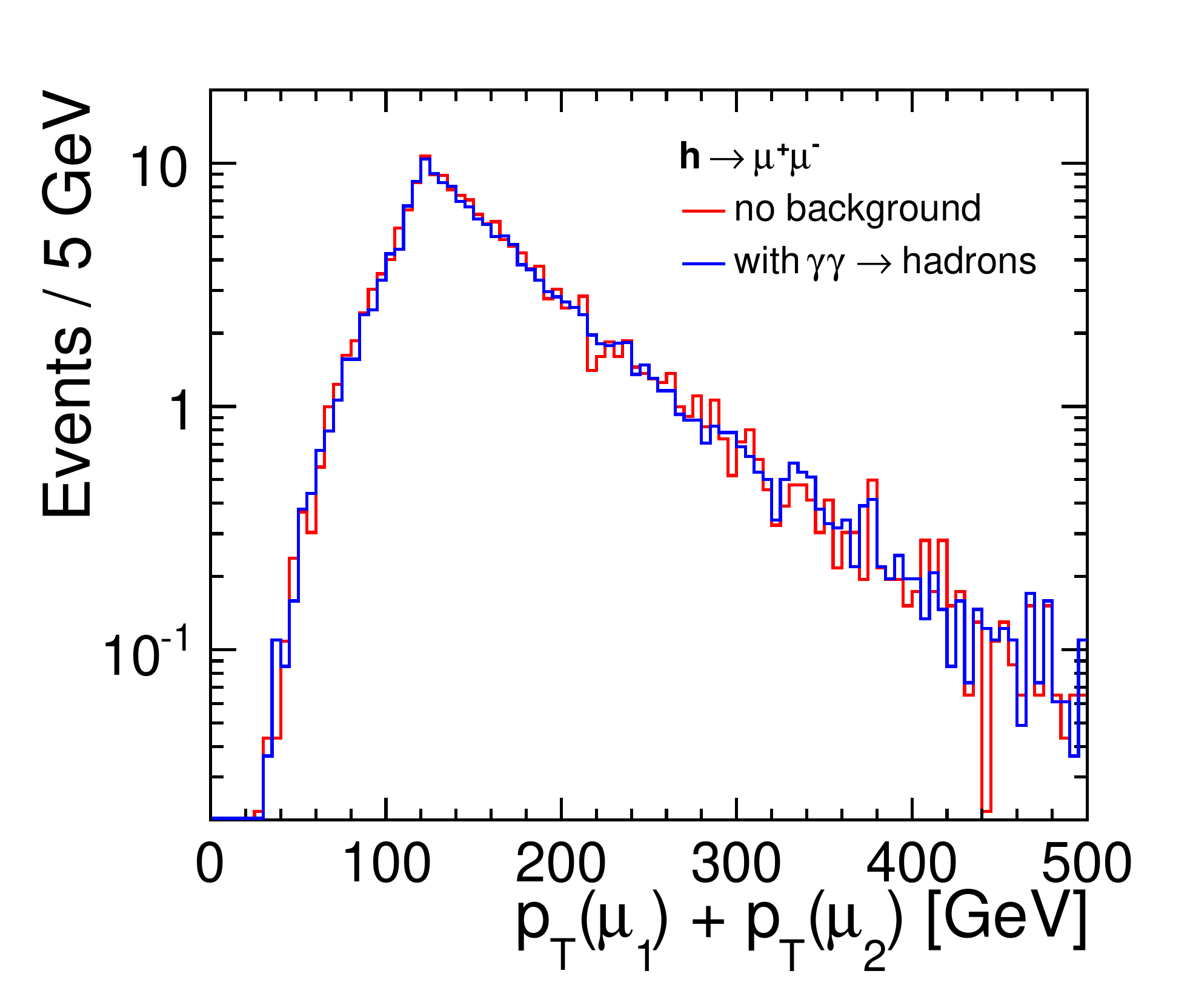}
 \end{subfigure}
 \begin{subfigure}[]{0.49\textwidth}
  \includegraphics[width=\textwidth]{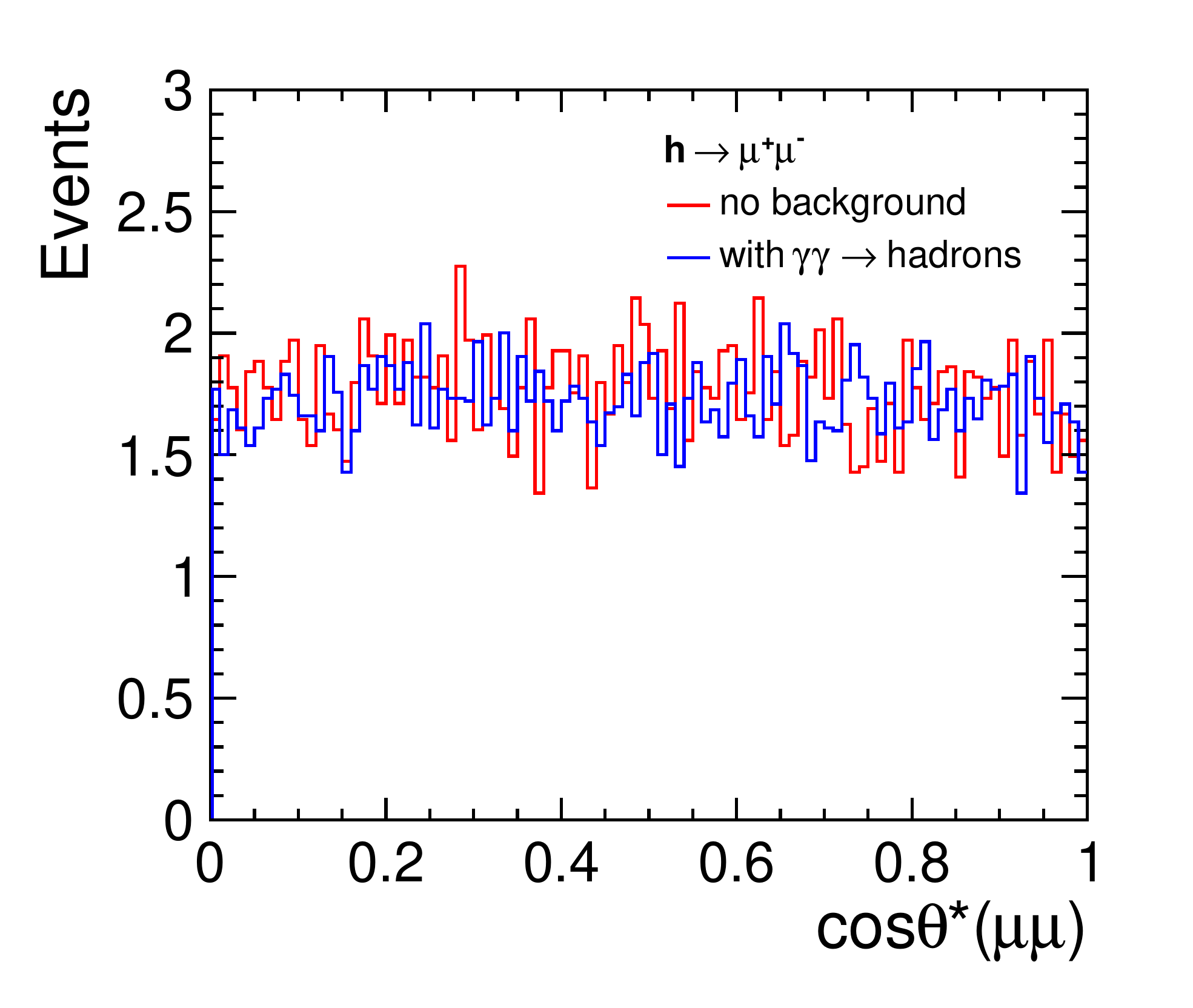}
 \end{subfigure}
\hfill
 \begin{subfigure}[]{0.49\textwidth}
  \includegraphics[width=\textwidth]{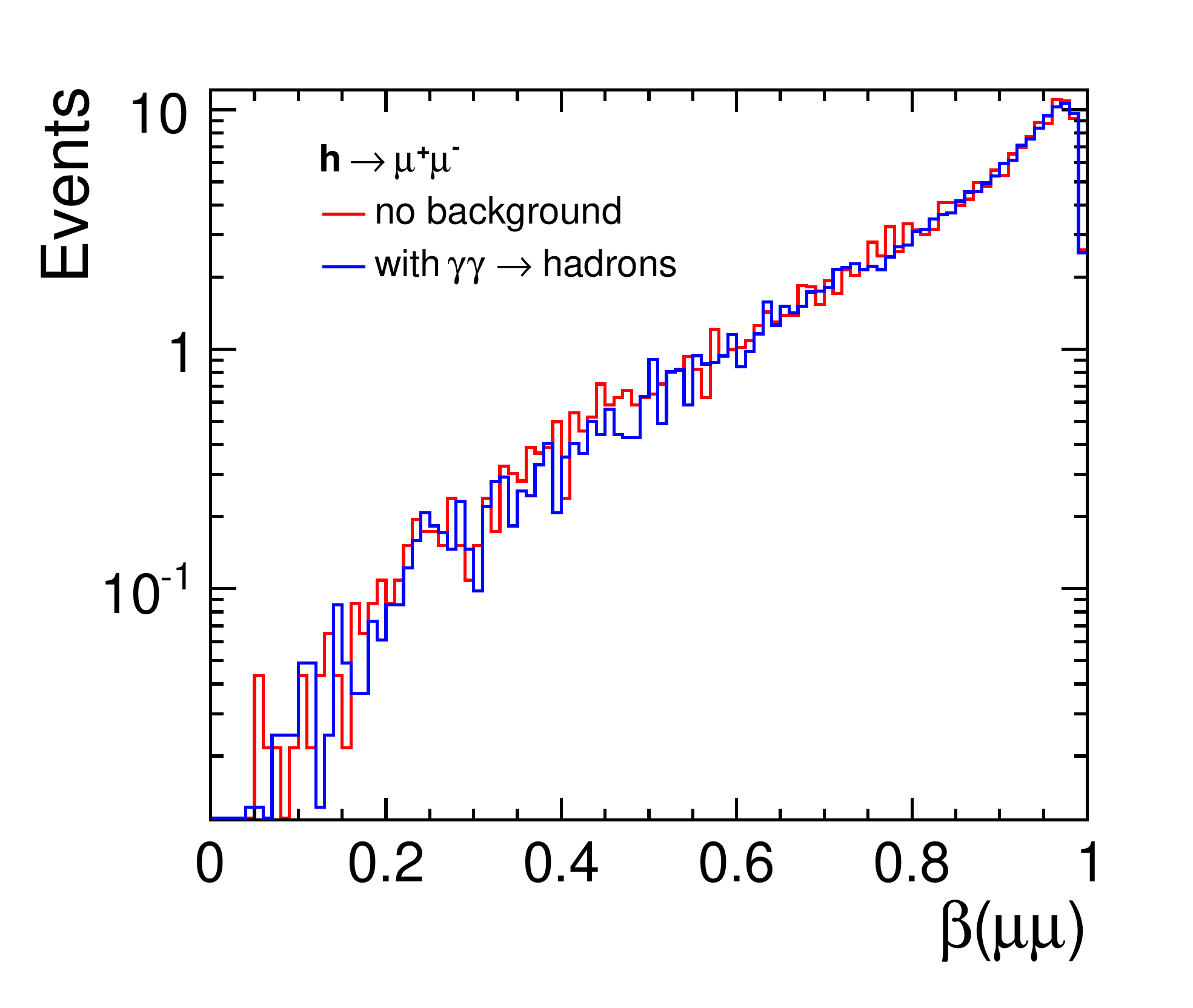}
 \end{subfigure}
 \begin{subfigure}[]{0.49\textwidth}
  \includegraphics[width=\textwidth]{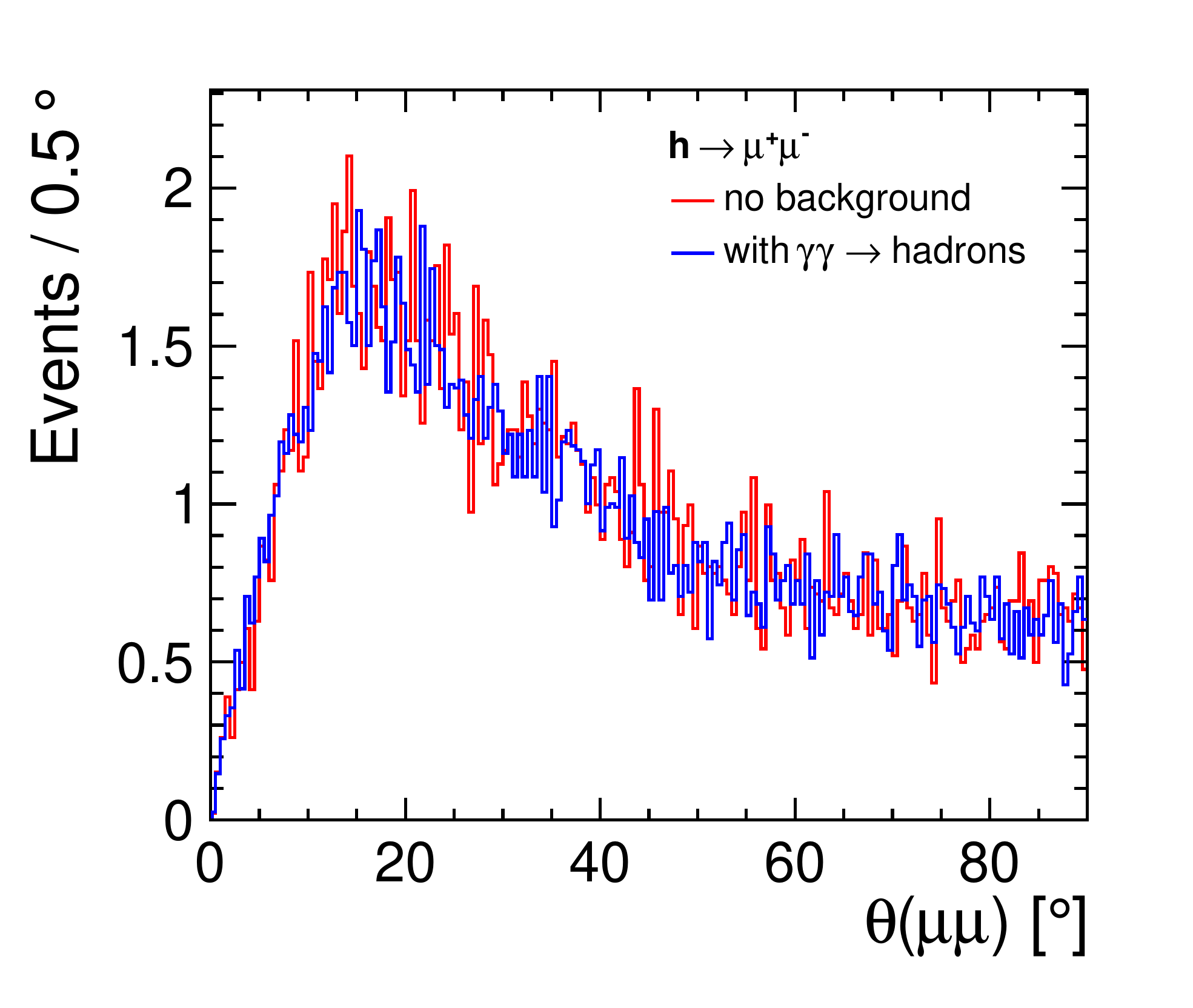}
 \end{subfigure}
\hfill
 \begin{subfigure}[]{0.49\textwidth}
  \includegraphics[width=\textwidth]{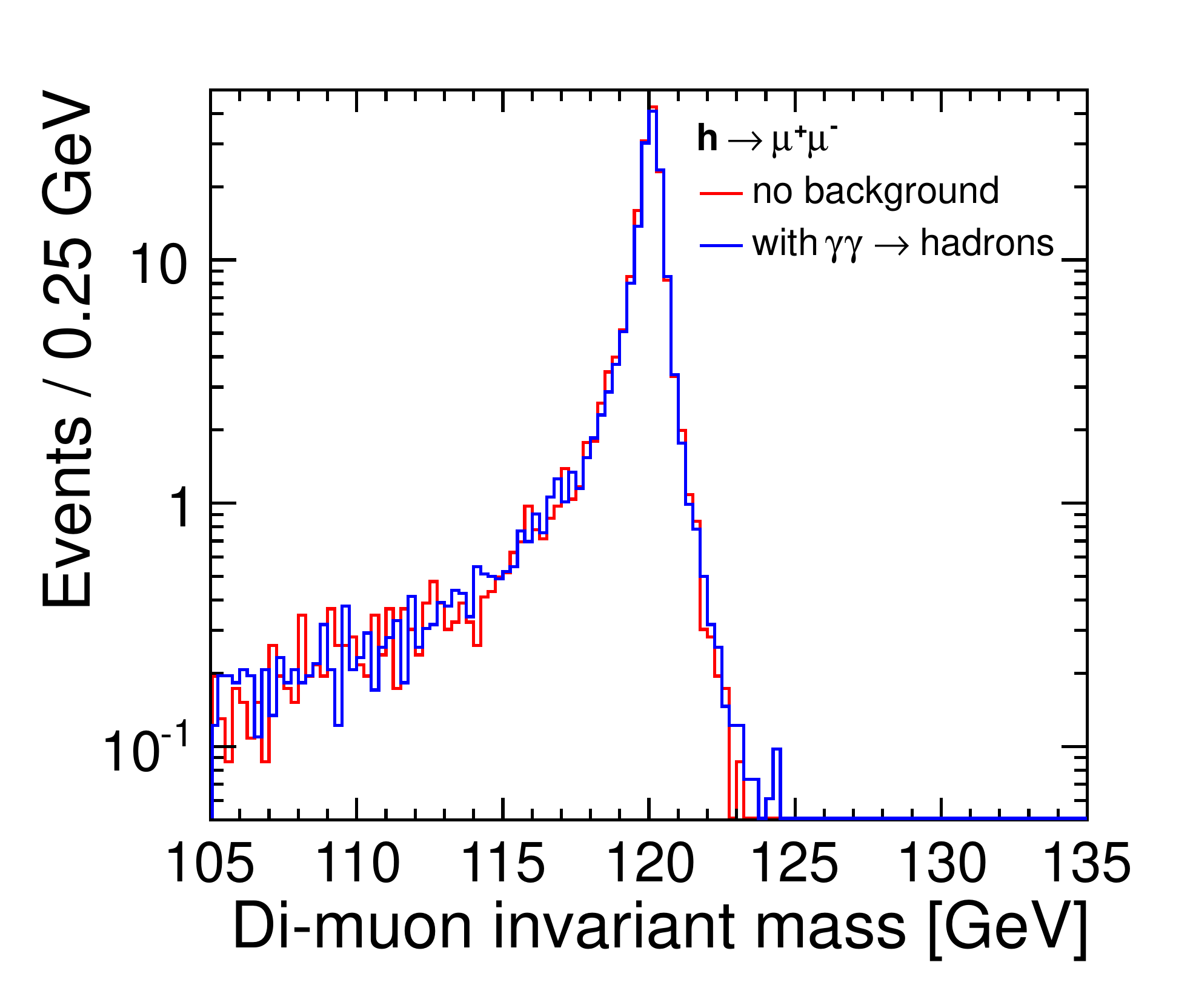}
 \end{subfigure}
\caption[Kinematic variables with and without overlay of the \gghad background.]{Impact of the \gghad background on the distributions of the kinematic variables used in the event selection in $\PSh \to \mpmm$ events. Histograms are normalized to a total integrated luminosity of \unit[2]{\abinv}.}
\label{fig:background_variables}
\end{figure}
\begin{figure}
 \begin{subfigure}[]{0.49\textwidth}
  \includegraphics[width=\textwidth]{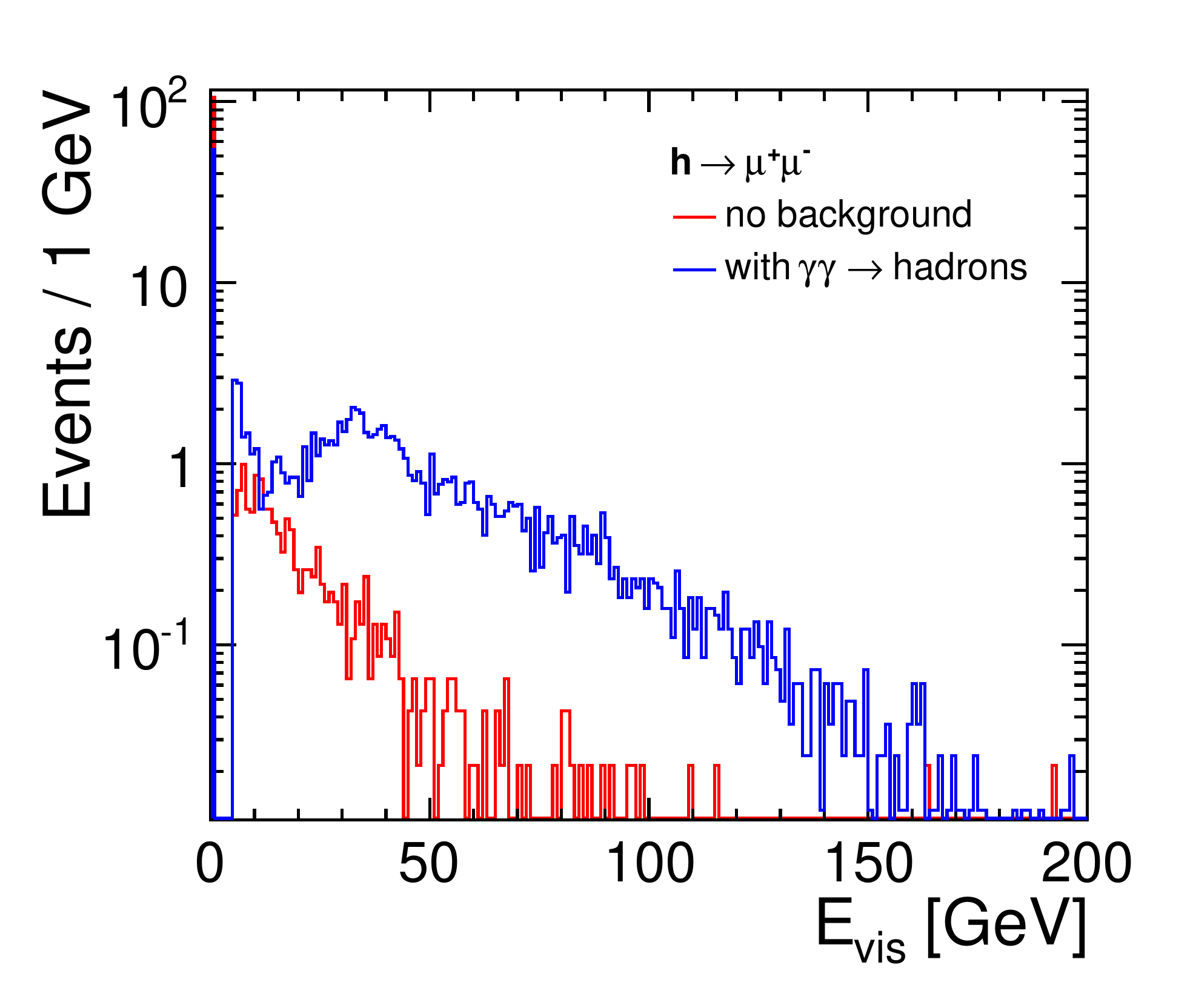}
 \end{subfigure}
\hfill
 \begin{subfigure}[]{0.49\textwidth}
  \includegraphics[width=\textwidth]{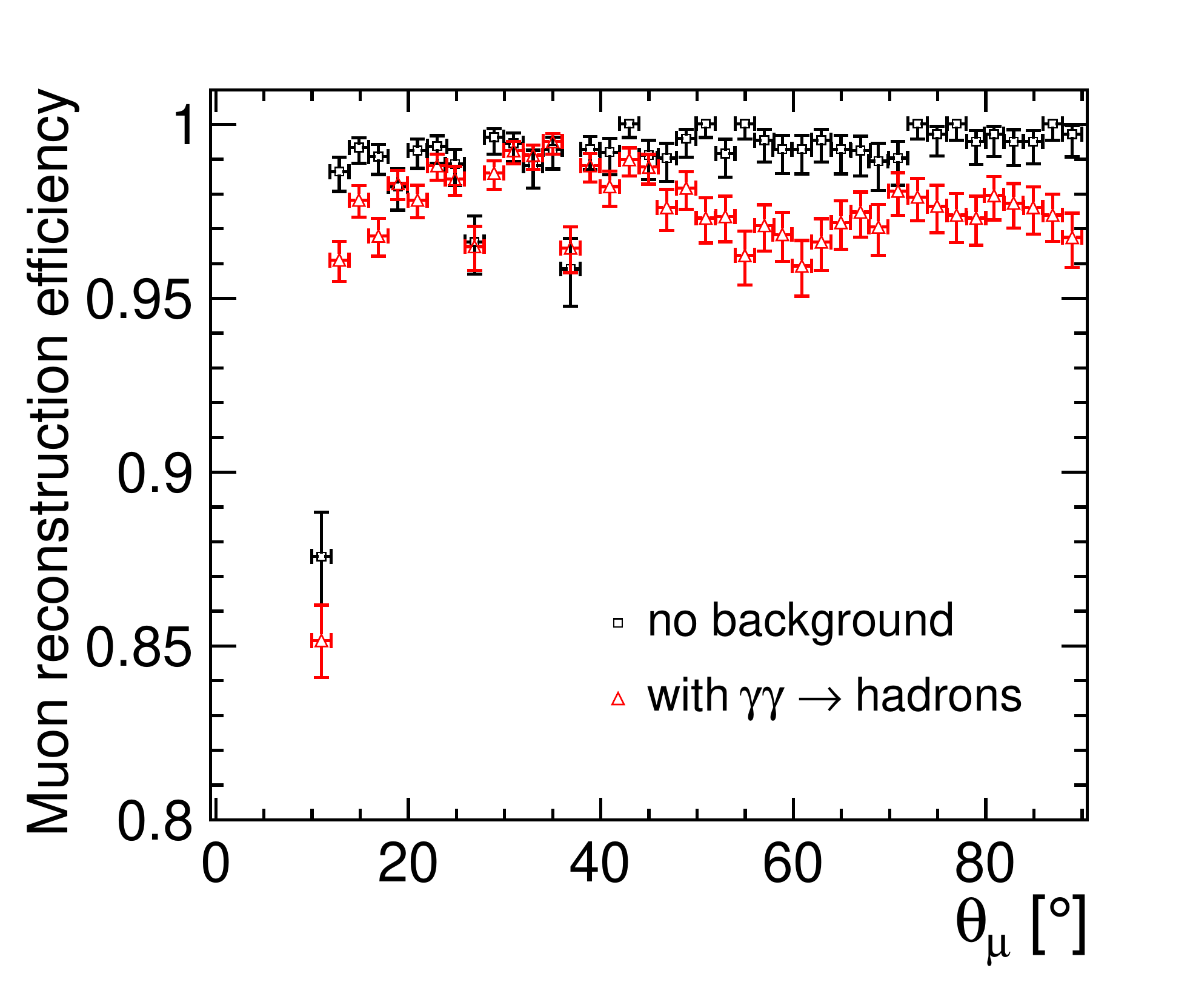}
 \end{subfigure}
\caption[Impact of the \gghad background on $E_\text{vis}$ and on the muon reconstruction efficiency.]{Impact of the \gghad background on the distribution of $E_\text{vis}$ of the signal sample, normalized to ${\cal L} = \unit[2]{\abinv}$ (left). Muon reconstruction efficiency for the signal sample with and without \gghad overlay (right).}
\label{fig:background_evis}
\end{figure}

In case of the signal sample, the reconstructed event topology of two muons with a typical transverse momentum of \unit[100]{GeV} and more (see \cref{fig:kinematic_variables_1}) is not affected by the addition of the \gghad background. Nevertheless a cut requiring a minimum transverse momentum of \unit[5]{GeV} for each reconstructed particle is introduced to remove low-\pT background particles and retain the clean event topology. None of the distributions of the kinematic variables of the muons and the di-muon system are affected by the addition of the \gghad background, see \cref{fig:background_variables}. On the other hand, the distribution of the visible energy, which is shown in \cref{fig:background_evis}~(left), is affected. Without background, non-zero $E_{\mathrm{vis}}$ is solely due to final state radiation. Despite the cut on the \pT of the reconstructed particles, the reconstructed energy is increased due to background, since the \gghad events can contain some particles with higher \pT that may add up to \unit[100]{GeV} to the total visible energy, after cuts. The fraction of $\PSh \to \mpmm$ events with $E_{\mathrm{vis}} > 0$ increases from 9\% to about 52\% when the background is added.

\begin{figure}
 \begin{subfigure}[]{0.49\textwidth}
  \includegraphics[width=\textwidth]{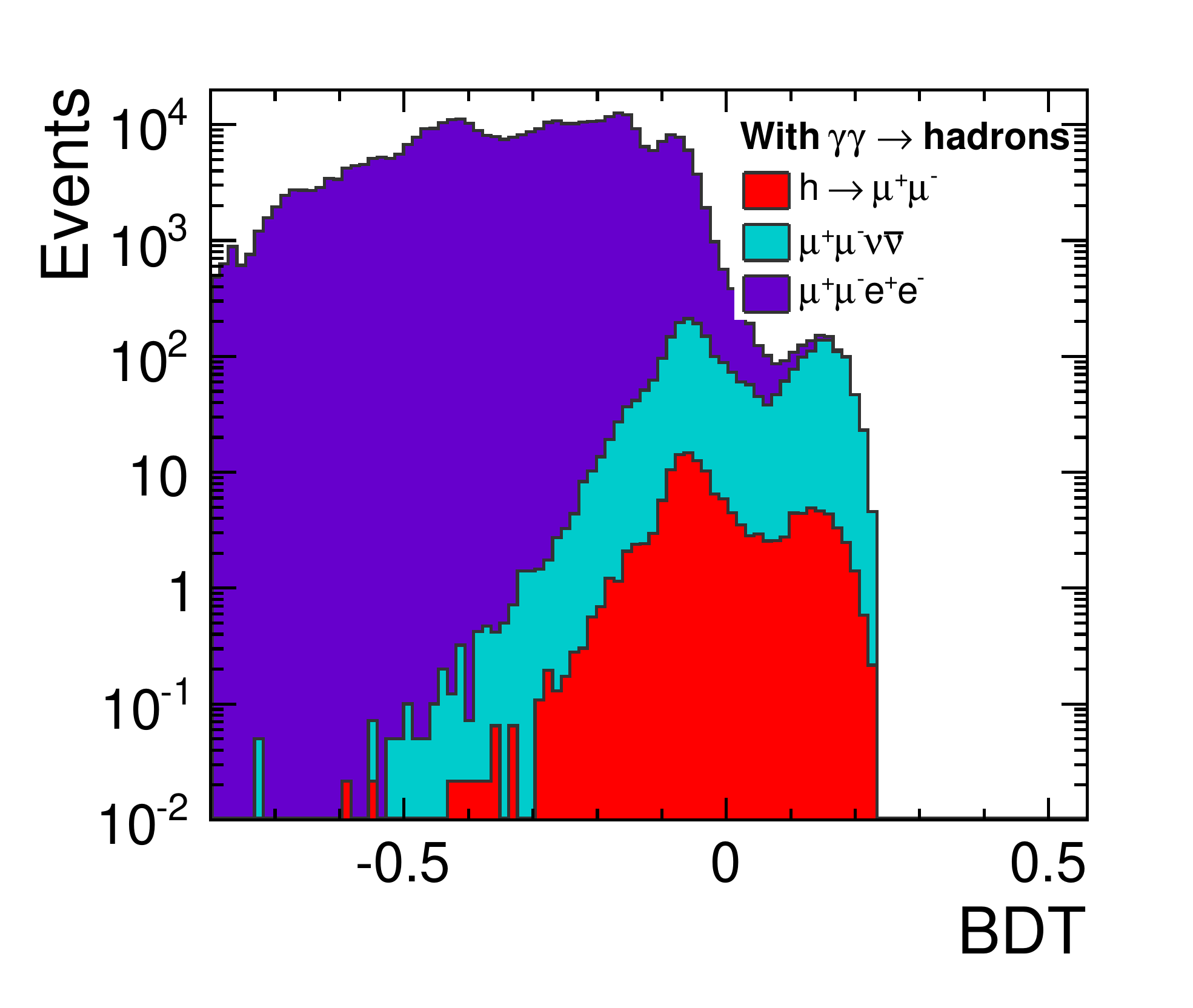}
 \end{subfigure}
\hfill
 \begin{subfigure}[]{0.49\textwidth}
    \includegraphics[width=\textwidth]{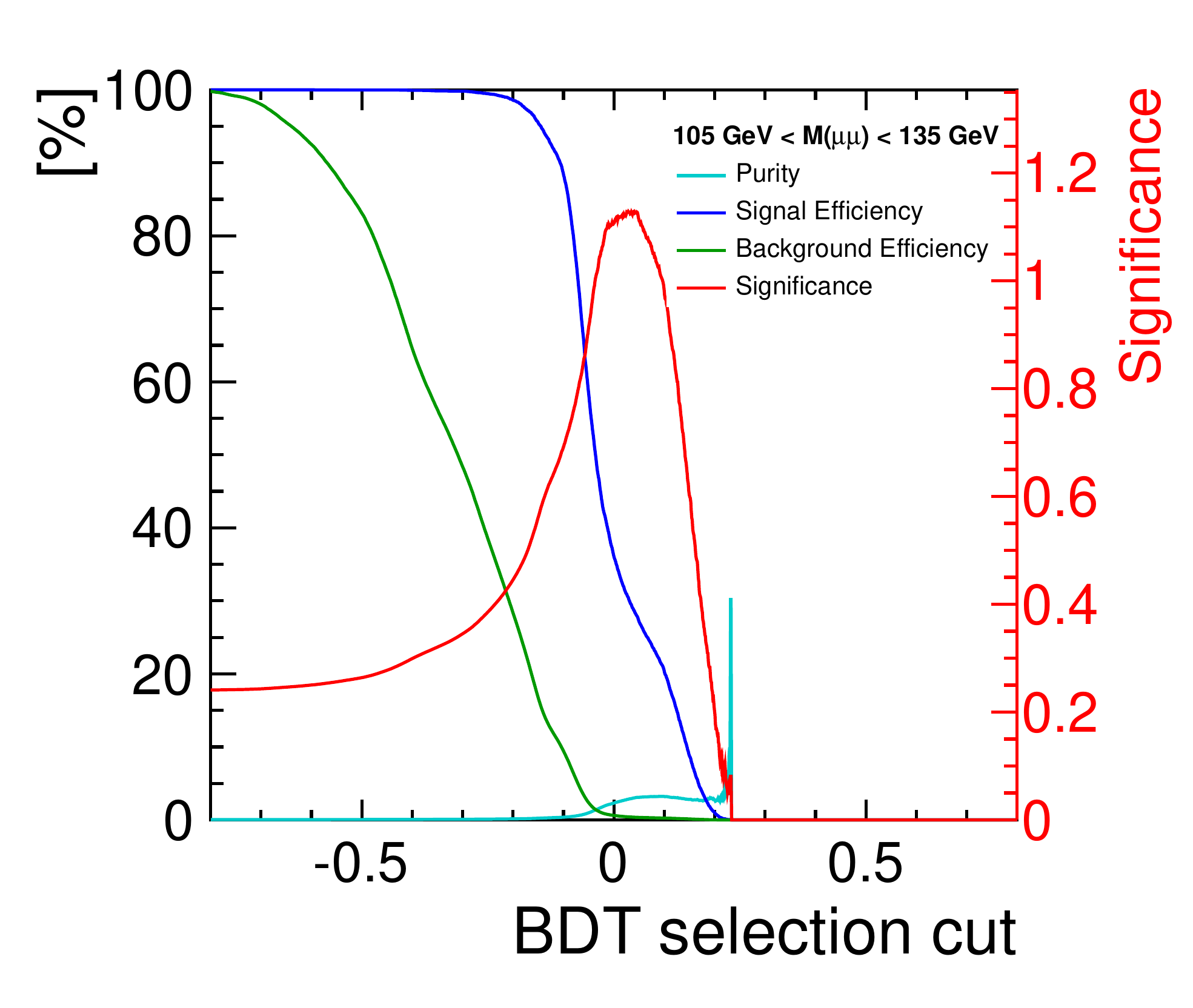}
 \end{subfigure}
\caption[Response of the boosted decision tree classifier when including the \gghad background.]{Response of the boosted decision tree classifier for the signal including the \gghad background and the two most important background processes (left) and the resulting significance, purity, signal efficiency and background efficiency (right).}
\label{fig:background_bdt}
\end{figure}

The muon reconstruction efficiency in the signal sample is slightly reduced in the barrel region, as shown in \cref{fig:background_evis}~(right). The average muon reconstruction efficiency for polar angles greater than 10\degrees is 98.4\% with background compared to 99.6\% without. The inefficiency is due to failed muon identification in the calorimeters which was tuned without background overlay and can most likely be improved. The total reconstruction efficiency, requiring two reconstructed muons with an invariant mass between \unit[105]{GeV} and \unit[135]{GeV}, is reduced to 72\% in the presence of background.

The event selection is performed with a \ac{BDT}, similar to the selection procedure described in \cref{sec:EventSelection}. Since the $E_\text{vis}$ distribution of the signal sample is modified by the addition of the \gghad background, which was not added to the other physics processes, this variable can not be used in the classifier. Instead, a pre-selection cut of $E_\text{vis} < \unit[150]{GeV}$ is used to reject $\epem \to \mpmm\epem$ events with electrons at large polar angles. This cut removes only 0.6\% of the $\PSh \to \mpmm$ events that passed the original pre-selection cut. Applying a pre-selection cut on a variable instead of using it in a multivariate classifier leads to a worse signal and background separation since correlations are no longer taken into account. In addition, the fact that $E_\text{vis}$ is underestimated for the background processes means that the number of selected background events is overestimated. This means that our estimate of the impact of the \gghad background is conservative.

\cref{fig:background_bdt} shows the response of the \ac{BDT} classifier including the \gghad background. The \ac{BDT} selection cut that yields the highest significance corresponds to a signal selection efficiency of 21.7\%. Using the fitting procedure explained in \cref{sec:MassFit} in 100 toy Monte~Carlo experiments results in an expected relative statistical uncertainty of 26.3\%, see \cref{tab:resultsgghad}. This is worse than the result obtained in \cref{sec:result} but probably too pessimistic due to the reasons stated above.

In addition to the systematic uncertainties discussed in \cref{sec:result}, no systematic uncertainties due to the \gghad background are expected. The huge statistics of all machine-induced backgrounds will allow to understand those process with almost unlimited precision and will allow to remove any systematic uncertainty on $E_\mr{vis}$. The impact of these backgrounds on the muon reconstruction efficiency can be studied in $\PZ$ boson production events. If the reduced reconstruction efficiency due to the presence of \gghad background can not be removed by an improved muon finding algorithm, a larger cross section of the \gghad process than expected here will lead to a lower significance of the branching ratio measurement. 

\begin{table}
 \centering
 \caption[Summary of the results for the $\PSh \to \mpmm$ branching ratio measurement with \gghad background overlaid.]{Summary of the results for the $\PSh \to \mpmm$ branching ratio measurement with \gghad background overlaid for an integrated luminosity of $\unit[2]{\abinv}$.}
 \label{tab:resultsgghad}
\begin{tabular}{l r r r}
\toprule
Signal events                                                 & $53 \pm 14$              \\
Signal efficiency                                             & 21.7\%                   \\
$\sigma_{\PSh\nuenuebar} \times \mathrm{BR}_{\PSh \to \mpmm}$ & \unit[0.121]{fb}         \\
Stat. uncertainty                                             & 26.3\%                   \\\bottomrule
\end{tabular}

\end{table}

\section{Impact of the Momentum Resolution}
\label{sec:MomentumResolution}

The momentum resolution is the most important detector parameter that directly influences the branching ratio measurement. It determines the width of the Higgs boson mass peak observed in the signal sample, which then translates into the significance of the observed peak over the background.

\begin{figure}
\begin{subfigure}[]{0.49\textwidth}
 \includegraphics[width=\textwidth]{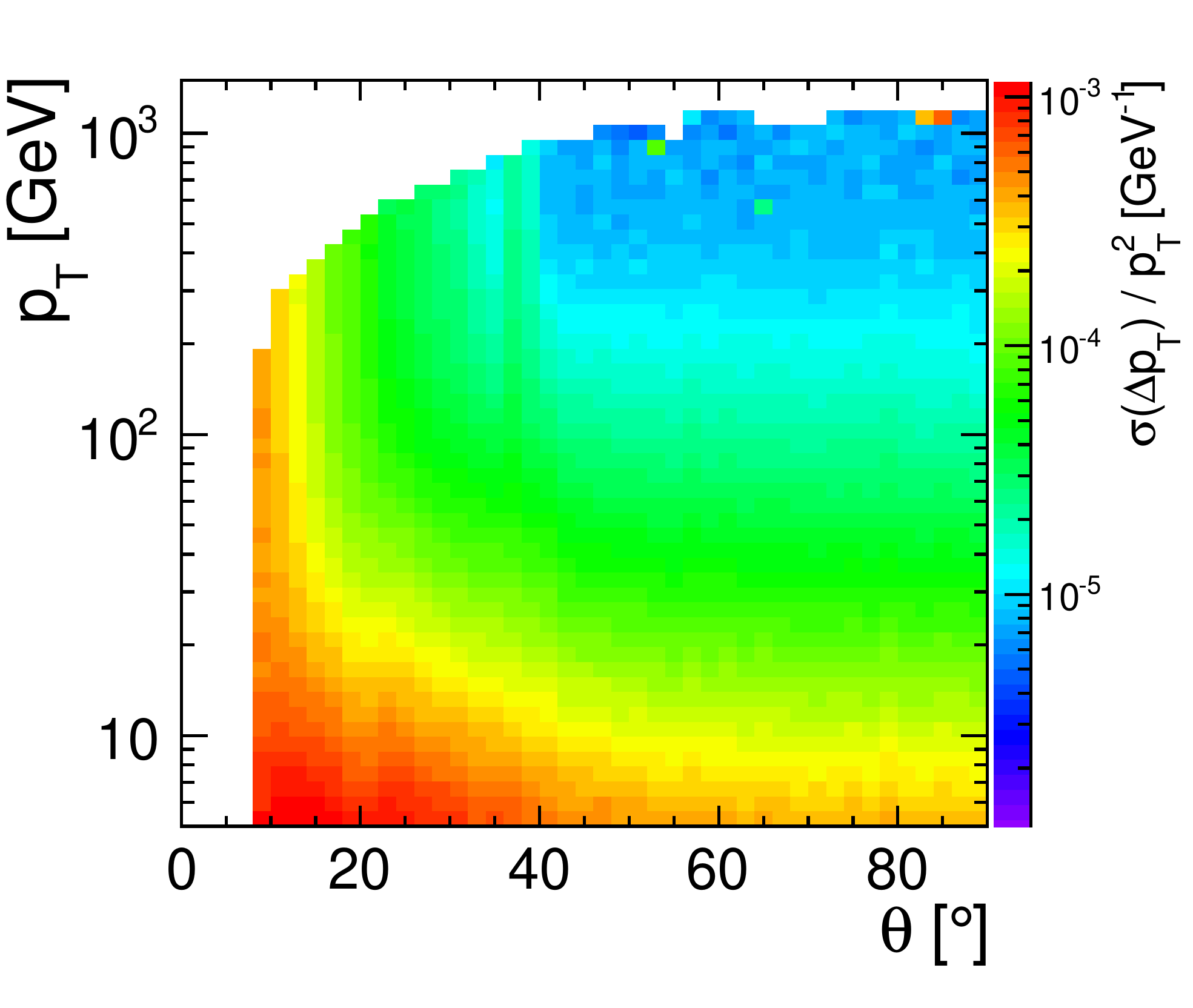}
\end{subfigure}
\hfill
\begin{subfigure}[]{0.49\textwidth}
 \includegraphics[width=\textwidth]{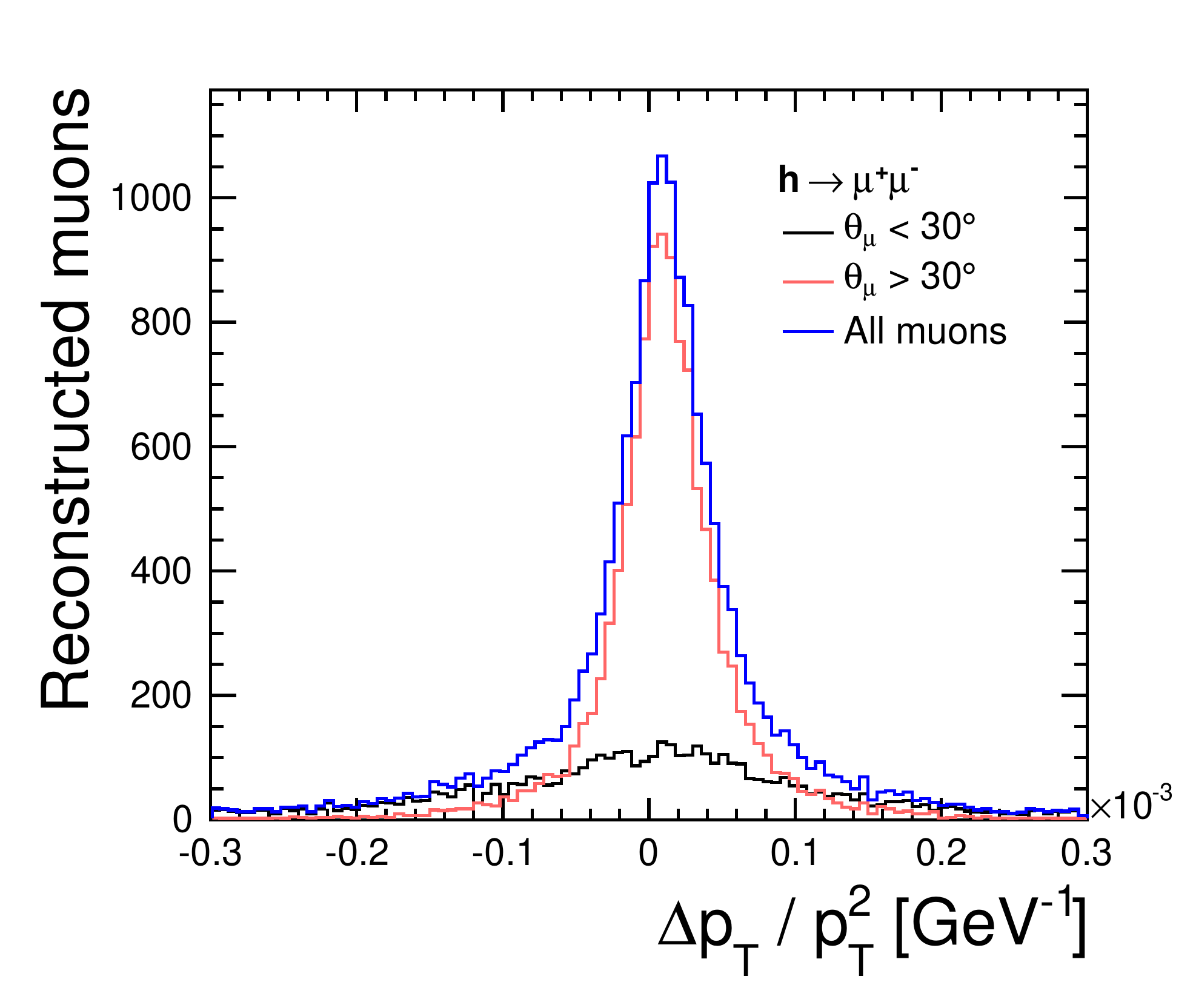}
\end{subfigure}
\caption[Momentum resolution of the muons depending on their polar angle and their transverse momentum.]{$\Delta\pT/\pT^2$ resolution of muons depending on the polar angle $\theta$ and the transverse momentum extracted from all fully simulated samples (left) and $\Delta\pT/\pT^2$ resolution in $\PSh \to \mpmm$ events for different regions in $\theta$ (right).}
\label{fig:momResFullReco}
\end{figure}

The momentum resolution $\sigma(\Delta(\pT)/\pT^2)$ depends strongly on the particle momentum and its polar angle $\theta$. \cref{fig:momResFullReco}~(left) shows the momentum resolution as a function of those two variables obtained from the muons in all of the fully simulated samples used in this analysis. For each bin in $\theta$ and $\pT$ the resulting $\Delta(\pT)/\pT^2$-distribution was fitted to a Gaussian distribution to obtain its width. Outliers are rejected by a cut at five times the RMS around the mean of the distribution prior to the fitting. As expected, the momentum resolution has a $1/\pT$-dependence due to multiple scattering. For transverse momenta above \unit[100]{GeV} the constant term driven by the magnetic field and the lever arm becomes more important and the distribution flattens out.  It can be seen that the momentum resolution throughout the barrel region, $\theta > \unit[40]{\degrees}$, only depends on the transverse momentum. It degrades sharply for polar angles below $30^\circ$. \cref{fig:momResFullReco}~(right) shows the momentum resolution of the reconstructed muons in the signal sample and for the two subsets where the polar angle of the muon is either above or below $30^\circ$. The average momentum resolution for reconstructed muons with $\theta > 30\degrees$ is \unit[$3.5\times10^{-5}$]{GeV$^{-1}$} but only \unit[$1.1\times10^{-4}$]{GeV$^{-1}$} for reconstructed muons with $\theta < 30\degrees$. The average momentum resolution for all reconstructed tracks in the signal sample is \unit[$5.2\times10^{-5}$]{GeV$^{-1}$}. Those events that pass the final event selection, as discussed in Section~\ref{sec:EventSelection}, are biased towards the central region, such that the average momentum resolution for the selected signal sample is \unit[$3.9\times10^{-5}$]{GeV$^{-1}$}.

\begin{figure}
\begin{subfigure}[]{0.49\textwidth}
 \includegraphics[width=\textwidth]{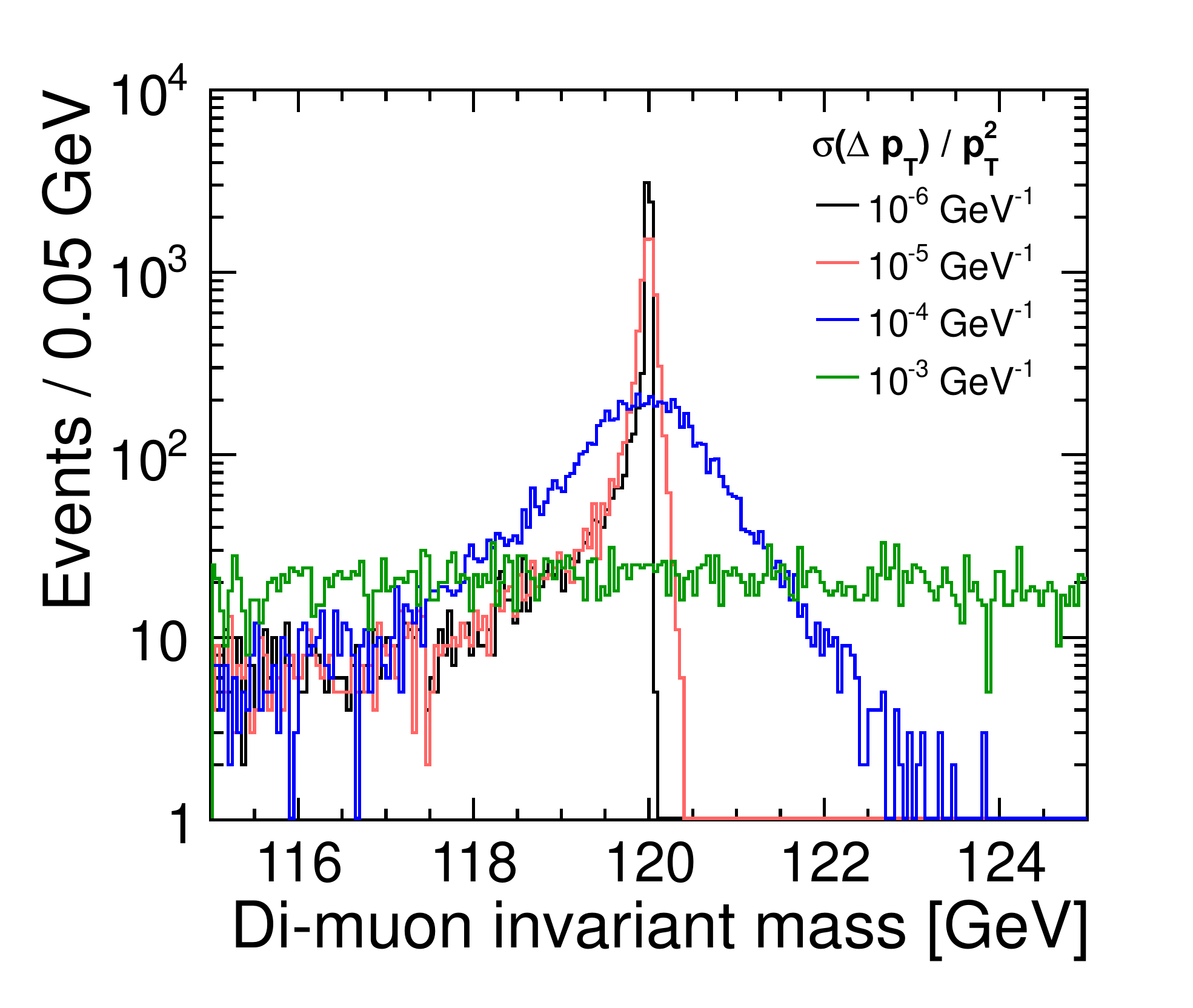}
\end{subfigure}
\hfill
\begin{subfigure}[]{0.49\textwidth}
 \includegraphics[width=\textwidth]{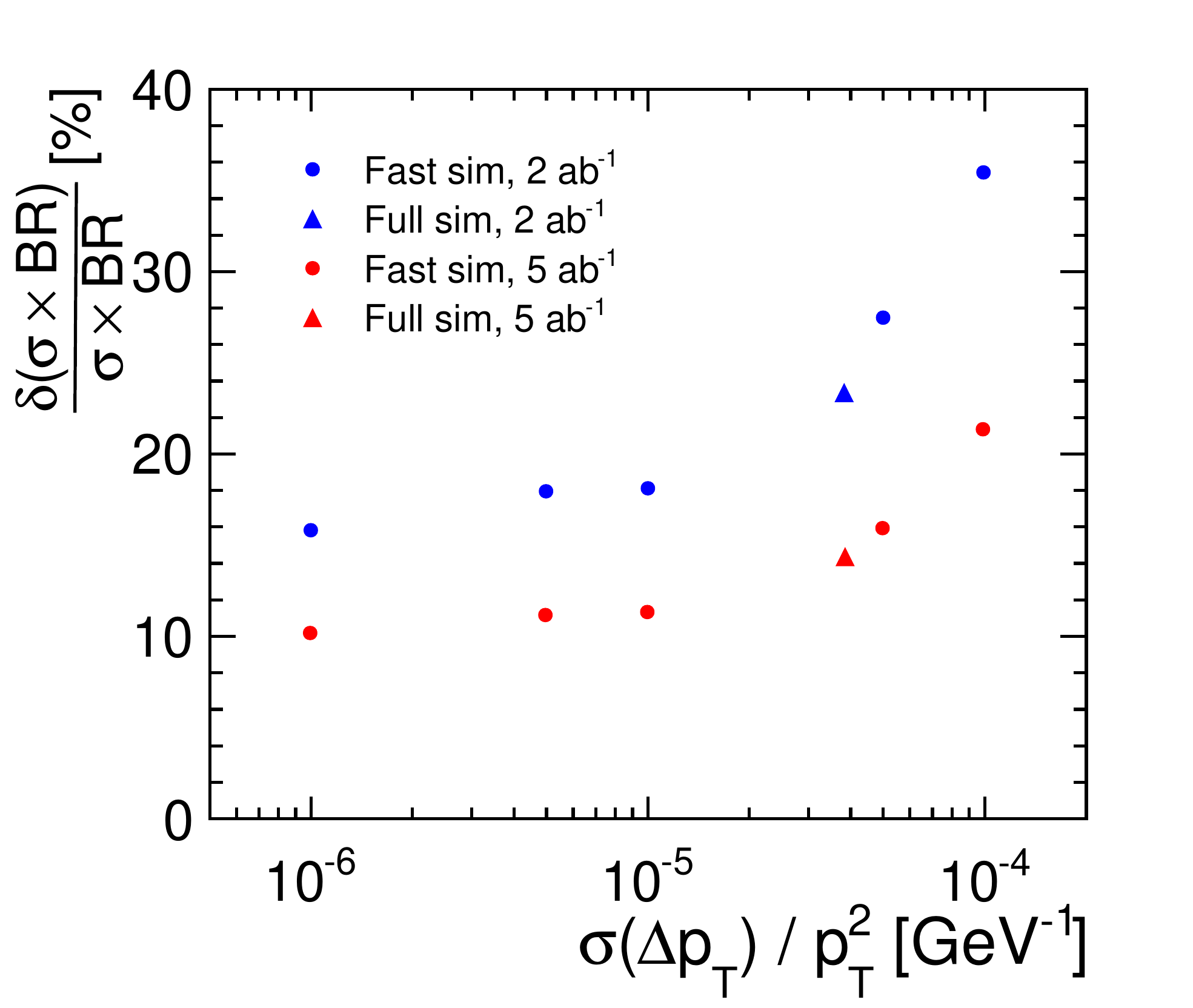}
\end{subfigure}
\caption[Di-muon invariant mass distribution and relative statistical uncertainty on the cross section times branching ratio measurement depending on the momentum resolution.]{Di-muon invariant mass distribution for different $\Delta\pT/\pT^2$ resolutions (left) and relative statistical uncertainty on the $\PSh \to \mpmm$ cross section times branching ratio measurement depending on the $\Delta\pT/\pT^2$ resolution for an integrated luminosity of 2 and \unit[5]{\abinv} (right).}
\label{fig:momResDependence}
\end{figure}

To assess the impact of the momentum resolution, various samples were created using fast simulation and assuming different widths of the $\Delta\pT/\pT^2$ distribution. For simplicity, flat momentum resolutions are assumed without any dependence on the transverse momentum or the polar angle of the particle. A Gaussian smearing with the width corresponding to the desired momentum resolution has been applied to the true momenta of the reconstructed muons to generate the reconstructed momenta. The reconstructed track angles are unchanged. The broadening of the Higgs peak due to increasing momentum resolution is shown in \cref{fig:momResDependence}~(left).

The event selection and fit of the invariant mass distribution follows the procedure explained in the previous sections using the same boosted decision tree. Only the \ac{BDT} selection cut has been changed to optimize the significance. The di-muon invariant mass distribution of the signal and background contributions are fitted to obtain their expected shapes and the cross section times branching ratio is extracted from 100 toy Monte~Carlo experiments. The impact of the \gghad background is not included but expected to be similar to the results obtained in \cref{sec:BeamInducedBackground}.

\cref{fig:momResDependence}~(right) shows the resulting statistical uncertainty on the cross section times branching ratio measurement depending on the momentum resolution, assuming a total integrated luminosity of \unit[2]{\abinv} and \unit[5]{\abinv}, respectively. The points corresponding to the result from the full simulation sample obtained in \cref{sec:result} are consistent with the results obtained from fast simulations. The relative statistical uncertainty degrades noticeably for momentum resolutions worse than a few times \unit[$10^{-5}$]{GeV$^{-1}$}. The measurement does not improve for better momentum resolutions because of the intrinsic statistical fluctuations in the number of selected signal events. The results of the cross section times branching ratio measurement for different momentum resolutions are summarized in \cref{tab:resultsMomentumResolution}.

\begin{table}
 \centering
 \caption[Summary of the results for the $\PSh \to \mpmm$ branching ratio measurement for different momentum resolutions.]{Summary of the results for the $\PSh \to \mpmm$ branching ratio measurement using fast simulation samples with different momentum resolutions $\sigma(\Delta\pT)/\pT^2$ assuming an integrated luminosity of $\unit[2]{\abinv}$. Given is the corresponding invariant mass resolution $\sigma(\Delta M(\mumu))$ and the resulting statistical uncertainty of the $\sigma_{\PSh\nuenuebar} \times \mathrm{BR}_{\PSh \to \mpmm}$ measurement.}
 \label{tab:resultsMomentumResolution}
\begin{tabular}{r r r}
\toprule
$\sigma(\Delta\pT)/\pT^2$    & $\sigma(\Delta M(\mumu))$ & Stat. uncertainty \\\midrule
\unit[$10^{-3}$]{GeV$^{-1}$} & \unit[6.5]{GeV}           &   -               \\
\unit[$10^{-4}$]{GeV$^{-1}$} & \unit[0.70]{GeV}          & 34.3\%            \\
\unit[$10^{-5}$]{GeV$^{-1}$} & \unit[0.068]{GeV}         & 18.2\%            \\
\unit[$10^{-6}$]{GeV$^{-1}$} & \unit[0.022]{GeV}         & 16.0\%            \\\bottomrule
\end{tabular}

\end{table}

\section{Impact of Forward Electron Tagging}
\label{sec:ElectronTagging}

The results presented in \cref{sec:result} can be improved further by rejecting $\mpmm\epem$ events through electron tagging in the forward calorimeters, thus improving the signal to background ratio. The forward calorimeters LumiCal and BeamCal were not part of the full simulation, since this would have required the addition of the incoherent pair background to every event in order to be realistic. To estimate the possible improvement by including this information, several electron tagging efficiencies are assumed and events are rejected based on the Monte~Carlo truth information. The electron tagging efficiency in the forward calorimeters at \ac{CLIC} in the presence of realistic beam-induced backgrounds is currently being investigated in a dedicated full simulation study.


\subsection{Event Rejection}
\begin{figure}
\begin{subfigure}[]{0.49\textwidth}
 \includegraphics[width=\textwidth]{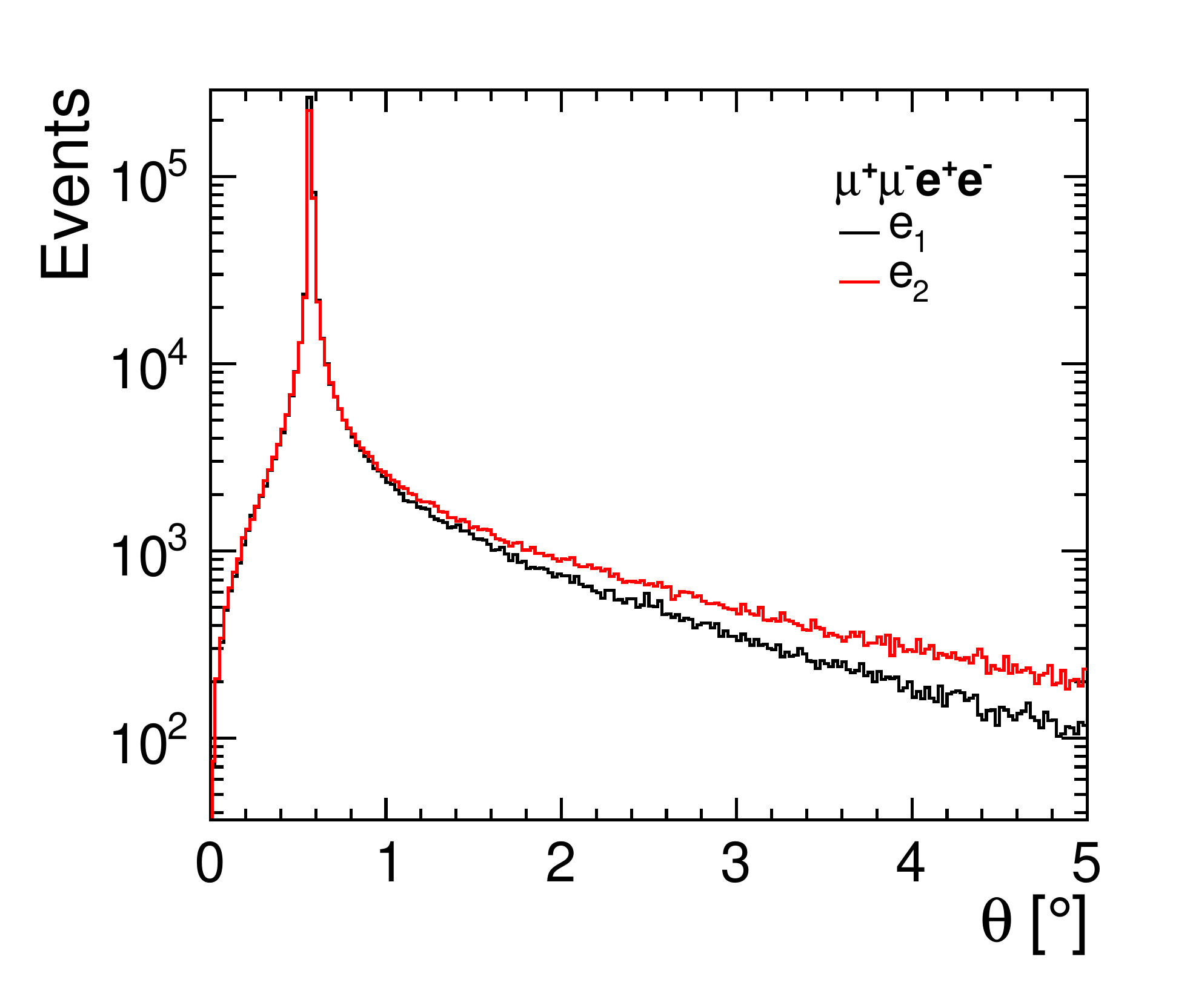}
\end{subfigure}
\hfill
\begin{subfigure}[]{0.49\textwidth}
 \includegraphics[width=\textwidth]{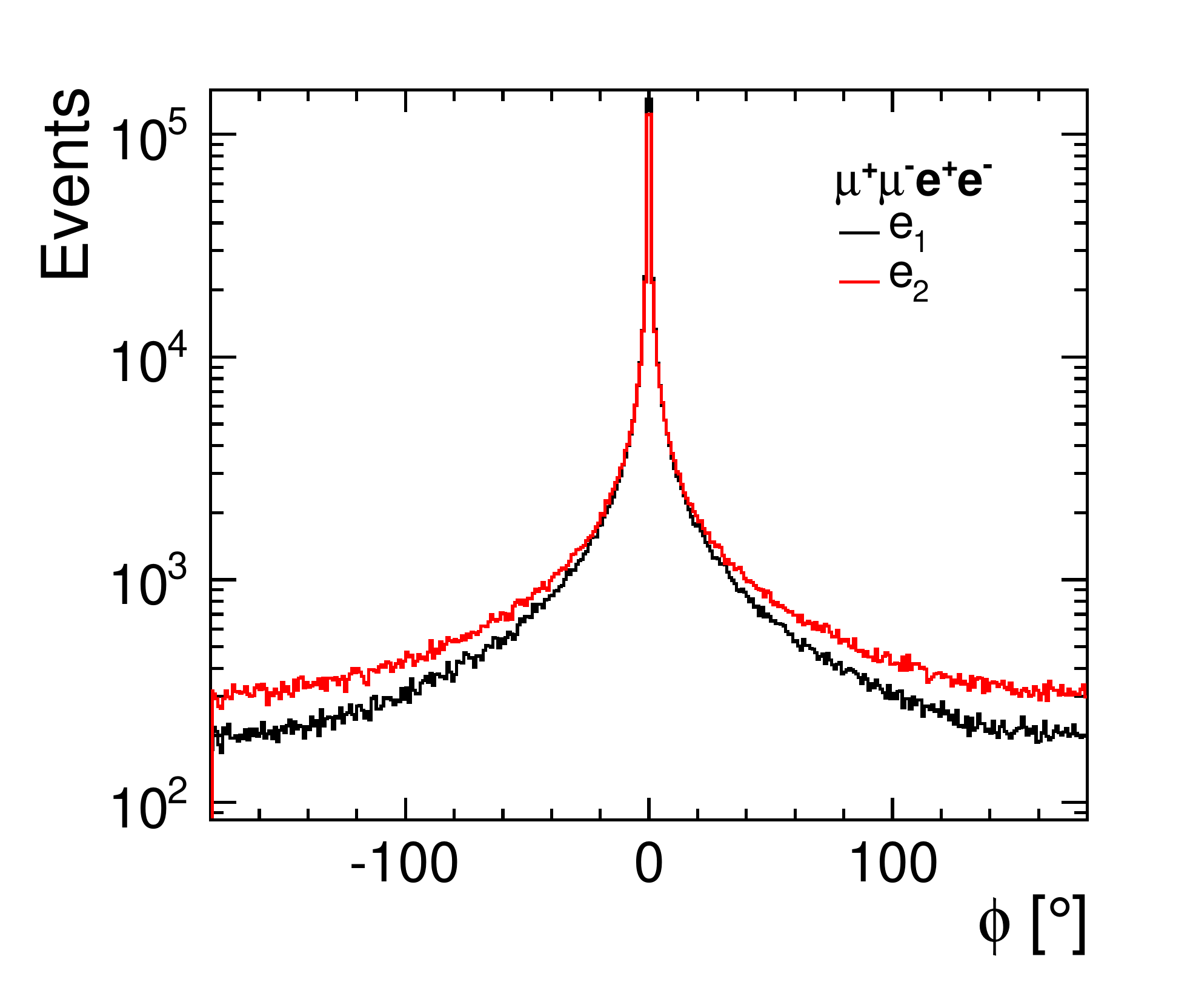}
\end{subfigure}
\caption[Angular distributions of the electrons in $\epem \to \mpmm\epem$.]{Distribution of the polar angle $\theta$ (left) and azimuthal angle $\phi$ (right) of the most energetic electron (e$_1$) and the second most energetic electron (e$_2$) in $\epem \to \mpmm\epem$ events.}
\label{fig:electronAngles}
\end{figure}

The event rejection of the $\epem \to \mpmm\epem$ events is based on the Monte~Carlo truth information of the two electrons, where
e$_1$ is the most energetic electron and e$_2$ is the second most energetic electron.
\cref{fig:electronAngles} shows the polar angle $\theta$ and the azimuthal angle $\phi$ of the two electrons in $\epem \to \mpmm\epem$ events. The electrons are mostly going very forward and peak at a polar angle of \unit[10]{mrad}, which is half of the crossing angle of the two beams. The $\phi$ distribution peaks at 0\degrees due to the inclusion of the crossing angle. The outgoing beam pipe is displaced in the $x$-$z$-plane, which corresponds to an azimuthal angle of 0\degrees.

\begin{figure}
\begin{subfigure}[]{0.49\textwidth}
 \includegraphics[width=\textwidth]{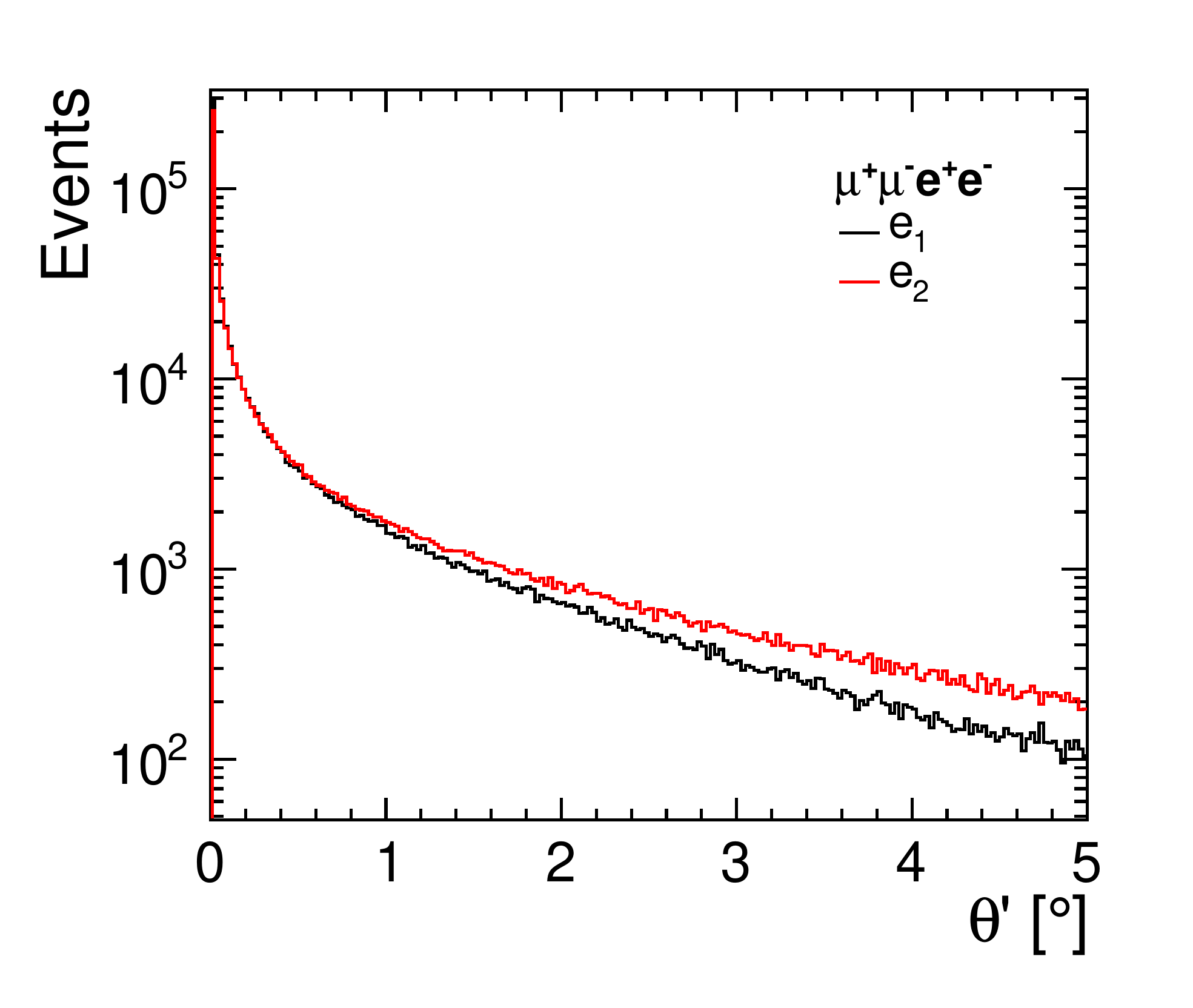}
\end{subfigure}
\hfill
\begin{subfigure}[]{0.49\textwidth}
 \includegraphics[width=\textwidth]{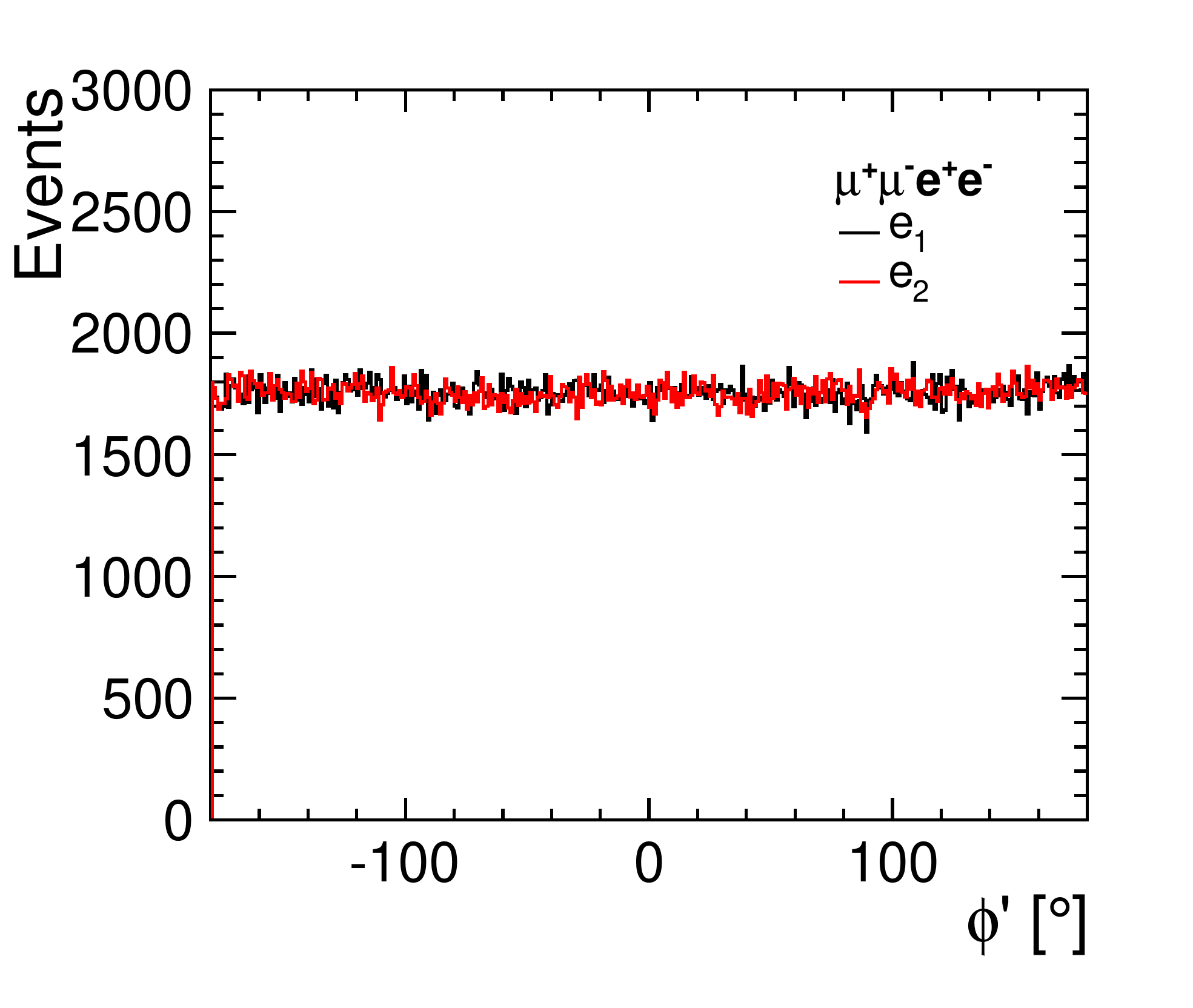}
\end{subfigure}
\caption[Angular distributions of the electrons in $\epem \to \mpmm\epem$ with respect to the outgoing beam axis.]{Distribution of the electron angles with respect to the outgoing beam axis. Polar angle $\theta'$ (left) and azimuthal angle $\phi'$ (right) of the most energetic electron (e$_1$) and the second most energetic electron (e$_2$) in $\epem \to \mpmm\epem$ events.}
\label{fig:electronAnglesTransformed}
\end{figure}

For this study we use the electron directions with respect to the outgoing beam axis instead of the directions with respect to the detector axis. This requires a rotation of the coordinate system around the $y$-axis by \unit[10]{mrad} for $z > 0$ and a rotation by \unit[-10]{mrad} for $z < 0$. The resulting angles are referred to as $\theta'$ for the polar angle and $\phi'$ for the azimuthal angle. Their distributions are shown in \cref{fig:electronAnglesTransformed}.

\begin{figure}
 \centering
 \includegraphics[width=0.49\textwidth]{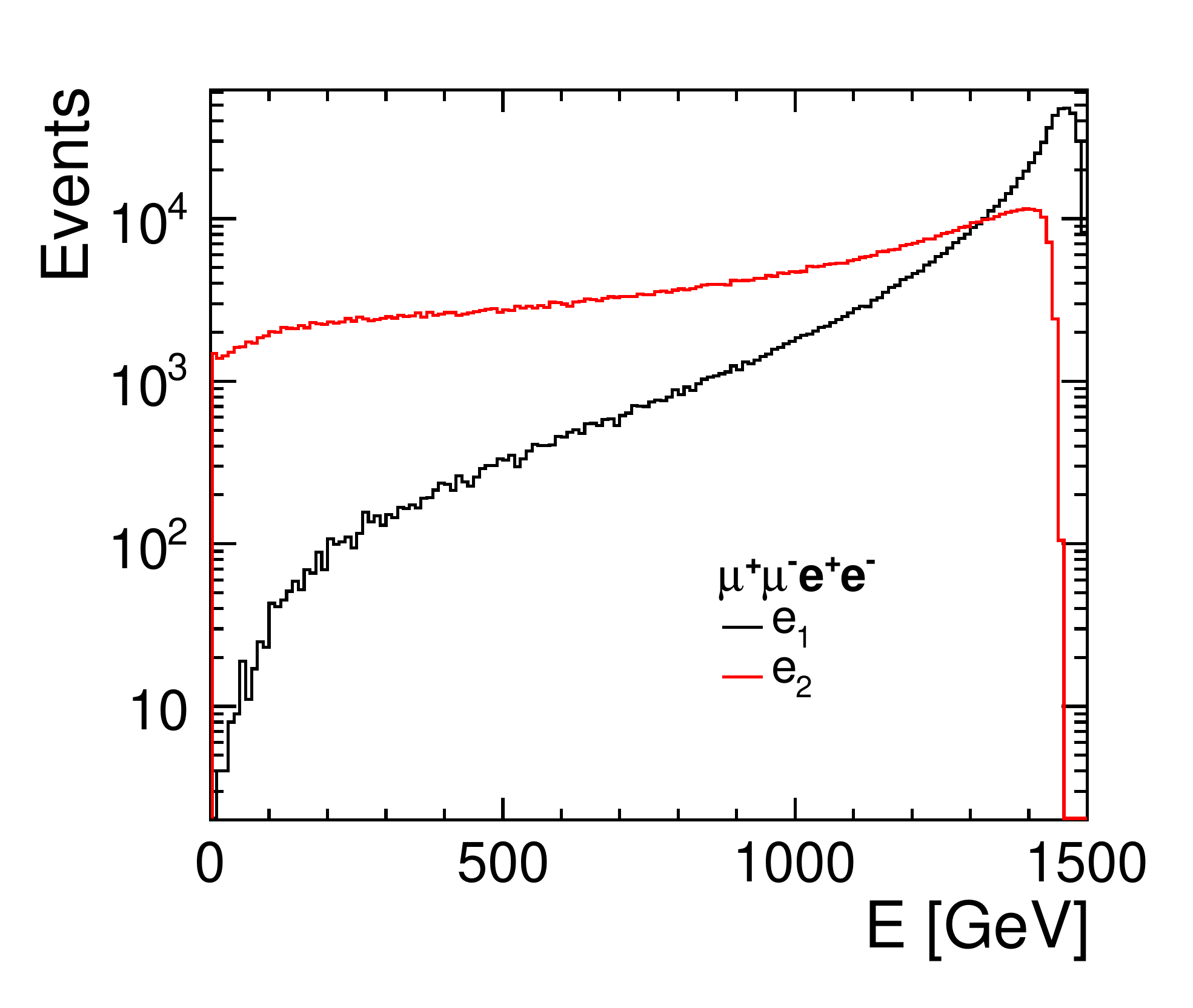}
 \hfill
 \includegraphics[width=0.49\textwidth]{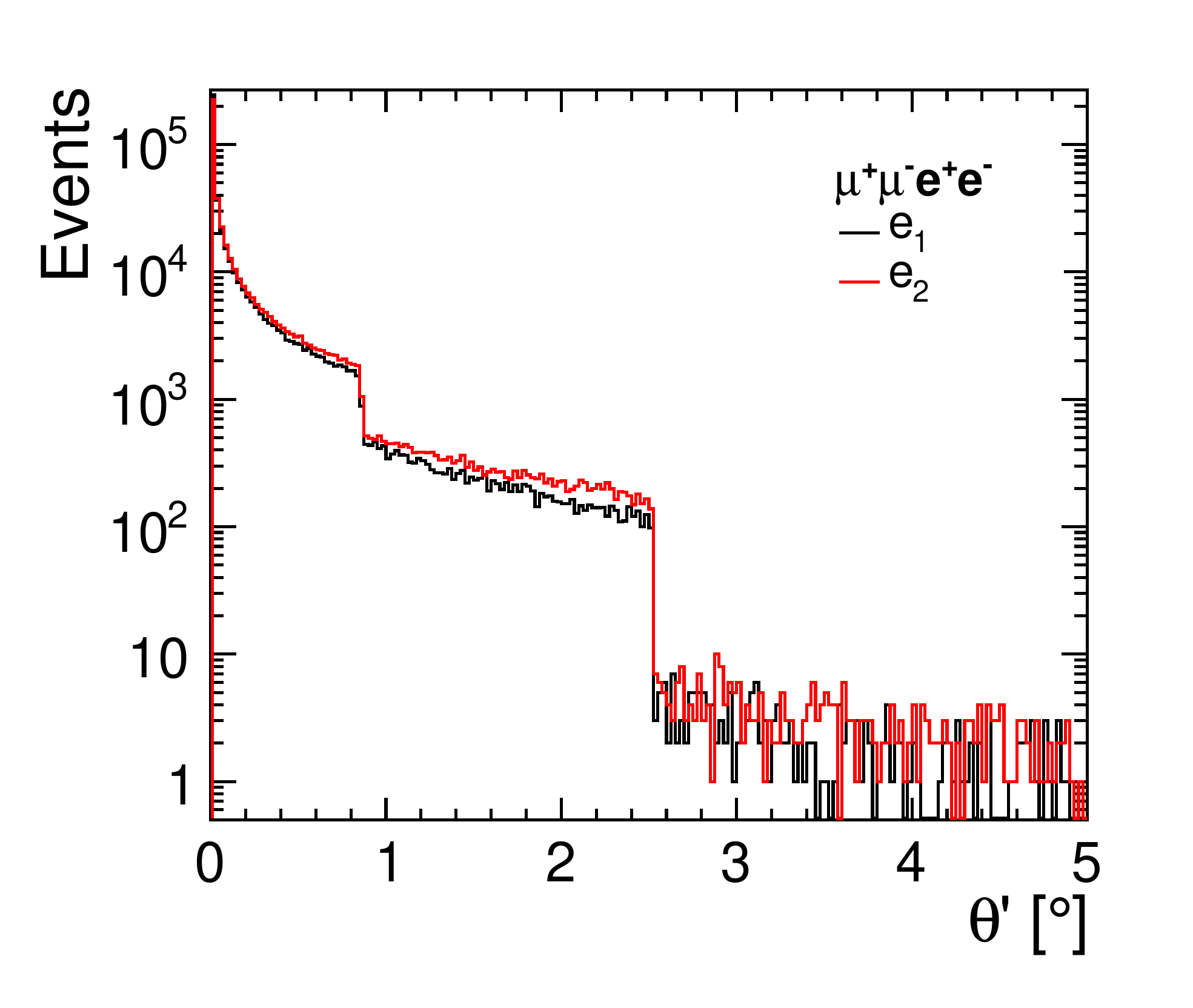}
\caption[Energy distributions of the electrons in $\epem \to \mpmm\epem$ events.]{Distribution of the energy of the most energetic electron (e$_1$) and the second most energetic electron (e$_2$) in $\epem \to \mpmm\epem$ events (left). Distribution of the polar angle $\theta'$ of the most energetic electron (e$_1$) and the second most energetic electron (e$_2$) in $\epem \to \mpmm\epem$ events after applying the BeamCalCut$_{70}$ selection cut (right).}
\label{fig:electronTaggingEnergy}
\end{figure}

In order to reject the $\epem \to \mpmm \epem$ background events, the electrons have to be tagged by identifying their energy deposition in presence of the beam-induced backgrounds. As shown in \cref{fig:electronTaggingEnergy}~(left), the energy of the electrons is usually of the order of several hundreds of GeV up to the full beam energy. We assume an ad-hoc tagging efficiency of 95\% or 99\% is for electrons with a polar angle greater then \unit[44]{mrad}, which corresponds to the inner radius of the fiducial volume of the LumiCal. The two cases are referred to as LumiCalCut$_{95}$, for 95\% efficiency, and LumiCalCut$_{99}$, for 99\% efficiency. The BeamCal on the other hand has a much higher occupancy and electron tagging at these polar angles is thus less efficient. Two cases have been studied, one assuming 30\% electron tagging efficiency for polar angles $\theta'$ greater than \unit[15]{mrad} and one assuming 70\% electron tagging efficiency in the same angular region. These selection cuts are referred to as BeamCalCut$_{30}$ and BeamCalCut$_{70}$, both implicitly including the LumiCalCut$_{99}$ to account for the higher tagging efficiency at larger polar angles. For simplicity, no further dependence of the electron tagging efficiency on the polar angle is assumed and no dependence of the tagging efficiency on the electron energy was introduced. A dedicated full simulation study of the electron tagging efficiency in the \ac{BeamCal}, including the incoherent pair background, found that tagging efficiencies of more than 80\% seem feasible for electron energies of \unit[1]{TeV} and higher~\cite{sailerphd}.

To model the tagging efficiency, two random numbers between 0 and 1 are assigned to each event, one for each electron. An event is rejected if any of the two most energetic electrons has a polar angle $\theta'$ greater than the assumed acceptance angle and the corresponding random number is above the assumed efficiency. \cref{fig:electronTaggingEnergy}~(right) shows the impact of the BeamCalCut$_{70}$ event selection on the distribution of the polar angles of the electrons. The number of events drops sharply at the polar angles corresponding to the assumed BeamCal and LumiCal acceptance, but most of the $\epem \to \mpmm\epem$ events are unaffected by these selection cuts, since in most cases both electrons are leaving the detector outside of the BeamCal acceptance. For example, more than 70\% of the events pass the BeamCalCut$_{70}$ rejection cut. The resulting total event rejection efficiencies for the different cuts are summarized in \cref{tab:electronTaggingCuts}.

\begin{table}
 \centering
 \caption[Event selection cuts for removing $\epem \to \mpmm\epem$ events based on the electron angles.]{Event selection cuts for removing $\epem \to \mpmm\epem$ events. The assumed single electron tagging efficiency $\epsilon_{\mathrm e}$, the minimum polar angle required to tag the electron $\theta'_{\mathrm{min}}$ and the resulting efficiency to reject events $\epsilon_{\mathrm{reject}}$ are given.}
 \label{tab:electronTaggingCuts}
\begin{threeparttable}
\begin{tabular}{l r r r}
\toprule
                  & $\epsilon_{\mathrm e}$ [\%] & $\theta'_{\mathrm{min}}$ [mrad] & $\epsilon_{\mathrm{reject}}$ [\%] \\\midrule
LumiCalCut$_{95}$ & 95                          & 44                              & 15.1                              \\
LumiCalCut$_{99}$ & 99                          & 44                              & 15.7                              \\
BeamCalCut$_{30}$\tnote{$\ast$} & 30            & 15                              & 21.5                              \\
BeamCalCut$_{70}$\tnote{$\ast$} & 70            & 15                              & 29.0                              \\\bottomrule
\end{tabular}
\begin{tablenotes}
 \item[$\ast$] Implicitely includes LumiCalCut$_{99}$.
\end{tablenotes}
\end{threeparttable}
\end{table}

\subsection{Coincidence with Bhabha Scattering Events}

Vetoing events based on low-angle electrons bears of course the danger of rejecting signal events if they coincide within the readout time window for example with a Bhabha scattering event. The total cross section for the $\epem \to \epem$ process with a minimum electron polar angle of \unit[44]{mrad} and a minimum electron energy of \unit[100]{GeV} is approximately \unit[60]{pb}. With a nominal luminosity of \unit[$5.9\times10^{34}$]{cm$^{-2}$s$^{-1}$} and a rate of 50 bunch trains per second, 0.071 events of this kind are expected per bunch train. Each bunch train consists of 312 bunches and has a length of \unit[156]{ns}. If a time stamping of \unit[10]{ns} is assumed in addition, the probability of coincidence with a signal event within this time window is less than 0.5\% and of no concern.

If a minimum electron polar angle of only \unit[15]{mrad} is required, the total cross section of the $\epem \to \epem$ process rises to approximately \unit[650]{pb}. This corresponds to 0.767 events per bunch train and 0.051 events within a \unit[10]{ns} time window. This means that a significant amount of signal events of about 5\% would be rejected.

The cross section for the $\epem \to \epem f\bar{f}$ process, with similar cuts on the electrons and a maximum fermion polar angle of \unit[100]{mrad}, is always more than one order of magnitude smaller than the cross section of the \epem process and can be neglected.

\subsection{Event Selection}

The final event selection is done similarly to the strategy introduced in \cref{sec:EventSelection} using boosted decision trees. Since the electron angles are correlated with the angle of the di-muon system as well as the muon angles, the distributions of the input variables used in the boosted decision tree change when the electron cuts are applied during the event pre-selection. Dedicated BDTs are thus trained for each of the electron cuts and the \ac{BDT} value of the final event selection cut is determined as the value with the highest significance $N_{\mathrm S}/\sqrt{N_{\mathrm S} + N_{\mathrm B}}$.

Since most of the $\epem \to \mpmm\epem$ events with a high \pT of the di-muon system are removed a priori by the electron rejection cuts, the background rejection by the BDT is largely improved. The signal selection efficiency increases to about 49\% with the LumiCalCut$_{95}$ and almost 56\% with the BeamCalCut$_{70}$ selection.



\subsection{Branching Ratio Measurement}
\label{sec:Higgs_taggingBR}

\begin{figure}
 \centering
 \includegraphics[width=0.49\textwidth]{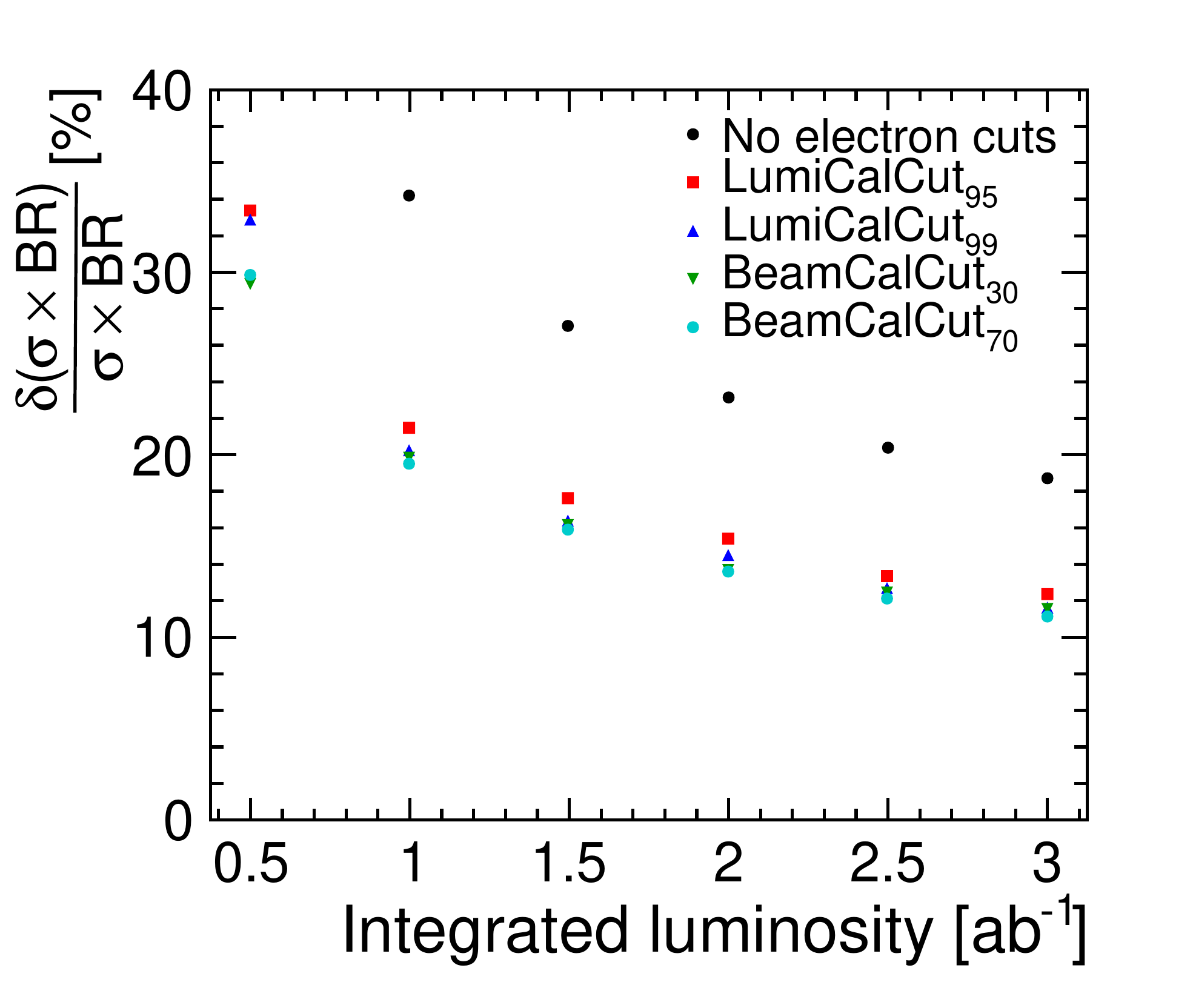}
\caption[Relative statistical uncertainty on the $\PSh \to \mpmm$ cross section times branching ratio measurement depending on the integrated luminosity.]{Relative statistical uncertainty on the $\PSh \to \mpmm$ cross section times branching ratio measurement depending on the integrated luminosity with and without assuming electron tagging in the forward calorimeters.}
\label{fig:electronTaggingBRMeasurement}
\end{figure}

The statistical uncertainty on the cross section times branching ratio measurement is determined similarly to \cref{sec:MassFit}. First the shapes of the di-muon invariant mass distributions for the individual signal and background contributions are determined. Then the signal plus background model is used to measure the number of signal events in 100 toy Monte~Carlo experiments and finally the cross section times branching ratio is calculated using the selection efficiency. \cref{fig:electronTaggingBRMeasurement} shows the uncertainty that can be achieved, depending on the integrated luminosity and the electron rejection cut applied. The results for an integrated luminosity of \unit[2]{\abinv} are given in \cref{tab:electronTaggingResults}.

\begin{table}
 \centering
 \caption[Summary of the results for the $\PSh \to \mpmm$ branching ratio measurement assuming different electron tagging efficiencies in the forward calorimeters.]{Summary of the results for the $\PSh \to \mpmm$ branching ratio measurement assuming different electron tagging efficiencies in the forward calorimeters for an integrated luminosity of $\unit[2]{\abinv}$.}
 \label{tab:electronTaggingResults}
\begin{tabular}{l r r r r}
\toprule
                                                              & LumiCalCut$_{95}$ & LumiCalCut$_{99}$ & BeamCalCut$_{30}$ & BeamCalCut$_{70}$ \\\midrule
Signal events                                                 & $120 \pm 17$      & $127 \pm 18$      & $130 \pm 18$      & $132 \pm 18$      \\
Signal efficiency                                             & 49.3\%            & 53.2\%            & 55.1\%            & 55.9\%            \\
$\sigma_{\PSh\nuenuebar} \times \mathrm{BR}_{\PSh \to \mpmm}$ & \unit[0.121]{fb}  & \unit[0.119]{fb}  & \unit[0.118]{fb}  & \unit[0.118]{fb}  \\
Stat. uncertainty                                             & 15.0\%            & 14.3\%            & 14.1\%            & 13.8\%            \\\bottomrule
\end{tabular}

\end{table}

\begin{figure}
 \begin{subfigure}[]{0.49\textwidth}
  \includegraphics[width=\textwidth]{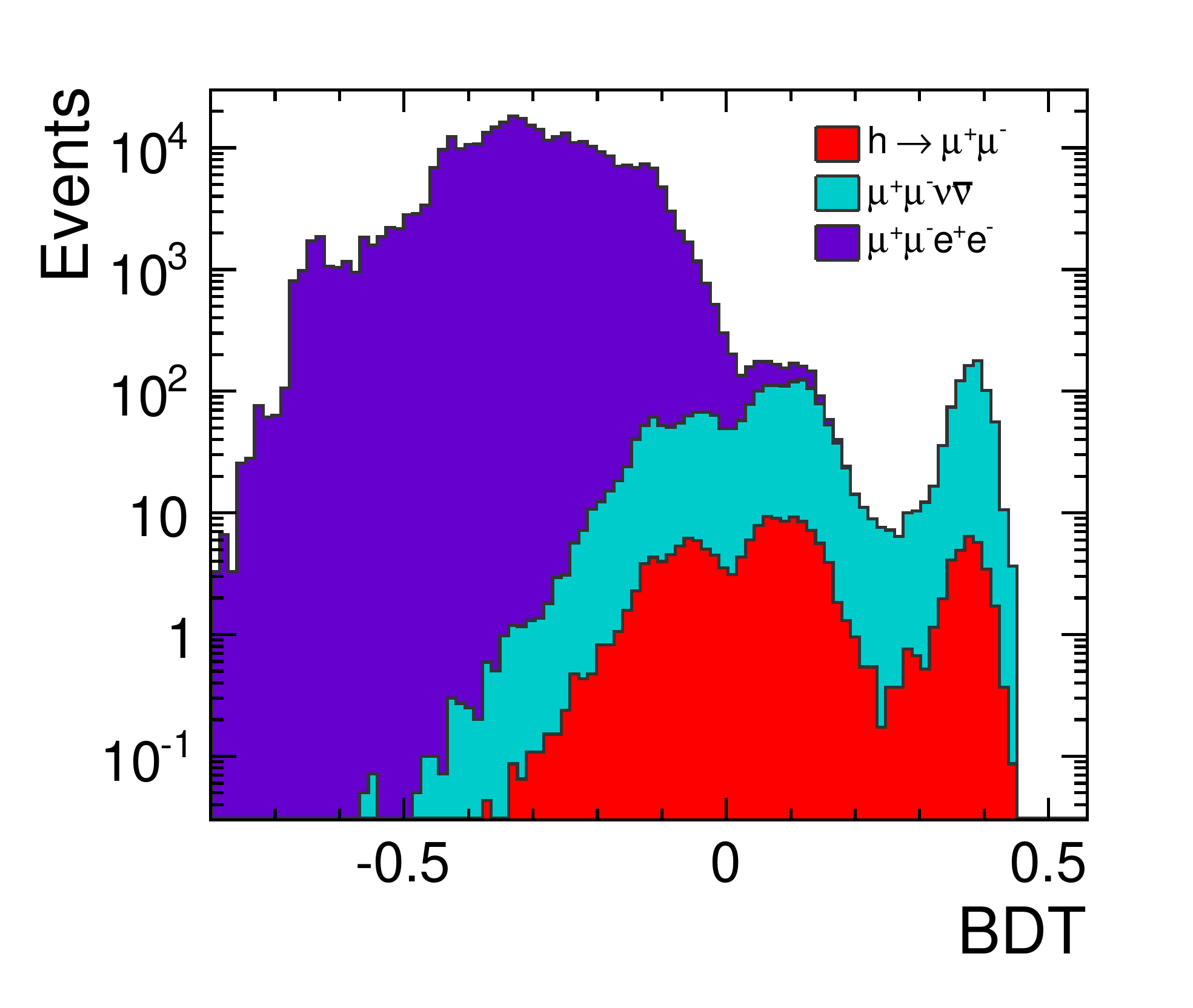}
 \end{subfigure}
\hfill
 \begin{subfigure}[]{0.49\textwidth}
    \includegraphics[width=\textwidth]{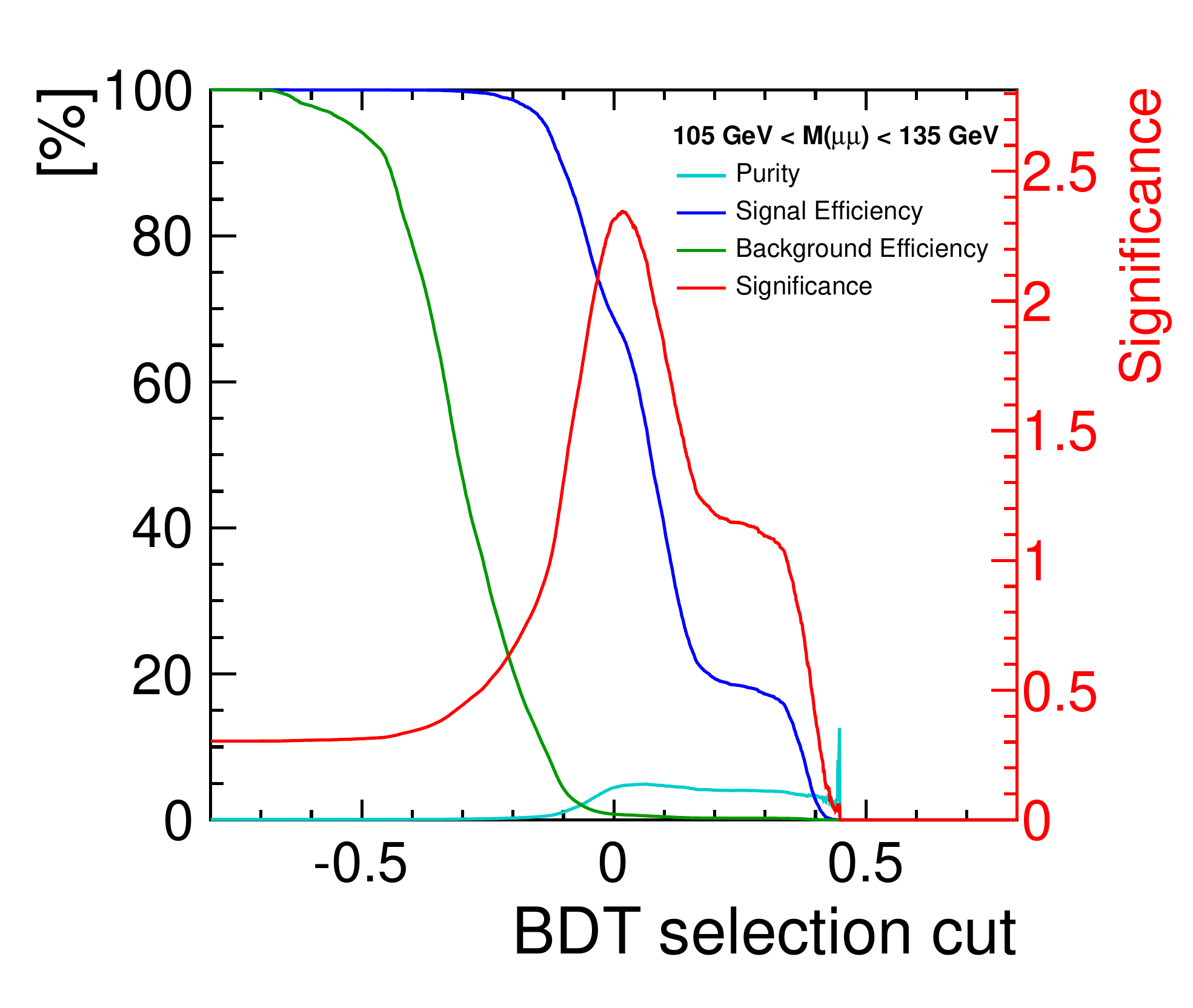}
 \end{subfigure}
\caption[Response of the boosted decision tree classifier in case of the LumiCalCut$_{95}$]{Response of the boosted decision tree classifier for the signal and the two most important background processes (left) and the resulting significance, purity, signal efficiency and background efficiency (right), in case of the LumiCalCut$_{95}$.}
\label{fig:bdt_lumi_95}
\end{figure}

\begin{figure}
 \begin{subfigure}[]{0.49\textwidth}
  \includegraphics[width=\textwidth]{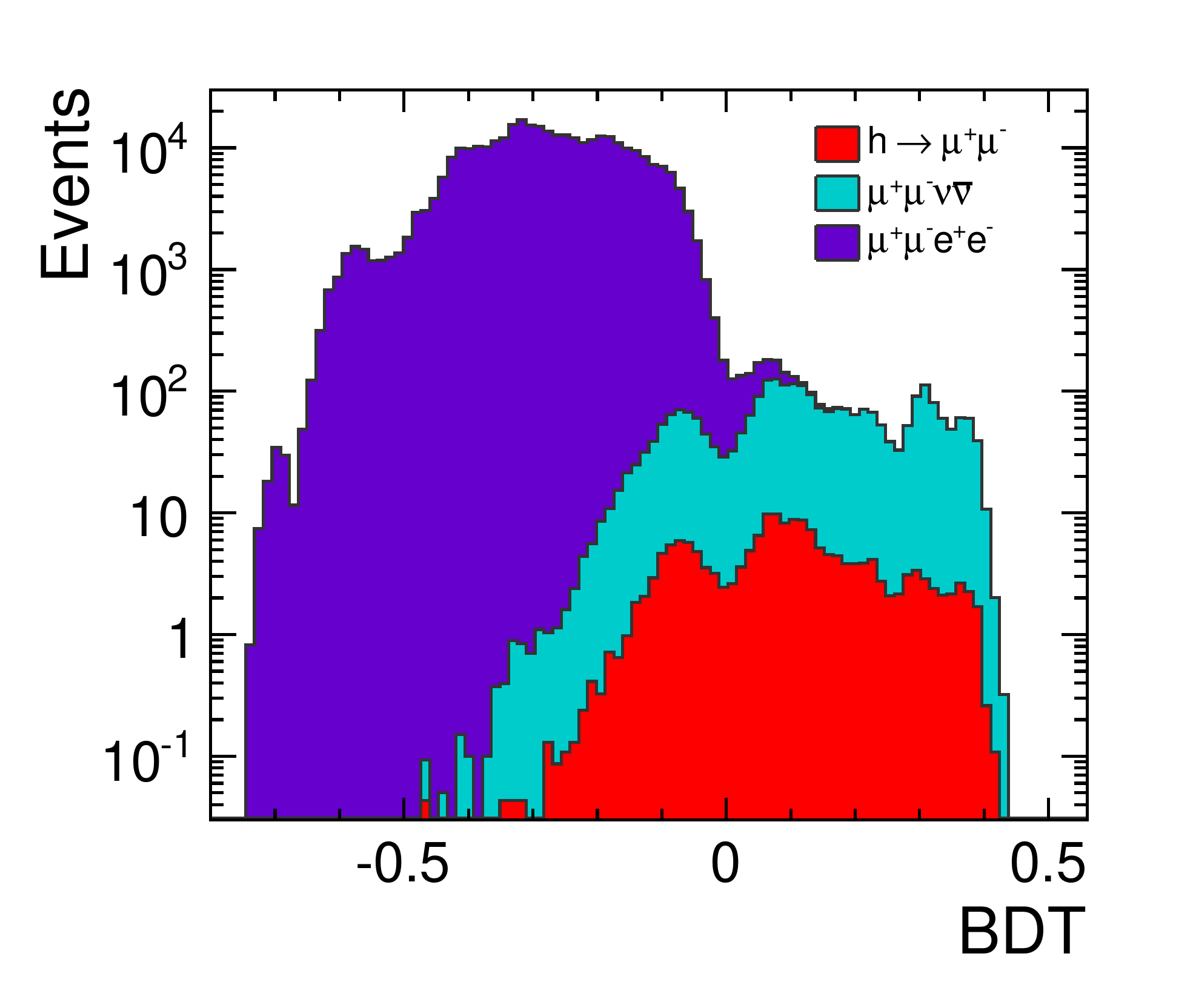}
 \end{subfigure}
\hfill
 \begin{subfigure}[]{0.49\textwidth}
    \includegraphics[width=\textwidth]{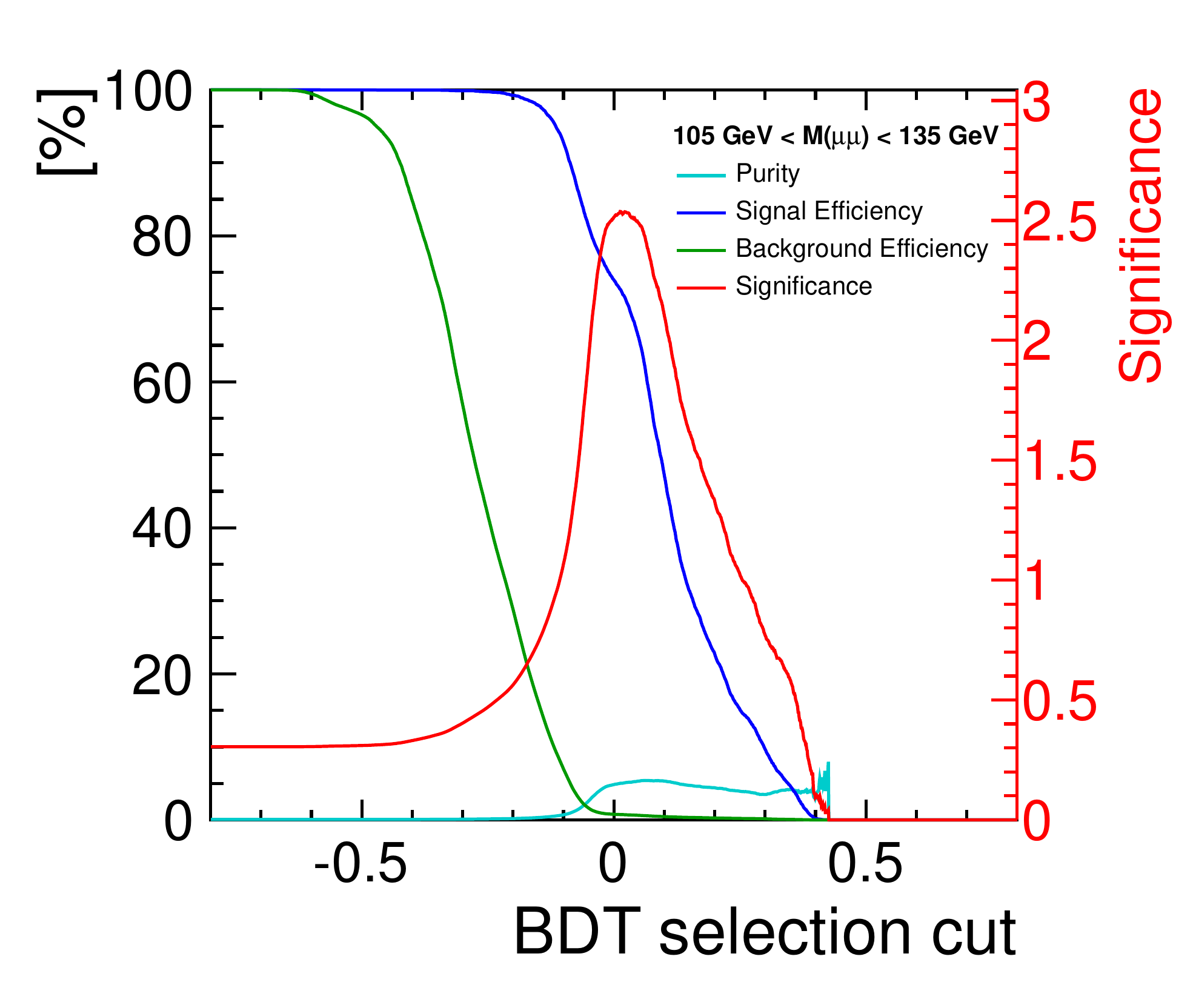}
 \end{subfigure}
\caption[Response of the boosted decision tree classifier in case of the LumiCalCut$_{99}$]{Response of the boosted decision tree classifier for the signal and the two most important background processes (left) and the resulting significance, purity, signal efficiency and background efficiency (right), in case of the LumiCalCut$_{99}$.}
\label{fig:bdt_lumi_99}
\end{figure}

\begin{figure}
 \begin{subfigure}[]{0.49\textwidth}
  \includegraphics[width=\textwidth]{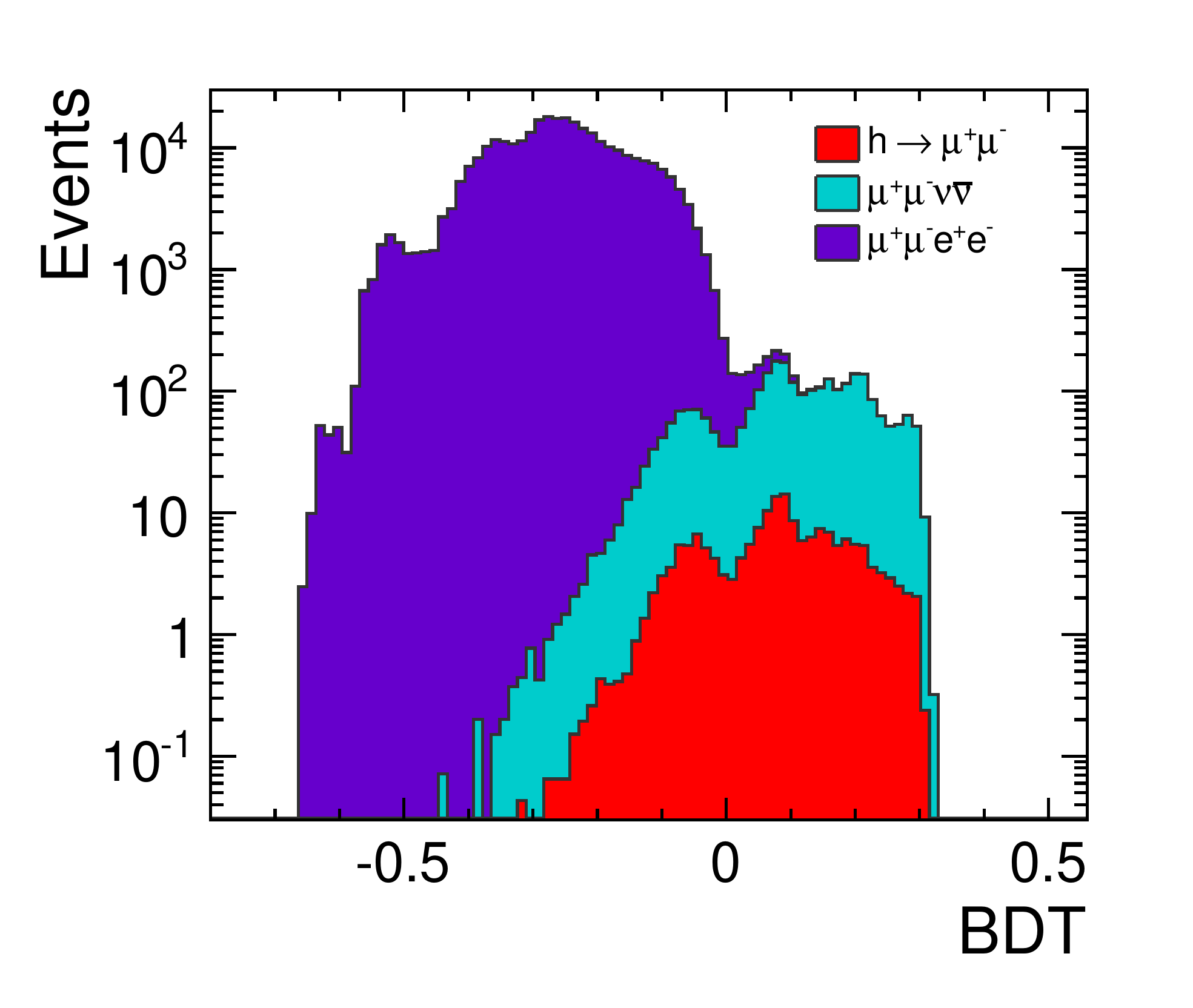}
 \end{subfigure}
\hfill
 \begin{subfigure}[]{0.49\textwidth}
    \includegraphics[width=\textwidth]{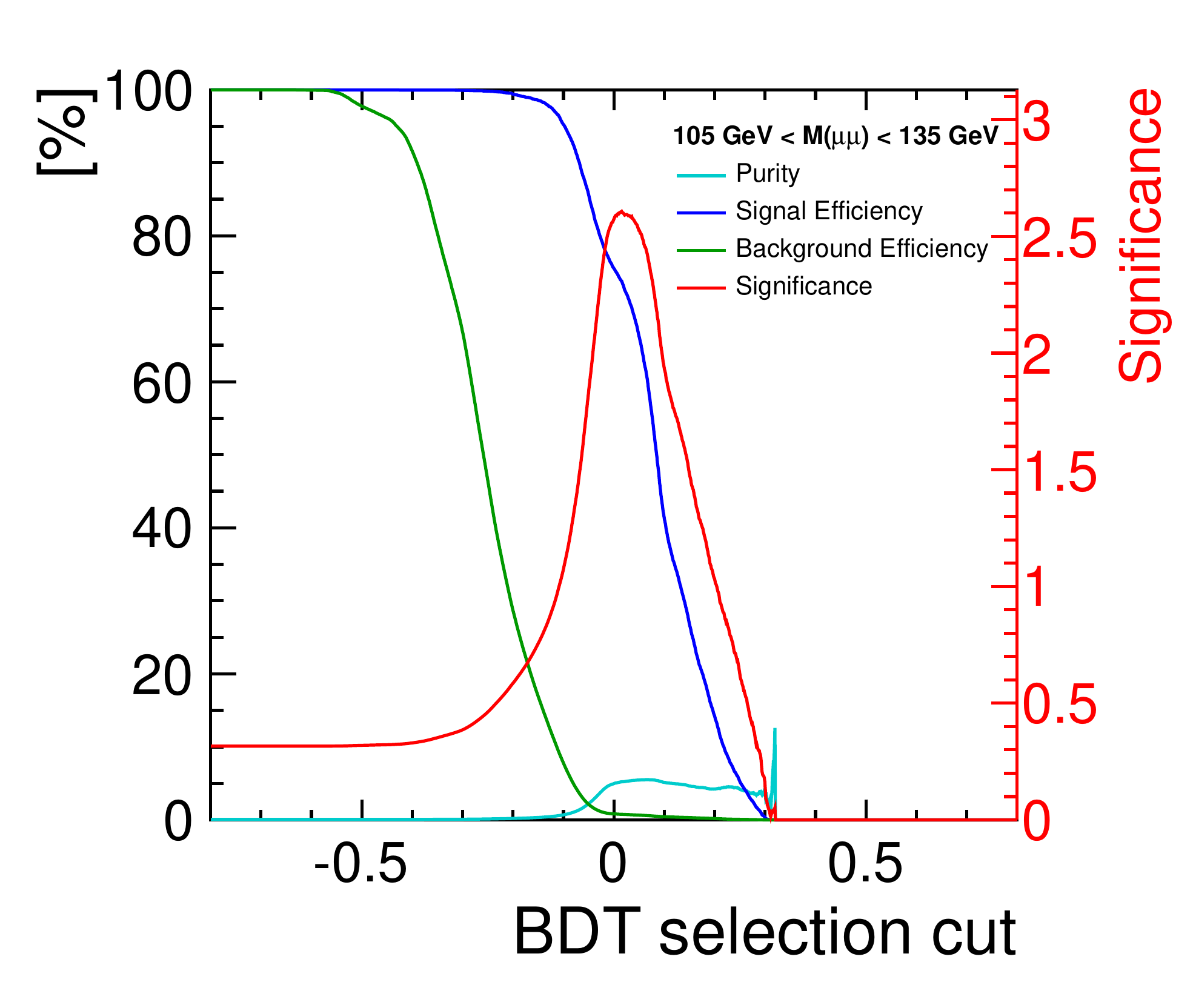}
 \end{subfigure}
\caption[Response of the boosted decision tree classifier in case of the BeamCalCut$_{30}$]{Response of the boosted decision tree classifier for the signal and the two most important background processes (left) and the resulting significance, purity, signal efficiency and background efficiency (right), in case of the BeamCalCut$_{30}$.}
\label{fig:bdt_beam_30}
\end{figure}

\begin{figure}
 \begin{subfigure}[]{0.49\textwidth}
  \includegraphics[width=\textwidth]{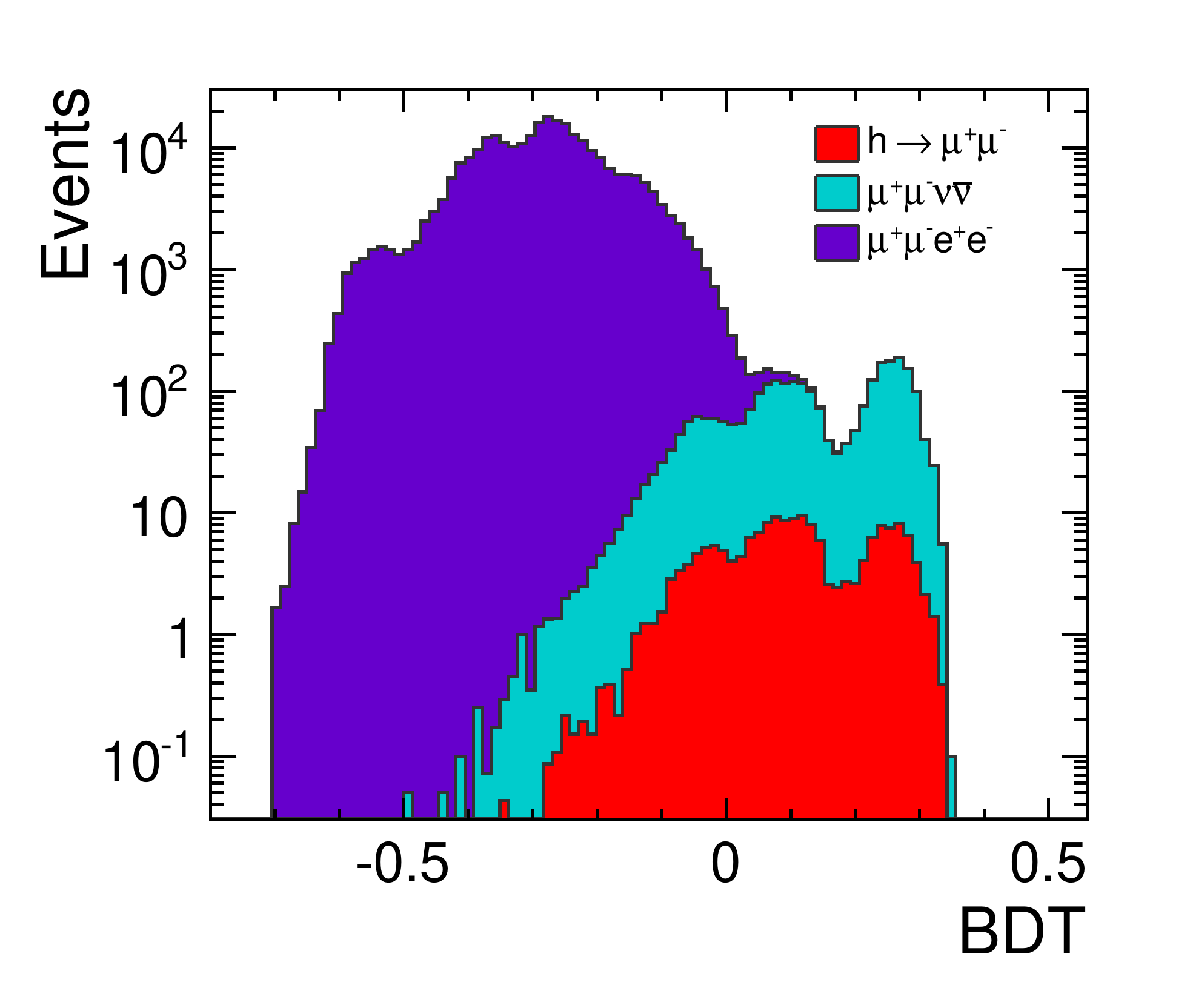}
 \end{subfigure}
\hfill
 \begin{subfigure}[]{0.49\textwidth}
    \includegraphics[width=\textwidth]{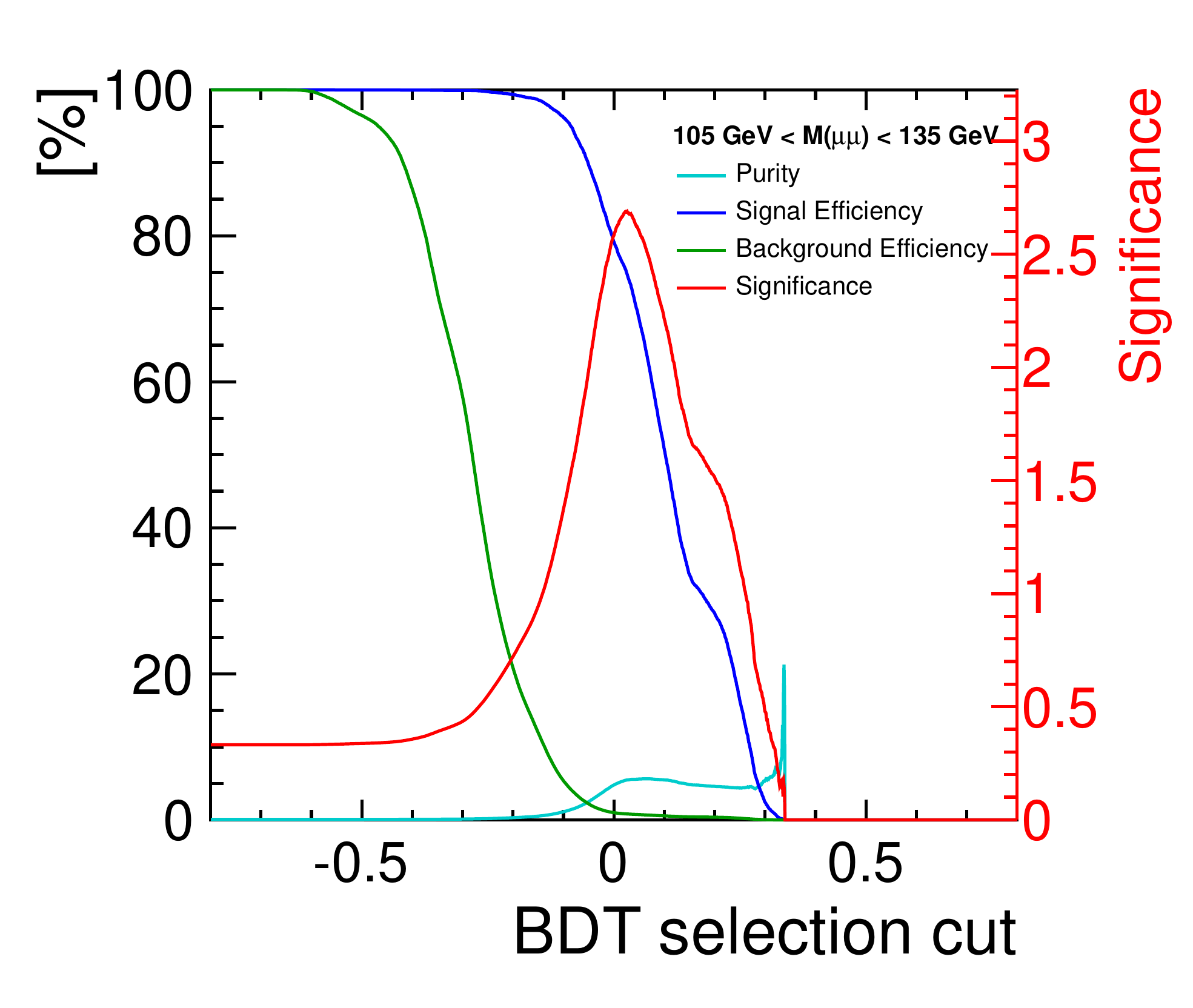}
 \end{subfigure}
\caption[Response of the boosted decision tree classifier in case of the BeamCalCut$_{70}$]{Response of the boosted decision tree classifier for the signal and the two most important background processes (left) and the resulting significance, purity, signal efficiency and background efficiency (right), in case of the BeamCalCut$_{70}$.}
\label{fig:bdt_beam_70}
\end{figure}

Since the statistical uncertainty is directly related to the number of signal events selected, the higher selection efficiency that can be achieved when rejecting some of the $\epem \to \mpmm\epem$ events a priori leads to a big improvement of the cross section times branching ratio measurement from a relative uncertainty of 23\% down to 15\% when using the LumiCalCut$_{95}$. Assuming even higher electron tagging efficiencies and also electron tagging in the BeamCal can improve the statistical uncertainety of the cross section times branching ratio measurement to less than 14\%. The additional improvement is only modest since the background rejection based on the transverse momentum of the di-muon system is already very efficient for low $\pT(\mumu)$. This is also reflected in the small differences in significance achievable by the \ac{BDT} event selection shown in Figures~\ref{fig:bdt_lumi_95}, \ref{fig:bdt_lumi_99}, \ref{fig:bdt_beam_30} and \ref{fig:bdt_beam_70}.

Similar to the study in \cref{sec:BeamInducedBackground}, the analysis was repeated with the signal sample that had been overlaid with the \gghad background for the case of the LumiCalCut$_{95}$ selection cut. A different \ac{BDT} without the $E_{\mathrm{vis}}$ variable was trained and instead a pre-selection cut of $E_{\mathrm{vis}} < \unit[150]{GeV}$ was used. The most significant selection cut yields a signal selection efficiency of 49.7\% and results in a relative statistical uncertainty on the $\sigma_{\PSh\nuenuebar} \times \mathrm{BR}_{\PSh \to \mpmm}$ measurement of 15.7\%. The result degrades only by a small amount since the $E_{\mathrm{vis}}$ variable is mostly used to reject background events with electrons within the detector acceptance which are mostly removed by the LumiCalCut$_{95}$ cut instead. This result would correspond to a significance of almost $6.4\sigma$.

These results show that a high electron tagging in the LumiCal can improve the measurement substantially. Efficient electron tagging in the BeamCal would improve the results only by a small amount. It also significantly increases the probability of rejecting signal events if they coincide with low-angle $\epem \to \epem$ events.

\section{Extracting the Higgs Coupling}
\label{sec:Higgs_extractCouplings}

Once the cross-section times branching ratio has been measured the Yukawa coupling constant $g_{\PGm}$ can be determined. In general this has to be done by performing a global fit to all branching ratio measurements involving the Higgs. This allows to determine all coupling constants simultaneously including their correlations. An example of such a global fit is presented in~\cite{Desch:2001xh} or more recently in~\cite{Rauch:2012wa}.

If we assume that all Higgs couplings are exactly as in the \ac{SM} except for the coupling to muons, we can estimate the sensitivity on this coupling just from the branching ratio measurement presented here. If the coupling to muons differs from the \ac{SM} expectation by a value $\delta_g$, \ie $g_{\PGm} / g_{\PGm \text{, SM}} = 1 + \delta_g$, the partial width can be written as $\Gamma_{\PGm} / \Gamma_{\PGm \text{, SM}} = \left(1 + \delta_g\right)^2$, using \cref{eq:higgsBranchinRatio}. The branching ratio is then given by
\begin{align}
 \text{BR}_{\PSh \to \mpmm} &= \frac{\Gamma_{\PGm}}{\Gamma} \\
                            &= \frac{\Gamma_{\PGm \text{, SM}}\left(1 + \delta_g\right)^2}{\Gamma_\text{SM} - \Gamma_{\PGm \text{, SM}} + \Gamma_{\PGm \text{, SM}}\left(1 + \delta_g\right)^2} \\
                            &= \frac{\left(1 + \delta_g\right)^2}{\frac{1}{\text{BR}_{\PGm \text{, SM}}} - 1 + \left(1 + \delta_g\right)^2}.
\label{eq:Higgs_BRmod}
\end{align}
Since we assumed all other couplings to be as in the \ac{SM}, the production cross section is also as in the \ac{SM} and thus
\begin{equation}
 \frac{\sigma_{\PSh\nuenuebar} \times \mathrm{BR}_{\PSh \to \mpmm}}{\left(\sigma_{\PSh\nuenuebar} \times \mathrm{BR}_{\PSh \to \mpmm}\right)_\text{SM}} = \frac{\text{BR}_{\PSh \to \mpmm}}{\text{BR}_{\PSh \to \mpmm \text{, SM}}}.
\end{equation}
If $\delta_g$ is small, we can use \cref{eq:Higgs_BRmod} and expand it around 0
\begin{equation}
 \frac{\text{BR}_{\PSh \to \mpmm}}{\text{BR}_{\PSh \to \mpmm \text{, SM}}} = 1 + 2(1 - \text{BR}_{\PSh \to \mpmm \text{, SM}})\delta_g + \mathcal{O}(\delta_g^2) + \dots
\end{equation}
The measured relative uncertainty on the cross section times ratio thus translates to a sensitivity on the coupling as
\begin{equation}
 \frac{\delta(\sigma_{\PSh\nuenuebar} \times \mathrm{BR}_{\PSh \to \mpmm})}{\sigma_{\PSh\nuenuebar} \times \mathrm{BR}_{\PSh \to \mpmm}} = 2(1 - \text{BR}_{\PSh \to \mpmm \text{, SM}})\delta_g \approx 2\ \delta_g.
\end{equation}
For example, the relative statistical uncertainty of 15.7\% obtained in \cref{sec:Higgs_taggingBR}, translates to an uncertainty on $g_{\PGm}$ of approximately 7.9\%.


\section{Measurement Potential at other Accelerators}
\label{sec:Higgs_otherPotential}

\subsection{Comparison with \acs{ILC} Potential}
\label{sec:Higgs_ILCpotential}
The measurement of Higgs decays into muons at the \ac{ILC} was studied in the \ac{SiD} detector concept in~\cite{Aihara:2009ad}. In that study a center-of-mass energy of \unit[250]{GeV} and an integrated luminosity of \unit[250]{\fbinv} is assumed. Like in our study, a \ac{SM} Higgs boson with a mass of \unit[120]{GeV} is assumed. At that $\sqrt{s}$ the Higgsstrahlung process is the dominating Higgs production process. In the hadronic decay channel of the \PZ boson, $\epem \to \PZ\PSh \to \qqbar \mpmm$, 7.7 signal events are left after the final event selection, resulting in a significance of only $1.1\sigma$ is achieved. The final state where the \PZ decays invisibly, $\epem \to \PZ\PSh \to \nunubar \mpmm$, has an even lower number of signal events left after the event selection. Nevertheless, these 2.7 events correspond to a significance of $1.8\sigma$, since the backgrounds can be removed very efficiently

A combination of these two analyses will improve the significance but ultimately more luminosity will be required. For example, in order to reach $5\sigma$ in the neutrino channel, 7.7 times more luminosity would be required, all collected at $\sqrt{s} = \unit[250]{GeV}$. For higher center-of-mass energies the Higgs production cross section drops significantly until the \ww fusion process becomes dominating. At around $\sqrt{s} = \unit[250]{GeV}$ the Higgs production cross section would be comparable to that at the $\PZ\PSh$ threshold, but the impact of the backgrounds at that energy can not be easily estimated.

From this comparison one can see that a multi-TeV \ac{CLIC} has two advantages over a low energy linear collider for measuring this rare decay. First, the Higgs production cross section is larger at high energies due to the $\sqrt{s}$ dependence. Secondly, \ac{CLIC} has a higher total luminosity which can, at least in this channel, be fully exploited despite the long tail in the luminosity spectrum.

\subsection{Comparison with \acs{LHC} Potential}
\label{sec:Higgs_LHCpotential}

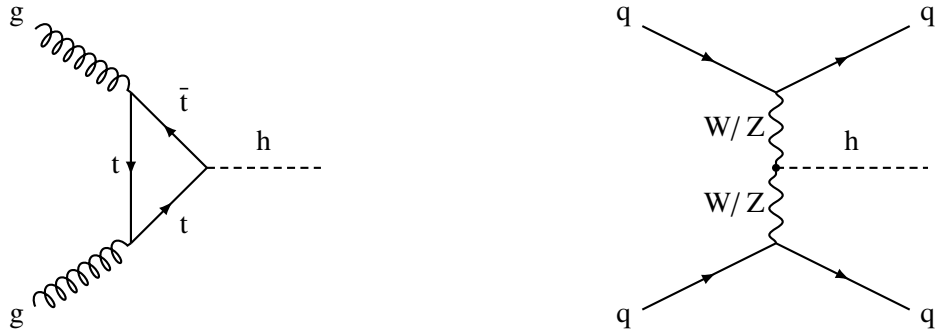
\begin{figure}
 \centering
 \begin{subfigure}[]{0.49\textwidth}
  \centering
  \begin{tikzpicture}[thick, >=latex]

   \node (gluon1) at (-2.0,2.0) {\Pg};
   \node (gluon2) at (-2.0,-2.0) {\Pg};
   \coordinate (top1) at (-0.5, 1.0);
   \coordinate (top2) at (-0.5, -1.0);
   \coordinate (higgsLeft) at (0.5,0.0);
   \coordinate (higgsRight) at (2.0,0.0);

   \draw[gluon] (gluon1) to (top1);
   \draw[fermion] (top1) to node[left] {\PQt} (top2);
   \draw[antifermion] (top1) to node[above right] {\PAQt} (higgsLeft);
   \draw[fermion] (top2) to node[below right] {\PQt} (higgsLeft);
   \draw[higgs] (higgsLeft) to node[above] {\PSh} (higgsRight);
   \draw[gluon] (gluon2) to (top2);
  \end{tikzpicture}
 \end{subfigure}
 \hfill
 \begin{subfigure}[]{0.49\textwidth}
  \centering
  \begin{tikzpicture}[thick, >=latex]

   \node (q1) at (-2.0,2.0) {\PQq};
   \node (q3) at (2.0,2.0) {\PQq};
   \node (q2) at (-2.0,-2.0) {\PQq};
   \node (q4) at (2.0,-2.0) {\PQq};
   \coordinate (boson1) at (0.0, 1.0);
   \coordinate (boson2) at (0.0, -1.0);
   \coordinate (higgsLeft) at (0.0,0.0);
   \coordinate (higgsRight) at (2.0,0.0);

   \draw[fermion] (q1) to (boson1);
   \draw[fermion] (boson1) to (q3);
   \draw[vectorboson] (boson1) to node[left] {\PW / \PZ} (higgsLeft);
   \draw[vectorboson] (boson2) to node[left] {\PW / \PZ} (higgsLeft);
   \fill[black] (higgsLeft) circle (1.5pt);
   \draw[higgs] (higgsLeft) to node[above] {\PSh} (higgsRight);
   \draw[fermion] (q2) to (boson2);
   \draw[fermion] (boson2) to (q4);
  \end{tikzpicture}
 \end{subfigure}

 \caption[Feynman diagrams of the Higgs production processes at the \ac{LHC} suitable for measuring Higgs decays into muons.]{Feynman diagrams of the Higgs production processes at the \ac{LHC} suitable for measuring Higgs decays into muons. Gluon fusion process (left) and vector boson fusion process (right).  }
 \label{fig:Higgs_productionLHC}
\end{figure}

The main production process for a light \ac{SM} Higgs boson at the \ac{LHC} is the gluon fusion process shown in \cref{fig:Higgs_productionLHC}~(left). This process and the vector boson fusion process shown in \cref{fig:Higgs_productionLHC}~(right) might allow for the measurement of the Higgs branching ratio into muons at the \ac{LHC}. The prospects of measuring the Higgs branching ratio into muons at the \ac{LHC} using these channels has been investigated in several studies~\cite{Plehn:2001qg,Han:2002gp,Cranmer:2006zs}. The different results are summarized below for a light \ac{SM} Higgs boson with a mass of \unit[120]{GeV}, assuming a center-of-mass energy of \unit[14]{TeV} and an integrated luminosity of \unit[300]{\fbinv}~\footnote{Expected total integrated luminosity per experiment for the \ac{LHC} without a possible luminosity upgrade.}.

The measurement in the \PW and \PZ boson fusion channel has been studied in~\cite{Plehn:2001qg}. The significance of the signal is estimated to be around 1.8 $\sigma$ per experiment, translating to an uncertainty on the cross section times branching ratio measurement of approximately 60\%.

The Higgs production through the gluon fusion channel is investigated in~\cite{Han:2002gp}. The estimated significance for measuring the Higgs decay into two muons is $2.51\sigma$ per experiment. They also find a better significance compared to~\cite{Plehn:2001qg} of $2.37\sigma$ using the vector boson fusion channel. The combined significance using both channels is given as $3.45\sigma$ per experiment, which corresponds to an uncertainty on the cross section times branching ratio measurement of about 29\%.

Ref.~\cite{Cranmer:2006zs} presents a method of calculating the maximum significance of a measurement using all phase space information on generator level. This method is applied to the Higgs into muons branching ratio measurement using the vector boson fusion channel. A large improvement compared to the cut based analysis presented in~\cite{Plehn:2001qg} is achieved. The maximum significance for an integrated luminosity of \unit[300]{\fbinv} is reported as $3.54\sigma$ per experiment. Any realistic analysis will necessarily result in a lower significance.

It should be noted that all \ac{LHC} studies listed above are done on generator level. Gaussian smearing is applied to the di-muon mass distribution to model the detector resolution and additional kinematic cuts on the muons are introduced to model the detector acceptance. The impact of pile-up and minimum bias events is not taken into account.

\section{Summary}
\label{sec:Higgs_summary}

It has been shown that the measurement of the cross section times branching ratio of a light \acl{SM} Higgs boson with a mass of \unit[120]{GeV} decaying into two muons can be performed at \ac{CLIC} with a center-of-mass energy of \unit[3]{TeV} and total integrated luminosity of \unit[2]{\abinv} with a statistical uncertainty of about 23\%\footnote{The results reported in \cite{Battaglia:2008aa}, scaled to the same integrated luminosity, quote a smaller error than is reported here, since the $\epem \to \mpmm\epem$ background was neglected in that study.}. Systematic uncertainties will be negligible compared to the statistical uncertainties. This result has been obtained using full detector simulation and taking into account all relevant \acl{SM} backgrounds.

The \gghad background present at \ac{CLIC} degrades the statistical uncertainty of the measurement to about 26\%. This result was obtained overlaying realistic amounts of \gghad events over the signal sample. The background samples were not overlayed with \gghad background events, meaning that background rejection based on the visible energy in the event was underestimated and the estimated uncertainty on the branching ratio measurement is conservative.

A fast simulation study was performed to assess the required momentum resolution and it was found that a resolution of a few \unit[$10^{-5}$]{GeV$^{-1}$} is required, otherwise the result will degrade significantly. The results from fast simulation are consistent with the results from full simulation, where an average momentum resolution of \unit[$3.9\times10^{-5}$]{GeV$^{-1}$} is achieved for the selected signal events.

The impact of forward electron tagging was studied by randomly rejecting $\epem \to \mpmm\epem$ background events based on the true electron angles. Assuming an electron tagging efficiency of 95\% in the LumiCal, the statistical uncertainty of the cross section times branching ratio measurement can be improved to about 15\% or 15.7\% when including the effect of the \gghad background. If we assume that all other Higgs couplings are as predicted by the \ac{SM}, this translates to an uncertainty on the coupling $g_{\PGm}$ of approximately 7.5\% or 7.9\%, when including the degradation due to \gghad.





\chapter{Conclusion and Outlook}
\label{cha:Conclusion}

The basic detector requirements of any experiment are derived from its physics goals. In addition, there are additional requirements that come from experimental conditions at the interaction region. In case of \ac{CLIC} the short time between two bunch crossings, the amount of beam-related backgrounds originating from beamstrahlung, and the requirements of the placement and stability of the \acl{QD0} have important consequences for the design of the detector as well as the choice of detector technologies. Based on the \ac{SiD} concept we have developed a simulation model that takes into account the requirements specific to \ac{CLIC} at a center-of-mass energy of \unit[3]{TeV} as well as the performance goals required for the desired precision measurements.

We have shown in simulation studies that the performance of the proposed all-silicon tracking system is robust even in the presence of the \gghad background. A transverse momentum resolution of $\sigma(\pT) = \unit[1.9\times10^{-5}]{\pT/GeV} \oplus 3.8\times10^{-3}/\sin\theta$ or better is achieved for polar angles $\theta > 30$. Although we assume a low material budget, it is still the limiting factor for the momentum resolution for track momenta of up to $\unit[\sim200]{GeV}$. The track finding efficiency in high energetic jets is 95--98\%, depending on the polar angle and the track momentum. The efficiency is lower in regions of very high local occupancies, \eg in the center of a jet. Increasing the segmentation in the barrel strip detectors is expected to improve this. In addition, we have discussed the total occupancies in the tracking detectors from the incoherent pair background and the \gghad background, and identified a possible problem with the occupancies in the innermost strip detectors. This originates from the long strip sizes and can be avoided by increasing the segmentation. Nevertheless, a time resolution of the order of \unit[5]{ns} and multi-hit capabilities will be necessary to mitigate the effects of beam-related backgrounds. Dedicated R\&D is necessary especially for fast low material pixel technologies for the vertex detectors, since no current technology fulfills all prerequisits for a \ac{CLIC} detector.

Tungsten was chosen as the absorber material in the \ac{HCal} barrel region to allow for sufficient shower containment for jet energies typical for a multi-TeV collider experiment. In this way, an \ac{HCal} corresponding to approximately \unit[7.5]{\lambdaint} can be accommodated within a free bore radius of the coil of \unit[2.9]{m}. A simulation study using a \acl{NN}, taking several shower shape variables as input and which was trained for various calorimeter configurations, has been performed to identify the optimal configuration. In the mean time, two test beam campaigns with a tungsten \ac{HCal} prototype using analog readout have been succesfully completed and preliminary results show promising results. Further test beams using other readout technologies will follow to validate the simulation results.

The current default for the \ac{SiD} concept is to use a linear energy reconstruction. We have presented the calibration method to obtain the correction factors used to reconstruct the energy. The energy resolution for single neutral particles was determined to be $\sigma(E)/E = 19.1\%/\sqrt{E/\mr{GeV}} \oplus 1.0\%$ in case of photons and $\sigma(E)/E = 50.3\%/\sqrt{E/\mr{GeV}} \oplus 6.5\%$ in case of \PKzL. The jet-energy resolution achievable with the current particle flow algorithm was studied in di-jet events and was found to be between $\sigma(E)/E = 5\%$ and $\sigma(E)/E = 3.5\%$, using \rmsn, for jet energies between \unit[45]{GeV} and \unit[1.5]{GeV}. The mean reconstructed jet energy is underestimated by approximately 5\% which can be corrected by adjusting the jet energy scale. It is assumed that this systematic bias is due to confusion, but further studies are required. 

Future simulation studies will have to be refined and made more realistic in order to arrive at an optimal detector concept for the next collider experiment. We have pointed out several issues that could be improved beyond those software tools that we have implemented for the studies presented here. For example the \pandora algorithms will have to be specifically tuned for \ac{SiD}-like detector geometries, and additional track finding algorithms to recover non-prompt tracks should be explored.

Finally, we have presented one of the benchmark analyses that have been selected for the \ac{CLIC} \ac{CDR}, the measurement of the decays of a \unit[120]{GeV} \ac{SM} Higgs boson into two muons. The study was performed in full simulation, taking into account all relevant \ac{SM} background processes as well as the most important beam-related background from \gghad processes. It was found that for an integrated luminosity of \unit[2]{\abinv} taken at \ac{CLIC} with a center-of-mass energy of \unit[3]{TeV} a statistical uncertainty of approximately 26\% can be achieved. The sensitivity of the result on the momentum resolution has been determined in a fast simulation study and it was found that an average momentum resolution of $\sigma(\pT)/\pT^2 < \unit[5\times10^{-5}]{GeV^{-1}}$ is desirable. If the forward calorimeters are used to reject the dominant reducible \ac{SM} background $\epem \to \mpmm \epem$, the statistical uncertainty can be improved to approximately 15\%. If all other branching ratios are assumed to be exactly as predicted by the \ac{SM}, this corresponds to an uncertainty on the Higgs Yukawa coupling to muons of approximately 7.5\%. This measurement would be significantly more accurate than what is achievable at the \ac{ILC} or the \ac{LHC}, at least without a possible high luminosity upgrade.

The results that the \ac{LHC} will produce in the next years will answer the question if the Higgs mechanism can be the reason for the electroweak symmetry breaking in the \ac{SM}. In addition, there might be hints for new physics beyond the \ac{SM}. A linear \epem collider like \ac{CLIC} will complement the measurements possible at the \ac{LHC} and will probably be necessary to decide between different concurrent models. If discovered, the energy scale of the new physics will guide the decision for the scope of the next linear collider.

\appendix
\part*{Appendix}
\setcounter{secnumdepth}{1}
%
%

\chapter{Units, Conventions and Notations}
\label{app:Convention}

\section{Units}
Throughout this thesis we use the \emph{natural} units for the speed of light, $c$, and the reduced Planck constant, $\hbar = \frac{h}{2\pi}$ with
\begin{equation*}
 c \equiv 1 \text{ and } \hbar \equiv 1.
\end{equation*}
With the respective values in SI units, taken from~\cite{PDG},
\begin{align*}
 c &= \unit[299792458]{m / s}, \\
 \hbar &= \unit[1.054571628(53) \times 10^{-34}]{\text{J} \cdot \text{s}}, \\
 \unit[1]{GeV} &= \unit[1.602176487(40) \times 10^{-10}]{J}
\end{align*}
we can translate the units of common dimensions:
\begin{align*}
 \unit[1]{GeV} &\equiv \unit[1]{GeV / \mathnormal{c}^2} \approx \unit[1.783 \times 10^{-27}]{kg} \\
 \unit[1]{GeV} &\equiv \unit[1]{GeV / \mathnormal{c}} \approx \unit[5.344 \times 10^{-19}]{kg \cdot m / s} \\
 \unit[1]{GeV^{-1}} &\equiv \unit[1]{\hbar \cdot \mathnormal{c} / GeV} = \unit[1.973 \times 10^{-15}]{m} \\
 \unit[1]{GeV^{-1}} &\equiv \unit[1]{\hbar / GeV} = \unit[6.582 \times 10^{-25}]{s}.
\end{align*}
The cross section of particle interactions is usually expressed in barn
\begin{equation*}
 \unit[1]{fb} = \unit[10^{-43}]{m^2}.
\end{equation*}

\section{Conventions}
\label{sec:Conventions}
For convenience we use $\oplus$ to denote quadratic addition, \eg in case of uncorrelated errors
\begin{equation*}
 a^2 + b^2 = c^2 \qquad \Leftrightarrow \qquad a \oplus b = c.
\end{equation*}

The following conventions are only relevant to \cref{cha:SM}.
Our metric is defined by the tensor
\begin{equation*}
 g_{\mu\nu} = g^{\mu\nu} = \begin{pmatrix} 1 & 0 & 0 & 0 \\
                                           0 &-1 & 0 & 0 \\
                                           0 & 0 &-1 & 0 \\
                                           0 & 0 & 0 &-1 \end{pmatrix}.
\end{equation*}
The Greek indices denote the four components $t$, $x$, $y$, $z$ and are lowered or raised by multiplication with the metric tensor
\begin{equation*}
 x_\mu = g_{\mu\nu}x^{\nu}.
\end{equation*}
The Latin indices denote the three spatial components.
The partial derivative $\partial_\mu$ is defined as
\begin{equation*}
 \partial_\mu = \frac{\partial}{\partial x^{\mu}}.
\end{equation*}
The three Pauli matrices $\sigma^a$ are defined as
\begin{equation*}
 \sigma^1 = \begin{pmatrix} 0 & 1 \\ 1 & 0 \end{pmatrix},\qquad
 \sigma^2 = \begin{pmatrix} 0 &-i \\ i & 0 \end{pmatrix},\qquad
 \sigma^3 = \begin{pmatrix} 1 & 0 \\ 0 &-1 \end{pmatrix}.
\end{equation*}
We use the chiral representation of the Dirac matrices
\begin{align*}
 \gamma^0 = \begin{pmatrix} 0 & 0 & 1 & 0 \\
                             0 & 0 & 0 & 1 \\
                             1 & 0 & 0 & 0 \\
                             0 & 1 & 0 & 0\end{pmatrix},\quad
 \gamma^1 = \begin{pmatrix} 0 & 0 & 0 & 1 \\
                             0 & 0 & 1 & 0 \\
                             0 &-1 & 0 & 0 \\
                            -1 & 0 & 0 & 0\end{pmatrix},\quad
 \gamma^2 = \begin{pmatrix} 0 & 0 & 0 &-i \\
                             0 & 0 & i & 0 \\
                             0 & i & 0 & 0 \\
                            -i & 0 & 0 & 0\end{pmatrix},\quad
 \gamma^3 = \begin{pmatrix} 0 & 0 & 1 & 0 \\
                             0 & 0 & 0 &-1 \\
                            -1 & 0 & 0 & 0 \\
                             0 & 1 & 0 & 0\end{pmatrix}.
\end{align*}
In addition we use the Einstein notation, \ie all terms with an index occurring twice implies a sum over the range of the index.

\section{Detector Coordinate Systems}
\label{sec:coordinateSystems}
Throughout this thesis we are using Cartesian, cylindrical and spherical coordinate systems which are defined as follows.

\subsection{Cartesian Coordinates}
\label{sec:carthesianCoordinates}
Right handed euclidean coordinate system defined by the orthogonal axes $x$, $y$ and $z$. The detector axis is identical to $z$, $x$ is in the horizontal plane and $y$ is in the vertical plane pointing upwards. The \acl{IP} in the center of the detector is located at the origin $(0,0,0)$.

\subsection{Cylindrical Coordinates}
\label{seccylindricalCoordinates}
Coordinate system defined by the detector axis $z$, the azimuthal angle $\phi$ and the distance from the detector axis $r$.
The azimuthal angle $\phi$ is defined in the interval $[0,2\pi] = [0\degrees,360\degrees]$.
The transformation from Cartesian coordinates are
\begin{align*}
 r &= \sqrt{x^2+y^2}, \\
 \phi &= \tan^{-1}\dfrac{y}{x}.
\end{align*}
The coordinate transformations to the Cartesian coordinates are
\begin{align*}
 x &= r\cos\phi, \\
 y &= r\sin\phi.
\end{align*}

\subsection{Spherical Coordinates}
\label{sec:sphericalCoordinates}
Coordinate system defined by the polar angle $\theta$, the azimuthal angle $\phi$ and the distance from the origin $R$.
The azimuthal angle is defined identical to the cylindrical case. The polar angle $\theta$ is defined in the interval $[0,\pi] = [0\degrees,180\degrees]$.
\begin{align*}
 R &= \sqrt{x^2+y^2+z^2} = \sqrt{r^2 + z^2}, \\
 \theta &= \arctan\dfrac{\sqrt{x^2+y^2}}{z} = \arctan\dfrac{r}{z}.
\end{align*}
Due to the symmetry of the detector it is often convenient to use a slightly different definition of the polar angle
\begin{align*}
 \theta &= \arctan\dfrac{\sqrt{x^2+y^2}}{|z|} = \arctan\dfrac{r}{|z|}
\end{align*}
in the interval $[0,\dfrac{\pi}{2}] = [0\degrees,90\degrees]$. We mostly use the latter definition. It should be evident from the context which definition was used.

\chapter{RMS and RMS$_{90}$}
\label{App:RMS90}

For a finite sample of $N$ values $x_1,...,x_N$, the arithmetic mean is defined as
\begin{equation}
 \mr{MEAN} = \overline{x} = \frac{1}{N}\sum\limits_{i=0}^N x_i.
 \label{eq:mean}
\end{equation}
We define the \ac{RMS} as
\begin{equation}
 \mathrm{RMS} = \sqrt{\frac{1}{N}\sum\limits_{i=0}^N \left( x_i - \overline{x}\right)^2},
 \label{eq:rms}
\end{equation}
which is identical to the standard deviation of the sample and the square root of the variance. When estimating the mean and the RMS of a larger sample from a small sample, the expressions given in \autoref{eq:mean} and \autoref{eq:rms} are biased. An unbiased estimate of the standard deviation is instead given by replacing $1/N$ with $1/(N-1)$ in \autoref{eq:rms}. This correction is negligible for sufficiently large samples, which is the case for all \ac{RMS} values calculated in this thesis.

The mean and RMS are very sensitive to strong outliers. Especially for an otherwise narrow distribution with several strong outliers they can lead to a poor description of the majority of the distribution. When characterizing for example the resolution of a measurement, the distribution of the majority is of most interest. In these cases it is useful to restrict the sample to the interval which contains the majority of the values. A commonly used quantity is the \rmsn which is defined as the standard deviation of the values in the smallest interval that contains at least 90\% of the values. The mean of this interval is denoted as $\mr{MEAN}_{90}$.

One has to keep in mind that the \rmsn is systematically smaller than the \ac{RMS}. The size of the difference of these two measures of the spread depends on the respective distribution. For a normal distribution the Gaussian width and the \ac{RMS} are identical. The \rmsn is then approximately 79\% of the \ac{RMS}.

In addition, we use the interval defined by the \rmsn method as the fit range for Gaussian fits. This ensures useful fit results in cases of long non-Gaussian tails.

\chapter{Multivariate Techniques}

\section{\aclp{NN}}

\section{\aclp{BDT}}

\subsection{Decision Trees}

\subsection{Boosting}

\chapter{Track Parametrization}
\label{App:trackParametrization}
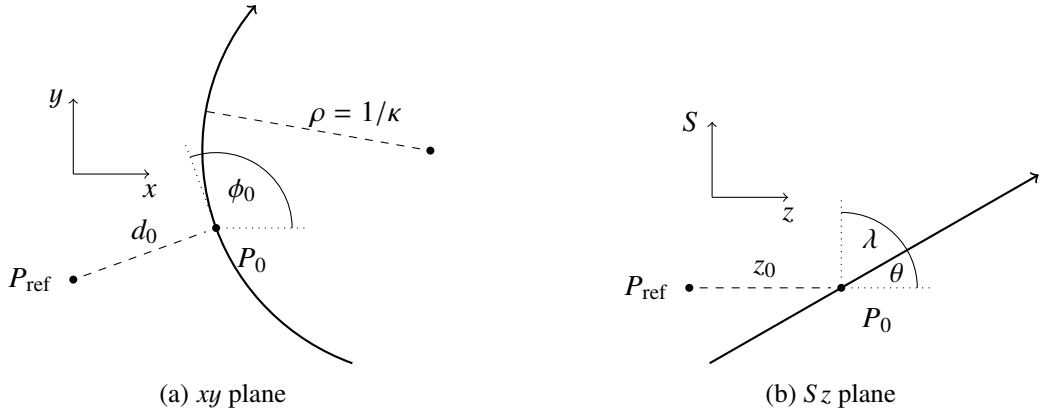
\begin{figure}
 \begin{subfigure}[b]{0.49\textwidth}
 \centering
 \begin{tikzpicture}
  \node (origin) [label={left:$P_\mathrm{ref}$}] at (0,0) {};
  \node (d0) [label={below right:$P_0$}] at (20:2cm) {};
  \node (center) at (20:5cm) {};
  \draw[dashed] (origin) -- (d0) node [pos=0.5, above] {$d_0$};
  \draw[thick, <-] (center) +(140:3cm) arc [start angle=140, delta angle=110, radius=3cm];
  \draw[dashed] (center) -- +(170:3cm) node [pos=0.3, above] {$\rho = 1/\kappa$};
  \draw[dotted] (d0) -- +(0:1.2cm);
  \draw[dotted] (d0) -- +(110:1.2cm);
  \draw (d0) +(0:1cm) arc [start angle=0, delta angle=110, radius=1cm];
  \path (d0) ++(55:0.6cm) node {$\phi_0$};
  \fill[black] (origin) circle (1.5pt);
  \fill[black] (d0) circle (1.5pt);
  \fill[black] (center) circle (1.5pt);
  \draw[->] (0.0,1.4) -- (0.0,2.4) node [left] {$y$};
  \draw[->] (0.0,1.4) -- (1.0,1.4) node [below] {$x$};
 \end{tikzpicture}
 \caption{$xy$ plane}
 \end{subfigure}
 \hfill
 \begin{subfigure}[b]{0.49\textwidth}
 \centering
 \begin{tikzpicture}
  \node (origin) [label={left:$P_\mathrm{ref}$}] at (0,0) {};
  \node (z0) [label={below right:$P_0$}] at (2,0) {};
  \draw[dashed] (origin) -- (z0) node [pos=0.5, above] {$z_0$};
  \draw[dotted] (z0) -- +(1.2,0);
  \draw[dotted] (z0) -- +(0,1.2);
  \draw (z0) +(1.0,0) arc [start angle=0, delta angle=90, radius=1cm];
  \path (z0) ++(15:0.75cm) node {$\theta$};
  \path (z0) ++(60:0.75cm) node {$\lambda$};
  \draw[thick, ->] (z0) ++(210:2cm) -- +(30:5cm);
  \fill[black] (origin) circle (1.5pt);
  \fill[black] (z0) circle (1.5pt);
  \draw[->] (0.3,1.2) -- (0.3,2.2) node [left] {$S$};
  \draw[->] (0.3,1.2) -- (1.3,1.2) node [below] {$z$};
 \end{tikzpicture}
 \caption{$Sz$ plane}
 \end{subfigure}
\caption{Illustration of the track parameters as defined in LCIO.}
\label{fig:trackParametrization}
\end{figure}

The tracks in LCIO are helices which are defined by 5 parameters~\cite{Kramer:2006zz}. This definition is based on the \acs{L3} convention~\cite{L3Helix}. For a uniform magnetic field, the axis of the helix is always parallel to the direction of the magnetic field. A solenoid field along $z$ with $\vec{B}=(0,0,B)$ is assumed. The parameters are defined with respect to an arbitrary reference point, which we define as the origin $P_\mathrm{ref} = (0,0,0)$. The point of closest approach of the track projection in the $xy$ plane to the reference point defines the starting point of the helix $P_0 = (x_0,y_0,z_0)$ . The arc length along the curvature of the helix in the $xy$ projection starting at $P_0$ is called $S$. The track projection into $Sz$ is a straight line. \autoref{fig:trackParametrization} gives an illustration of the track parameters which are defined as follow.
\begin{itemize}
 \item The distance of closest approach to the reference point in the $xy$ plane, $d_0 = \sqrt{x_0 + y_0}$.
 \item The track curvature in the $xy$ plane, $\kappa$, which is the inverse of the radius $\rho$ of the track projection in the $xy$ plane. The sign of $\kappa$ gives the orientation of the helix.
 \item The track direction at $P_0$ in the $xy$ plane, $\phi_0$.
 \item The slope of the particle trajectory in the $Sz$ projection, $\tan\lambda = \cot\theta$.
 \item The $z$ coordinate of the point of closest approach to the reference point in the $xy$ plane, $z_0$.
\end{itemize}

\section{Converting Track Parameters}
Assuming a constant magnetic field along $z$ with $\vec{B}=(0,0,B)$ and a particle with a charge of $q = \pm e$, the track parameters can be translated into physical quantities using the following formulas
\begin{align}
 \begin{split}
 \pT &= \frac{k\ B}{|\kappa|} \\
 p_x &= \pT\ \cos\phi_0 , \\
 p_y &= \pT\ \sin\phi_0 , \\
 p_z &= \pT\ \tan\lambda , \\
 p   &= \frac{\pT}{\cos\lambda} = \pT\sqrt{1 + \tan^2\lambda}, \\
 q   &= \frac{\kappa}{|\kappa|} .
 \end{split}
\label{eq:track_physicalQuantities}
\end{align}
The constant $k$ is introduced to absorb the units and is defined as
\begin{equation}
 k \approx 0.3 \left[\frac{\mathrm{GeV}}{\mathrm{T\ m}}\right].
\end{equation}

\section{Track Parameter Uncertainties}
If we define a vector $\vec{x} = (d_0,\kappa,\phi_0,\tan\lambda,z_0)$ the corresponding uncertainties from the track fit are given by the symmetric covariance matrix
\begin{equation}
 C = \begin{pmatrix}
       \sigma^2(d_0)            & \sigma(d_0, \kappa)         & \sigma(d_0, \phi_0)         & \sigma(d_0, \tan\lambda)    & \sigma(d_0, z_0)         \\
       \sigma(\kappa, d_0)      & \sigma^2(\kappa)            & \sigma(\kappa, \phi_0)      & \sigma(\kappa, \tan\lambda) & \sigma(\kappa, z_0)      \\
       \sigma(\phi_0, d_0)      & \sigma(\phi_0, \kappa)      & \sigma^2(\phi_0)            & \sigma(\phi_0, \tan\lambda) & \sigma(\phi_0, z_0)      \\
       \sigma(\tan\lambda, d_0) & \sigma(\tan\lambda, \kappa) & \sigma(\tan\lambda, \phi_0) & \sigma^2(\tan\lambda)       & \sigma(\tan\lambda, z_0) \\
       \sigma(z_0, d_0)         & \sigma(z_0, \kappa)         & \sigma(z_0, \phi_0)         & \sigma(z_0, \tan\lambda)    & \sigma^2(z_0)
     \end{pmatrix}.
\end{equation}
The off-diagonal elements are the covariances which are defined as
\begin{equation}
 \sigma(a,b) = \sigma(a)\sigma(b)\delta(a,b),
\end{equation}
where $\delta(a,b)$ is the correlation coefficient which is defined in the interval $[-1, 1]$. Estimations for the diagonal elements of the covariance matrix from detector resolutions are given in \autoref{sec:SiD_momentumMeasurement}. For a full derivation of all terms in the covariance matrix we refer to \cite{Karimaki:1997ff, Regler:2008zza, Valentan:2009zz}.

The track fitting that is explained in \autoref{sec:Software_trackReconstruction} does an independent circle fit to obtain $d_0$, $\kappa$ and $\phi_0$ followed by a linear fit to obtain $\tan\lambda$ and $z_0$. The correlation coefficients between the parameters of the two fits are thus 0 and the covariance matrix is given by two independent sub-matrices of dimensions $3\times3$ and $2\times2$
\begin{equation}
 C = \begin{pmatrix}
       \sigma^2(d_0)            & \sigma(d_0, \kappa)         & \sigma(d_0, \phi_0)         & 0                           & 0                        \\
       \sigma(\kappa, d_0)      & \sigma^2(\kappa)            & \sigma(\kappa, \phi_0)      & 0                           & 0                        \\
       \sigma(\phi_0, d_0)      & \sigma(\phi_0, \kappa)      & \sigma^2(\phi_0)            & 0                           & 0                        \\
       0                        & 0                           & 0                           & \sigma^2(\tan\lambda)       & \sigma(\tan\lambda, z_0) \\
       0                        & 0                           & 0                           & \sigma(z_0, \tan\lambda)    & \sigma^2(z_0)
     \end{pmatrix}.
\end{equation}

\section{Propagation of Track Parameter Uncertainties}
For a linear combinations of the track parameters
\begin{equation}
 f(d_0,\kappa,\phi_0,\tan\lambda,z_0) = A\vec{x}
\end{equation}
the uncertainty is given by
\begin{equation} 
 \sigma^2(f) = A\ C\ A^\mathrm{T}.
\end{equation}
For a non-linear combination of variables, which is needed to calculate the uncertainties of the quantities described in \autoref{eq:track_physicalQuantities}, the function has to be expanded into a Taylor series. The first order approximation for the expansion around the vector $\vec{x}_0$ is then given by
\begin{equation}
 f(d_0,\kappa,\phi_0,\tan\lambda,z_0) \approx f(\vec{x}_0) + J(\vec{x}-\vec{x}_0),
\end{equation}
where $J$ is the Jacobian matrix
\begin{equation}
 J = \left(\frac{\partial f}{\partial d_0},\frac{\partial f}{\partial \kappa},\frac{\partial f}{\partial \phi_0},\frac{\partial f}{\partial \tan\lambda},\frac{\partial f}{\partial z_0}\right).
\end{equation}
It should be noted that the first order approximation is only valid for small uncertainties. The term $f(\vec{x}_0)$ is constant and does not contribute to the uncertainty. The uncertainty for $f$ is then given by
\begin{equation}
 \sigma^2(f) \approx J\ C\ J^\mathrm{T}.
\end{equation}
As an example, for an arbitrary function of two parameters, $f(a,b)$,  the uncertainty is given by
\begin{equation}
 \sigma^2(f) \approx \left(\frac{\partial f(a,b)}{\partial a}\right)^2\sigma^2(a) + \left(\frac{\partial f(a,b)}{\partial b}\right)^2\sigma^2(b) + 2\frac{\partial f(a,b)}{\partial a}\frac{\partial f(a,b)}{\partial b}\sigma(a,b).
 \label{eq:twoParameterPropagation}
\end{equation}
Using \autoref{eq:twoParameterPropagation} we can now estimate the uncertainties of the quantities given in \autoref{eq:track_physicalQuantities}:
\begin{align}
 \sigma^2(\pT) &\approx \left(\frac{k\ B}{\kappa^2}\right)^2 \sigma^2(\kappa) \\
 \sigma^2(p_x) &\approx \left(\frac{k\ B}{\kappa^2}\right)^2 \left( \left(\frac{\cos\phi_0}{\kappa}\right)^2 \sigma^2(\kappa) + \sin^2\phi_0 \sigma^2(\phi_0) + \frac{2\ \sin\phi_0\ \cos\phi_0}{\kappa} \sigma(\kappa,\phi_0) \right)\\
 \sigma^2(p_y) &\approx \left(\frac{k\ B}{\kappa^2}\right)^2 \left( \left(\frac{\sin\phi_0}{\kappa}\right)^2 \sigma^2(\kappa) + \cos^2\phi_0 \sigma^2(\phi_0) + \frac{2\ \sin\phi_0\ \cos\phi_0}{\kappa} \sigma(\kappa,\phi_0) \right)\\
 \sigma^2(p_z) &\approx \left(\frac{k\ B}{\kappa^2}\right)^2 \left( \left(\frac{\tan\lambda}{\kappa}\right)^2 \sigma^2(\kappa) + \sigma^2(\tan\lambda) + \frac{2\ \tan\lambda}{\kappa} \sigma(\kappa,\tan\lambda) \right)\\
 \sigma^2(p)   &\approx \left(\frac{k\ B}{\kappa^2}\right)^2 \left( \frac{1 + \tan^2\lambda}{\kappa^2} \sigma^2(\kappa) + \frac{\tan^2\lambda}{1 + \tan\lambda}\sigma^2(\tan\lambda) + \frac{2\ \tan\lambda}{\kappa\ \sqrt{1 + \tan^2\lambda}} \sigma(\kappa,\tan\lambda) \right)
\end{align}

\chapter{Tracking Efficiency in Top Pair Production}
\label{cha:Appendix_ttTracking}

Here we want to briefly discuss the tracking efficiencies and fake rates observed in simulations of $\epem \to \ttbar$ at $\sqrt{s} = \unit[3]{TeV}$ and compare them to the results found for the di-jet events discussed in \cref{cha:Tracking}. 10000 simulated events with and without background are used for the study.

\begin{figure}[htpb]
 \includegraphics[width=0.49\textwidth]{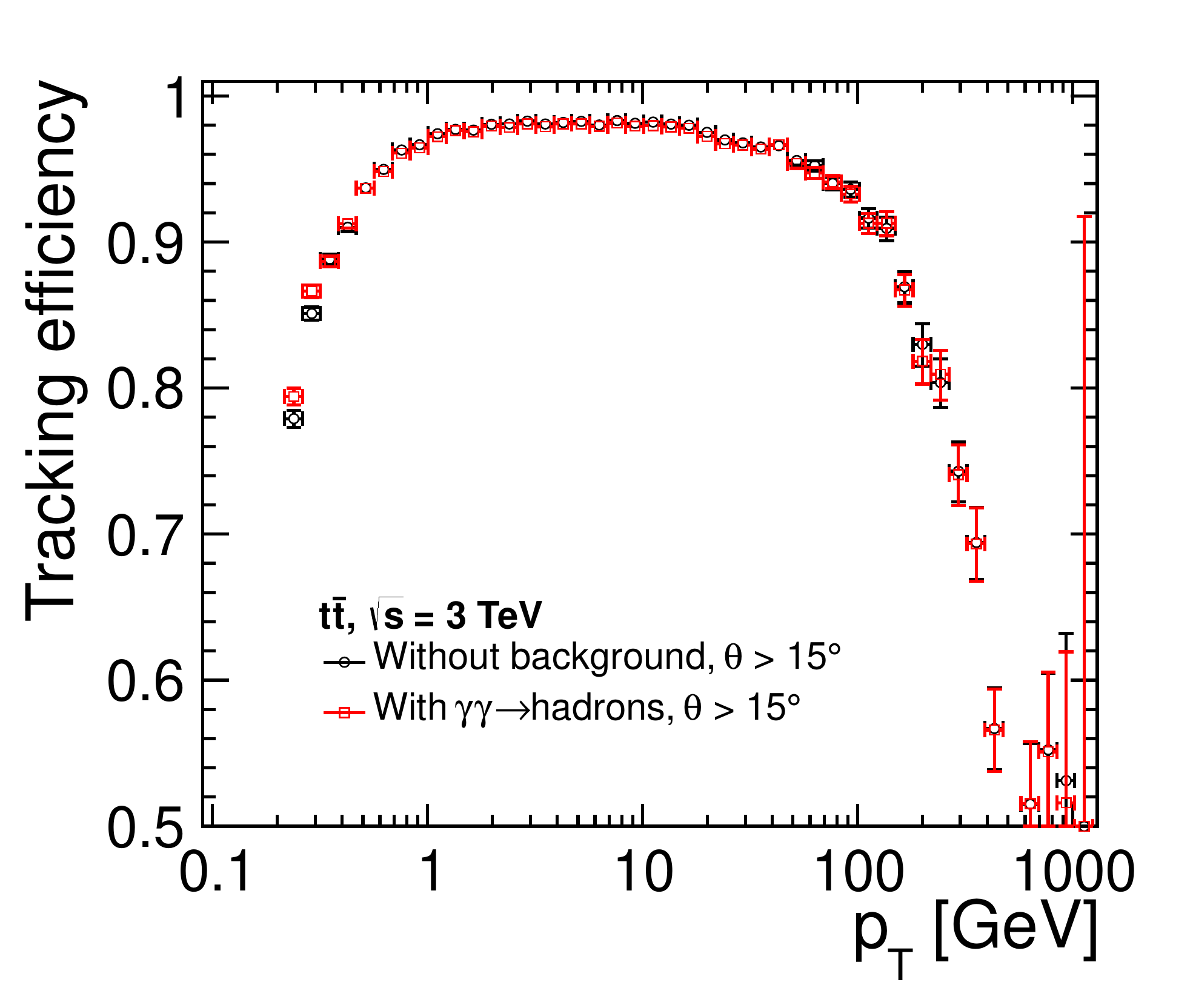}
 \hfill
 \includegraphics[width=0.49\textwidth]{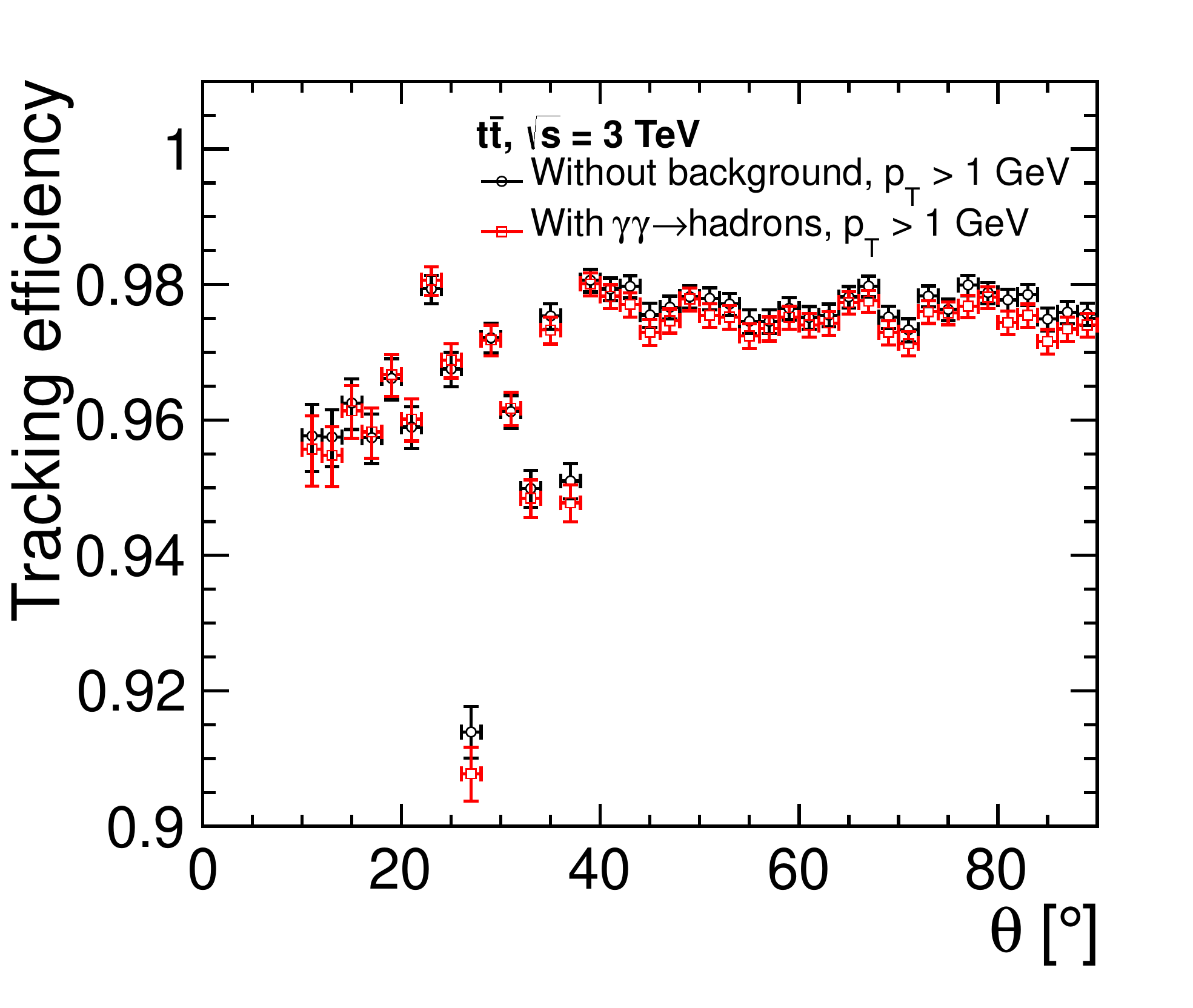}
\caption[Fake rate depending on the transverse momentum and the polar angle in \ttbar events.]{Tracking efficiency depending on the transverse momentum \pT (left) and the polar angle $\theta$ (right) in \ttbar events at a center-of-mass energy of \unit[3]{TeV} with and without \gghad background.}
\label{tteff}
\end{figure}

The highest tracking efficiency of 98\% is achieved for tracks of intermediate momentum below \unit[100]{GeV} which is very similar to what is achieved in di-jet events. The dependency towards higher transverse momenta is very different though, as shown in \cref{tteff}. The drop in efficiency is less sharp for momenta below \unit[100]{GeV} but becomes stronger than in the case of the di-jet events for higher energies dropping as low as only 50\% for transverse momenta of \unit[500--1000]{GeV}. This is due to the different average particle moment in the \ttbar events which contain more lower energetic tracks. The angular dependence of the tracking efficiency is very similar between the two topologies.

\begin{figure}[htpb]
 \includegraphics[width=0.49\textwidth]{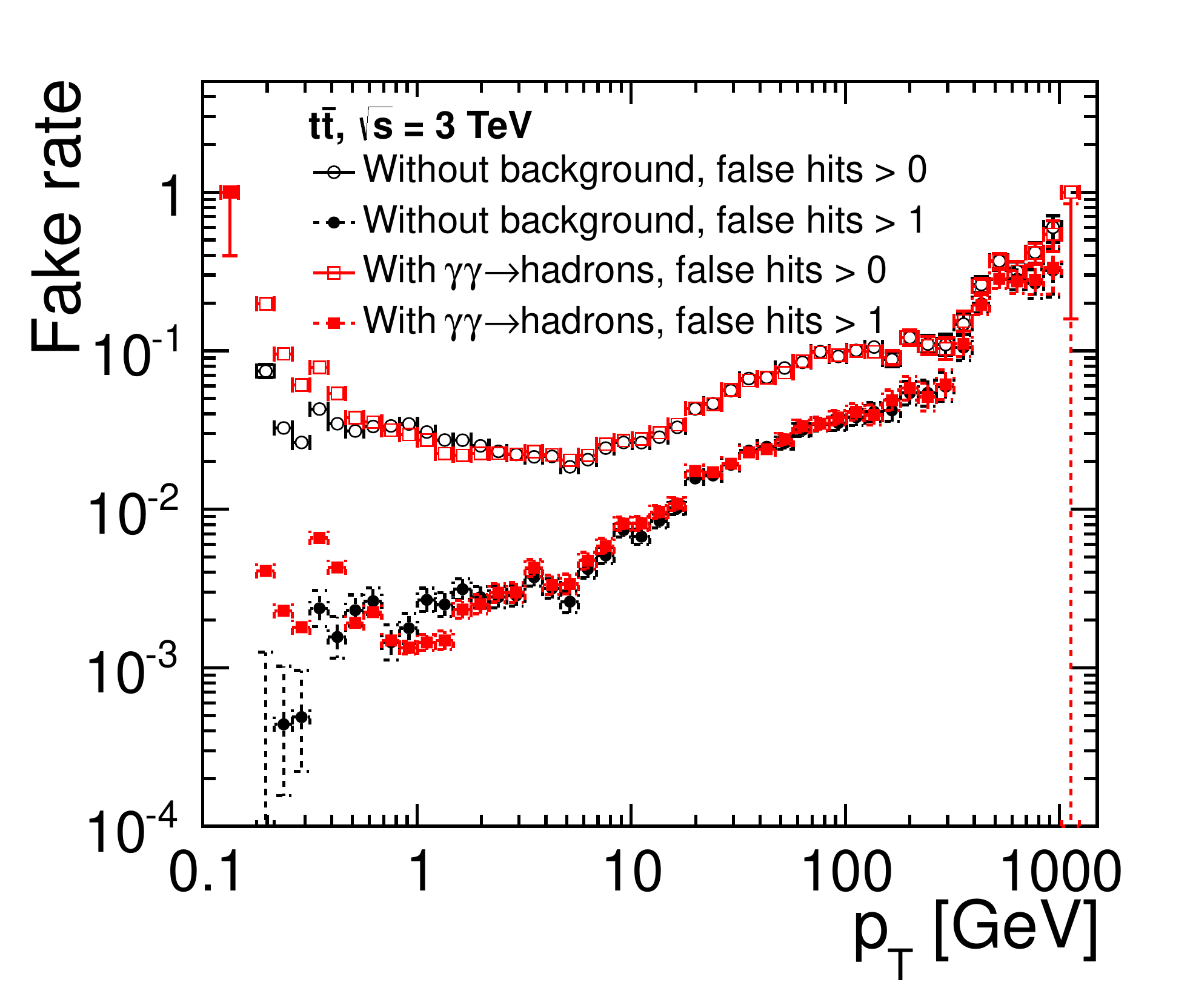}
 \hfill
 \includegraphics[width=0.49\textwidth]{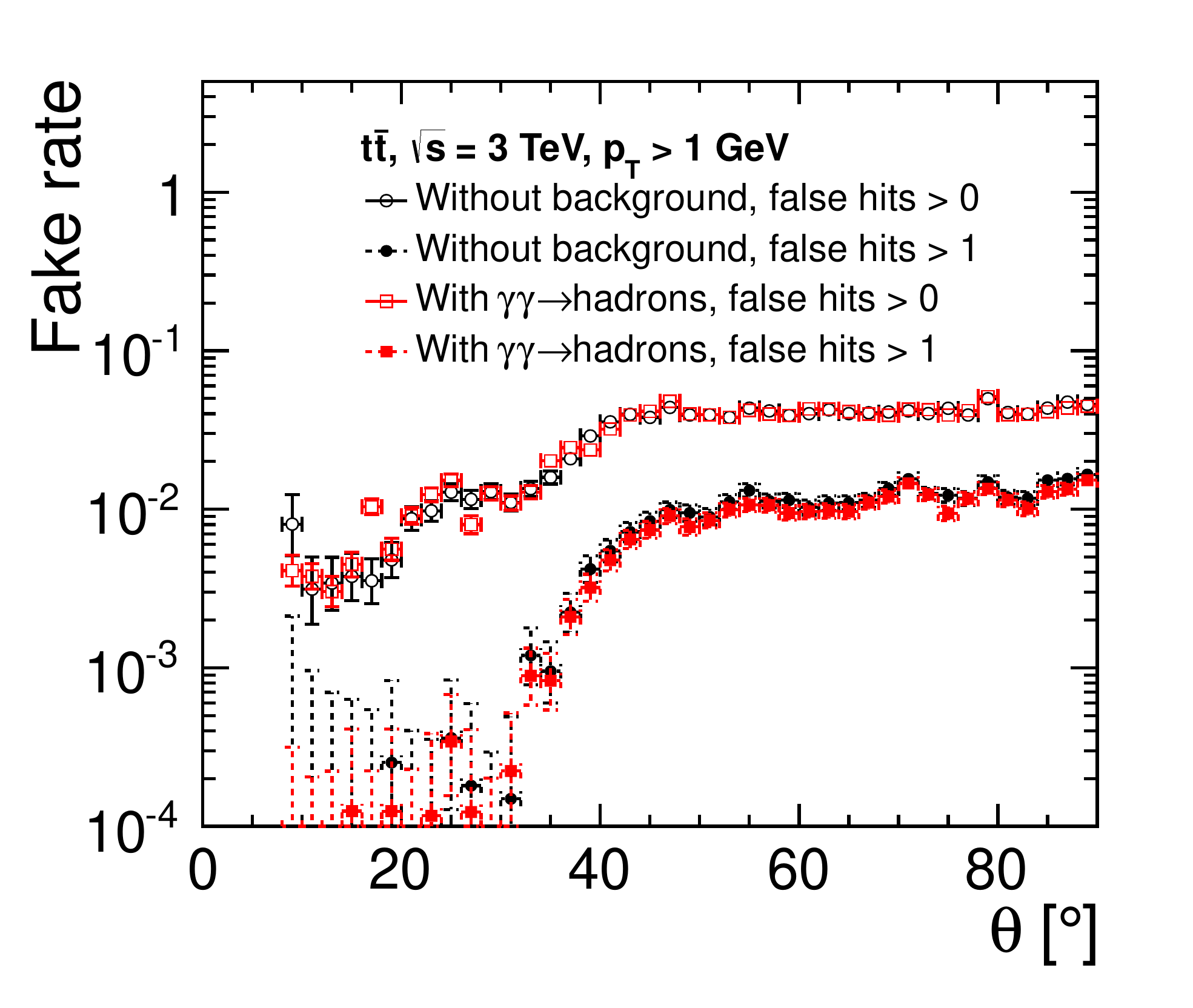}
\caption[Fake rate depending on the transverse momentum and the polar angle in \ttbar events.]{Fake rate depending on the transverse momentum \pT (left) and the polar angle $\theta$ (right) in \ttbar events at a center-of-mass energy of \unit[3]{TeV} with and without \gghad background.}
\label{ttfake}
\end{figure}

The results for the fake rate are also very similar, as shown in \cref{ttfake}. The only difference between di-jet and \ttbar events is for tracks of medium transverse momentum. At these momenta the fake rates in \ttbar are slightly lower. In general, like in case of the di-jet events, the impact of the \gghad background is negligible except for lowest track momenta.

This comparison shows that there is no strong dependency of the results presented in \cref{cha:Tracking} on the event topology.

\backmatter
\chapter{List of Acronyms}

\begin{acronym}[BeamCal]
 \acro{ADC}{Analog to Digital Converter}
 \acro{ALEPH}{\acroextra{Experiment at LEP}}
 \acro{ALICE}{\acroextra{Experiment at the LHC}}
 \acro{ASIC}{Application-Specific Integrated Circuit}
 \acro{ATLAS}{\acroextra{Experiment at the LHC}}

 \acro{BeamCal}{Beam Calorimeter}
 \acro{BDS}{Beam Delivery System}
 \acro{BDT}{Boosted Decision Tree\acroextra{ - a multi variate classifier}}
 \acro{BSM}{Beyond the Standard Model}
 \acro{BX}{Bunch Crossing}

 \acro{CDR}{Conceptual Design Report}
 \acro{CERN}{European Organization for Nuclear Research\acroextra{, Geneva, Switzerland}}
 \acro{CHIPS}{Chiral Invariant Phase Space model}
 \acro{CKM}{Cabibbo-Kobayashi-Maskawa\acroextra{- name of the quark mixing matrix in the \ac{SM} in flavor changing electroweak interactions}}
 \acro{CLIC}{Compact Linear Collider}
 \acro{CMS}{Compact Muon Solenoid}
 \acro{CTF3}{CLIC test facility 3}

 \acro{DELPHI}{\acroextra{Experiment at LEP}}
 \acro{DESY}{Deutsches Elektronen Synchrotron\acroextra{, Hamburg, Germany}}

 \acro{ECal}{Electromagnetic Calorimeter}


 \acro{GPS}{General Particle Source}

 \acro{H1}{\acroextra{experiment at HERA}}
 \acro{HERA}{Hadron Elektron Ring Anlage\acroextra{ - lepton-proton collider at DESY}}
 \acro{HCal}{Hadronic Calorimeter}
 \acro{HEP}{High Energy Physics}

 \acro{ILC}{International Linear Collider\acroextra{ - concept for a future \epem linear collider}}
 \acro{ILD}{International Large Detector\acroextra{ - concept for an experiment at the ILC or CLIC}}
 \acro{IP}{Interaction Point}



 \acro{L3}{\acroextra{Experiment at LEP}}
 \acro{LCIO}{Linear Collider Input/Output}
 \acro{LEP}{Large Electron-Positron Collider}
 \acro{LHC}{Large Hadron Collider\acroextra{ - proton-proton collider at CERN}}
 \acro{LHCb}{LHC beauty}
 \acro{LINAC}{Linear Accelerator}
 \acro{LumiCal}{Luminosity Calorimeter}
 \acro{LSP}{lightest supersymmetric particle}

 \acro{MIP}{Minimum Ionizing Particle}
 \acro{MPGD}{Micro-Pattern Gas Detectors}
 \acro{MSSM}{Minimal Supersymmetric Standard Model}

 \acro{NN}{Neural Network}

 \acro{OPAL}{\acroextra{Experiment at LEP}}

 \acro{PDF}{Probability Density Function}
 \acro{PETS}{Power Extraction and Transfer System}
 \acro{PFA}{Particle Flow Algorithm}
 \acro{PFO}{Particle Flow Object}
 \acro{PS}{Proton Synchrotron\acroextra{ - accelerator at CERN}}

 \acro{QCD}{Quantum Chromodynamics}
 \acro{QD0}{final focusing quadrupole}
 \acro{QED}{Quantum Electrodynamics}
 \acro{QFT}{Quantum Field Theory}

 \acro{RF}{Radio Frequency}
 \acro{RMS}{Root Mean Square}
 \acro{RPC}{Resistive Plate Chamber}

 \acro{SiD}{Silicon Detector\acroextra{ - concept for an experiment at the ILC or CLIC}}
 \acro{SiPM}{Silicon Photomultiplier}
 \acro{SLAC}{National Accelerator Laboratory\acroextra{, Stanford, USA}}
 \acro{SLC}{Stanford Linear Collider\acroextra{ - \epem collider at SLAC}}
 \acro{SLD}{SLAC Large Detector\acroextra{ - experiment at the SLC}}
 \acro{SLIC}{Simulator for the Linear Collider}
 \acro{SM}{Standard Model}
 \acro{SPS}{Super Proton Synchrotron\acroextra{ - accelerator at CERN}}
 \acro{SUSY}{Supersymmetry}
 \acro{SUGRA}{Supergravity}

 \acro{TESLA}{\acroextra{Concept for a future \epem linear collider}}
 \acro{Tevatron}{\acroextra{Proton-antiproton collider at FermiLab}}
 \acro{TMVA}{Toolkit for Multivariate Data Analysis}

 \acro{UA1}{\acroextra{Experiment at the SPS}}
 \acro{UA2}{\acroextra{Experiment at the SPS}}



 \acro{XML}{Extensible Markup Language}



\end{acronym}

%
%
%
%
%
\printbibliography[heading=bibintoc]

\listoffigures
\listoftables

%
%

\end{document}